\documentclass[11pt,a4paper]{article}
\pdfoutput=1

\usepackage{jheppub-nosort,amsthm}

\usepackage{amsmath,amssymb,amsthm,amscd,wasysym}
\usepackage{tikz}
\tikzset{node distance=2cm, auto}
\usetikzlibrary{arrows}
\usetikzlibrary{decorations.markings}
\usepackage{xcolor}
\usepackage{epsfig}
\usepackage{fancybox}
\usepackage{psfrag,comment}
\usepackage{graphicx}
\usepackage{mathdots}

\newcommand{\CB}{{\cal B}}

\newcommand{\CF}{{\cal F}}

\newcommand{\CJ}{{\cal J}}
\newcommand{\CK}{{\cal K}}
\newcommand{\CL}{{\cal L}}

\newcommand{\CN}{{\cal N}}
\newcommand{\CO}{{\cal O}}
\newcommand{\CP}{{\cal P}}

\newcommand{\CS}{{\cal S}}
\newcommand{\CT}{{\cal T}}

\newcommand{\CV}{{\cal V}}

\newcommand{\CX}{{\cal X}}
\newcommand{\CY}{{\cal Y}}
\newcommand{\CZ}{{\cal Z}}

\def\BN{{\mathbb N}}
\def\BZ{{\mathbb Z}}
\def\BR{{\mathbb R}}
\def\BC{{\mathbb C}}
\def\BP{{\mathbb P}}
\def\BT{{\mathbb T}}
\def\BS{{\mathbb S}}

\def\BL{{\mathbb L}}
\def\BK{{\mathbb K}}
\def\BQ{{\mathbb Q}}
\def\BF{{\mathbb F}}

\newcommand{\be}{\begin{equation}}
\newcommand{\ee}{\end{equation}}
\newcommand{\bea}{\begin{eqnarray}}
\newcommand{\eea}{\end{eqnarray}}

\newcommand{\nres}{\boldsymbol{\mathfrak{n}}_0}

\def\r{\right\rangle}

\def\1{\mathbf{1}}
\def\0{|\1\r}
\def\im{{\mathbb{I}}{\mathrm{m}}}
\def\re{{\mathbb{R}}{\mathrm{e}}}

\def\disc{{\mathrm{Disc}}}

\newcommand{\rme}{{\mathrm{e}}}
\newcommand{\rmi}{{\mathrm{i}}}
\newcommand{\rmd}{{\mathrm{d}}}

\def\XXint#1#2#3{{\setbox0=\hbox{$#1{#2#3}{\int}$}
     \vcenter{\hbox{$#2#3$}}\kern-.5\wd0}}

\newdimen\tableauside\tableauside=1.0ex
\newdimen\tableaurule\tableaurule=0.4pt
\newdimen\tableaustep
\def\phantomhrule#1{\hbox{\vbox to0pt{\hrule height\tableaurule width#1\vss}}}
\def\phantomvrule#1{\vbox{\hbox to0pt{\vrule width\tableaurule height#1\hss}}}
\def\sqr{\vbox{%
  \phantomhrule\tableaustep
  \hbox{\phantomvrule\tableaustep\kern\tableaustep\phantomvrule\tableaustep}%
  \hbox{\vbox{\phantomhrule\tableauside}\kern-\tableaurule}}}
\def\squares#1{\hbox{\count0=#1\noindent\loop\sqr
  \advance\count0 by-1 \ifnum\count0>0\repeat}}
\def\tableau#1{\vcenter{\offinterlineskip
  \tableaustep=\tableauside\advance\tableaustep by-\tableaurule
  \kern\normallineskip\hbox
    {\kern\normallineskip\vbox
      {\gettableau#1 0 }%
     \kern\normallineskip\kern\tableaurule}%
  \kern\normallineskip\kern\tableaurule}}
\def\gettableau#1{\ifnum#1=0\let\next=\null\else
\squares{#1}\let\next=\gettableau\fi\next}
\subheader{\hfill
{\small{\textsf{NSF-ITP-17-153}}}
}

\title{A Primer on Resurgent Transseries and Their Asymptotics}

\author[a,b]{In\^es~Aniceto,}
\affiliation[a]{Kavli Institute for Theoretical Physics, University of California,\\ Santa Barbara, CA 93106, United States of America}
\affiliation[b]{Institute of Physics, Jagiellonian University, ul. {\L}ojasiewicza 11, 30-348 Krak\'ow, Poland}
\emailAdd{ines@th.if.uj.edu.pl}

\author[c]{G\"ok\c ce~Ba\c sar,}
\affiliation[c]{Department of Physics, University of Illinois, Chicago, IL 60607, United States of America}
\emailAdd{gbasar@uic.edu}

\author[a,d]{Ricardo~Schiappa\,}
\affiliation[d]{CAMGSD, Departamento de Matem\'atica, Instituto Superior T\'ecnico,\\ Universidade de Lisboa, 1049-001 Lisboa, Portugal\\}
\emailAdd{schiappa@math.tecnico.ulisboa.pt}

\abstract{
The computation of observables in general interacting theories, be them quantum mechanical, field, gauge or string theories, is a non-trivial problem which in many cases can only be addressed by resorting to perturbative methods. In most physically interesting problems these perturbative expansions result in asymptotic series with zero radius of convergence. These asymptotic series then require the use of resurgence and transseries in order for the associated observables to become nonperturbatively well-defined. Resurgence encodes the complete large-order asymptotic behaviour of the coefficients from a perturbative expansion, generically in terms of (multi) instanton sectors and for each problem in terms of its Stokes constants. Some observables arise from linear problems, and have a finite number of instanton sectors and associated Stokes constants; some other observables arise from nonlinear problems, and have an infinite number of instanton sectors and Stokes constants. By means of two very explicit examples, and with emphasis on a pedagogical style of presentation, this work aims at serving as a primer on the aforementioned resurgent, large-order asymptotics of general perturbative expansions. This includes discussions of transseries, Stokes phenomena, generalized steepest-descent methods, Borel transforms, nonlinear resonance, and alien calculus. Furthermore, resurgent properties of transseries---usually described mathematically via alien calculus---are recast in equivalent physical languages: either a ``statistical mechanical'' language, as motions in chains and lattices; or a ``conformal field theoretical'' language, with underlying Virasoro-like algebraic structures.
}


\arxivnumber{1802.10441}

\begin{document}

\maketitle

\vfill

\eject

\allowdisplaybreaks

\section{Introduction}\label{sec:intro}

Exactly-solvable interacting quantum theories, from quantum mechanics and field theory through gauge and string theories, are sparse. This naturally implies that, in practice, most problems are hard to solve. In principle, one would like to compute (generating functions for) all possible correlation functions, involving both local and non-local observables. But, generically, even focusing on the simplest local observables such as the partition function or the free energy may already present hard and daunting calculations if not for resorting to perturbation theory. At first this may strike as a reasonable course of action, after all perturbation theory can be applied to a broad range of different systems. The non-trivial question, however, asks what exactly is computable from the resulting series (in the interaction weak coupling-constant). This, of course, will very much depend on the \textit{convergence} properties of such series, determining if any results may be extracted from it, and how far may one go into the strong-coupling region. Unfortunately, in most interesting interacting quantum theories, perturbative expansions for large classes of observables have \textit{vanishing} radius of convergence (and are dubbed \textit{asymptotic}).

With hindsight this should not come as a surprise. As discussed long ago by Dyson \cite{d52} a given theory may have dramatically different physics when considering positive versus negative coupling; in particular being unstable at negative coupling---and thus the physical requirement of finding a null radius of convergence. To be more specific, let us consider some observable $\Phi := \langle \CO \rangle$ and denote by $\Phi^{(0)}_{k}$ the $k$-loop coefficients of its perturbative expansion (in some adequate coupling constant). That this perturbative series is asymptotic translates to a (leading) factorial growth of these coefficients, $\Phi^{(0)}_{k} \sim k!$ at large order $k$. In many cases this is precisely due to the factorial growth of Feynman diagrams relevant at each order $k$ of perturbation theory, see, \textit{e.g.}, \cite{z81b, lz90}. But what can be said about the next-to-leading growth of the coefficients $\Phi^{(0)}_{k}$?

This is where the theory of resurgence first appears in full glory. Of course early research on instantons had already made clear that at next-to-leading order one finds $\Phi^{(0)}_{k} \sim k!\, A^{-k}$, with $A$ the instanton action, and at next-to-next-to-leading order one will start seeing the appearance---the resurgence---of the perturbative series \textit{around} the \textit{one-instanton} background, $\Phi^{(1)}_{\ell}$ (a list of references will follow shortly). But one has to deploy resurgence in order to take this story to its full conclusion; where it tells us exactly what are \textit{all} the different components appearing in the \textit{exact} growth/behaviour of the $k$-sequence $\Phi^{(0)}_{k}$. The remarkable feature is that \textit{only}, and in fact \textit{all}, multi-instanton sequences appear, $\Phi^{(n)}_{\ell}$ (where here $\Phi^{(n)}_{\ell}$ are the coefficients of the perturbative loop-expansions around the $n$-instanton sectors). Schematically, the corresponding resurgence large-order relation is of the form
\be
\label{resurgence_predicting}
\begin{tikzpicture}[>=latex]
  \node (Phi0) {$\Phi^{(0)}_{k \gg 1}$};
  \node (equiv) [right of=Phi0] {$\longleftrightarrow$};  
  \node (Phi1) [right of=equiv] {$\left[ \Phi^{(1)}_\ell \right]$};
  \node (+1) [node distance=1cm, right of=Phi1] {$+$};
  \node (Phi2) [node distance=1cm, right of=+1] {$\left[ \Phi^{(2)}_\ell \right]$};
  \node (+2) [node distance=1cm, right of=Phi2] {$+$};
  \node (Phi3) [node distance=1cm, right of=+2] {$\left[ \Phi^{(3)}_\ell \right]$};
  \node (+3) [node distance=1cm, right of=Phi3] {$+$};
  \node (dots) [node distance=0.75cm, right of=+3] {$\cdots$};
  \node (comma) [node distance=0.5cm, right of=dots] {,};
  \draw[->, bend right] (Phi1) to node [swap] {} (Phi0);
  \draw[->, bend right] (Phi2) to node [swap] {} (Phi0);
  \draw[->, bend right] (Phi3) to node [swap] {``predicting''} (Phi0);
\end{tikzpicture}
\ee
\noindent
where we illustrate via the solid arrows how resurgence works: if one knows all multi-instanton data---including the instanton action $A$ and all perturbative coefficients around all instanton sectors $\Phi^{(n)}_{\ell}$---then one can predict the \textit{exact} behaviour of the perturbative coefficients. Initially written as a large-order relation (thus the subscript $k \gg 1$), the more nonperturbative sectors one gets a handle of, the less of a large-order and the more of an exact relation one obtains.

While the relations implied by \eqref{resurgence_predicting} are usually taken as expectable, it is sometimes the case that one forgets that the converse is also true. One can in fact invert the direction of the arrows in \eqref{resurgence_predicting} and extract the complete nonperturbative content of a given observable out of its exact perturbative expansion, \textit{i.e.}, the perturbative series \textit{already encodes} all other nonperturbative series, even if deep in its large-order behaviour. As such, decomposing the sequence $\Phi^{(0)}_{k}$ into its different components, be them polynomial $\sim k^{-\ell}$ or exponential $\sim n^{-k}$ or else, one can read the resurgence large-order relation instead as
\be
\label{resurgence_decoding}
\begin{tikzpicture}[>=latex]
  \node (Phi0) {$\Phi^{(0)}_{k \gg 1}$};
  \node (equiv) [right of=Phi0] {$\longleftrightarrow$};  
  \node (Phi1) [right of=equiv] {$\left[ \Phi^{(1)}_\ell \right]$};
  \node (+1) [node distance=1cm, right of=Phi1] {$+$};
  \node (Phi2) [node distance=1cm, right of=+1] {$\left[ \Phi^{(2)}_\ell \right]$};
  \node (+2) [node distance=1cm, right of=Phi2] {$+$};
  \node (Phi3) [node distance=1cm, right of=+2] {$\left[ \Phi^{(3)}_\ell \right]$};
  \node (+3) [node distance=1cm, right of=Phi3] {$+$};
  \node (dots) [node distance=0.75cm, right of=+3] {$\cdots$};
  \node (fullstop) [node distance=0.5cm, right of=dots] {.};
  \draw[->, bend right, dashed] (Phi0) to node [swap] {} (Phi1);
  \draw[->, bend right, dashed] (Phi0) to node [swap] {} (Phi2);
  \draw[->, bend right, dashed] (Phi0) to node [swap] {``decoding''} (Phi3);
\end{tikzpicture}
\ee
\noindent
This makes it clear how the asymptotics of perturbative expansions encode a lot of physics!

Resurgence relations such as \eqref{resurgence_predicting} or \eqref{resurgence_decoding} may in fact be written for \textit{any} instanton\footnote{While in this paper we shall mainly stick to the use of the word ``instanton'' to denote generic nonperturbative (exponentially non-analytic) contributions, and will thus be a bit loose on its use, of course in many quantum field theory examples this should be more precisely replaced by the word ``renormalon''; see, \textit{e.g.}, \cite{b98}.} sector, where on the left-hand-side one now finds the growth of any (multi) instanton $k$-sequence $\Phi^{(n)}_{k}$ and, on the right-hand-side, the different components describing its complete behaviour; of course now themselves expressible in terms of all other instanton sectors alongside the perturbative expansion. This is in fact the reason for the name \textit{resurgence}: picking any sector of the theory, by digging deep enough into its large-order behaviour one will eventually find the ``resurgence'' of all other sectors. Having seen some initial physical developments within quantum mechanics, \textit{e.g.}, \cite{bw69, bw73, bow77, cs78, sr78, bpv78a, bpv78b, v80, b80, z81a, z81c, hdgclc82, z83, v83, z84, shcp85, s85, ddp90, ddp93, v93, v94a, v94b, ddp97, dp97, v00} (see \cite{z81b} for further earlier references) and (non-critical) string theory, \textit{e.g.}, \cite{gp88, bmp90, s90, gz91, ez92, ez93} (see \cite{fgz93} for further earlier references), the modern (physical) version of resurgence theory has itself recently resurged in many areas, from quantum mechanics \cite{ac00a, ac00b, ahs02, z03, zj04a, zj04b, jz04, a04, jsz10, jz11, du13, bdu13, as13, ggs13, du14, mns15, gt15, fkmns16, gt16, su16, du16c, kstu16, ssv16, cm16, bdu17, fkmns17a, ssv17, fkmns17b, gs17, cms17, h1712} through gauge and field theories \cite{s02, msw07, m08, msw08, ps09, mpp09, sw09, asv11, r12, au12a, au12b, m10, du12a, du12b, sv13, cddu13, as13, cdu14, mns14a, mns14b, cku14, ars14, bc14, mns14c, csv15, bd15, dsu15, hs15, hj15, kt15, du15, a15, dh15, bd15b, bdssu15, as15, k16, h16a, mns16, ddt16, h16b, gmp16, fkmns16, ads16, g16, du16c, rt16, bc16, m17, b17, c1701, m_b17, yy17, cd17, acpy17, ad17, h1710, dg17, bcgmp17, hy17, cgm17, mv18, mpn18} to string theory \cite{m06, msw07, m08, glm08, msw08, em08, gm08, ps09, mpp09, gikm10, kmr10, dmp11, asv11, m10, sv13, as13, cesv13, gmz14, cesv14, v15, ho15, c15, csv16, cms16, cm16, gs17, cms17, c1712} and cosmology \cite{bhn16, flt17a, flt17b, dhhhj17, flt17c, bccs17}. The modern (mathematical) version of the theory of resurgence (and name) arises in the seminal work of \'Ecalle \cite{e81}. For earlier reviews on resurgence see, \textit{e.g.}, \cite{cnp93a, cnp93b, c95, c98, c98b, b99, dp99, ss03, kt05}. More modern reviews include, \textit{e.g.}, \cite{o05a, o05b, d06, s07, m10, s14, d14, d2014, du15b, m15, du16a, du16b, ms16, lr16, d16, p16, mvb17, bh17}; see also \cite{e08} for a review specifically on transseries. Our (novel) aim in this work sets on an elementary introduction to these mathematical ideas, within physical contexts, and with an emphasis on their relation to asymptotics (itself mostly absent in the aforementioned reviews).

While in the main body of the paper we shall be quantitatively precise on the exact nature of the resurgence relations \eqref{resurgence_predicting} or \eqref{resurgence_decoding}, for the moment let us note that they have a somewhat universal character. This universal structure depends upon the nature of the \textit{transseries} and of the \textit{bridge equations}. Briefly, transseries enlarge  power series by the inclusion of non-analytic terms and, as such, allow for the construction of a single object (the transseries itself) which gathers all different nonperturbative contributions to whatever observable we are computing (where by ``all'' we mean all contributions having already appeared in the resurgence relations). The simplest way to think about this is to imagine the ``trans'' in transseries as encompassing all powers of the non-analytic monomial $\exp \left( - \frac{1}{x} \right)$, and generalizations thereof, with $x$ the (small) coupling constant of the problem under consideration. Further, the bridge equations essentially encode the resurgent nature of each transseries component, \textit{i.e.}, they make precise the resurgence link between a chosen sector $(n)$, and all other sectors $(m)$ associated to $(n)$ via resurgence as in \eqref{resurgence_predicting} or \eqref{resurgence_decoding}. In this way, many different problems may have the same transseries structure and the same bridge equations. So what tells them apart? For once, obviously, the perturbative and multi-instanton data appearing in \eqref{resurgence_predicting} or \eqref{resurgence_decoding}. But this is not all. There is also a (possibly infinite) set of constants, the \textit{Stokes constants}, $S_n$, which are distinct for each different problem. They already appear in the resurgence relations: to be more precise \eqref{resurgence_predicting} and \eqref{resurgence_decoding} above should have been written in the schematic form
\be
\label{resurgence_stokes}
\Phi^{(0)}_{k \gg 1} \quad \longleftrightarrow \quad \boldsymbol{S_1} \left[ \Phi^{(1)}_\ell \right] + \boldsymbol{S_2} \left[ \Phi^{(2)}_\ell \right] + \boldsymbol{S_3} \left[ \Phi^{(3)}_\ell \right] + \cdots.
\ee
\noindent
In this way it should be clear that, without having the Stokes constants, any resurgence statement will be quantitatively limited. So how do we compute them? This is generically a hard problem, specially for nonlinear systems, and one which we shall discuss at length in this paper.

The origin of these Stokes constants predates resurgence by over one hundred years: they of course arise with the discovery of Stokes phenomena \cite{s64} (see, \textit{e.g.}, the excellent review \cite{b91} in a closely related context). When the transseries is first assembled, all its nonperturbative content $\Phi^{(n)}$ is exponentially suppressed; in fact the higher the instanton number the greater the exponential damping. This means that already the one-instanton contribution is invisible from the point-of-view of the perturbative power series---at least as long as the (overall) argument of the exponential is negative. But as one changes the couplings or other parameters in the theory along the complex plane, this argument may vary and consequentially these exponentially suppressed contributions may change dominance. In fact they can become of the same order as the perturbative series and, eventually, even exponentially \textit{enhanced} as compared to $\Phi^{(0)}$. This is the essence of Stokes phenomena: small, suppressed exponentials, hidden behind perturbation theory,  may actually grow to become the dominant contribution. The transseries formalism already incorporates Stokes phenomena in the sense that it assembles all non-analytic exponential contributions together. The Stokes constants then precisely control how the transseries changes as one varies the couplings across their complex domains and different types of Stokes phenomena may occur. As we shall show later, this is accomplished by having the \textit{transseries parameters} $\boldsymbol{\sigma}$ transform at the Stokes lines, with these transformations explicitly dependent upon the Stokes constants ``attached'' to the corresponding Stokes line (in the simplest cases these are just shifts $\boldsymbol{\sigma} \to \boldsymbol{\sigma} + \boldsymbol{S_1}$, such that if one starts off with the perturbative series alone, $\boldsymbol{\sigma}=\boldsymbol{0}$, upon crossing a Stokes line the exponential contributions become ``turned on'' with their weight measured by $\boldsymbol{S_1}$). How general transseries incorporate Stokes phenomena was recently discussed in \cite{as13}.

The fundamental role played by the Stokes constants in the nonperturbative construction of a given observable should be now clear. They allow for precise, quantitative statements concerning the large-order resurgent behaviour of its different sectors; and they precisely quantify Stokes phenomena, allowing for a fully nonperturbative exploration of the whole parameter (coupling) space. Yet, computing them may be very hard. For instance, in the examples of the Painlev\'e~I \cite{msw07, gikm10, asv11} or the Painlev\'e~II \cite{m08, sv13} equations, out of a possibly infinite set of Stokes constants only one (in each problem) was accessible to analytic calculation. All others had to be determined numerically. Even worse, most of these resulting numbers seemed to be transcendental. Having a first principles guide to the computation of Stokes constants is thus a critical step towards achieving analytical control over this type of nonperturbative constructions.

There is one more feature which may be found in general resurgent asymptotics and we still wish to mention: \textit{resonance}. While we have been talking about a single instanton action, in many scenarios one may in fact have \textit{several} instanton (or renormalon) actions. The number of nonperturbative sectors then multiplies accordingly. For simplicity, let us assume there are only two instanton actions, $A_1$ and $A_2$, and the nonperturbative sectors are organized as $(n,m)$ corresponding to an exponential factor of the form $\sim n A_1 + m A_2$. The reader immediately realizes that a door is now open towards new phenomena: indeed, if the two actions happen to be $\BZ$-linearly dependent, then there is $n^\star$, $m^\star$ such that $n^\star A_1 + m^\star A_2 = 0$ and the nonperturbative exponential contributions ``collapse'' to order one, no longer suppressed or enhanced. Such phenomena have appeared recurrently in recent string theoretic examples \cite{gikm10, asv11, sv13}. In these cases both the large-order resurgence relations as well as the transseries structure will change, in order to incorporate new (nonperturbative) resonant sectors, and logarithms may appear throughout. Because general physical problems may depend upon many parameters, it is conceivable that resonance is ``tunable'' and may generically appear in many examples. We shall later discuss in a very explicit example how this may come about and what are its implications.

\textbf{These lectures are organized as follows}. We begin in \textbf{section~\ref{sec:quartic}} by considering a toy model for an interacting quantum theory: a one-dimensional integral representing the partition function of a system with quartic potential. Because this partition function satisfies a ``Schwinger--Dyson'' linear differential equation, this allows us to start the analysis from simple saddle-point arguments and, in the process of computation, introduce many basic resurgence methods and ideas. As we turn to the free energy of this system, the problem becomes nonlinear and full-fledged resurgence is then required. We discuss its asymptotics, both analytically and via high-precision numerical studies, and show how Stokes constants from the linear problem may be used to find the Stokes constants of the nonlinear problem. Most numerical tools used in the asymptotic analyses are always introduced as needed. \textbf{Section~\ref{sec:quartic}} also aims at being \textit{self-contained}, and enough material for the \textit{first-time reader} who is only interested in the basics of resurgence. Having grasped the basics of resurgent analysis, we then start addressing its diverse generalizations. \textbf{Section~\ref{sec:lefschetz}} discusses a generalization of saddle-point analysis (previously used in section~\ref{sec:quartic}) to higher dimensions, replacing one-dimensional steepest-descent contours with multi-dimensional Lefschetz thimbles, and how these are adequate tools to use when addressing linear problems. Nonlinear problems require heavier machinery and in \textbf{section~\ref{sec:borel}} we discuss it by addressing general results on Borel transforms and their associated resurgence singularities. These linear and nonlinear discussions lead up to \textbf{section~\ref{sec:physics}}, where we describe the (more complicated) \textit{general} structure of resurgent transseries, involving arbitrary instanton actions, and always having in mind a physical point-of-view. In particular, ``physical interpretations'' of alien calculus are suggested, either via statistical mechanics or via conformal field theory. Ensemble, these sections yield an overview of many technical ideas and concepts pertaining to resurgence and transseries. \textbf{Section~\ref{sec:elliptic}} then reconsiders a more complicated toy model where resonance may also appear: a one-dimensional integral representing the partition function of a system with an elliptic-function potential. Modularity now enters the game and the instanton actions depend upon a modulus, which may be chosen so that the system resonates. In both linear (partition function) and nonlinear (free energy) systems we analyze how resonance appears in the asymptotics, and how it changes as the elliptic modulus is varied. \textbf{Sections~\ref{sec:quartic}} and~\textbf{\ref{sec:elliptic}} thus introduce and review, in very explicit examples, most working methods of resurgent asymptotics. There are also a few appendices, collecting some technicalities and directions of further study. A more mathematical approach to alien calculus, alongside further technical details, appears in \textbf{appendix~\ref{app:alien-calculus}}. Then, \textbf{appendix~\ref{app:stokes}} discusses (anti) Stokes lines and their wall crossing in quantum mechanics, while \textbf{appendix~\ref{app:strongcoupling}} discusses resummations and strong-coupling expansions---together, these two appendices complement some discussions from \textbf{section~\ref{sec:quartic}}. Technicalities concerning generating high-order data, used in \textbf{sections~\ref{sec:quartic}} and~\textbf{\ref{sec:elliptic}}, are collected in \textbf{appendix~\ref{app:recursive-free-en}}; while some other further details on Borel transforms, complementing \textbf{section~\ref{sec:borel}}, may be found in \textbf{appendix~\ref{app:borel}}. Finally, it is worth pointing out two interesting topics which are \textit{not} included in the present work. One of these topics is the resummation of transseries (albeit briefly discussed in appendix~\ref{app:strongcoupling}). This involves not only the classical idea of Borel resummation, alongside its practical implementation as Borel--Pad\'e resummation, but mainly its newer version---as it applies to transseries with multi-instanton sectors and occurrence of Stokes phenomena---in the form of Borel--Pad\'e--\'Ecalle resummation. We refer the reader to, \textit{e.g.}, the recent papers \cite{m08, gmz14, csv15, a15, cms16, cms17, c1712} for nice illustrations of resummations within gauge and string theoretic contexts. The one other topic which is left absent deals with reorganizations of transseries, by exchanging the order of the sums in the perturbative and multi-instanton indexes. This actually leads to a myriad of interesting results connecting to, \textit{e.g.}, modular properties \cite{bde00, b_e08, em08, mpp09, asv18a, asv18b}, transasymptotics \cite{cc01, cch13} and its ``nonlinear'' generalizations \cite{asv18a, asv18b} (all relating back to modularity \cite{bde00, b_e08, em08}), and (generalized) partition-function phases \cite{asv18a, asv18b}. The interested reader is encouraged to proceed into the aforementioned references, given the basis acquired in the present lectures.

\section{Resurgent Analysis of a Quartic-Potential Integral}\label{sec:quartic}

Let us begin by setting up some notation and definitions which will be needed in order to develop the resurgence analyses that follow. As mentioned earlier, there are a few reviews on the theory of resurgence and its associated transseries and alien calculus, but on what concerns an introduction to resurgent \textit{asymptotics} we are not aware of much beyond section~2 of \cite{asv11}. We begin precisely by recalling a few formulae from that paper. Consider an observable $\Phi := \langle \CO \rangle$ and its perturbative computation in some coupling constant, denoted by $x$, which as usual we shall take around $x \sim 0$.

The perturbative series\footnote{Note that at this stage we are not concerned on the details of how the perturbative series for the observable $\Phi$ was obtained, nor on whether this observable may also have a nonperturbative definition beyond perturbation theory. The machinery of resurgence simply starts from such a perturbative expansion, and works regardless of the aforementioned possibilities. As such, this series may equally arise from, \textit{e.g.}, Rayleigh--Schr\"odinger perturbation theory in quantum mechanics, Feynman diagrams in field theory, or a perturbative solution to some differential equation. In the following we shall be more specific on how to compute perturbative series in examples.}
\be
\label{divergent}
\Phi^{(0)} (x) \simeq \sum_{g=0}^{+\infty} \Phi^{(0)}_g\, x^{g+1}
\ee
\noindent
is asymptotic with \textit{zero radius of convergence} if its coefficients grow as $\Phi^{(0)}_g \sim g!$. But if this power-series does not converge for any non-zero value of $x$, then how can one compute the observable $\Phi$ as a function of the coupling? One of the most powerful frameworks to extract information out of this type of asymptotic series is Borel analysis, where one starts with the \textit{Borel transform}; a linear map acting on formal power series
\begin{align}
\CB \colon \BC [[x]] &\rightarrow \BC [[s]] \\
x^{\alpha+1} &\mapsto \CB \left[ x^{\alpha+1} \right](s) = \frac{s^{\alpha}}{\Gamma(\alpha+1)}.
\label{boreltrans}
\end{align}
\noindent
This produces a new series, $\CB [\Phi^{(0)}] (s)$, out of the asymptotic series \eqref{divergent}, which has finite \textit{non-zero} convergence radius, as its coefficients no longer grow factorially but instead only exponentially fast. It may thus be analytically continued throughout $s \in \BC$, into a \textit{function}, albeit having singularities and branch cuts. Once this is done, and chosen a direction $\theta$ in the complex $s$-plane along which $\CB [\Phi^{(0)}] (s)$ has no singularities, the \textit{Borel resummation} (or inverse Borel transform)
\be
\label{borelresum}
\CS_\theta \Phi^{(0)} (x) = \int_0^{\rme^{\rmi\theta}\infty} \rmd s\, \CB[\Phi^{(0)}](s)\, \rme^{-\frac{s}{x}}
\ee
\noindent
finally associates a finite value to the divergent sum \eqref{divergent}, for each $x$. If the Borel transform was an entire function this would pretty much be the end of the story. But that can only happen if the original series was \textit{not} asymptotic to begin with; being asymptotic, then $\CB [\Phi^{(0)}] (s)$ \textit{will} have singularities---and this is instead the beginning of a fascinating story. Indeed, if $\CB [\Phi^{(0)}] (s)$ has singularities along $\theta$, this (singular) direction is now known as a \textit{Stokes line}, where \eqref{borelresum} is no longer well-defined: there is an \textit{ambiguity} concerning along which direction should these singularities be avoided by the  integration contour. But if the end-result of this process of resummation might be ambiguous, it would seem we still cannot compute the observable $\Phi$ as a function of $x$? In order to solve this issue, one first defines \textit{lateral} Borel resummations, $\CS_{\theta^{\pm}} \Phi^{(0)} (x)$, which avoid all singularities via a contour just to the left (to the right) of the direction $\theta$. They naturally lead to different results and the one seemingly simple question which actually sparks many features of resurgence is: are these left- and right-lateral resummations related somehow?

As it turns out, and as we shall pedagogically discuss at greater length in this section---and, in a more rigorous mathematical language, in appendix~\ref{app:alien-calculus}---, the answer to the above question is ``yes''. The lateral Borel resummations lead to \textit{distinct} sectorial resummations of our original asymptotic series, but they are nonetheless \textit{connected} via the so-called \textit{Stokes automorphism} $\underline{\mathfrak{S}}_\theta$,
\be\label{stokesauto}
\CS_{\theta^+} = \CS_{\theta^-} \circ \underline{\mathfrak{S}}_\theta.
\ee
\noindent
In words, there is an operator, $\underline{\mathfrak{S}}_\theta$, relating both resummations (and essentially encoding all the singular structure along the Stokes direction $\theta$), and which we shall explain below how to construct starting from a simple example. This is also the point where \textit{alien calculus} kicks in, as the Stokes automorphism may be computed in terms of the \textit{alien derivative}, $\Delta_\omega$, a differential operator encoding the singular behaviour of the Borel transform at the point $\omega$. We shall also explain in the following how this alien derivative may be very simply read from the Borel structure of a given asymptotic series. Furthermore, it turns out that $\underline{\mathfrak{S}}_\theta \sim \exp \left( \rme^{-\frac{\omega}{x}} \Delta_\omega \right)$ with $\omega$ in the $\theta$--direction (more on all this below), a relation which also implies that the aforementioned ambiguity is nonperturbative (or non-analytic), essentially of order $\sim \rme^{-\frac{1}{x}}$. Clearly, if we know all possible Stokes automorphisms associated to all Stokes lines, or, equivalently, all possible alien derivatives along these same singular directions, then we will know how to match all possible sectorial resummations and it is thus possible to reconstruct the nonperturbative solution of the problem at hand, anywhere in the coupling-constant complex-plane. Because each of these matches is non-analytic in $\sim \rme^{-\frac{1}{x}}$, this further implies that the nonperturbative solution we are trying to construct must involve all possible such (exponential-type) instanton contributions. This naturally leads to our next ingredient. 

As we shall see, the calculation of alien derivatives is undissociated from \textit{transseries}, general solutions to nonlinear problems, going beyond the realm of power series. For example, a transseries \textit{ansatz} for our resurgent function, depending on a single parameter $\sigma$, could be
\be
\label{Phitransseries}
\Phi (x,\sigma) = \sum_{n=0}^{+\infty} \sigma^n\, \Phi^{(n)} (x),
\ee
\noindent
where $\Phi^{(0)} (x)$ is the formal asymptotic power series \eqref{divergent}, and where the $\Phi^{(n)} (x)$ are the $n$-instanton contributions,
\be
\label{Phi(n)}
\Phi^{(n)} (x) \simeq \rme^{-n \frac{A}{x}}\, \sum_{g=1}^{+\infty} \Phi_g^{(n)}\, x^{g + \beta_n},
\ee
\noindent
with $A$ the instanton action and $\beta_n$ a characteristic exponent. The exponentials in \eqref{Phi(n)} are \textit{non-analytic}. In this way, indeed the transseries  \eqref{Phitransseries} includes all possible (exponential-type) instanton contributions, going beyond the realm of power-series and organizing itself as a double series: for each $n$ as an asymptotic series, and for the sum in $n$ as a ``power series'' in the non-analytic term $\rme^{-\frac{A}{x}}$. The transseries parameter $\sigma$ is an instanton counting parameter, associated to the instanton action $A$, and it parameterizes different choices of boundary conditions for whatever problem one is addressing (\textit{e.g.}, \eqref{Phitransseries} would be an appropriate solution \textit{ansatz} for a first-order nonlinear ordinary differential equation, representing a one-parameter \textit{family} of solutions for unspecified $\sigma$). Generically one should consider transseries depending on multiple parameters; \textit{e.g.}, a two-parameter transseries was fundamental in the solution to the Painlev\'e~I equation and the quartic matrix model in \cite{gikm10, asv11}, or the solution to the Painlev\'e~II equation in \cite{sv13}. Even more complicated multi-parameter transseries solutions seem to appear within the topological string theory context, \textit{e.g.}, \cite{dmp11, cesv13, cesv14, c15}. In particular, note that a problem whose transseries solution involves $k$-parameters will have multiple instanton sectors, labeled by a set of integers $\boldsymbol{n} = (n_1, \ldots, n_k) \in \BN^k$, as $\Phi^{(\boldsymbol{n})} (x)$. The $\boldsymbol{n} = (0, \ldots, 0)$ sector is the perturbative sector, the $\boldsymbol{n} = (n_1, 0, \ldots, 0)$ sector is a typical instanton sector associated to the instanton action $A_1$, but most sectors will be mixed and one may even find new (``unusual'') sectors associated to complex instantons as in, \textit{e.g.}, \cite{sr78, bpv78a, bpv78b, v83, gikm10, asv11, sv13, bdu13, gmz14, cesv14, cms16, cms17}.

Finally, let us mention that the reason why the automorphism $\underline{\mathfrak{S}}_\theta$ in \eqref{stokesauto} is known as the \textit{Stokes} automorphism is, of course, because it precisely describes \textit{Stokes} phenomena---which we already explained qualitatively in the introduction. We may now be a bit more quantitative by considering the one-parameter transseries \eqref{Phitransseries}, for which $\theta=0$ is a singular direction of the Borel transform, \textit{i.e.}, it is a Stokes line (for $A>0$). As we shall explain later, applying the Stokes automorphism \eqref{stokesauto} one finds
\be\label{stokespheno}
\CS_{+} \Phi (x,\sigma) = \CS_{-} \Phi \left( x, \sigma + S_1 \right),
\ee
\noindent
where $S_1$ is a Stokes constant as in \eqref{resurgence_stokes}. One immediately realizes that this expression, \eqref{stokespheno}, precisely describes the Stokes phenomena of classical asymptotics within the resurgence framework: starting out with $\sigma = 0$, all that the transseries \eqref{Phitransseries} includes is the original perturbative series \eqref{divergent}. But once we cross a Stokes line, signaled by the fact that the Borel resummation \eqref{borelresum} is no longer yielding correct  results, then \eqref{stokespheno} above will make appear a series of exponentially suppressed terms, as in \eqref{Phi(n)}, which correct the Borel resummation of the perturbative expansion and which may eventually grow and take over the dominance of the transseries \eqref{Phitransseries}.

\subsection{Phase Diagram from Stokes Phenomena}\label{subsec:4intphases}

Let us now describe these ideas in greater detail as they appear in a simple example, that of a partition-function toy-model described by a one-dimensional integral with quartic potential. We will set up this analysis within the resurgence framework, identifying Stokes and anti-Stokes lines, and constructing all different asymptotic expansions which hold on different sides of the Stokes lines. For pedagogical purposes, we shall start over familiar ground by considering simple saddle-point analysis and steepest-descent contours in the present subsection.

\begin{figure}[t!]
\begin{center}
\includegraphics[width=6cm]{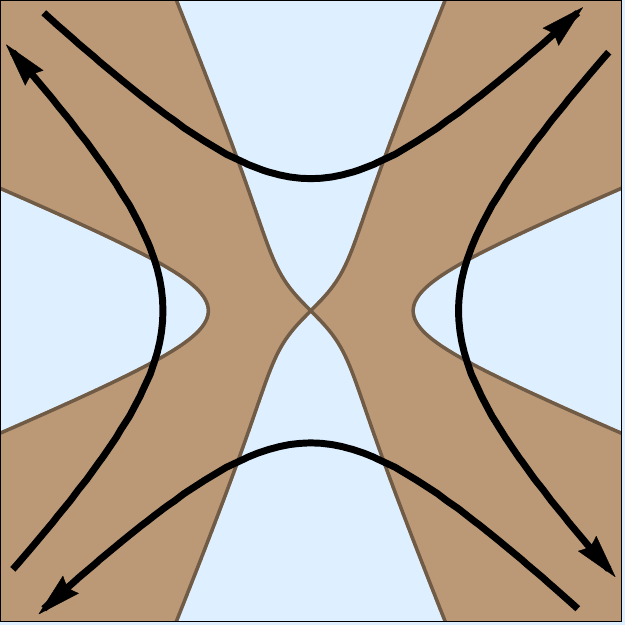}
\end{center}
\caption{The quartic potential \eqref{quarticpotential} in the complex $z$-plane, with $\re\, \lambda > 0$ ($\lambda=1$ in the plot). The brown (darker) region is where $\re\, V(z) > 0$, while the blue (lighter) region is where $\re\, V(z) < 0$. The four contours are the admissible contours, leading to three homologically independent integration paths. The $\BZ_2$ symmetry $V(z) = V(-z)$ of the quartic potential \eqref{quarticpotential} further narrows down the number of independent results one may compute over homologically distinct contours.}
\label{quarticpotentialfig}
\end{figure}

Consider the ``quartic partition function'', described by
\be
\label{quarticintegral}
Z (\hbar) = \frac{1}{2\pi} \int_{\Gamma} \rmd z\, \rme^{- \frac{1}{\hbar} V(z)},
\ee
\noindent
where $V(z)$ is the quartic potential
\be
\label{quarticpotential}
V(z) = \frac{1}{2} z^2 - \frac{\lambda}{24} z^4,
\ee
\noindent
with coupling constant $\lambda$, and $\Gamma$ is a contour to be specified. Not every contour is admissible: there are only four directions going off to infinity with $\re\, V(z) > 0$ as $z \to \infty$, naturally leading to three homologically-independent admissible integration paths. This is illustrated in figure~\ref{quarticpotentialfig} when both $\re\, \hbar > 0$ and $\re\, \lambda > 0$. As these conditions change, so will the contours.

\begin{figure}[t!]
\begin{center}
\includegraphics[width=3cm]{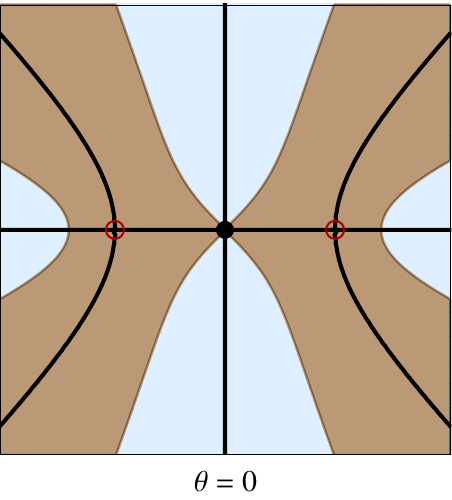}
$\qquad$
\includegraphics[width=3cm]{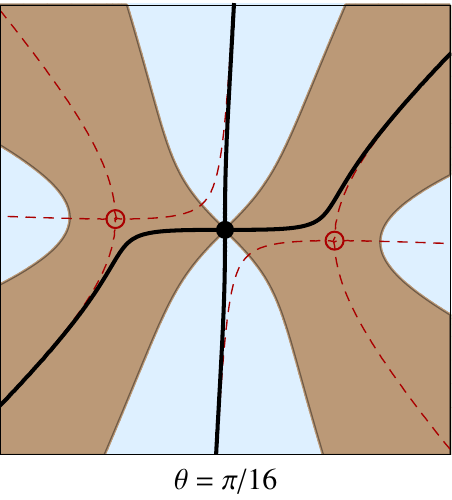}
$\qquad$
\includegraphics[width=3cm]{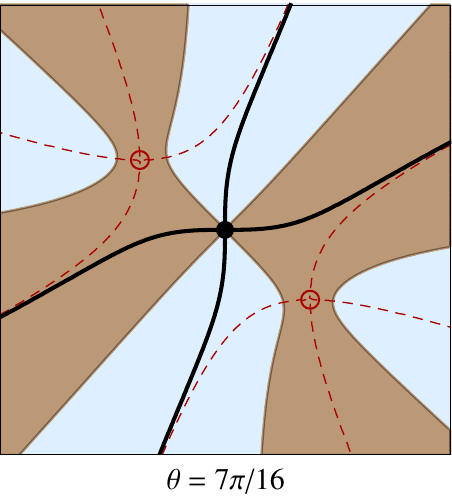}
$\qquad$
\includegraphics[width=3cm]{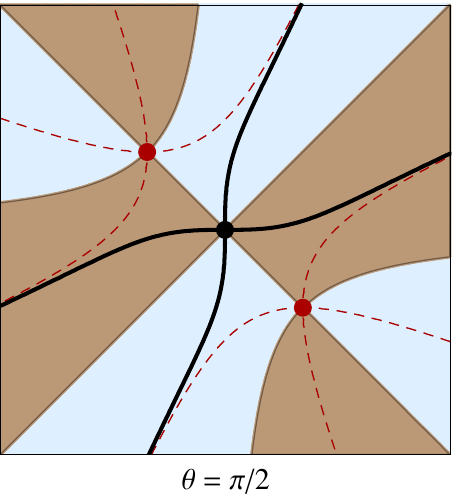}
\break
\includegraphics[width=3cm]{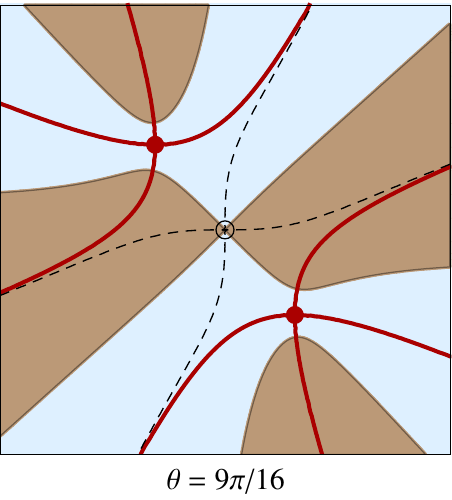}
$\qquad$
\includegraphics[width=3cm]{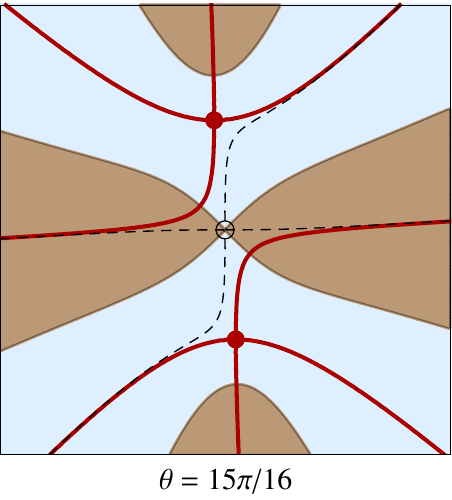}
$\qquad$
\includegraphics[width=3cm]{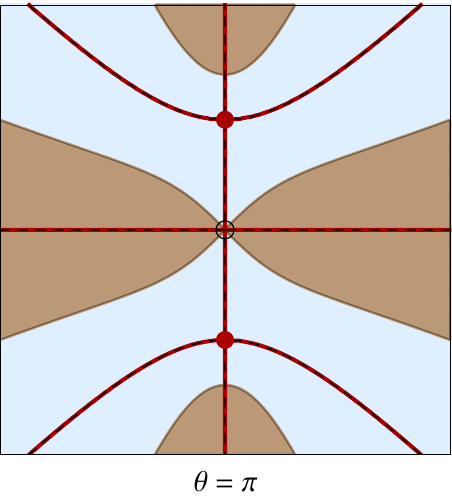}
$\qquad$
\includegraphics[width=3cm]{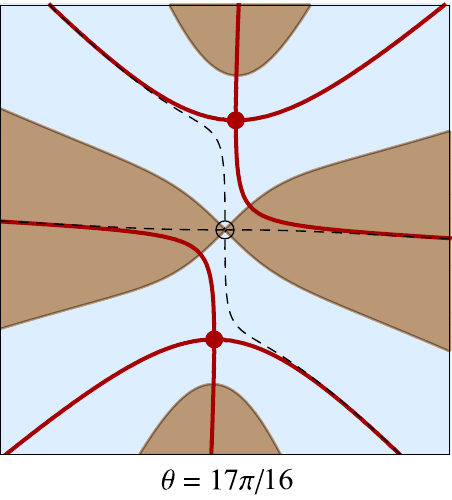}
\break
\includegraphics[width=3cm]{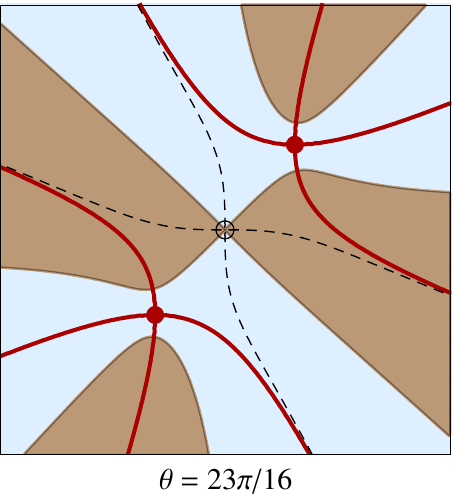}
$\qquad$
\includegraphics[width=3cm]{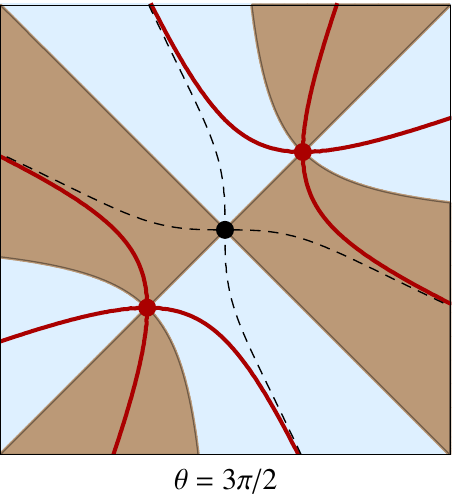}
$\qquad$
\includegraphics[width=3cm]{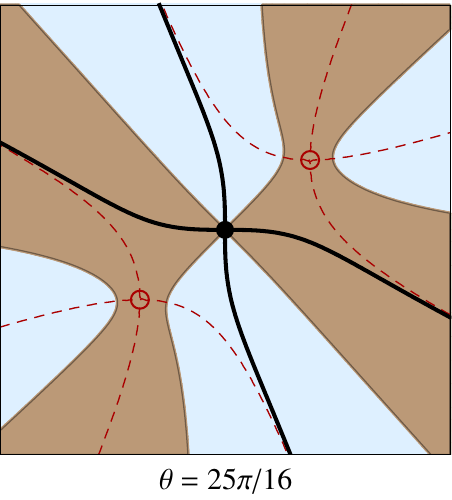}
$\qquad$
\includegraphics[width=3cm]{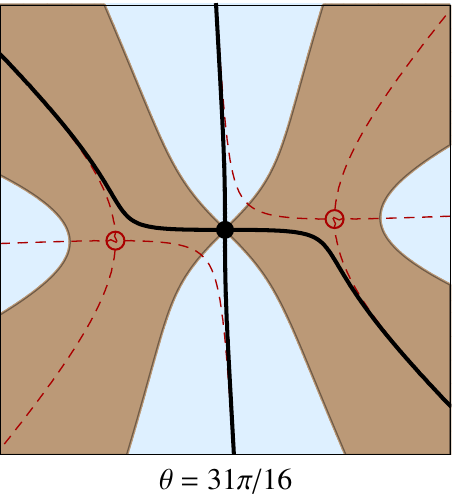}
\end{center}
\caption{Steepest descent (ascent) contours for the quartic integral, through the several saddles (black for the ``perturbative'' saddle and red for the ``instanton'' saddle), along with the positivity of the potential, for different values of $\theta = \arg x$ ($|x|=1$ in the plot). This makes clear which are Stokes and anti-Stokes lines: Stokes lines at $\arg x = 0, \pi$ when the steepest descent contours hit the three saddles; anti-Stokes lines at $\arg x = \frac{\pi}{2}, \frac{3\pi}{2}$. See the main text for a full discussion of the plot sequence.}
\label{steepest}
\end{figure}

With a simple change of variables the partition function \eqref{quarticintegral} may be written as
\be
\label{quarticintegralX}
Z (x) = \frac{\sqrt{\hbar}}{2\pi} \int_{\Gamma} \rmd z\, \exp \left( - \frac{1}{2} z^2 + \frac{x}{24} z^4 \right),
\ee
\noindent
which explicitly illustrates how all non-trivial dependence occurs through the variable $x \equiv \lambda \hbar$, and this is what we shall study in the following. One way to evaluate this integral as an asymptotic expansion is through the method of steepest descents (see, \textit{e.g.}, \cite{bo78}). First, one computes the saddle-points $z^*_0 = 0$, $z^*_{\pm} = \pm \sqrt{\frac{6}{x}}$, with $V(z^*_0) = 0$ and $V(z^*_{\pm}) = \frac{3}{2x}$. Their leading contribution to the quartic partition function is thus given by $\exp \left( - V(z^*) \right)$ and which is the dominant saddle will of course depend on the value of $\arg x$. Next, chosen a reference saddle $z^*$, one deforms the contour of integration into the infinite, oriented path of steepest descent through $z^*$, defined as
\be
\label{steepestdescentcontour-def}
\im \left( V(z) - V(z^*) \right) = 0,
\ee
\noindent
and increasing away from $z^*$. For both ``perturbative'' saddle at the origin, $z^*_0$, and ``instanton'' saddles, $z^*_{\pm}$, steepest descent (and steepest ascent) contours are depicted in figure~\ref{steepest}, for different values of $\theta = \arg x$. We have also plotted the regions where the quartic potential is positive (negative): comparing with figure~\ref{quarticpotentialfig}, we now see how these regions change with $\theta$. In particular, this allows us to distinguish steepest descent\footnote{Departing off to infinity along brown (darker) regions, ensuring convergence of the integral.} from steepest ascent contours. Let us explain the plotting details of figure~\ref{steepest}: for each value of $\theta$ we have plotted the leading saddle(s) with a solid disk, while the subleading saddle(s) were plotted with a circle. Regardless of dominance, the ``perturbative'' saddle is always plotted in black, with the ``instanton'' saddles in red. Then, solid contours are steepest descent (ascent) contours associated with the leading saddle(s), while dashed contours are steepest descent (ascent) contours associated with the subleading saddle(s).

\begin{figure}[t!]
\begin{center}
\includegraphics[width=6cm]{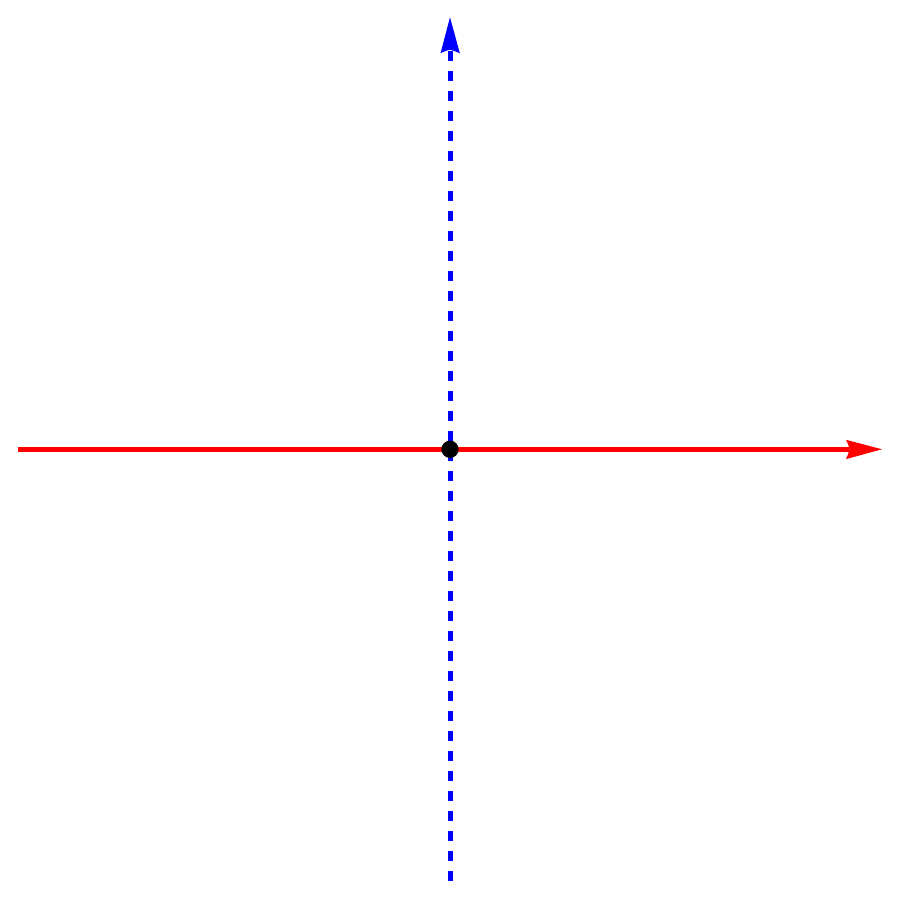}
$\qquad\qquad$
\includegraphics[width=6cm]{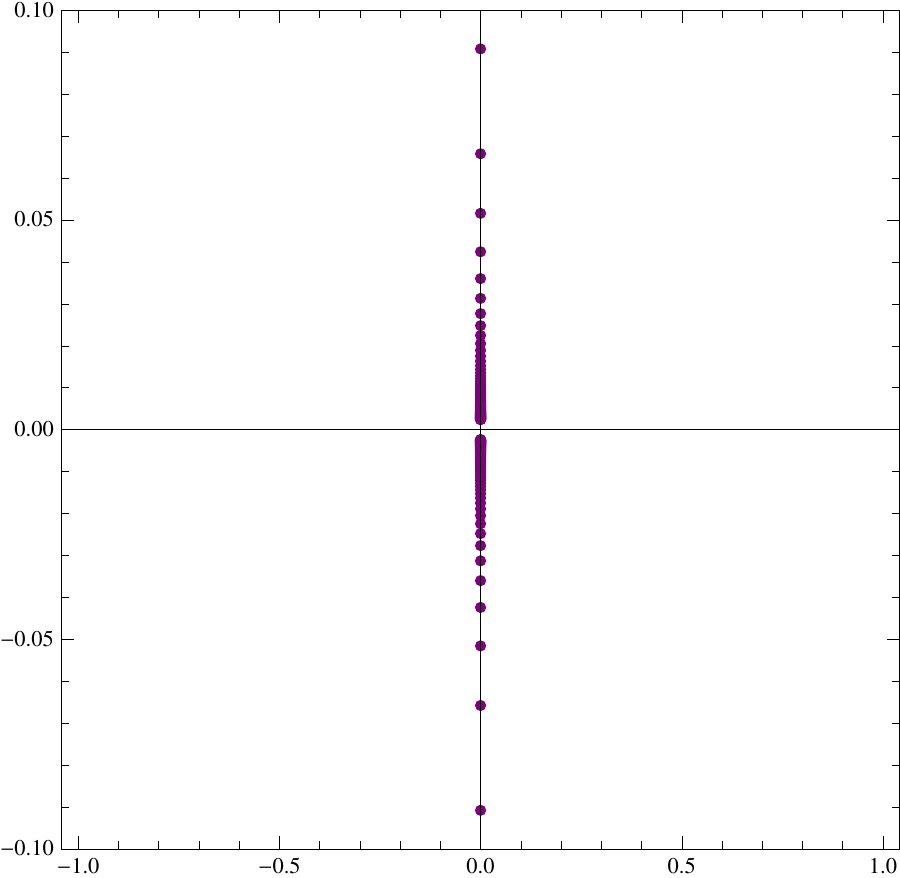}
\end{center}
\caption{The first image shows the coupling constant $x$-plane, representing the phase diagram of the quartic partition function. Stokes lines are in red (solid), at $\arg x = 0,\pi$, while anti-Stokes lines are in blue (dashed), at $\arg x = \frac{\pi}{2}, \frac{3\pi}{2}$. The second image shows the Lee--Yang accumulation of zeros of the modified Bessel function describing the quartic partition-function, signaling a phase transition. One can clearly see that indeed the anti-Stokes lines correspond to phase boundaries.}
\label{quarticphase}
\end{figure}

What we immediately learn from figure~\ref{steepest} is that, generically, a steepest-descent contour goes through a single saddle---but this is not always the case. Indeed, this procedure yields well-defined asymptotic expansions as long as the path of steepest descent used in their calculation does not pass through a second saddle. This will occur when
\be
\im \left( V(z^*_{\pm}) - V(z^*_0) \right) = 0 \quad \Leftrightarrow \quad \im\ \frac{1}{x} = 0,
\ee
\noindent
corresponding to $\arg x = 0, \pi$ (these lines are very distinctive in figure~\ref{steepest}). These are\footnote{Note that the definitions of Stokes and anti-Stokes lines are usually reversed between the mathematical and physical literatures. In this work we are using the appropriate definitions within the resurgence framework. As such, the reader who has been exposed to these ideas before should pay attention to our precise definitions.} the \textit{Stokes lines} for the quartic partition-function, where, because other (suppressed) saddles start contributing to the evaluation of the integral, subleading exponentials likewise start contributing to the asymptotics. At $\theta=0$ the leading saddle is $z^*_0$, while $z^*_{\pm}$ lead to exponentially suppressed contributions (in fact, \textit{purely} exponential contributions along the Stokes direction). As $\theta$ keeps increasing, these exponentially subleading contributions grow and, when they becomes of the same magnitude as the leading term, we reach what is known as an \textit{anti-Stokes line}. Different exponential contributions are of the same magnitude when both saddles have the same real part,
\be
\re \left( V(z^*_{\pm}) - V(z^*_0) \right) = 0 \quad \Leftrightarrow \quad \re\ \frac{1}{x} = 0,
\ee
\noindent
corresponding to $\arg x = \frac{\pi}{2}, \frac{3\pi}{2}$ (these lines are also shown in figure~\ref{steepest}). At $\theta=\frac{\pi}{2}$ the subleading saddles $z^*_{\pm}$ have become of the same magnitude as $z^*_0$ and one thus finds an \textit{oscillatory} asymptotic behaviour along this direction. As we keep increasing $\theta$ beyond $\frac{\pi}{2}$ one sees that the different saddles have reversed roles: it is now the $z^*_0$ contribution which is exponentially suppressed with respect to that of $z^*_{\pm}$. This situation is maintained all the way through the Stokes line at $\theta=\pi$, where the $z^*_0$ contribution, albeit suppressed, starts ``growing'' again (as expected upon crossing a Stokes line), until we reach the anti-Stokes line at $\theta=\frac{3\pi}{2}$. Here all saddles recovered the same magnitude and one finds oscillatory asymptotic behaviour once again. Past this line, the leading contribution is again given by $z^*_0$ all the way back to $\theta=2\pi$. The (very simple) ``phase diagram'' of the quartic partition function, describing the aforementioned behaviour, is shown in figure~\ref{quarticphase}.

What we have just described is the Stokes phenomenon: in an asymptotic approximation, \textit{different} asymptotic formulae hold, on different sectors, for the \textit{same} analytic function. As we shall see, transseries precisely capture all these possible asymptotic formulae in a single go. In the meantime, it is also interesting to understand the physical interpretation where Stokes phenomenon describes phase transitions \cite{ps93}, building upon the Lee--Yang construction \cite{ly52a, ly52b} (see \cite{gg07} as well). In the Lee--Yang set-up, the finite-size grand-canonical partition function is first written as a polynomial. A phase transition then arises if, in the thermodynamic limit, the zeros of this polynomial condense onto lines describing phase boundaries. In the case of the quartic partition function (where the asymptotic approximation plays the role of the thermodynamic limit), it is simple to see that phase transitions occur at the anti-Stokes lines, from the ``one-saddle'' phase to the ``two-saddle'' phase and back again. This follows from an explicit Lee--Yang analysis: one can actually exactly compute the  integral \eqref{quarticintegralX} as a modified Bessel function of the first kind (up to an exponential piece and a change of variables), whose zeros\footnote{Rigorously, this depends upon a choice of contour, yielding appropriate linear combinations of two Bessel functions. We are plotting the zeros of each of these building blocks (not exactly the same, but extremely similar).} precisely condense upon the imaginary axis as shown in figure~\ref{quarticphase}. We shall next describe how resurgent analysis unfolds, but for the moment just note that the picture we have unveiled for the phase diagram of the quartic partition function\footnote{A short extension of this analysis, from the present realm of partition-function toy models to the full framework of one-dimensional quantum mechanics, is included in appendix~\ref{app:stokes} for the benefit of the interested reader.} is the following: in the complex $x$-plane, the real axis is a Stokes line, while the imaginary axis is an anti-Stokes line/phase boundary.

\subsection{Basic Formulae for Resurgent Analysis}\label{subsec:basicresurgence}

Having understood the steepest-descent contour structure of the integral \eqref{quarticintegralX} we still need to explicitly evaluate it, as an asymptotic series. This is most simply done by first noticing that this integral satisfies a second-order \textit{linear} ordinary differential equation\footnote{As mentioned, peeling off an exponential piece and changing variables, this is the modified Bessel equation.} (ODE),
\be
\label{quarticODE}
16 x^2\, Z'' (x) + \left( 32 x - 24 \right) Z' (x) + 3\, Z (x) = 0.
\ee
\noindent
An \textit{ansatz} of the form $Z (x) = \rme^{-A/x}\, \Phi (x)$, with $\Phi (x)$ a power series, perturbatively yields (algebraic and recursive) equations for both $A$ and the coefficients of the power series $\Phi (x)$. The quadratic equation for $A$ yields the two solutions
\be
\label{instactions}
A = 0 \qquad \text{or} \qquad A = \frac{3}{2}
\ee
\noindent
as expected from our earlier saddle-point analysis; while the recursive equations for the corresponding power-series coefficients yield the asymptotic series
\bea
\label{coeff-quartic}
\Phi_{0} (x) &\simeq& \sum_{n=0}^{+\infty} Z^{(0)}_n\, x^n, \qquad Z^{(0)}_n = \left( \frac{2}{3} \right)^n \frac{(4n)!}{2^{6n}\, (2n)!\, n!}\, Z^{(0)}_0, \\
\label{coeff-quartic-inst}
\Phi_{1} (x) &\simeq& \sum_{n=0}^{+\infty} Z^{(1)}_n\, x^n, \qquad Z^{(1)}_n = (-1)^n \left( \frac{2}{3} \right)^n \frac{(4n)!}{2^{6n}\, (2n)!\, n!}\, Z^{(1)}_0.
\eea
\noindent
The ``instanton action'' $A=\frac{3}{2}$ is evident in the growth of the coefficients $Z^{(0)}_n, Z^{(1)}_n \sim A^{-n} n!$. Notice that one still has to evaluate the \textit{constant} ($x$-independent) contributions, $Z^{(0)}_0$ and $Z^{(1)}_0$, sometimes denoted by \textit{residual coefficients} in the resurgence literature. This is done by associating each of the above asymptotic expansions to distinct steepest-descent contours. For the perturbative saddle, in the ``zero-instanton'' sector, it is simple to evaluate $Z (0) = \sqrt{\frac{\hbar}{2\pi}} \equiv Z^{(0)}_0$ from \eqref{quarticintegralX}; while for the two subleading saddles, in the ``one-instanton'' sector, the integral over the appropriate contour\footnote{This is one of the ``side'' contours in figure~\ref{quarticpotentialfig}, only picking-up one ``instanton'' saddle.} may be evaluated in terms of modified Bessel functions whose asymptotic expansion then yields $Z^{(1)}_0 = -\frac{\rmi}{\sqrt{2}}\, Z^{(0)}_0$. The asymptotic evaluation of the quartic partition function thus boils down to
\bea
\label{Z010}
Z^{(0)} (x) &\simeq& \sqrt{\frac{\hbar}{2\pi}}\, \sum_{n=0}^{+\infty} \left( \frac{2}{3} \right)^n \frac{(4n)!}{2^{6n}\, (2n)!\, n!}\, x^n, \\
\label{Z101}
Z^{(1)} (x) &\simeq& -\frac{\rmi}{\sqrt{2}}\, \sqrt{\frac{\hbar}{2\pi}}\, \rme^{-\frac{3}{2x}}\, \sum_{n=0}^{+\infty} (-1)^n \left( \frac{2}{3} \right)^n \frac{(4n)!}{2^{6n}\, (2n)!\, n!}\, x^n.
\eea
\noindent
Note how one may also obtain the Stokes and anti-Stokes lines directly from the exponential structure of the above solutions; by looking at their instanton structure. Stokes lines are given by
\be
\im \left( \frac{3}{2 x} \right) = 0,
\ee
\noindent
such that the exponential in \eqref{Z101} is as damped (enhanced) as possible; while anti-Stokes lines correspond to
\be
\re \left( \frac{3}{2 x} \right) = 0,
\ee
\noindent
such that the exponential in \eqref{Z101} is of order one, and \eqref{Z010} and \eqref{Z101} are of the same order. This obviously yields the precise same results as discussed earlier and depicted in figure~\ref{quarticphase}.

Having constructed two linearly independent solutions\footnote{Respectively associated to each of the two linearly/homologically independent steepest-descent contours of integration in figure~\ref{quarticpotentialfig}; \textit{i.e.}, one for the perturbative saddle and another for the nonperturbative saddle (due to the $\BZ_2$ symmetry of the potential, both $\pm \sqrt{6/x}$ saddles obviously yield the same result).} \eqref{Z010} and \eqref{Z101} to the linear second-order ODE \eqref{quarticODE}, the general solution follows as their linear combination. This introduces two integration constants, parameterizing boundary conditions. One finds:
\be
\label{quartictransseries}
\CZ (x, \sigma_{0}, \sigma_{1}) = \sigma_{0}\, Z^{(0)}(x) + \sigma_{1}\, Z^{(1)}(x).
\ee
\noindent
This is in fact the \textit{transseries solution} to the problem, in the sense that it incorporates all saddles. The transseries parameters are $\sigma_{0}$ and $\sigma_{1}$, and it is also clear how they are parameterizing boundary conditions or contours of integration, depending upon the point-of-view. This is, of course, a trivial statement in elementary (linear) differential equations theory. Further, when comparing this explicit result \eqref{quartictransseries} with the general form of a (one-parameter) transseries \eqref{Phitransseries}, in here we only find a \textit{finite} number of instanton sectors. Again, this is due to \textit{linearity} of the present problem. As we shall explicitly see in an upcoming subsection, only when considering \textit{nonlinear} problems do we find an infinite number of (multi) instanton sectors.

In order to illustrate the ideas of resurgence in an elementary setting, we now turn to the computation of the Borel transforms \eqref{boreltrans}, for the asymptotic power series $\Phi_{0} (x)$ and $\Phi_{1} (x)$. It is straightforward to obtain\footnote{Note that the Borel transform \eqref{boreltrans} is undefined for the residual coefficient, with $\alpha+1=0$. This is simply handled by dropping this coefficient and only adding it back-in at the end---its contribution is trivial to include at any stage. Our Borel transforms thus follow from \eqref{coeff-quartic} and \eqref{coeff-quartic-inst} with the asymptotic series starting at $n=1$.
\label{foot:residual-coefficients}}
\bea 
\label{borel-quartic-int}
\CB [\Phi_{0}] (s) &=& \frac{1}{8}\, \sqrt{\frac{\hbar}{2\pi}}\, {}_2 F_{1} \left( \frac{5}{4}, \frac{7}{4}, 2 \right. \left| \frac{2s}{3} \right), \\
\label{borel-quartic-int-inst}
\CB [\Phi_{1}] (s) &=& \frac{\rmi}{8 \sqrt{2}}\, \sqrt{\frac{\hbar}{2\pi}}\, {}_2 F_{1} \left( \frac{5}{4}, \frac{7}{4}, 2 \right. \left| - \frac{2s}{3} \right),
\eea
\noindent
where ${}_2 F_{1} \left( a, b, c \right. \left| z \right)$ is the hypergeometric function, with a branch cut on the complex $z$-plane running from $z=1$ to infinity. Thus, $\CB [\Phi_{0}] (s)$ has a singularity at $s = \frac{3}{2} \equiv A$ and $\CB [\Phi_{1}] (s)$ has a singularity at $s = - \frac{3}{2} \equiv - A$, implying that both have non-zero radius of convergence around $s=0$. If one next asks what is the nature of these singularities, it is not too hard to find the following expansions\footnote{Note how the aforementioned residual coefficients are again visible in here; attached to the simple poles.} near $s=\pm A$,
\bea
\label{borel-quartic-int-expanded}
\CB [\Phi_{0}] (s) &=& \left( + 2 \right) \times \frac{Z_{0}^{(1)}}{2\pi\rmi \left( s-A \right)} + \left( + 2 \right) \times \CB [\Phi_{1}] \left( s-A \right) \frac{\log \left( s-A \right)}{2\pi\rmi} + \text{holomorphic}, \\
\label{borel-quartic-int-expanded-inst}
\CB [\Phi_{1}] (s) &=& \left( - 1 \right) \times \frac{Z_{0}^{(0)}}{2\pi\rmi \left( s+A \right)} + \left( - 1 \right) \times \CB [\Phi_{0}] \left( s+A \right) \frac{\log \left( s+A \right)}{2\pi\rmi} + \text{holomorphic}.
\eea
\noindent
The first thing one notices in these results is the origin of the name ``resurgence'': in fact, at the singularity of $\CB [\Phi_{0}] (s)$ one finds the ``reappearance'' or \textit{resurgence} of the $\Phi_{1}$ sector, via its Borel transform; while at the singularity of $\CB [\Phi_{1}] (s)$ one finds the resurgence of the $\Phi_{0}$ sector. This reappearance includes the full asymptotic series, with the residual coefficient attached to the simple pole and the rest of the coefficients appearing next to the logarithmic branch-cut.

This property of ``reappearance'' turns out to be rather generic and it is in fact at the basis of the theory of resurgence. If we forget for one moment that it was uncovered by looking at the functional nature of the Borel singularities, or even that these singularities may be poles or logarithmic branch-cuts or else, what \eqref{borel-quartic-int-expanded} and \eqref{borel-quartic-int-expanded-inst} are expressing are relations in-between the asymptotic series $\Phi_{0} (x)$ and $\Phi_{1} (x)$, the building blocks of the transseries \eqref{quartictransseries}. Such a level of abstraction may be easily reachable from \eqref{borel-quartic-int-expanded} and \eqref{borel-quartic-int-expanded-inst} by first considering the difference between left and right Borel resummations, \eqref{stokesauto}, along the appropriate Stokes directions, $\theta=0$ and $\theta=\pi$ respectively,
\bea
\label{disc-to-alien-Phi0}
\left( \CS_{0^+} - \CS_{0^-} \right) \Phi_0 (x) &=& \left( + 2 \right) \times \left( - Z_{0}^{(1)} - \Phi_1^\star (x) \right) \rme^{-\frac{A}{x}} = \left( - 2 \right) \times \Phi_1 (x) \times \rme^{-\frac{A}{x}},
\\
\label{disc-to-alien-Phi1}
\left( \CS_{\pi^+} - \CS_{\pi^-} \right) \Phi_1 (x) &=& \left( - 1 \right) \times \left( - Z_{0}^{(0)} - \Phi_0^\star (x) \right) \rme^{+\frac{A}{x}} = \left( + 1 \right) \times \Phi_0 (x) \times \rme^{+\frac{A}{x}}.
\eea
\noindent
After the first equality, we momentarily used the notation $\Phi^\star (x)$ to explicitly denote the asymptotic series which follow from $\Phi (x)$ by dropping their residual coefficients, \textit{i.e.}, $\Phi_i^\star := \Phi_i - Z^{(i)}_0$ (recall the discussion leading up to \eqref{borel-quartic-int} and \eqref{borel-quartic-int-inst}; in particular footnote~\ref{foot:residual-coefficients}). It is very interesting to observe how precisely these coefficients are being added back-in by the simple-pole contributions in \eqref{borel-quartic-int-expanded} and \eqref{borel-quartic-int-expanded-inst}, which then lead to the very neat final equalities. As just discussed, we are aiming at a level of abstraction where we can do without looking at the particular functional nature of the Borel transforms. That is finally obtained by defining a linear operator, $\Delta_\omega$, with $\omega$ a point in the complex $s$-plane, which acts on the transseries building blocks $\Phi_{n}$ (the asymptotic series), and which describes the basic structure encoded in \eqref{borel-quartic-int-expanded} and \eqref{borel-quartic-int-expanded-inst}. In this particular linear example, a definition for such an operator naturally follows from the previous two equations, as\footnote{It is very important to keep in mind that this definition \textit{only holds} in the present \textit{linear} case. It will \textit{not} hold for \textit{nonlinear} problems, where it will require (substantial) modification. More on this in the following.}
\be
\label{def-linear-alien}
\Delta_\omega := \rme^{\frac{\omega}{x}} \left( \CS_{\theta^+} - \CS_{\theta^-} \right),
\ee
\noindent
if the na\"\i ve action of $\Delta_\omega$ results in a ``pure'' power-series (\textit{i.e.}, another transseries block, without any exponential factors), and $\Delta_\omega := 0$ otherwise (\textit{i.e.}, if the na\"\i ve action of $\Delta_\omega$ still includes exponential factors\footnote{These are the so-called ``transmonomials'' factors, to be discussed in section~\ref{sec:physics}.}). This definition leads to the (for the moment) abstract structure formalizing the property of resurgence:
\bea
\label{Delta+A}
\Delta_{+A} \Phi_{0} &=& \left( - 2 \right) \times \Phi_{1}, \\
\label{Delta-A}
\Delta_{-A} \Phi_{1} &=& \Phi_{0}.
\eea
\noindent
Our departure equations may here be recovered by reading these expressions as follows: the Borel transform of $\Phi_{0}$ has a singularity at $\omega=+A$ which is given by the \textit{Stokes coefficient} $S_1 = -2$, multiplied by the Borel transform of $\Phi_{1}$ (and ``attached'' to the logarithmic\footnote{If all singularities are simple poles alongside logarithmic branch-cuts, this implies we are in the class of so-called \textit{simple} resurgent functions, but we will get back to this point in section~\ref{sec:borel}.} branch-cut). The second line is read similarly. Furthermore, this algebraic structure closes upon itself,
\be
\label{linear-algebraic-structure}
\begin{tikzpicture}[>=latex,decoration={markings, mark=at position 0.6 with {\arrow[ultra thick]{stealth};}} ]
\begin{scope}[node distance=2.5cm]
  \node (Phi0) [draw] at (0,0) {$\Phi_0$};
  \node (Phi1) [right of=Phi0] [draw] {$\Phi_1$};
\end{scope}
   \draw [thick,postaction={decorate},-,>=stealth,shorten <=2pt,shorten >=2pt]  (Phi0.north) -- +(0.3,0.4) -- ++(2.2,0.4) -- (Phi1.north);
   \draw [thick,postaction={decorate},-,>=stealth,shorten <=2pt,shorten >=2pt] (Phi1.south) -- +(-0.3,-0.4) -- ++(-2.2,-0.4) -- (Phi0.south);
  \node (+A) at (1.3,1) {\small{$\Delta_{+A}$}};
  \node (-A) at (1.3,-1) {\small{$\Delta_{-A}$}};
\end{tikzpicture}
\ee
\noindent
as, in particular, $\Delta_\omega \Phi_{n} = 0$ if the Borel transform of $\Phi_{n}$ is \textit{not} singular at $\omega$. This linear operator is known as the \textit{alien derivative} as it is in fact a linear \textit{differential} operator, satisfying the Leibniz rule (see appendix~\ref{app:alien-calculus}). We refer the reader to the reviews \cite{cnp93a, cnp93b, dp99, ss03, s07, s14} and references therein for the proof that $\Delta_\omega$ is a derivation, and subsequently the construction of \textit{alien calculus}.

In our spirit of introducing the main ideas of resurgence in a simple example, let us proceed and address the Stokes automorphism \eqref{stokesauto}. One of the reasons to introduce alien calculus, as sketched above, is that the Stokes automorphism along some direction $\theta$ is essentially given by the exponential of the alien derivative (why this is so will hopefully be motivated by the end result that follows),
\be
\label{Stokes-aut-as exponential-singularities}
\underline{\mathfrak{S}}_\theta = \exp \left( \sum_{\left\{ \omega_\theta \right\}} \rme^{-\frac{\omega_\theta}{x}} \Delta_{\omega_\theta} \right).
\ee
\noindent
Herein, the set $\left\{ \omega_\theta \right\}$ corresponds to all possible Borel singularities along the direction $\theta$ (in the present linear case there is only one such possibility for either $\theta=0$ or $\theta=\pi$, and the summation is immaterial, but in the nonlinear case there may be infinite possibilities and this action will be much more intricate). Now, from \eqref{Delta+A} and \eqref{Delta-A}, it is easy to see that $\Delta_{+A} \Delta_{+A} \Phi_{0} = 0 = \Delta_{-A} \Delta_{-A} \Phi_{1}$, implying that the above exponential collapses to a finite sum for both the $\theta=\arg \left( +A \right)=0$ and $\theta=\arg \left( -A \right)=\pi$ Stokes rays. The only non-trivial relations are thus:
\be
\label{quarticStokes}
\underline{\mathfrak{S}}_0 \Phi_{0} = \exp \left( \rme^{-\frac{A}{x}} \Delta_{+A} \right) \Phi_{0} = \Phi_{0} - 2\, \rme^{-\frac{3}{2x}}\, \Phi_{1} \qquad \text{and} \qquad \underline{\mathfrak{S}}_\pi \Phi_{1} = \Phi_{1} + \rme^{\frac{3}{2x}}\, \Phi_{0},
\ee
\noindent
or, equivalently,
\be
\label{StokesZ0andZ1}
\underline{\mathfrak{S}}_0 Z^{(0)} (x) = Z^{(0)} (x) - 2\, Z^{(1)} (x), \qquad \text{and} \qquad \underline{\mathfrak{S}}_\pi Z^{(1)} (x) = Z^{(1)} (x) + Z^{(0)} (x).
\ee
\noindent
That $\Phi_{0}$ only has one Stokes line along $\theta=0$ on the Borel plane, and $\Phi_{1}$ only one along $\theta=\pi$, is of course already dictated by the alien calculus \eqref{Delta+A} and \eqref{Delta-A} (or, earlier on, by \eqref{borel-quartic-int-expanded} and \eqref{borel-quartic-int-expanded-inst}). The above expressions now completely encode the Stokes phenomena \eqref{stokespheno} across all Stokes lines, where, in this problem, singular directions on the Borel plane equally relate with Stokes lines on the complex $x$-plane. Using \eqref{quartictransseries} one thus finds that the Stokes automorphism \eqref{stokesauto} at $\theta=0$ is
\be\label{crossingthetazero}
\CS_{0^+} \CZ (x, \sigma_0, \sigma_1) = \CS_{0^-} \CZ (x, \sigma_0, \sigma_1-2\sigma_0),
\ee
\noindent
while at $\theta=\pi$ it is
\be\label{crossingthetapi}
\CS_{\pi^+} \CZ (x, \sigma_0, \sigma_1) = \CS_{\pi^-} \CZ (x, \sigma_0+\sigma_1, \sigma_1).
\ee
\noindent
It is interesting to notice that if we move twice around the origin, $\left( \underline{\mathfrak{S}}_\pi \underline{\mathfrak{S}}_0 \right)^2$, the transseries parameters change as $(\sigma_0, \sigma_1) \to (-\sigma_0, -\sigma_1)$, while if we move four times around the origin we recover the identity,
\be\label{quarticmonodromy}
\left( \underline{\mathfrak{S}}_\pi \underline{\mathfrak{S}}_0 \right)^4 = \boldsymbol{1}.
\ee
\noindent
This expression tells us what is the monodromy encoded in the transseries solution to the quartic integral, which is thus best understood as a multi-sheeted Riemann surface\footnote{See \cite{as13} for a discussion on the relation between Stokes transitions and the monodromy of transseries solutions.}.

At this stage, it is instructive to take a step back and re-derive the above monodromy relation \eqref{quarticmonodromy} directly from our discussion of steepest-descent contours in the previous subsection. Recall that we had three saddle-points, $z^*_0$ and $z^*_{\pm}$, which led to three steepest-descent contours for $\Gamma$ in the evaluation of the partition function \eqref{quarticintegralX}, and which we shall now respectively denote by $\CJ_0$ and $\CJ_\pm$ (we shall discuss the nature of such steepest-descent contours in further detail later-on, in section~\ref{sec:lefschetz}). The asymptotic series $\Phi_{0}$ and $\Phi_{1}$, which we computed in \eqref{coeff-quartic} and \eqref{coeff-quartic-inst}, are obviously also generated by the integration \eqref{quarticintegralX} along these contours. Now, what we wish to point out is that these contours will acquire \textit{discontinuities} at the Stokes rays $\theta=0$ and $\pi$, as shown in figure~\ref{steepest}. In fact, in order to instead ensure the continuity of each asymptotic series $\Phi_i$ along these Stokes rays, one needs to make sure that the contour has the same asymptotics for, \textit{e.g.}, $\theta=0^-$ and $\theta=0^+$ (\textit{i.e.}, the tails of the contour lie in the same brown regions of figure~\ref{steepest} for both $\theta=0^-$ and $\theta=0^+$). This \textit{continuity condition} implies that the steepest-descent contours need to change as $\CJ_0 \to \CJ_0 + \CJ_- - \CJ_+$ as one rotates from $\theta=0^-$ to $0^+$ (compare the last and second plots in figure~\ref{steepest}). On the other hand, $\CJ_\pm$ need not acquire any jumps at $\theta=0$. These conditions may be compactly expressed as
\begin{equation}
\left(
\begin{array}{ccc}
\CJ_0 \\
\CJ_+ \\
\CJ_-
\end{array}
\right) \to S_{\circlearrowleft}^{(0)} \left(
\begin{array}{ccc}
\CJ_0 \\
\CJ_+ \\
\CJ_-
\end{array}
\right) = \left(
\begin{array}{ccc}
1 & -1 & 1 \\
0 &  1 & 0 \\
0 &  0 & 1
\end{array}
\right) \left(
\begin{array}{ccc} 
\CJ_0 \\
\CJ_+ \\
\CJ_-
\end{array}
\right),
\end{equation}
\noindent
where $S_{\circlearrowleft}^{(0)}$ is the ``Stokes matrix'' associated to $\theta=0$. Essentially the same reasoning goes through around $\theta=\pi$, where the continuity conditions may now be obtained by comparing the sixth and eight plots in figure~\ref{steepest}. The reader can check that one will now obtain
\begin{equation}
\left(
\begin{array}{ccc}
\CJ_0 \\
\CJ_+ \\
\CJ_-
\end{array}
\right) \to S_{\circlearrowleft}^{(\pi)} \left(
\begin{array}{ccc}
\CJ_0 \\
\CJ_+ \\
\CJ_-
\end{array}
\right) = \left(
\begin{array}{ccc}
 1 & 0 & 0 \\
 1 & 1 & 0 \\
-1 & 0 & 1
\end{array}
\right) \left(
\begin{array}{ccc} 
\CJ_0 \\
\CJ_+ \\
\CJ_-
\end{array}
\right).
\end{equation}
\noindent
If we finally define the \textit{monodromy matrix} $\mathfrak{M} := S_{\circlearrowleft}^{(\pi)} \cdot S_{\circlearrowleft}^{(0)}$ it is simple to check that $\mathfrak{M}^4 = \boldsymbol{1}$, which is the steepest-descent equivalent of the monodromy relation \eqref{quarticmonodromy}.

Having understood the basics of the Stokes automorphism, and how Stokes transitions may be implemented, let us ask what occurs at a Stokes line, say at $\arg x = 0$ (equivalently, along the singular direction $\theta = 0$). For our transseries \eqref{quartictransseries}, one starts off at $\arg x = 0^-$ with $\sigma_1 = 0$ and $\sigma_0 \not = 0$, which we may always normalize to $\sigma_0=1$, \textit{i.e.}, we start off with $\CZ (x, 1, 0)$ which is simply an adequate Borel resummation of the (perturbative) asymptotic expansion \eqref{Z010}. As we cross the Stokes line via \eqref{crossingthetazero} we obtain the transseries
\be
\CZ (x, 1, 2) = \CS_{\theta>0} \Phi_0 (x) + 2\, \rme^{-\frac{3}{2x}}\, \CS_{\theta>0} \Phi_{1} (x).
\ee
\noindent
Increasing $\theta$ until we reach the positive imaginary axis where $x = + \rmi |x|$, this becomes, to leading order,
\be
\CZ (\rmi |x|, 1, 2) \simeq \sqrt{\frac{\hbar}{2\pi}} \left( 1 + \sqrt{2} \sin \left( \frac{3}{2 |x|} \right) - \rmi \sqrt{2} \cos \left( \frac{3}{2 |x|} \right) \right) + \cdots,
\ee
\noindent
yielding the expected asymptotic \textit{oscillatory} behaviour to be found at anti-Stokes lines.

\begin{figure}[ht!]
\begin{center}
\includegraphics[width=6.5cm]{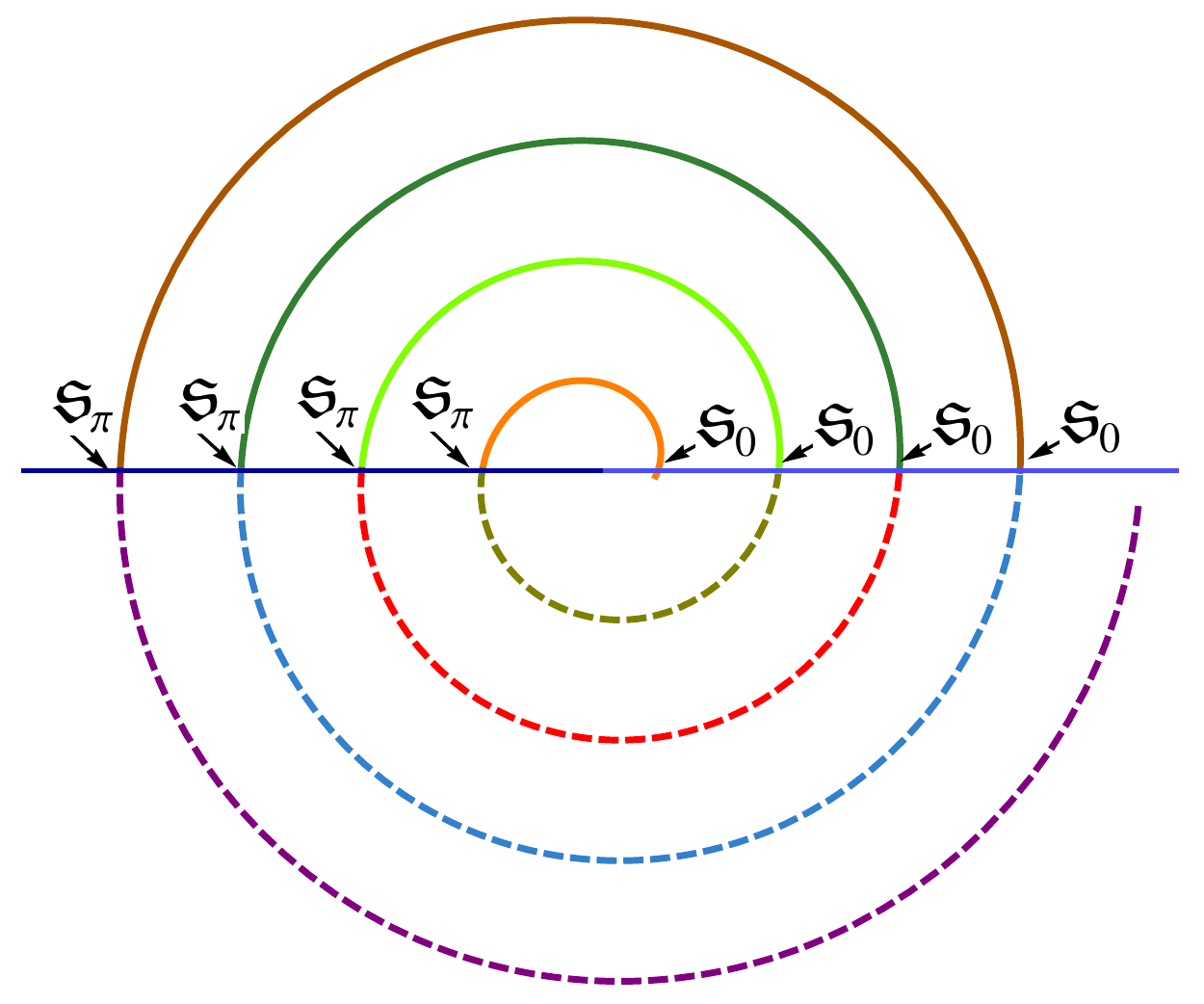}
\break
\break
\includegraphics[width=15.5cm]{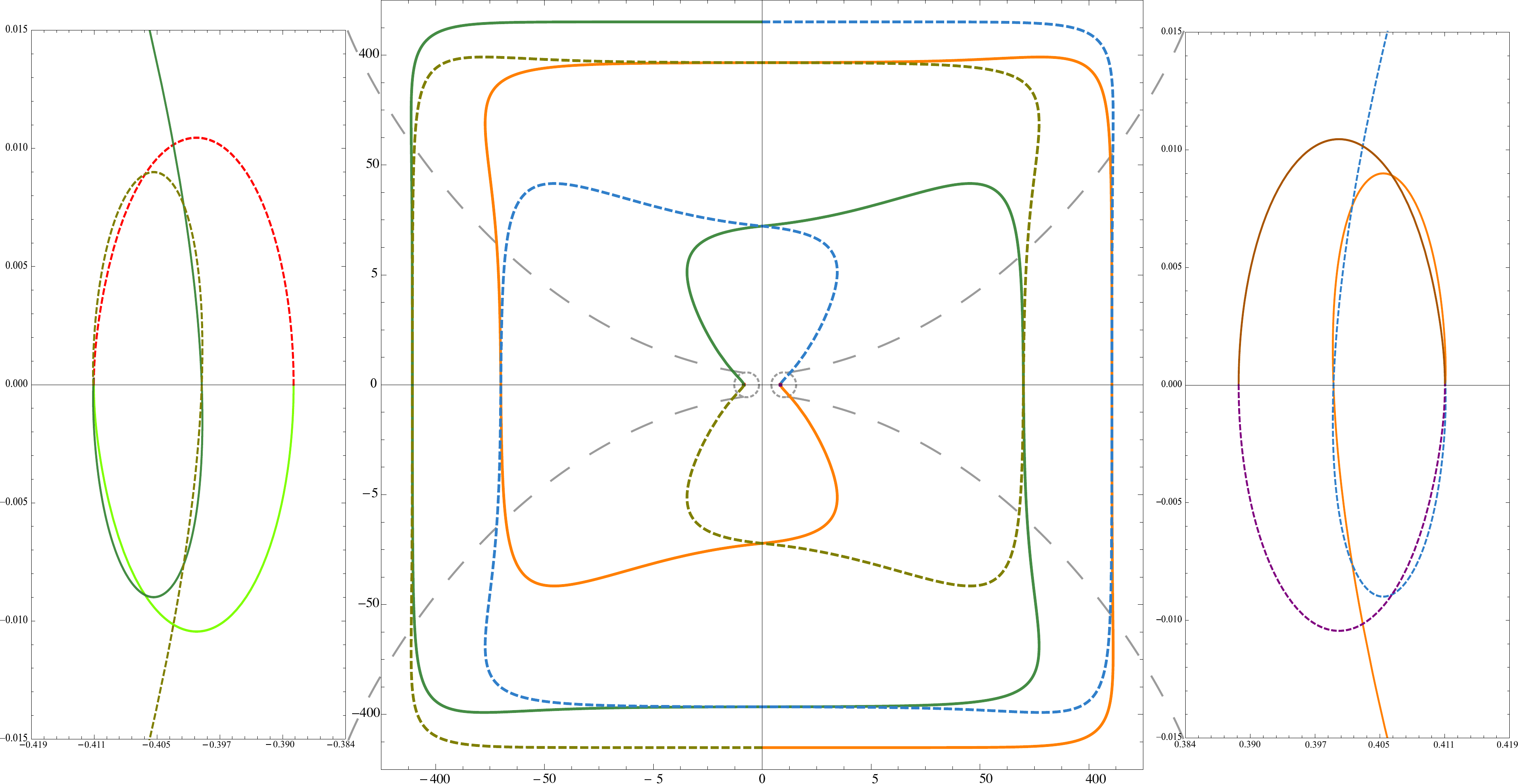}
\end{center}
\caption{The monodromy of the transseries solution to the quartic integral. We start at $\arg x = 0^-$ with $\CZ (0.21\, \rme^{\rmi \theta}, 1, 0)$, and rotate $\theta$ four times around the origin according to \eqref{quarticmonodromy}. This four-fold rotation is \textit{abstractly} represented in the upper plot, where the different colors/types are signaling different regions (on the Borel plane) in-between Stokes transitions. For instance, at $2\pi n$, the Stokes transition \eqref{crossingthetazero} takes place, while at $\pi + 2\pi n$ it is \eqref{crossingthetapi} instead ($n=0,1,2,3$). In this way, each color/type occurs in an interval of length $\pi$ and the complete sequence yields the full $8\pi$ rotation as shown. As we represent this four-fold rotation back into the $x$-plane (illustrated in the lower plots, with matching colors/types as in the upper plot), it is clear how the transseries trajectory closes upon itself, fulfilling the predicted monodromy and showing how indeed the transseries encodes the full nonperturbative solution. The axes in the lower plots correspond to real and imaginary parts of the partition function. A word on numerics: we have resummed the asymptotic series with Pad\'e approximants, and chosen relatively small $|x|$ for improved accuracy. This, however, results in the different scales appearing in the lower plots due to the exponential factor. The central figure is plotted with a logarithmic scale and also gives an (almost complete) overview of the trajectory. The exponential factor makes it hard to understand what happens closer to the origin and, hence, the lateral images zoom-in onto the small ``loops'' which are already visible in the central plot.}
\label{monodromyplot}
\end{figure}

Of course the transseries construction includes much more information beyond finding this oscillatory behaviour. As one keeps rotating in the complex plane, successively crossing Stokes lines, the transseries solution displays its full multi-sheeted structure encoded in \eqref{quarticmonodromy}. Let us explore this particular point. An exact evaluation of the quartic-potential integral around the ``elementary'' contours of figure~\ref{quarticpotentialfig} is possible, with the different results being written in terms of modified Bessel functions of the first kind; concretely as linear combinations of\footnote{One way to obtain this result is via a strong-coupling analysis, as in appendix~\ref{app:strongcoupling}.}
\be
\label{bessel-mentioned}
\sqrt{\frac{3}{4x}}\, \rme^{-\frac{3}{4x}}\, I_{\pm \frac{1}{4}} \left( \frac{3}{4x} \right).
\ee
\noindent
However, all such results have branch cuts along the negative real axis and the question remains on how to explicitly implement the appropriate analytic continuation yielding a fully nonperturbative solution to our problem, in its correct domain of validity. This is one of the points encoded in the Stokes transitions \eqref{crossingthetazero} and \eqref{crossingthetapi}: they already \textit{automatically} yield this analytic continuation with the prescribed monodromy---the only point left to address being an adequate resummation of the different asymptotic series, which is now just a numerical question. To illustrate this point, let us again start off at $\arg x = 0^-$ with $\CZ (x, 1, 0)$, let us fix $|x|$, and let us rotate $\arg x$ around the origin four times, according to the monodromy \eqref{quarticmonodromy}. The result is shown in figure~\ref{monodromyplot}, where we have set $|x|=0.21$, resummed the asymptotic series with (appropriately chosen) Pad\'e approximants, and rotated as described: in this way, via Stokes transitions, the transseries describes a smooth trajectory over the multi-sheeted solution to our problem.

This brief overview of the quartic-potential integral introduced some of the ideas of resurgence in a very simple setting, and also illustrated how resurgent analysis is particularly straightforward for linear systems (\textit{i.e.}, with a \textit{finite} number of instanton sectors and Stokes constants). It is of course important to point out that the story is very different for generic nonlinear systems, with multi-instanton sectors and even resonance, where trying to build the correct analytic continuation as described above is already an extremely non-trivial problem (see \cite{as13} for the full Stokes transitions generalizing \eqref{crossingthetazero} and \eqref{crossingthetapi} and a glimpse at the general monodromy operator \eqref{quarticmonodromy}). The interested reader may find more details on resurgent calculus in appendix~\ref{app:alien-calculus}.

\subsection{Asymptotics and Large-Order Behaviour: Partition Function}\label{subsec:Zasymptotics}

\begin{figure}[t!]
\begin{center}
\includegraphics[width=6cm]{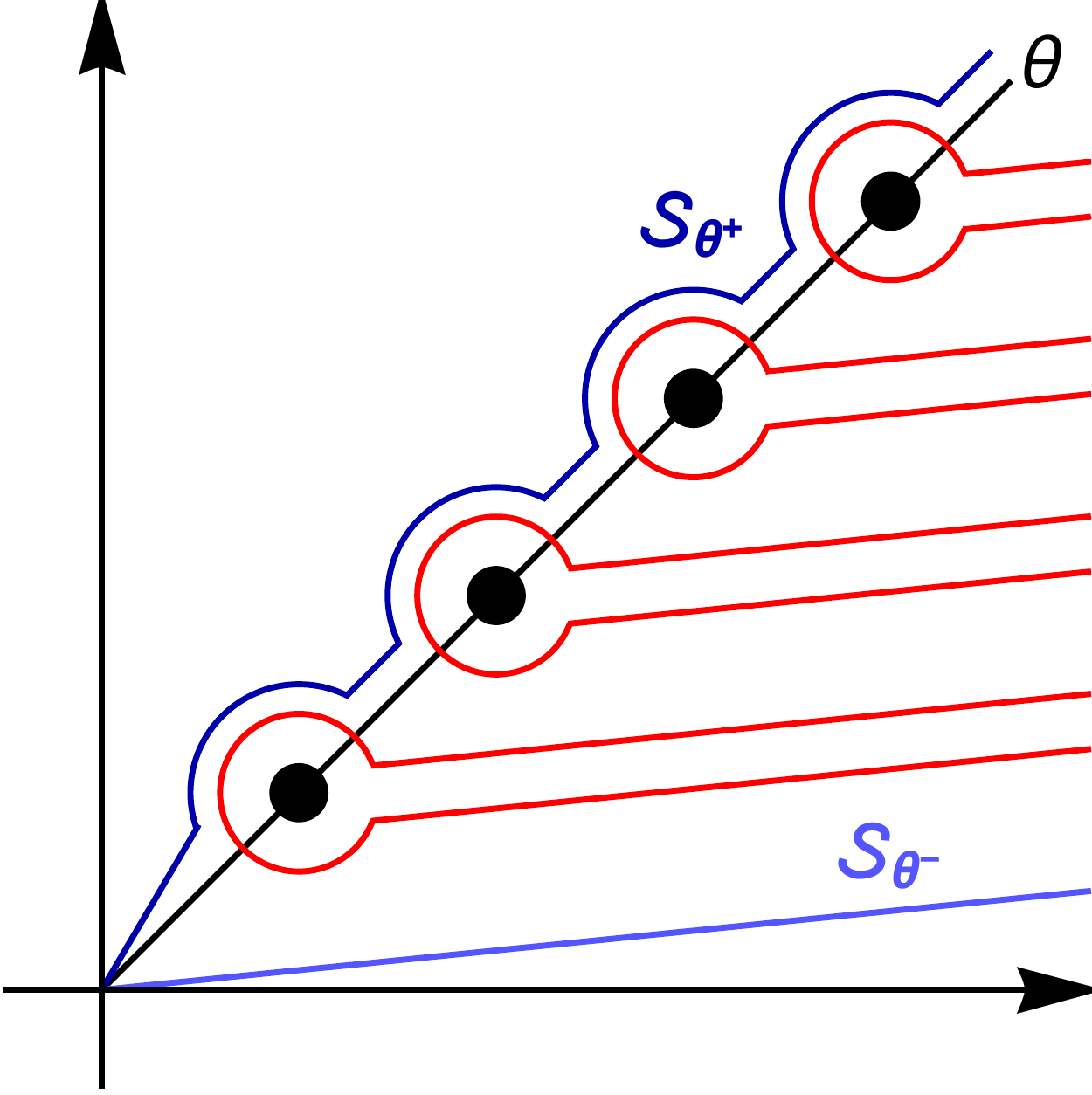}
\end{center}
\caption{The left Borel resummation $\CS_{\theta^+}$ crosses the singular Stokes line along $\theta$, to yield the right Borel resummation $\CS_{\theta^-}$ alongside the Stokes discontinuity $\disc_\theta$. This discontinuity has a geometrical representation as sum over the (red) Hankel contours associated to each (multi-instanton) singularity.}
\label{stokescrossingfig}
\end{figure}

As we already made clear in the introduction, one very important physical outcome of resurgence is the capability it gives us to \textit{completely} understand the asymptotics of some given perturbative expansion. Let us go back to \eqref{stokesauto},
\be
\CS_{\theta^+} = \CS_{\theta^-} \circ \underline{\mathfrak{S}}_\theta,
\ee
\noindent
which quantifies how different asymptotic expansions hold, on different sides of a Stokes line, for the same function. This implies that there is a \textit{discontinuity} of the asymptotic expansions\footnote{So that the nature of this discontinuity is completely clear, let us repeat what we wrote earlier concerning Stokes phenomenon: in an asymptotic approximation, \textit{different} asymptotic formulae hold, on different sectors, for the \textit{same} analytic function. In this way, it is this discontinuity which then relates them to each other.}, $\disc_\theta$, which must relate to the Stokes automorphism as
\be
\label{StokesDisc}
\disc_\theta = \1 - \underline{\mathfrak{S}}_\theta,
\ee
\noindent
so that $\CS_{\theta^+} - \CS_{\theta^-} = - \CS_{\theta^-} \circ \disc_\theta$ (see figure~\ref{stokescrossingfig} for a sketch of this discontinuity).

\begin{figure}[t!]
\begin{center}
\includegraphics[width=10.5cm]{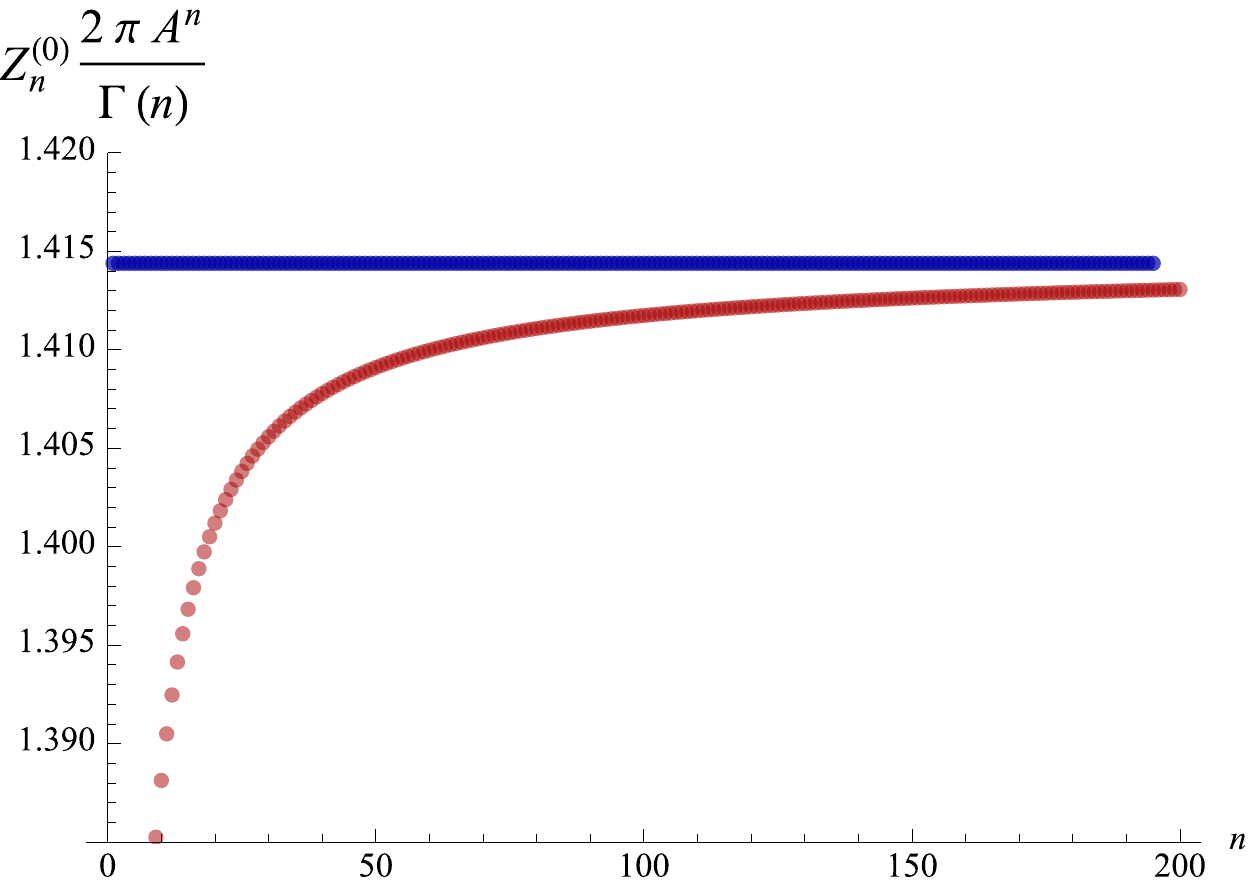}
\end{center}
\caption{Numerical plot of the coefficients $Z_{n}^{(0)}$, weighted by the growth factor $\frac{2\pi A^n}{\Gamma(n)}$ and up to 200 terms. The coefficients of the original series are shown in red, and in blue we plot their corresponding fifth Richardson transform, denoted by $\mathrm{RT}_{0}(0,n,5)$ (details on these numerical acceleration schemes will appear in a following subsection). The value to which the sequence is converging can be read directly from \eqref{Z0largeorder} to be $2\mathrm{i}Z_{0}^{(1)}=\sqrt{2}$ (where in this test we have set $\hbar=2\pi$). The numerics fully validate the (leading) resurgence large-order relation \eqref{Z0largeorder}, up to an extremely small error of order $\sim \CO ( 10^{-12} )$. Similar tests may be done to validate the subleading components of the growth, as predicted by resurgence.}
\label{fig:quarticpartfunc-largeorder}
\end{figure}

Now given a function $F(x)$, with a discontinuity along some ray starting from the origin on the complex plane along the direction $\theta$, Cauchy's theorem translates to 
\be
F(x) = - \frac{1}{2\pi\rmi} \int_{0}^{\rme^{\rmi\theta} \infty} \rmd w\, \frac{\disc_\theta F(w)}{w-x},
\ee
\noindent
at least if there is no contribution around infinity---which is usually the case in the large-order context; \textit{e.g.}, \cite{bw69, bw73, cs78, z81b}. If we apply this relation to the perturbative asymptotic expansion for $\Phi_0 (x)$, given in \eqref{coeff-quartic}, which will in fact have a discontinuity along $\theta=0$ due to the non-triviality of the Stokes automorphism in \eqref{quarticStokes}, it first follows\footnote{Using asymptotic series on both sides of the equality, and changing summation and integration signs, the resulting expression only holds formally and thus the ``asymptotic equality'' sign meaning it holds at large order.}
\be
\sum_{n=0}^{+\infty} Z^{(0)}_n\, x^n \simeq -\frac{1}{2\pi\rmi} \int_{0}^{+ \infty} \rmd w\, \frac{2}{w-x}\, \rme^{-\frac{3}{2w}} \sum_{k=0}^{+\infty} Z^{(1)}_k\, w^k.
\ee
\noindent
This immediately implies\footnote{Note that convergence of the integral requires $n>k$, again signaling the large-order nature of this relation.}
\be
Z^{(0)}_n \simeq -\frac{2}{2\pi\rmi}\, \frac{\Gamma\left(n\right)}{A^n}\, \sum_{k=0}^{+\infty} \frac{\Gamma\left(n-k\right)}{\Gamma\left(n\right)}\, Z^{(1)}_k A^k.
\ee
\noindent
One may regard this expression as a large-order relation (see figure~\ref{fig:quarticpartfunc-largeorder}),
\be
\label{Z0largeorder}
Z^{(0)}_n \simeq -\frac{2}{2\pi\rmi}\, \frac{\left(n-1\right)!}{A^n} \left( Z^{(1)}_0 + \frac{Z^{(1)}_1 A}{n-1} + \frac{Z^{(1)}_2 A^2}{\left( n-1 \right) \left( n-2 \right)} + \cdots \right);
\ee
\noindent
or else make use of the exact coefficients $Z^{(1)}_k$, given in \eqref{coeff-quartic-inst}, to reobtain (without surprise) the exact coefficients \eqref{coeff-quartic}. Indeed, replacing \eqref{coeff-quartic-inst} in the above expression,
\be
Z^{(0)}_n \simeq \frac{1}{\sqrt{2}\pi} \left( \frac{2}{3} \right)^n \sum_{k=0}^{n-1} (-1)^k \frac{(4k)!\, (n-k-1)!}{2^{6k}\, (2k)!\, k!}\, Z^{(0)}_0,
\ee
\noindent
and recalling that it only holds as a large-order expression, \textit{i.e.}, in a regime where $n \gg k$ and where subsequently one has
\be
\lim_{n \to +\infty} \frac{2^{6n}\, (2n)!\, n!}{(4n)!} \left(n-k-1\right)! = \sqrt{2}\pi\, \delta_{k0},
\ee
\noindent
one then directly recovers \eqref{coeff-quartic}. Naturally, an exactly analogue procedure goes through when analyzing the large-order behaviour of the instanton asymptotic expansion \eqref{coeff-quartic-inst} instead.

\subsection{First Steps Towards Nonlinear Resurgent Analysis}\label{subsec:Fnonlinear}

Resurgence only shows its full powers when addressing nonlinear problems, \textit{i.e.}, problems where most popular (nonperturbative) techniques might be of little use. At the same time, in many quantum theoretic settings, it is often not the partition function which is most amenable to calculations, but rather the \textit{free energy}. With this dual motivation in mind, let us now address the ``quartic free energy'', $F = \log Z$. Unlike the quartic-potential partition function in \eqref{quarticODE}, this time around the quartic-potential free energy satisfies a \textit{nonlinear} second-order ODE,
\be
\label{quarticNLODE}
16 x^2\, F'' (x) + 16 x^2 \left( F'(x) \right)^2 + \left( 32 x - 24 \right) F' (x) + 3 = 0.
\ee
\noindent
An asymptotic perturbative solution $F (x) \simeq \sum_{n=0}^{+\infty} F_n^{(0)} x^n$ yields the recursion (determining it)
\be\label{quarticFrecursion}
F_{n+1}^{(0)} = \frac{2}{3} \left( n\, F_n^{(0)} + \frac{1}{n+1} \sum_{k=1}^{n-1} k \left(n-k\right) F_k^{(0)}\, F_{n-k}^{(0)} \right),
\ee
\noindent
with $F_1^{(0)} = \frac{1}{8}$ and $F_0^{(0)}$ an integration constant. Imagine we knew nothing about $F$ (although we can know a lot, as we have already analyzed the partition function thoroughly!), and let us thus proceed with a large-order analysis of this recursion\footnote{For the reader who is less familiar with such techniques, we shall discuss in detail such type of analyses in a following subsection. For the moment, take this in the spirit of what was just done in the previous subsection.}. To very high numerical precision (for instance, using Richardson transforms---which we will describe in detail in a following subsection) it is possible to show that the leading large-order behaviour of the perturbative coefficients in the quartic free energy is 
\be
\label{quarticFdivergesn!}
F^{(0)}_n \sim \frac{S_1 F^{(1)}_0}{2\pi\rmi}\, \frac{\Gamma \left( n \right)}{A^n} + \cdots,
\ee
\noindent
with\footnote{Where $S_1 F^{(1)}_0$ is the ``scale invariant'' combination carrying physical information; see \cite{asv11}.}
\be
\label{freeen-initial-large-order}
A \approx \frac{3}{2} \qquad \text{and} \qquad S_1 F^{(1)}_0 \approx \rmi \sqrt{2}.
\ee

This behaviour points to the existence of instanton corrections, with action $A=\frac{3}{2}$. Now, in the study of the partition function, we saw that we only needed two sectors: the perturbative expansion and a single nonperturbative instanton sector, both solutions to a linear differential equation. On the other hand, in the present case the free energy obeys a \textit{nonlinear} differential equation, and we thus expect to have a \textit{tower} of multi-instanton sectors alongside the perturbative series\footnote{This is intuitively obvious. When searching for solutions, the nonlinear term in \eqref{quarticNLODE} essentially grabs any (non-analytic) exponential contribution and iteratively induces all possible powers of this exponential. In this way, a solution \textit{ansatz} involving a single-instanton must then necessarily involve all possible multi-instanton contributions.}. Let us try to derive the above ``experimental'' results by considering an \textit{ansatz} for the solution to the quartic free-energy equation \eqref{quarticNLODE}, of the form
\be\label{oneparametertransseriesquartic}
F (x) \simeq \rme^{-\frac{\ell A}{x}}\, x^{\ell \beta}\, \sum_{n=0}^{+\infty} F_n^{(\ell)}\, x^n,
\ee
\noindent
where, besides the multi-loop multi-instanton coefficients $F_n^{(\ell)}$, we also have to determine the instanton action $A$ and a characteristic exponent $\beta$. Plugging this \textit{ansatz} back into \eqref{quarticNLODE}, one finds that a solution exists if one chooses
\be
\label{freeen-initial-transansatz}
A = 0,\, \frac{3}{2}, \qquad \beta = 0.
\ee
\noindent
We find two possible values for the instanton action, but only one non-zero. The first reasonable guess in order to compute the free energy is then to assume that the above zero instanton action is in fact part of the perturbative sector, and to expect to describe the complete (and most general!) solution to the free energy of the quartic-potential as the following one-parameter transseries
\be
\label{one-param-transseries}
F (x, \sigma) = \sum_{\ell=0}^{+\infty} \sigma^\ell\, \rme^{-\frac{\ell A}{x}}\, \Phi_{\ell} (x).
\ee
\noindent
Here $\sigma$ is the transseries instanton-counting parameter (an integration constant parameterizing boundary conditions), and the $\Phi_{\ell} (x)$ are the transseries building blocks, \textit{i.e.}, the asymptotic expansions associated to each (multi) instanton sector,
\be
\label{one-param-inst-asympt}
\Phi_{\ell} (x) \simeq \sum_{n=0}^{+\infty} F_n^{(\ell)}\, x^n.
\ee
\noindent
Just like for the perturbative sector in \eqref{quarticFrecursion}, the coefficients of these instanton expansions also obey recursive relations, obtained by keeping track of equal powers of $\rme^{-\frac{\ell A}{x}}$ and of $x^n$ when substituting the full transseries \textit{ansatz}, \eqref{one-param-transseries} and \eqref{one-param-inst-asympt}, into the differential equation \eqref{quarticNLODE}. These recursive relations are now more intricate\footnote{In particular, they keep on demanding for the need of extra (higher) multi-instanton sectors, coupled to the previous ones (and eventually to the perturbative series), in such a way that in the end one indeed needs for an \textit{infinite} number of multi-instanton sectors as initially put forward in the general transseries \textit{ansatz} \eqref{one-param-transseries}.} than \eqref{quarticFrecursion}, and they may be found in appendix~\ref{app:recursive-free-en}. From these recursive relations one can determine enough coefficients in the expansions of any (multi) instanton sector to find that the expansions $\Phi_{\ell}$ are indeed asymptotic, as they diverge factorially\footnote{The simplest way to find this factorial growth is to plot $\frac{F_{n}^{(\ell)}}{F_{n+1}^{(\ell)}}$, and notice that this quotient converges to \textit{zero}. Had it converged to any other value than zero, one would need to extend the \textit{ansatz} used, beyond the one-parameter transseries, and this will be thoroughly discussed below.}
\be
\label{quarticmultiFdivergesn!}
F_n^{(\ell)} \sim \frac{\Gamma(n)}{A^n}.
\ee

We shall now illustrate how resurgence extends to this \textit{one-parameter transseries} solution of a \textit{nonlinear} differential equation. Akin to what was done in earlier sections, within the linear problem, one starts by addressing the Borel transforms of each multi-instanton sector $\Phi_{\ell}$,
\be
\CB [\Phi_{\ell}] (s) = \sum_{n=1}^{+\infty} \frac{F_n^{(\ell)}}{\Gamma(n)}\, s^{n-1}.
\ee
\noindent
Of course one immediately stumbles upon a major difference as compared to the linear case: now, one no longer has closed-form expressions for the multi-loop multi-instanton coefficients, $F_n^{(\ell)}$, and so it is no longer possible to analytically compute the Borel transforms (as was done in the linear problem in \eqref{borel-quartic-int} and \eqref{borel-quartic-int-inst}). Recall that such calculations led us to understand the singularity structure of the Borel transforms, \eqref{borel-quartic-int-expanded} and \eqref{borel-quartic-int-expanded-inst}, which opened the door to the introduction of the resurgence framework in \eqref{Delta+A} and \eqref{Delta-A}. Nevertheless, given their recursive construction, it is straightforward to computationally calculate several of these coefficients---for any sector $\Phi_{(\ell)}$ of the one-parameter transseries---and one can then numerically approximate the analytic function we are interested in, \textit{e.g.}, via Pad\'{e} approximants (see, \textit{e.g.}, \cite{bo78}, for a classical treatment, or, \textit{e.g.}, \cite{as17}, for very recent applications; but more on this below). Being applied to the computation of Borel transforms, $\CB [\Phi_n]$, we shall denote this approximation scheme as the Borel--Pad\'{e} approximation method, written as $\mathrm{BP}_{\ell} [\Phi_n]$ for a (diagonal) order-$\ell$ Pad\'e approximant. For the moment all one needs to know is that this method approximates the Borel transform by a rational function---in fact a ratio of two polynomials---whose singularity structure is then encoded in its poles (immediately computed as the zeroes of the polynomial in the denominator). As it turns out, in the case of resurgent functions these poles condense to form cuts which start at the positions of the instanton actions on the Borel plane. 

\begin{figure}[t!]
\begin{center}
\includegraphics[width=7.4cm]{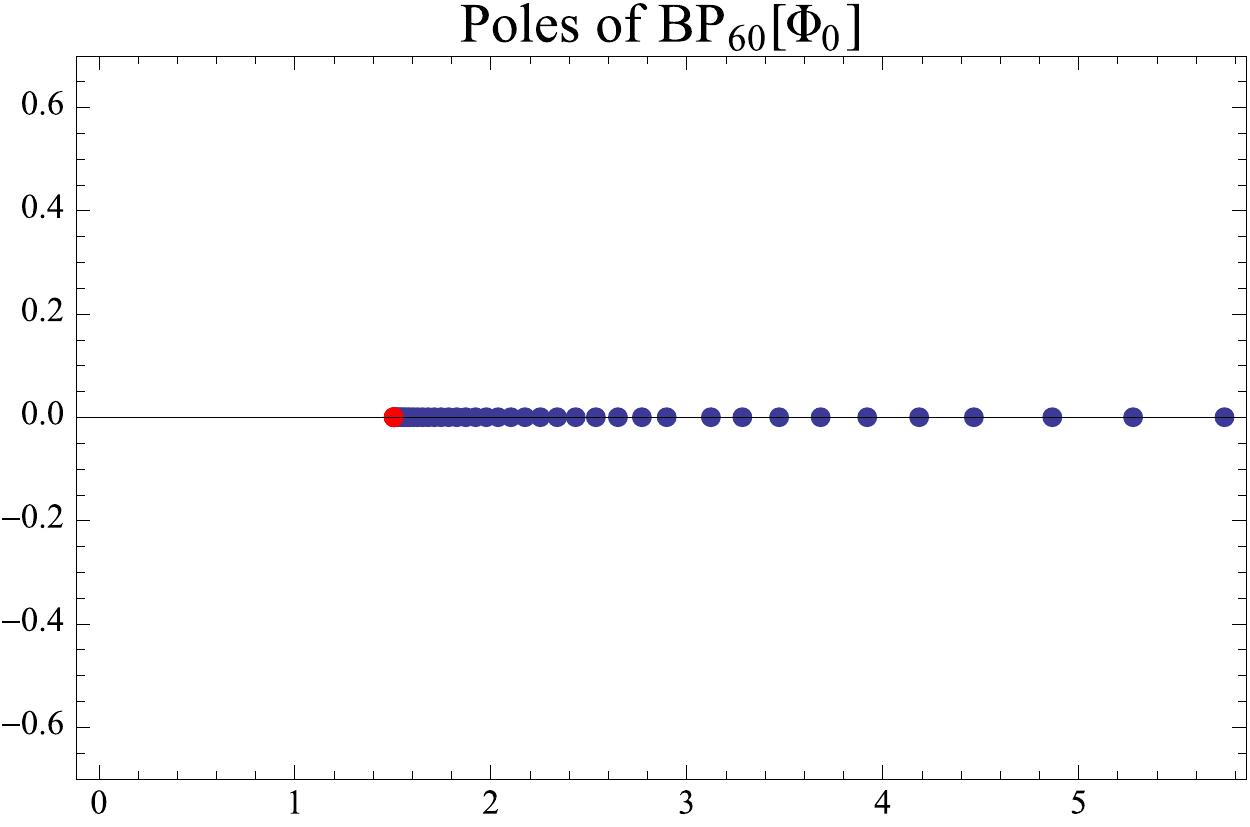}
$\qquad$
\includegraphics[width=7.4cm]{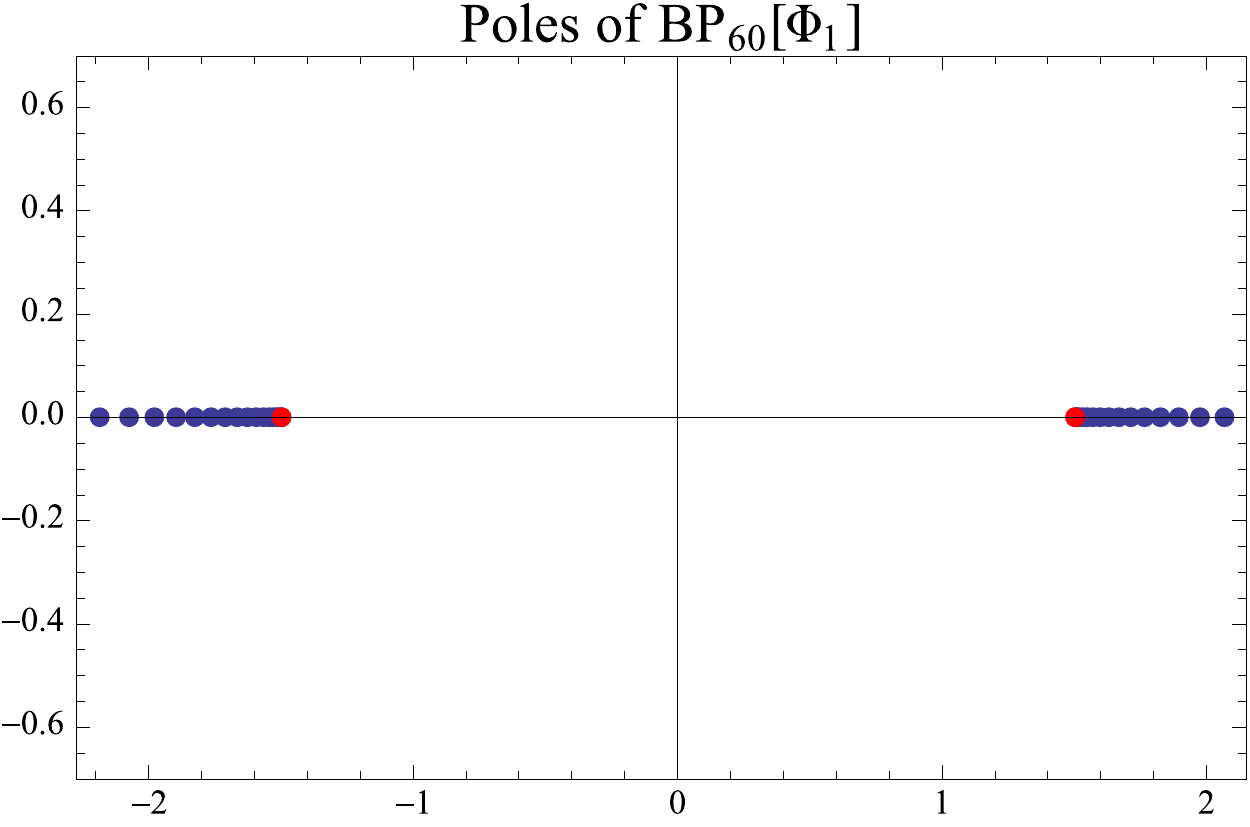}
\end{center}
\caption{Poles of the (diagonal) order-60 Pad\'e approximants for the Borel transforms of perturbative ($\mathrm{BP}_{60} [\Phi_0]$) and one-instanton ($\mathrm{BP}_{60} [\Phi_1]$) sectors in the free energy of the quartic potential. We have truncated the series $\Phi_0$ and $\Phi_1$ at order $N=120$, and used the built-in \textit{Mathematica} command \texttt{PadeApproximant} at order 60 (typically the order of the approximant is half of the cut-off order, $N$, for a diagonal Pad\'e approximant). One can easily see that the poles are all on the positive real line; for the perturbative expansion $\CB [\Phi_0]$ condensing upon a branch-cut starting at $A=\frac{3}{2}$ (shown in red); and for the one-instanton series $\CB [\Phi_1]$ condensing upon two separate branch-cuts, starting at either $\pm A$ (also in red).}
\label{fig:Pade-poles-quartic-pert-series}
\end{figure}

As an example, we show in figure~\ref{fig:Pade-poles-quartic-pert-series} the poles of the Pad\'e approximants for the Borel transforms of perturbative, $\Phi_0$, and one-instanton, $\Phi_1$, asymptotic series in the quartic free energy. It is rather evident how these poles condense into either one or two branch-cuts, starting at the values of the instanton actions on the Borel plane, $s=\pm \frac{3}{2}$. The respective Borel transforms thus have a non-zero radius of convergence around $s=0$. Using the recursions in appendix~\ref{app:recursive-free-en} it is straightforward to plot further  examples and to study numerically the nature of the resulting singularities and branch-cuts, which we leave as an exercise for the reader.

One very important feature of resurgence is its \textit{universality}: it is a rather general property of arbitrary complex functions. It is thus quite reasonable to expect that the nature of the Borel singularities associated to each multi-instanton sector of our one-parameter transseries \eqref{one-param-transseries} ought to be of similar nature to what was found in the linear problem, \eqref{borel-quartic-int-expanded} and \eqref{borel-quartic-int-expanded-inst}, and we shall take this as granted in the following. The novelty of the nonlinear problem will amount to an \textit{infinite} number of (multi) instanton sectors, implying that the Borel transforms may generically have \textit{many more} singularities than it was the case in the linear problem. Indeed, and to make a long-story short, it turns out that for the present one-parameter transseries, at a given fixed sector with instanton number $n$, the Borel transform will have singularities at $s = k A$, with $k \neq 0$ and $k \ge -n$ (the transseries does not include any sectors $\Phi_n$ with $n < 0$). Up to holomorphic content, the expansions near these singularities are of the form\footnote{One could also easily include a simple pole in this expression; the discussion that follows would remain unchanged. But, as seen before, the role of the simple poles is solely to carry through with the residual coefficients and, as such, they are a bit secondary to the main analysis. We leave it to the reader to include them as fit.} \cite{e81}
\be
\label{Borel-transf-expanded-one-param}
\CB [\Phi_n] (s) \Big|_{s=kA} = \mathsf{S}_{n\to n+k} \times \CB [\Phi_{n+k}] \left( s-kA\right)\, \frac{\log \left( s-kA \right)}{2\pi\rmi}, \qquad k\neq0.
\ee
\noindent
Near the singularity $s=kA$ of $\CB [\Phi_n] (s)$ one finds the \textit{resurgence} of the $\Phi_{n+k}$ sector, through its corresponding Borel transform $\CB [\Phi_{n+k}] (s)$. These singularities are logarithmic\footnote{This is not necessarily always the case. However, as we shall later discuss in section~\ref{sec:borel}, generically there is an adequate ``representative'' of the asymptotic series where indeed the branch cuts \textit{are} logarithmic, as in the equation above (this is known as the \textit{simple} representative). The main interest of using the representative with logarithms is that the properties of resurgence are most easily read using the logarithms branch points, as opposed to other types of singularities (but, again, this will be discussed in detail in section~\ref{sec:borel}).} branch-cuts (for \textit{simple} resurgent functions), while the Borel transform of the ``reappearing'' sector is being evaluated around the origin---thus being locally regular. One also finds a proportionality constant $\mathsf{S}_{n\to n+k}$, the \textit{``Borel residue''}, numerically characterizing each distinct logarithmic singularity.

\begin{figure}[t!]
\begin{center}
\includegraphics[width=6cm]{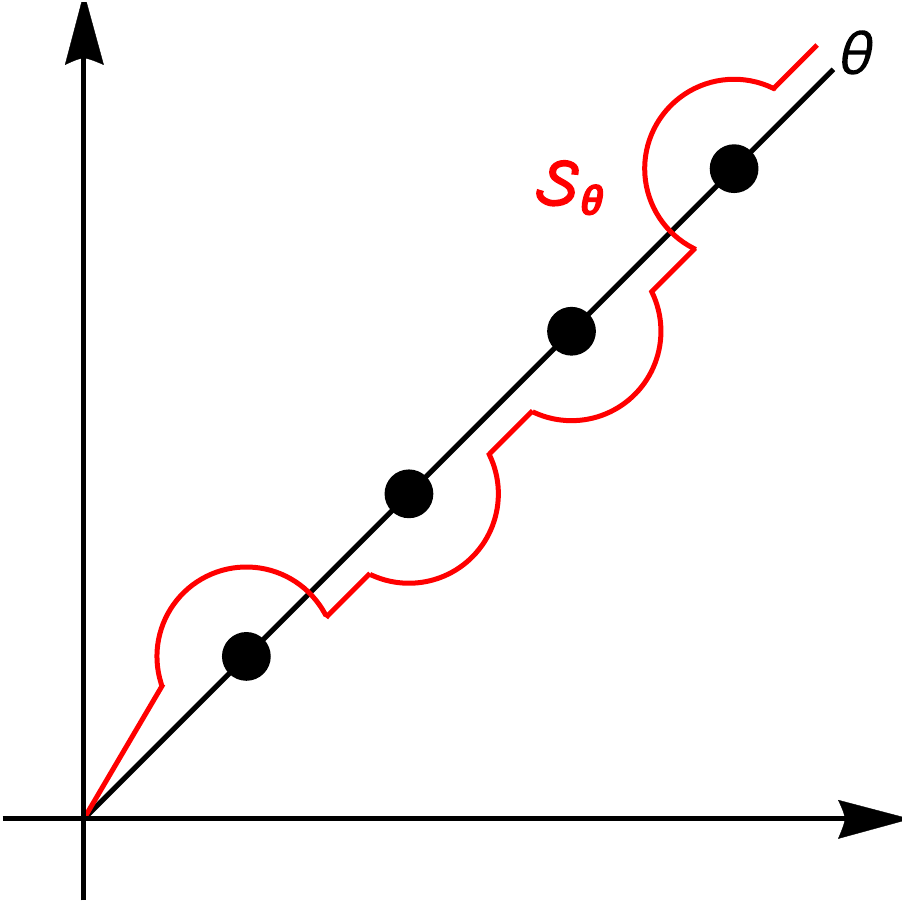}
\end{center}
\caption{One of the many convoluted paths that we could choose for resummation, when in the presence of multiple instanton singularities. The alien derivative is a weighted-average over all these possible paths.}
\label{stokesweirdcrossingfig}
\end{figure}

Being a very general statement about complex functions, and as we have already discussed in the linear case, resurgence is best expressed at an algebraic level of abstraction. In order to uncover algebraic properties behind the singularity structure encoded in \eqref{Borel-transf-expanded-one-param}, let us thus move beyond the functional nature of the Borel transforms. As discussed earlier, this is possible via the use of a linear differential operator, the \textit{alien derivative} $\Delta_{\omega}$, which directly express any relations in-between different sectors in the transseries. However, unfortunately, the simple definition of alien derivative that we used in the linear case, \eqref{def-linear-alien}, no longer works in the present nonlinear scenario. The reason why is simple. Indeed, while in the linear case there were only two possible distinct resummations to consider in \eqref{def-linear-alien}, in the current nonlinear situation we will have many more due to the proliferation of Borel singularities. Of course that the original paths for lateral Borel resummations in \eqref{def-linear-alien} are still present, as in, \textit{e.g.}, figure~\ref{stokescrossingfig}; but now many others (and some quite convoluted) may be considered---see figure~\ref{stokesweirdcrossingfig} for one such example. In the nonlinear case, the definition of alien derivative \eqref{def-linear-alien} must thus be \textit{extended} to accommodate all these new possibilities: its definition \cite{e81} turns out to be an adequate, weighted-average over all these possible paths---but it is also too technical for the introductory and pedagogical goals of these lectures (nonetheless, see appendix~\ref{app:alien-calculus} for further details and references). Note, however, how this caveat is also a blessing: one reason why the alien derivative is such a powerful concept is because it includes, inside it, all possible paths of resummation.

The one remarkable result in alien calculus is that, in spite of its intricate definition, the final result for the alien derivative is surprisingly simple. To some extent, $\Delta_{\omega}$ acts on the transseries building blocks, the asymptotic series $\Phi_n$, simply according to what is encoded in \eqref{Borel-transf-expanded-one-param}: it is only non-zero if the corresponding Borel transform has a singularity at $\omega$, in which case its action is schematically given by
\be
\Delta_{k A} \Phi_n \propto \Phi_{n+k}.
\ee
\noindent
For any other value of $\omega \in \mathbb{C}$, $\Delta_{\omega} \Phi_n=0$ (\textit{e.g.}, as aforementioned the transseries does not include any sectors $\Phi_n$ with $n<0$, in which case if $n+k<0$ then $\Delta_{kA}\Phi_n=0$). At this stage, we are still lacking the proportionality coefficients, \textit{i.e.}, the \textit{Stokes constants}. In the linear case, \eqref{Delta+A} and \eqref{Delta-A}, they matched the Borel residues up to an overall sign (see \eqref{disc-to-alien-Phi0} and \eqref{disc-to-alien-Phi1}). But this can no longer hold true in the nonlinear setting. In fact, due to the intertwining paths of resummation, Borel residues and Stokes constants, albeit obviously related, will not be equal. For the moment, we shall conveniently parameterize the Stokes constants $S_k$ as\footnote{At this stage, the reader may want to compare back to \eqref{Delta+A} and \eqref{Delta-A} in the linear case.}:
\be
\label{Delta-kA-one-param}
\Delta_{k A} \Phi_n = S_{k} \left(n+k\right) \Phi_{n+k}, \qquad k \le 1 \text{  and  } k \neq 0.
\ee
\noindent
These \textit{resurgence relations} follow directly from the so-called \textit{bridge equations} of \cite{e81}, and that is one way to derive them (in fact, contrary to what we have done here, they are usually the starting point in the discussions of resurgence). These bridge equations, which we describe at greater length in appendix~\ref{app:alien-calculus}, are also behind the reason why\footnote{The reason being intimately related to the one-parameter \textit{ansatz} for the transseries \eqref{one-param-transseries}.} alien-derivative singularities truncate to $k \le 1$ in \eqref{Delta-kA-one-param} (while Borel singularities did not in \eqref{Borel-transf-expanded-one-param}). 

The resulting structure in \eqref{Delta-kA-one-param} is clearly the same as in the Borel singularities \eqref{Borel-transf-expanded-one-param}, but the proportionality constants $\mathsf{S}_{n\to n+k}$ and $S_k$ must still be related to each other. Hopefully, an intuition for the relations expressed in the following formulae will be developed by the reader as we discuss ``alien chains'' and Stokes automorphisms in the next couple of pages. Let us display some of these relations. When moving forward in the ``chain'' of singularities, one finds for example
\bea
\label{stokesS-to-borelS+1}
\mathsf{S}_{n\to n+1} &=& - \left(n+1\right) S_1, \\
\label{stokesS-to-borelS+2}
\mathsf{S}_{n\to n+2} &=& - \frac{1}{2} \left(n+1\right) \left(n+2\right) S_1^2, \\
\label{stokesS-to-borelS+3}
\mathsf{S}_{n\to n+3} &=& - \frac{1}{6} \left(n+1\right) \left(n+2\right) \left(n+3\right) S_1^3,
\eea
\noindent
to which corresponds the closed-form expression\footnote{While this expression could be rewritten using a binomial coefficient, the present form is more natural in light of the upcoming vectorial generalization for multi-parameter transseries (to be discussed in section~\ref{sec:physics}).}:
\be
\label{stokesS-to-borelS+GEN}
\mathsf{S}_{n\to n+k} = - \frac{1}{k!}\, \frac{\left(n+k\right)!}{n!}\, S_1^k.
\ee
\noindent
These formulae are also simply invertible, allowing to write the ``more abstract'' Stokes constant $S_1$ as a function of the ``more numerical'' Borel residue $\mathsf{S}_{n\to n+1}$. When, instead, moving backwards along the ``chain'' of singularities, one finds for example
\bea
\label{stokesS-to-borelS-1}
\mathsf{S}_{n\to n-1} &=& -\left(n-1\right) S_{-1}, \\
\label{stokesS-to-borelS-2}
\mathsf{S}_{n\to n-2} &=& -\left(n-2\right) \left( S_{-2} + \frac{1}{2} \left(n-1\right) S_{-1}^2 \right), \\
\label{stokesS-to-borelS-3}
\mathsf{S}_{n\to n-3} &=& -\left(n-3\right) \left( S_{-3} + \frac{1}{2} \left(2n-3\right) S_{-1} S_{-2} + \frac{1}{6} \left(n-2\right) \left(n-1\right) S_{-1}^3 \right),
\eea
\noindent
to which corresponds the closed-form expression\footnote{This expression could be rewritten using sums over partitions (many closed-form formulae of this type may be found in \cite{asv11, as13}). However, in light of the upcoming vectorial generalization for multi-parameter transseries (to be discussed in section~\ref{sec:physics}), it is more natural not to embark in such discussion at this stage.}
\be
\label{stokesS-to-borelS-GEN}
\mathsf{S}_{n\to n-k} = - \sum_{s=1}^{k} \frac{1}{s!}\, \sum_{\substack{\left\{ \ell_1,\ldots,\ell_s\ge 1 \right\} \\ \vphantom{\frac{1}{2}} \sum \ell_i=k}}\, \prod_{a=1}^{s} \left( n - \sum_{j=1}^{a} \ell_j \right)\, \prod_{b=1}^{s} S_{-\ell_b}, \qquad  1\le k < n.
\ee
\noindent
Although more intricate, also in this case inverse formulae are available; allowing to write the ``more abstract'' Stokes constants as functions of the ``more numerical'' Borel residues. For example, inverting equations \eqref{stokesS-to-borelS-1} through \eqref{stokesS-to-borelS-3} yields
\bea
\label{borelS-to-stokesS-1}
S_{-1} &=& - \frac{1}{n-1}\, \mathsf{S}_{n\to n-1}, \\
\label{borelS-to-stokesS-2}
S_{-2} &=& - \frac{1}{n-2} \left( \mathsf{S}_{n\to n-2} + \frac{1}{2}\, \mathsf{S}_{n\to n-1}\, \mathsf{S}_{n-1\to n-2} \right), \\
\label{borelS-to-stokesS-3}
S_{-3} &=& - \frac{1}{n-3} \left( \mathsf{S}_{n\to n-3} + \frac{1}{2}\, \mathsf{S}_{n\to n-1}\, \mathsf{S}_{n-1\to n-3} + \frac{1}{2}\, \mathsf{S}_{n\to n-2}\, \mathsf{S}_{n-2\to n-3} + \right. \nonumber \\
&&
\left. + \frac{1}{3}\, \mathsf{S}_{n\to n-1}\, \mathsf{S}_{n-1\to n-2}\, \mathsf{S}_{n-2\to n-3} \right),
\eea
\noindent
to which corresponds the closed-form expression:
\be
\label{borelS-to-stokesS-GEN}
S_{-k} = - \frac{1}{n-k}\, \sum_{s=1}^{k} \frac{1}{s}\, \sum_{\substack{\left\{ \ell_1,\ldots,\ell_s\ge 1 \right\} \\ \vphantom{\frac{1}{2}} \sum \ell_i=k}} \mathsf{S}_{n\to n-\ell_1}\, \mathsf{S}_{n-\ell_1\to n-\ell_1-\ell_2}\, \cdots\, \mathsf{S}_{n-k+\ell_s\to n-k}, \qquad 1\le k < n.
\ee

Note how the Borel residues $\mathsf{S}_{n\to n+k}$ depend on ``departure'' and ``arrival'' sectors, but how the Stokes constants $S_k$ only depend on the specific singularity being probed on the complex Borel plane. Furthermore, because $\Delta_{-n A} \Phi_n = 0$, one can show that
\be
\mathsf{S}_{n\to 0} = 0.
\ee
\noindent
Finally, the truncation $k \le 1$ in \eqref{Delta-kA-one-param}, or, equivalently, $\Delta_{k A} \Phi_n = 0$ for $k \ge 2$, which does not occur in \eqref{Borel-transf-expanded-one-param}, implies that the Borel residues must satisfy the following consistency condition:
\be
\label{borelS-consistency}
\mathsf{S}_{n\to n+k} = \frac{(-1)^{k-1}}{k!}\, \prod_{\ell=0}^{k-1} \mathsf{S}_{n+\ell\to n+\ell+1}.
\ee
\noindent
In the following we shall mostly work with the Stokes constants (attached to the specific $s = kA$ singularity, regardless of ``departure'' $\Phi_n$ and ``arrival'' $\Phi_{n+k}$ sectors), but sometimes the Borel residues turn out to be more useful, simplifying some formulae.

\begin{figure}[t!]
\begin{center}
\begin{tikzpicture}[>=latex,decoration={
markings, mark=at position 0.6 with {\arrow[ultra thick]{stealth};}} ]
\begin{scope}[node distance=2.5cm]
  \node (Phi0) [draw] at (0,0) {$\Phi_0$};
  \node (Phi1) [right of=Phi0] [draw] {$\Phi_1$};
  \node (Phi2) [right of=Phi1] [draw] {$\Phi_2$};
  \node (Phi3) [right of=Phi2] [draw] {$\Phi_3$};
\end{scope}
\begin{scope}[node distance=3.5cm]
  \node (Phidots) [right of=Phi3] [draw] {$\Phi_k$};
  \node (Phin) [right of=Phidots] [draw] {$\Phi_n$};
\end{scope}
  \draw [thick,postaction={decorate},-,>=stealth,shorten <=2pt,shorten >=2pt] (Phi0.north) -- +(0.3,0.4) -- ++(2.2,0.4) -- (Phi1.north);
  \draw [thick,postaction={decorate},-,>=stealth,shorten <=2pt,shorten >=2pt] (Phi1.north) -- +(0.3,0.4) -- ++(2.2,0.4) -- (Phi2.north);
  \draw [thick,postaction={decorate},-,>=stealth,shorten <=2pt,shorten >=2pt] (Phi2.north) -- +(0.3,0.4) -- ++(2.2,0.4) -- (Phi3.north);
  \draw [thick,->,>=stealth,shorten <=2pt,shorten >=2pt] (Phidots.north) -- +(0.3,0.4) -- ++(0.6,0.4) -- ++(0.3,-0.4);
  \draw [thick,->,>=stealth,shorten <=2pt,shorten >=2pt] (Phidots.north)+(0.9,0) -- ++(1.1,0.4) -- ++(0.3,0) -- ++(0.3,-0.4);
  \draw [thick,->,>=stealth,shorten <=2pt,shorten >=2pt] (Phidots.north)+(1.7,0) -- ++(2.,0.4) -- ++(0.3,-0) -- ++(0.3,-0.4);
  \draw [thick,<-,>=stealth,shorten <=2pt,shorten >=2pt] (Phin.north) -- +(-0.3,0.4) -- ++(-0.6,0.4) -- ++(-0.3,-0.4);  
  \draw [thick,->,>=stealth,shorten <=2pt,shorten >=2pt] (Phi3.north) -- +(0.3,0.4) -- ++(0.6,0.4) -- ++(0.3,-0.4);
  \draw [thick,->,>=stealth,shorten <=2pt,shorten >=2pt] (Phi3.north)+(0.9,0) -- ++(1.1,0.4) -- ++(0.3,0) -- ++(0.3,-0.4);
  \draw [thick,->,>=stealth,shorten <=2pt,shorten >=2pt] (Phi3.north)+(1.7,0) -- ++(2.,0.4) -- ++(0.3,-0) -- ++(0.3,-0.4);
  \draw [thick,<-,>=stealth,shorten <=2pt,shorten >=2pt] (Phidots.north) -- +(-0.3,0.4) -- ++(-0.6,0.4) -- ++(-0.3,-0.4);
  \draw [thick,postaction={decorate},-,>=stealth,shorten <=2pt,shorten >=2pt] (Phi1.south) -- +(-0.3,-0.4) -- ++(-2.2,-0.4) -- (Phi0.south);
  \draw [thick,postaction={decorate},-,>=stealth,shorten <=2pt,shorten >=2pt] (Phi2.south) -- +(-0.3,-0.4) -- ++(-2.2,-0.4) -- (Phi1.south);
  \draw [thick,postaction={decorate},-,>=stealth,shorten <=2pt,shorten >=2pt] (Phi3.south) -- +(-0.3,-0.4) -- ++(-2.2,-0.4) -- (Phi2.south);
  \draw [thick,->,>=stealth,shorten <=2pt,shorten >=2pt] (Phidots.south) -- +(-0.3,-0.4) -- ++(-0.6,-0.4) -- ++(-0.3,0.4);
  \draw [thick,->,>=stealth,shorten <=2pt,shorten >=2pt] (Phidots.south)+(-0.9,0) -- ++(-1.1,-0.4) -- ++(-0.3,-0) -- ++(-0.3,0.4);
  \draw [thick,->,>=stealth,shorten <=2pt,shorten >=2pt] (Phidots.south)+(-1.7,0) -- ++(-2.,-0.4) -- ++(-0.3,-0) -- ++(-0.3,0.4);
  \draw [thick,<-,>=stealth,shorten <=2pt,shorten >=2pt] (Phi3.south) -- +(0.3,-0.4) -- ++(0.6,-0.4) -- ++(0.3,0.4);  
  \draw [thick,->,>=stealth,shorten <=2pt,shorten >=2pt] (Phin.south) -- +(-0.3,-0.4) -- ++(-0.6,-0.4) -- ++(-0.3,0.4);
  \draw [thick,->,>=stealth,shorten <=2pt,shorten >=2pt] (Phin.south)+(-0.9,0) -- ++(-1.1,-0.4) -- ++(-0.3,-0) -- ++(-0.3,0.4);
  \draw [thick,->,>=stealth,shorten <=2pt,shorten >=2pt] (Phin.south)+(-1.7,0) -- ++(-2.,-0.4) -- ++(-0.3,-0) -- ++(-0.3,0.4);
  \draw [thick,<-,>=stealth,shorten <=2pt,shorten >=2pt] (Phidots.south) -- +(0.3,-0.4) -- ++(0.6,-0.4) -- ++(0.3,0.4);  
  \draw[thick,color=blue!80!black,postaction={decorate},-,>=stealth,shorten <=2pt,shorten >=2pt] (Phi3.south) .. controls (7.,-2.7) and (3.,-1.7)  ..  (Phi1.south);
  \draw[thick,color=blue!80!black,postaction={decorate},-,>=stealth,shorten <=2pt,shorten >=2pt] (Phi2.south) .. controls (4.5,-1.7) and (0.5,-2.7)  ..  (Phi0.south);
   \draw[thick,color=red!70!black,postaction={decorate},-,>=stealth,shorten <=2pt,shorten >=2pt] (Phi3.south) .. controls (7.5,-3.6) and (-0.3,-3.6)  ..  (Phi0.south);
   \draw[thick,color=red!70!black,postaction={decorate},-,>=stealth,shorten <=2pt,shorten >=2pt] (Phin.south) .. controls +(0.4,-3.6) and +(0.3,-3.6)  ..  (Phi3.south) ;
   \draw[thick,color=orange!80!black,postaction={decorate},-,>=stealth,shorten <=2pt,shorten >=2pt] (Phin.south) .. controls +(-0.1,-1.9) and +(0.4,-1.9)  ..  (Phidots.south);
   \draw[thick,color=orange!80!black,postaction={decorate},-,>=stealth,shorten <=2pt,shorten >=2pt] (Phidots.south) .. controls +(-0.1,-1.9) and +(0.4,-1.9)  ..  (Phi3.south);
   \draw[color=green!40!black,postaction={decorate},-,>=stealth,shorten <=2pt,shorten >=2pt] (Phidots.south) .. controls +(-0.3,-3.7) and +(0.8,-3.5)  ..  (Phi2.south);
   \draw[color=green!40!black,postaction={decorate},-,>=stealth,shorten <=2pt,shorten >=2pt] (Phidots.south) .. controls +(-0.5,-5.1) and +(0.8,-4.7)  ..  (Phi1.south);
   \draw[color=green!40!black,postaction={decorate},-,>=stealth,shorten <=2pt,shorten >=2pt] (Phidots.south) .. controls +(-0.55,-6.5) and +(0.,-6.1)  ..  (Phi0.south);
   \draw[color=white!30!black,postaction={decorate},-,>=stealth,shorten <=2pt,shorten >=2pt] (Phin.south) .. controls +(0.3,-6.6) and +(1.2,-5.8)  .. (Phi2.south);
   \draw[color=white!30!black,postaction={decorate},-,>=stealth,shorten <=2pt,shorten >=2pt] (Phin.south) .. controls +(0.8,-8.) and +(1.1,-8.)  ..  (Phi1.south);
   \draw[color=white!30!black,postaction={decorate},-,>=stealth,shorten <=2pt,shorten >=2pt] (Phin.south) .. controls +(1.5,-9.8) and +(-1,-9.8)  ..  (Phi0.south);
  \node (+A) at (1.2,1) {\small{$\Delta_{+A}$}};
  \node (+A) at (3.7,1) {\small{$\Delta_{+A}$}};
  \node (+A) at (6.2,1) {\small{$\Delta_{+A}$}};
  \node (+A) at (9.3,1) {\small{$\Delta_{+A} \cdots \Delta_{+A}$}};
  \node (+A) at (12.8,1) {\small{$\Delta_{+A} \cdots \Delta_{+A}$}};
  \node (dots) at (9.3,0) {$\cdots$};
  \node (dots) at (12.8,0) {$\cdots$};
  \node (-A) at (6.2,-1) {\small{$\Delta_{-A}$}};
  \node (-A) at (9.3,-1) {\small{$\Delta_{-A} \cdots \Delta_{-A}$}};
  \node (-A) at (12.8,-1) {\small{$\Delta_{-A} \cdots \Delta_{-A}$}};
  \node (-A) at (3.7,-1) {\small{$\Delta_{-A}$}};
  \node (-A) at (1.2,-1) {\small{$\Delta_{-A}$}};
  \node (-2A) at (1.7,-1.95) {\textcolor{blue!80!black}{\small{$\Delta_{-2A}$}}};
  \node (-2A) at (4.7,-1.95) {\textcolor{blue!80!black}{\small{$\Delta_{-2A}$}}};
  \node (-3A) at (2.1,-2.85) {\textcolor{red!70!black}{\small{$\Delta_{-3A}$}}};
  \node (-nm3A) at (12.,-3.25) {\textcolor{red!70!black}{\small{$\Delta_{-(n-3)A}$}}};
  \node (-nmkA) at (12.7,-1.9) {\textcolor{orange!80!black}{\small{$\Delta_{-(n-k)A}$}}};
   \node (-km3A) at (9.35,-1.95) {\textcolor{orange!80!black}{\small{$\Delta_{-(k-3)A}$}}};
   \node (-km2A) at (7.9,-3.2) {\textcolor{green!30!black}{\small{$\Delta_{-(k-2)A}$}}};
   \node (-km1A) at (5.9,-4.15) {\textcolor{green!40!black}{\small{$\Delta_{-(k-1)A}$}}};
   \node (-kA) at (3.6,-5.05) {\textcolor{green!40!black}{\small{$\Delta_{-kA}$}}};     
   \node (-nm2A) at (10.5,-5.25) {\textcolor{white!30!black}{\small{$\Delta_{-(n-2)A}$}}};
   \node (-nm1A) at (8.3,-6.55) {\textcolor{white!30!black}{\small{$\Delta_{-(n-1)A}$}}};
   \node (-nA) at (7,-7.9) {\textcolor{white!30!black}{\small{$\Delta_{-nA}$}}};
\end{tikzpicture}
\end{center}
\vspace*{-50pt}
\caption{The \textit{alien chain}: a pictorial representation of the (nonlinear) transseries \eqref{one-param-transseries}, where each node in the chain corresponds to each of the transseries building blocks (each of the asymptotic series); and of the resurgence relations \eqref{Delta-kA-one-param}, herein represented as possible motions in-between the nodes of this infinite one-dimensional chain. The arrows in the figure represent allowed actions of the alien derivative on the (multi) instanton sectors $\Phi_k$, with different colors encoding different possible actions. Forward motions along the chain (\textit{i.e.}, motions from left to right) can only be achieved by the action of $\Delta_{+A}$, in which case to connect two different instanton sectors $\Phi_m$ and $\Phi_{\ell}$ with $\ell>m$ we need to act with this alien derivative $\ell-m$ times. But due to \eqref{Delta-kA-one-param}, backward motions have much more freedom. To move from right to left, one finds several more ways to connect $\Phi_{\ell}$ to $\Phi_m$: in fact any left-moving paths are allowed with any combination of instanton sectors $\Phi_k$ with $m<k<\ell$ as mid-nodes. For example, to connect $\Phi_3$ to $\Phi_0$ we have four possible paths: act with $\Delta_{-A}$ three times (the black links $\Phi_3 \rightarrow \Phi_2  \rightarrow \Phi_1  \rightarrow \Phi_0$); act with $\Delta_{-A}$ (the black link $\Phi_3 \rightarrow \Phi_2$) followed by $\Delta_{-2A}$ (the blue link $\Phi_2 \rightarrow \Phi_0$); act with $\Delta_{-2A}$ (the blue link $\Phi_3 \rightarrow \Phi_1$) followed by $\Delta_{-A}$ (the black link $\Phi_1 \rightarrow \Phi_0$); or act with $\Delta_{-3A}$ (the red link $\Phi_3 \rightarrow \Phi_0$).}
\label{fig:alien-action-1-param}
\end{figure}
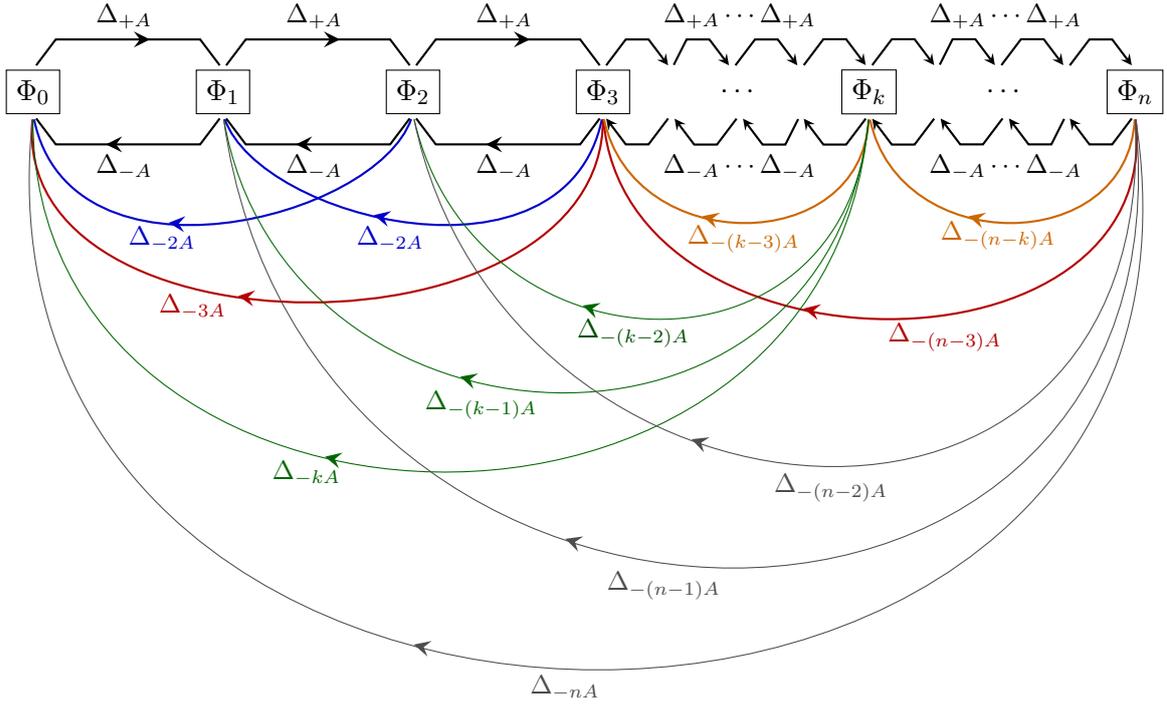

Let us briefly summarize what we have uncovered so far. For the nonlinear one-parameter transseries \eqref{one-param-transseries}, the Borel transform of each of its asymptotic series $\Phi_n$ has logarithmic branch-cut singularities at $s = k A$, which are essentially given by the corresponding Borel residue $\mathsf{S}_{n\to n+k}$, multiplied by the Borel transform of $\Phi_{n+k}$. This is completely analogous to what happened for the linear transseries \eqref{quartictransseries}, the only difference being that we have now moved from a finite to an \textit{infinite} number of building blocks or \textit{nodes} in the transseries. This will have a dramatic consequence: while the resurgence relations in the linear problem, \eqref{Delta+A} and \eqref{Delta-A}, led to a simple algebraic structure which closed upon itself, \eqref{linear-algebraic-structure}, the resurgence relations for the nonlinear problem, \eqref{Delta-kA-one-param}, will instead lead to a much more intricate algebraic structure. In fact, one can now keep on connecting different transseries nodes by acting several times with the alien derivative, just like in a one-dimensional chain as shown in figure~\ref{fig:alien-action-1-param}.

It is at this point that we depart from the standard descriptions or introductions to resurgent analysis. In fact, either having arrived at the resurgence relations \eqref{Delta-kA-one-param} or starting out from them, most texts on this subject proceed into rather abstract algebraic discussions of their implications, fully developing alien calculus in this process. Herein, we wish to use a more physical language and pathway towards resurgence, and as such will use the one-dimensional alien chain of figure~\ref{fig:alien-action-1-param}, with its set of nodes and allowed motions in-between these nodes, in order to proceed with a more physical, ``statistical mechanical'' approach to resurgence. Besides being a new point-of-view on resurgence, we also believe it is more intuitive and pedagogical.

The first obvious step is to reinterpret the (nonlinear) \textit{one}-parameter \textit{transseries} \eqref{one-param-transseries} as an (infinite) \textit{one}-dimensional \textit{chain}, with the asymptotic series which make up the transseries understood as the \textit{nodes} of this chain. The resurgence relations \eqref{Delta-kA-one-param} are then reinterpreted as defining an allowed set of \textit{resurgence motions} in-between nodes of the chain. In this language, the content of figure~\ref{fig:alien-action-1-param} is obvious. The resurgence-relations constraint $k \le 1$ (and $k \neq 0$) implies that the alien derivatives at each node $\Phi_n$ only have one singularity on the \textit{positive} real axis, at $s=+A$, allowing for a single type of \textit{forward} motions on the chain (iterations still allowed). On the other hand, upon the \textit{negative} real axis one finds logarithmic branch-cuts starting at every point $s = - |k| A$, with $k\ge1$ (but truncating at $k=n$ as one cannot go further left in the chain than $\Phi_0$), allowing for many different types of \textit{backward} motions on the chain. For example, to arrive at a node $\Phi_{\ell}$, starting at another node $\Phi_{m}$, the resurgence relations tell us that one needs to act with alien derivatives $\Delta_{\ell_i A}$, $\ell_i \neq 0$, consecutively as in $\Delta_{\ell_N A} \cdots \Delta_{\ell_2 A}\Delta_{\ell_1 A}$, and such that $\sum_{i=1}^{N}\ell_i = (\ell-m)$. Now if we are moving forward along the chain, \textit{i.e.}, $\ell > m$, there is only the $\Delta_{+A}$ action, and one must thus use this operator $\ell-m$ times. But if on the other hand we are moving backwards along the chain, $\ell < m$, there are now several possible paths to arrive at the desired node. Such paths can be composed of many steps, and go through any number of intermediary nodes. Each \textit{step} is defined by a single \textit{link}, one arrow connecting two nodes. In other words, in this situation one may use all possible combinations $\Delta_{-\ell_i A}$, $\ell_i>0$, with the same starting and ending nodes. One example of a choice of paths is given in the caption of figure~\ref{fig:alien-action-1-param}.

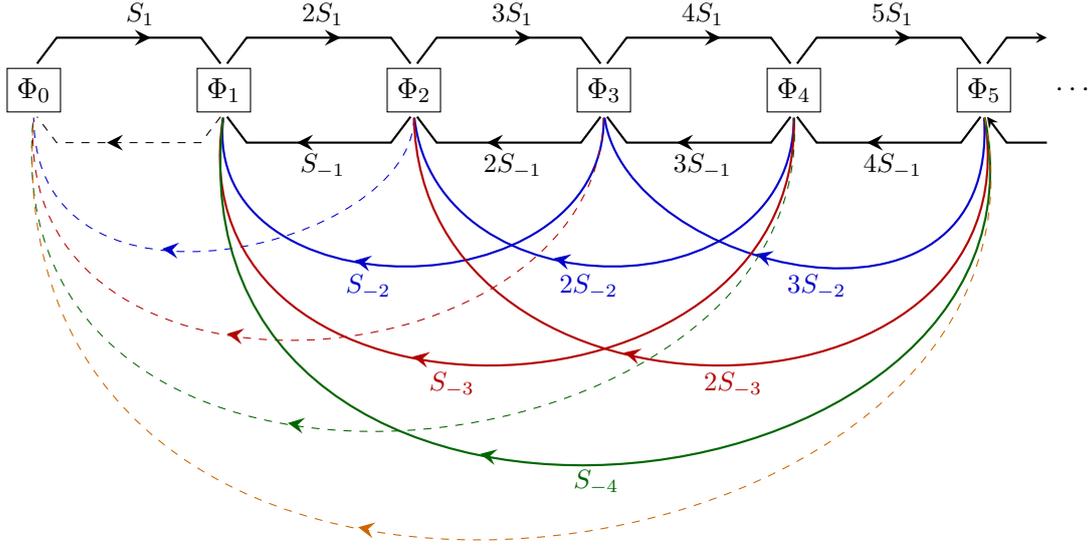
\begin{figure}[t!]
\begin{center}
\begin{tikzpicture}[>=latex,decoration={markings, mark=at position 0.6 with {\arrow[ultra thick]{stealth};}} ]
\begin{scope}[node distance=2.5cm]
  \node (Phi0) [draw] at (0,0) {$\Phi_0$};
  \node (Phi1) [right of=Phi0] [draw] {$\Phi_1$};
  \node (Phi2) [right of=Phi1] [draw] {$\Phi_2$};
  \node (Phi3) [right of=Phi2] [draw] {$\Phi_3$};
  \node (Phi4) [right of=Phi3] [draw] {$\Phi_4$};
  \node (Phi5) [right of=Phi4] [draw] {$\Phi_5$};
\end{scope}
  \draw [thick,postaction={decorate},-,>=stealth,shorten <=2pt,shorten >=2pt] (Phi0.north) -- +(0.3,0.4) -- ++(2.2,0.4) -- (Phi1.north);
  \draw [thick,postaction={decorate},-,>=stealth,shorten <=2pt,shorten >=2pt] (Phi1.north) -- +(0.3,0.4) -- ++(2.2,0.4) -- (Phi2.north);
  \draw [thick,postaction={decorate},-,>=stealth,shorten <=2pt,shorten >=2pt] (Phi2.north) -- +(0.3,0.4) -- ++(2.2,0.4) -- (Phi3.north);
  \draw [thick,postaction={decorate},-,>=stealth,shorten <=2pt,shorten >=2pt] (Phi3.north) -- +(0.3,0.4) -- ++(2.2,0.4) -- (Phi4.north);
  \draw [thick,postaction={decorate},-,>=stealth,shorten <=2pt,shorten >=2pt] (Phi4.north) -- +(0.3,0.4) -- ++(2.2,0.4) -- (Phi5.north);
  \draw [thick,->,>=stealth,shorten <=2pt,shorten >=2pt] (Phi5.north) -- +(0.3,0.4) -- +(0.9,0.4);
  \draw [dashed,postaction={decorate},-,>=stealth,shorten <=2pt,shorten >=2pt] (Phi1.south) -- +(-0.3,-0.4) -- ++(-2.2,-0.4) -- (Phi0.south);
  \draw [thick,postaction={decorate},-,>=stealth,shorten <=2pt,shorten >=2pt] (Phi2.south) -- +(-0.3,-0.4) -- ++(-2.2,-0.4) -- (Phi1.south);
  \draw [thick,postaction={decorate},-,>=stealth,shorten <=2pt,shorten >=2pt] (Phi3.south) -- +(-0.3,-0.4) -- ++(-2.2,-0.4) -- (Phi2.south);
  \draw [thick,postaction={decorate},-,>=stealth,shorten <=2pt,shorten >=2pt] (Phi4.south) -- +(-0.3,-0.4) -- ++(-2.2,-0.4) -- (Phi3.south);
  \draw [thick,postaction={decorate},-,>=stealth,shorten <=2pt,shorten >=2pt] (Phi5.south) -- +(-0.3,-0.4) -- ++(-2.2,-0.4) -- (Phi4.south);
  \draw [thick,<-,>=stealth,shorten <=2pt,shorten >=2pt] (Phi5.south) -- +(0.3,-0.4) -- ++(0.9,-0.4);
  \draw[thick,blue!80!black,postaction={decorate},-,>=stealth,shorten <=2pt,shorten >=2pt] (Phi3.south) .. controls +(0,-2.7) and +(-0.3,-2.7)  ..  (Phi1.south);
  \draw[dashed,blue!80!black,postaction={decorate},-,>=stealth,shorten <=2pt,shorten >=2pt] (Phi2.south) .. controls +(0,-1.8) and +(0,-3.)  ..  (Phi0.south);
  \draw[thick,blue!80!black,postaction={decorate},-,>=stealth,shorten <=2pt,shorten >=2pt] (Phi4.south) .. controls +(0,-2.7) and +(0.3,-2.7)  ..  (Phi2.south);
  \draw[thick,blue!80!black,postaction={decorate},-,>=stealth,shorten <=2pt,shorten >=2pt] (Phi5.south) .. controls +(0.2,-3.3) and +(0.3,-2.1)  ..  (Phi3.south);
   \draw[dashed,red!70!black,postaction={decorate},-,>=stealth,shorten <=2pt,shorten >=2pt] (Phi3.south) .. controls +(-0.1,-3.7) and +(-0.3,-4.3)  ..  (Phi0.south);
   \draw[thick,red!70!black,postaction={decorate},-,>=stealth,shorten <=2pt,shorten >=2pt] (Phi4.south) .. controls +(0.1,-4.4) and +(-0.7,-4.5)  ..  (Phi1.south);
   \draw[thick,red!70!black,postaction={decorate},-,>=stealth,shorten <=2pt,shorten >=2pt] (Phi5.south) .. controls +(0.7,-4.4) and +(0.1,-4.5)  ..  (Phi2.south);
   \draw[dashed,green!40!black,postaction={decorate},-,>=stealth,shorten <=2pt,shorten >=2pt] (Phi4.south) .. controls +(0.3,-5) and +(-0.6,-6.2)  ..  (Phi0.south);
   \draw[thick,green!40!black,postaction={decorate},-,>=stealth,shorten <=2pt,shorten >=2pt] (Phi5.south) .. controls +(1.1,-5.6) and +(-0.8,-6.8)  ..  (Phi1.south);
   \draw[dashed,orange!80!black,postaction={decorate},-,>=stealth,shorten <=2pt,shorten >=2pt] (Phi5.south) .. controls +(1.3,-6.5) and +(-0.6,-8.5)  ..  (Phi0.south);
  \node (+A) at (1.4,1) {\small{$S_1$}};
  \node (+A) at (3.8,1) {\small{$2S_1$}};
  \node (+A) at (6.3,1) {\small{$3S_1$}};
  \node (+A) at (8.8,1) {\small{$4S_1$}};
  \node (+A) at (11.3,1) {\small{$5S_1$}};
  \node (-A) at (11.3,-1) {\small{$4S_{-1}$}};
  \node (-A) at (8.8,-1) {\small{$3S_{-1}$}};
  \node (-A) at (6.3,-1) {\small{$2S_{-1}$}};
  \node (-A) at (3.8,-1) {\small{$S_{-1}$}};
  \node (dots) at (13.7,0) {$\cdots$};
  \node (-2A) at (4.4,-2.6) {\textcolor{blue!80!black}{\small{$S_{-2}$}}};
  \node (-2A) at (7.3,-2.6) {\textcolor{blue!80!black}{\small{$2S_{-2}$}}};
  \node (-2A) at (10.3,-2.6) {\textcolor{blue!80!black}{\small{$3S_{-2}$}}};
  \node (-3A) at (5.5,-3.9) {\textcolor{red!70!black}{\small{$S_{-3}$}}}; 
  \node (-3A) at (9.2,-3.9) {\textcolor{red!70!black}{\small{$2S_{-3}$}}}; 
  \node (-4A) at (7.4,-5.2) {\textcolor{green!40!black}{\small{$S_{-4}$}}};
\end{tikzpicture}
\end{center}
\vspace*{-65pt}
\caption{The \textit{alien chain} revisited: pictorial representation of the \textit{action} of the alien derivative (the right-hand-side of the resurgence relations \eqref{Delta-kA-one-param}) upon the perturbative and first five instanton sectors of the alien chain. Different single arrows correspond to different steps, where each step has an associated \textit{weight} as dictated by \eqref{Delta-kA-one-param}. The black links corresponds to the action of $\Delta_{+A}$ if the arrow is pointing forward (to the right); and to the action of $\Delta_{-A}$ if the arrow is pointing backward (to the left). Then the blue links correspond to the action of $\Delta_{-2A}$, the red links to the action of $\Delta_{-3A}$, the green ones to $\Delta_{-4A}$, and finally the orange link refers to $\Delta_{-5A}$. All colored links are backward motions. The dashed links refer to cases where the action of the alien derivative precisely yields zero, and thus invalidates such would-be motion. As one moves backward along the chain one finds different paths connecting the same chosen two nodes, each composed of different steps. The weight of each step is written next to the corresponding arrow. Further note that the action of two different alien derivatives does not commute (the mid-steps will be different), and this is seen explicitly as one takes weights into account. For example, if one starts off at $\Phi_4$ and acts with $\Delta_{-2A}$ followed by $\Delta_{-A}$ (the blue link followed by the black one, with mid-step on $\Phi_2$), one obtains a total path weight of $2 S_{-2} S_{-1} \Phi_1$. But if instead one first acts with $\Delta_{-A}$ followed by $\Delta_{-2A}$ (the black link followed by the blue one, with mid-step on $\Phi_3$), one rather obtains $3 S_{-1} S_{-2} \Phi_{1}$.}
\label{fig:first-5-sector-alien-1-param}
\end{figure}

Above we were mostly interested in understanding resurgence motions on the alien chain as dictated by the left-hand-side of \eqref{Delta-kA-one-param}. We will now further include the information coming from its right-hand-side, in particular make the Stokes data explicit in all these motions\footnote{This is another reason why to work with Stokes constants more often than with Borel residues. In some sense, the Stokes constants $S_k$ live on the complex Borel plane, but they are attached to the singular points \textit{regardless} of which (instanton) sector we are considering. Borel residues $\mathsf{S}_{n\to n+k}$, on the other hand, live on the \textit{specific} complex Borel plane describing the singular nature of each particular (instanton) sector. This is already made clear by their labeling. But now, there are also subsequent differences in the language of motions on the alien chain. Stokes constants are ``translational invariant'' on the chain, in the sense that they only depend on the ``distance vector'' $k$ in-between nodes. But Borel residues are \textit{not}: they need clear specification of both departure and arrival nodes.}. This simply leads to the rewriting of the alien chain as shown in figure~\ref{fig:first-5-sector-alien-1-param}. With this in mind, we may finally introduce a set of ``statistical mechanical'' definitions, quantitatively describing motions on the alien chain and allowing for a more physical understanding of many upcoming resurgence formulae:

\begin{itemize}
\item \textit{Step $\mathcal{S}$}: any single link connecting two nodes on the chain
\be
\label{DEF-stepS}
\mathcal{S}\,\,=\,\,\,
\begin{tikzpicture}[>=latex,decoration={
markings, mark=at position 0.6 with {\arrow[ultra thick]{stealth};}} ]  \draw[thick,black,postaction={decorate},-,>=stealth,shorten <=2pt,shorten >=2pt] (0,0) .. controls +(0.3,0.8) and +(-0.3,0.8)  ..  (2.5,0);
\end{tikzpicture}
\ee
\item \textit{Weight of a step $\mathcal{S}$, $w (\mathcal{S})$}: the product of Stokes coefficient and numerical factor associated with the step, which can be read directly from \eqref{Delta-kA-one-param}. In figure~\ref{fig:first-5-sector-alien-1-param} the weights of the steps are written right next to the colored arrows. For a step connecting nodes $\Phi_m$ and $\Phi_k$
\be
\label{equation:weight1DIM}
w \left( \mathcal{S} \left( m \rightarrow k \right) \right) = k\, S_{k-m}.
\ee
\item \textit{Path $\mathcal{P}$}: a trajectory composed by any number of steps, connecting two nodes $\Phi_m$ to $\Phi_k$
\be
\label{DEF-path}
\mathcal{P} = \mathcal{S}_1 \cup \mathcal{S}_2 \cup \cdots \cup \mathcal{S}_{\ell} \,\, =\,\,\,
\begin{tikzpicture}[>=latex,decoration={markings, mark=at position 0.6 with {\arrow[ultra thick]{stealth};}} ]  
\begin{scope}[node distance=2.5cm]
  \node (Phim) [draw] at (0,0) {$\Phi_{\phantom{k}\hspace{-4pt}m}$};
  \node (mid) [right of=Phim]  {$\cdots$};
  \node (Phik) [right of=mid] [draw] {$\Phi_{\phantom{m}\hspace{-6pt}k}$};
\end{scope}
  \draw[thick,black,postaction={decorate},-,>=stealth,shorten <=2pt,shorten >=2pt] (Phim.north) .. controls +(0.3,0.8) and +(-0.3,0.8)  ..  (mid.north);
  \draw[thick,black,postaction={decorate},-,>=stealth,shorten <=2pt,shorten >=2pt] (mid.north) .. controls +(0.3,0.8) and +(-0.3,0.8)  ..  (Phik.north);
\end{tikzpicture}
\ee
\item \textit{Length of a path $\mathcal{P}$, $\ell (\mathcal{P})$}: the number of steps composing the path (where any step is always considered to have length one)
\be
\label{DEF-ell-path}
\ell (\mathcal{P}) = \# \left\{ \mathcal{S}_i \in \mathcal{P} \right\}.
\ee
\item \textit{Weight of a path $\mathcal{P}$, $w (\mathcal{P})$}: the product of the weights of every step composing the path
\be
\label{DEF-w-path}
w (\mathcal{P}) = \prod_{i=1}^{\ell (\CP)} \,w (\mathcal{S}_i).
\ee
\item \textit{Combinatorial factor of path $\CP$, $\mathrm{CF} (\mathcal{P})$}: the division by the permutations of $\left\{\mathcal{S}_i\in\mathcal{P}\right\}$
\be
\label{DEF-CF-path}
\mathrm{CF} (\mathcal{P}) = \frac{1}{\left(\ell (\mathcal{P})\right)!}.
\ee
\end{itemize}

The last two points to notice concerning figure~\ref{fig:first-5-sector-alien-1-param} are the following. First, we have plotted some dashed backward steps. While these are in principle allowed, they show no singularity, \textit{i.e.}, have zero weight. For example, the node $\Phi_5$ does not have a singularity at $s=-5A$, as the alien derivative vanishes $\Delta_{-5A}\Phi_5=0$. More generally, the Borel transform of $\Phi_n$ does not have a singularity at $s=-nA$, \textit{i.e.}, $\Delta_{-nA}\Phi_n=0$. Second, also noteworthy is the non-commutativity of alien derivatives which translates to a non-commutativity of different paths. Take for example $m, k>0$ and $m+k<n$ to compute:
\begin{eqnarray}
\Delta_{-mA} \Delta_{-kA} \Phi_n &=& S_{-k} \left(n-k\right) \Delta_{-mA} \Phi_{n-k} = S_{-k}\, S_{-m} \left(n-k\right) \left(n-k-m\right) \Phi_{n-k-m}, \\
\Delta_{-kA} \Delta_{-mA} \Phi_n &=& S_{-m} \left(n-m\right) \Delta_{-kA} \Phi_{n-m} = S_{-m}\, S_{-k} \left(n-m\right) \left(n-m-k\right) \Phi_{n-m-k}.
\end{eqnarray}
\noindent
The start and end nodes are the same, but the multiplicative factors are different as long as $k\ne m$. This happens because the mid-step is different ($\Phi_{n-k}$ for the first case and $\Phi_{n-m}$ for the second). In particular,
\bea
\left[ \Delta_{-mA}, \Delta_{-kA} \right] \Phi_n &=& S_{-m}\, S_{-k} \left( m-k \right) \left(n-m-k\right) \Phi_{n-m-k} = \nonumber \\
&=& \frac{S_{-m}\, S_{-k}}{S_{-m-k}} \left( m-k \right) \Delta_{-(m+k)A} \Phi_{n}.
\label{virasoro-1d-teaser}
\eea
\noindent
This means that two given paths made of smaller steps will not commute, as long as there is a single-step path between the departure and arrival nodes.

Recall that the discussion of the partition function in subsection~\ref{subsec:basicresurgence} was much simpler. In fact, the alien chain in that linear problem only had two nodes, \eqref{linear-algebraic-structure}, and the corresponding algebraic structure was elementary. It was nonetheless required in order to study Stokes discontinuities and large-order asymptotics later on. The same is true in the present nonlinear case. Having discussed the alien chain for the transseries describing nonlinear problems, let us next see how to calculate the Stokes automorphism \eqref{stokesauto}, or, more specifically, the related Stokes discontinuities \eqref{StokesDisc} (illustrated in figure~\ref{stokescrossingfig}) for each of the asymptotic expansions (chain nodes) $\Phi_n$. This is a required step, in the present one-parameter transseries set-up, in order to finally fully understand the asymptotics of all multi-instanton sectors in the quartic free energy.

The Stokes discontinuities along some direction $\theta$ on the Borel plane, defined in \eqref{StokesDisc} and depicted in figure~\ref{stokescrossingfig}, arise from the Stokes automorphism, which itself relates to the alien derivative via exponentiation \eqref{Stokes-aut-as exponential-singularities}. The motions on the alien chain we have just discussed dictate that each node of the chain, $\Phi_n$, only has two singular directions (\textit{i.e.}, Stokes lines; directions with Borel singularities), forward and backward or $\theta=0$ and $\pi$. Now, along the positive real line $\theta=0$ the alien derivative of any node $\Phi_n$ only has one single singularity, $s=A$; while along the negative real line $\theta=\pi$ there are many singularities at $s=-kA$ for $0<k<n$ (recall \eqref{Delta-kA-one-param}). Via \eqref{Stokes-aut-as exponential-singularities} applied to these directions, one finds
\begin{eqnarray}
\label{one-param-stokes-0}
\underline{\mathfrak{S}}_0 \Phi_n &=& \mathrm{exp} \left( \rme^{-\frac{A}{x}} \Delta_{A} \right) \Phi_n, \\
\label{one-param-stokes-pi}
\underline{\mathfrak{S}}_{\pi} \Phi_n &=& \mathrm{exp} \left( \sum_{k=1}^{n} \rme^{\frac{kA}{x}} \Delta_{-kA} \right) \Phi_n.
\end{eqnarray}
\noindent
The discontinuities for each node then immediately follow as $\disc_{\theta} \Phi_n = \Phi_n - \underline{\mathfrak{S}}_{\theta} \Phi_n$.

In order to understand the Stokes discontinuities in terms of motions and data on the alien chain, let us first discuss the above two formulae, \eqref{one-param-stokes-0} and \eqref{one-param-stokes-pi}, in a couple of examples. Stokes discontinuities will have several terms, each with two components: a ``functional'' component, comprising of the non-analytic exponential factor $\rme^{kA/x}$ alongside the nodes (which are themselves asymptotic series as in \eqref{one-param-inst-asympt}); and a ``statistical'' component, arising from a subset of paths on the alien chain in figure~\ref{fig:first-5-sector-alien-1-param}, and comprising their weights and combinatorial factors.
\begin{itemize}
\item \textit{What is the Stokes automorphism on the positive real line, when acting upon the node $\Phi_2$?} One simply expands the exponential \eqref{one-param-stokes-0} to obtain\footnote{Note that the alien derivative does \textit{not} act on the (non-analytic) exponential; but \textit{only} upon asymptotic series.}:
\begin{eqnarray}
\underline{\mathfrak{S}}_0 \Phi_2 &=& \left( 1 + \rme^{-\frac{A}{x}}\, \Delta_{A} + \frac{1}{2!} \left( \rme^{-2\frac{A}{x}}\, \Delta_{A} \Delta_{A} \right) + \frac{1}{3!} \left( \rme^{-3\frac{A}{x}}\, \Delta_{A} \Delta_{A} \Delta_{A} \right) + \cdots \right)\Phi_2 = \nonumber \\
&=&
\Phi_2 + 3 S_1\, \rme^{-\frac{A}{x}}\, \Phi_3 + \frac{12}{2!} \left( S_1 \right)^2 \rme^{-2\frac{A}{x}}\, \Phi_4 + \frac{60}{3!} \left( S_1 \right)^3 \rme^{-3\frac{A}{x}}\, \Phi_5 + \cdots \nonumber \\
\label{stokes-zero-one-param-expanded}
&=&
\Phi_2 - \mathsf{S}_{2\to3}\, \rme^{-\frac{A}{x}}\, \Phi_3 - \mathsf{S}_{2\to4}\, \rme^{-2\frac{A}{x}}\, \Phi_4 - \mathsf{S}_{2\to5}\, \rme^{-3\frac{A}{x}}\, \Phi_5 - \cdots \\
&=&
\Phi_2 - \mathsf{S}_{2\to3}\, \rme^{-\frac{A}{x}}\, \Phi_3 + \frac{1}{2!} \mathsf{S}_{2\to3} \mathsf{S}_{3\to4}\, \rme^{-2\frac{A}{x}}\, \Phi_4 - \frac{1}{3!} \mathsf{S}_{2\to3} \mathsf{S}_{3\to4} \mathsf{S}_{4\to5}\, \rme^{-3\frac{A}{x}}\, \Phi_5 + \cdots. \nonumber
\end{eqnarray}
\noindent
The result may be written with either Stokes constants or, equivalently, Borel residues, via\footnote{Or else, if we did not know equations \eqref{stokesS-to-borelS+GEN} and \eqref{borelS-consistency}, one could also take each exponential term in \eqref{stokes-zero-one-param-expanded} as the \textit{defining equations} for the Borel residues as functions of the Stokes constants.} \eqref{stokesS-to-borelS+GEN} and \eqref{borelS-consistency}. Let us compare this action of $\underline{\mathfrak{S}}_0$ on $\Phi_2$ with the allowed \textit{forward} resurgence motions out of the node $\Phi_2$, on the alien chain of figure~\ref{fig:first-5-sector-alien-1-param}. The coefficients of the several terms in the expansion above, \eqref{stokes-zero-one-param-expanded}, have a simple ``physical'' origin:
\begin{itemize}
\item $\Phi_2\rightarrow\Phi_2$: this simply leaves the node invariant;
\item $\Phi_2\rightarrow\Phi_3$: there is a single path, of length $\ell=1$ and weight $w = 3 S_1$, leading to a combinatorial factor $\mathrm{CF}=\frac{1}{1!}$ and nonperturbative contribution $\rme^{-\frac{A}{x}}$;
\item $\Phi_2\rightarrow\Phi_4$: there is a single path (made of two steps), with length $\ell=2$ and weight $w = 12 \left( S_1 \right)^2$, leading to $\mathrm{CF}=\frac{1}{2!}$ and nonperturbative contribution $\rme^{-2\frac{A}{x}}$;
\item $\Phi_2\rightarrow\Phi_5$: again there is a single path (with three steps), of length $\ell=3$ and weight $w = 60 \left( S_1 \right)^3$, leading to $\mathrm{CF}=\frac{1}{3!}$ and nonperturbative contribution $\rme^{-3\frac{A}{x}}$.
\end{itemize}
\item \textit{What is the Stokes automorphism on the negative real line, when acting upon the node $\Phi_4$?} Now we expand the exponential \eqref{one-param-stokes-pi} instead:
\begin{eqnarray}
\underline{\mathfrak{S}}_{\pi} \Phi_4 &=& \left( 1 + \sum_{k=1}^4 \rme^{\frac{kA}{x}}\, \Delta_{-kA} + \frac{1}{2!} \left( \sum_{k=1}^4 \rme^{\frac{kA}{x}}\, \Delta_{-kA} \right)^2 + \frac{1}{3!} \left( \sum_{k=1}^4 \rme^{\frac{kA}{x}}\, \Delta_{-kA} \right)^3 + \cdots \right) \Phi_4 = \nonumber \\
&=&
\Phi_4 + 3 S_{-1}\, \rme^{\frac{A}{x}} \Phi_3 + \left( 2 S_{-2} + \frac{6}{2!} S_{-1}^2 \right) \rme^{\frac{2A}{x}} \Phi_2 + \left( S_{-3} + \frac{5}{2!} S_{-1} S_{-2} + \frac{6}{3!} S_{-1}^3 \right) \rme^{\frac{3A}{x}} \Phi_1 \nonumber \\
&=&
\Phi_4 - \mathsf{S}_{4\to3}\, \rme^{\frac{A}{x}}\, \Phi_3 - \mathsf{S}_{4\to2}\, \rme^{\frac{2A}{x}}\, \Phi_2 - \mathsf{S}_{4\to1}\, \rme^{\frac{3A}{x}}\, \Phi_1.
\label{stokes-pi-one-param-expanded}
\end{eqnarray}
\noindent
The result may be written with either Stokes constants or, equivalently, Borel residues, via\footnote{Again, if we did not know equation \eqref{stokesS-to-borelS-GEN}, one could also take each exponential term in \eqref{stokes-pi-one-param-expanded} as the \textit{defining equations} for the Borel residues as functions of the Stokes constants.} \eqref{stokesS-to-borelS-1}, \eqref{stokesS-to-borelS-2} and \eqref{stokesS-to-borelS-3} (or, more generally, via \eqref{stokesS-to-borelS-GEN}). This action of $\underline{\mathfrak{S}}_{\pi}$ on $\Phi_4$ should now be compared with the allowed \textit{backward} resurgence motions out of the node $\Phi_4$, on the alien chain of figure~\ref{fig:first-5-sector-alien-1-param}. As already mentioned, we now have many more paths to consider. But, once again, the coefficients of the several terms in the expansion \eqref{stokes-pi-one-param-expanded} have a simple ``physical'' origin:
\begin{itemize}
\item $\Phi_4\rightarrow\Phi_4$: this simply leaves the node invariant;
\item $\Phi_4\rightarrow\Phi_3$: there is a single path, of length $\ell=1$ and weight $w = 3 S_{-1}$, leading to a combinatorial factor $\mathrm{CF}=\frac{1}{1!}$ and nonperturbative contribution $\rme^{+\frac{A}{x}}$;
\item $\Phi_4\rightarrow\Phi_2$: there are now two paths. One consists of a single step (the blue link in figure~\ref{fig:first-5-sector-alien-1-param}), with length $\ell=1$, weight $w = 2 S_{-2}$ and $\mathrm{CF}=\frac{1}{1!}$; the other consists of two steps (the black links in figure~\ref{fig:first-5-sector-alien-1-param}), with length $\ell=2$, weight $w = 6 \left( S_{-1} \right)^2$ and $\mathrm{CF}=\frac{1}{2!}$. They both lead to a nonperturbative contribution $\rme^{+2\frac{A}{x}}$;
\item $\Phi_4\rightarrow\Phi_1$: there are now four paths. One consists of a single step (the red link in figure~\ref{fig:first-5-sector-alien-1-param}), with length $\ell=1$, weight $w = S_{-3}$ and $\mathrm{CF}=\frac{1}{1!}$; the second consists of two steps (first the blue link, followed by the black one), with length $\ell=2$, weight $w = 2  S_{-1} S_{-2}$ and $\mathrm{CF}=\frac{1}{2!}$;  the third similarly consisting of two steps (now first a black link, followed by a blue one), with length $\ell=2$, $w = 3 S_{-1} S_{-2}$ and $\mathrm{CF}=\frac{1}{2!}$; and the fourth consists of three steps (the black links in figure~\ref{fig:first-5-sector-alien-1-param}), with length $\ell=3$, weight $w = 6 \left( S_{-1} \right)^3$ and $\mathrm{CF}=\frac{1}{3!}$. They all have nonperturbative contribution $\rme^{+3\frac{A}{x}}$;
\item $\Phi_4\rightarrow\Phi_0$: all paths arriving at node $\Phi_0$ have zero weight. Further, one cannot move backwards from this node, in which case the expansion necessarily truncates here.
\end{itemize}
\end{itemize}
\noindent
The pattern should be rather clear. In fact, one may now define the Stokes discontinuities, $\disc_{\theta}$, purely in terms of motions and data on the alien chain\footnote{One may also think of the statistical component below, $\mathrm{SF}_{(n \rightarrow m)}$, as a ``statistical-mechanical'' \textit{definition} of the Borel residues (up to a sign).}:

\vspace{10pt}
\setlength{\fboxsep}{10pt}
\Ovalbox{%
\parbox{14cm}{%
\vspace{5pt}

\textbf{Stokes discontinuity on positive real line:} $\disc_{0}\Phi_n$ is given by a \textit{sum} over \textit{all} forward paths, linking nodes to the right $\Phi_{m>n}$. 
\vspace{5pt}

\textbf{Stokes discontinuity on negative real line:} $\disc_{\pi}\Phi_n$ is given by a \textit{sum} over \textit{all} backward paths, linking nodes to the left $\Phi_{m<n}$. 
\vspace{5pt}

In both cases each term in this sum ($\Phi_n\rightarrow\Phi_m$) can be decomposed into two factors:
\begin{itemize}
\item Functional component, dictated solely by beginning and end nodes:
\begin{equation*}
-\rme^{-\left(m-n\right) \frac{A}{x}}\, \Phi_m.
\end{equation*}
\item Statistical component, sum over all the allowed paths $\mathcal{P}(n \rightarrow m)$ linking the nodes as in figure~\ref{fig:first-5-sector-alien-1-param}:
\begin{equation*}
\mathrm{SF}_{(n \rightarrow m)} \equiv \sum_{\mathcal{P}(n \rightarrow m)} \mathrm{CF} (\CP)\, w (\mathcal{P}).
\end{equation*}
\end{itemize}
\vspace{-5pt}
}}
\vspace{15pt}

\noindent
For our present purposes, this will be our definition of Stokes discontinuities (and automorphism). It is a simple graphical and combinatorial procedure to understand how the structure of Borel singularities in \eqref{Borel-transf-expanded-one-param} translates to discontinuities of the asymptotic expansions within our transseries. One can also write general closed-form formulae, for all these quantities, based on sums over partitions, but we refer the interested reader to section~2 of \cite{asv11} for those.

Let us now get back into our starting problem in this subsection. Recall that in trying to compute the quartic free energy (assuming no knowledge of the partition function), we were faced with a nonlinear differential equation \eqref{quarticNLODE} whose perturbative series diverged factorially \eqref{quarticFdivergesn!}. Trying to solve this \textit{nonlinear} differential equation with a transseries \textit{ansatz} led to an \textit{infinite} number of multi-instanton sectors, as described in appendix~\ref{app:recursive-free-en}, where the multi-loop expansions of all these nonperturbative sectors also diverged factorially, \eqref{quarticmultiFdivergesn!}. These growths translate to intricate singularity structures on the Borel plane, which may be studied numerically as in figure~\ref{fig:Pade-poles-quartic-pert-series}, and such analyses motivate general resurgence relations at the level of Borel transforms as in \eqref{Borel-transf-expanded-one-param}. The Stokes discontinuities which are created by this singular Borel structure then may be described as above, as motions and data upon the alien chain representing the transseries, and depicted in figure~\ref{fig:first-5-sector-alien-1-param}. Many of the results described above are in fact valid for the resurgent structure of \textit{general} nonlinear problems solved by one-parameter transseries. But as we focus back upon the quartic free energy, and proceeding in analogy to what was earlier done for the (linear) partition function in subsection~\ref{subsec:Zasymptotics}, we would now want to use the above Stokes discontinuities in order to write down large-order resurgent relations which would allow us to check (or predict) the multi-instanton content of the free energy. However, there is still one subtlety, particular to the quartic free energy and having its origin in \eqref{freeen-initial-transansatz}, that we have to consider.

Take a closer look at the first two sectors in the transseries; the $\Phi_0$ and $\Phi_1$ nodes in the chain. As shown in figure~\ref{fig:Pade-poles-quartic-pert-series}, the Borel transform of the perturbative sector has a branch-cut starting at $s=A$. Figure~\ref{fig:first-5-sector-alien-1-param} then shows how this is correctly reproduced by our transseries \textit{ansatz}: there are no singularities along the negative real axis because $\Phi_0$ is the first node in the chain, and there is indeed one singularity at $s=A$ because $\Delta_A \Phi_0 \propto \Phi_1$. But figure~\ref{fig:Pade-poles-quartic-pert-series} also illustrates how the one-instanton sector has \textit{two} branch-cuts, one similarly starting at $s=A$, but another at $s=-A$ along the \textit{negative} real axis. As is made very clear in figure~\ref{fig:first-5-sector-alien-1-param}, this \textit{cannot} happen within the one-parameter transseries set-up: in this case there should be a single singularity at $s=A$ and none at $s=-A$, because $\Delta_{-A}\Phi_1=0$ (\textit{i.e.}, one has $\mathsf{S}_{1\to0}=0$)! We are then forced to conclude that the one-parameter transseries \textit{ansatz} is \textit{not enough} to solve our problem, and more transseries parameters may be required. This is in fact a generic feature as many systems have diverse instanton (or renormalon) actions; even by simply considering higher-order differential equations one will need to parameterize a larger space of boundary conditions with more transseries parameters.

As we already alluded to, the origin of this problem may be found in \eqref{freeen-initial-transansatz}. Indeed when we plugged the multi-instanton \textit{ansatz} \eqref{oneparametertransseriesquartic} into the nonlinear differential equation for the quartic free energy, we obtained \textit{two} possible values for the instanton action: $A=0$ and $A=\frac{3}{2}$. The one-parameter \textit{ansatz} we then chose was implicitly assuming that the solution $A=0$ was just part of the perturbative solution. What we are now realizing is that this is not the case, the action $A=0$ must be considered as part of a truly independent sector, beyond the perturbative $\ell=0$ sector of the $A = \frac{3}{2}$ solution. In this case, the \textit{two}-parameter transseries that fully describes the quartic free energy should be
\be
\label{two-param-quartic}
F (x, \sigma_1, \sigma_2) = \sum_{n=0}^{+\infty} \sum_{m=0}^{+\infty} \sigma_1^n\, \sigma_2^m\, \rme^{-\frac{n A}{x}}\, \Phi_{(n,m)} (x).
\ee
\noindent
Now do note that this is actually a simplified version of what a general two-parameter transseries would look like, as it might allow for two (independent) non-zero actions and thus a generic exponential contribution of the form $\sim \rme^{- \left( n\, A_1 + m\, A_2 \right) \frac{1}{x}}$ (we shall discuss such systems later on). As usual, the sectors $\Phi_{(n,m)}$ are themselves asymptotic series,
\be
\label{two-param-sectors-quartic}
\Phi_{(n,m)} (x) \simeq \sum_{k=0}^{+\infty} F_{k}^{(n,m)}\, x^{k},
\ee
\noindent
with their coefficients growing factorially. Plugging this two-parameter transseries \textit{ansatz} as a solution to the free-energy nonlinear differential equation, we obtain new recursive relations for the multi-loop multi-instanton coefficients $F_{k}^{(n,m)}$. Fortunately, given that one of the instanton actions is actually \textit{zero}, we will be able to reuse much of what was previously done for the one-parameter case. Indeed, looking back at \eqref{two-param-quartic}, it turns out that one can ``hide'' the second parameter into the definition of new ``hatted'' coefficients as
\be
\label{two-into-one-param-quartic}
F (x, \sigma_1, \sigma_2) = \sum_{n=0}^{+\infty} \sigma_1^n\, \rme^{-\frac{n A}{x}}\, \widehat{\Phi}_{n} (x),
\ee
\noindent
where
\be
\label{two-into-one-param-sectors-quartic}
\widehat{\Phi}_{n}(x) \simeq \sum_{k=0}^{+\infty} \widehat{F}_{k}^{(n)}\, x^{k}, \qquad\widehat{F}_{k}^{(n)} := \sum_{m=0}^{+\infty} \sigma_2^m\, F_{k}^{(n,m)}.
\ee
\noindent
Above, we have rewritten our new \textit{ansatz} such as to look like the earlier one-parameter transseries, where the newly defined multi-loop multi-instanton coefficients now obey the \textit{same} recursive relations as those found in appendix~\ref{app:recursive-free-en}. To reiterate, this is only possible within our example as one of the instanton actions \textit{identically vanishes}; we shall loosely refer to this as having a ``one-and-a-half'' transseries. Of course that in a general two-parameter transseries this would not be the case.

But having rewritten the two-parameter \textit{ansatz} as a one-parameter transseries, one is led to ask why did out first attempt at using a one-parameter transseries did not work? The rough answer is that even though we were able to reorganize our \textit{ansatz} as to look like a one-parameter transseries, the ``hidden'' second parameter will still be seen by the resurgence relations.

Let us see how this comes about, for our ``1.5''-parameter transseries. As it will turn out, the full transseries \eqref{two-param-quartic} is not even needed; restricting to the sectors $\Phi_{(n,0)}$, $n\ge0$, and $\Phi_{(0,1)}$ is enough to (finally!) fully describe the quartic free energy . Moreover, this last sector is not even an asymptotic series; it is just a constant:
\be
\label{01nonasymptsector}
\Phi_{(0,1)}(x) = F_0^{(0,1)}.
\ee
\noindent
The main reason why one can restrict our two-parameter transseries into the form of a one-parameter transseries with an added single sector is the same reason why we were able to ``hide'' the second parameter in \eqref{two-into-one-param-sectors-quartic}: one can check that the sectors associated to $\Phi_{(n,m)}$ with $m$ non-zero will \textit{not} be asymptotic; with all asymptotic behaviour being restricted to the $\Phi_{(n,0)}$. Still, our solution to the ODE does tell us that one\footnote{In principle one could include several sectors $\Phi_{(0,m)}$, but they would all be constant and indistinguishable, which in the end does amount to a single contribution.} extra sector needs to be included (albeit non-asymptotic), associated to the action $A=0$---and \eqref{01nonasymptsector} is it. Notice that these asymptotic statements have a rather natural explanation\footnote{By this we mean an explanation going beyond the fact that resurgence \textit{dictates} them (our emphasis in these lectures), \textit{i.e.}, that our \textit{ansatz} is indeed strongly validated by the associated large-order relations arising from resurgence---which we shall get to discuss in detail a bit later on, in subsection~\ref{sec:large-order-F}.} by looking at the quartic free-energy second-order ODE \eqref{quarticNLODE}. Indeed, this equation has \textit{no} dependence on $F(x)$ itself, but only on \textit{derivatives} of $F(x)$. It can then be trivially integrated into a \textit{first}-order ODE by the redefinition $\mathfrak{F} (x) \equiv F' (x)$, where the new sector which we have just uncovered can then be identified with the integration constant associated to this redefinition. Another natural explanation will soon be obtained when we directly compare linear and nonlinear problems (\textit{i.e.}, the quartic partition function and free energy) in the next subsection. To make all these statements a bit more clear, let us finally rewrite the transseries solution as
\be
\label{one-point-five-transseries}
F (x, \sigma_1, \sigma_2) = \sum_{n=0}^{+\infty} \sigma_1^n\, \rme^{-\frac{n A}{x}}\, \Phi_{(n,0)} (x) + \sigma_2\, F_0^{(0,1)}.
\ee

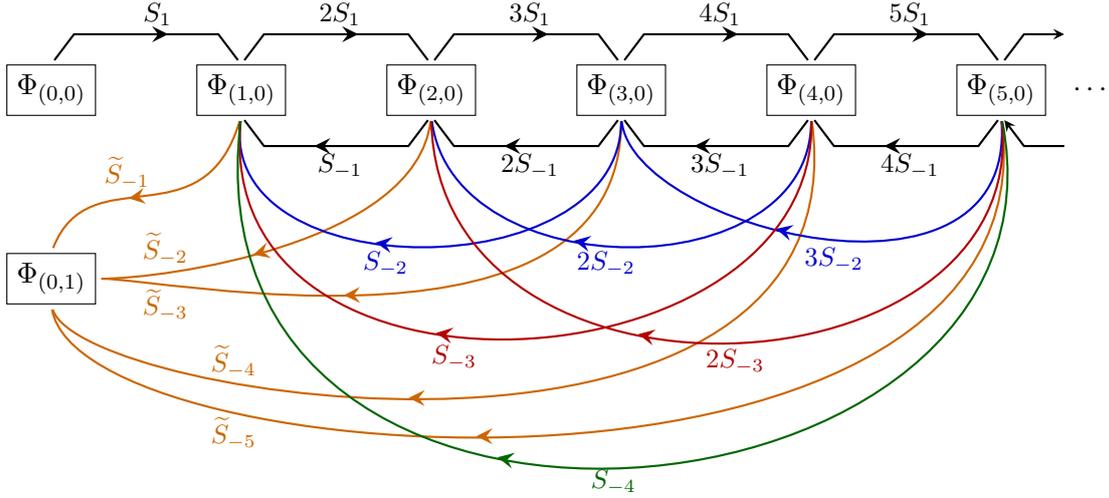
\begin{figure}[t!]
\begin{center}
\begin{tikzpicture}[>=latex,decoration={markings, mark=at position 0.6 with {\arrow[ultra thick]{stealth};}} ]
\begin{scope}[node distance=2.5cm]
  \node (Phi0) [draw] at (0,0) {$\Phi_{(0,0)}$};
  \node (Phi1) [right of=Phi0] [draw] {$\Phi_{(1,0)}$};
  \node (Phi2) [right of=Phi1] [draw] {$\Phi_{(2,0)}$};
  \node (Phi3) [right of=Phi2] [draw] {$\Phi_{(3,0)}$};
  \node (Phi4) [right of=Phi3] [draw] {$\Phi_{(4,0)}$};
  \node (Phi5) [right of=Phi4] [draw] {$\Phi_{(5,0)}$};
  \node (Phi01) [below of=Phi0] [draw] {$\Phi_{(0,1)}$};
\end{scope}
  \draw [thick,postaction={decorate},-,>=stealth,shorten <=2pt,shorten >=2pt] (Phi0.north) -- +(0.3,0.4) -- ++(2.2,0.4) -- (Phi1.north);
  \draw [thick,postaction={decorate},-,>=stealth,shorten <=2pt,shorten >=2pt] (Phi1.north) -- +(0.3,0.4) -- ++(2.2,0.4) -- (Phi2.north);
  \draw [thick,postaction={decorate},-,>=stealth,shorten <=2pt,shorten >=2pt] (Phi2.north) -- +(0.3,0.4) -- ++(2.2,0.4) -- (Phi3.north);
  \draw [thick,postaction={decorate},-,>=stealth,shorten <=2pt,shorten >=2pt] (Phi3.north) -- +(0.3,0.4) -- ++(2.2,0.4) -- (Phi4.north);
  \draw [thick,postaction={decorate},-,>=stealth,shorten <=2pt,shorten >=2pt] (Phi4.north) -- +(0.3,0.4) -- ++(2.2,0.4) -- (Phi5.north);
  \draw [thick,->,>=stealth,shorten <=2pt,shorten >=2pt] (Phi5.north) -- +(0.3,0.4) -- +(0.9,0.4);
  \draw [thick,postaction={decorate},-,>=stealth,shorten <=2pt,shorten >=2pt] (Phi2.south) -- +(-0.3,-0.4) -- ++(-2.2,-0.4) -- (Phi1.south);
  \draw [thick,postaction={decorate},-,>=stealth,shorten <=2pt,shorten >=2pt] (Phi3.south) -- +(-0.3,-0.4) -- ++(-2.2,-0.4) -- (Phi2.south);
  \draw [thick,postaction={decorate},-,>=stealth,shorten <=2pt,shorten >=2pt] (Phi4.south) -- +(-0.3,-0.4) -- ++(-2.2,-0.4) -- (Phi3.south);
  \draw [thick,postaction={decorate},-,>=stealth,shorten <=2pt,shorten >=2pt] (Phi5.south) -- +(-0.3,-0.4) -- ++(-2.2,-0.4) -- (Phi4.south);
  \draw [thick,<-,>=stealth,shorten <=2pt,shorten >=2pt] (Phi5.south) -- +(0.3,-0.4) -- ++(0.9,-0.4);
  \draw [thick,orange!80!black,postaction={decorate},-,>=stealth,shorten <=2pt,shorten >=2pt] (Phi1.south) .. controls +(-0.5,-1.7) and +(0.5,1.3)  ..  (Phi01.north);
  \draw [thick,orange!80!black,postaction={decorate},-,>=stealth,shorten <=2pt,shorten >=2pt] (Phi2.south) .. controls +(-0.2,-1.7) and +(0.9,0)  ..  (Phi01.east);
  \draw [thick,orange!80!black,postaction={decorate},-,>=stealth,shorten <=2pt,shorten >=2pt] (Phi3.south) .. controls +(-0.,-3.4) and +(0.9,0)  ..  (Phi01.east);
  \draw [thick,orange!80!black,postaction={decorate},-,>=stealth,shorten <=2pt,shorten >=2pt] (Phi4.south) .. controls +(0.7,-5.3) and +(0.4,-1.3)  ..  (Phi01.south);
  \draw [thick,orange!80!black,postaction={decorate},-,>=stealth,shorten <=2pt,shorten >=2pt] (Phi5.south) .. controls +(0.7,-5.6) and +(0.4,-2.3)  ..  (Phi01.south);
  \draw[thick,blue!80!black,postaction={decorate},-,>=stealth,shorten <=2pt,shorten >=2pt] (Phi3.south) .. controls +(0,-2.3) and +(-0.3,-2.3)  ..  (Phi1.south);
  \draw[thick,blue!80!black,postaction={decorate},-,>=stealth,shorten <=2pt,shorten >=2pt] (Phi4.south) .. controls +(0,-2.3) and +(0.3,-2.3)  ..  (Phi2.south);
  \draw[thick,blue!80!black,postaction={decorate},-,>=stealth,shorten <=2pt,shorten >=2pt] (Phi5.south) .. controls +(0.2,-2.8) and +(0.15,-1.5)  ..  (Phi3.south);
   \draw[thick,red!70!black,postaction={decorate},-,>=stealth,shorten <=2pt,shorten >=2pt] (Phi4.south) .. controls +(0.1,-3.65) and +(-0.4,-4.2)  ..  (Phi1.south);
   \draw[thick,red!70!black,postaction={decorate},-,>=stealth,shorten <=2pt,shorten >=2pt] (Phi5.south) .. controls +(0.4,-3.8) and +(0.1,-4.2)  ..  (Phi2.south);
   \draw[thick,green!40!black,postaction={decorate},-,>=stealth,shorten <=2pt,shorten >=2pt] (Phi5.south) .. controls +(1.1,-5.6) and +(-0.8,-6.8)  ..  (Phi1.south);
  \node (+A) at (1.4,1) {\small{$S_1$}};
  \node (+A) at (3.8,1) {\small{$2S_1$}};
  \node (+A) at (6.3,1) {\small{$3S_1$}};
  \node (+A) at (8.8,1) {\small{$4S_1$}};
  \node (+A) at (11.3,1) {\small{$5S_1$}}; 
  \node (-A) at (11.3,-1) {\small{$4S_{-1}$}};
  \node (-A) at (8.8,-1) {\small{$3S_{-1}$}};
  \node (-A) at (6.3,-1) {\small{$2S_{-1}$}};
  \node (-A) at (3.8,-1) {\small{$S_{-1}$}};
  \node (dots) at (13.7,0) {$\cdots$};
  \node (-nA) at (1.0,-1.1) {\textcolor{orange!80!black}{\small{$\widetilde{S}_{-1}$}}};
  \node (-nA) at (1.5,-2.1) {\textcolor{orange!80!black}{\small{$\widetilde{S}_{-2}$}}};
  \node (-nA) at (1.5,-2.85) {\textcolor{orange!80!black}{\small{$\widetilde{S}_{-3}$}}};
  \node (-nA) at (2.4,-3.6) {\textcolor{orange!80!black}{\small{$\widetilde{S}_{-4}$}}};
  \node (-nA) at (2.4,-4.55) {\textcolor{orange!80!black}{\small{$\widetilde{S}_{-5}$}}};
  \node (-2A) at (4.4,-2.3) {\textcolor{blue!80!black}{\small{$S_{-2}$}}};
  \node (-2A) at (7.3,-2.3) {\textcolor{blue!80!black}{\small{$2S_{-2}$}}};
  \node (-2A) at (10.3,-2.25) {\textcolor{blue!80!black}{\small{$3S_{-2}$}}};
  \node (-3A) at (5.3,-3.55) {\textcolor{red!70!black}{\small{$S_{-3}$}}}; 
  \node (-3A) at (9.,-3.6) {\textcolor{red!70!black}{\small{$2S_{-3}$}}};   
  \node (-4A) at (7.4,-5.2) {\textcolor{green!40!black}{\small{$S_{-4}$}}};
\end{tikzpicture}
\end{center}
\vspace*{-60pt}
\caption{The \textit{alien chain} for the ``$1.5$''-parameter transseries: a pictorial representation of the action of the alien derivative upon the perturbative and first five instanton sectors, $\Phi_{(k,0)}$, with $k=0,\ldots,5$; alongside the extra ``orthogonal'' sector $\Phi_{(0,1)}$. As before, different single arrows correspond to different steps, and each step has an associated weight as dictated by the new resurgence relations \eqref{Delta-ka-one-point-five}. For simplicity, we have only drawn steps which have non-zero weight. One obtains Stokes discontinuities for this problem applying the same rules as before, now having in mind the set of motions illustrated above.}
\label{fig:first-5-sector-alien-one-point-five}
\end{figure}

The local structure of Borel singularities may likewise be obtained having resurgence in mind. Although the sector $\Phi_{(0,1)}$ is not asymptotic, all the other sectors in \eqref{one-point-five-transseries} are, and their Borel transforms will have associated singularities and branch-cuts. One expects only mild corrections to \eqref{Borel-transf-expanded-one-param} and performing similar calculations as in, \textit{e.g.}, \cite{asv11}, it is simple to obtain
\be
\label{Borel-transf-expanded-one-point-five}
\CB [\Phi_{(n,0)}] (s+kA) = \left( \mathsf{S}_{n\to n+k} \times \CB [\Phi_{(n+k,0)}] (s) + \delta_{n+k}\, \widetilde{\mathsf{S}}_{n\to n+k} \times \CB [\Phi_{(0,1)}] (s) \right) \frac{\log s}{2\pi\rmi}, \qquad k \neq 0,
\ee
\noindent
where now $\widetilde{\mathsf{S}}_{n\to0} \neq 0$. Of course as $\Phi_{(0,1)}(x)$ is just a constant (and not an asymptotic series), its Borel transform will not have any singularities\footnote{To be precise, Borel transforms are not defined for constant functions (residual coefficients). Nevertheless one may define them as an identity on the Borel plane and perform calculations in this way; see \cite{d14} for some examples.}. The resurgence relations which follow from this resurgent singularity-structure are simply
\be
\label{Delta-ka-one-point-five}
\Delta_{k A} \Phi_{(n,0)} = S_k \left(n+k\right) \Phi_{(n+k,0)} + \delta_{n+k}\, \widetilde{S}_{-n}\, \Phi_{(0,1)}, \qquad k \le 1 \text{  and  } k \neq 0,
\ee
\noindent
all other combinations vanishing\footnote{Recall that we are taking $\Phi_{(n,m)}=0$ if either $n$ or $m$ are negative.} (in particular, $\Delta_{kA} \Phi_{(0,1)} = 0$). The pictorial representation of these algebraic equations, as motions on the alien chain associated to the new ``one-and-a-half'' transseries, are depicted in figure~\ref{fig:first-5-sector-alien-one-point-five}. Do notice that this time around the chain is not strictly one-dimensional, having an extra ``orthogonal'' node which we placed just below the perturbative node $\Phi_{(0,0)}$. On what concerns forward motions, the chain still starts at the original perturbative series $\Phi_{(0,0)}$, but on what concerns backward motions, now the ``last'' node becomes $\Phi_{(0,1)}$.

Having a clear picture of this ``1.5-dimensional'' alien chain, describing the transseries solution to the quartic free energy, we can proceed and compute the Stokes discontinuities along the singular directions $\theta=0,\pi$ just like before (the rules are the same, either for calculating functional or statistical factors for each term; only the set of allowed motions is different). For example when evaluating $\disc_{\pi} \Phi_{(n,0)}$, which in the present case should be done with the motions depicted in figure~\ref{fig:first-5-sector-alien-one-point-five}, the last node on backward motions to the left is now $\Phi_{(0,1)}$. Further, this node $\Phi_{(0,1)}$ does not correspond to an asymptotic series, in which case
\be
\label{stokes-zeropi-0|1-quartic}
\disc_{0,\pi}\Phi_{(0,1)}=0.
\ee
\noindent
Let us next write a few examples of discontinuities one may compute (we choose them as they will be useful in what follows). These should not be too hard to obtain, and we leave it as an exercise for the reader to prove them.
\begin{itemize}
\item The Stokes discontinuity along the positive real line is unchanged in this new chain, as there are no new forward motions. It is actually straightforward to find a general expression for all sectors, as
\bea
\label{stokes-zero-pert-quartic}
\disc_{0} \Phi_{(n,0)} (x) &=& - \sum_{k=1}^{+\infty} \binom{n+k}{n}\, S_1^k\, \rme^{-\frac{kA}{x}}\, \Phi_{(n+k,0)} (x) \\
&=& \sum_{k=1}^{+\infty} \mathsf{S}_{n\to n+k}\, \rme^{-\frac{kA}{x}}\, \Phi_{(n+k,0)} (x). \nonumber
\eea
\item The Stokes discontinuities along the negative real line will now have contributions from paths ending on the new node $\Phi_{(0,1)} = F_0^{(0,1)}$. A few examples are:
\begin{eqnarray}
\label{stokes-pi-1inst-quartic}
\disc_{\pi} \Phi_{(1,0)}(x) &=& - \widetilde{S}_{-1}\, \rme^{\frac{A}{x}}\, \Phi_{(0,1)} = \widetilde{\mathsf{S}}_{1\to0}\, \rme^{\frac{A}{x}}\, \Phi_{(0,1)}, \\
\label{stokes-pi-2inst-quartic}
\disc_{\pi} \Phi_{(2,0)}(x) &=& - S_{-1}\, \rme^{\frac{A}{x}}\, \Phi_{(1,0)}(x) - \left( \widetilde{S}_{-2} + \frac{1}{2!} S_{-1} \widetilde{S}_{-1} \right) \rme^{\frac{2A}{x}}\, \Phi_{(0,1)} \\
&=& \mathsf{S}_{2\to1}\, \rme^{\frac{A}{x}}\, \Phi_{(1,0)}(x) + \widetilde{\mathsf{S}}_{2\to0}\, \rme^{\frac{2A}{x}}\, \Phi_{(0,1)}, \nonumber \\
\label{stokes-pi-3inst-quartic}
\disc_{\pi} \Phi_{(3,0)}(x) &=& - 2S_{-1}\, \rme^{\frac{A}{x}}\, \Phi_{(2,0)}(x) - \left( \frac{2}{2!} S_{-1}^2 + S_{-2} \right) \rme^{\frac{2A}{x}}\, \Phi_{(1,0)}(x) - \\
&&
- \left( \frac{2}{3!} S_{-1}^2 \widetilde{S}_{-1} + \frac{1}{2!} S_{-2} \widetilde{S}_{-1} + \frac{2}{2!} S_{-1} \widetilde{S}_{-2} + \widetilde{S}_{-3} \right) \rme^{\frac{3A}{x}}\, \Phi_{(0,1)} \nonumber \\
&=& \mathsf{S}_{3\to2}\, \rme^{\frac{A}{x}}\, \Phi_{(2,0)}(x) + \mathsf{S}_{3\to1}\, \rme^{\frac{2A}{x}}\, \Phi_{(1,0)}(x) + \widetilde{\mathsf{S}}_{3\to0}\, \rme^{\frac{3A}{x}}\, \Phi_{(0,1)}.
\nonumber
\end{eqnarray}
\end{itemize}

At this stage, let us make a few remarks. First, this may seem like an awful lot of work to finally construct a complete solution to the quartic free energy, when we had already computed its partition function and it simply would have followed that $F = \log Z$. Of course we took this route in choosing a \textit{very simple example} to illustrate what resurgence is and how it works. There are of course much more complicated examples where all one has in hand is a nonlinear system, without any possibility to bypass the nonlinearities via an easier problem. For example, this is the case concerning Painlev\'e equations which connect to 2d (super) gravity. In fact, a similar albeit more involved analysis was followed for the Painlev\'e~I equation, describing 2d gravity, in \cite{msw07, msw08, gikm10, asv11} and for the Painlev\'e~II equation, describing 2d supergravity, in \cite{m08, sv13}.

Further, one may wonder if this resurgent transseries we have constructed is in fact the \textit{correct} and \textit{complete} solution to the (nonlinear) quartic free energy. One interesting feature of these resurgence constructions is that they may be \textit{checked} via large-order analyses: one may use the recursive relations to compute multi-loop multi-instanton data in the transseries, $F_k^{(n,0)}$, and then resurgence yields very intricate relations in-between all these coefficients and their asymptotics which may be tested numerically. However, there is one point to handle before this may be done. The construction above was rather general on Stokes constants; in fact it had nothing to say about what are their exact, numerical values. If one wants to be more specific and \textit{quantitative}, one still needs to determine the Stokes constants for the quartic free energy, and this is what we shall turn to now.

\subsection{Stokes Constants of Partition Function versus Free Energy}\label{subsec:stokesZvsF}

Now we wish to make contact between the results in the previous subsection, concerning the quartic free energy $F$, and our earlier (analytical) results on the quartic partition function $Z$. This will serve a dual purpose of both checking the general structure of the two-parameter transseries for the quartic free energy, alongside determining its Stokes constants from first principles (rather than numerically, as in many cases; \textit{e.g.}, \cite{gikm10, asv11, sv13}). For example, recalling the (exact) perturbative expansion of the partition function \eqref{Z010}, it is immediate to compute (and check) the perturbative expansion of the free energy. Using $F = \log Z$ one simply obtains
\be
\label{F0fromZ0}
F^{(0)} = \log Z^{(0)} \simeq \frac{1}{2} \log \frac{\hbar}{2\pi} + \frac{1}{8}\, x + \frac{1}{12}\, x^2 + \frac{11}{96}\, x^3 + \frac{17}{72} x^4 + \frac{619}{960}\, x^5 + \cdots,
\ee
\noindent
which is exactly the same expansion as the one produced by \eqref{quarticFrecursion} (and further fixing the integration constant $F_0^{(0)}$). Similar checks may be done concerning the nonperturbative sectors. For example, crossing a Stokes line to turn-on $Z^{(1)}$, as in \eqref{crossingthetazero}, will induce a subsequent change in \eqref{F0fromZ0} above as
\be
\CF \equiv \log \CZ = \log \left( Z^{(0)}(x) + 2\, Z^{(1)}(x) \right).
\ee
\noindent
Expanding the ratio of \eqref{Z101} with \eqref{Z010}, one obtains
\be
\CF \simeq F^{(0)} - \sum_{n=1}^{+\infty} \left( \rmi \sqrt{2} \right)^n \rme^{- n \frac{3}{2x}} \left( \frac{1}{n} - \frac{1}{4}\, x + \frac{n}{32}\, x^2 + \cdots \right).
\ee
\noindent
One immediately finds $A = \frac{3}{2}$ and $S_1 F^{(1)}_0 = \rmi \sqrt{2}$, as expected from earlier numerical results. In particular, this expression also fixes \textit{all} coefficients in the one-parameter transseries \eqref{oneparametertransseriesquartic}.

Let us make these checks more precise and more systematic, by translating resurgent properties of the partition function into resurgence results for the free energy. In particular, let us compute the Stokes constants of the free energy, given the Stokes constants of the partition function. The comparison of the structure of the transseries for partition function and free energy starts as above, but which we now write a bit more formally. First, the partition function transseries \eqref{quartictransseries} is slightly rewritten as
\be
\CZ (x, \sigma_{0}, \sigma_{1}) = \sigma_{0}\, \Phi_0 (x) + \sigma_{1}\, \rme^{-\frac{3}{2x}}\, \Phi_1 (x) = \sigma_{0}\, \Phi_0 (x) \left( 1 + \zeta\, \rme^{-\frac{3}{2x}}\, \frac{\Phi_1 (x)}{\Phi_0 (x)} \right),
\ee
\noindent
where we have introduced $\zeta \equiv \frac{\sigma_1}{\sigma_0}$. Next, we write the free energy as the logarithm of the partition function, understood as a \textit{formal} expansion of the function $\log\left(1+z\right)$. It follows:
\begin{equation}
\label{eq:Free-en-from-part-func}
\CF (x, \sigma_{0}, \sigma_{1}) = \log \sigma_{0} + \log \Phi_{0} + \sum_{n=1}^{+\infty} \zeta^n\, \rme^{-n \frac{A}{x}}\, F^{(n)}(x),
\end{equation}
\noindent
where the nonperturbative sectors $F^{(n)}$ may be written in terms of the original sectors of the partition function as
\be
\label{eq:Free-en-sectors-instantons}
F^{(n)} := \frac{(-1)^{n+1}}{n} \left( \frac{\Phi_1}{\Phi_0} \right)^n.
\ee
\noindent
On the other hand, and as discussed at length in the previous subsection, we also expect that the free energy will be given by a two-parameter\footnote{Here we have changed notation for the parameters of the transseries, as compared to \eqref{one-point-five-transseries}, in order not to confuse transseries parameters of the partition function and of the free energy.} transseries,
\begin{equation}
\label{eq:Free-en-two-param-transs}
F (x, \tau_1, \tau_2) = \sum_{n=0}^{+\infty} \sum_{m=0}^{+\infty} \tau_1^n\, \tau_2^m\, \rme^{-\frac{n A_1 + m A_2}{x}}\, \Phi_{(n,m)} (x),
\end{equation}
\noindent
where $A_1 = A \equiv \frac{3}{2}$ and $A_2 = 0$, and the nonperturbative sectors $\Phi_{(n,m)} (x)$ are given by the asymptotic series \eqref{two-param-sectors-quartic}. If we now compare this expression for the free energy with the previous one, \textit{i.e.}, compare the two-parameter transseries \eqref{eq:Free-en-two-param-transs} with the transseries following from the partition function \eqref{eq:Free-en-from-part-func}, one may obtain the following identifications\footnote{Obviously assuming $\sigma_{0}\ne0$.}
\bea
\label{transseries-sigma-vs-tau}
\tau_1 &=& \frac{\sigma_1}{\sigma_{0}}, \qquad \tau_2 = \log \sigma_0, \\
\label{transseries-Phi00-vs-logPhi0}
\Phi_{(0,0)} &=& \log \Phi_0, \\
\label{transseries-Phin0-vs-Fn}
\Phi_{(n,0)} &=& F^{(n)}, \\
\label{transseries-Phi01-vs-1}
\Phi_{(0,1)} &=& 1, \\
\Phi_{(n,m)} &=& 0, \qquad \text{if }\, m>1 \vee \left( m=1 \wedge n>0 \right).
\eea
\noindent
In particular, this identification is now validating our ``1.5''-parameter transseries \textit{ansatz} in \eqref{one-point-five-transseries}.

Let us next see if we can make similar identifications at the level of resurgence relations and, eventually, at the level of Stokes data. Recall that the resurgence relations for the partition function, with sectors $\Phi_{0}$ and $\Phi_{1}$, were very simple, \eqref{Delta+A} and \eqref{Delta-A} (implying Stokes constants\footnote{In this subsection we shall label Stokes constants with a superscript $Z$ or $F$, depending on whether they are associated to the quartic partition function or to the quartic free energy, respectively.} $S_{1}^{Z}=-2$ and $S_{-1}^{Z}=1$, respectively). The Stokes automorphism \eqref{Stokes-aut-as exponential-singularities} was then very simply written as \eqref{quarticStokes}, implying \eqref{crossingthetazero} and \eqref{crossingthetapi}, which we reproduce in here as\footnote{Notice in this simple example how \textit{alien} derivatives, which act upon transseries \textit{nodes}, may be translated to \textit{regular} derivatives, which are now acting upon transseries \textit{parameters}. This is the essence of the \textit{bridge equations}: establishing a ``bridge'' between alien and ordinary calculus; see appendix~\ref{app:alien-calculus} for further details.}
\begin{eqnarray}
\underline{\mathfrak{S}}_{0} \CZ (x, \sigma_{0}, \sigma_{1}) &:=& \exp \left( \rme^{-\frac{A}{x}} \Delta_{+A} \right) \CZ (x, \sigma_{0}, \sigma_{1}) = \CZ (x, \sigma_{0}, \sigma_{1} + S_{1}^{Z} \sigma_{0} ) = \nonumber \\
&\equiv& \exp \left\{ S_{1}^{Z}\, \sigma_{0} \frac{\partial}{\partial\sigma_{1}} \right\} \CZ (x, \sigma_{0}, \sigma_{1}), 
\label{stokes-zero-sigmas}\\
\underline{\mathfrak{S}}_{\pi} \CZ (x, \sigma_{0}, \sigma_{1}) &:=& \exp \left( \rme^{+\frac{A}{x}} \Delta_{-A} \right) \CZ (x, \sigma_{0}, \sigma_{1}) = \CZ (x, \sigma_{0} + S_{-1}^{Z} \sigma_{1}, \sigma_{1} ) = \nonumber \\
&\equiv& \exp \left\{ S_{-1}^{Z}\, \sigma_{1} \frac{\partial}{\partial\sigma_{0}} \right\} \CZ (x, \sigma_{0}, \sigma_{1}).
\label{stokes-pi-sigmas}
\end{eqnarray}
\noindent
As already briefly discussed earlier, one way to summarize the action of the Stokes automorphisms is to say that transseries parameters \textit{jump} when crossing Stokes lines (see, \textit{e.g.}, \cite{as13} for further details). In this way, crossing the Stokes line at $\theta = 0$ entails $\sigma_0 \to \sigma_0$ and $\sigma_1 \to \sigma_{1} + S_{1}^{Z} \sigma_{0}$; while crossing the Stokes line at $\theta = \pi$ entails $\sigma_0 \to \sigma_{0} + S_{-1}^{Z} \sigma_{1}$ and $\sigma_1 \to \sigma_1$. We may now use these results to directly determine the Stokes discontinuities associated to the free-energy transseries \eqref{eq:Free-en-two-param-transs}. Its associated transseries parameters need to satisfy \eqref{transseries-sigma-vs-tau}, which immediately implies their ``jumps''. As such, crossing the Stokes line at $\theta = 0$ will now entail $\tau_1 \to \tau_1 + S_{1}^{Z}$ and $\tau_2 \to \tau_2$; while crossing the Stokes line at $\theta = \pi$ will entail $\tau_1 \to \frac{\tau_1}{1 + S_{-1}^{Z} \tau_1}$ and $\tau_2 \to \tau_2 + \log \left( 1 + S_{-1}^{Z} \tau_1 \right)$ (again, see \cite{as13} for first-principle derivations of formulae like these). It is then straightforward to write the analogues of \eqref{stokes-zero-sigmas} and \eqref{stokes-pi-sigmas} for the free energy as
\begin{eqnarray}
\label{eq:Stokes-aut-zero-v1}
\underline{\mathfrak{S}}_{0} F (x, \tau_1, \tau_2) &=& F \left(x, \tau_1+S_{1}^{Z}, \tau_{2}\right), \\
\label{eq:Stokes-aut-pi-v1}
\underline{\mathfrak{S}}_{\pi} F (x, \tau_1, \tau_2) &=& F \left(x, \frac{\tau_1}{1 + S_{-1}^{Z} \tau_{1}}, \tau_{2} + \log \left( 1 + S_{-1}^{Z} \tau_{1} \right) \right).
\end{eqnarray}

The two equations above, \eqref{eq:Stokes-aut-zero-v1} and \eqref{eq:Stokes-aut-pi-v1}, should be equally obtainable by direct action of the Stokes automorphism \eqref{Stokes-aut-as exponential-singularities} upon the free-energy transseries \eqref{eq:Free-en-two-param-transs}, computed via the resurgence relations \eqref{Delta-ka-one-point-five} (\textit{i.e.}, via explicit use of alien derivatives). This calculation is rather straightforward. Along $\theta=0$ the only non-vanishing alien derivative is $\Delta_{+A}$, in which case\footnote{Recall that $\Delta_{+A} \Phi_{(0,1)} = 0$, as this sector is not even asymptotic (it is just a constant).}
\begin{eqnarray}
\rme^{-\frac{A}{x}} \Delta_{+A} F &=& \sum_{n=0}^{+\infty} \tau_1^n\, \rme^{- \left( n+1 \right) \frac{A}{x}}\, \Delta_{+A} \Phi_{(n,0)} = S_{1}^{F}\, \sum_{n=0}^{+\infty} \tau_{1}^{n} \left( n+1 \right) \rme^{- \left( n+1 \right) \frac{A}{x}}\, \Phi_{(n+1,0)} = \nonumber \\
&=& S_{1}^{F} \frac{\partial}{\partial\tau_{1}} F,
\end{eqnarray}
\noindent
immediately leading to the action of the Stokes automorphism \eqref{one-param-stokes-0} as
\begin{equation}
\label{eq:Stokes-aut-zero-v2}
\underline{\mathfrak{S}}_{0} F (x, \tau_{1}, \tau_{2}) = \exp \left\{ S_{1}^F \frac{\partial}{\partial\tau_{1}} \right\} F (x, \tau_{1}, \tau_{2}).
\end{equation}
\noindent
A completely analogous calculation holds along the direction $\theta=\pi$. One finds (with $k>0$)
\begin{eqnarray}
\rme^{k\frac{A}{x}} \Delta_{-kA} F &=& S_{-k}^F\, \sum_{n=k}^{+\infty} \tau_1^n \left(n-k\right) \rme^{- \left( n-k \right) \frac{A}{x}}\, \Phi_{(n-k,0)} + \widetilde{S}_{-k}^F\, \tau_1^k\, \Phi_{(0,1)} = \nonumber \\
&=& \left( S_{-k}^{F}\, \tau_1^{k+1} \frac{\partial}{\partial\tau_1} + \widetilde{S}_{-k}^F\, \tau_1^k \frac{\partial}{\partial\tau_2} \right) F,
\label{Delta-kAF:generic}
\end{eqnarray}
\noindent
now leading to the action of the Stokes automorphism \eqref{one-param-stokes-pi} as
\begin{equation}
\label{eq:Stokes-aut-pi-v2}
\underline{\mathfrak{S}}_{\pi} F (x, \tau_{1}, \tau_{2}) = \exp \left\{ \sum_{k=1}^{+\infty} \left( S_{-k}^{F}\, \tau_1^{k+1} \frac{\partial}{\partial\tau_1} + \widetilde{S}_{-k}^F\, \tau_1^k \frac{\partial}{\partial\tau_2} \right) \right\} F (x, \tau_{1}, \tau_{2}).
\end{equation}

This is all the needed information in order to compare Stokes constants associated to the partition function and to the free energy. In fact, \eqref{eq:Stokes-aut-zero-v1} and \eqref{eq:Stokes-aut-zero-v2} should yield the same result, as should \eqref{eq:Stokes-aut-pi-v1} and \eqref{eq:Stokes-aut-pi-v2}. Making sure such identifications hold will automatically compute the Stokes constants. For instance, from \eqref{eq:Stokes-aut-zero-v2} follows
\begin{equation}
\underline{\mathfrak{S}}_{0} F (x, \tau_1, \tau_2) = F \left(x, \tau_1+S_{1}^{F}, \tau_{2}\right), 
\end{equation}
\noindent
which, when compared to \eqref{eq:Stokes-aut-zero-v1}, immediately yields the Stokes constant $S_{1}^{F}$
\begin{equation}
\label{eq:Stokes-consts-relation-1}
S_{1}^{F} = S_{1}^{Z} \quad (\, = - 2\, ).
\end{equation}

For the second comparison, that of \eqref{eq:Stokes-aut-pi-v1} and \eqref{eq:Stokes-aut-pi-v2}, the action of the exponential in \eqref{eq:Stokes-aut-pi-v2} is now more intricate and we begin by expanding it as\footnote{Note that $\widetilde{S}_{-k}^F\, \tau_1^k \frac{\partial}{\partial\tau_2}$ only acts once upon the transseries, as its $\tau_2$-dependence is purely linear.} (using the explicit transseries in, \textit{e.g.}, \eqref{eq:Free-en-two-param-transs})
\bea
\exp \left\{ \sum_{k=1}^{+\infty} \left( S_{-k}^{F}\, \tau_1^{k+1} \frac{\partial}{\partial\tau_1} + \widetilde{S}_{-k}^F\, \tau_1^k \frac{\partial}{\partial\tau_2} \right) \right\} F (x, \tau_{1}, \tau_{2}) &=& \tau_2 + \sum_{\ell=1}^{+\infty} \frac{1}{\ell!} \left( \sum_{k=1}^{+\infty} S_{-k}^{F}\, \tau_1^{k+1} \frac{\partial}{\partial\tau_1} \right)^{\ell-1} \times \nonumber \\
&&
\hspace{-200pt}
\times \sum_{k'=1}^{+\infty}\widetilde{S}_{-k'}^F\, \tau_1^{k'} + \sum_{n=0}^{+\infty} \sum_{\ell=0}^{+\infty} \frac{1}{\ell!} \left( \sum_{k=1}^{+\infty} S_{-k}^{F}\, \tau_1^{k+1} \frac{\partial}{\partial\tau_1}\right)^{\ell} \tau_1^n\, \rme^{-\frac{n A}{x}}\, \Phi_{(n,0)} (x).
\eea
\noindent
Clearly, in order to compare back to the resulting (shifted) transseries in \eqref{eq:Stokes-aut-pi-v1}, we will also need to expand that expression. One simply obtains
\bea
\label{Stokes-freeen-pi-from-partfunc}
F \left(x, \frac{\tau_1}{1 + S_{-1}^{Z} \tau_{1}}, \tau_{2} + \log \left( 1 + S_{-1}^{Z} \tau_{1} \right) \right) &=& \\
&&
\hspace{-100pt}
= \tau_2 + \log \left( 1 + S_{-1}^{Z} \tau_{1} \right) + \sum_{n=0}^{+\infty} \left( \frac{\tau_1}{1 + S_{-1}^{Z} \tau_{1}} \right)^n \rme^{-\frac{n A}{x}}\, F^{(n)} (x). \nonumber
\eea
\noindent
The above two expressions will match if, upon the standard identification \eqref{transseries-Phin0-vs-Fn}, one further has
\begin{eqnarray}
\label{Stokes-comp-1st}
\log \left( 1 + S_{-1}^{Z} \tau_{1} \right) &=& \sum_{k=1}^{+\infty} \widetilde{S}_{-k}^F \left(\, \sum_{\ell=1}^{+\infty} \frac{1}{\ell!} \left\{ \sum_{k'=1}^{+\infty} S_{-k'}^{F}\, \tau_1^{k'+1} \frac{\partial}{\partial\tau_1} \right\}^{\ell-1}\, \right) \tau_1^{k}, \\
\frac{\tau_1^n}{\left( 1 + S_{-1}^{Z} \tau_{1} \right)^n} &=& \left(\, \sum_{\ell=0}^{+\infty} \frac{1}{\ell!} \left\{ \sum_{k=1}^{+\infty} S_{-k}^{F}\, \tau_1^{k+1} \frac{\partial}{\partial\tau_1} \right\}^{\ell}\, \right) \tau_1^n, \qquad \forall n\ge1.
\label{Stokes-comp-2nd}
\end{eqnarray}
\noindent
While at first these may seem as intricate conditions, we shall see in the following that there is a very simple set of values for the Stokes constants $S_{-k}^F$ and $\widetilde{S}_{-k}^F$, $k\ge1$, which solve both conditions. Let us exemplify how this works starting with \eqref{Stokes-comp-2nd} and computing the $S_{-k}^F$. For $n=1$, a power-series expansion of the left-hand side, together with the recursive action of the differential operator on the right-hand side, leads to
\begin{eqnarray}
\label{Stokes-comp-powers-tau1}
\sum_{\ell=0}^{+\infty} \left(-S_{-1}^{Z}\right)^{\ell} \tau_{1}^{\ell+1} & = & \tau_{1} + \sum_{k_{1}=1}^{+\infty} S_{-k_{1}}^{F}\, \tau_{1}^{k_{1}+1} + \frac{1}{2!} \sum_{k_{1},k_{2}=1}^{+\infty} \left( k_{1}+1 \right) S_{-k_{1}}^{F} S_{-k_{2}}^{F}\, \tau_{1}^{k_{1}+k_{2}+1} + \\
&+&
\frac{1}{3!} \sum_{k_{1},k_{2},k_{3}=1}^{+\infty} \left( k_{1}+1 \right) \left( k_{1}+k_{2}+1 \right) S_{-k_{1}}^{F} S_{-k_{2}}^{F} S_{-k_{3}}^{F}\, \tau_{1}^{\sum_{i=1}^{3}k_{i}+1} + \cdots.
\nonumber
\end{eqnarray}
\noindent
Let us compare equal powers of $\tau_1^{\ell+1}$. For $\ell=0$ we get an identity; for the next few $\ell$'s we find
\begin{eqnarray}
\label{Stokes-comp-one}
\ell = 1: & -S_{-1}^{Z} & = S_{-1}^{F}, \\
\label{Stokes-comp-two}
\ell = 2: & \left( -S_{-1}^{Z} \right)^{2} & = S_{-2}^{F} + \frac{1}{2!} \left( 1+1 \right) \left( S_{-1}^{F} \right)^{2}, \\
\label{Stokes-comp-three}
\ell = 3: & \left( -S_{-1}^{Z} \right)^{3} & = S_{-3}^{F} + \frac{1}{2!} \left( 2+3 \right) S_{-1}^{F} S_{-2}^{F} + \frac{1}{3!} \left( 2 \times 3 \right) \left( S_{-1}^{F} \right)^{3}.
\end{eqnarray}
\noindent
The first equation above, \eqref{Stokes-comp-one}, immediately yields $S_{-1}^{F} = -S_{-1}^{Z}$. Recursively solving the following equations one sees that \eqref{Stokes-comp-two} then yields $S_{-2}^{F} =0$ and \eqref{Stokes-comp-three} $S_{-3}^{F} =0$. This is a pattern which generalizes to higher powers of $\tau_1$. Indeed, the first time the Stokes constant $S_{-k}^F$ will enter the game is when we write down the equation associated to $\ell=k$. But, at this stage, such equation will also include a term just in $\left(S_{-1}^F\right)^k$, while all other terms will be products of Stokes constants which always include at least one\footnote{The $m$-term contribution to the sum in \eqref{Stokes-comp-2nd} is given by
\begin{equation}
\frac{1}{m!} \prod_{i=1}^{m} \left(\, \sum_{k_{i}=1}^{+\infty} \left(\, \sum_{j=1}^{i-1}k_{j}+1\, \right) S_{-k_{i}}^{F}\, \right) \tau_{1}^{\sum_{j=1}^{m}k_{j}+1}.
\end{equation}
} 
$S_{-k'}^F$ with $1<k'<k$. For a general $\ell$ this looks like
\begin{equation}
\left( -S_{-1}^{Z} \right)^{\ell} = S_{-\ell}^{F} + \cdots + \frac{1}{\ell!}\, \ell!\left( S_{-1}^{F} \right)^{\ell},
\end{equation}
\noindent
where the dots indicate terms involving Stokes constants $S_{-k}^F$, $1<k<\ell$, which were already set to zero from previous conditions. By induction, it follows that $S_{-k}^F=0$, for all $k>1$. Finally, it is straightforward to show that this solution also solves the initial constraint \eqref{Stokes-comp-2nd} for any other value of $n\ge1$. This is indeed the case as
\begin{equation}
\sum_{\ell=0}^{+\infty} \frac{1}{\ell!} \left( S_{-1}^{F} \right)^{\ell} \left( \tau_{1}^{2} \frac{\partial}{\partial\tau_{1}} \right)^{\ell} \tau_{1}^{n} = \tau_{1}^{n}\, \sum_{\ell=0}^{+\infty} \frac{\left( n + \ell - 1 \right)!}{\ell! \left( n - 1 \right)!} \left( S_{-1}^{F} \tau_{1} \right)^{\ell} = \frac{\tau_{1}^{n}}{\left( 1 - S_{-1}^{F} \tau_{1} \right)^{n}}.
\end{equation}
\noindent
A completely analogous procedure can be used to solve the remaining constraint \eqref{Stokes-comp-1st} (where one also has to use the above solution for the $S_{-k}^F$), which we leave as an exercise for the reader. At the end of the day, one finally obtains
\begin{align}
\label{eq:Stokes-consts-relation-3}
S_{-1}^{F} &= - S_{-1}^{Z} &(\, &= - 1\,\, ), \\
\label{eq:Stokes-consts-relation-2}
\widetilde{S}_{-1}^{F} &= S_{-1}^{Z} &(\, &= 1\,\, ), \\
\label{eq:Stokes-consts-relation-4}
S_{-m}^{F} &= 0, &&m\ge2, \\
\label{eq:Stokes-consts-relation-5}
\widetilde{S}_{-m}^{F} &= 0, &&m\ge2.
\end{align}

These results tell us that a large number of Stokes constants for the free energy actually vanish. This will be numerically checked via large-order analysis in the next subsection. For the moment let us just note that this implies that some of the resurgence relations allowed for in \eqref{Delta-ka-one-point-five} are in fact trivial, \textit{e.g.},
\begin{equation}
\Delta_{nA}\Phi_{(m,0)}(x) \equiv 0, \qquad n\ne-1,1.
\end{equation}
\noindent
That this is the case could have been somewhat expected from the results concerning the linear problem, where the only non-zero alien operators already were $\Delta_{\pm A}$. In fact, from the transseries comparisons of \eqref{eq:Free-en-from-part-func} and \eqref{eq:Free-en-two-param-transs} we learnt that the $\Phi_{(n,0)} (x)$ sectors are given by the combinations $F^{(n)} (x)$ in \eqref{eq:Free-en-sectors-instantons}, and these are ratios of two asymptotic series. Now, Borel transforms for these types of contributions are known to be given by convolutions of the Borel transforms of the original asymptotic series: as the inverse Borel transform is a Laplace transform, it is simple to show that the Borel of the product is a convolution, \textit{i.e.}, that $\CB [ \Phi_1 \times \Phi_2 ] = \CB [\Phi_1] * \CB [\Phi_2]$, with $*$ the convolution of functions. This allows us to relate Borel singularities of both $\Phi_1$ and $\Phi_2$ with Borel singularities of their product. In fact (see, \textit{e.g,} \cite{s14}), the convolution will inherit the original singularities of each of the asymptotic series in the principal branch of the Borel plane, together with combinations of singularities, but which will fall on different sheets of the Borel plane, thus not contributing directly to the Stokes automorphism\footnote{If $\Phi_{1}$ and $\Phi_{2}$ are asymptotic series with singularities on the Borel plane at $s=\omega_{1}$ and $s=\omega_{2}$, respectively, then the product $\Phi_{1} \times \Phi_{2}$ will have singularities on the Borel plane at $s=\omega_{1},\omega_{2},\omega_{1}+\omega_{2}$. Nevertheless, only the original singularities $s=\omega_{1},\omega_{2}$ will be on the principal branch in the Borel plane; the one at $s=\omega_{1}+\omega_{2}$ will fall on a different sheet, and thus the alien derivative acting on the product will not see this extra singularity \cite{s14}.}.

Another way to understand the discussion in the above paragraph is the following. As we have the solution for the non-vanishing Stokes constants in hand, we may now rewrite the Stokes automorphism \eqref{eq:Stokes-aut-pi-v2} in a simpler form, as
\bea
\label{eq:Stokes-aut-pi-v3}
\underline{\mathfrak{S}}_{\pi} F (x, \tau_{1}, \tau_{2}) &=& \exp \left\{S_{-1}^{F}\, \tau_1^{2} \frac{\partial}{\partial\tau_1} + \widetilde{S}_{-1}^F\, \tau_1 \frac{\partial}{\partial\tau_2} \right\} F (x, \tau_{1}, \tau_{2}) = \\
&=& \exp \left\{ - \tau_1^{2} \frac{\partial}{\partial\tau_1} + \tau_1 \frac{\partial}{\partial\tau_2} \right\} F (x, \tau_{1}, \tau_{2}).
\eea
\noindent
Alongside \eqref{eq:Stokes-aut-zero-v2}, this equation is telling us something very interesting. They show how the Stokes automorphisms for the free energies, \eqref{eq:Stokes-aut-zero-v2} and \eqref{eq:Stokes-aut-pi-v3}, turn out to be the \textit{very same operators} as those acting at the level of the partition function, \eqref{stokes-zero-sigmas} and \eqref{stokes-pi-sigmas}. This is immediately made explicit once one uses the respective relation between transseries parameters \eqref{transseries-sigma-vs-tau}, to find
\begin{eqnarray}
\sigma_{0} \frac{\partial}{\partial\sigma_{1}} &=& \frac{\partial}{\partial\tau_1}, \\
\sigma_{1} \frac{\partial}{\partial\sigma_{0}} &=& - \tau_{1}^2 \frac{\partial}{\partial\tau_{1}} + \tau_{1} \frac{\partial}{\partial\tau_{2}}.
\end{eqnarray}

There is yet another way to see how all this is not a big surprise, but rather a straightforward consequence\footnote{In all rigour, this is more of a consistency check that $\Delta_\omega$ is a derivation, as we have not addressed such proof.} of the fact that $\Delta_\omega$ is a derivation. If $\Delta_\omega$ is a derivation and at some specific point $\omega$ one has $\Delta_\omega Z = 0$, then it must be the case that at this same point one finds $\Delta_\omega F = 0$, as $F = \log Z$. But then all the possibilities encoded in \eqref{Delta-kAF:generic} must immediately simplify, as $\Delta_{-kA} Z = 0$ for $k\ge2$. In fact, this condition sets $\Delta_{-kA} F = 0$ for $k\ge2$, which further yields $S_{-k}^{F} = 0$ and $\widetilde{S}_{-k}^{F} = 0$ for $k\ge2$ (indeed, as was later found in \eqref{eq:Stokes-consts-relation-4} and \eqref{eq:Stokes-consts-relation-5}). In other words, the resurgence relations for the partition function, \eqref{Delta+A} and \eqref{Delta-A}, alongside the fact that $\Delta_\omega$ is a derivation, are enough to tell us that the only non-zero result in \eqref{Delta-kAF:generic} must be
\be
\rme^{\frac{A}{x}} \Delta_{-A} F = \left( S_{-1}^{F}\, \tau_1^{2} \frac{\partial}{\partial\tau_1} + \widetilde{S}_{-1}^F\, \tau_1 \frac{\partial}{\partial\tau_2} \right) F.
\ee
\noindent
But this is precisely what we just had in \eqref{eq:Stokes-aut-pi-v3} above, as expected.

\subsection{Asymptotics and Large-Order Behaviour: Free Energy}\label{sec:large-order-F}

Having discussed all basic ideas and constructed a transseries solution for the free energy, displaying an infinite number of (multi) instanton sectors, we may finally turn to a rather interesting feature of resurgent transseries: the fact that these constructions may be very explicitly \textit{tested}, by means of resurgent large-order numerical analysis. In particular the tests we shall present next will very precisely confirm the structure we have put forward in the previous subsections.

The numerical tests we wish to perform are tests on the asymptotics of the different mutli-instantonic sectors, as predicted by the resurgent structure of the transseries we have proposed, and given known coefficients associated to \textit{other} sectors than the one we are testing (\textit{e.g.}, computed by the recursive formulae in appendix~\ref{app:recursive-free-en}). But before diving into numerics, one must pay attention to a couple of details concerning these coefficients. Comparison of the actual quartic free-energy transseries \eqref{one-point-five-transseries} with its ``one-parameter'' rewriting \eqref{two-into-one-param-quartic} and \eqref{two-into-one-param-sectors-quartic}, tells that the multi-loop multi-instanton coefficients $F_k^{(n,0)}$ will obey the same recursive equations as the $F_k^{(n)}$ coefficients. However, for this match to remain precise, one also has to keep in mind that for instance the coefficient $F_0^{(0)}$ should actually be thought of as $F_0^{(0,0)} + \sigma_2\, F_0^{(0,1)}$, and that none of these will be fixed by the recursion relations. Instead, they belong to the set of initial conditions one needs to account for, in order to specify a solution to our differential equation \eqref{quarticNLODE}. It turns out things are slightly simpler as we may directly compare free energy and partition function. For instance, the comparisons \eqref{transseries-Phi00-vs-logPhi0} to \eqref{transseries-Phi01-vs-1} immediately yield a relation between initial data for the free energies and the coefficients $Z_{0}^{(0)}$ and $Z_{0}^{(1)}$ of the partition function. In particular,
\be
F_0^{(1,0)} = \frac{Z_0^{(1)}}{Z_0^{(0)}} = -\frac{\rmi}{\sqrt{2}},
\ee
\noindent
where we used the relation $Z_0^{(1)} = -\frac{\rmi}{\sqrt{2}}\, Z_0^{(0)}$. We shall use this value for the coefficient $F_0^{(1,0)}$ in all the large-order tests carried out in this subsection.

Now, in order to understand the nature of the aforementioned resurgent, large-order, numerical tests, let us go back to the Stokes discontinuities \eqref{StokesDisc} depicted in figure~\ref{stokescrossingfig}. As discussed in subsection~\ref{subsec:Fnonlinear}, the resurgent structure of the quartic transseries \eqref{one-point-five-transseries} tells us that the sectors which build-it-up will have either: no Stokes discontinuities (the case of the sector $\Phi_{(0,1)}$, as in \eqref{stokes-zeropi-0|1-quartic}); or else one single Stokes discontinuity along $\theta=0$ (the case of $\Phi_{(0,0)}$); or else two Stokes discontinuities, along both $\theta=0$ and $\pi$ (for every other sector, as in \eqref{stokes-zero-pert-quartic} and, \textit{e.g.}, \eqref{stokes-pi-1inst-quartic} or \eqref{stokes-pi-2inst-quartic} or \eqref{stokes-pi-3inst-quartic}). Next, as already discussed in subsection~\ref{subsec:Zasymptotics}, given a function $F(x)$ with discontinuities along some rays starting from the origin on the complex plane, say along directions $\theta=0$ and $\theta=\pi$, Cauchy's theorem translates to (assuming no contributions around infinity, also as already discussed in subsection~\ref{subsec:Zasymptotics})
\be
\label{Cauchy-two-disc}
F(x) = - \frac{1}{2\pi\rmi} \int_{0}^{+\infty} \rmd w\, \frac{\disc_{0}F(w)}{w-x} - \frac{1}{2\pi\rmi} \int_{0}^{-\infty} \rmd w\, \frac{\disc_{\pi}F(w)}{w-x}.
\ee
\noindent
For the function $F$ above one selects any sector within the transseries, $\Phi_{(n,m)}$ (of course in the case of the perturbative sector there is only one discontinuity, and the second term disappears). As should be very clear by now, we only know these sectors expressed as asymptotic series. We do know their Stokes discontinuities, but also these are given as sums over weighted sectors (\textit{e.g.}, \eqref{stokes-zero-pert-quartic}--\eqref{stokes-pi-3inst-quartic}), themselves only known as asymptotic series. So when comparing both sides of \eqref{Cauchy-two-disc} we will be comparing asymptotic series, resulting in \textit{large-order} relations.

Let us begin with the perturbative sector $F=\Phi_{(0,0)}$. Using the Stokes discontinuity \eqref{stokes-zero-pert-quartic} in the Cauchy formula \eqref{Cauchy-two-disc}, and making asymptotic series explicit, 
\be
\sum_{k=0}^{+\infty} F_{k}^{(0,0)}\, x^k  \simeq \sum_{n=1}^{+\infty}\frac{S_1^n}{2\pi\rmi}\, \int_{0}^{+\infty} \rmd w\, \frac{\rme^{-n \frac{A}{w}}}{w-x}\, \sum_{h=0}^{+\infty} F_{h}^{(n,0)}\, w^h.
\ee
\noindent
Formally expanding the denominator on the right-hand-side in powers of $x$, and performing the resulting $w$-integrations, one may compare equal powers of $x$ on both sides to obtain (after a slight rearrangement)\footnote{In here Stokes coefficients and Borel residues are essentially the same, up to a trivial transformation \eqref{stokesS-to-borelS+GEN}.}
\begin{eqnarray}
\label{large-order-pert-quartic}
F_{k}^{(0,0)}\, \frac{2\pi\rmi A^k}{\Gamma(k)} &\simeq& S_1 \left( F_0^{(1,0)} + \frac{A}{k-1}\, F_1^{(1,0)} + \frac{A^2}{(k-1)(k-2)}\, F_2^{(1,0)} + \cdots \right) + \\
&& + 2^{-k}\, S_1^2 \left( F_0^{(2,0)} + \frac{2A}{k-1}\, F_1^{(2,0)} + \frac{(2A)^2}{(k-1)(k-2)}\, F_2^{(2,0)} + \cdots \right) + \mathcal{O}(3^{-k}). \nonumber
\end{eqnarray}
\noindent
Holding at \textit{large} $k$, this expression\footnote{It is also simple to obtain the closed-form expression
\be
F_{k}^{(0,0)} \simeq \sum_{n=1}^{+\infty} \frac{\Gamma(k)}{\left( n A \right)^k}\, \frac{S_1^n}{2\pi\rmi}\, \sum_{h=0}^{+\infty} \frac{\Gamma(k-h)}{\Gamma(k)}\, F_{h}^{(n,0)} \left( n A \right)^{h}.
\ee
\noindent
Validity for $k\gg h$ reveals the \textit{large-order nature} of this relation. Note that the ratio of Gamma-functions appearing in the second sum is subleading: it grows as $\sim k^{-h}$. See, \textit{e.g.}, \cite{asv11} for more closed-form large-order relations.} is a \textit{large}-order relation. With the aforementioned slight rearrangement, we made explicit both the leading factorial growth and the subleading exponential growths. In fact a very distinctive feature in this type of expressions are the $\sim n^{-k}$ contributions: in the first line above one finds the leading (one-instanton) terms with $n=1$, while in the second line one already finds exponentially-suppressed contributions with the two-instanton damping $\sim 2^{-k}$, and then on. In this way, large-order power-law corrections are associated to multi-loops, at fixed instanton number; while large-order exponentially-suppressed corrections are associated to the different multi-instanton numbers.

As they follow from the Stokes discontinuities via simple application of Cauchy's theorem, these large-order relations may also be directly written down starting-off by analyzing allowed motions on our alien chain (which dictated the resurgent structure of the transseries). In fact the ``statistical mechanical'' rules for writing down large-order relations will be very similar to the ones for writing down Stokes discontinuities in subsection~\ref{subsec:Fnonlinear}, and equally rather general. In our present example we are dealing with the chain depicted in figure~\ref{fig:first-5-sector-alien-one-point-five}, and the ``functional component'' appearing in the Stokes discontinuity will now be essentially replaced by its integral, which we shall denote as the \textit{large-order factor} (the ``statistical component'' will be essentially the same). For a path on the chain, connecting the nodes $\Phi_{(n,0)}\rightarrow \Phi_{(m,c_m)}$, this is
\be
\label{large-order-factor}
\chi_{(n\rightarrow m)} (k) \equiv \frac{1}{(m-n)^k}\, \sum_{h=0}^{+\infty} \frac{\Gamma(k-h)}{\Gamma(k)}\, F_h^{(m,c_m)} \left( \left(m-n\right) A\right)^h, \qquad c_m = \left\{ \begin{array}{l} 1, \quad m=0 \\ 0, \quad m>0 \end{array} \right. .
\ee
\noindent
Using this definition, one may now write down generic large-order relations purely in terms of motions and data on the alien chain:

\vspace{10pt}
\setlength{\fboxsep}{10pt}
\Ovalbox{%
\parbox{14cm}{%
\vspace{5pt}

\textbf{Large-order relations:}  The large-order (large $k$) behaviour of the coefficients $F_k^{(n,0)}$, associated to the node $\Phi_{(n,0)}$, is given by a \textit{sum} over \textit{all} forward \textit{and} backward paths, linking nodes to the right $\Phi_{(m>n,0)}$ and to the left $\Phi_{(m<n,0)}$, \textit{in addition} to a sum over paths to the extra ``orthogonal'' node in the chain, $\Phi_{(0,1)}$.
        
\vspace{10pt}        
Each term in this sum ($\Phi_{(n,0)} \rightarrow \Phi_{(m,\ell=0,1)}$) can be decomposed into three factors:
\begin{itemize}
\item Leading growth-factor, dictated solely by the order $k$ of the coefficient:
\begin{equation*}
\frac{\Gamma(k)}{2\pi\rmi\, A^k}.
\end{equation*}
\noindent
\item Large-order factor \eqref{large-order-factor}, dictated by beginning and end nodes:
\begin{equation*}
\chi_{(n\rightarrow m)}(k).
\end{equation*}
\noindent
\item Statistical factor, sum over all the allowed paths $\mathcal{P}(n \rightarrow m)$ linking the nodes as in figure~\ref{fig:first-5-sector-alien-one-point-five}:
\begin{equation*}
\mathrm{SF}_{(n \rightarrow m)} \equiv \sum_{\mathcal{P}(n \rightarrow m)} \mathrm{CF} (\CP)\, w (\mathcal{P}).
\end{equation*}
\end{itemize}
\vspace{-5pt}
}}
\vspace{15pt}

\noindent
Spelled out in writing, the result is:
\be
\label{large-order-one-point-five}
F_k^{(n,0)}\, \frac{2\pi\rmi\, A^k}{\Gamma(k)} \simeq \sum_{m\ne n} \mathrm{SF}_{(n\rightarrow m)}\, \chi_{(n\rightarrow m)}(k).
\ee

Let us make a few remarks concerning the above large-order relation. First, as the statistical factors do not depend upon the order, $k$, and once the leading growth is factored out, the (subleading) large-order behaviour will be fully dictated by the large-order factors \eqref{large-order-factor}---implying that in some sense this is where all the non-trivial large-order growth is included. Now equation \eqref{large-order-factor} very sharply tells us that the further away two nodes are, the more exponentially suppressed the corresponding large-order factor will be, due to the term $\left( m-n \right)^{-k}$. One other interesting feature is how forward and backward motions appear at large order. When walking forward along the chain, $m-n>0$, but when walking backwards $m-n<0$ leading to an oscillatory behaviour as $(-1)^k$ at large $k$. Finally, note that the ratio of Gamma functions in \eqref{large-order-factor} may be easily expanded in powers of $k^{-1}$. This precisely corresponds to expanding the fractions inside the brackets in \eqref{large-order-pert-quartic}. In fact, rewriting the large-order relation for the perturbative series in this light, we find
\begin{eqnarray}
\label{large-order-pert-quartic-2}
F_{k}^{(0,0)}\, \frac{2\pi\rmi A^k}{\Gamma(k)} &\simeq& \mathrm{SF}_{(0\rightarrow1)} \left( F_0^{(1,0)} + \frac{F_1^{(1,0)} A}{k} + \frac{F_1^{(1,0)} A + F_2^{(1,0)} A^2}{k^2} + \cdots \right) + \\
&&
\hspace{-30pt}
+ 2^{-k}\, \mathrm{SF}_{(0\rightarrow2)} \left( F_0^{(2,0)} + \frac{F_1^{(2,0)} \left( 2A \right)}{k} + \frac{F_1^{(2,0)} \left( 2A \right) + F_2^{(2,0)} \left( 2A \right)^2}{k^2} + \cdots \right) + \mathcal{O}(3^{-k}). \nonumber
\end{eqnarray}
\noindent
The statistical factors are read directly from figure~\ref{fig:first-5-sector-alien-one-point-five}, as $\mathrm{SF}_{(0\rightarrow1)} = S_1$ and $\mathrm{SF}_{(0\rightarrow2)} = S_1^2$. The agreement with \eqref{large-order-pert-quartic} is then exact.

We may now go back to the discussion at the beginning of the present subsection, in particular the two main questions which are still open. First, having introduced so many mathematical tools to calculate discontinuities and such, how can we be absolutely sure that the structures we have put forward precisely describe our problem? Second, all statistical factors, defined using weights for the steps in each path, they all depend on a set of \textit{a priori} unknown coefficients---the Stokes coefficients---and we still need to make sure that they may be determined (either via the partition function as in the previous subsection, or else). As already suggested, one may now settle both these issues by performing \textit{numerical checks} of the previous asymptotic relations. Because such formulae have their origin in the resurgent structure associated to the transseries we started off from, then, if they hold, so will all that stands behind them hold. In order to understand what types of numerical tests may be performed, let us consider the expression above, \eqref{large-order-pert-quartic-2}. Having moved the leading growth-factor to the left-hand side, the right-hand side is now composed of a series of sequences, each one more and more exponentially suppressed. Each of these sequences is given by a number (the statistical factor) times the actual sequence itself,  in powers of $\frac{1}{k}$. At very large order, both exponential and power suppressed contributions fade away, and the right-hand side above approaches a constant value (which in here is $\mathrm{SF}_{(0\rightarrow1)} F_0^{(1,0)}$). Can we test this?

\begin{figure}[t!]
\begin{center}
\includegraphics[height=5.1cm]{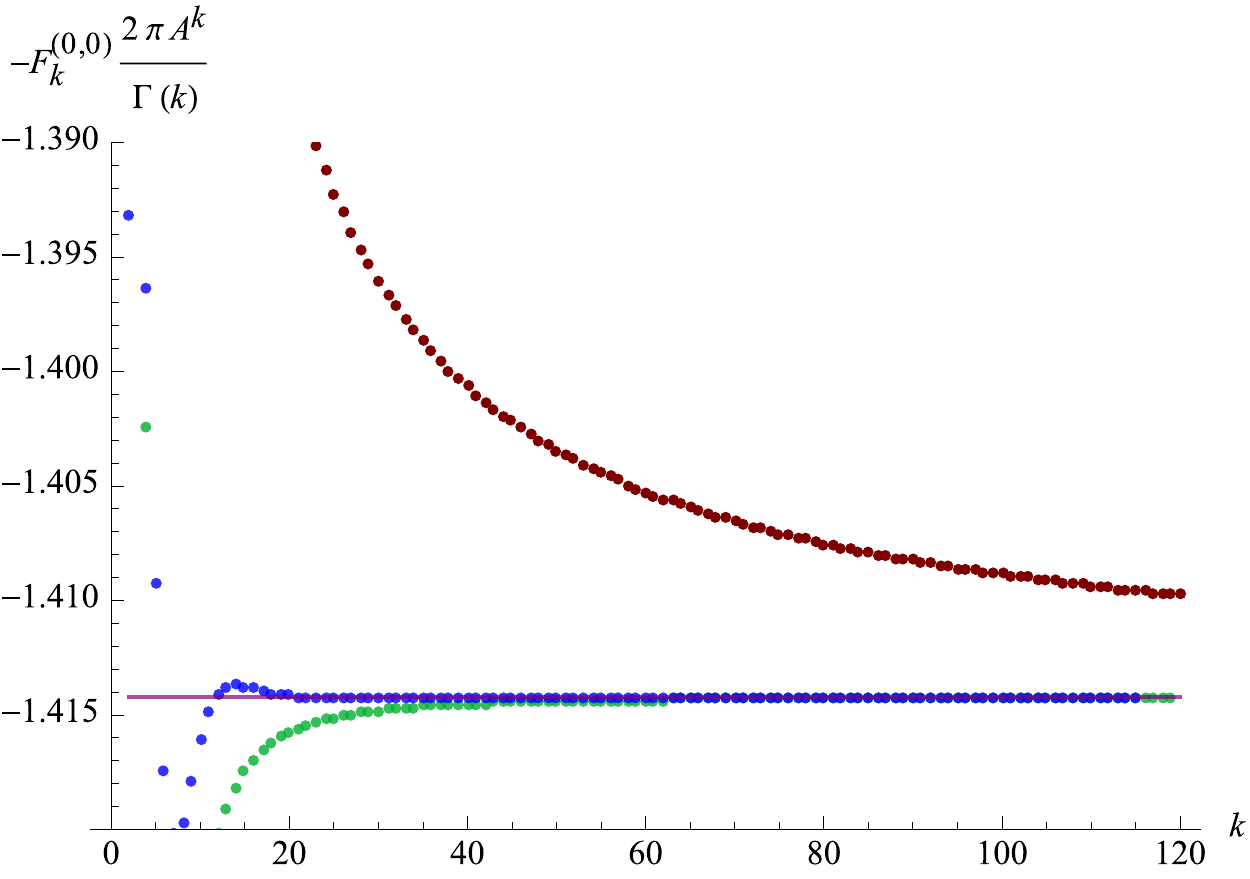}
$\qquad$
\includegraphics[height=5.1cm]{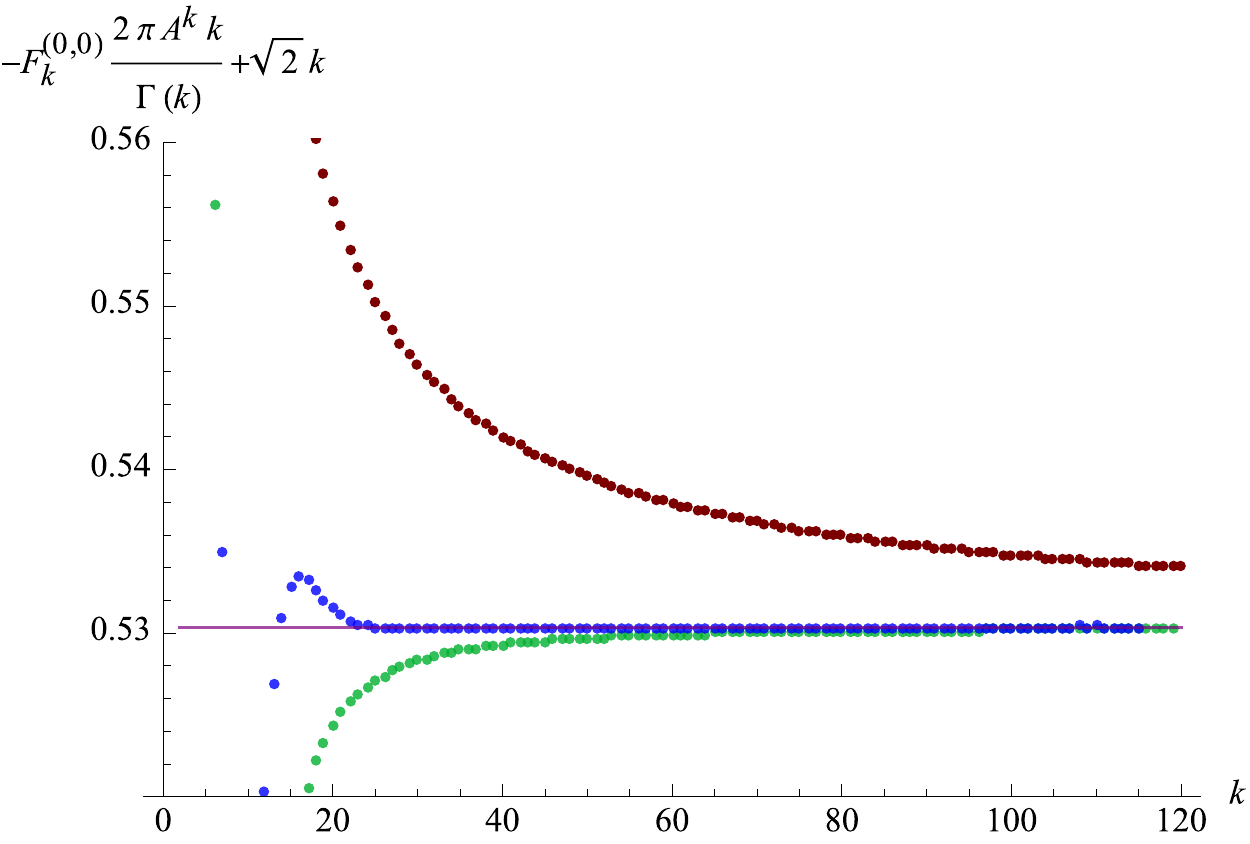}
\end{center}
\caption{Plot of the coefficients of leading (left) and $1/k$ (right) behaviour of the perturbative sector, $F_k^{(0,0)}$, weighted by the leading growth-factor. In red we plot the original sequences, while in green and blue we plot Richardson transforms $\mathrm{RT}_0 (r,k,1)$ and $\mathrm{RT}_0 (r,k,5)$, respectively ($r=0$ for the left plot and $r=1$ for the right plot). The purple lines show the constant values that these sequences are converging to: for the case of the left plot $\rmi S_1 F_0^{(1,0)} = - \sqrt{2}$; while for the case of the right plot $\rmi S_1 F_1^{(1,0)} A = \frac{3}{8} \sqrt{2}$ (both correct up to extremely small numerical errors). These plots also nicely illustrate the improved rate of convergence of each initial sequence, the higher the Richardson transform which was subsequently used.}
\label{fig:large-order-pert-series}
\end{figure}

In order to check \eqref{large-order-pert-quartic-2}, one first needs data. Using the recursive equations listed in the appendix, it is computationally straightforward to generate coefficients for both perturbative and one-instanton sectors, $F_k^{(0,0)}$ and $F_k^{(1,0)}$, and we have done this for $k\le 120$. Let us begin with the perturbative data. Plotting the left-hand side of \eqref{large-order-pert-quartic-2}, the larger the value $k$ the closer we expect that this sequence will approach the constant\footnote{Recall that even though $F_0^{(1,0)}$ is unknown from the perspective of the differential equation (it belongs to the initial conditions), we have set its value, via the partition function coefficients, to be $F_0^{(1,0)} = -\frac{\rmi}{\sqrt{2}}$.} $S_1 F_0^{(1,0)}$. In turn, this should allow us to determine, at least numerically, the value of the Stokes coefficient $S_1$. This is depicted in figure~\ref{fig:large-order-pert-series}, where the red curve shows the coefficients of the left-hand side of \eqref{large-order-pert-quartic-2} for growing $k$, and the purple line shows the constant value they are converging towards.

The first thing we notice is that this convergence is very slow! This brings us to our first parenthesis in this subsection, where we wish to digress a bit and discuss a simple numerical method to accelerate convergence of sequences. The type of sequences we are interested in are generically of the type appearing in \eqref{large-order-pert-quartic-2}, \textit{i.e.}, a leading constant term followed by a whole series in powers of $\frac{1}{k}$. A simple and powerful numerical method to accelerate convergence of this sort of sequences is that of \textit{Richardson transforms} (mentioned earlier; but see, \textit{e.g.}, \cite{bo78}). Richardson extrapolation starts with a $k$-sequence
\be
\label{sequence-large-order}
\BS (k) \simeq s_0 + \frac{s_1}{k} + \frac{s_2}{k^2} + \cdots,
\ee
\noindent
from where its $N^{\text{th}}$-Richardson transform, herein denoted by $\mathrm{RT}_{\BS} (0,k,N)$, is a new $k$-sequence which can be recursively defined as\footnote{A closed form expression for the $N^{\text{th}}$-Richardson transform is
\be
\mathrm{RT}_{\BS} (0,k,N) = \sum_{n=0}^N (-1)^{n+N}\, \frac{\left(k+n\right)^N}{n! \left(N-n\right)!}\, \BS(k+n).
\ee
}
\begin{eqnarray}
\mathrm{RT}_{\BS} (0,k,0) &=& \BS(k), \\
\mathrm{RT}_{\BS} (0,k,N) &=& \mathrm{RT}_{\BS} (0,k+1,N-1) + \frac{k}{N}\, \Big(\, \mathrm{RT}_{\BS} (0,k+1,N-1) - \mathrm{RT}_{\BS} (0,k,N-1)\, \Big), \\
&&
\hspace{+340pt} N \ge 1. \nonumber
\end{eqnarray}
\noindent
Loosely speaking, the Richardson transforms cancel the \textit{subleading} terms in the original sequence $\BS(k)$ up to order $k^{-N}$. This results in a much better numerical convergence towards the \textit{first} term of the original sequence (denoted by $s_0$ above), the larger both $k$ and $N$ are. A couple of words on notation. First, the apparently ``silent'' zero appearing in the definition above simply refers to the fact that the Richardson transforms we have set up are zooming-in on the $s_0$ coefficient of the original sequence. Second, we will be using Richardson extrapolation to check large-order relations involving multi-instanton coefficients, $F_k^{(n,0)}$. It will thus prove convenient to make explicit which series are we addressing, by using the notation $\mathrm{RT}_n (0,k,N)$ to signal Richardson transforms of the leading large-order of the $n$-instanton sector.

Let us now go back to the analysis of the perturbative sector, considering the case $\BS(k) \equiv F_k^{(0,0)}\, \frac{2\pi\rmi A^k}{\Gamma(k)}$, as in the left-hand side of \eqref{large-order-pert-quartic-2}. The left image in figure~\ref{fig:large-order-pert-series} depicts the first and fifth Richardson transforms (RTs) of this sequence, closing-in on the constant $\mathrm{SF}_{(0\rightarrow1)} F_0^{(1,0)} = S_1 F_0^{(1,0)}$. It is rather clear how the more RTs one considers, the faster the rate of convergence one observes. The best estimate we have produced for the value of this constant was upon taking the 5-RT, at order $k=115$. It so happens that in this case the value of the constant is also very easy to pinpoint as $\mathrm{SF}_{(0\rightarrow1)} F_0^{(1,0)} = S_1 F_0^{(1,0)} = \rmi \sqrt{2}$. Comparison with $\mathrm{RT}_0 (0,115,5)$ yields an extremely small error,
\be
\frac{\mathrm{RT}_0 (0,115,5) - \rmi \sqrt{2}}{\rmi \sqrt{2}} \approx 1.211978 \times 10^{-10}.
\ee
\noindent
Finally note that this value of $S_1 F_0^{(1,0)} = \rmi \sqrt{2}$ precisely corresponds to a value for the Stokes coefficient of $S_1^{F}=-2$, which is in complete agreement with the predicted value \eqref{eq:Stokes-consts-relation-1}.

This numerical check has confirmed the first (leading) term in the large-order expansion for the perturbative sector \eqref{large-order-pert-quartic-2}. But we would further like to check the subsequent terms---for the moment, at least at leading exponential order. This would verify that the exponentially-leading large-order behaviour is indeed dictated by \textit{all} the coefficients of the one instanton sector, $F_h^{(1,0)}$, as predicted by resurgence. Such checks follow in very similar fashion to the one above. In fact the first line of \eqref{large-order-pert-quartic-2} is precisely of the form \eqref{sequence-large-order}, and it is rather simple to check the value of $s_1$ in this latter sequence. All one has to do is to repeat the procedure above, this time applied to the (trivially) modified sequence
\be
\BS_1(k) \equiv \left( \BS(k) - s_0 \right) k \simeq s_1 + \frac{s_2}{k} + \frac{s_3}{k^2} + \cdots.
\ee
\noindent
The RT is defined in a completely similar way. More generally, we can recursively check every coefficient in the sequence \eqref{sequence-large-order} simply by defining
\be
\BS_r (k) \equiv \left( \BS_{r-1}(k) - s_{r-1} \right) k \simeq s_r + \frac{s_{r+1}}{k} + \frac{s_{r+2}}{k^2} + \cdots, \qquad r \ge 1,
\ee
\noindent
and the corresponding RTs
\begin{eqnarray}
\mathrm{RT}_{\BS} (r,k,0) &=& \BS_r (k), \\
\mathrm{RT}_{\BS} (r,k,N) &=& \mathrm{RT}_{\BS} (r,k+1,N-1) + \frac{k}{N}\, \Big(\, \mathrm{RT}_{\BS} (r,k+1,N-1) - \mathrm{RT}_{\BS} (r,k,N-1)\, \Big), \\
&&
\hspace{+340pt} N \ge 1. \nonumber
\end{eqnarray}

One may now proceed in the exact same way as above, checking each coefficient in \eqref{large-order-pert-quartic-2}. The right image in figure~\ref{fig:large-order-pert-series} is checking the coefficient $r=1$, where we have obtained a convergence towards the expected value of $s_1 \equiv \mathrm{SF}_{(0\rightarrow1)} F_1^{(1,0)} A = -\frac{3\rmi}{8}\sqrt{2}$, with extremely small error
\be
\frac{\mathrm{RT}_0 (1,115,5) - s_1}{s_1} \approx -3.142630 \times 10^{-9}.
\ee
\noindent
Checking the subsequent coefficients may be easily done in the very same way, albeit sometimes one does need to use a higher RT to keep good accuracy. For instance, at six-loops the convergence towards $s_{5} = -\frac{6767133\rmi}{131072\sqrt{2}}$ (do recall we are starting with the sequence $\BS(k) \equiv F_k^{(0,0)}\, \frac{2\pi\rmi A^k}{\Gamma(k)}$) again yields an extremely small error
\be
\frac{\mathrm{RT}_0 (5,110,8) - s_{5}}{s_{5}} \approx 1.818376 \times 10^{-8}.
\ee
\noindent
One word of caution on the nature of these numerical tests should still be said. As per definition, the coefficients in the sequence $\BS(k)$ from \eqref{large-order-pert-quartic-2} grow factorially---this is simply due to the appearance of the one-instanton series, $F_h^{(1,0)}$. As such, this can lead to large numerical errors when checking higher multi-loop coefficients, unless one is careful from scratch in implementing an elevated numerical accuracy in the calculations.

Proceeding along these lines, one eventually arrives at a stage where one has very strong checks on the validity of the first line in \eqref{large-order-pert-quartic-2}, and thus the fact that it was correctly predicted by the resurgent structure of the transseries we started off with. This check corresponds to verifying the correctness of the term $\mathrm{SF}_{(0\rightarrow1)}\, \chi_{(0\rightarrow1)}$ in \eqref{large-order-one-point-five}, which is the exponentially leading motion in the large-order relations of the perturbative sector. But one may likewise naturally wonder if it is possible to check all other motions, even if exponentially suppressed? In other words, may one also check, say, the (exponentially suppressed) second line in \eqref{large-order-pert-quartic-2}? The first thing to notice is that in order to probe the $2^{-k}$ behaviour, we first need to subtract out of the original series all contribution coming from the $\mathrm{SF}_{(0\rightarrow1)}\, \chi_{(0\rightarrow1)}$ motion, \textit{i.e.}, to subtract out the first line in \eqref{large-order-pert-quartic-2}. But this line corresponds to an \textit{asymptotic} expansion on its own (more generically, do recall \eqref{large-order-factor}), and in order to subtract out something which is actually meaningful one first needs to re-sum this asymptotic series into some \textit{number}.

This brings us to our second parenthesis in this subsection, where we wish to digress a bit and discuss resummations of asymptotic series. The most straightforward method to re-sum asymptotic series is that of \textit{optimal truncation}, where the series is cut-off essentially at the point where its coefficients start growing. While this method may sometimes yield reasonable accuracy, for the purposes of this review it does not reach the required accuracy for our tests, and we will not dwell on it any further\footnote{The reader who might be interested to see this in an example may take a look at the study of Painlev\'e~I solutions in \cite{asv11}. In this reference and within that example it was shown how the accuracy of the optimal truncation method was exactly of the same order as the first exponentially-suppressed term, and would thus be blind to the full large-order structure of the $k$-sequences associated to any exponentially suppressed contributions.}. A more intricate resummation method, which in fact reaches the accuracy we are looking for, is the method of \textit{Borel--Pad\'e} approximants. As already discussed in an earlier subsection, given an asymptotic series $\BS(k)$, with zero radius of convergence, its Borel transform
\be
\CB [\BS] (t) = \sum_{m=0}^{+\infty} \frac{s_m}{\Gamma(m+1)}\, t^m
\ee
\noindent
will instead have a (finite) non-zero radius of convergence. Of course as long as there are no closed-form expressions for all the coefficients $s_m$, one cannot precisely pinpoint the above analytic function. However, using the recursion relations for, \textit{e.g.}, the free energy coefficients listed in the appendix, we have computationally calculated them in our example up to $2N=120$. If we make sole use of these resulting coefficients, what we have managed to compute is
\be
\label{borel-S}
\CB [\BS] (t) \approx \sum_{m=0}^{2N} \frac{s_m}{\Gamma(m+1)}\, t^m.
\ee
\noindent
Now instead of immediately ``inverting'' this result via Borel resummation \eqref{borelresum} (which at this point would only allow us to recover these first $2N$ terms of the asymptotic expansion $\BS(k)$), we shall first approximate the above \textit{truncated} Borel transform by an order-$N$ diagonal Pad\'e approximant, as in:
\be
\label{pade-S}
\mathrm{BP}_N [\BS] (t) = \frac{\sum_{\ell=0}^{N} a_n\, t^n}{\sum_{\ell=0}^{N} b_n\, t^n},
\ee
\noindent
where the coefficients $a_n$, $b_n$ are chosen to match the ones in the expansion of the Borel transform in a small-$t$ expansion. The main reason for choosing this rational function instead of the expansion \eqref{borel-S} is that, while for small $t$ they are basically the same, for large $t$ the rational function converges to a constant and is much better behaved\footnote{When performing Borel resummation \eqref{borelresum}, we need to integrate over the range of both $t$ small \textit{and} large!}. Moreover, the poles of this rational function will also prove to be very good approximations (upon ``condensation'') to the branch-cuts of the Borel transform. It is then to the rational function \eqref{pade-S} which we apply Borel resumation, along the positive real axis (as the original variable $k$ is the order of the coefficients, and is thus positive),
\be
\label{SBP-Laplace}
\CS_0 \mathrm{BP}_{N} [\BS] (k) = \int_0^{+\infty} \rmd t\, \mathrm{BP}_N [\BS] \left( \frac{t}{k} \right) \rme^{-t}.
\ee
\noindent
As should be familiar by now, this integration can only be performed when there are no poles of $\mathrm{BP}_N [\BS] (t)$ along the positive real axis of the Borel $t$-plane. However, this is very many times the case, and in such situations one needs to shift the integration contour off the real axis, say by performing a lateral Borel resummation, \textit{e.g.}, $\CS_{0^{-}} \mathrm{BP}_{N} [\BS] (k)$. Imaginary contributions may then appear, and one has to be careful handling them as we shall further discuss below.

\begin{figure}[t!]
\begin{center}
\includegraphics[height=4.9cm]{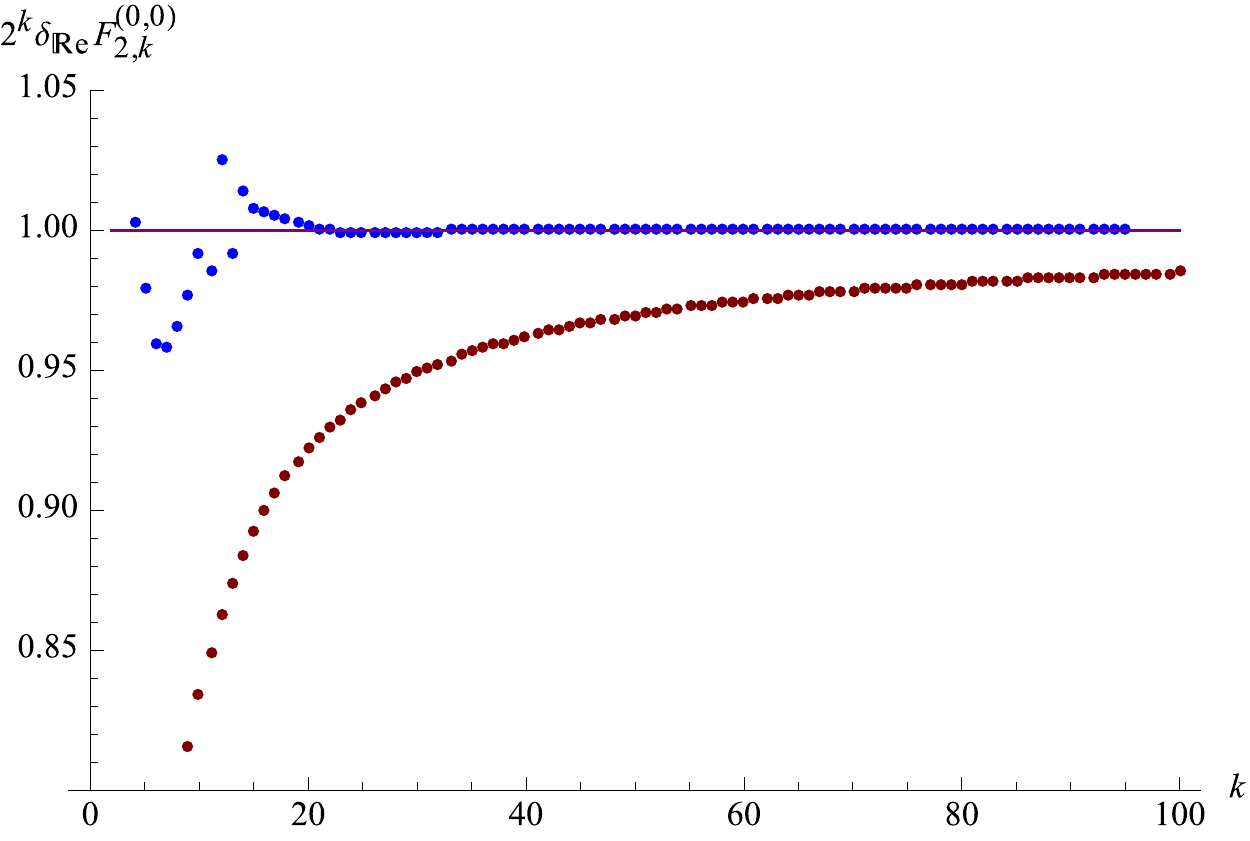}
$\qquad$
\includegraphics[height=4.9cm]{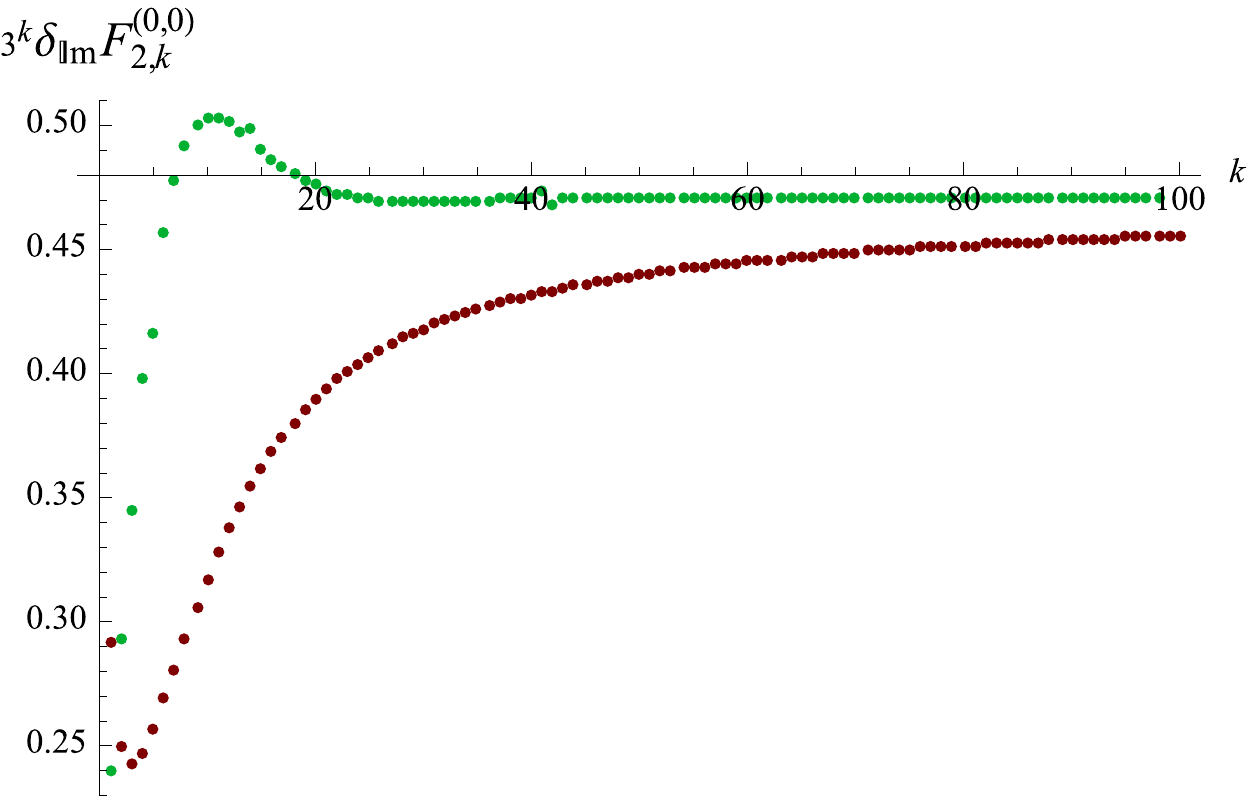}
\end{center}
\caption{The image on the left, plots the leading behaviour of the coefficients $2^k\, \delta_{\re} F^{(0,0)}_{2,k} \equiv - 2^{k}\,\mathrm{SF}_{(0\rightarrow1)} \times \CS_{\re} \mathrm{BP}_{60} [\chi_{(0\rightarrow1)}] (k)$. In red we plot the original sequence, while in blue we plot its $\mathrm{RT}_{0,2} (0,k,5)$ Richardson transform. The purple line shows the constant value that this sequence is converging towards, $S_1^2 F_0^{(2,0)}= 1$. The image on the right, plots the leading behaviour of the coefficients $3^k\, \delta_{\im} F^{(0,0)}_{2,k} \equiv \im \left( F_k^{(0,0)}\, \frac{2\pi\rmi \left(3A\right)^k}{\Gamma(k)} \right) - 3^{k}\, \mathrm{SF}_{(0\rightarrow1)} \times \CS_{\im} \mathrm{BP}_{60} [\chi_{(0\rightarrow1)}] (k)$, in red, and their first Richardson transform, in green. It is very clear how this sequence is converging towards a constant value of order $\sim \CO (1)$, which immediately validates the expected $3^{-k}$ decay.}
\label{fig:large-order-pert-series-1inst-resum}
\end{figure}

Let us now go back to our ongoing analysis of the perturbative large-order growth, and use the method described above to check exponentially-suppressed contributions to its asymptotics. If we start by assuming that there will be no poles upon the real axis, and that the Laplace-type integral \eqref{SBP-Laplace} can thus be evaluated without any shift in the integration contour, then the procedure should be rather clear. Applying the Borel--Pad\'e (BP) resummation to the asymptotic series $\chi_{(0\rightarrow1)} (k)$, we expect to obtain a resulting meaningful quantity which may be \textit{subtracted} out of the original perturbative sequence $F_k^{(0,0)}\, \frac{2\pi\rmi A^k}{\Gamma(k)}$. Looking back at \eqref{large-order-pert-quartic-2}, one writes\footnote{Recall that $\mathrm{SF}_{(0\rightarrow2)} = S_1^2$.}
\bea
\label{Fk00BP-1}
F_k^{(0,0)}\, \frac{2\pi\rmi A^k}{\Gamma(k)} - \mathrm{SF}_{(0\rightarrow1)} \times  \CS_0 \mathrm{BP}_{N} [\chi_{(0\rightarrow1)}] (k) &\simeq& \\
&&
\hspace{-200pt}
\simeq 2^{-k}\, \mathrm{SF}_{(0\rightarrow2)} \left( F_0^{(2,0)} + \frac{F_1^{(2,0)} \left( 2A \right)}{k} + \frac{F_1^{(2,0)} \left( 2A \right) + F_2^{(2,0)} \left( 2A \right)^2}{k^2} + \cdots \right) + \mathcal{O}(3^{-k}), \nonumber
\eea
\noindent
and the analysis would now proceed akin to what we did earlier. But, as we said, in many situations, including the case we are now addressing, the BP approximant to $\chi_{(0\rightarrow1)}$ will indeed have poles along the positive real axis. We shall deal with them by simply performing a lateral resummation, $\CS_{0^{-}} \mathrm{BP}_{N} [\chi_{(0\rightarrow1)}] (k)$, which integrates right \textit{below} the real line. This will then produce a result with both real \textit{and} imaginary components. As it turns out, the real contribution will be similar to above, and can be compared to the exponential order $2^{-k}$ of the large-order behaviour of the perturbative series, just as in \eqref{Fk00BP-1}. But the imaginary contribution, on the other hand, cancels with the perturbative coefficients up to order $3^{-k}$ and is thus irrelevant at the order we are interested in (see also \cite{asv11}). Let us make this explicit. Writing
\be
\label{S-BP=Re+Im-compare}
\CS_{0^{-}} \mathrm{BP}_{N} [\chi_{(0\rightarrow1)}] (k) = \CS_{\re} \mathrm{BP}_{N} [\chi_{(0\rightarrow1)}] (k) + \rmi\, \CS_{\im} \mathrm{BP}_{N} [\chi_{(0\rightarrow1)}] (k),
\ee
\noindent
then, splitting imaginary (first line) and real (second and third lines) components, what we find is
\begin{eqnarray}
F_k^{(0,0)}\, \frac{2\pi A^k}{\Gamma(k)} - \mathrm{SF}_{(0\rightarrow1)} \times \CS_{\im} \mathrm{BP}_{N} [\chi_{(0\rightarrow1)}] (k) &\simeq& \mathcal{O}(3^{-k}), \\
- \mathrm{SF}_{(0\rightarrow1)} \times \CS_{\re} \mathrm{BP}_{N} [\chi_{(0\rightarrow1)}] (k) &\simeq& \\
&&
\hspace{-188pt}
\simeq 2^{-k}\, \mathrm{SF}_{(0\rightarrow2)} \left( F_0^{(2,0)} + \frac{F_1^{(2,0)} \left( 2A \right)}{k} + \frac{F_1^{(2,0)} \left( 2A \right) + F_2^{(2,0)} \left( 2A \right)^2}{k^2} + \cdots \right) + \mathcal{O}(3^{-k}).
\nonumber
\end{eqnarray} 
\noindent
That this is indeed the resulting behaviour can be tested numerically, and is illustrated in figure~\ref{fig:large-order-pert-series-1inst-resum} for a BP of $N=60$. On its left image, we plot the convergence of the real part of the coefficients (do note that they are already properly weighted by the exponential term $2^k$)
\be
2^k\, \delta_{\re} F^{(0,0)}_{2,k} \equiv - 2^{k}\,\mathrm{SF}_{(0\rightarrow1)} \times \CS_{\re} \mathrm{BP}_{N} [\chi_{(0\rightarrow1)}] (k)
\ee
\noindent
towards the leading term within the two-instanton contribution, $\mathrm{SF}_{(0\rightarrow2)} F_0^{(2,0)} = S_1^2 F_0^{(2,0)} = 1$, which is achieved with great precision. Let us use the notation $\mathrm{RT}_{n,\ell}(0,k,N)$ to denote the $N^{\text{th}}$-RT associated to the $\ell^{-k}$ exponentially-suppressed behaviour of the large-order relation obeyed by the $n$-instanton coefficients $F_k^{(n,0)}$, when zooming-in on the leading contribution, $s_0$, and as obtained via the resummation procedure described above (thus $\mathrm{RT}_{n}(0,k,N)$ defined previously is just $\mathrm{RT}_{n,1}(0,k,N)$ in the new notation). Then, in the present case, we find that the convergence error is once again extremely small,
\be
\frac{\mathrm{RT}_{0,2}(0,95,5) - 1}{1} \approx 3.40814 \times 10^{-8}.
\ee
\noindent
On the right image of figure~\ref{fig:large-order-pert-series-1inst-resum}, we plot the imaginary part (again, already properly weighted by the corresponding exponential term $3^k$)
\be
3^k\, \delta_{\im} F^{(0,0)}_{2,k} \equiv F_k^{(0,0)}\, \frac{2\pi \left(3A\right)^k}{\Gamma(k)} - 3^{k}\, \mathrm{SF}_{(0\rightarrow1)} \times \CS_{\im} \mathrm{BP}_{N} [\chi_{(0\rightarrow1)}] (k).
\ee
\noindent
The plot clearly shows that indeed the imaginary part of this resummation is of order $3^{-k}$, and will mix with the \textit{next} exponentially-suppressed order in the large-order relations.

\begin{figure}[t!]
\begin{center}
\includegraphics[height=7cm]{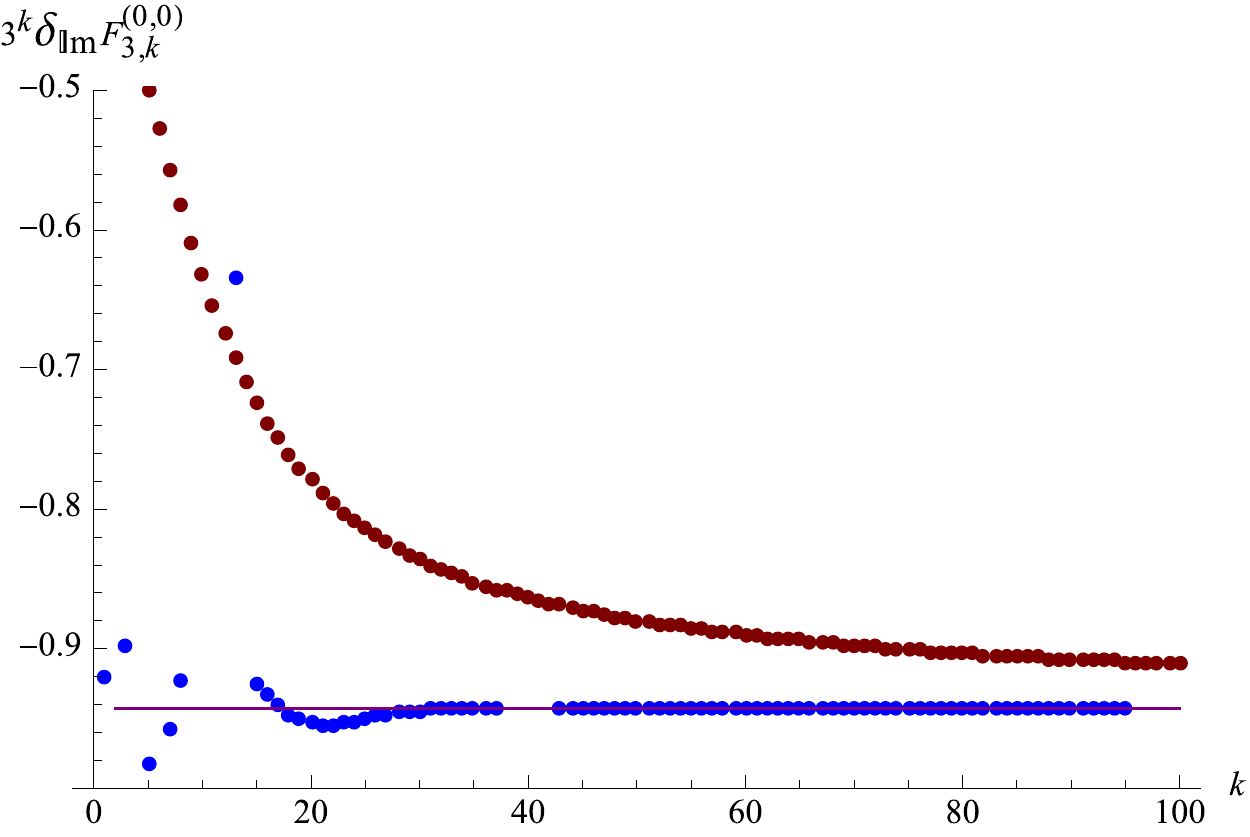}
\end{center}
\caption{Plot of the large-order behaviour of the coefficients $3^k\, \delta_{\im} F^{(0,0)}_{3,k}$. The original sequence is shown in red, while in blue we plot its $\mathrm{RT}_{0,3} (0,k,5)$ Richardson transform. The purple line shows the constant value this sequence is converging towards, $- \rmi S_1^3 F_0^{(3,0)} = -\frac{2\sqrt{2}}{3}$.}
\label{fig:large-order-pert-series-2inst-resum}
\end{figure}

The procedure above can of course keep being applied to study more and more exponentially-suppressed contributions to the perturbative asymptotics, in a completely straightforward fashion. Nonetheless, due to the aforementioned mixing of real and imaginary contributions arising upon BP resummation, let us still schematically work out the next step, illustrating how to re-sum the order $2^{-k}$ contributions to the large-order behaviour of the perturbative sequence $F_k^{(0,0)}$ in \eqref{large-order-pert-quartic-2}. As expected, one may once again verify that the BP of $\chi_{(0\rightarrow2)}$ will have poles upon the positive real axis, and we need to perform a lateral resummation. We shall do this \textit{consistently} with the previous choice of resummation direction, and will thus re-sum \textit{below} the poles obtaining both a real and an imaginary contribution, as in
\be
\CS_{0^{-}} \mathrm{BP}_{N} [\chi_{(0\rightarrow2)}] (k) = \CS_{\re} \mathrm{BP}_{N} [\chi_{(0\rightarrow2)}] (k) + \rmi\, \CS_{\im} \mathrm{BP}_{N} [\chi_{(0\rightarrow2)}] (k).
\ee
\noindent
The real and imaginary parts of the resummed result are defined as above,
\begin{eqnarray}
\delta_{\re} F^{(0,0)}_{3,k} + \rmi\, \delta_{\im} F^{(0,0)}_{3,k} &\equiv& F_k^{(0,0)}\, \frac{2\pi\rmi A^k}{\Gamma(k)} - \mathrm{SF}_{(0\rightarrow1)}\, \CS_{\re} \mathrm{BP}_{N} [\chi_{(0\rightarrow1)}] (k) - \nonumber \\
&&
\hspace{-110pt}
- \rmi \mathrm{SF}_{(0\rightarrow1)}\, \CS_{\im} \mathrm{BP}_{N} [\chi_{(0\rightarrow1)}] (k) - \mathrm{SF}_{(0\rightarrow 2)}\, \CS_{\re} \mathrm{BP}_{N} [\chi_{(0\rightarrow2)}] (k) - \rmi \mathrm{SF}_{(0\rightarrow2)}\, \CS_{\im} \mathrm{BP}_{N} [\chi_{(0\rightarrow2)}] (k) = \nonumber \\
&=& F_k^{(0,0)}\, \frac{2\pi\rmi A^k}{\Gamma(k)} + 2\, \CS_{\re} \mathrm{BP}_{N} [\chi_{(0\rightarrow1)}] (k) + 2 \rmi\, \CS_{\im} \mathrm{BP}_{N} [\chi_{(0\rightarrow1)}] (k) - \nonumber \\
&&
- (-2)^2\, \CS_{\re} \mathrm{BP}_{N} [\chi_{(0\rightarrow2)}] (k) - (-2)^2 \rmi\, \CS_{\im} \mathrm{BP}_{N} [\chi_{(0\rightarrow2)}] (k).
\end{eqnarray}
\noindent
It turns out that this time around it will be the imaginary part, which was earlier of order $3^{-k}$, which will yield the leading behaviour at this order; while the real part will now be of order $4^{-k}$ and only contribute at next order, \textit{i.e.},
\begin{eqnarray}
\rmi \delta_{\im} F^{(0,0)}_{3,k} &\simeq& 3^{-k}\, \mathrm{SF}_{(0\rightarrow3)} \left( F_0^{(3,0)} + \cdots \right) + \mathcal{O}(4^{-k}), \\
\delta_{\re} F^{(0,0)}_{3,k} &\simeq& \mathcal{O}(4^{-k}).
\end{eqnarray} 
\noindent
Herein, the coefficients $\rmi \delta_{\im} F^{(0,0)}_{3,k}$ are expected to converge, at leading order in $1/k$, towards $\mathrm{SF}_{(0\rightarrow3)} F_0^{(3,0)} = - \frac{2\sqrt{2}\rmi}{3}$. This is tested in figure~\ref{fig:large-order-pert-series-2inst-resum} for a BP of $N=60$, with the usual extremely small error,
\be
\frac{\mathrm{RT}_{0,3}(0,95,5) - \left( -\frac{2\sqrt{2}\rmi}{3} \right)}{\left( -\frac{2\sqrt{2}\rmi}{3} \right)} \approx 9.017 \times 10^{-7}.
\ee

\begin{figure}[t!]
\begin{center}
\includegraphics[height=7cm]{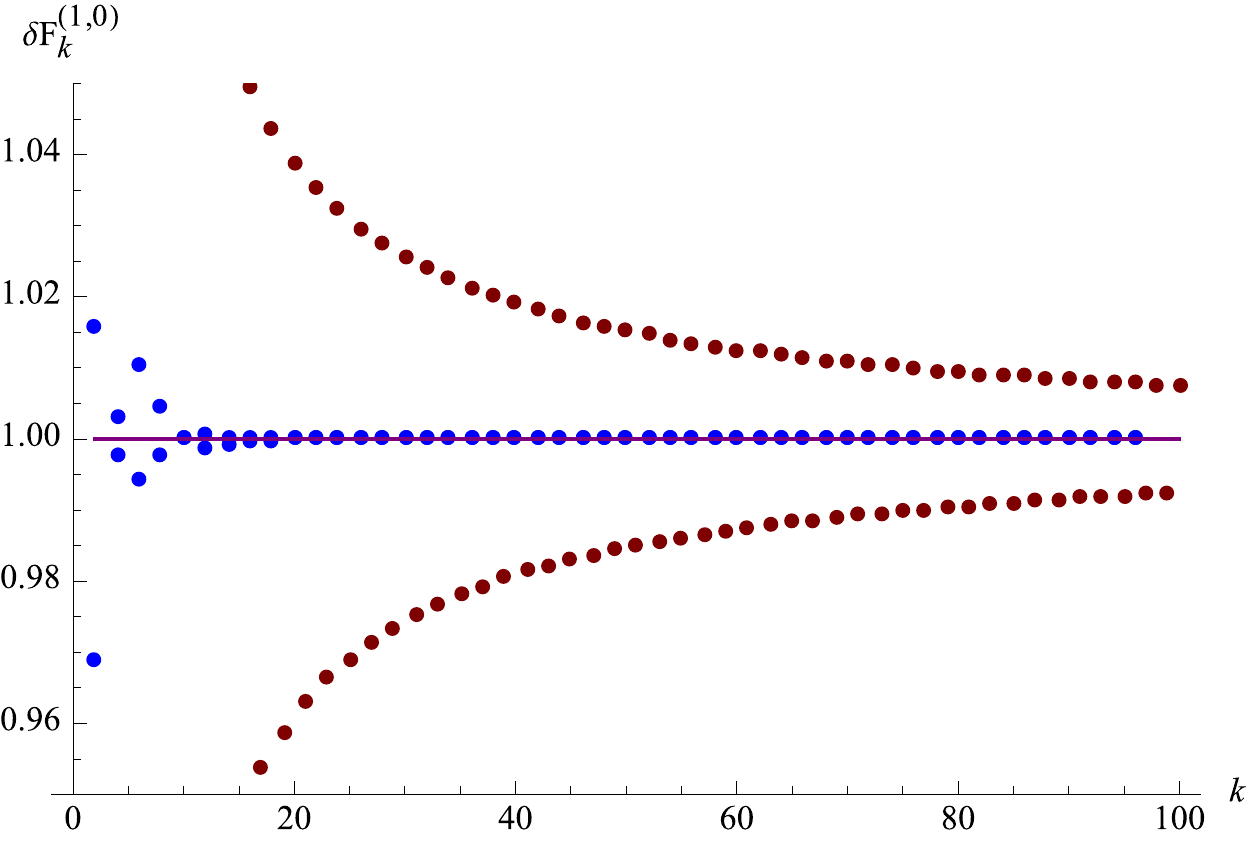}
\end{center}
\caption{Plot of the large-order behaviour of the coefficients $\delta F_k^{(1,0)}$. The original sequence is shown in red, while in blue we plot its $\mathrm{RT}_{1} (0,k,5)$ Richardson transform. The purple line shows the constant value towards which this sequence is converging, $\widetilde{S}_{-1}\, F_0^{(0,1)}= 1$.}
\label{fig:Large-order-quartic-freeen-1inst}
\end{figure}

Having thoroughly tested the resurgent structure of the perturbative sector, let us now turn to the analysis of the large-order behaviour of higher (multi) instanton sectors. In particular, we shall next focus upon the one and two-instanton sectors. Starting with the asymptotic expansion around the one-instanton sector, $\Phi_{(1,0)}$, we proceed in complete analogy with what was done for the perturbative sector. Using both Stokes discontinuities\footnote{Once again, Stokes coefficients and Borel residues are trivially related, via \eqref{stokesS-to-borelS+GEN}.} \eqref{stokes-zero-pert-quartic} and \eqref{stokes-pi-1inst-quartic} alongside Cauchy's theorem \eqref{Cauchy-two-disc}, one immediately obtains the large-order relation\footnote{For completeness, let us also include the respective closed-form expression, valid for $k>h$,
\be
F_{k}^{(1,0)} \simeq \sum_{n=1}^{+\infty} \frac{\Gamma(k)}{\left( n A \right)^k} \left( n+1 \right) \frac{S_1^n}{2\pi\rmi}\, \sum_{h=0}^{+\infty} \frac{\Gamma(k-h)}{\Gamma(k)}\, F_{h}^{(n+1,0)} \left( n A \right)^{h} + \frac{\Gamma(k)}{\left( - A \right)^k}\, \frac{\widetilde{S}_{-1}}{2\pi\rmi}\, F_0^{(0,1)}.
\ee
}
\begin{eqnarray}
F_{k}^{(1,0)}\, \frac{2\pi\rmi A^k}{\Gamma(k)} &\simeq& 2S_1 \left( F_0^{(2,0)} + \frac{A}{k-1}\, F_1^{(2,0)} + \frac{A^2}{(k-1)(k-2)}\, F_2^{(2,0)} + \cdots \right) + \frac{\widetilde{S}_{-1}}{(-1)^k}\, F_0^{(0,1)} + \nonumber\\
&&
\hspace*{-30pt}
+ 2^{-k}\, 3 S_1^2 \left( F_0^{(3,0)} + \frac{2A}{k-1}\, F_1^{(3,0)} + \frac{(2A)^2}{(k-1)(k-2)}\, F_2^{(3,0)} + \cdots \right) + \mathcal{O}(3^{-k}).
\label{large-order-1-inst-quartic}
\end{eqnarray}
\noindent
The main novelty as compared back to \eqref{large-order-pert-quartic} is the appearance of ``sideways'' resurgence, via the Stokes constant $\widetilde{S}_{-1}$, which we analyze in the following. In any case, the whole procedure is very similar to what we did before. We first divide this large-order relation into its leading asymptotics and the exponentially-suppressed contributions at order $2^{-k}$ and smaller. Then, we plot the leading $1/k$ coefficients, and their respective RTs, to numerically determine the Stokes constant $\widetilde{S}_{-1}$. This is shown in figure~\ref{fig:Large-order-quartic-freeen-1inst}, where the coefficients
\be
\delta F_k^{(1,0)} \equiv \left(-1\right)^k \left( F_{k}^{(1,0)}\, \frac{2\pi\rmi A^k}{\Gamma(k)} - 2S_1\, F_0^{(2,0)}\right)
\ee
\noindent
are plotted, alongside their fifth RT. From \eqref{large-order-1-inst-quartic} above, these coefficients are predicted to converge to $\widetilde{S}_{-1}\, F_0^{(0,1)} = 1$ (where we use the predicted value for the Stokes constant $\widetilde{S}_{-1}=1$). The figure very clearly shows how this is indeed the case, with the usual extremely small error
\be
\frac{\mathrm{RT}_{1}(0,95,5) - 1}{1} \approx 8.870 \times 10^{-10}.
\ee
\noindent
The procedure is entirely analogous to what we did earlier for the perturbative sector (only the resurgent structure of the large-order relations changes), and one could thus proceed into performing BP resummations of the leading behaviour and eventually test exponentially-suppressed predictions in the exact same fashion. We leave such tests as an exercise for the reader.

\begin{figure}[t!]
\begin{center}
\includegraphics[height=7cm]{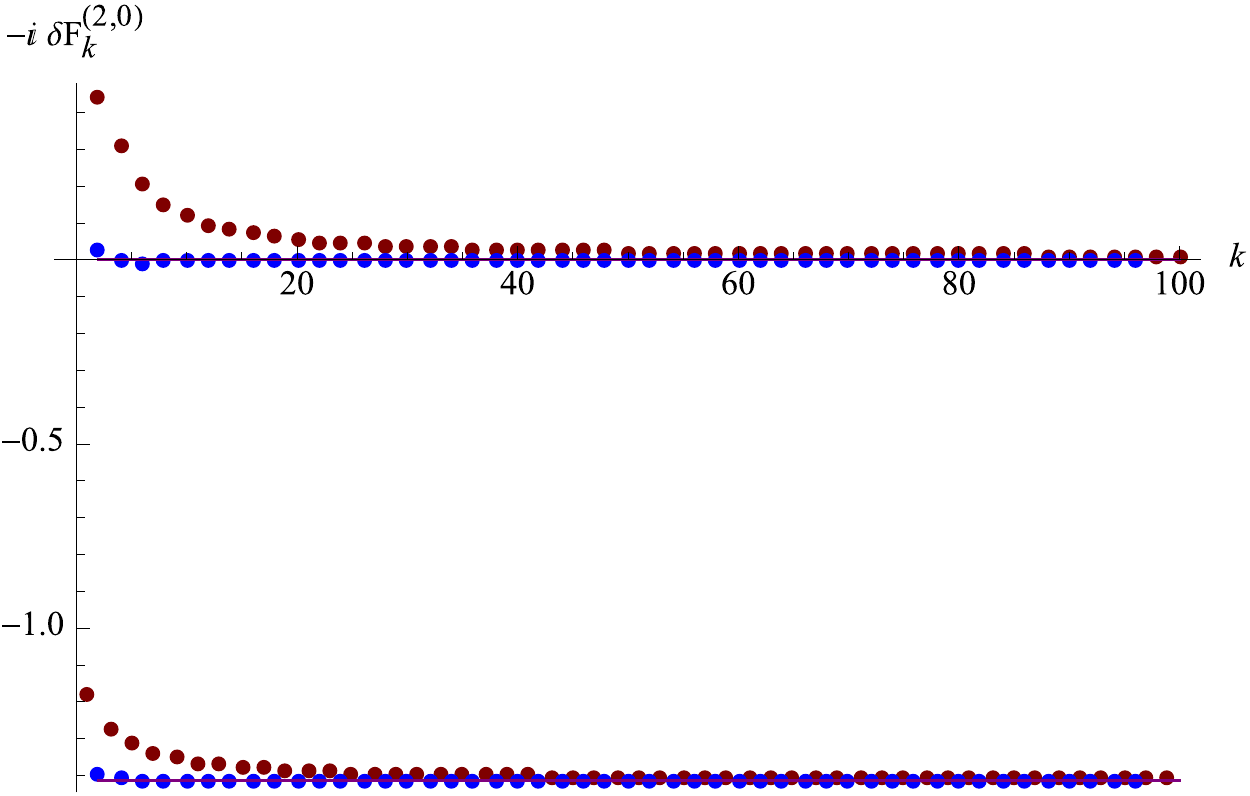}
\end{center}
\caption{Plot of the large-order behaviour of the two-instanton coefficients $-\rmi\, \delta F_{k}^{(2,0)} \equiv F_{k}^{(2,0)}\, \frac{2\pi A^k}{\Gamma(k)}$. As usual the original sequence is shown in red, while in blue we plot the Richardson transforms for each branch, $k$ even or odd, $\mathrm{RT}_{2,\mathrm{even/odd}} (0,k,5)$. The purple lines show the constant values towards which each of these sequences is converging to, depending on parity, which is either $0$ or $-\sqrt{2}$.}
\label{fig:Large-order-quartic-freeen-2inst}
\end{figure}

The one final test we still wish to discuss in here deals with the large-order behaviour of the two-instanton sector. Using the Stokes discontinuities \eqref{stokes-zero-pert-quartic} and \eqref{stokes-pi-2inst-quartic}, the large-order relation of this sector is:
\begin{eqnarray}
F_{k}^{(2,0)}\, \frac{2\pi\rmi A^k}{\Gamma(k)} &\simeq& 3 S_1 \left( F_0^{(3,0)} + \frac{A}{k-1}\, F_1^{(3,0)} + \frac{A^2}{(k-1)(k-2)}\, F_2^{(3,0)} + \cdots \right) + \nonumber \\
&&
+ \frac{1}{(-1)^k}\, S_{-1} \left( F_0^{(1,0)} + \frac{\left( -A \right)}{k-1}\, F_1^{(1,0)} + \frac{\left( -A \right)^2}{(k-1)(k-2)}\, F_2^{(1,0)} + \cdots \right) + \nonumber \\
&&
+ \frac{1}{2^k}\, 6 S_1^2 \left( F_0^{(4,0)} + \frac{\left( 2A \right)}{k-1}\, F_1^{(4,0)} + \frac{\left( 2A \right)^2}{(k-1)(k-2)}\, F_2^{(4,0)} + \cdots \right) + \nonumber \\
&&
+ \frac{1}{(-2)^k} \left( \widetilde{S}_{-2} + \frac{1}{2} \widetilde{S}_{-1} S_{-1}  \right) F_0^{(0,1)} + \mathcal{O} (3^{-k}),
\label{large-order-2-inst-quartic}
\end{eqnarray}
\noindent
or, equivalently but with slightly simpler combinatorial factors, with Borel residues:
\begin{eqnarray}
F_{k}^{(2,0)}\, \frac{2\pi\rmi A^k}{\Gamma(k)} &\simeq& - \mathsf{S}_{2\to3} \left( F_0^{(3,0)} + \frac{A}{k-1}\, F_1^{(3,0)} + \frac{A^2}{(k-1)(k-2)}\, F_2^{(3,0)} + \cdots \right) - \nonumber \\
&&
- \frac{1}{(-1)^k}\, \mathsf{S}_{2\to1} \left( F_0^{(1,0)} + \frac{\left( -A \right)}{k-1}\, F_1^{(1,0)} + \frac{\left( -A \right)^2}{(k-1)(k-2)}\, F_2^{(1,0)} + \cdots \right) - \nonumber \\
&&
- \frac{1}{2^k}\, \mathsf{S}_{2\to4} \left( F_0^{(4,0)} + \frac{\left( 2A \right)}{k-1}\, F_1^{(4,0)} + \frac{\left( 2A \right)^2}{(k-1)(k-2)}\, F_2^{(4,0)} + \cdots \right) - \nonumber \\
&&
- \frac{1}{(-2)^k}\, \widetilde{\mathsf{S}}_{2\to0}\, F_0^{(0,1)} + \mathcal{O} (3^{-k}).
\end{eqnarray}
\noindent
This two-instanton sector is particularly interesting as it is the first large-order relation which displays \textit{all} types of resurgence in this problem; ``forward'', via $S_+$, ''backward'', via $S_-$, and ``sideway'', via $\widetilde{S}_-$. Much of the analysis of this relation now follows the same steps as before. For example, one may firstly check the leading large-order behaviour in the same manner as we did earlier with RTs. Then one can also re-sum the exponentially leading sectors, associated to the motions $\chi_{(2\rightarrow3)}$ and $\chi_{(2\rightarrow1)}$, in order to study the exponentially suppressed $2^{-k}$ contribution and, in this way, check the value of the Stokes constant $\widetilde{S}_{-2}$. Let us first say a few words about the leading large-order behaviour. From the expression above a novelty readily emerges, as this leading behaviour will now depend upon the \textit{parity} of $k$. In fact, one has at leading order in $1/k$
\be
F_{k}^{(2,0)}\, \frac{2\pi\rmi A^k}{\Gamma(k)} \approx 3 S_1\, F_0^{(3,0)} + \left(-1\right)^k S_{-1}\, F_0^{(1,0)} + \mathcal{O}(1/k).
\ee
\noindent
Making use of $F_0^{(3,0)} = \frac{\rmi}{6\sqrt{2}}$, it then follows that when $k$ is odd the sequence above converges towards $3 S_1\, F_0^{(3,0)} - S_{-1}\, F_0^{(1,0)} = - \rmi \sqrt{2}$; while when $k$ is even the sequence converges towards $3 S_1\, F_0^{(3,0)} + S_{-1}\, F_0^{(1,0)} = 0$ instead. It should be rather evident that one may actually perform convergence acceleration for either $k$ even or odd, separately, by simple application of RTs on these sub-sequences. This is illustrated in figure~\ref{fig:Large-order-quartic-freeen-2inst}, where we can perfectly see the convergence towards each of the above values separately. The error associated with the convergence is  extremely small, as before; it is about $10^{-9}$. Moving onwards, resumming the sectors $\chi_{(2\rightarrow3)}$ and $\chi_{(2\rightarrow1)}$ and subtracting them from the original two-instanton sequence will then allow us to directly access the $2^{-k}$ behaviour of the large-order relations. In particular, this exercise (which we leave for the reader) directly confirms that $\widetilde{S}_{-2}=0$ as expected. Of course we could keep on analyzing higher and higher sectors, and given enough time and patience numerically check all the predicted Stokes constants alongside the transseries resurgent structure.

After all this, we may finally conclude that all these tests allowed us to very precisely verify the structure of the resurgent transseries we had put forward earlier, as a solution to the quartic free energy. In general, in harder problems (and without a linear ``companion-problem'' to help us bypass some of the nonlinear difficulties), the procedure develops in the opposite direction, or as a mix in-between. One puts forward a reasonable \textit{ansatz} for a transseries solution, then checks it via large-order relations as above. Some aspects may be validated, others not---and require for extra structure to be added to the transseries \textit{ansatz}. So one does, and repeats the different analyses and checks. Iteratively, one eventually constructs the full resurgent transseries solution to the problem at hand (also sometimes involving numerical determination of many Stokes constants). Recent examples where such procedure was applied include, \textit{e.g.}, the cases of the first Painlev\'e equations relating to 2d (super) gravity and addressed in \cite{msw07, m08, msw08, gikm10, asv11, sv13, asv18a}. Again, we have here chosen such a simple example as the quartic integral in order to introduce and develop resurgence and transseries in a ``friendly environment'', for pedagogical purposes and the ease of the first-time reader. Let us say that one issue we have not covered concerns transseries resummation, \textit{i.e.}, how to use the transseries to obtain nonperturbative results (numbers!) to whatever problem at hand. Such a discussion---alongside an analysis of strong-coupling expansions (as opposed to the present weak-coupling asymptotic expansions)---is briefly addressed in appendix~\ref{app:strongcoupling}. We have nonetheless decided not to include this topic in the main part of this review (where we chose to focus more on asymptotics), and rather refer the interested reader to the general theory of \textit{median resummation} as described in, \textit{e.g.}, \cite{as13}, and to the (numerical) examples of, \textit{e.g.}, \cite{m08, gmz14, csv15, a15, cms16, cms17, c1712}, which very nicely illustrate the procedure of resummation and Stokes phenomena within the contexts of string theory and the large $N$ approximation.

\section{Lefschetz Thimbles for Linear Problems}\label{sec:lefschetz}

Having understood most of the basics concerning resurgence---within a very simple toy-model given by an ordinary one-dimensional integral---one would next like to bring these powerful techniques into the realm of quantum field theory. Following in analogy with what was done at the beginning of section~\ref{sec:quartic}, let us first consider the generalization of the saddle-point method into higher (but still finite) dimensions, which will provide a basis for a more systematic approach. This generalization entails higher-dimensional analogues of steepest-descent contours, which go by the name of \textit{Lefschetz thimbles} (essentially, steepest-descent hypersurfaces), and the whole mathematical machinery surrounding it is known as Picard--Lefschetz theory (which has been considered since long ago within the mathematical-physics literature \cite{p83, f77, dh02}). These methods relate to the theory of resurgence via \textit{hyperasymptotics}, see, \textit{e.g.}, \cite{bh90, bh91, b91, h97, b99, hlo04, o04, chko07}. Further, given their natural use in the evaluation of multi-dimensional integrals, they have also entailed recent developments and activity with the hope of opening a door towards a better understanding of both quantum-theoretic functional integrals, \textit{e.g.}, \cite{w10a, w10b, hmw11, cdu14, tk14, t14, kt14, bssu15, tnk15, bpsu15, ft15, du15b, tnv17}, and lattice field theory, \textit{e.g.}, \cite{cdms13, mcs13, fhkkks13, abb15, abbrw15, abbvw16, abbrw16, ns17}. For most of this section we shall restrict to finite-dimensional integrals.

Let us begin with a $d$-dimensional ``partition function'', given by the integral
\be
\label{eq:nd_integral}
Z (\hbar) = \frac{1}{\left( 2\pi \right)^d} \int_{\boldsymbol{\Gamma}} \rmd z^1 \wedge \cdots \wedge \rmd z^d\, \rme^{- \frac{1}{\hbar} V(\boldsymbol{z})}.
\ee
\noindent
Herein, the potential $V(\boldsymbol{z})$ is a polynomial\footnote{In section~\ref{sec:elliptic} we shall discuss an example involving a meromorphic function rather than a polynomial potential. Nevertheless, most of the constructions in this section will apply to that example, with only slight modifications.} in the $d$ complex variables which we have collectively denoted by $\boldsymbol{z} \equiv ( z^1, \ldots, z^d ) \in \BC^d$, and $\hbar \in \BC$ as before. We shall further assume that the critical points of the potential, $V$, are isolated and non-degenerate, \textit{i.e.}, the  determinant of the Hessian matrix of $V$, $\text{Hess} \equiv \det (\partial_a\partial_b V)$, at these critical points is non-zero.

Now, in order to give a precise meaning to the partition function \eqref{eq:nd_integral}, we still need to know over which (real middle-dimensional) integration cycle(s), $\boldsymbol{\Gamma}$, is the integral convergent. Further, we would like to know how this cycle changes as one changes the phase of $\hbar$, a feature which was already discussed in detail in the simpler context of section~\ref{sec:quartic} (recall figure~\ref{steepest}). Clearly, a cycle over which \eqref{eq:nd_integral} is convergent for, say, $\hbar>0$, will be different from a convergent cycle with $\hbar<0$. Without surprise, we shall see how the change which the integration cycle undergoes, as one varies the phase of $\hbar$, is intimately related to Stokes phenomenon. Finally, we also want to learn how to explicitly evaluate the partition function, which usually can only be done as an asymptotic expansion of $Z (\hbar)$ for small coupling $\hbar$. As already advertised, we shall use the mathematical machinery of Picard--Lefschetz theory to answer these questions, which by generalizing the notion of steepest-descent contours to higher complex-dimensions further gives a \textit{topological interpretation} to the possible integration cycles which ensure convergence of the partition-function integral.

\subsection*{Constructing the Lefschetz Thimbles}

Our first goal is thus to classify the set of integration cycles over which the integral \eqref{eq:nd_integral} is convergent. Convergence can be achieved by simply requiring $\re \left( - \frac{1}{\hbar} V(\boldsymbol{z}) \right) \rightarrow -\infty$ as $|\boldsymbol{z}| \rightarrow +\infty$. Of course that, depending on the form of the potential $V(\boldsymbol{z})$, there may be distinct regions in $\BC^d$ which satisfy this requirement. To be more more precise, let us introduce the set
\be
\label{eq:chi}
\CX_{\epsilon} = \left\{ \boldsymbol{z} \in \BC^d\, \left|\, \re \left( - \frac{1}{\hbar} V (\boldsymbol{z}) \right) = - \frac{1}{\epsilon} \right. \right\},
\ee
\noindent
with $\epsilon$ a small (but positive) number. Generically this set will be made of disjoint subsets, with these different subsets corresponding to the different asymptotic regions that a convergent integration cycle may have. Recalling the example of the quartic potential, discussed in section~\ref{sec:quartic}, the set $\CX_\epsilon$ then had four disjoint subsets, each of which a region far away from the origin inside each of the four sectors associated to the four brown regions in figures~\ref{quarticpotentialfig} and~\ref{steepest}. The non-trivial and convergent integration contours thus became those which join two distinct such regions. In fact, the integral will depend solely on which regions does the integration cycle join, and not on the specific (local) details of the contour itself (as long as, along its path, this contour does not cross any other critical point than the one which was initially attached to it). This is the key point which will assign a \textit{topological interpretation} to our multi-dimensional integration cycles. 

In fact, the main result we now wish to discuss is that any integration cycle, $\boldsymbol{\Gamma}$, over which \eqref{eq:nd_integral} is convergent, is a $d$-\textit{cycle} with \textit{integer coefficients}. Furthermore, the partition-function integral will subsequently only depend upon the \textit{homology class} of $\boldsymbol{\Gamma}$ in the relative\footnote{A $d$-cycle $\gamma$ in $\mathbb C^d$ is called a \textit{relative cycle} mod $\CX$ if its boundary, $\partial \gamma$, lies in $\CX$ (if we think of $\BC^d$ as a $2d$-dimensional real manifold, then $\CX$ is a $(2d-1)$-dimensional real manifold). A natural equivalence class of relative $d$-cycles is simply constructed by saying that two cycles $\gamma$ and $\gamma'$ are equivalent iff $\gamma-\gamma^{\prime}$ is homologous to zero. Then the relative homology group $H_d (\BC^d, \CX; \BZ )$ is the set of these equivalence classes with integer coefficients.} homology group $H_d ( \BC^d, \CX_\epsilon; \BZ )$ \cite{p83, f77, w10a}. Now, the properties of this relative homology group depend only upon the local properties of the potential, $V$, near its critical points. Moreover, a basis for this homology group may be identified with the equivalence classes of integration cycles which intersect only \textit{one}\footnote{The violation of this requirement concerning critical points leads to Stokes phenomenon, to be discussed later.} critical point, $\boldsymbol{z}^*_i$, and which we shall denote by $\left\{ \CJ_i \right\}$. It then turns out that, in this case, the original cycle $\boldsymbol{\Gamma}$ can be decomposed as
\be
\label{eq:decompositionG}
\boldsymbol{\Gamma} = \sum_{i=1}^N n_i\, \CJ_i,
\ee
\noindent
leading to
\be
\label{eq:decompositionI}
\int_{\boldsymbol{\Gamma}} \rmd z^1 \wedge \cdots \wedge \rmd z^d = \sum_{i=1}^N n_i\, \int_{\CJ_i} \rmd z^1 \wedge \cdots \wedge \rmd z^d,
\ee
\noindent
where the $n_i$'s are integers and $N$ is the total number of critical points. The cycles $\CJ_i$ are what is known as the \textit{Lefschetz thimbles}. Let us justify this construction in the following.

In order to construct the thimbles, start by labeling the critical points of the potential $V$ as\footnote{In order to avoid notational confusion, we shall denote the critical points with a subscript $\{ i,j,... \}$, and the coordinate vector index, whenever present, with a superscript $\{ a,b,... \}$.} $\boldsymbol{z}^*_i$ with $i \in \{ 1, 2, \dots, N \}$; satisfying
\be
\left. \frac{\partial V}{\partial z^a} \right|_{\boldsymbol{z} = \boldsymbol{z}^*_i} = 0.
\ee
\noindent
Intuitively, if we start integrating out from a critical point $\boldsymbol{z}^*_i$, convergence of the partition-function integral can be ensured by requiring
\begin{equation}
\rho (\boldsymbol{z}) \equiv \re \left( - \frac{1}{\hbar}\, \Big( V (\boldsymbol{z}) - V (\boldsymbol{z}^*_i) \Big) \right) \leq 0
\end{equation}
\noindent
along the integration cycle. Moreover, this convergence will be fastest along that collection of curves (parameterized by $\tau$) which flow along the \textit{gradient} of the above function, \textit{i.e.}, along
\be
\label{eq:gradient_flow}
\frac{\rmd\boldsymbol{z}}{\rmd\tau} = - 2\, \frac{\partial\rho}{\partial\bar{\boldsymbol{z}}}.
\ee
\noindent
This set of equations \eqref{eq:gradient_flow} is known as the \textit{gradient flow equations} \cite{f77, k94}. It is then straightforward to show that $\rho \left( \boldsymbol{z}(\tau) \right)$ \textit{decreases} along any curve with increasing $\tau$:
\begin{equation}
\frac{\rmd \rho \left( \boldsymbol{z}(\tau) \right)}{\rmd\tau} = \frac{\partial\rho}{\partial\boldsymbol{z}} \cdot \frac{\rmd\boldsymbol{z}}{\rmd\tau} = - 2 \left| \frac{\partial\rho}{\partial\boldsymbol{z}} \right|^2 \leq 0.
\end{equation}
\noindent
Such solutions define \textit{downward flows} out from a given critical point. In this context, the function $\rho (\boldsymbol{z}) < \rho (\boldsymbol{z}^*_i)$ may be further interpreted as a \textit{height function}, in the Morse-theory sense.

It is the set of all points that may be reached out from a given critical point, $\boldsymbol{z}^*_i$, via downward flow as above, which is called a \textit{Lefschetz thimble} (associated with the critical point\footnote{Following our assumption that there is only \textit{one} critical point associated with a given thimble---and deferring the discussion of Stokes phenomenon, which occurs when this assumption is violated, to the end of this section.} $\boldsymbol{z}^*_i$). Expanding the potential around the critical point, one can use Morse theory to show that there are $d$ real, orthogonal directions along which one can flow downwards. Therefore, the thimble generated by the collection of all such flows is a $d$-real-dimensional cycle. To be more precise, consider a $(d-1)$-sphere encircling the critical point, and pick any point on this sphere. Following the trajectory of such point along the gradient flow \eqref{eq:gradient_flow} will generate a curve in $\BC^d$. The collection of \textit{all} curves that originate from \textit{all} points on the $(d-1)$-sphere is the $d$-dimensional real manifold which is the Lefschetz thimble. This $(d-1)$-sphere is known as the \textit{vanishing cycle}; vanishing at the critical point (for $d=1$, the vanishing cycle is just two points).

Another important feature of the gradient flow curves is that the imaginary part of the potential is \textit{constant} along them, and hence remains unchanged on the whole thimble. This implies that the thimble may be viewed as the natural higher-dimensional generalization of the stationary-phase contour (recall \eqref{steepestdescentcontour-def}), with
\be
\label{saddlegeneratethimble}
\im \left( - \frac{1}{\hbar} \left( V(\boldsymbol{z}) - V(\boldsymbol{z}^*_i) \right) \right) = 0
\ee
\noindent
holding true everywhere on the thimble\footnote{Of course that generically \eqref{saddlegeneratethimble} alone is not enough to define the thimble, unlike in our earlier one-dimensional stationary-phase contours. This is essentially because \eqref{saddlegeneratethimble} defines a real-codimension-one surface, whereas the thimble is in fact a real-codimension-$d$ surface, defined via the collection of gradient-flow curves described above.}. To see how this comes about, let us split the complex coordinates $\boldsymbol{z}$ into their real and imaginary parts, $\boldsymbol{z} = \boldsymbol{x} + \rmi \boldsymbol{y}$, and use the Cauchy--Riemann equations to obtain
\begin{eqnarray}
\label{eq:hamilton_flow-a}
\frac{\rmd\boldsymbol{x}}{\rmd\tau} &=& - \frac{\partial\rho}{\partial\boldsymbol{x}} = - \frac{\partial}{\partial\boldsymbol{y}}\, \im \left( - \frac{1}{\hbar} V (\boldsymbol{z}) \right), \\
\label{eq:hamilton_flow-b}
\frac{\rmd\boldsymbol{y}}{\rmd\tau} &=& - \frac{\partial\rho}{\partial\boldsymbol{y}} = \frac{\partial}{\partial\boldsymbol{x}}\, \im \left( - \frac{1}{\hbar} V (\boldsymbol{z}) \right),
\end{eqnarray}
\noindent
from where it follows
\begin{equation}
\frac{\rmd}{\rmd\tau}\, \im \left( - \frac{1}{\hbar} V (\boldsymbol{z}) \right) = \frac{\rmd\boldsymbol{x}}{\rmd\tau} \cdot \frac{\partial}{\partial\boldsymbol{x}}\, \im \left( - \frac{1}{\hbar} V (\boldsymbol{z}) \right) + \frac{\rmd\boldsymbol{y}}{\rmd\tau} \cdot \frac{\partial}{\partial\boldsymbol{y}}\, \im \left( - \frac{1}{\hbar} V (\boldsymbol{z}) \right) = 0.
\end{equation}
\noindent
In fact the gradient flow equations, combined with the Cauchy--Riemann equations \eqref{eq:hamilton_flow-a} and \eqref{eq:hamilton_flow-b}, define a \textit{Hamiltonian flow} where the imaginary part $\im \left( - \frac{1}{\hbar} V (\boldsymbol{z}) \right)$ is the \textit{conserved quantity} (the ``Hamiltonian'') associated with this flow. One can combine the gradient flow equations for the real part, and the Hamiltonian flow equations for the imaginary part, into a single equation in complex variables (where we also make explicit the phase of $\hbar$, as in $\hbar = |\hbar|\, \rme^{\rmi\theta}$)
\be
\label{eq:gradient_flow_complex}
\frac{\rmd z^a}{\rmd\tau} = \frac{\rme^{\rmi\theta}}{|\hbar|}\, \frac{\partial\bar{V}}{\partial\bar{z}^a},
\ee
\noindent
with the boundary condition $z^a (\tau\to-\infty) = (z^*_i)^a$. This is the \textit{complex} gradient flow equation, also known as \textit{Picard--Lefschetz equation}, whose solutions then naturally satisfy \eqref{saddlegeneratethimble}.

In complete analogy to the downward flow, one may likewise define \textit{upward flow} as originating from a critical point and defined by the equations
\be
\frac{\rmd z^a}{\rmd\tau} = - \frac{\rme^{\rmi\theta}}{|\hbar|}\, \frac{\partial\bar{V}}{\partial\bar{z}^a}.
\ee
\noindent
Along the upward flow the imaginary part of the potential still remains constant, but the real part $\rho (\boldsymbol{z})$ now increases. Morse theory again tells us that there will be $d$ real directions along which one can flow upwards. As such, and akin to the Lefschetz thimble construction, all the points in $\BC^d$ which may be reached via upward flow originating on an $(d-1)$-sphere surrounding the critical point $\boldsymbol{z}^*_i$, will construct a $d$-dimensional cycle which we denote by $\CK_i$. The set of all these cycles (upon adequate equivalence classes) constitute a basis for a dual homology class $H_d ( \BC^d, \CY_\epsilon; \BZ )$, where
\be
\CY_{\epsilon} = \left\{ \boldsymbol{z} \in \BC^d\, \left|\, \re \left( - \frac{1}{\hbar} V (\boldsymbol{z}) \right) = + \frac{1}{\epsilon} \right. \right\}. 
\ee

Note that since the imaginary part of the potential associated with each critical point is a conserved quantity, and generically the imaginary parts of $V$ at different critical points are distinct, this implies that the downward/upward flows originating from different critical points do not generically intersect each other. However, when this does happen to be the case, these cycles form a good basis for the relative homology groups we started off with, above. Given two cycles $\boldsymbol{\Gamma}$ and $\boldsymbol{\Gamma}^\prime$, one can define an inner product via the intersection number
\be
\label{eq:intersection_norm}
\left\langle\, \CK_n, \CJ_m\, \right\rangle = \delta_{nm}\,.
\ee
\noindent
This inner product is a map $H_d ( \BC^d, \CY_\epsilon; \BZ ) \otimes H_d ( \BC^d, \CX_\epsilon; \BZ ) \to \BZ$ which tells us when does a given downward-flowing thimble intersects some upward-flowing thimble. In the generic case described above, where all the imaginary parts of the potential at different critical points are distinct, such an intersection never happens---except at the critical point itself. Moreover, this intersection occurs between the upward and downward flows originating from that same critical point leading to \eqref{eq:intersection_norm}. Consequently, any integration cycle $\boldsymbol{\Gamma}$ such that \eqref{eq:nd_integral} is convergent, can be decomposed as a chain \cite{p83}
\be
\label{eq:decomposition}
\boldsymbol{\Gamma} = \sum_{i=1}^N n_i\, \CJ_i,
\ee
\noindent
where $n_i= \left\langle\, \CK_i, \boldsymbol{\Gamma}\, \right\rangle$. This conclusion finally justifies our starting main equations, \eqref{eq:decompositionG} and \eqref{eq:decompositionI}.

\subsection*{Evaluation of the Asymptotic Expansion}

Having understood the integration cycles \eqref{eq:decompositionI}, let us now try to explicitly evaluate the partition-function integral \eqref{eq:nd_integral}. Generically, this will entail an asymptotic series expansion in powers of $\hbar$. To start off with, let us impose a constraint regarding the critical points; which is that no pair of critical values has the same imaginary part, \textit{i.e.},
\be
\label{impotentialij}
\im \left( \frac{1}{\hbar} V (\boldsymbol{z}^*_i) \right) \neq \im \left( \frac{1}{\hbar} V (\boldsymbol{z}^*_j) \right) \quad \text{for} \quad i\neq j.
\ee
\noindent
Since the imaginary part is a conserved quantity along the thimble, this assumption guarantees that there are no intersections between thimbles associated with different critical points, and the decomposition of the integration cycle \eqref{eq:decomposition} is thus well defined. In fact, this condition generically holds except for a finite set of particular values that $\theta = \arg (\hbar)$ takes. These are the points where Stokes phenomenon occurs, to be discussed a bit later on. But for the moment let us simply assume a generic value of $\theta$ for which there is no Stokes phenomenon (and thus also suppress the dependence on $\theta$ from our notation).

Let us consider the partition-function integral over some particular thimble, $\CJ_n$, associated with the critical point $\boldsymbol{z}^*_n$. Introducing the variable $s \equiv V (\boldsymbol{z}) - V (\boldsymbol{z}^*_n)$, we can write this integral as
\be
\label{eq:cov}
Z^{(n)} (\hbar) = \frac{1}{\left( 2\pi \right)^d} \int_{\CJ_n} \rme^{- \frac{1}{\hbar} V(\boldsymbol{z})}\, \Omega = \rme^{- \frac{1}{\hbar} V^*_n} \int_{0}^{\rme^{\rmi\theta}\infty} \rmd s\, \CB[\Psi_n](s)\, \rme^{-\frac{s}{\hbar}},
\ee
\noindent
where $\Omega = \rmd z^1 \wedge \cdots \wedge \rmd z^d$ is the volume form and $V^*_n \equiv V (\boldsymbol{z}^*_n)$ is the value of the potential at the critical point $\boldsymbol{z}^*_n$. Remarkably, this change of variables is actually converting our original integral into a Borel-type integral, as in \eqref{borelresum} (hence the notation!). In fact, it will turn out that the function
\be
\label{eq:Borel_van}
\CB[\Psi_n](s) \equiv \frac{1}{\left( 2\pi \right)^d} \int_{\CV_n (s)} \left. \frac{\Omega}{\rmd s} \right|_{V (\boldsymbol{z}) = V^*_n + s}
\ee
\noindent
is precisely a \textit{generalized Borel transform}, of the type we shall discuss in the upcoming section~\ref{sec:borel}, and later denoted by $\CB[\Psi_{[d/2]}](s)$ (but herein let us keep denoting it by just $\CB[\Psi](s)$). The final integral in \eqref{eq:cov}, over $s$, is along a ray which we shall refer to as $\Sigma_n$, such that the exponent $s/\hbar$ is positive and bounded from below ensuring convergence. The remaining $(d-1)$-dimensional integral in \eqref{eq:Borel_van} is over a $(d-1)$-cycle, $\CV_n(s)$, which is nothing but the vanishing cycle transported to the point $s$ along $\Sigma_n$. As such, the thimble $\CJ_n$ can be thought of as a $d$-dimensional cycle generated by the orbit of the vanishing cycle along $\Sigma_n$.

Around the origin, $\CB[\Psi_n](s)$ has a \textit{convergent} series expansion\footnote{This expansion can be generalized to include the cases where the critical points are degenerate and/or the integration cycles have boundaries. For the degenerate case, one ``Morsifies'' the critical point by perturbing it away from the degenerate case. The perturbation deforms the degenerate critical point into a $\mu_n$ number of non-degenerate critical points, where $\mu_n$ is the Milnor number of the original critical point. In this case the Lefschetz thimble $\CJ_n$ develops $\mu_n$ foldings in the limit where the perturbation vanishes. Furthermore, $\CB[\Psi_n](s)$ is analytic over the universal covering of the punctured disk $\BC \setminus \{0\}$ and satisfies an ordinary differential equation, known as a Picard--Fuchs equation or a Gauss--Manin connection, which has at most a regular singularity at the origin. The order of this Gauss--Manin connection can be greater than one, but is bounded by $\mu_n$ and the expansion of $\CB[\Psi_n]$ near the origin will have powers of $\log s$ in addition to $s$. For more details, we refer the reader to \cite{m74, p83, p85, dh02}.} of the form
\be
\CB[\Psi_n](s) = s^{\frac{d}{2}-1} \sum_{k=0}^{+\infty} B^{(n)}_k s^k.
\ee
\noindent
Notice that for odd dimensions there is a square root branch-cut, whereas for even dimensions the function is analytic near the origin. The above leading term is easy to calculate. Around the critical point (\textit{i.e.}, $s=0$), we can expand $V (\boldsymbol{z}^*_n + \boldsymbol{z}) \approx V^*_n + \frac{1}{2}\, H_{ab}^{n}\, z^a z^b \equiv V^*_n + s$, where $H_{ab}^{n} \equiv \text{Hess} [V] (\boldsymbol{z}^*_n)$ is the Hessian matrix of the potential, $V$, at the critical point $\boldsymbol{z}^*_n$. Therefore, the vanishing cycle $\CV_n (s)$ is a $(d-1)$-sphere whose radius is $\sqrt{|s|}$ and \eqref{eq:Borel_van} reduces to a surface integral over this sphere, yielding
\be
\CB[\Psi_n](s) \approx \frac{1}{\sqrt{\left( 2\pi \right)^d \det \big(\, \text{Hess} [V] (\boldsymbol{z}^*_n)\, \big)}}\, \frac{s^{\frac{d}{2}-1}}{\Gamma \left( \frac{d}{2} \right)}.
\ee
\noindent
The rest of the coefficients may be computed depending on the specific details of the potential $V$. Evaluating the remaining $s$-integral then leads to an \textit{asymptotic}\footnote{This is actually made \textit{explicit} in the formula, due to the $\Gamma \left(k+d/2\right)$ factor.} expansion for $Z^{(n)} (\hbar)$
\bea
Z^{(n)} (\hbar) &\simeq& \rme^{- \frac{1}{\hbar} V^*_n}\, \hbar^{\frac{d}{2}}\, \sum_{k=0}^{+\infty} \Gamma \left( k + \frac{d}{2} \right) B^{(n)}_k \hbar^{k} \\
&\simeq& \frac{\rme^{- \frac{1}{\hbar} V^*_n}}{\sqrt{(2 \pi)^{d}\, \det \big(\, \text{Hess} [V] (\boldsymbol{z}^*_n)\, \big)}}\, \hbar^{\frac{d}{2}} + \rme^{- \frac{1}{\hbar} V^*_n}\, \hbar^{\frac{d}{2}}\, \sum_{k=1}^{+\infty} Z^{(n)}_k \hbar^{k},
\eea
\noindent
where we have redefined $Z^{(n)}_k \equiv \Gamma \left(k+d/2 \right) B^{(n)}_k$. 

Now that we have all the basic building blocks of the partition function \eqref{eq:nd_integral}---\textit{i.e.}, the asymptotic expansions of the thimble integrations \eqref{eq:cov} as written above---we may construct the final asymptotic expansion of $Z (\hbar)$ over the cycle $\boldsymbol{\Gamma}$. Using the decomposition of $\boldsymbol{\Gamma}$ into thimbles \eqref{eq:decomposition}, we find that the partition function in fact has a \textit{transseries} form, as
\be
\label{eq:thimble_asymp}
Z (\hbar) = \frac{1}{\left( 2\pi \right)^d} \int_{\boldsymbol{\Gamma}} \rme^{- \frac{1}{\hbar} V(\boldsymbol{z})}\, \Omega = \hbar^{\frac{d}{2}}\, \sum_{i=1}^N n_i\, \rme^{- \frac{1}{\hbar} V^*_i} \Phi_i(\hbar),
\ee
\noindent
where each $\Phi_i(\hbar)$ is an asymptotic series
\be
\Phi_i (\hbar) \simeq \sum_{k=0}^{+\infty} Z^{(i)}_k \hbar^{k}
\ee
\noindent
whose coefficients are obtained by the local expansion of the vanishing-cycle integral \eqref{eq:Borel_van} around the critical point $\boldsymbol{z}^*_i$, which is a solution of the Gauss--Manin connection. To sum up, each term $\rme^{- \frac{1}{\hbar} V^*_i} \Phi_i(\hbar)$ in the transseries is associated with an element of a basis of a homology group, where the topology is induced by restricting the allowed integration cycles into regions in $\BC^n$ that ensure the convergence of the integral $Z(\hbar)$. The asymptotic expansion of a given term is constructed \textit{locally} by the geometric properties of the vanishing cycle.

With the total number of critical points $N$ a finite number, our final solution \eqref{eq:thimble_asymp} also has a finite number of multi-instanton sectors and is thus an appropriate solution to some \textit{linear} problem (in the distinction already discussed in section~\ref{sec:quartic}). One could now be tempted to think that extending this formalism to situations with an infinite number of multi-instanton sectors simply amounts to taking the limit $N \to + \infty$ in \eqref{eq:thimble_asymp}. This would cover nonlinear problems and eventually quantum field theoretic functional-integrals. But while for the former there may not necessarily exist any adequate integral representation of solutions, invalidating the whole procedure, for the latter one needs to be careful with subtleties due to renormalization \cite{k16}.

Indeed, let us consider the partition function of some scalar quantum field theory,
\be
\label{eq:path_integral}
Z (\lambda) = \int \left[ \rmd \varphi \right] \rme^{- S(\varphi)},
\ee
\noindent
with action $S(\varphi)$ and coupling-constant $\lambda$. In order to use the toolbox described in this section, one would first want to factor out the coupling constant, as in
\be
S (\varphi) \to \frac{1}{\lambda} S (\widetilde{\varphi} ),
\ee
\noindent
and then compute (functional) saddle points for the functional-integral partition function \eqref{eq:path_integral}. But this factoring can only be done when considering \textit{bare} quantities. As renormalization kicks in, the coupling runs as $\lambda \equiv \lambda (\mu)$ at energy-scale $\mu$, further inducing mass renormalization and wave-function renormalization. For our purposes, the overall effect of such renormalization procedure is to \textit{invalidate} the above factoring. One rather finds (very schematically---see \cite{k16} for a full discussion)
\be
S (\varphi) \to \frac{1}{\lambda} S_{-1} (\widetilde{\varphi} ) + S_{0} (\widetilde{\varphi} ) + \lambda S_{1} (\widetilde{\varphi} ) + \lambda^2 S_{2} (\widetilde{\varphi} ) + \cdots.
\ee
\noindent
This automatically implies that in order to address renormalizable quantum field theories, one will need more sophisticated saddle-point methods and answers than just \eqref{eq:thimble_asymp}, and likely more complicated transseries structures than the ones we have been addressing so far (see section~\ref{sec:physics} or, \textit{e.g.}, \cite{e08}, for some such structures; and \cite{acpy17} for recent results pointing towards the need to include some of them within Yang--Mills/QCD theories). Do notice that this caveat is very much attached to any renormalization needs. In quantum field theoretic contexts where such needs do not arise, the transseries structures addressed so far are certainly enough; see, \textit{e.g.}, \cite{ars14, h16a, h16b, h1710, hy17} for supersymmetric gauge theories, \textit{e.g.}, \cite{msw07, m08, msw08, kmr10, dmp11, asv11, sv13, csv15} for large $N$ gauge theories, and, \textit{e.g.}, \cite{m06, ps09, mpp09, gikm10, cesv13, gmz14, cesv14, c15, cms16, cm16, cms17} for string theories.

\subsection*{Dealing with Stokes Phenomena}

The transseries construction of the partition function in \eqref{eq:thimble_asymp} arises from a \textit{local} analysis on the Lefschetz thimbles, near their corresponding critical points. But, as we have been postponing for a while, there are also \textit{global} aspects on the thimbles (and hence of the transseries) which need to be considered; most notably those leading to Stokes phenomena. In order to understand how global issues may come about, let us go back to the evaluation of the partition-function integral over some specific thimble $\CJ_n$, denoted by $Z^{(n)} (\hbar)$ in \eqref{eq:cov}. Its asymptotic expansion was generated by a Laplace-type integration of $\CB[\Psi_n](s)$ (defined in \eqref{eq:Borel_van}) along some ray $\Sigma_n$ of direction $\theta$ on the complex $s$-plane. However, the function $\CB[\Psi_n](s)$ may, in general, have singularities, which one can hit by varying the ray of integration as one varies $\theta$. These singularities of $\CB[\Psi_n](s)$ occur precisely at the critical points; which follows from the fact that
\be
\rmd s = \sum_{a=1}^d \frac{\partial V}{\partial z^a}\, \rmd z^a
\ee
\noindent
vanishes at a critical point. Previously, we have required that no two critical values of the potential had the same imaginary part; recall \eqref{impotentialij}. It was this assumption which saved us so far from hitting a singularity along the $s$-integration: because the imaginary part of the potential is conserved on a thimble, integration along a fixed thimble could not hit another critical point---with the same holding true for the projections of the Lefschetz thimbles onto rays of integration as in \eqref{eq:cov}. Let us now relax this assumption and discuss what happens as we cross a singularity.

For simplicity, let us begin by assuming that there is only \textit{one} critical point to be crossed, which we shall label by~$m$. The location of this singularity is $s = V (\boldsymbol{z}^*_m) - V (\boldsymbol{z}^*_n) = V^*_m - V^*_n \equiv V^*_{nm}$, which is nothing but the relative potential of the critical points $n$ and $m$. In order for the integration in \eqref{eq:cov} to be well defined, the integration contour needs to avoid this singularity. This can be done by avoiding it either from above or from below, by simply shifting to either $\theta^+$ or $\theta^-$. Each of these shifts will yield a different result for the partition function $Z^{(n)} (\hbar)$, which we shall denote by $Z_{\pm}^{(n)} (\hbar)$, respectively. Their difference, $Z_{+}^{(n)} (\hbar) - Z_{-}^{(n)} (\hbar)$, essentially yields the monodromy that $\CB[\Psi_n](s)$ acquires as one moves from $\theta^-$ to $\theta^+$. This monodromy is induced by the monodromy of the vanishing cycle, $\CV_n (s)$, as ones goes around the singularity $V^*_{nm}$ in a counter-clockwise fashion, and which is captured by the \textit{Picard--Lefschetz formula} \cite{l24}
\be
\label{eq:picard_lefschetz}
\CV_n ( \mathsf{r}_{nm}\, \rme^{\rmi \left( \phi + 2\pi \right)} ) = \CV_n ( \mathsf{r}_{nm}\, \rme^{\rmi\phi} ) + \left(-1\right)^{\frac{d \left(d-1\right)}{2}}\, \mathsf{S}_{nm}\, \CV_m ( \mathsf{r}_{nm}\, \rme^{\rmi\phi} ).
\ee
\noindent
Here, $\mathsf{r}_{nm}\, \rme^{\rmi\phi}$ parameterizes an infinitesimal circle around $V^*_{nm}$, and the coefficient $\mathsf{S}_{nm}$ is the intersection number of the two vanishing cycles, $\mathsf{S}_{nm} \equiv \langle \CV_n, \CV_m \rangle \in \BZ$. Formula \eqref{eq:picard_lefschetz} shows how when the vanishing cycle $\CV_n$, associated to the critical point~$n$, is moved along a counter-clockwise closed loop around the projection of the critical point~$m$ (\textit{i.e.}, around the singular point at $s = V^*_{nm}$), then the resulting cycle will receive an additional contribution given by $\CV_m$, the vanishing cycle associated with the critical point~$m$.   

From the Picard--Lefschetz expression \eqref{eq:picard_lefschetz}, one can further deduce the monodromy of $\CB[\Psi_n](s)$ around the singularity $V^*_{nm}$. Simply plugging the monodromy \eqref{eq:picard_lefschetz} in \eqref{eq:Borel_van}, one obtains
\be
\label{eq:borel_pl}
\CB_-[\Psi_n](s) = \CB_+[\Psi_n](s) + \left(-1\right)^{\frac{d \left(d-1\right)}{2}}\, \mathsf{S}_{nm}\, \CB_+[\Psi_m](s).
\ee
\noindent
Note that, since the $s$-integration in \eqref{eq:cov} basically corresponds to a Borel resummation, what we are describing here by tilting $\theta$ to $\theta^{\pm}$ in order to avoid the singularity is essentially the lateral Borel resummations which were discussed at the beginning of section~\ref{sec:quartic} (recall \eqref{stokesauto}). The resulting monodromy between the different lateral Borel resummations describes the \textit{Stokes phenomenon} \eqref{stokespheno}, where the intersection numbers $\mathsf{S}_{nm}$ relate to the \textit{Stokes constants}. The computation of Stokes constants within this framework is a technically challenging problem, and for an extensive discussion along the present context we refer the reader to, \textit{e.g.}, \cite{h97, bh90, bh91}.

It is worth mentioning that the resurgent relations take a standard form in this setting \cite{dh02}
\begin{eqnarray}
\label{eq:B_Psi_n:d=even}
\CB[\Psi_n](s) \Big|_{s=V^*_{nm}} &=& \mathsf{S}_{nm}\, \CB[\Psi_m] (s-V^*_{nm})\, \frac{\log (s-V^*_{nm})}{2\pi\rmi} + \text{holomorphic}, \qquad d = \text{even}, \\
\label{eq:B_Psi_n:d=odd}
\CB[\Psi_n](s) \Big|_{s=V^*_{nm}} &=& - \frac{1}{2}\, \mathsf{S}_{nm}\, \CB[\Psi_m] (s-V^*_{nm}) + \text{holomorphic}, \qquad\qquad\qquad\, d = \text{odd}.
\end{eqnarray}
\noindent
As comparing to \eqref{Borel-transf-expanded-one-param}---where we were dealing with $\Phi$-sectors, which relate to the $\Psi$-sectors herein as $\Psi = \hbar^{\frac{d}{2}}\, \Phi$---one should match $V^*_{nm} \leftrightarrow k A$ and $\left(-1\right)^d \mathsf{S}_{nm} \leftrightarrow - \mathsf{S}_{n\to n+k} + \cdots$ (so that the $\mathsf{S}_{nm}$ intersection numbers essentially are Borel residues). An example illustrating how to carry through this match will be discussed in the upcoming section~\ref{sec:borel}. Furthermore, the generalization of the above discussion to the case of multiple singularities is conceptually straightforward. One just has to use \eqref{eq:borel_pl}, but where one should now sum over \textit{all} the critical points~$m$ that are found along the path. Finally, all the critical points whose singularities show up in the path of~$n$, are below~$n$ in the Morse-theory sense, \textit{i.e.}, $\re\, ( - \frac{1}{\hbar} V^*_n ) > \re\, ( - \frac{1}{\hbar} V^*_m ) > \cdots$.

It is also interesting to understand the relation between Stokes phenomenon and the thimble decomposition. The rays of integration in \eqref{eq:cov}, $\Sigma_n$, are the projections of the Lefschetz thimbles upon the $s$-plane. Therefore, the aforementioned monodromy structure on the $s$-plane can be lifted onto a monodromy structure for the thimbles. In particular from \eqref{eq:borel_pl} and \eqref{eq:cov} one gets\footnote{At this stage, the reader may want to review the steepest-descent discussion of Stokes phenomena from subsection~\ref{subsec:basicresurgence}, which took place in the paragraph after equation \eqref{quarticmonodromy}.}
\be
\CJ_n (\theta^-) = \CJ_n (\theta^+) + \left(-1\right)^{\frac{d \left(d-1\right)}{2}} \sum_{m=1}^N \mathsf{S}_{nm}\, \CJ_m (\theta^+).
\ee
\noindent
As a result, our original thimble decomposition \eqref{eq:decomposition} ceases to be well-defined at $\theta$, precisely because the intersection numbers for the thimbles, the $n_i$'s, are not well defined at these values. Instead, the intersection numbers acquire a $\theta$-dependence, as in $n_i \equiv n_i(\theta)$, and become piecewise-constant functions of $\theta$. Integer-valued jumps occur for certain values of $\theta$, which are encoded by the intersection numbers of the vanishing cycles, or ``Stokes constants'', $\mathsf{S}_{nm}$. Consequently we have to refine the decomposition of the chain as
\be
\label{eq:decomposition2}
\Gamma (\theta) = \sum_{i=1}^N n_i(\theta)\, \CJ_i (\theta),
\ee
\noindent
which eventually finds its way back onto our transseries construction of the partition function, in \eqref{eq:thimble_asymp}. This finally becomes
\be
\label{eq:thimble_integral}
Z (|\hbar|\, \rme^{\rmi\theta}) = \hbar^{\frac{d}{2}}\, \sum_{i=1}^N n_i (\theta)\, \rme^{- \frac{1}{\hbar} V^*_i} \Phi_i(\hbar).
\ee
\noindent
When Stokes phenomenon occurs, the intersection numbers $n_i(\theta)$ jump in integer amounts. These jumps are precisely balanced by the jumps that arise from the lateral Borel resummations of $\Phi_i(\hbar)$, rendering \eqref{eq:thimble_integral} unambiguous. With the knowledge of the Stokes constants, we can therefore give a precise meaning to this expansion as a well-defined, \textit{continuous} function of $\hbar$.

\section{Borel Transforms for Nonlinear Problems}\label{sec:borel}

Let us now go back to Borel analysis, as discussed throughout section~\ref{sec:quartic}, in order to give it a more systematic basis. This is a more general and powerful tool than steepest-descent analysis (and its generalizations): all it requires to start-off is some asymptotic series, \textit{regardless} of where it came from or how was it computed originally\footnote{It even works for sequences with rather abstract origins, as for example sequences of enumerative invariants in algebraic geometry \cite{csv16}, or even more abstract combinatorial sequences \cite{m_b16}.}. Its blindness to the origin of the asymptotic sequences at play implies that it addresses linear and nonlinear problems equally well (as introduced in section~\ref{sec:quartic}). Furthermore, from a quantum field theoretic standpoint, it \textit{always} works for \textit{any} observable, not being limited to the partition function, or to functional integrals or functional differential equations, or whatever else; and not caring if the nonperturbative effects at hand are multi-instantons, or multi-renormalons, or whatever else. Following a pragmatic point-of-view, one need not dwell much on the origins of any given (perturbative) factorially-divergent sequences, but rather simply focus on how to use them as an \textit{input} in order to start the resurgence program, building towards finding their (nonperturbative) transseries completions.

In fact, the Borel transform of any such asymptotic sequence is the first step towards implementing the resurgence program: as discussed in section~\ref{sec:quartic}, resurgence emerges out of an analysis of Borel singularities, with its algebraic formulation following from an abstraction of the behaviour of these very same Borel transforms near their respective singularities. How Borel transforms are defined, and how resurgence is already evident from their singular structure on the Borel plane, is what we shall discuss in the following and place on firmer grounds. A better understanding of resurgence, as we wish to discuss in the upcoming section~\ref{sec:physics}, implies that we first need to understand Borel analysis in a bit more detail than we have done up to this point.

\subsection*{The Borel Integral Representation of a Partition Function}

In retrospect, our definition of a Borel transform in \eqref{boreltrans} might now seem a bit vague or imprecise: why were the powers of the $x$ and $s$ variables related the way they were?, or why was the argument of the Gamma-function chosen the way it was? In order to properly define a Borel transform, and understand what kind of freedom such definition will nonetheless still entail, we shall start by going back to our earlier discussion on the Borel-like rewriting of a partition function, in \eqref{eq:cov} and \eqref{eq:Borel_van} (albeit for the present purposes it will be enough to restrict to one-dimensional integrals). This will illustrate the aforementioned freedom in a (by now) familiar setting.

Consider as usual a ``partition function'', described by a one-dimensional integral\footnote{Throughout this section we shall suppress $\hbar$ and only focus on the coupling constant, for the sake of notational simplicity. Of course it can be added back-in straightforwardly, or attached to the coupling as in section~\ref{sec:quartic}.}
\begin{equation}
\label{partfunctatn}
Z^{(n)}(\lambda) = \frac{1}{2\pi} \int_{\Gamma_{n}} \rmd z\, \rme^{-V_\lambda(z)}.
\end{equation}
\noindent
Herein we are considering a general analytic potential $V_\lambda (z)$, depending on some coupling constant $\lambda$, with a set of saddle-points labelled by $z^*_{n} (\lambda)$ and given by its critical points. The integration contour $\Gamma_{n}$ is the steepest descent-contour, or Lefschetz thimble, through the saddle $z^*_{n}$. Thus, the above integration defines the contribution to the partition function from each saddle. To make it a bit more explicit, let us expand the potential around this $n$th saddle as\footnote{Herein, $N$ could in principle be infinite.}
\be
V_\lambda (z) = V_\lambda (z^*_{n}) + \sum_{k=1}^N \frac{\left( z - z^*_{n} \right)^k}{k!}\, \left. \frac{\rmd^k V_\lambda}{\rmd z^k} \right|_{z= z^*_{n}} \equiv V^*_n (\lambda) + V_\lambda^{(n)} (z).
\ee
\noindent
Rewriting the integral above, \eqref{partfunctatn}, we then have
\begin{equation}
Z^{(n)} (\lambda) = \frac{\rme^{- V^*_n (\lambda)}}{2\pi} \int_{\Gamma_{n}} \rmd z\, \rme^{- V_\lambda^{(n)} (z)}.
\end{equation}

The coupling constant can take any complex value, $\lambda = \left|\lambda\right| \rme^{\rmi\theta}$, which implies that the steepest-descent contour itself will depend on its phase, $\theta$, as $\Gamma_{n} \equiv \Gamma_{n} (\theta)$. Having fixed $\theta$, one can then perform the familiar (asymptotic expansion) evaluation of the integral, in powers of the coupling. This results in the usual
\begin{equation}
Z^{(n)} (\lambda) = \rme^{- V^*_n (\lambda)}\, \Phi_{n} (\lambda),
\end{equation}
\noindent
where\footnote{In this expression we are allowing for the leading term to start at some general power, $\beta_{n}$, of the coupling.}
\begin{equation}
\label{eq:Borel-Phi-n-exp}
\Phi_{n} (\lambda) \simeq \lambda^{\beta_{n}}\, \sum_{k=0}^{+\infty} Z_{k}^{(n)} \lambda^{k},
\end{equation}
\noindent
with the coefficients $Z_{k}^{(n)}$ given by
\begin{equation}
Z_{k}^{(n)} \equiv \frac{1}{k!} \int_{\Gamma_{n}} \rmd z \left. \frac{\rmd^k}{\rmd\lambda^k} \rme^{- V_\lambda^{(n)} (z)} \right|_{\lambda=0}.
\end{equation}
\noindent
Of course that in all cases of interest to us, this expansion is \textit{asymptotic} with the standard large-order behaviour
\begin{equation}
Z_{k}^{(n)} \sim k!.
\end{equation}
\noindent
$\Phi_{n} (\lambda)$ is thus the asymptotic perturbative expansion of the original partition function around the $z^*_{n}$ saddle; an analysis which may be repeated for all saddle-points of the potential.

Following the discussion at the beginning of section~\ref{sec:quartic}, one should now Borel transform the asymptotic expansion \eqref{eq:Borel-Phi-n-exp} by means of \eqref{boreltrans}. This uses the (still not fully justified) assignment where $\lambda^{\alpha} \to \frac{s^{\alpha-1}}{\Gamma (\alpha)}$ (with the exception of the residual coefficient $Z_{0}^{(n)}$ when there is one, \textit{i.e.}, the leading constant term when $\beta_{n}=0$; and while assuming $\beta_n \ge 0$). The Borel transform is thus written as
\begin{equation}
\label{eq:Borel-Phi-n-transf-exp}
\CB[\Phi_n] (s) = \sum_{k=0}^{+\infty} \left( 1 - \delta_{k,0} \delta_{\beta_{n},0} \right) Z_{k}^{(n)}\, \frac{s^{k+\beta_n-1}}{\Gamma (k+\beta_n)}.
\end{equation}
\noindent
This series has a finite, non-zero radius of convergence, whose analytic-continuation yields a structure of singularities and branch-cuts which will give rise to the resurgence properties of the partition function. This partition function can then be properly defined via Borel resummation, \eqref{borelresum}, which is given by a Laplace-type\footnote{In fact, the map $\lambda^{\alpha} \to \frac{s^{\alpha-1}}{\Gamma (\alpha)}$ is an inverse Laplace transform on each term of the asymptotic series.} integral as (for the moment considering the direction of integration along the positive real axis)
\begin{equation}
\label{eq:Borel-Laplace-at-zero}
Z^{(n)} (\lambda) = \rme^{- V^*_n (\lambda)} \left( Z_{0}^{(n)}\, \delta_{\beta_{n},0} + \int_{0}^{+\infty} \rmd s\, \CB[\Phi_{n}](s)\, \rme^{-\frac{s}{\lambda}} \right).
\end{equation}

This is not the only way to find a Borel-like integral-representation of our partition function. For example, one could have instead followed the strategy outlined in the previous section~\ref{sec:lefschetz}, changing the variable of integration from $z$ to the potential as in \eqref{eq:cov}. Let us next check whether this leads to the same result as \eqref{eq:Borel-Laplace-at-zero}. First, one needs to assume that there exists a change of variables, from our current integration variable $z$ to a new one $x$, as in $z=z(x,\lambda)$, such that the potential can be rewritten factoring out the coupling constant,
\begin{equation}
V_\lambda (z) \equiv \frac{1}{\lambda}\, v (x).
\end{equation}
\noindent
In this case, the partition function may be written as
\begin{equation}
Z^{(n)} (\lambda) = \frac{\rme^{- \frac{1}{\lambda} v^*_n}}{2\pi} \int_{\Gamma_{n}} \rmd x\,\frac{\rmd z (x,\lambda)}{\rmd x}\, \rme^{- \frac{1}{\lambda} v^{(n)} (x)}.
\end{equation}
\noindent
Of course that the contour $\Gamma_n$ also changes, as to become a steepest-descent contour in the new variable $x$. We shall proceed by following the discussion in \cite{bh91}, addressing the slightly more general integration problem
\begin{equation}
\label{eq:Partition-function-with Sn(x)}
Z^{(n)} (\lambda) = \frac{\rme^{- \frac{1}{\lambda} v^*_n}}{2\pi} \int_{\Gamma_{n}} \rmd x\, g (x,\lambda)\, \rme^{- \frac{1}{\lambda} v^{(n)} (x)},
\end{equation}
\noindent
for some function $g (x,\lambda)$ with an explicit coupling dependence. But we shall nonetheless focus on the situation of interest to the present discussion, where this coupling dependence may be factored out in the form
\begin{equation}
g(x,\lambda)=\frac{\widetilde{g}(x)}{\sqrt{\lambda}}.
\end{equation}
\noindent
Having this in mind, the adequate rewriting of the partition function becomes
\begin{equation}
\label{Psidefinition}
Z^{(n)} (\lambda) = \rme^{- \frac{1}{\lambda} v^*_n}\, \frac{1}{\sqrt{\lambda}}\, \Psi_{n} (\lambda),
\end{equation}
\noindent
where
\be
\label{Psiintegration}
\Psi_{n} (\lambda) = \frac{1}{2\pi} \int_{\Gamma_n} \rmd x\, \widetilde{g}(x)\, \rme^{- \frac{1}{\lambda} v^{(n)} (x)}.
\ee

First note that \eqref{Psiintegration} is already of similar form to our earlier Borel resummation in \eqref{eq:Borel-Laplace-at-zero}. To try to make it equivalent, and identify the Borel transform in the process, one is tempted to define the Borel plane as
\begin{equation}
\label{eq:Change-of-variables-to-Borel}
s \equiv v^{(n)}(x).
\end{equation}
\noindent
Given that the integration contour $\Gamma_n$ in \eqref{Psiintegration} is a steeepest-descent contour, this seems to guarantee that the corresponding integration upon the $s$-plane, via \eqref{eq:Change-of-variables-to-Borel}, is in fact an integration over the interval $(0,+\infty)$ as one would expect from a Borel resummation (recall \eqref{borelresum}). However, this change of variables is not without a few caveats, which have to be taken into account in order to find the correct integration contour and corresponding Borel transform. The first thing to notice is that upon inversion of \eqref{eq:Change-of-variables-to-Borel}, to find $x \equiv x(s)$, and due to the general nature of the potential, there can in fact be \textit{many} roots $\left\{ x_{\alpha}(s) \right\}$ which solve this equation. In particular, for a general polynomial potential there will be as many roots as the order of the potential, all of them seemingly resulting in the same integration contour $s \in (0,+\infty)$. Closer inspection reveals that this is not quite so \cite{bh91}: in fact, it turns out that \textit{only two} of these roots will pass through the saddle-point $x(z^*_n)$ when $s=0$, while having the required limits at infinity as $s \to +\infty$ (and which are determined by the steepest-descent contour $\Gamma_n$). These two solutions will be labeled as $x_{+}(s)$, corresponding to the half of the contour $\Gamma_n$ which emerges from the saddle point and \textit{goes off} to infinity; and $x_{-}(s)$, which corresponds to the second half of the contour $\Gamma_n$ which \textit{comes in} from infinity and into the saddle point. The resulting integration \eqref{Psiintegration}, after the change of variables \eqref{eq:Change-of-variables-to-Borel}, then becomes
\begin{equation}
\label{eq:BorelINV-Psi-n}
\Psi_{n} (\lambda) = \frac{1}{2\pi} \int_{0}^{+\infty} \rmd s \left\{ \frac{\widetilde{g} \left( x_{+}(s) \right)}{\frac{\rmd v^{(n)}}{\rmd x} \left( x_{+}(s) \right)} - \frac{\widetilde{g} \left( x_{-}(s) \right)}{\frac{\rmd v^{(n)}}{\rmd x} \left( x_{-}(s) \right)}\right\} \rme^{- \frac{s}{\lambda}}.
\end{equation}
\noindent
The integrand inside the wavy-brackets is the Borel transform of $\Psi_{n} (\lambda)$. Following \citep{bh91}, it can also be written as a contour integral\footnote{Notice that this calculation also helps making \eqref{eq:Borel_van} a bit more explicit.},
\begin{equation}
\label{eq:Borel-Psi-n}
\CB[\Psi_{n}](s) = \frac{1}{2\pi} \left\{ \frac{\widetilde{g} \left( x_{+}(s) \right)}{\frac{\rmd v^{(n)}}{\rmd x} \left( x_{+}(s) \right)} - \frac{\widetilde{g} \left( x_{-}(s) \right)}{\frac{\rmd v^{(n)}}{\rmd x} \left( x_{-}(s) \right)}\right\} = \frac{1}{4\pi^{2}\rmi}\, \frac{1}{\sqrt{s}} \oint_{\Gamma_{\circlearrowleft}} \rmd x\, \frac{\sqrt{v^{(n)} (x)}}{v^{(n)} (x) - s}\, \widetilde{g}(x).
\end{equation}
\noindent
The contour $\Gamma_{\circlearrowleft}$ is an anti-clockwise contour enclosing the original steepest descent contour $\Gamma_n$. This expression makes explicit the appearance of the $\sqrt{s}$ factor, crucial to understanding the origin of the double-valuedness of the map between the $s$ and $x$ variables. One may then think of this map as defined on a two-sheeted Riemann surface, connected at the saddle-point $s=0$, and where it has zero-phase on the contour associated to $x_{+}(s)$ and acquiring a phase $\pi$ on the contour associated to $x_{-}(s)$ (see \cite{bh91} for figures and further details).

At this point, we have described two routes towards computing the Borel transform of our original partition function \eqref{partfunctatn}. The first approach directly uses the asymptotic expansion of $\Phi_{n}(\lambda)$, alongside our initial definition of a Borel transform, \eqref{boreltrans}. The second approach changes variables inside the integral for $\Psi_{n}(\lambda)$, following the strategy in section~\ref{sec:lefschetz}, in particular \eqref{eq:cov} and \eqref{eq:Borel_van}. The results are \eqref{eq:Borel-Phi-n-transf-exp} and \eqref{eq:Borel-Psi-n}, respectively, but these two functions are \textit{distinct}. This is rather obvious once one notices that the starting asymptotic expansions differ by a factor of $\sqrt{\lambda}$,
\begin{equation}
\sqrt{\lambda}\, \Phi_{n}(\lambda) \simeq \Psi_{n}(\lambda).
\end{equation}
\noindent
Of course their respective Borel transforms might still be related. Using the asymptotic expansion of $\Phi_{n}$ in \eqref{eq:Borel-Phi-n-exp}, and rewriting it with this extra $\sqrt{\lambda}$ factor, essentially only changes $\beta_n$ as $\beta_n \to \beta_n+\frac{1}{2}$. Then the power-series definition for the Borel transform of $\Psi_{n}(\lambda)$ is
\be
\label{eq:borelwithsquareroot}
\CB[\Psi_{n}](s) = \CB[\sqrt{\lambda}\, \Phi_{n}](s) = \sum_{k=0}^{+\infty} \left( 1 - \delta_{k,0} \delta_{\beta_n,-\frac{1}{2}} \right) Z_k^{(n)}\, \frac{s^{k+\beta_n-\frac{1}{2}}}{\Gamma (k+\beta_n+\frac{1}{2})}.
\ee

As such---and in spite of the fact that $\CB[\Phi_{n}]$ and $\CB[\Psi_{n}]$ are different functions, possibly with different singularity structure upon the complex plane---the question we have left to answer is whether the resurgence information encoded in $\CB[\Phi_{n}]$, \eqref{eq:Borel-Phi-n-transf-exp}, is \textit{the same} as that encoded in $\CB[\Psi_{n}]$, \eqref{eq:borelwithsquareroot}, and how can it be read-off easily in these two different cases. We shall carry out this analysis in our favourite (and simplest!) example so far, the quartic partition-function \eqref{quarticintegral}.

But before we do that, let us make two comments. First, as already mentioned, Borel transforms are naturally attached to any asymptotic (perturbative) expansions, being completely independent of partition functions, or integrals, or steepest-descent methods. We just chose the above example of a partition-function integral, evaluated via saddle-points, in order to illustrate this apparent ambiguity concerning the definition of a Borel transform in a simple and familiar setting. Second, expecting that $\CB[\Phi_{n}]$ and $\CB[\Psi_{n}]$ yield the same resurgence information, \textit{i.e.}, that if the initial asymptotic sequences are somehow related then so should their Borel transforms be, is an extremely natural expectation: as the inverse Borel transform is essentially a Laplace transform, under reasonable assumptions the basic properties of this linear integral operator imply elementary, straightforward relations such as, \textit{e.g.},
\be
\CB [ \lambda^{-1} \Phi ] (s) = \frac{\partial}{\partial s} \CB [\Phi] (s).
\ee
\noindent
Relations such as this strongly hint that very likely also the resurgence structure will be translated along, and we shall make the required dictionary precise in the following.

\subsection*{Constructing Borel Transforms: Some Examples}

Now that we have realized that there is some ambiguity, or \textit{freedom}, in the definition of what a Borel transform is, the reader might be asking: then what is the correct definition? As we shall see, this question is somewhat irrelevant for one should ask beforehand: does this freedom matter? To anticipate the upcoming conclusion, the answer is that in fact \textit{it does not matter}, as the resurgence properties may be read-off from any ``representative'' of the Borel transform.

The quartic partition-function \eqref{quarticintegralX} was thoroughly analyzed in section~\ref{sec:quartic}, with Borel transforms computed straight out from the (perturbative) asymptotic expansions \eqref{coeff-quartic} and \eqref{coeff-quartic-inst}; recall the resulting \eqref{borel-quartic-int} and \eqref{borel-quartic-int-inst}. Let us next try to obtain these same Borel transforms directly out from the partition function, as described above. Recall that the quartic potential \eqref{quarticpotential} has three saddle points, located at $z^*_0 = 0$, $z^*_{\pm} = \pm \sqrt{\frac{6}{\lambda}}$. Rescaling out the coupling constant\footnote{Further recall that, as compared to section~\ref{sec:quartic}, we have now set $\hbar=1$ and only use $\lambda$ for the overall coupling.} in \eqref{quarticintegral} by changing variables inside the integral as\footnote{Do not mistake the (silent) integrand variable $x$ in this section with the (overall) coupling $x$ of section~\ref{sec:quartic}!} $z = \frac{x}{\sqrt{\lambda}}$, one can rewrite the quartic partition-function around each saddle, following \eqref{eq:Partition-function-with Sn(x)} and \eqref{Psidefinition}, as
\begin{eqnarray}
Z^{(0)} (\lambda) &=& \frac{1}{2\pi \sqrt{\lambda}}\, \int_{\Gamma_{0}} \rmd x\, \rme^{- \frac{v^{(0)}(x)}{\lambda}} \equiv \frac{1}{\sqrt{\lambda}}\, \Psi_{0} (\lambda), \\
Z^{(\pm)} (\lambda) &=& \frac{\rme^{- \frac{3}{2\lambda}}}{2\pi \sqrt{\lambda}}\, \int_{\Gamma_{\pm}} \rmd x\, \rme^{- \frac{v^{(\pm)}(x)}{\lambda}} \equiv \rme^{- \frac{3}{2\lambda}}\, \frac{1}{\sqrt{\lambda}}\, \Psi_{\pm} (\lambda),
\end{eqnarray}
\noindent
where the rescaled saddles are now $x^*_0 = 0$ and $x^*_{\pm} = \pm \sqrt{6}$. The ``local'' potentials above are
\begin{equation}
v^{(0)} (x) = \frac{1}{2} x^2 - \frac{1}{24} x^{4} \qquad \text{and} \qquad 
v^{(\pm)} (x) = - \frac{3}{2} + \frac{1}{2} x^2 - \frac{1}{24} x^4,
\end{equation}
\noindent
while the steepest-descent contours of integration can be read-off from figure~\ref{steepest}; which for $\arg\lambda=0$ are
\begin{eqnarray}
\Gamma_{0} &=& \left( - \infty\, \rme^{\rmi\frac{\pi}{4}}, + \infty\, \rme^{\rmi \frac{\pi}{4}} \right), \\
\Gamma_{-} &=& \left( - \infty\, \rme^{\rmi \frac{\pi}{4}}, + \infty\, \rme^{\rmi \frac{3\pi}{4}} \right), \\
\Gamma_{+} &=& \left( - \infty\, \rme^{\rmi \frac{5\pi}{4}}, + \infty\, \rme^{\rmi \frac{7\pi}{4}} \right).
\end{eqnarray}
\noindent
Do note that each of these contours naturally passes through its respective saddle. 

We proceed with the change of variables \eqref{eq:Change-of-variables-to-Borel}, defining the Borel plane as
\begin{equation}
s = v^{(n)} (x) \qquad \text{for} \qquad n=0,\pm.
\end{equation}
\noindent
Inverting this map yields $x \equiv x(s)$. As we are dealing with a polynomial potential of fourth degree, one finds four different solutions to this equation. They are:
\begin{eqnarray}
\text{Saddle }\, x_{0}^{*}\, : &\qquad& x_{\pm}^{(\pm)} (s) = (\pm) \sqrt{ 6 \left( 1 \pm \sqrt{1-\frac{2s}{3}} \right)}, \\
\text{Saddle }\, x_{\pm}^{*}\, : &\qquad& x_{\pm}^{(\pm)} (s) = (\pm) \sqrt{6 \left( 1 \pm \sqrt{-\frac{2s}{3}} \right)}.
\end{eqnarray}
\noindent
Note that, with a slight abuse of notation, in here the $\pm$ and $(\pm)$ signs in the solutions are \textit{labeling the solutions} for each saddle, and are \textit{not related} to the $\pm$ signs labeling the saddles---which hopefully will be clear from context. All these maps will result in a variable $s$ which is varying in $(0,+\infty)$, but, as we discussed earlier, only \textit{two} for each saddle will correspond to the original steepest-descent contours $\Gamma_n$. After some careful examination of the signs of the square roots, one finds that as $s$ varies from zero to positive infinity for the saddle $x_0^*$ one needs to choose
\begin{equation}
x_{-}^{(-)} (s) \in \left( 0, + \infty\, \rme^{\rmi \frac{\pi}{4}} \right), \qquad x_{-}^{(+)} (s) \in \left( 0, - \infty\, \rme^{\rmi \frac{\pi}{4}} \right).
\end{equation}
\noindent
In this way $x_{-}^{(-)}$ corresponds to the half-contour leaving the saddle-point off to infinity, while $x_{-}^{(+)}$ corresponds to the half-contour coming into the saddle-point from infinity. In a self-explanatory notation, we can thus write $\Gamma_{0} \equiv \Gamma_{-}^{(-)} - \Gamma_{-}^{(+)}$. A completely analogous analysis may be done for the other saddles, $x^*_{\pm}$, where from the four solutions previously found, two go through one saddle, the other two through the other. In particular, as one varies $s \in (0,+\infty)$, one has:
\begin{eqnarray}
\text{Saddle }\, x^*_+\, : &\qquad& \Gamma_+ = \Gamma_+^{(+)} - \Gamma_-^{(+)}, \\
\text{Saddle }\, x^*_-\, : &\qquad& \Gamma_- = \Gamma_-^{(-)} - \Gamma_+^{(-)}.
\end{eqnarray}

One can proceed by using the Laplace integral-representation for $\Psi_n (\lambda)$ in \eqref{eq:BorelINV-Psi-n} in order to compute the Borel transforms of $\Psi_{0} (\lambda)$ and $\Psi_{\pm} (\lambda)$, via \eqref{eq:Borel-Psi-n}. For the ``perturbative saddle'',
\bea
\Psi_0 (\lambda) &=& \frac{1}{2\pi} \int_{0}^{+\infty} \rmd s\, \left\{ \frac{1}{\frac{\rmd v^{(0)}}{\rmd x} \left( x_-^{(-)} (s) \right)} - \frac{1}{\frac{\rmd v^{(0)}}{\rmd x} \left( x_-^{(+)} (s) \right)} \right\} \rme^{-\frac{s}{\lambda}} = \nonumber \\
&=& \frac{1}{2\pi} \int_{0}^{+\infty} \rmd s\, \left\{ \frac{2}{\frac{\rmd v^{(0)}}{\rmd x} \left( x_-^{(-)} (s) \right)} \right\} \rme^{-\frac{s}{\lambda}},
\label{Psi0lambdadefined}
\eea
\noindent
as $\frac{\rmd v^{(0)}}{\rmd x} \left( x_-^{(-)} (s) \right) = - \frac{\rmd v^{(0)}}{\rmd x} \left( x_-^{(+)} (s) \right) = \sqrt{6 ( 1 - \sqrt{1-2s/3} )}\, \sqrt{1-2s/3}$. It then follows:
\be
\label{eq:Borel-Z0-after-change-var}
\CB [\Psi_0] (s) = \frac{1}{2\pi \sqrt{s}}\, \frac{\sqrt{1+\sqrt{1-\frac{2s}{3}}}}{\sqrt{1-\frac{2s}{3}}}.
\ee
\noindent
In much the same way we can calculate
\begin{equation}
\label{borel1/2factor}
\Psi_+ (\lambda) = - \Psi_- (\lambda) \equiv \int_{0}^{+\infty} \rmd s\, \CB [\Psi_1] (s)\, \rme^{-\frac{s}{\lambda}},
\end{equation}
\noindent
leading to the Borel transform
\begin{equation}
\label{eq:Borel-Z1-after-change-var}
\CB [\Psi_1] (s) = - \frac{\rmi}{2\pi \sqrt{2s}}\, \frac{\sqrt{1+\sqrt{1+\frac{2s}{3}}}}{\sqrt{1+\frac{2s}{3}}}.
\end{equation}

At first sight the Borel transforms we have just computed, $\CB [\Psi_0] (s)$ and $\CB [\Psi_1] (s)$, look very different from the Borel transforms computed in subsection~\ref{subsec:basicresurgence}, \eqref{borel-quartic-int} and \eqref{borel-quartic-int-inst}. This should not come as a surprise: indeed, as discussed earlier in this section, the asymptotic expansions for the $\Phi$'s and the $\Psi$'s differ by a factor of $\sqrt{\lambda}$. But as a consistency check on our calculations, we may now show how to obtain $\CB [\Psi_0] (s)$ and $\CB [\Psi_1] (s)$ \textit{directly} from the asymptotic expansions for $\Phi_0$ and $\Phi_1$, written in \eqref{coeff-quartic} and \eqref{coeff-quartic-inst}. Recalling that $\Phi_0$ is the expansion around the saddle $x^*_0$, and that $\Phi_1$ the expansion around either saddle $x^*_{\pm}$ , we are expecting to find:
\bea
\CB[\Psi_0](s) &=& \CB[\sqrt{\lambda}\, \Phi_0](s), \\
\CB[\Psi_1](s) &=& \CB[\sqrt{\lambda}\, \Phi_1](s).
\eea
\noindent
Let us check that this is in fact verified. Start with the asymptotic expansion for $\Phi_0 (\lambda)$, in \eqref{coeff-quartic}, which we reproduce in here
\be
\Phi_{0} (\lambda) \simeq \sum_{k=0}^{+\infty} Z^{(0)}_k\, \lambda^k, \qquad Z^{(0)}_k = \left( \frac{2}{3} \right)^k \frac{(4k)!}{2^{6k}\, (2k)!\, k!}\, Z^{(0)}_0,
\ee
\noindent
and where in the present conventions $Z^{(0)}_0 = \frac{1}{\sqrt{2\pi}}$. Computing the Borel transform above, using \eqref{eq:borelwithsquareroot}, it follows
\be
\CB[\sqrt{\lambda}\, \Phi_0](s) = \sum_{k=0}^{+\infty} Z^{(0)}_k\, \frac{s^{k-\frac{1}{2}}}{\Gamma \left( k+\frac{1}{2} \right)} = \frac{1}{2\pi \sqrt{s}}\, \frac{\sqrt{1+\sqrt{1-\frac{2s}{3}}}}{\sqrt{1-\frac{2s}{3}}},
\ee
\noindent
which is precisely $\CB[\Psi_0](s)$ as expected. Performing the same analysis using the asymptotic expansion for $\Phi_1 (\lambda)$, in \eqref{coeff-quartic-inst}, one likewise obtains
\be
\CB[\sqrt{\lambda}\, \Phi_1](s) = \sum_{k=0}^{+\infty} Z^{(1)}_k\, \frac{s^{k-\frac{1}{2}}}{\Gamma \left( k+\frac{1}{2} \right)} = - \frac{\rmi}{2\pi \sqrt{2s}}\, \frac{\sqrt{1+\sqrt{1+\frac{2s}{3}}}}{\sqrt{1+\frac{2s}{3}}},
\ee
\noindent
which precisely matches $\CB[\Psi_1](s)$. This explicitly checks our calculations, verifying how there is more than one way to compute a Borel transform, leading to \textit{distinct, albeit related} results.

At this stage we are back to our main question. The Borel transforms computed in section~\ref{sec:quartic}, \eqref{borel-quartic-int} and \eqref{borel-quartic-int-inst}, had a singularity structure with poles and logarithms, \eqref{borel-quartic-int-expanded} and \eqref{borel-quartic-int-expanded-inst}, which immediately made explicit the resurgence properties of the quartic partition-function. In fact, the abstraction of these resurgence properties out from their associated Borel singularities, realized in the resurgence relations \eqref{Delta+A} and \eqref{Delta-A}, or in the (finite) alien chain \eqref{linear-algebraic-structure}, was what then allowed to jump-start the whole resurgence program. But will the same remain true with the Borel transforms we have just computed, \eqref{eq:Borel-Z0-after-change-var} and \eqref{eq:Borel-Z1-after-change-var}? How can we recover the (algebraic) resurgence structure encoded in \eqref{linear-algebraic-structure} out of these ``new'' Borel transforms?

The first thing to notice is that the Borel transforms \eqref{eq:Borel-Z0-after-change-var} and \eqref{eq:Borel-Z1-after-change-var} both have a square-root singularity at the origin, \textit{i.e.}, $\sim 1/\sqrt{s}$ near $s=0$. This is a somewhat artificial singularity from the resurgence standpoint, and is essentially related to the coupling square-root in the initial asymptotic expansion. In order to infer whether there is any clean resurgent behaviour in our ``new'' Borel transforms, we need to analyze them away from this singularity. In fact, we shall need to focus on the singularity at $s=\frac{3}{2}$ of the Borel transform $\CB[\Psi_0](s)$, and on the singularity at $s=-\frac{3}{2}$ of the Borel transform $\CB[\Psi_1](s)$. If resurgence is also at work in here, near each of these singularities we should see the ``reappearance'' of the other sector.

Expanding the Borel transforms \eqref{eq:Borel-Z0-after-change-var} and \eqref{eq:Borel-Z1-after-change-var} near their singularities at $s = \pm A \equiv \pm \frac{3}{2}$, we find
\begin{eqnarray}
\label{expansion-Borel-sqrt}
\CB[\Psi_0](s) &=& \sum_{k=-1}^{+\infty} B_k^{(0)} \left( s - A \right)^{\frac{k}{2}}, \qquad B_k^{(0)} = \frac{\rme^{\rmi \frac{k\pi}{2}}}{2\pi^{3/2}}\, \frac{\Gamma \left( k+\frac{3}{2} \right)}{\Gamma \left( k+2 \right)}\, A^{-\frac{k+1}{2}}, \\
\CB[\Psi_1](s) &=& \sum_{k=-1}^{+\infty} B_k^{(1)} \left( s + A \right)^{\frac{k}{2}}, \qquad B_k^{(1)} = \frac{\rme^{\rmi\pi}}{\left(2\pi\right)^{3/2}}\, \frac{\Gamma \left( k+\frac{3}{2} \right)}{\Gamma \left( k+2 \right)}\, A^{-\frac{k+1}{2}}.
\end{eqnarray}
\noindent
Recall that back in \eqref{borel-quartic-int-expanded} and \eqref{borel-quartic-int-expanded-inst} the singularities were restricted to a simple pole and a logarithm branch-cut. Things here are a bit ``messier'', as there is now an \textit{infinite number} of branch-cuts, one per fractional power in the above expansions (\textit{i.e.}, whenever $k$ is odd). Thus, at first sight, it now seems much harder to find the resurgence of, say, the $\Psi_1$ sector inside the expansion of $\CB[\Psi_0](s)$ around $s=A$, due to this plethora of branch-cuts. This scenario quickly becomes more amenable once one computes the expansion of $\CB[\Psi_1](s-A)$ near $s=A$ (which anyways is what we hope to find inside $\CB[\Psi_0](s)$ near $s=A$ in the first place). One easily obtains
\begin{equation}
\label{psi1borelHAT}
\CB[\Psi_1] \left(s-A\right) = \sum_{k=-1}^{+\infty} \widehat{B}_k^{(1)} \left( s - A \right)^{k+\frac{1}{2}}, \qquad \widehat{B}_k^{(1)} = \frac{\rmi\, \rme^{\rmi\pi k}}{2\pi^{3/2}}\, \frac{\Gamma \left( 2k+\frac{5}{2} \right)}{\Gamma \left( 2k+3 \right)}\, A^{- \left( k+1 \right)}.
\end{equation}
\noindent
Comparing the coefficients $\widehat{B}_k^{(1)}$ with those particular coefficients $B_k^{(0)}$ which were attached to a \textit{singular} fractional-power, \textit{i.e.}, with the coefficients $B_{2k+1}^{(0)}$, one finds
\be
B_{2k+1}^{(0)} = \widehat{B}_k^{(1)}.
\ee
\noindent
In this way, one may \textit{split} the power-series expansion \eqref{expansion-Borel-sqrt} into its regular \textit{holomorphic} half and its \textit{singular} half, both of which are \textit{infinite series} on their own. Then, the singular series will match \eqref{psi1borelHAT} above in such a way that one can write the resurgence-like relation
\be
\label{resurgentPsi0}
\CB[\Psi_0](s) = \left( + 2 \right) \times \frac{1}{2}\, \CB[\Psi_1] \left(s-A\right) + \text{holomorphic}.
\ee
\noindent
The exact same analysis equally applies to $\CB[\Psi_1](s)$. This time around
\be
\CB[\Psi_0] \left(s+A\right) = \sum_{k=-1}^{+\infty} \widehat{B}_k^{(0)} \left( s + A \right)^{k+\frac{1}{2}}, \qquad \widehat{B}_k^{(0)} = \frac{1}{2^{1/2}\pi^{3/2}}\, \frac{\Gamma \left( 2k+\frac{5}{2} \right)}{\Gamma \left( 2k+3 \right)}\, A^{- \left( k+1 \right)},
\ee
\noindent
in such a way that
\be
B_{2k+1}^{(1)} = - \frac{1}{2}\, \widehat{B}_k^{(0)},
\ee
\noindent
from where the resurgence-like relation emerges
\be
\label{resurgentPsi1}
\CB[\Psi_1](s) = \left( - 1 \right) \times \frac{1}{2}\, \CB[\Psi_0] \left(s+A\right) + \text{holomorphic}.
\ee 
\noindent
One may compare these results, \eqref{resurgentPsi0} and \eqref{resurgentPsi1}, to \eqref{borel-quartic-int-expanded} and \eqref{borel-quartic-int-expanded-inst}, respectively. Recall that one needs to use the ``dictionary'' from \eqref{eq:B_Psi_n:d=odd} to read Stokes constants. In this way, via \eqref{eq:B_Psi_n:d=odd}, \eqref{resurgentPsi0} yields the Stokes constant $S_1 = - 2$, which is the precise constant obtained in \eqref{Delta+A}. Likewise, \eqref{resurgentPsi1} yields the Stokes constant $S_{-1} = 1$, which matches the one obtained in \eqref{Delta-A}.

We may finally conclude that the resurgence structure is seen \textit{both} by the $\Psi$-``representative'' which was used above, as well as by the $\Phi$-``representative'' used in section~\ref{sec:quartic}. Further, as expected, \textit{both} these descriptions lead to the exact \textit{same} structure, \textit{i.e.}, they are in the same ``class''. This means that, starting-off from our ``new'' resurgence relations \eqref{resurgentPsi0} and \eqref{resurgentPsi1}, we may once again forget about any functional details concerning Borel singularities, and express resurgence solely at the level of the asymptotic series $\Psi_0 (\lambda)$ and $\Psi_1 (\lambda)$ (our transseries building blocks). Herein, resurgence is likewise formalized at an abstract level using alien derivatives and bridge equations in the \textit{exact same fashion} as before, \textit{i.e.}, using the same bridge equations \eqref{Delta+A} and \eqref{Delta-A} and the same (finite) alien chain \eqref{linear-algebraic-structure} with its associated allowed motions. The main difference, of course, is in the calculation process leading up to \eqref{resurgentPsi0} and \eqref{resurgentPsi1}. Unlike \eqref{borel-quartic-int-expanded} and \eqref{borel-quartic-int-expanded-inst}, we no longer have just a \textit{finite} number of singular structures (in fact, \textit{two} singular contributions, one pole and one logarithmic branch-cut), but rather an \textit{infinite} number of singular fractional powers, with the ``reappearance'' properties intertwined inside this series. While in the former case resurgence is seen directly in the functional coefficients of the singularities, in the latter case resurgence is only seen once the numerical coefficients of \textit{all} singular powers are hold \textit{together}. In other words, although (as before) we still need to take into account \textit{all} singular structures, now this amounts to an \textit{infinite} number of terms---just a few will not be enough!

This implies that using the $\Phi$-``representative'' rather than the $\Psi$-``representative'' allows for a more transparent reading of the encoded resurgence structure. For convenience, one should thus select the $\Phi$-``representative'' as the ``cleanest'' representative, leading up to the definition of a \textit{simple resurgent function}, see, \textit{e.g.}, \cite{e81, cnp93a, cnp93b, ss03}. A \textit{simple resurgent function} is a resurgent function (asymptotic series) for which its Borel singularities solely reduce to a simple pole alongside a logarithmic branch-cut. To be fully precise, we should not say ``\textit{simple} resurgent \textit{function}'', but rather ``\textit{simple} resurgent \textit{Borel representative}''---out of the (possibly) infinite number of distinct but related Borel representatives, for the (possibly) infinite number of distinct organizations of the asymptotic series we started off with. This will be made precise below.

Do note that our example above corresponds to a linear problem, where we were able to compute everything exactly. In general, nonlinear problems this will \textit{not} be the case. This implies that in such cases one can only access partial information concerning Borel singularities and, if not dealing with a \textit{simple representative}, the resurgence properties of our asymptotic sequence will be masked and may be very difficult to retrieve. It then becomes of utmost importance to know how to reach a simple Borel representative, independent of our starting point. As we shall discuss next, there is a very precise way of relating Borel transforms associated to asymptotic series which differ by factors of the coupling, and all members of such a family of sequences will be equally well described by their single \textit{simple representative}.

This analysis makes two things clear. The first is that the definition of a Borel transform does not need to be very precise, and may allow for some freedom as discussed. It does not really matter, as at the end of the day one reaches the same resurgence structure. The second is that precisely because one always ends up with the same algebraic structure (upon abstraction from the functional nature of Borel singularities), the true fundamental structure to work with is that of abstract bridge equations and alien chains, from where all nonperturbative results may follow.

\subsection*{On the Class of Simple Resurgent Functions}

Let us consider any observable $\Phi$, typically arising from some nonlinear problem, with an asymptotic perturbative expansion in the coupling $\lambda$ given by
\be
\label{simpleasymptoticsequence}
\Phi (\lambda) \simeq \sum_{k=0}^{+\infty} \Phi_k\, \lambda^{\alpha k + \beta},
\ee
\noindent
with $\alpha, \beta \in \BR$. Let us further assume that the large-order behaviour of the (factorially divergent) coefficients above is of the form
\be
\Phi_k \sim ( \alpha k + \delta - 1 )!,
\ee
\noindent
with some other $\delta \in \BR$. Given this data, we wish to compute the Borel transform of $\Phi$, $\CB [\Phi] (s)$, in such a way as to obtain a \textit{simple representative}, \textit{i.e.}, in such a way that $\CB [\Phi] (s)$ only has simple poles and logarithmic branch-points as singularities (this is known as a \textit{simple singularity}\footnote{All which is required is that the function in front of the logarithm is holomorphic. This requirement allows this function to be rewritten---or understood---as the Borel transform of \textit{some other} asymptotic series (evaluated around the origin of \textit{its own} complex Borel plane), as given in the expression above. In this setting, $\CB [\Psi_\omega]$ is usually denoted as the \textit{minor} of $\CB [\Phi]$ (at $\omega$, along the ray $\arg \omega$).}). That is to say
\be
\label{simpleBorelsingularities}
\CB [\Phi] (s) = \frac{\mathsf{S}'_\omega}{2\pi\rmi \left( s-\omega \right)} + \mathsf{S}_\omega \times \CB [\Psi_\omega] \left( s-\omega \right) \frac{\log \left( s-\omega \right)}{2\pi\rmi} + \text{holomorphic},
\ee
\noindent
near each singular point $\omega$, with\footnote{To be precise, $\mathsf{S}'_\omega = \mathsf{S}_\omega \times \text{residual coefficient}$. In other words, up to the residual coefficient (and a factor of $2\pi\rmi$), $\mathsf{S}_\omega$ is the \textit{residue} of $\CB [\Phi] (s)$ at $\omega$ (hence the nomenclature ``Borel residue'' used throughout the text).} $\mathsf{S}'_\omega, \mathsf{S}_\omega \in \BC$ and $\Psi_\omega$ some other asymptotic sector (such that $\CB [\Psi_\omega]$ is holomorphic near the origin, \textit{i.e.}, it introduces no further singularities in \eqref{simpleBorelsingularities} above).

When all that is available to us is an asymptotic sequence as in \eqref{simpleasymptoticsequence}, one might be tempted to use our initial definition of Borel transform, \eqref{boreltrans}, and thus write
\be
\CB [\Phi] (s) = \sum_{k=0}^{+\infty} \Phi_k\, \frac{s^{\alpha k + \beta - 1}}{\Gamma \left( \alpha k + \beta \right)}.
\ee
\noindent
The problem, of course, is the \textit{off-set} $\delta$. If $\delta \neq \beta$, we are not removing the precise factorial growth and this will result in more intricate Borel transforms than one would wish for. The solution is nonetheless simple. Introducing
\be
\label{Phi_[gamma]-DEF}
\Phi_{[\gamma]} (\lambda) \equiv \lambda^\gamma\, \Phi (\lambda) \simeq \sum_{k=0}^{+\infty} \Phi_k\, \lambda^{\alpha k + \beta + \gamma}
\ee
\noindent
and choosing $\gamma = \delta-\beta$, then the Borel transform of $\Phi_{[\gamma]}$ will remove the precise factorial growth leading to the simple representative,
\be
\CB [\Phi_{[\delta-\beta]}] (s) = \sum_{k=0}^{+\infty} \Phi_k\, \frac{s^{\alpha k + \delta - 1}}{\Gamma \left( \alpha k + \delta \right)}.
\ee

All asymptotic expansions of the type $\Phi_{[\gamma]} (\lambda) = \lambda^\gamma\, \Phi (\lambda)$, with $\gamma \in \BR$, are said to be in the same \textit{resurgence class} (or family), meaning that their resurgence structure is equivalent for all values of $\gamma$ and that they may all be related to the one particular choice $\bar{\gamma}$ where the Borel transform $\CB [\Phi_{[\bar{\gamma}]}] (s)$ will have the singularity structure \eqref{simpleBorelsingularities} of a \textit{simple resurgent function}. We shall denote this particular choice, $\Phi_{[\bar{\gamma}]}$, as the \textit{standard representative}, the representative where resurgent analysis is most straightforward. One may reach this standard representative \textit{directly} at the level of the Borel transforms: in appendix~\ref{app:borel} we show how to relate Borel transforms of distinct members in the family $\{ \Phi_{[\gamma]} \}$, leading up to\footnote{Residual coefficients, which need to be removed upon Borel transformation, must be considered separately.}
\begin{equation}
\label{eq:finding-Borel-rep-from-other}
\CB [\Phi_{[\gamma']}] (s) = D_s^{\gamma-\gamma'} \CB [\Phi_{[\gamma]}] (s).
\end{equation}
\noindent
In here, $D_s^{\alpha}$ is the order-$\alpha$ (\textit{fractional}) derivative with respect to $s$, when $\alpha>0$; and it is the (\textit{fractional}) $|\alpha|$-integration with respect to $s$, when $\alpha<0$ (but see appendix~\ref{app:borel} and, \textit{e.g.}, \cite{lot76} for further details). Having this relation in mind, once one computes the Borel transform of some non-standard representative $\Phi_{[\gamma]}$, the only difficulty in subsequently finding the standard-representative Borel transform $\CB [\Phi_{[\bar{\gamma}]}] (s)$ is how to find $\bar{\gamma}$. At the level of Borel transforms this is done via an analysis of the original $\CB [\Phi_{[\gamma]}] (s)$ around its singular points, which however is not necessarily always straightforward as we shall illustrate next.

\subsection*{Another Look at the Quartic Partition-Function}

Let us illustrate how \eqref{eq:finding-Borel-rep-from-other} works in practice, back in the example of the quartic partition-function. We shall focus on the saddle $x^*_0$ and, for the sake of notational simplicity, suppress the $0$ subscript in the asymptotic series around this saddle, say $\Phi_0 (\lambda)$ in \eqref{coeff-quartic} or $\Psi_0 (\lambda)$ in \eqref{Psi0lambdadefined}, from now on. In this example we already know which is the standard representative; it is the original asymptotic series \eqref{coeff-quartic} with (simple) Borel transform \eqref{borel-quartic-int-expanded}, which we shall denote by $\Phi_{[0]}$. All other family members within this resurgence class may be obtained via
\be
\Phi_{[\gamma]} (\lambda) = \lambda^\gamma\, \Phi_{[0]} (\lambda).
\ee
\noindent
For example, $\Psi = \sqrt{\lambda}\, \Phi$ in \eqref{Psi0lambdadefined} may now be written as $\Phi_{[1/2]} (\lambda)$. Equation \eqref{eq:finding-Borel-rep-from-other} then explicitly relates both these Borel transforms (which we computed earlier) as
\be
\label{semiderivativeBorel}
\CB [\Phi_{[0]}] (s) = D_s^{1/2}\, \CB [\Phi_{[1/2]}] (s).
\ee
\noindent
Here we have to deal with a \textit{non-local} fractional operator, the semi-derivative $D_s^{1/2}$, which is defined by (see, \textit{e.g.}, \cite{lot76} or appendix~\ref{app:borel})
\be
D_s^{1/2} f(s) = \frac{1}{\Gamma \left(1/2\right)}\, \frac{\rmd}{\rmd s} \int_0^s \rmd t\, \frac{f(t)}{\sqrt{s-t}}.
\ee
\noindent
It is this non-locality of the semi-derivative which explains why all the resurgence information which is ``localized'', as functional coefficients, in the two singularities of $\CB [\Phi_{[0]}] (s)$ becomes ``spread-out'' over the infinite number of singular branch-cuts in $\CB [\Phi_{[1/2]}] (s)$, as discussed earlier. The formal derivation that \eqref{semiderivativeBorel} above holds, at the level of series expansions, may be found in appendix~\ref{app:borel}. Here, we shall instead focus on this relation at the level of \textit{Borel singularities}, \textit{i.e.}, we shall analyze the validity of \eqref{semiderivativeBorel} around $s=A$ (we are addressing the perturbative saddle).

Recall that the expansion of $\CB [\Phi_{[1/2]}] (s)$ around $s=A$ is given by \eqref{expansion-Borel-sqrt}, which we (partially) recall in here as
\be
\CB [\Phi_{[1/2]}] (s) = \sum_{k=-1}^{+\infty} B_k^{(0)} \left( s - A \right)^{\frac{k}{2}}.
\ee
\noindent
In order to find the action of the semi-derivative on this function, let us first compute the action of the semi-derivative on each fractional-power term. Using its definition, one finds\footnote{Below, the $(-1)$ factor multiplying $A$ is actually $\rme^{-\rmi\pi}$, while we change signs as $(A-s) = \rme^{\rmi\pi} (s-A)$.}
\begin{eqnarray}
D_s^{1/2} \left(s-A\right)^{\frac{n}{2}} &=& \frac{\left( -A \right)^{\frac{n}{2}+1}}{\left( n-1 \right) \sqrt{\pi\, s^3}} - \frac{\left( -A \right)^{\frac{n}{2}}\left( s-A \right)}{\left( n-1 \right) \sqrt{\pi\, s^{3}}}\, {}_2 F_{1} \left( 1, -\frac{n}{2}, -\frac{1}{2} \right. \left| \frac{s}{A} \right), \qquad n\ge2, \\
D_s^{1/2} \left(s-A\right)^{\frac{1}{2}} &=& - \rmi \sqrt{\frac{A}{\pi\, s}} + \frac{\rmi}{\sqrt{\pi}}\, \log \left( \sqrt{s}+\sqrt{A} \right) - \frac{\rmi}{2\sqrt{\pi}}\, \log\left( s-A \right) + \frac{\sqrt{\pi}}{2}, \\
D_s^{1/2} \left(s-A\right)^0 &=& \frac{1}{\sqrt{\pi\, s}}, \\
D_s^{1/2} \left(s-A\right)^{-\frac{1}{2}} &=& \sqrt{\frac{A}{\pi\, s}}\, \frac{\left( -\rmi \right)}{s-A}.
\end{eqnarray}
\noindent
The hypergeometric function is easily expanded using its standard properties; see, \textit{e.g.}, \cite{olbc10}. In particular, this directly shows how when $n \ge 2$ and $n$ is \textit{even} there are no logarithmic contributions; while when $n \ge 3$ and $n$ is \textit{odd}, there will be. In the latter case, for $z \equiv \frac{s}{A} \to 1^{-}$,
\begin{eqnarray}
{}_2 F_{1} \left( 1, -\frac{n}{2}, \left. -\frac{1}{2}\, \right| z \right) &=& \frac{\frac{3}{2}}{\Gamma \left(\frac{n-1}{2}\right) \Gamma \left(-\frac{n}{2}\right)}\,
\left( z-1 \right)^{\frac{n-3}{2}}\, \log \left(1-z\right) \times \\
&&
\times
\sum_{k=0}^{+\infty} \frac{\Gamma \left( k-\frac{3}{2} \right)}{\Gamma \left( k+1 \right)}\, \left( 1-z \right)^k + \text{regular (non-logarithmic) terms}. \nonumber
\end{eqnarray}
\noindent
Using this expansion alongside the semi-derivatives above, one can compute, term by term,
\bea
D_s^{1/2}\, \CB [\Phi_{[1/2]}] (s) &=& \sum_{k=-1}^{+\infty} B_k^{(0)}\, D_s^{1/2} \left( s - A \right)^{\frac{k}{2}} = -\frac{\rmi B^{(0)}_{-1}}{\sqrt{\pi}}\, \frac{1}{s-A} - \frac{\rmi B^{(0)}_1}{2\sqrt{\pi}}\, \log \left( s-A \right) + \\
&&
+ \rmi \log \left( s-A \right)\, \sum_{m=1}^{+\infty} \frac{\left(-1\right)^m B^{(0)}_{2m+1}}{\Gamma \left( m+1 \right) \Gamma \left( -m-\frac{1}{2} \right)} \left(s-A\right)^m + \text{regular terms}. \nonumber
\eea
\noindent
Recall how we are only interested in the singular terms. Next, in order to compare to the expansion of $\CB [\Phi_{[0]}] (s)$ around the same point and check if we do obtain the same result, we also need to expand \eqref{borel-quartic-int-expanded}. Using the exact information encoded in \eqref{borel-quartic-int-inst} and expanding the hypergeometric function akin to what was done above, it follows
\bea
\CB [\Phi_{[0]}] (s) &=& - \frac{1}{2\pi^{\frac{3}{2}}}\, \frac{1}{s-A} + \frac{1}{16\pi^{\frac{3}{2}}}\, \log \left( s-A \right) + \\
&&
+ \frac{1}{3 \sqrt{2} \pi^{\frac{5}{2}}} \log \left( s-A \right)\, \sum_{m=1}^{+\infty} \frac{\left(-1\right)^m}{A^m}\, \frac{\Gamma \left( m+\frac{5}{4} \right) \Gamma \left( m+\frac{7}{4} \right)}{\Gamma \left( m+1 \right) \Gamma \left( m+2 \right)} \left(s-A\right)^m + \text{regular terms}. \nonumber
\eea
\noindent
It is now a matter of some simple algebra to check that we have obtained the very same result and thus checked the validity of \eqref{semiderivativeBorel}. A completely analogous comparison may be done for the nonperturbative saddles, matching the Borel transforms of $\Phi_1$ and $\Psi_1$ close to the singularity at $s = - A$. In this way, these Borel transforms encode the same resurgence information and thus belong to the same  resurgence family. Further, it should be clear that the resurgence information which is somewhat ``locally'' encoded in $\CB [\Phi] (s)$, becomes ``non-locally'' encoded in $\CB [\Psi] (s)$ due to the non-local nature of the (fractional) semi-derivative.

\section{Physical Resurgence: From Lattices to Virasoro Algebras}\label{sec:physics}

Transseries structures are rather simple yet very general constructions. They prove extremely useful in finding or (iteratively) constructing \textit{explicit} solutions to a wide range of linear and nonlinear problems. For instance, they can give rise to general solutions of differential equations, algebraic equations, finite difference equations, integrals or integral equations, functional equations, and so on (see, \textit{e.g.}, \cite{e08} for a review with many such explicit examples). Yet, so far we have been solely constructing transseries within very specific and simple examples. We shall now be more precise on how they come about, and how general their building-blocks might be.

In this process, it is important to have two points in mind. The first point was already discussed in the previous section, where explicit Borel calculations taught us how, at the end of the day, all one should care about are the \textit{abstract resurgence relations} in-between the different transseries building-blocks, or \textit{nodes}. In other words, the specific details concerning which kind of Borel singularities show-up are somewhat irrelevant, as any ``representative'' will essentially yield the very same resurgence relations. But these \textit{nodes} are asymptotic series which are ``attached'' to specific (and distinct) non-analytic transmonomials, which begs for some sort of \textit{classification} of different types of transmonomials. Such structures will surely play a role in generalizing the alien \textit{chains} we studied in section~\ref{sec:quartic} to more general (and \textit{higher-dimensional}) \textit{lattices}.

The second point has to do with the resurgent nature of a general transseries. Resurgence induces relations between the many nodes that build-up a transseries, as was already schematically depicted early-on, in \eqref{resurgence_predicting} and \eqref{resurgence_decoding}. One can imagine an example where these resurgence relations \textit{directly} connect the perturbative expansion to \textit{all} other possible nonperturbative sectors. In this case, one may ask how exactly should \eqref{resurgence_decoding} be fully exploited to decode all multi-instanton towers---and thus construct the complete transseries solution---out of the (also complete!) perturbative expansion. For this strategy to be fully implemented, one must first understand what is the full \textit{set of all allowed motions} in-between transseries nodes, in the case of general, higher-dimensional alien lattices. Let us explore these several issues in the following.

\subsection*{General Transseries from Ordinary Differential Equations}

A natural and illustrative set-up where transseries appear is that of ordinary differential equations (ODEs). Suppose we are given a $k$th-order nonlinear ODE, for the (yet unknown) function $u(x)$ in the variable $x$ (for instance, recall the second-order ODE for the free energy $F(x)$, \eqref{quarticNLODE}, which was our main example back in subsection~\ref{subsec:Fnonlinear}). If there are no singularities at $x=0$ one may first try to solve this ODE for $u(x)$ perturbatively, in powers of the monomial $x$ (again, as we did back in subsection~\ref{subsec:Fnonlinear}, where a perturbative \textit{ansatz} for $F(x)$ led to the recursion \eqref{quarticFrecursion}, yielding the perturbative coefficients $F_n^{(0)}$). Now, if the resulting perturbative coefficients $u_n^{(0)}$ grow at most exponentially with $n$, then the perturbative series has a finite, non-zero radius of convergence and our power-series solution may be studied with standard tools from classical analysis. Unfortunately, this is generically \textit{not} the case, with the perturbative coefficients $u_n^{(0)}$ instead diverging factorially fast. What this divergence essentially means is that $x=0$ is sitting right next to a branch-cut of the solution, and thus it cannot possibly be made of powers of $x$ alone---one needs extra \textit{non-analytic} contributions at $x=0$ in order to properly characterize $u(x)$. Such contributions, which add to the \textit{monomial} $M \equiv x$, will be denoted by \textit{transmonomials}, $T$, and different sets of transmonomials may be required for different problems.

One basic transmonomial we have already found several times before is $\exp \left( - \frac{1}{x} \right)$; but we might naturally ask what other types of transmonomials may one usually encounter? To address this question in a concrete example, let us go back to our $k$th-order nonlinear ODE and try to solve it. The first (rather general) step is to rewrite it vectorially, as a rank-$k$ system of first-order (nonlinear) ODEs; schematically:
\be
\label{generalNLODE}
\frac{\rmd \boldsymbol{u}}{\rmd x} (x) = \boldsymbol{\varphi} \left( x, \boldsymbol{u} (x) \right).
\ee
\noindent
Under broad conditions, equations of this type may be further rewritten in a rather useful form, known as the \textit{prepared form}\footnote{This prepared form \eqref{preparedNLODE} might look a bit odd at first. This is because we are working in the variable $x$, with perturbative expansions constructed around $x \sim 0$. This, however, is not the most standard variable in the  resurgence literature. Therein one frequently uses $z = \frac{1}{x}$, with perturbative expansions built around $z \sim \infty$ and with the usual non-analytic transmonomial now being $\rme^{-z}$ (non-analytic around $z \sim \infty$). In this variable, \eqref{preparedNLODE} looks a bit more ``prepared'', as
\be
\frac{\rmd \boldsymbol{u}}{\rmd z} (z) = - \mathbb{A} \cdot \boldsymbol{u} (z) - \frac{1}{z}\, \mathbb{B} \cdot \boldsymbol{u} (z) + \CO \left( z^{-2} \boldsymbol{u}, \| \boldsymbol{u} \|^2 \right).
\ee
\noindent
Appendix~\ref{app:alien-calculus} also presents a ``notational bridge'' from our present discussions to the standard mathematical literature on resurgence, and many resurgence formulae in this variable $z$ will appear in there.} (see, \textit{e.g.}, \cite{m10} for an introduction to such procedure, and, \textit{e.g.}, \cite{c95, c98} for further details), which in this case is\footnote{We are slightly abusing notation, as the prepared form is obtained after possibly more than one change-of-variables; while in this expression we are using the exact same notation as in the previous equation, \eqref{generalNLODE}.}:
\be
\label{preparedNLODE}
\frac{\rmd \boldsymbol{u}}{\rmd x} (x) = \frac{1}{x^2}\, \mathbb{A} \cdot \boldsymbol{u} (x) + \frac{1}{x}\, \mathbb{B} \cdot \boldsymbol{u} (x) + \CO \left( \boldsymbol{u}, x^{-2} \| \boldsymbol{u} \|^2 \right),
\ee
\noindent
where the $k \times k$ matrices $\mathbb{A}$ and $\mathbb{B}$ are \textit{diagonal}; $\mathbb{A} = \text{diag} \left( A_1, \cdots, A_k \right)$ and $\mathbb{B} = \text{diag} \left( \beta_1, \cdots, \beta_k \right)$. The reason why this prepared form (or, sometimes denoted, normal form) is useful is that if we neglect the higher-order and nonlinear terms, and just focus on the ``linearized'' version of \eqref{preparedNLODE}, then the solution is immediate, as, component wise,
\be
\label{preparedNLODE:linsol}
u_i (x) = \sigma_i\, x^{\beta_i}\, \rme^{-\frac{A_i}{x}},
\ee
\noindent
with $\boldsymbol{\sigma}$ a $k$-dimensional vector of coefficients (the boundary conditions). It is now very explicit how the transmonomial $\exp \left( - \frac{1}{x} \right)$ first appears, and how it must play a central role in the construction of possible solutions. In fact, the next step in constructing a solution to \eqref{generalNLODE} will be to try using a transseries \textit{ansatz} precisely based upon the solution \eqref{preparedNLODE:linsol}. In other words, in the same way that powers of the monomial $M = x$ were used as building blocks of the (incomplete) perturbative, power-series solution to the original nonlinear equation; one may now use powers of the transmonomials $T_i = \rme^{-\frac{A_i}{x}}$ as new building blocks for the (eventually complete) nonperturbative, transseries solution to our ODE. Then, the typical transseries \textit{ansatz} for such a solution is, without surprise,
\be
\label{vec-k-parameter-TS}
\boldsymbol{u} \left( x, \boldsymbol{\sigma} \right) = \sum_{\boldsymbol{n} \in \BN_0^k} \boldsymbol{\sigma}^{\boldsymbol{n}}\, x^{\boldsymbol{n} \cdot \boldsymbol{\beta}}\, \rme^{- \frac{\boldsymbol{n} \cdot \boldsymbol{A}}{x}}\, \boldsymbol{u}^{(\boldsymbol{n})} (x),
\ee
\noindent
where $\boldsymbol{\sigma}^{\boldsymbol{n}} = \prod_{i=1}^k \sigma_i^{n_i}$ are the $k$ independent transseries parameters. Further, there still are asymptotic series in powers of $x$ ``attached'' to each transmonomial, as
\be
\boldsymbol{u}^{(\boldsymbol{n})} (x) \simeq \sum_{g=0}^{+\infty} \boldsymbol{u}^{(\boldsymbol{n})}_g\, x^g.
\ee

What other transmonomials could one find? Imagining we were forced to go beyond prepared form, it is simple to glare into the diversity of some of these new structures. Let us focus on the one-dimensional case for simplicity. Then, including for instance different powers of the monomial $x$ into the ODE \eqref{preparedNLODE}, one finds terms in, \textit{e.g.},
\be
\frac{\rmd u}{\rmd x} + \alpha\, n\, x^{n-1} u = 0 \qquad \Rightarrow \qquad u (x) = \sigma\, \rme^{- \alpha\, x^n}.
\ee
\noindent
When $n>0$ (these would be higher-order \textit{regular} terms in \eqref{preparedNLODE}) the corresponding solutions are analytic. But when $n<-1$ (these would be higher-order \textit{irregular} terms in \eqref{preparedNLODE}, going beyond prepared form) the corresponding solutions yield \textit{new} transmonomials. One could further envisage finding known transmonomials already in the ODE. For instance,
\bea
\frac{\rmd u}{\rmd x} - \frac{1}{x^2}\, \rme^{\frac{1}{x}}\, u (x) = 0 \qquad &\Rightarrow& \qquad u (x) = \sigma\, \rme^{- \rme^{\frac{1}{x}}}, \\
\frac{\rmd u}{\rmd x} + \frac{\log x}{x^2}\, u (x) = 0 \qquad &\Rightarrow& \qquad u (x) = \sigma\, \rme^{- \frac{1}{x} \left( 1 + \log x \right)},
\eea
\noindent
which immediately leads to yet some \textit{new} transmonomials as solutions. The diversity is enormous.

Let us next be more precise on definitions and nomenclature associated to transseries (now following \cite{e08}). A \textit{monomial} is typically a term of the form
\be
M \equiv x^\alpha,
\ee
\noindent
where $\alpha \in \BR$. In this context, one defines a \textit{power-series} $\boldsymbol{\CS}$ as a (possibly infinite) \textit{formal}\footnote{As we shall discuss shortly, issues of convergence need not concern us.} sum of monomials (and their respective powers). Monomials near $x \sim 0$ are easily ordered as
\be
x^\alpha \ll x^\beta \qquad \text{iff} \qquad \alpha > \beta,
\ee
\noindent
where the power-series $\boldsymbol{\CS}$ then collects monomials ordered in this way. In a similar fashion, a \textit{transmonomial} is typically a  term of the form\footnote{To be precise this is a so-called \textit{log-free} transmonomial; we will include logarithms shortly.}
\be
T \equiv x^{\alpha}\, \rme^{\boldsymbol{\CS}(x)},
\ee
\noindent
where $\alpha \in \BR$ and $\boldsymbol{\CS}(x)$ may be a function---it may be a (convergent) power-series, or it may even be a transseries itself (now we are running the risk of falling upon a slightly circular definition, so we refer the interested reader to \cite{e08} for further details). These transmonomials may be either analytic (in which case they essentially lead back to our earlier power-series) or non-analytic (where they finally lead to transseries!). In this language, it is straightforward to define a (log-free) \textit{transseries} $\boldsymbol{\CT}$ as a (possibly infinite) \textit{formal} sum of transmonomials (and their respective powers). Ordering transmonomials near $x \sim 0$ is now done \textit{lexicographically}, \textit{i.e.},
\be
x^{\alpha}\, \rme^{- A(x)} \ll x^{\beta}\, \rme^{- B(x)} \qquad \text{iff} \qquad A(x) \gg B(x) \qquad \text{or} \qquad \left\{ A(x) = B(x) \text{ and } \alpha > \beta \right\},
\ee
\noindent
and the transseries $\boldsymbol{\CT}$ then collects transmonomials ordered in this way. The fact that the definitions above only require ``formal sums'' naturally leads to rather simple constructions, as we are now allowed to do essentially any \textit{formal} (nonlinear) operations on transseries we may wish, from algebraic to differential to integral, without worrying about any convergence issues (we nonetheless refer back to \cite{e08} for precise definitions of all such basic manipulations).

In order to define a \textit{general} transmonomial and a \textit{general} transseries, one further needs to include \textit{logarithms}. This is done quite straightforwardly. Define
\be
\log_m x := \underbrace{\log \log \cdots \log }_{m} x,
\ee
\noindent
with $m \in \BN$. A general transmonomial is then obtained by replacing \textit{some} of the $x$'s in a log-free transmonomial by $\log_m x$. The exact same definition holds for a general transseries. These compositions also lead to a natural \textit{classification} of transmonomials and transseries. On one side, composing exponentials as in
\be
\rme^{\rme^{\rme^{\cdots}}},
\ee
\noindent
one can define the (positive) \textit{exponential height} of a transmonomial (or of a transseries) in the obvious way. Likewise, this is easily generalized to \textit{negative} exponential height, more commonly denoted by exponential \textit{depth} (or even ``logarithmic depth'' if to make it more transparent), where a transmonomial is obtained by composing a log-free transmonomial with $\log_m$ (on the right). So far, most fully worked-out physical problems seem to remain within height one and depth one\footnote{It is not hard to find contributions with logarithmic depth greater than one in several (higher-loop) Feynman diagrammatic calculations in field theory. But those are just contributions appearing at \textit{fixed} loop-order, \textit{not} contributions of higher logarithmic depth \textit{in} the coupling constant (which would change the transseries nature).}, but there are recent results pointing towards the need to include (at the very least) height-two contributions in the Sachdev--Ye--Kitaev model describing AdS black holes \cite{cghpsssst16}, and in Yang--Mills/QCD theories \cite{acpy17}. These are very interesting venues for future study.

One rather interesting consequence of all these constructions is the following. Whenever our transseries construction (as a solution to some problem) is associated to convergent series, then we are essentially assembling elementary functions. But whenever these transseries constructions become associated to \textit{divergent} series, then this means that upon (Borel \textit{and} \'Ecalle!) resummation we are obtaining functions which are \textit{not} elementary functions \cite{e08}.

\subsection*{Multi-Parameter Transseries and their Resurgence Relations}

Having discussed some basic generalities on transseries, we may now turn to their resurgence properties. For the discussion that follows we shall go back to the original $k$th-order nonlinear ODE (\textit{i.e.}, go back to addressing the original unknown function $u(x)$, rather than its vectorial rewriting $\boldsymbol{u} (x)$ from \eqref{generalNLODE}), and we shall further assume a simple $k$-parameter transseries with only transmonomials of the type $\sim \rme^{-\frac{A_i}{x}}$. This obviously implies there are $k$ distinct instanton actions, which we may again assemble vectorially as $\boldsymbol{A} \equiv \left( A_1, \ldots, A_k \right)$. We shall further assume the so-called \textit{non-resonant} case\footnote{Resonance will be thoroughly discussed in the upcoming section~\ref{sec:elliptic}.}, \textit{i.e.}, that these instanton actions are $\BZ$-\textit{linearly independent},
\be
\label{non-resonant-case}
\left. \nexists \,\, \boldsymbol{n} \neq \boldsymbol{0} \in \BZ^k \,\, \right| \,\, \boldsymbol{n} \cdot \boldsymbol{A} = 0.
\ee
\noindent
In this case, and much like \eqref{vec-k-parameter-TS}, the typical transseries \textit{ansatz} for our solution is
\be
\label{eq:sec5-transseries-vec}
u \left( x, \boldsymbol{\sigma} \right) = \sum_{\boldsymbol{n} \in \BN_0^k} \boldsymbol{\sigma}^{\boldsymbol{n}}\, \rme^{- \frac{\boldsymbol{n} \cdot \boldsymbol{A}}{x}}\, \Phi_{\boldsymbol{n}} (x),
\ee
\noindent
with the $\Phi_{\boldsymbol{n}} (x)$ asymptotic series\footnote{As before, we could consider $\Phi_{\boldsymbol{n}} (x) \simeq x^{\beta_{\boldsymbol{n}}}\, \sum_{g=0}^{+\infty} u_g^{(\boldsymbol{n})}\, x^g$, but this will not influence the discussion below.} in $x$
\begin{equation}
\label{eq:sec5-asympt-sectors-vec}
\Phi_{\boldsymbol{n}} (x) \simeq \sum_{g=0}^{+\infty} u_g^{(\boldsymbol{n})}\, x^g.
\end{equation}
\noindent
For example, the one-parameter transseries studied in subsection~\ref{subsec:Fnonlinear} is just the particular case where $k=1$, while the ``1.5''-parameter transseries also studied in subsection~\ref{subsec:Fnonlinear} just corresponds to taking $k=2$, but with $\boldsymbol{A}=(A,0)$ and where the only surviving sectors are $\Phi_{(n,0)}$, $n\ge0$, and $\Phi_{(0,1)}$ (this one not asymptotic; in fact it was just a constant).

The $\Phi_{\boldsymbol{n}}$ sectors in \eqref{eq:sec5-transseries-vec} are labeled by $\boldsymbol{n} \in \BN_0^k$ and thus live on a $k$-dimensional (semi-positive) lattice. What should we expect concerning their Borel singularities? If the structure is to be similar to the one-dimensional case, then they are located on the complex plane but must also be related to the aforementioned transseries sectors living on this $k$-dimensional lattice. Recalling the one-dimensional case, described by \eqref{Borel-transf-expanded-one-param}, it is reasonable to expect that the singularities of $\CB [\Phi_{\boldsymbol{n}}] (s)$ will be determined by the argument of the (non-analytic) exponential transmonomials in the transseries, \textit{i.e.}, they will be located at $s = \boldsymbol{\ell} \cdot \boldsymbol{A}$. This certainly satisfies the previous requirement. Further, due to possible ``backwards'' motions out from $\Phi_{\boldsymbol{n}}$, their location should be said more precisely as $s = \boldsymbol{\ell} \cdot \boldsymbol{A}$ \textit{and with} $\boldsymbol{\ell} \in \BZ^k$. Finally, we must certify that $\Phi_{\boldsymbol{n}+\boldsymbol{\ell}}$ is still a valid transseries sector, and thus can ``resurge'' at this $s = \boldsymbol{\ell} \cdot \boldsymbol{A}$ Borel singularity.

Let us make these requirements more precise. When starting off at $\Phi_{\boldsymbol{n}}$, asking whether $\Phi_{\boldsymbol{n}+\boldsymbol{\ell}}$ is still a transseries sector is a question on a $\BZ^k$ transseries grid. But if it is still a sector, and the transseries is resurgent, this may generically imply a possible Borel singularity of $\CB [\Phi_{\boldsymbol{n}}] (s)$ at $s = \boldsymbol{\ell} \cdot \boldsymbol{A}$, which effectively projects the $\BZ^k$ transseries grid into the complex Borel plane $\BC$. This defines a projection map\footnote{In the present \textit{non-resonant} case this linear map is one-to-one, \textit{i.e.}, $\ker \mathfrak{P} = \boldsymbol{0}$.}
\begin{align}
\mathfrak{P} : \BZ^k & \to \BC \nonumber \\
\boldsymbol{\ell} & \mapsto \boldsymbol{A} \cdot \boldsymbol{\ell}  
\label{projectionmapZkC}
\end{align}
\noindent
which specifies singularity locations and resurgent ``motions'' upon the Borel plane. In this way, the natural $k$-dimensional generalization of \eqref{Borel-transf-expanded-one-param} becomes\footnote{Recalling the discussion in section~\ref{sec:borel}, we only need to consider the standard representative, \textit{i.e.}, we only need to focus upon the singularity structure of a simple resurgent function. Further, this expression may be derived (rather than motivated, as done here for pedagogical reasons) by starting-off with the \textit{bridge equations}, and the interested reader may consult our appendix~\ref{app:alien-calculus} on these technicalities.
\label{footnoteBEink-transeries}}
\begin{equation}
\label{eq:sec5-Borel-transf}
\CB [\Phi_{\boldsymbol{n}}] (s) \Big|_{s=\boldsymbol{\ell}\cdot\boldsymbol{A}} = \mathsf{S}_{\boldsymbol{n} \to \boldsymbol{n}+\boldsymbol{\ell}} \times \CB [\Phi_{\boldsymbol{n}+\boldsymbol{\ell}}] (s-\boldsymbol{\ell}\cdot\boldsymbol{A})\, \frac{\log \left(s-\boldsymbol{\ell}\cdot\boldsymbol{A}\right)}{2\pi\rmi}, \qquad \boldsymbol{\ell} \neq \boldsymbol{0},
\end{equation}
\noindent
where each singular point $s=\boldsymbol{\ell}\cdot\boldsymbol{A}$ is uniquely\footnote{Again, this is due to presently addressing the non-resonant case.} determined by the vector $\boldsymbol{\ell} \in \BZ^k$.

As already extensively discussed throughout these lectures, resurgence is best expressed at an algebraic level of abstraction, based on the alien derivative $\Delta_{\omega}$. In this way, we may now extract the algebraic structure encoded in \eqref{eq:sec5-Borel-transf} and write it in terms of $\Delta_{\omega}$. One may proceed very much in analogy with the reasoning that led to \eqref{Delta-kA-one-param} (or one may also easily generalize the corresponding bridge-equation calculations in appendix~\ref{app:alien-calculus} to the present $k$-dimensional setting). In any case, the natural extension of the resurgence relations is found to be:
\begin{equation}
\label{eq:sec5-bridge-eqs}
\Delta_{\boldsymbol{\ell} \cdot \boldsymbol{A}} \Phi_{\boldsymbol{n}} = \boldsymbol{S}_{\boldsymbol{\ell}} \cdot \left(\boldsymbol{n}+\boldsymbol{\ell}\right) \Phi_{\boldsymbol{n}+\boldsymbol{\ell}},
\end{equation}
\noindent
where we now have to deal with a $k$-dimensional \textit{vector} of Stokes coefficients $\boldsymbol{S}_{\boldsymbol{\ell}} \equiv ( S_{\boldsymbol{\ell}}^{(1)}, \ldots, S_{\boldsymbol{\ell}}^{(k)} )$. Notice that, much like in \eqref{Delta-kA-one-param}, we labeled these Stokes vectors with subscripts $\boldsymbol{\ell}$, and think of them as living on lattice sites. We could have equally labeled them with subscripts $\boldsymbol{\ell} \cdot \boldsymbol{A}$, in which case they would be living on the complex Borel plane, attached to each singular point. As long as the projection \eqref{projectionmapZkC} is one-to-one, as it presently is, this choice does not really matter. Further, of course not \textit{every} point of the form $s = \boldsymbol{\ell} \cdot \boldsymbol{A}$ should correspond to a new alien-derivative singularity (this was certainly not the case back in \eqref{Delta-kA-one-param}), and, as such, at this stage one must still allow for some possible extra constraints or ``limitations'' either on the vector $\boldsymbol{\ell}$ or on the Stokes vectors (\textit{e.g.}, this was the requirement that $k \leq 1$ in \eqref{Delta-kA-one-param}). These are probably best understood at the level of motions on the ``transseries lattice'', to which we turn next. Akin to what happened in subsection~\ref{subsec:Fnonlinear}, \eqref{eq:sec5-bridge-eqs} leads to a set of allowed motions, this time around defined on an \textit{alien lattice}. Again, this interpretation will naturally lead to a ``statistical mechanical'' approach to $k$-dimensional resurgence. To understand these motions, first recall the one-dimensional case illustrated in figures~\ref{fig:alien-action-1-param} and~\ref{fig:first-5-sector-alien-1-param}. In that case there was a \textit{single} type of  forward motion, essentially dictated by the constraint that the only non-vanishing ``positive'' Stokes constant was $S_1$. At the same time, there were many different types of backward motions. Let us consider $k=2$ to illustrate the higher-dimensional cases in the following.

\begin{figure}[t!]
\begin{center} 
\begin{tikzpicture}[>=latex,decoration={markings, mark=at position 0.6 with {\arrow[ultra thick]{stealth};}} ]
\begin{scope}[node distance=2.5cm]
  \node (Phi00) [draw] at (0,0) {$\Phi_{(0,0)}$};
  \node (Phi10) [right of=Phi00] [draw] {$\Phi_{(1,0)}$};
  \node (Phi20) [right of=Phi10] [draw] {$\Phi_{(2,0)}$};
  \node (Phi30) [right of=Phi20] [draw] {$\Phi_{(3,0)}$};
  \node (Phi40) [right of=Phi30] [draw] {$\Phi_{(4,0)}$};
  \node (Phi50) [right of=Phi40] [draw] {$\Phi_{(5,0)}$};
  \node (cdots50) [right of=Phi50]  {$\cdots$}; 
  \node (Phi01) [above of=Phi00] [draw] {$\Phi_{(0,1)}$};
  \node (Phi02) [above of=Phi01] [draw] {$\Phi_{(0,2)}$};
  \node (Phi03) [above of=Phi02] [draw] {$\Phi_{(0,3)}$};
  \node (Phi04) [above of=Phi03] [draw] {$\Phi_{(0,4)}$};
  \node (cdots05) [above of=Phi04]  {$\vdots$};  
  \node (Phi11) [above of=Phi10] [draw] {$\Phi_{(1,1)}$};
  \node (Phi21) [above of=Phi20] [draw] {$\Phi_{(2,1)}$};
  \node (Phi31) [above of=Phi30] [draw] {$\Phi_{(3,1)}$};
  \node (Phi41) [above of=Phi40] [draw] {$\Phi_{(4,1)}$};
  \node (Phi51) [above of=Phi50] [draw] {$\Phi_{(5,1)}$};
  \node (cdots51) [above of=cdots50]  {$\cdots$};
  \node (Phi12) [above of=Phi11] [draw] {$\Phi_{(1,2)}$};
  \node (Phi22) [above of=Phi21] [draw] {$\Phi_{(2,2)}$};
  \node (Phi32) [above of=Phi31] [draw] {\color{red}$\Phi_{(3,2)}$};
  \node (Phi42) [above of=Phi41] [draw] {$\Phi_{(4,2)}$};
  \node (Phi52) [above of=Phi51] [draw] {$\Phi_{(5,2)}$};
  \node (cdots52) [above of=cdots51]  {$\cdots$};  
  \node (Phi13) [above of=Phi12] [draw] {$\Phi_{(1,3)}$};
  \node (Phi23) [above of=Phi22] [draw] {$\Phi_{(2,3)}$};
  \node (Phi33) [above of=Phi32] [draw] {$\Phi_{(3,3)}$};
  \node (Phi43) [above of=Phi42] [draw] {$\Phi_{(4,3)}$};
  \node (Phi53) [above of=Phi52] [draw] {$\Phi_{(5,3)}$};
  \node (cdots53) [above of=cdots52]  {$\cdots$};  
  \node (Phi14) [above of=Phi13] [draw] {$\Phi_{(1,4)}$};
  \node (Phi24) [above of=Phi23] [draw] {$\Phi_{(2,4)}$};
  \node (Phi34) [above of=Phi33] [draw] {$\Phi_{(3,4)}$};
  \node (Phi44) [above of=Phi43] [draw] {$\Phi_{(4,4)}$};
  \node (Phi54) [above of=Phi53] [draw] {$\Phi_{(5,4)}$};
  \node (cdots54) [above of=cdots53]  {$\cdots$};
  \node (cdots15) [above of=Phi14] {$\vdots$};
  \node (cdots25) [above of=Phi24] {$\vdots$};
  \node (cdots35) [above of=Phi34] {$\vdots$};
  \node (cdots45) [above of=Phi44] {$\vdots$};
  \node (cdots55) [above of=Phi54]  {$\vdots$};  
  \node (cdotsbb) [above of=cdots54]  {$\iddots$};
\end{scope}
 \draw [dotted,thick,blue!80!black,postaction={decorate},-,>=stealth,shorten <=2pt,shorten >=2pt] (Phi32.north)  -- (Phi33.south);
 \draw [thick,postaction={decorate},-,>=stealth,shorten <=2pt,shorten >=2pt] (Phi32.east)  -- (Phi42.west);
 \draw [thick,postaction={decorate},-,>=stealth,shorten <=2pt,shorten >=2pt] (Phi32.west)  -- (Phi22.east);
 \draw [dotted,thick,blue!80!black,postaction={decorate},-,>=stealth,shorten <=2pt,shorten >=2pt] (Phi32.south)  -- (Phi31.north); 
 \draw [dash pattern={on 4pt off 2pt on 1pt off 2pt},thick,red!70!black,postaction={decorate},-,>=stealth,shorten <=2pt,shorten >=2pt] (Phi32.north) .. controls +(-0.1,1.7) and +(0.5,0)  ..  (Phi23.east);
  \draw [dash pattern={on 4pt off 2pt on 1pt off 2pt},thick,green!40!black,postaction={decorate},-,>=stealth,shorten <=2pt,shorten >=2pt] (Phi32.north) .. controls +(-0.2,1.5) and +(0.,-0.3)  ..  (Phi13.east);
  \draw [dash pattern={on 4pt off 2pt on 1pt off 2pt},thick,orange!80!black,postaction={decorate},-,>=stealth,shorten <=2pt,shorten >=2pt] (Phi32.north) .. controls +(-0.2,1.3) and +(0.1,-0.5)  ..  (Phi03.east);
  \draw [dash pattern={on 4pt off 2pt on 1pt off 2pt},thick,red!70!black,postaction={decorate},-,>=stealth,shorten <=2pt,shorten >=2pt] (Phi32.east) .. controls +(1.5,-0.1) and +(0,0.5)  ..  (Phi41.north);
  \draw [dash pattern={on 4pt off 2pt on 1pt off 2pt},thick,orange!80!black,postaction={decorate},-,>=stealth,shorten <=2pt,shorten >=2pt] (Phi32.east) .. controls +(1.5,-0.3) and +(-1.0,0.7)  ..  (Phi40.north);  
 \draw[dotted,thick,green!40!black,postaction={decorate},-,>=stealth,shorten <=2pt,shorten >=2pt] (Phi32.east) .. controls +(1.3,-0.7) and +(1.2,0.9)  ..  (Phi30.east);
 \draw[thick,blue!80!black,postaction={decorate},-,>=stealth,shorten <=2pt,shorten >=2pt] (Phi32.north) .. controls +(-1.3,0.7) and +(1.2,0.9)  ..  (Phi12.north);
 \draw[thick,red!70!black,postaction={decorate},-,>=stealth,shorten <=2pt,shorten >=2pt] (Phi32.north) .. controls +(-1.,1.3) and +(0.7,1.5)  ..  (Phi02.north); 
 \draw[dash pattern={on 4pt off 2pt on 1pt off 2pt},thick,red!70!black,postaction={decorate},-,>=stealth,shorten <=2pt,shorten >=2pt] (Phi32.south) .. controls +(-1.1,-1.6) and +(1.3,1.3)  ..  (Phi20.east);
 \draw[dash pattern={on 4pt off 2pt on 1pt off 2pt},thick,postaction={decorate},-,>=stealth,shorten <=2pt,shorten >=2pt] (Phi32.south) .. controls +(-1.3,-0.5) and +(0.6,0.5)  ..  (Phi21.east);
 \draw[dash pattern={on 4pt off 2pt on 1pt off 2pt},thick,yellow!60!black,postaction={decorate},-,>=stealth,shorten <=2pt,shorten >=2pt] (Phi32.west) .. controls +(-1.,-0.9) and +(0.3,1.)  ..  (Phi11.north);
 \draw[dash pattern={on 4pt off 2pt on 1pt off 2pt},thick,postaction={decorate},-,>=stealth,shorten <=2pt,shorten >=2pt] (Phi32.west) .. controls +(-1.,-0.95) and +(0.3,2.5)  ..  (Phi01.north);
\draw[dash pattern={on 4pt off 2pt on 1pt off 2pt},thick,blue!80!black,postaction={decorate},-,>=stealth,shorten <=2pt,shorten >=2pt] (Phi32.south) .. controls +(-0.9,-0.8) and +(1.7,0)  ..  (Phi10.east);
 \node (m3-1) at (1.3,8.5) {\textcolor{orange!80!black}{\small{$w_{-3A_1 +A_2}$}}};
 \node (1-arrow) at (1.3,7.7) {\textcolor{orange!80!black}{$\left\downarrow\rule{0cm}{0.4cm}\right.$}};
 \node (m2-1) at (3.7,8.5) {\textcolor{green!40!black}{\small{$w_{-2A_1 + A_2}$}}};
 \node (1-arrow) at (3.7,7.7) {\textcolor{green!40!black}{$\left\downarrow\rule{0cm}{0.4cm}\right.$}}; 
 \node (m1-1) at (6.3,8.4) {\textcolor{red!70!black}{\small{$w_{-A_1 + A_2}$}}};
 \node (1-arrow) at (6.3,7.7) {\textcolor{red!70!black}{$\left\downarrow\rule{0cm}{0.35cm}\right.$}};
 \node (1-m2) at (11.1,1.3) {\textcolor{orange!80!black}{\small{$w_{A_1 - 2A_2}$}}};
 \node (1-arrow) at (10.0,1.3) {\textcolor{orange!80!black}{$\longleftarrow$}}; 
 \node (1-m1) at (11.5,3.6) {\textcolor{red!70!black}{\small{$w_{A_1 - A_2}$}}};
 \node (1-arrow) at (10.4,3.6) {\textcolor{red!70!black}{$\longleftarrow$}};
 \node (m3-0) at (0.4,6.4) {\textcolor{red!70!black}{\small{$w_{-3A_1 }$}}};
\node (0-1) at (8.1,6.7) {\textcolor{blue!80!black}{\small{$w_{A_2}$}}};
\node (1-0) at (8.9,5.4) {\small{$w_{A_1}$}};
 \node (0-m2) at (8.3,1.4) {\textcolor{green!40!black}{\small{$w_{-2A_2}$}}}; 
 \node (m1-m2) at (7.1,0.8) {\textcolor{red!70!black}{\small{$w_{-A_1 - 2A_2}$}}};
 \node (m3-m1) at (0.1,4.25) {\small{$w_{-3A_1- A_2}$}}; 
 \node (m2-m2) at (3.7,1.0) {\textcolor{blue!80!black}{\small{$w_{-2A_1 - 2A_2}$}}};
\node (0-m1) at (8.1,3.6) {\textcolor{blue!80!black}{\small{$w_{-A_2}$}}};
\node (m1-0) at (6.2,5.3) {\small{$w_{-A_1}$}};
\node (m2-0) at (4.0,5.6) {\textcolor{blue!80!black}{\small{$w_{-2A_1}$}}};
\node (m2-m1) at (1.95,3.4) {\textcolor{yellow!60!black}{\small{$w_{-2A_1 - A_2}$}}};
\node (m1-m1) at (5.35,3.4) {\small{$w_{-A_1 - A_2}$}};
\end{tikzpicture}
\end{center} 
\caption {The 2d \textit{alien lattice}: a pictorial representation of the action of the alien derivative upon the sector $\Phi_{\boldsymbol{n}}$, with lattice-node $\boldsymbol{n}=(3,2)$ (now regarded as the origin for the $\BZ^2$ lattice of Stokes vectors, thus dividing the alien lattice into quadrants). Different single arrows correspond to different single steps, and each step has an associated weight, as dictated by the right-hand-side of the resurgence relations \eqref{eq:sec5-bridge-eqs-COMPLETE}. The weight of each step---written next to the corresponding arrow---is given in terms of Stokes vectors as explained in the text. The solid links refer to the action of the alien derivative along the grid direction $\boldsymbol{e}_1$ (with instanton action $A_1$), depicting both forward and backwards motions. The dotted links correspond to the action along the grid direction $\boldsymbol{e}_2$ (now with instanton action $A_2$). The mixed dashed/dotted links are motions which mix both directions. In the plot, colors are used solely to help differentiate distinct motions. The steps shown are \textit{all} the single steps which are allowed starting at the $(3,2)$ lattice-site (where \textit{paths} can then be constructed with different middle nodes, and corresponding products of weights).
}
\label{fig:sec5-2d-lattice}
\end{figure}
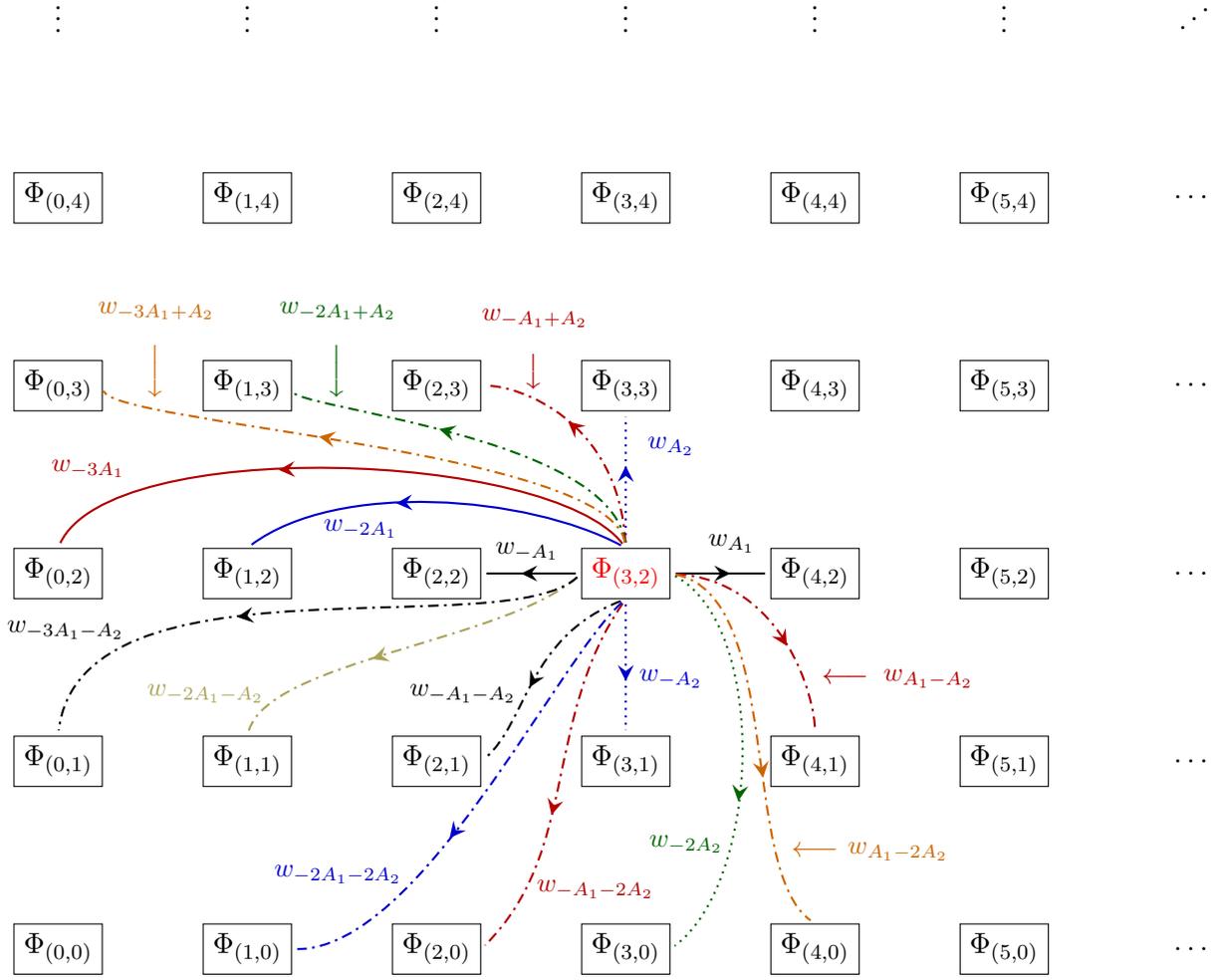

In the two-dimensional case the nodes $\Phi_{\boldsymbol{n}}$ live on a two-dimensional lattice, as $\boldsymbol{n} \in \BN^2_0$. The alien derivative \eqref{eq:sec5-bridge-eqs} will then induce different motions on this lattice, as illustrated in figure~\ref{fig:sec5-2d-lattice}. Let $\boldsymbol{e}_1 \equiv (1,0)$ and $\boldsymbol{e}_2 \equiv (0,1)$ be the canonical basis of $\BZ^2$ and write $\boldsymbol{\ell} = \ell_{1}\, \boldsymbol{e}_1 + \ell_{2}\, \boldsymbol{e}_2$ and $\boldsymbol{S}_{\boldsymbol{\ell}} = S_{\boldsymbol{\ell}}^{(1)}\, \boldsymbol{e}_1 + S_{\boldsymbol{\ell}}^{(2)}\, \boldsymbol{e}_2$. Starting-off at the lattice node $\Phi_{\boldsymbol{n}}$, strictly-backward motions---described by vectors with components $\ell_{1}, \ell_{2} \le 0$---are allowed, and lead to corresponding singularities of $\mathcal{B} [\Phi_{\boldsymbol{n}}] (s)$ at $s=\boldsymbol{\ell}\cdot\boldsymbol{A}$, as long as $\left|\ell_{i}\right| \leq n_{i}$ (recall that transseries sectors $\Phi_{\boldsymbol{m}}$ are set to vanish---do not exist---whenever one $m_j<0$), \textit{i.e.}, one can move backwards up to the boundary of the transseries $\BN^2_0$ semi-positive grid. This is a familiar scenario also from the one-dimensional case, \textit{e.g.}, recall the backward motions in figure~\ref{fig:first-5-sector-alien-1-param}. On the other hand, one expects forward motions to be more restricted. In the one-dimensional case, the only\footnote{As mentioned, this was the requirement that $k \leq 1$ back in \eqref{Delta-kA-one-param}.} allowed forward motion on the $\BZ$-chain was the one generated by $\boldsymbol{\ell} = \boldsymbol{e}_1$ (here $\boldsymbol{e}_1$ the canonical basis of $\BZ$). A similar two-dimensional scenario would entail that the only allowed strictly-forward motions, $\ell_{1}, \ell_{2} \ge 0$, would be associated to either $\boldsymbol{\ell} = \boldsymbol{e}_1$ (setting $\boldsymbol{S}_{\boldsymbol{\ell}} = ( S_{\boldsymbol{e}_1}^{(1)}, 0 )$ to be as close to the one-dimensional example as possible) or $\boldsymbol{\ell} = \boldsymbol{e}_2$ (now setting $\boldsymbol{S}_{\boldsymbol{\ell}} = ( 0, S_{\boldsymbol{e}_2}^{(2)} )$). This statement may be generalized to $k$ dimensions by only allowing strictly-forward motions which are \textit{elementary}, \textit{i.e.}, given by some basis vector $\boldsymbol{e}_i$ within the canonical basis of $\BZ^k$. In this way, an elementary forward motion $\boldsymbol{\ell} = \boldsymbol{e}_i$ would create a singularity at $s = \boldsymbol{\ell}\cdot \boldsymbol{A} = A_i$ with Stokes vector $\boldsymbol{S}_{\boldsymbol{e}_i} = ( 0, \ldots, 0, S_{\boldsymbol{e}_i}^{(i)}, 0, \ldots, 0 )$, and the resurgent structure \eqref{eq:sec5-bridge-eqs} becomes 
\begin{equation}
\label{AlienA_jonPhin}
\Delta_{A_i} \Phi_{\boldsymbol{n}} = S_{\boldsymbol{e}_i}^{(i)} \left(n_i+1\right) \Phi_{\boldsymbol{n}+\boldsymbol{e}_i}.
\end{equation}
\noindent
Consequentially, note that for forward motions along orthogonal directions, say $\boldsymbol{e}_i$ and $\boldsymbol{e}_j$ with $i \neq j$, the action of the alien derivatives commutes:
\begin{equation}
\label{eq:sec5-commutator-orthogonal-forward}
\left[ \Delta_{A_i}, \Delta_{A_j} \right] \Phi_{\boldsymbol{n}} = 0.
\end{equation}
\noindent
The above general expectations may be written as general constraints\footnote{Again, these conditions may be derived rather than just motivated; see footnote~\ref{footnoteBEink-transeries}.} on Stokes vectors,
\begin{equation}
\label{eq:sec5-Stokes-const-condition}
S_{\boldsymbol{\ell}}^{(j)}=0 \quad \text{ if } \quad \ell_{i} \ge 1 + \delta_{ij}, \quad \forall_{i \in \left\{ 1,\ldots,k \right\}};
\end{equation}
\noindent
\textit{i.e.}, the $k$-dimensional extension of \eqref{Delta-kA-one-param}, or, the complete formulation of \eqref{eq:sec5-bridge-eqs}, is
\be
\label{eq:sec5-bridge-eqs-COMPLETE}
\Delta_{\boldsymbol{\ell} \cdot \boldsymbol{A}} \Phi_{\boldsymbol{n}} = \boldsymbol{S}_{\boldsymbol{\ell}} \cdot \left(\boldsymbol{n}+\boldsymbol{\ell}\right) \Phi_{\boldsymbol{n}+\boldsymbol{\ell}}, \qquad \ell_i \leq \delta_{ij} \quad \text{and} \quad \boldsymbol{\ell} \neq \boldsymbol{0}.
\ee

Besides many strictly-backwards and few strictly-forward motions, of course one may also find a couple of \textit{mixed} forward/backward motions on the lattice; where along one (and only one) direction one has an elementary forward motion, and along all others one has either no motion or some backwards motion. A few of these possible motions are illustrated in figure~\ref{fig:sec5-2d-lattice}. Note that the ``statistical mechanical'' language from section~\ref{sec:quartic} essentially holds in here as well, with the straightforward $k$-dimensional adaptations (instead of motions on a chain we now have motions on a lattice, with transseries nodes at each grid-site, and so on). In this way, the \textit{weight} $w$ associated to a \textit{step} $\CS$ is defined analogously to \eqref{equation:weight1DIM}, and as read directly from \eqref{eq:sec5-bridge-eqs-COMPLETE}. For a step connecting the lattice nodes $\Phi_{\boldsymbol{n}}$ and $\Phi_{\boldsymbol{m}}$ one has
\be
\label{k-dim-weight!}
w \left( \mathcal{S} \left( \boldsymbol{n} \rightarrow \boldsymbol{m} \right) \right) = \boldsymbol{m} \cdot \boldsymbol{S}_{\boldsymbol{m} - \boldsymbol{n}}.
\ee
\noindent
In figure~\ref{fig:sec5-2d-lattice} we illustrate all motions out from $\boldsymbol{n} \equiv (n_{1},n_{2}) = (3,2)$. There are two strictly-forward motions, with $\boldsymbol{\ell} = \boldsymbol{e}_1, \boldsymbol{e}_2$. One can use them to either reach the $(4,2)$ node, in which case\footnote{To make notation clearer, we are labeling the weights with the subscript $s = \boldsymbol{\ell} \cdot \boldsymbol{A}$.} $w_{A_1} = (4,2) \cdot \boldsymbol{S}_{(4,2)-(3,2)} = (4,2) \cdot \boldsymbol{S}_{\boldsymbol{e}_1} = 4 S^{(1)}_{\boldsymbol{e}_1}$; or to reach the node $(3,3)$, in which case $w_{A_2} = (3,3) \cdot \boldsymbol{S}_{(3,3)-(3,2)} = (3,3) \cdot \boldsymbol{S}_{\boldsymbol{e}_2} = 3 S^{(2)}_{\boldsymbol{e}_2}$. There are also many strictly-backwards motions, for example the one that takes us back to node $(1,1)$. In this case, $w_{-2A_1-A_2} = (1,1) \cdot \boldsymbol{S}_{(1,1)-(3,2)} = (1,1) \cdot \boldsymbol{S}_{(-2,-1)} = S^{(1)}_{-2 \boldsymbol{e}_1-\boldsymbol{e}_2} + S^{(2)}_{-2 \boldsymbol{e}_1-\boldsymbol{e}_2}$. Finally, there are a few ``mixed'' motions; consider for example the one taking us to node $(2,3)$. In this case, $w_{-A_1+A_2} = (2,3) \cdot \boldsymbol{S}_{(2,3)-(3,2)} = (2,3) \cdot \boldsymbol{S}_{(-1,1)} = 2 S^{(1)}_{-\boldsymbol{e}_1+\boldsymbol{e}_2} + 3 S^{(2)}_{-\boldsymbol{e}_1+\boldsymbol{e}_2}$. All these, and many other, are illustrated in figure~\ref{fig:sec5-2d-lattice}.

The $k$-dimensional case follows in a straightforward fashion. The alien geometry induced on the lattice based upon a given fixed node, $\boldsymbol{n}$, naturally splits $\BZ^k$ (with origin at $\boldsymbol{n}$) into $k$-orthants (\textit{i.e.}, $k$-dimensional orthants or $k$-hyperoctants, generalizing the $2$-orthants or quadrants in figure~\ref{fig:sec5-2d-lattice}). These $k$-orthants have distinct features on what concerns distributions of Stokes vectors---in fact, generically, most $k$-orthants will be \textit{empty} of any Stokes vectors. The $\BZ^k$-axes will be populated with Stokes vectors much like the one-dimensional case was, with Stokes constants. The ``all-negative'' $(-,\cdots,-)$ $k$-orthant will be fully populated (at least up to the boundary of the original transseries semi-positive grid---now shifted by $\boldsymbol{n}$). The ``single-plus'' $(-,\cdots,-,+,-,\cdots,-)$ $k$-orthants will only be partially populated, much like the second and fourth quadrants in figure~\ref{fig:sec5-2d-lattice} are. But any ``double-plus'' $(\cdots,+,\cdots,+,\cdots)$ or higher $k$-orthant will be \textit{empty} of any Stokes vectors. This split of the $\BZ^k$ alien lattice into distinct orthants is solely dictated by the constraint $\ell_i \leq \delta_{ij}$ on geometric motions given in \eqref{eq:sec5-bridge-eqs-COMPLETE}.

As in the one-dimensional case, the structures encoded in \eqref{eq:sec5-Borel-transf} and \eqref{eq:sec5-bridge-eqs-COMPLETE} must be the same, upon relating the proportionality factors to each other. The $k$-dimensional novelty is that while Borel residues $\mathsf{S}_{\boldsymbol{n} \to \boldsymbol{n}+\boldsymbol{\ell}}$ are still \textit{scalars}, the Stokes coefficients $\boldsymbol{S}_{\boldsymbol{\ell}}$ have now been ``upgraded'' to \textit{vectors}. This obviously implies that relating them to each other will become a bit more intricate. Generalizing the one-dimensional calculations in appendix~\ref{app:alien-calculus}, a few of these results follow. Let us split them according to the $k$-orthant classification we discussed above:
\begin{itemize}
\item \textit{Forward Axes:} Let us pick any axis, generated by the $\BZ^k$ canonical-basis vector $\boldsymbol{e}_i$, and consider its \textit{forward direction}. This projects into a \textit{direction angle} of $\theta_i \equiv \arg A_i$ on the complex Borel plane. For this forward direction, the $k$-dimensional generalization of \eqref{stokesS-to-borelS+1} through \eqref{stokesS-to-borelS+3} is, very naturally,
\bea
\label{kd-FA-stokesS-to-borelS+1}
\mathsf{S}_{\boldsymbol{n}\to\boldsymbol{n}+\boldsymbol{e}_i} &=& -\, \left( \boldsymbol{n}+\boldsymbol{e}_i \right) \cdot \boldsymbol{S}_{\boldsymbol{e}_i}, \\
\label{kd-FA-stokesS-to-borelS+2}
\mathsf{S}_{\boldsymbol{n}\to\boldsymbol{n}+2\boldsymbol{e}_i} &=& - \frac{1}{2}\, \left( \boldsymbol{n}+\boldsymbol{e}_i \right) \cdot \boldsymbol{S}_{\boldsymbol{e}_i}\,\, \left( \boldsymbol{n}+2\boldsymbol{e}_i \right) \cdot \boldsymbol{S}_{\boldsymbol{e}_i}, \\
\label{kd-FA-stokesS-to-borelS+3}
\mathsf{S}_{\boldsymbol{n}\to\boldsymbol{n}+3\boldsymbol{e}_i} &=& - \frac{1}{6}\, \left( \boldsymbol{n}+\boldsymbol{e}_i \right) \cdot \boldsymbol{S}_{\boldsymbol{e}_i}\,\, \left( \boldsymbol{n}+2\boldsymbol{e}_i \right) \cdot \boldsymbol{S}_{\boldsymbol{e}_i}\,\, \left( \boldsymbol{n}+3\boldsymbol{e}_i \right) \cdot \boldsymbol{S}_{\boldsymbol{e}_i}.
\eea
\noindent
A closed-form expression is also simple to obtain (and comparing back to \eqref{stokesS-to-borelS+GEN}, it should now be clear why we wrote that equation without the binomial coefficient, back in section~\ref{sec:quartic}),
\be
\label{kd-FA-stokesS-to-borelS+GEN}
\mathsf{S}_{\boldsymbol{n}\to\boldsymbol{n}+\ell\boldsymbol{e}_i} = - \frac{1}{\ell!} \prod_{m=1}^{\ell} \left( \boldsymbol{n}+m\boldsymbol{e}_i \right) \cdot \boldsymbol{S}_{\boldsymbol{e}_i}.
\ee
\noindent
Now, along the positive direction $\boldsymbol{e}_i$ there is a single Stokes vector, with components $\boldsymbol{S}_{\boldsymbol{e}_i} = S_{\boldsymbol{e}_i}^{(i)}\, \boldsymbol{e}_i$, in which case the inverse map follows immediately from \eqref{kd-FA-stokesS-to-borelS+1} above,
\be
S_{\boldsymbol{e}_i}^{(i)} = - \frac{1}{n_i+1}\, \mathsf{S}_{\boldsymbol{n}\to\boldsymbol{n}+\boldsymbol{e}_i}.
\ee
\noindent
Finally, the fact that there are no further Stokes vectors along this positive direction results in the set of constraints generalizing \eqref{borelS-consistency} to the $k$-dimensional setting,
\be
\label{kd-FA-borelS-consistency}
\mathsf{S}_{\boldsymbol{n}\to\boldsymbol{n}+\ell\boldsymbol{e}_i} = \frac{(-1)^{\ell-1}}{\ell!}\, \prod_{m=0}^{\ell-1} \mathsf{S}_{\boldsymbol{n}+m\boldsymbol{e}_i\to\boldsymbol{n}+\left(m+1\right)\boldsymbol{e}_i}.
\ee
\item \textit{Backward Axes:} Let us remain along the same axis as in the previous case, but now consider its \textit{backward direction}, \textit{i.e.}, the one associated to the direction vector $-\boldsymbol{e}_i$ and projecting into a \textit{direction angle} of $\theta_i+\pi$ on the complex Borel plane. For this backward direction, the $k$-dimensional generalization of \eqref{stokesS-to-borelS-1} through \eqref{stokesS-to-borelS-3} is
\bea
\label{kd-BA-stokesS-to-borelS-1}
\mathsf{S}_{\boldsymbol{n}\to\boldsymbol{n}-\boldsymbol{e}_i} &=& - \left( \boldsymbol{n}-\boldsymbol{e}_i \right) \cdot \boldsymbol{S}_{-\boldsymbol{e}_i}, \\
\label{kd-BA-stokesS-to-borelS-2}
\mathsf{S}_{\boldsymbol{n}\to\boldsymbol{n}-2\boldsymbol{e}_i} &=& - \left( \boldsymbol{n}-2\boldsymbol{e}_i \right) \cdot \left\{ \boldsymbol{S}_{-2\boldsymbol{e}_i} + \frac{1}{2} \left( \boldsymbol{n}-\boldsymbol{e}_i \right) \cdot \boldsymbol{S}_{-\boldsymbol{e}_i}\,\, \boldsymbol{S}_{-\boldsymbol{e}_i} \right\}, \\
\label{kd-BA-stokesS-to-borelS-3}
\mathsf{S}_{\boldsymbol{n}\to\boldsymbol{n}-3\boldsymbol{e}_i} &=& - \left( \boldsymbol{n}-3\boldsymbol{e}_i \right) \cdot \left\{ \boldsymbol{S}_{-3\boldsymbol{e}_i} + \frac{1}{2} \left( \boldsymbol{n}-\boldsymbol{e}_i \right) \cdot \boldsymbol{S}_{-\boldsymbol{e}_i}\,\, \boldsymbol{S}_{-2\boldsymbol{e}_i} + \right. \nonumber \\
&&
\left. + \frac{1}{2} \left( \boldsymbol{n}-2\boldsymbol{e}_i \right) \cdot \left( \boldsymbol{S}_{-2\boldsymbol{e}_i} + \frac{1}{3} \left( \boldsymbol{n}-\boldsymbol{e}_i \right) \cdot \boldsymbol{S}_{-\boldsymbol{e}_i}\,\, \boldsymbol{S}_{-\boldsymbol{e}_i} \right) \boldsymbol{S}_{-\boldsymbol{e}_i} \right\}.
\eea
\noindent
A closed-form expression is now a bit more intricate to obtain than in the earlier forward-case. But, when comparing back to \eqref{stokesS-to-borelS-GEN}, it should also be clear how the present vectorial form is somewhat cleaner than the one-dimensional (scalar-like) case:
\bea
\mathsf{S}_{\boldsymbol{n}\to\boldsymbol{n}-\ell\boldsymbol{e}_i} &=& - \sum_{s=1}^{\ell} \frac{1}{s!}\, \sum_{\substack{\left\{ \ell_1,\ldots,\ell_s\ge 1 \right\} \\ \vphantom{\frac{1}{2}} \sum \ell_i=\ell}} \left( \boldsymbol{n}- \ell_1 \boldsymbol{e}_i \right) \cdot \boldsymbol{S}_{- \ell_1\, \boldsymbol{e}_i}\,\, \left( \boldsymbol{n}- \left(\ell_1+\ell_2\right) \boldsymbol{e}_i \right) \cdot \boldsymbol{S}_{- \ell_2 \boldsymbol{e}_i}\,\, \cdots \nonumber\\
&&
\cdots\,\, \left( \boldsymbol{n}- \left(\ell_1+\ell_2+\cdots+\ell_{s-1}\right) \boldsymbol{e}_i \right) \cdot \boldsymbol{S}_{- \ell_{s-1} \boldsymbol{e}_i}\,\, \left( \boldsymbol{n}- \ell\, \boldsymbol{e}_i \right) \cdot \boldsymbol{S}_{- \ell_s \boldsymbol{e}_i}.
\label{kd-BA-stokesS-to-borelS-GEN}
\eea
\noindent
Inverse maps follow as usual. A few examples, generalizing \eqref{borelS-to-stokesS-1} through \eqref{borelS-to-stokesS-3} to the present $k$-dimensional setting, are
\bea
\label{kd-BA-borelS-to-stokesS-1}
\left( \boldsymbol{n}-\boldsymbol{e}_i \right) \cdot \boldsymbol{S}_{-\boldsymbol{e}_i} &=& - \mathsf{S}_{\boldsymbol{n}\to\boldsymbol{n}-\boldsymbol{e}_i}, \\
\label{kd-BA-borelS-to-stokesS-2}
\left( \boldsymbol{n}-2\boldsymbol{e}_i \right) \cdot \boldsymbol{S}_{-2\boldsymbol{e}_i} &=& - \mathsf{S}_{\boldsymbol{n}\to\boldsymbol{n}-2\boldsymbol{e}_i} - \frac{1}{2}\, \mathsf{S}_{\boldsymbol{n}\to\boldsymbol{n}-\boldsymbol{e}_i}\, \mathsf{S}_{\boldsymbol{n}-\boldsymbol{e}_i\to\boldsymbol{n}-2\boldsymbol{e}_i}, \\
\left( \boldsymbol{n}-3\boldsymbol{e}_i \right) \cdot \boldsymbol{S}_{-3\boldsymbol{e}_i} &=& - \mathsf{S}_{\boldsymbol{n}\to\boldsymbol{n}-3\boldsymbol{e}_i} - \frac{1}{2}\, \mathsf{S}_{\boldsymbol{n}\to\boldsymbol{n}-\boldsymbol{e}_i}\, \mathsf{S}_{\boldsymbol{n}-\boldsymbol{e}_i\to\boldsymbol{n}-3\boldsymbol{e}_i} - \frac{1}{2}\, \mathsf{S}_{\boldsymbol{n}\to\boldsymbol{n}-2\boldsymbol{e}_i}\, \mathsf{S}_{\boldsymbol{n}-2\boldsymbol{e}_i\to\boldsymbol{n}-3\boldsymbol{e}_i} - \nonumber \\
&&
- \frac{1}{3}\, \mathsf{S}_{\boldsymbol{n}\to\boldsymbol{n}-\boldsymbol{e}_i}\, \mathsf{S}_{\boldsymbol{n}-\boldsymbol{e}_i\to\boldsymbol{n}-2\boldsymbol{e}_i}\, \mathsf{S}_{\boldsymbol{n}-2\boldsymbol{e}_i\to\boldsymbol{n}-3\boldsymbol{e}_i},
\label{kd-BA-borelS-to-stokesS-3}
\eea
\noindent
leading up to the vectorial closed-form expression (in fact very similar to its one-dimensional analogue \eqref{borelS-to-stokesS-GEN})
\be
\label{kd-BA-borelS-to-stokesS-GEN}
\left( \boldsymbol{n}- \ell\, \boldsymbol{e}_i \right) \cdot \boldsymbol{S}_{- \ell \boldsymbol{e}_i} = - \sum_{s=1}^{\ell} \frac{1}{s}\, \sum_{\substack{\left\{ \ell_1,\ldots,\ell_s\ge 1 \right\} \\ \vphantom{\frac{1}{2}} \sum \ell_i=\ell}} \mathsf{S}_{\boldsymbol{n}\to\boldsymbol{n}-\ell_1\boldsymbol{e}_i}\, \mathsf{S}_{\boldsymbol{n}-\ell_1\boldsymbol{e}_i\to\boldsymbol{n}- \left(\ell_1+\ell_2\right) \boldsymbol{e}_i}\, \cdots\, \mathsf{S}_{\boldsymbol{n}- \left(\ell-\ell_s\right) \boldsymbol{e}_i\to\boldsymbol{n}-\ell\boldsymbol{e}_i}.
\ee
\noindent
This equation is valid as long as $1 \leq \ell < n_i$. Together with what we have just discussed in the forward-axes case, this tells us that for $\boldsymbol{\ell} = \ell\, \boldsymbol{e}_i$ essentially the familiar one-dimensional constraints on the resurgence relations (from \eqref{Delta-kA-one-param}) applies; \textit{i.e.}, $\ell \leq 1$, $\ell \neq 0$.
\item \textit{``Single-Plus'' $k$-Orthant:} We shall illustrate this case with the $(+,-,\cdots,-)$ $k$-orthant (this is the so-called ``last'', or $2^k$-th $k$-orthant), but the results are completely generic. In this $k$-orthant, a specific \textit{direction vector} on the alien grid is given by, \textit{e.g.},
\be
\boldsymbol{v} = \boldsymbol{e}_1 - \sum_{i=2}^k \nu_i\, \boldsymbol{e}_i, \qquad \nu_i \in \BN,
\ee
\noindent
which projects into a \textit{direction angle} of $\theta_{\boldsymbol{v}} \equiv \arg \boldsymbol{A} \cdot \boldsymbol{v}$ on the complex Borel plane. Any other directions are not allowed as they would lead to vanishing Stokes vectors. For these directions, the $k$-dimensional generalization of \eqref{stokesS-to-borelS+1} through \eqref{stokesS-to-borelS+3} is in fact completly equivalent to the forward-axes case, \eqref{kd-FA-stokesS-to-borelS+1} through \eqref{kd-FA-stokesS-to-borelS+3},
\bea
\label{kd-SPO-stokesS-to-borelS+1}
\mathsf{S}_{\boldsymbol{n}\to\boldsymbol{n}+\boldsymbol{v}} &=& -\, \left( \boldsymbol{n}+\boldsymbol{v} \right) \cdot \boldsymbol{S}_{\boldsymbol{v}}, \\
\label{kd-SPO-stokesS-to-borelS+2}
\mathsf{S}_{\boldsymbol{n}\to\boldsymbol{n}+2\boldsymbol{v}} &=& - \frac{1}{2}\, \left( \boldsymbol{n}+\boldsymbol{v} \right) \cdot \boldsymbol{S}_{\boldsymbol{v}}\,\, \left( \boldsymbol{n}+2\boldsymbol{v} \right) \cdot \boldsymbol{S}_{\boldsymbol{v}}, \\
\label{kd-SPO-stokesS-to-borelS+3}
\mathsf{S}_{\boldsymbol{n}\to\boldsymbol{n}+3\boldsymbol{v}} &=& - \frac{1}{6}\, \left( \boldsymbol{n}+\boldsymbol{v} \right) \cdot \boldsymbol{S}_{\boldsymbol{v}}\,\, \left( \boldsymbol{n}+2\boldsymbol{v} \right) \cdot \boldsymbol{S}_{\boldsymbol{v}}\,\, \left( \boldsymbol{n}+3\boldsymbol{v} \right) \cdot \boldsymbol{S}_{\boldsymbol{v}},
\eea
\noindent
leading up to the expression which generalizes \eqref{stokesS-to-borelS+GEN} and \eqref{kd-FA-stokesS-to-borelS+GEN},
\be
\label{kd-SPO-stokesS-to-borelS+GEN}
\mathsf{S}_{\boldsymbol{n}\to\boldsymbol{n}+\ell\boldsymbol{v}} = - \frac{1}{\ell!} \prod_{m=1}^{\ell} \left( \boldsymbol{n}+m\boldsymbol{v} \right) \cdot \boldsymbol{S}_{\boldsymbol{v}} \qquad \text{for} \quad 1 \leq \ell < \min \left\{ \left\lfloor \frac{n_i}{\nu_i} \right\rfloor \right\}_{i=2,3,\ldots,k}
\ee
\noindent
and zero otherwise. Except for the constraint of reaching the transseries-grid boundary, this is essentially the same result as in the forward-axes case. Further, because along the direction $\boldsymbol{v}$ there is a single non-vanishing Stokes vector $\boldsymbol{S}_{\boldsymbol{v}}$, the inverse map is already implicit in \eqref{kd-SPO-stokesS-to-borelS+1} above. The fact that there are no further Stokes vectors along this direction, results in the set of constraints generalizing \eqref{borelS-consistency} and \eqref{kd-FA-borelS-consistency},
\be
\label{kd-SPO-borelS-consistency}
\mathsf{S}_{\boldsymbol{n}\to\boldsymbol{n}+\ell\boldsymbol{v}} = \frac{(-1)^{\ell-1}}{\ell!}\, \prod_{m=0}^{\ell-1} \mathsf{S}_{\boldsymbol{n}+m\boldsymbol{v}\to\boldsymbol{n}+\left(m+1\right)\boldsymbol{v}}.
\ee
\noindent
Again, this is essentially the same result as in the earlier forward-axes case.
\item \textit{``All-Negative'' $k$-Orthant:} This is the $(-,\cdots,-)$ $k$-orthant. In this $k$-orthant, a specific \textit{direction vector} on the alien grid is conventionally given by $-\boldsymbol{v}$, where, \textit{e.g.},
\be
\boldsymbol{v} = \sum_{i=1}^k \nu_i\, \boldsymbol{e}_i, \qquad \nu_i \in \BN,
\ee
\noindent
so that $\boldsymbol{v}$ belongs to the ``all-positive'' $(+,\cdots,+)$ $k$-orthant. Notice that the choice of coefficients $\nu_i$ is not completely arbitrary, as $\boldsymbol{v}$ must also be a \textit{lattice versor}. This does not imply that this vector is normalized to unit length (it would not be on the lattice), but that it is the \textit{smallest} lattice vector defining the said direction. The vector $\boldsymbol{v}$ projects  into a direction angle of $\theta_{\boldsymbol{v}} \equiv \arg \boldsymbol{A} \cdot \boldsymbol{v}$, and thus the ``all-negative'' $k$-orthant projection yields a \textit{direction angle} of $\theta_{\boldsymbol{v}}+\pi$ on the complex Borel plane. For this direction, the $k$-dimensional generalization of equations \eqref{stokesS-to-borelS-1}, \eqref{stokesS-to-borelS-2} and \eqref{stokesS-to-borelS-3}, is completely equivalent to the backward-axes case, \eqref{kd-BA-stokesS-to-borelS-1}, \eqref{kd-BA-stokesS-to-borelS-2} and \eqref{kd-BA-stokesS-to-borelS-3}, as
\bea
\label{kd-ANO-stokesS-to-borelS-1}
\mathsf{S}_{\boldsymbol{n}\to\boldsymbol{n}-\boldsymbol{v}} &=& - \left( \boldsymbol{n}-\boldsymbol{v} \right) \cdot \boldsymbol{S}_{-\boldsymbol{v}}, \\
\label{kd-ANO-stokesS-to-borelS-2}
\mathsf{S}_{\boldsymbol{n}\to\boldsymbol{n}-2\boldsymbol{v}} &=& - \left( \boldsymbol{n}-2\boldsymbol{v} \right) \cdot \left\{ \boldsymbol{S}_{-2\boldsymbol{v}} + \frac{1}{2} \left( \boldsymbol{n}-\boldsymbol{v} \right) \cdot \boldsymbol{S}_{-\boldsymbol{v}}\,\, \boldsymbol{S}_{-\boldsymbol{v}} \right\}, \\
\label{kd-ANO-stokesS-to-borelS-3}
\mathsf{S}_{\boldsymbol{n}\to\boldsymbol{n}-3\boldsymbol{v}} &=& - \left( \boldsymbol{n}-3\boldsymbol{v} \right) \cdot \left\{ \boldsymbol{S}_{-3\boldsymbol{v}} + \frac{1}{2} \left( \boldsymbol{n}-\boldsymbol{v} \right) \cdot \boldsymbol{S}_{-\boldsymbol{v}}\,\, \boldsymbol{S}_{-2\boldsymbol{v}} + \right. \nonumber \\
&&
\left. + \frac{1}{2} \left( \boldsymbol{n}-2\boldsymbol{v} \right) \cdot \left( \boldsymbol{S}_{-2\boldsymbol{v}} + \frac{1}{3} \left( \boldsymbol{n}-\boldsymbol{v} \right) \cdot \boldsymbol{S}_{-\boldsymbol{v}}\,\, \boldsymbol{S}_{-\boldsymbol{v}} \right) \boldsymbol{S}_{-\boldsymbol{v}} \right\}.
\eea
\noindent
It is tempting to notice how the content inside the curly brackets can be thought-of as a projection into the linear subspace spanned by $\CL \left\{ \boldsymbol{S}_{-\ell\boldsymbol{v}}, \boldsymbol{S}_{-\left(\ell-1\right)\boldsymbol{v}}, \ldots, \boldsymbol{S}_{-\boldsymbol{v}} \right\}$ (where, however, no statement may be made concerning dimension as, in general, nothing is said about the linear (in)dependence of these Stokes vectors). However, it turns out that this reasoning is not the best way to write a general formula (it leads to rather intricate coefficients at higher orders). Instead, the good generalization of \eqref{stokesS-to-borelS-GEN} is essentially the one already used in the backward-axes case \eqref{kd-BA-stokesS-to-borelS-GEN},
\bea
\mathsf{S}_{\boldsymbol{n}\to\boldsymbol{n}-\ell\boldsymbol{v}} &=& - \sum_{s=1}^{\ell} \frac{1}{s!}\, \sum_{\substack{\left\{ \ell_1,\ldots,\ell_s\ge 1 \right\} \\ \vphantom{\frac{1}{2}} \sum \ell_i=\ell}} \left( \boldsymbol{n}- \ell_1 \boldsymbol{v} \right) \cdot \boldsymbol{S}_{- \ell_1\, \boldsymbol{v}}\,\, \left( \boldsymbol{n}- \left(\ell_1+\ell_2\right) \boldsymbol{v} \right) \cdot \boldsymbol{S}_{- \ell_2 \boldsymbol{v}}\,\, \cdots \nonumber\\
&&
\cdots\,\, \left( \boldsymbol{n}- \left(\ell_1+\ell_2+\cdots+\ell_{s-1}\right) \boldsymbol{v} \right) \cdot \boldsymbol{S}_{- \ell_{s-1} \boldsymbol{v}}\,\, \left( \boldsymbol{n}- \ell\, \boldsymbol{v} \right) \cdot \boldsymbol{S}_{- \ell_s \boldsymbol{v}}.
\label{kd-ANO-stokesS-to-borelS-GEN}
\eea
\noindent
The inverse mapping also follows in a similar fashion to the backward-axes case. A few examples, generalizing \eqref{borelS-to-stokesS-1} through \eqref{borelS-to-stokesS-3} and completely equivalent to \eqref{kd-BA-borelS-to-stokesS-1} through \eqref{kd-BA-borelS-to-stokesS-3}, are 
\bea
\label{kd-ANO-borelS-to-stokesS-1}
\left( \boldsymbol{n}-\boldsymbol{v} \right) \cdot \boldsymbol{S}_{-\boldsymbol{v}} &=& - \mathsf{S}_{\boldsymbol{n}\to\boldsymbol{n}-\boldsymbol{v}}, \\
\label{kd-ANO-borelS-to-stokesS-2}
\left( \boldsymbol{n}-2\boldsymbol{v} \right) \cdot \boldsymbol{S}_{-2\boldsymbol{v}} &=& - \mathsf{S}_{\boldsymbol{n}\to\boldsymbol{n}-2\boldsymbol{v}} - \frac{1}{2}\, \mathsf{S}_{\boldsymbol{n}\to\boldsymbol{n}-\boldsymbol{v}}\, \mathsf{S}_{\boldsymbol{n}-\boldsymbol{v}\to\boldsymbol{n}-2\boldsymbol{v}}, \\
\left( \boldsymbol{n}-3\boldsymbol{v} \right) \cdot \boldsymbol{S}_{-3\boldsymbol{v}} &=& - \mathsf{S}_{\boldsymbol{n}\to\boldsymbol{n}-3\boldsymbol{v}} - \frac{1}{2}\, \mathsf{S}_{\boldsymbol{n}\to\boldsymbol{n}-\boldsymbol{v}}\, \mathsf{S}_{\boldsymbol{n}-\boldsymbol{v}\to\boldsymbol{n}-3\boldsymbol{v}} - \frac{1}{2}\, \mathsf{S}_{\boldsymbol{n}\to\boldsymbol{n}-2\boldsymbol{v}}\, \mathsf{S}_{\boldsymbol{n}-2\boldsymbol{v}\to\boldsymbol{n}-3\boldsymbol{v}} - \nonumber \\
&&
- \frac{1}{3}\, \mathsf{S}_{\boldsymbol{n}\to\boldsymbol{n}-\boldsymbol{v}}\, \mathsf{S}_{\boldsymbol{n}-\boldsymbol{v}\to\boldsymbol{n}-2\boldsymbol{v}}\, \mathsf{S}_{\boldsymbol{n}-2\boldsymbol{v}\to\boldsymbol{n}-3\boldsymbol{v}}.
\label{kd-ANO-borelS-to-stokesS-3}
\eea
\noindent
For the closed-form expression one finds
\be
\left( \boldsymbol{n}- \ell\, \boldsymbol{v} \right) \cdot \boldsymbol{S}_{- \ell \boldsymbol{v}} = - \sum_{s=1}^{\ell} \frac{1}{s}\, \sum_{\substack{\left\{ \ell_1,\ldots,\ell_s\ge 1 \right\} \\ \vphantom{\frac{1}{2}} \sum \ell_i=\ell}} \mathsf{S}_{\boldsymbol{n}\to\boldsymbol{n}-\ell_1\boldsymbol{v}}\, \mathsf{S}_{\boldsymbol{n}-\ell_1\boldsymbol{v}\to\boldsymbol{n}- \left(\ell_1+\ell_2\right) \boldsymbol{v}}\, \cdots\, \mathsf{S}_{\boldsymbol{n}- \left(\ell-\ell_s\right) \boldsymbol{v}\to\boldsymbol{n}-\ell\boldsymbol{v}},
\label{kd-ANO-borelS-to-stokesS-GEN}
\ee
\noindent
which generalizes \eqref{borelS-to-stokesS-GEN} to the present $k$-dimensional setting, in complete analogy to \eqref{kd-BA-borelS-to-stokesS-GEN}. In both \eqref{kd-ANO-stokesS-to-borelS-GEN} and \eqref{kd-ANO-borelS-to-stokesS-GEN} one has to constrain
\be
1 \leq \ell < \min \left\{ \left\lfloor \frac{n_i}{\nu_i} \right\rfloor \right\}_{i=1,\ldots,k},
\ee
\noindent
in order not to go past beyond the transseries-grid boundary.
\end{itemize}

So far the above considerations were rather generic. Furthermore, they were all \textit{motivated}, none was really \textit{derived}. Short of checking the derivation of these results (\textit{e.g.}, in appendix~\ref{app:alien-calculus}), how can we test if these considerations are actually correct, and realized in examples? If the reader recalls subsection~\ref{subsec:Fnonlinear}, one way to (numerically) probe the complex Borel plane of some specific example was via Pad\'e approximants to the Borel transforms, as in \eqref{pade-S}. Back then, while addressing the nonlinear second-order ODE \eqref{quarticNLODE}, we used this method to (numerically) learn about the nature of Borel singularities associated to different asymptotic sectors in its transseries solution \eqref{one-param-transseries}. In particular, we found out about the existence of the ``$1.5$''-parameter transseries due to the structure and accumulation of Borel (pole) singularities for the one-instanton sector; recall figure~\ref{fig:Pade-poles-quartic-pert-series}. Within our present context, is it also possible to carry through an analogous procedure which, eventually, could validate the previous general expectations, equations \eqref{eq:sec5-Borel-transf} and \eqref{eq:sec5-bridge-eqs-COMPLETE}? This is what we shall investigate next, numerically analyzing the singular structure of the Borel transforms \eqref{eq:sec5-Borel-transf} via Pad\'e approximants, within one example with $k>1$.

\begin{figure}[t!]
\begin{center}
\includegraphics[width=7.4cm]{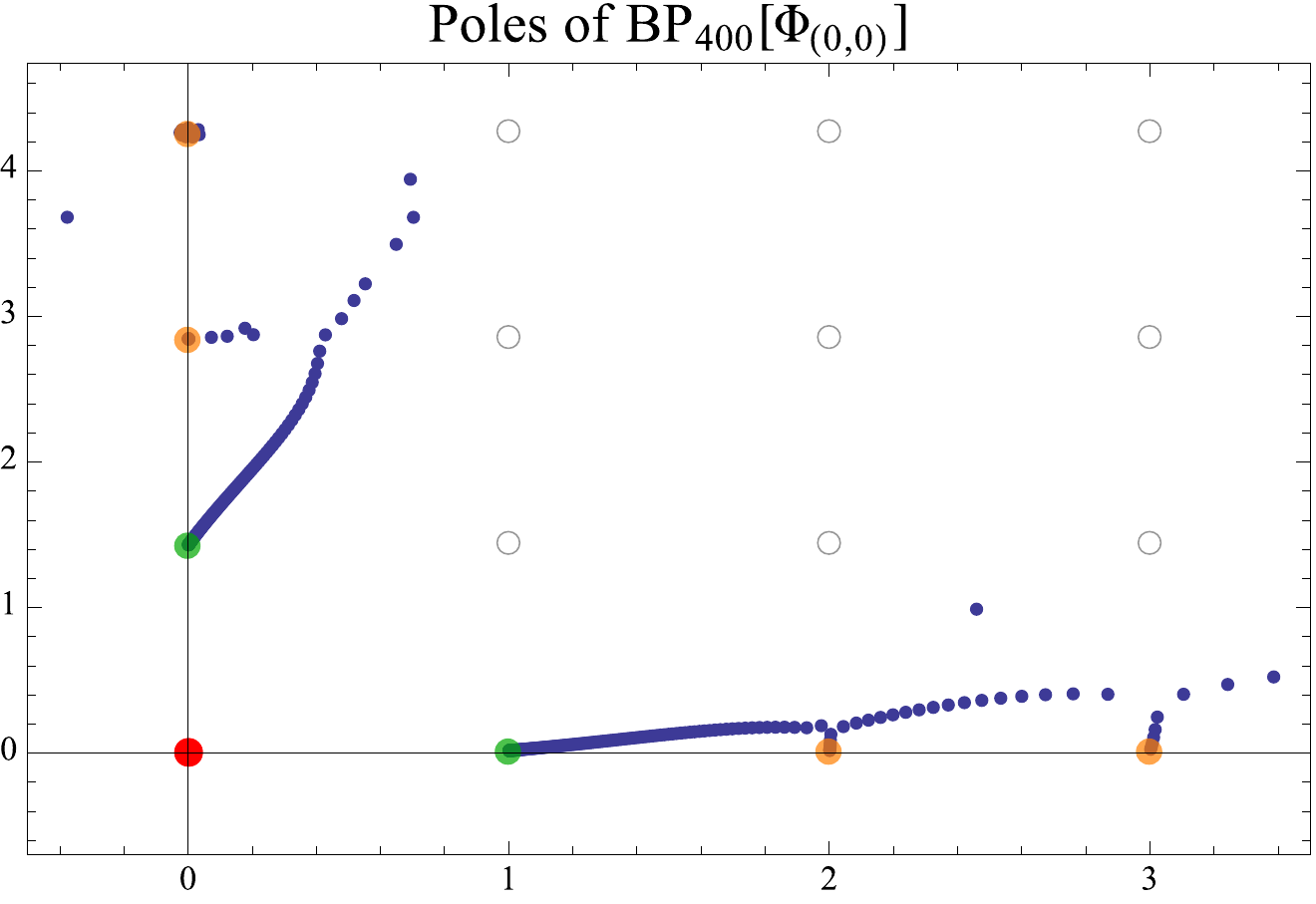}
$\qquad$
\includegraphics[width=7.4cm]{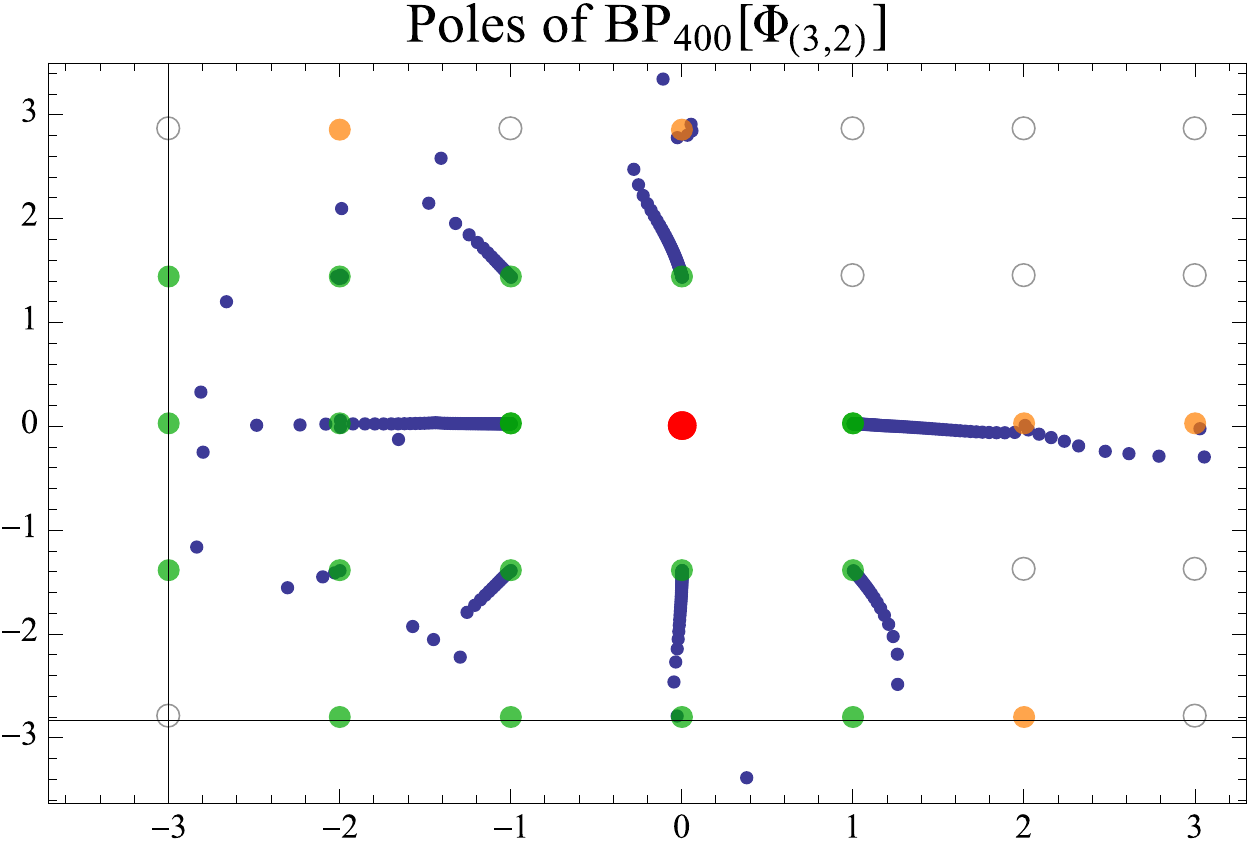}
\end{center}
\caption{Poles of the (diagonal) order-400 Pad\'e approximants for the Borel transforms of perturbative ($\mathrm{BP}_{400} [\Phi_{(0,0)}]$) and $(3,2)$-instanton ($\mathrm{BP}_{400} [\Phi_{(3,2)}]$) sectors in a modified, second-order nonlinear Riccati ODE. The particular perturbative (left image) or $(3,2)$-instanton (right image) sector which is under analysis is plotted with a red disk, whereas its ``basic'' singularities (\textit{i.e.}, the ones corresponding to the single-steps of figure~\ref{fig:sec5-2d-lattice}) are plotted by green disks. Singularities corresponding to multi-step paths are then plotted as orange disks. The grey circles mark points on the Borel plane where no singularities are to be found (see the corresponding discussion in the main text). The condensation of poles to form branch-cuts is very clear---in fact, close to many of the singular points---and the more data available the more visible this would become in other (\textit{i.e.}, further ``far away'') singularities.
}
\label{fig:sec5-BP-Riccati}
\end{figure}

For simplicity and clarity of comparison, we shall choose an example with $k=2$ and such that it reproduces figure~\ref{fig:sec5-2d-lattice} as much as possible (in order to validate---or not---the structure we have put forward). In other words, we want the projection map \eqref{projectionmapZkC}, from $\BZ^2$ to $\BC$, to be as close as possible to the identity map (but without any risk to induce resonance!). In this way, we chose the nonlinear second-order Riccati-type ODE from \cite{o05b}, which is solved by a two-parameter transseries, but we \textit{modified it} such that one of the instanton actions, $A_{1}$, is purely real, while the other, $A_{2}$, is purely imaginary. We also made sure their ratio is irrational. Following along similar steps to the ones in section~\ref{sec:quartic}, one now puts forward a two-parameter transseries \textit{ansatz} and after some work eventually finds\footnote{Our modified Ricatti-type ODE is (but see also \cite{o05b} for the original, unmodified version):
\begin{equation}
u'' (z) + \left( 1+\mathrm{i}\sqrt{2} \right) \left( u' (z) + u^2 (z) \right) + \mathrm{i}\sqrt{2}\, u (z) + \mathrm{i}\sqrt{2}\, u (z)\, u' (z) + u^3 (z) + \frac{1}{z} = 0,
\end{equation}
\noindent
where we are now using the variable $z$, instead of the usual $x = \frac{1}{z}$ from the main text (but see appendix~\ref{app:alien-calculus}). The solution to this differential equation, for $z\gg1$, is a two-parameter transseries with instanton actions $A_{1}=1$ and $A_{2}=\mathrm{i}\sqrt{2}$. We are not going to give any further details on how to solve this equation, or on how to plot figure~\ref{fig:sec5-BP-Riccati}. Instead, we leave it as an exercise to the reader to use the techniques learnt in section~\ref{sec:quartic}, alongside some elementary \textit{Mathematica} coding, in order to iteratively construct the transseries solution to the above equation, and reproduce the numerical plots in figure~\ref{fig:sec5-BP-Riccati}. Hopefully this will be a very illustrative exercise!} recursive equations for the coefficients in the asymptotic sectors $\Phi_{(n,m)}$. Once these coefficients have been computed, one can write down Borel transforms for each of the asymptotic sectors, determine their Pad\'e approximants, and thus finally analyze their singularity (pole) structure. These numerical results are summarized in the plots of figure~\ref{fig:sec5-BP-Riccati}, for the cases of perturbative, $\Phi_{(0,0)}$, and $(3,2)$-instanton, $\Phi_{(3,2)}$, sectors. Let us first consider the singularity structure of $\CB [\Phi_{(0,0)}] (s)$, depicted on the left plot of figure~\ref{fig:sec5-BP-Riccati}. It is simple to identify the expected branch-cuts starting at both $A_{1}=1$ and $A_{2}=\mathrm{i}\sqrt{2}$; these correspond to the Stokes \textit{single-step} elementary forward motions $\boldsymbol{e}_1$ and $\boldsymbol{e}_2$ from figure~\ref{fig:sec5-2d-lattice}, and are plotted in green. Apart from these branch-cuts, we further find the appearance of cuts at $2A_{1}$ and $3A_{1}$, as well as $2A_{2}$ and $3A_{2}$, plotted in orange. This is also precisely as expected, now corresponding to \textit{multi-step} paths (not plotted in figure~\ref{fig:sec5-2d-lattice}). In fact, recall that while figure~\ref{fig:sec5-2d-lattice} plotted (single-step) locations of (non-vanishing) \textit{Stokes vectors}, figure~\ref{fig:sec5-BP-Riccati} is now (numerically) identifying locations of \textit{all} possible \textit{Borel residues}, via the projection \eqref{projectionmapZkC}. So while in the former case the only non-vanishing possibility was, \textit{e.g.}, $\boldsymbol{S}_{\boldsymbol{e}_1} = S_{\boldsymbol{e}_1}^{(1)}\, \boldsymbol{e}_1$; in the latter case we have an infinite array of possibilities as listed in \eqref{kd-FA-stokesS-to-borelS+GEN} and, \textit{e.g.}, now translating to
\be
\mathsf{S}_{\boldsymbol{0}\to\ell\boldsymbol{e}_1} = - \left( S_{\boldsymbol{e}_1}^{(1)} \right)^{\ell}.
\ee
\noindent
The same structure appears when addressing the singularity structure of $\CB [\Phi_{(3,2)}] (s)$ (shown on the right plot of figure~\ref{fig:sec5-BP-Riccati}). On one side we find the branch-cuts corresponding to the expected Stokes \textit{single-step} backwards and forward motions depicted in figure~\ref{fig:sec5-2d-lattice}; they are plotted in green. Apart from these, we further find poles gathering around the remaining \textit{multi-step} singularities (and plotted in orange). The first quadrant is a ``double-plus'' 2-orthant and, as such, it is without surprise that we find it empty of any Borel singularities. All these results are exactly what was expected, based on our earlier $k$-orthant classification of singularities, thus yielding an excellent numerical support for the resurgent structure of Borel singularities which we have put forward in equations \eqref{eq:sec5-Borel-transf} and \eqref{eq:sec5-bridge-eqs-COMPLETE}.

In conclusion, this time around numerical tests did not produce any new features as compared to our original expectations---unlike what happened back in subsection~\ref{subsec:Fnonlinear}, where indeed they led to new results. Instead, on second thought, they helped us realize how the surprisingly simple resurgence relations \eqref{eq:sec5-bridge-eqs-COMPLETE} encode a vast amount of non-trivial information (the one displayed in figures~\ref{fig:sec5-2d-lattice} and~\ref{fig:sec5-BP-Riccati}). This should give the reader a glimpse into the fundamental role that alien derivatives play in this game, and the associated power of working at an algebraic level.

\subsection*{Stokes Discontinuities and Transseries Asymptotics}

Following along the steps from subsection~\ref{subsec:Fnonlinear}, our next goal is to compute Stokes discontinuities, $\disc_\theta = \1 - \underline{\mathfrak{S}}_\theta$, having in sight obtaining large-order formulae for the asymptotics of general transseries such as \eqref{eq:sec5-transseries-vec}. These are given in terms of the Stokes automorphism \eqref{Stokes-aut-as exponential-singularities}, but where $\theta$ may now take multiple values, depending on the projections \eqref{projectionmapZkC} of the several possible lattice directions (it was essentially restricted to being either $0$ or $\pi$ back in subsection~\ref{subsec:Fnonlinear}). Recall:
\be
\label{Stokes-aut-as exponential-singularities-AGAIN}
\underline{\mathfrak{S}}_\theta = \exp \left( \sum_{\left\{ \omega_\theta \right\}} \rme^{-\frac{\omega_\theta}{x}} \Delta_{\omega_\theta} \right).
\ee
\noindent
Although we are labeling this operator by $\theta$, on the complex Borel plane, it will be computed out of information on the alien lattice via the alien derivatives. This implies the ``statistical mechanical'' language we have used before holds generically. In particular, all concepts from subsection~\ref{subsec:Fnonlinear} pertaining to a step, $\CS$, depicted in \eqref{DEF-stepS}; the weight of a step $\CS$, $w(\CS)$, now given by \eqref{k-dim-weight!}; a path, $\CP$, defined by \eqref{DEF-path}; the length of a path, $\CP$, $\ell(\CP)$, defined in \eqref{DEF-ell-path}; the weight of a path, $\CP$, $w(\CP)$, defined in \eqref{DEF-w-path}; and the combinatorial factor of a path $\CP$, $\text{CF}(\CP)$, defined in \eqref{DEF-CF-path}; they all have completely clear analogues in the present higher-dimensional case.

Let us illustrate the general principles by computing Stokes discontinuities within a two-dimensional example, as usual, where $k=2$ and where we assume there is no resonance. We shall denote the two instanton actions by $A_1$, $A_2$, and will illustrate the computation of Stokes discontinuities along different singular directions by focusing on the  $\Phi_{(3,2)}$ asymptotic sector shown in figure~\ref{fig:sec5-2d-lattice}. The first question to address is how many different singular directions does this sector have, on the complex Borel plane (also have figure~\ref{fig:sec5-BP-Riccati} in mind in the following). This question was already answered in full generality when we classified the distribution of Stokes vectors along $k$-orthants, but let us now see it in practice. For the moment, we simply split the set of singular directions into directions including \textit{one forward} motion (including both the ``forward axes'' and ``single-plus $k$-orthant'' cases), and directions consisting \textit{only} of \textit{backward} motions (including both the ``backward axes'' and ``all-negative $k$-orthant'' cases). This implies the following, momentary, classification of singular directions (and their singular branch-points):
\begin{itemize}
\item \textit{Forward motions}: forward resurgence motions which connect the original sector $\Phi_{(3,2)}$ with the sectors $\Phi_{(3-\ell_{1},3)}$, $\ell_{1}=0,1,2,3$ (the alien-lattice vectors defining these motions are $\boldsymbol{\ell}=\left(-\ell_{1},1\right)$, \textit{i.e.}, the ``upwards'' motions of figure~\ref{fig:sec5-2d-lattice}); and iterations thereof (meaning we can still reach sectors $\Phi_{(1,4)}$ and $\Phi_{(0,5)}$ by iterating the $\boldsymbol{\ell}=\left(-1,1\right)$ motion, and indefinitely reach sectors $\Phi_{(3,m)}$ with $m>3$ by iterating the $\boldsymbol{\ell}=\left(0,1\right)$ motion). Or else, forward resurgence motions which connect the original sector $\Phi_{(3,2)}$ with the sectors $\Phi_{(4,2-\ell_{2})}$, $\ell_{2}=0,1,2$ (the alien-lattice vectors defining these motions are $\boldsymbol{\ell}=\left(1,-\ell_{2}\right)$, \textit{i.e.}, the ``rightwards'' motions of figure~\ref{fig:sec5-2d-lattice}); and iterations thereof (meaning we can still reach sector $\Phi_{(5,0)}$ by iterating the $\boldsymbol{\ell}=\left(1,-1\right)$ motion, and indefinitely reach sectors $\Phi_{(n,2)}$ with $n>4$ by iterating the $\boldsymbol{\ell}=\left(1,0\right)$ motion). The angle $\theta_{\boldsymbol{\ell}}$ on the complex Borel plane, associated to each of the corresponding projected directions, is simply $\theta_{\boldsymbol{\ell}} = \arg \left( \boldsymbol{\ell} \cdot \boldsymbol{A} \right)$. 
\item \textit{Backward motions}: all other resurgence motions from figure~\ref{fig:sec5-2d-lattice}, now connecting the original sector $\Phi_{(3,2)}$ with the  sectors $\Phi_{(3-\ell_{1},2-\ell_{2})}$, where $\boldsymbol{\ell}=\left(-\ell_{1},-\ell_{2}\right)$ with $0 \le \ell_1 \le 3$, $0 \le \ell_2 \le 2$ but $\boldsymbol{\ell} \ne \boldsymbol{0}$ and $\boldsymbol{\ell} \ne (3,2)$; and iterations thereof (in this case, only the motions $\boldsymbol{\ell}=\left(-1,0\right)$, $\boldsymbol{\ell}=\left(0,-1\right)$, and $\boldsymbol{\ell}=\left(-1,-1\right)$ may be iterated, at least up to the boundary of the transseries grid). Again, the angle $\theta_{\boldsymbol{\ell}}$ on the complex Borel plane, associated to each of the corresponding projected directions, is $\theta_{\boldsymbol{\ell}} = \arg \left( \boldsymbol{\ell} \cdot \boldsymbol{A} \right)$. 
\end{itemize}

Let us proceed with some examples, and first focus on the above class of \textit{forward motions}. Fix a lattice direction $\boldsymbol{\ell}$, corresponding to a singular direction $\theta_{\boldsymbol{\ell}} = \arg \left( \boldsymbol{\ell} \cdot \boldsymbol{A} \right)$ on the complex Borel plane. In this case, what will the Stokes automorphism \eqref{Stokes-aut-as exponential-singularities-AGAIN} look like? It should be clear by now that we need to sum over all \textit{singular points} along this \textit{fixed direction} $\theta_{\boldsymbol{\ell}}$, which may be reachable via (lattice) \textit{trajectories} along this very same (projected) \textit{fixed direction} $\theta_{\boldsymbol{\ell}}$ (recall from figure~\ref{stokescrossingfig} that the Stokes automorphism, being the operator behind Stokes phenomena as in, \textit{e.g.}, \eqref{stokespheno}, is \textit{local} in the singular direction of the complex Borel plane). It should also be clear that there is a \textit{single} Stokes vector associated to each of these forward directions. This means that the Stokes automorphism for \textit{forward motions} will be very similar to \eqref{one-param-stokes-0}: akin to back then, there is basically one single forward motion which needs to be iterated in order to reach each of the other singular points along the chosen fixed direction. The straightforward $k$-dimensional generalization of \eqref{one-param-stokes-0} is then, very simply\footnote{Recall that here, when selecting directions, $\boldsymbol{\ell}$ is lattice versor.},
\begin{equation}
\label{eq:sec5-Stokes-forward-direction-theta}
\underline{\mathfrak{S}}_{\theta_{\boldsymbol{\ell}}} \Phi_{\boldsymbol{n}} = \exp \left( \rme^{-\frac{\boldsymbol{\ell}\cdot\boldsymbol{A}}{x}} \Delta_{\boldsymbol{\ell}\cdot\boldsymbol{A}} \right) \Phi_{\boldsymbol{n}}.
\end{equation}
\noindent
Considering our ``example-sector'' $\boldsymbol{n} = (3,2)$, one finds\footnote{Just like in the one-dimensional case, if we did not know equation \eqref{kd-SPO-stokesS-to-borelS+GEN}, one could also take each exponential term in the equality \eqref{eq:sec5-forward-Stokes-expanded} as the \textit{defining equations} for the Borel residues as functions of the Stokes constants.}
\begin{eqnarray}
\underline{\mathfrak{S}}_{\theta_{\boldsymbol{\ell}}} \Phi_{(3,2)} &=& \left( 1 + \rme^{-\frac{\boldsymbol{\ell} \cdot \boldsymbol{A}}{x}}\, \Delta_{\boldsymbol{\ell} \cdot \boldsymbol{A}} + \frac{1}{2!} \rme^{-2 \frac{\boldsymbol{\ell} \cdot \boldsymbol{A}}{x}}\, \Delta_{\boldsymbol{\ell} \cdot \boldsymbol{A}}^2 + \frac{1}{3!} \rme^{-3 \frac{\boldsymbol{\ell} \cdot \boldsymbol{A}}{x}}\, \Delta_{\boldsymbol{\ell} \cdot \boldsymbol{A}}^3 + \cdots \right) \Phi_{(3,2)} = \nonumber \\
&=& \Phi_{(3,2)} + \boldsymbol{S}_{\boldsymbol{\ell}} \cdot \left( 3-\ell_{1}, 3 \right) \rme^{-\frac{\boldsymbol{\ell} \cdot \boldsymbol{A}}{x}}\, \Phi_{\left( 3-\ell_{1}, 3 \right)} + \frac{1}{2!}\, \Big( \boldsymbol{S}_{\boldsymbol{\ell}} \cdot \left( 3-\ell_{1}, 3 \right) \Big)\, \Big( \boldsymbol{S}_{\boldsymbol{\ell}} \cdot \left( 3-2\ell_{1}, 4 \right) \Big) \times \nonumber \\
&&
\times\, \mathrm{e}^{-2 \frac{\boldsymbol{\ell}\cdot\boldsymbol{A}}{x}}\, \Phi_{\left( 3-2\ell_{1}, 4 \right)} + \frac{1}{3!} \prod_{m=1}^{3} \Big( \boldsymbol{S}_{\boldsymbol{\ell}} \cdot \left( 3-m\,\ell_{1}, 2+m \right) \Big)\, \rme^{-3\frac{\boldsymbol{\ell}\cdot\boldsymbol{A}}{x}}\, \Phi_{\left( 3-3\ell_{1}, 5 \right)} + \cdots \nonumber \\
&=& \Phi_{(3,2)} - \mathsf{S}_{(3,2)\to\left( 3-\ell_{1}, 3 \right)}\, \rme^{-\frac{\boldsymbol{\ell} \cdot \boldsymbol{A}}{x}}\, \Phi_{\left( 3-\ell_{1}, 3 \right)} - \mathsf{S}_{(3,2)\to\left( 3-2\ell_{1}, 4 \right)}\, \mathrm{e}^{-2 \frac{\boldsymbol{\ell}\cdot\boldsymbol{A}}{x}}\, \Phi_{\left( 3-2\ell_{1}, 4 \right)} - \nonumber \\
&&
- \mathsf{S}_{(3,2)\to\left( 3-3\ell_{1}, 5 \right)}\, \rme^{-3\frac{\boldsymbol{\ell}\cdot\boldsymbol{A}}{x}}\, \Phi_{\left( 3-3\ell_{1}, 5 \right)} - \cdots,
\label{eq:sec5-forward-Stokes-expanded}
\end{eqnarray}
\noindent
where we chose $\boldsymbol{\ell} = \left(-\ell_{1},1\right)$ and, as we already said above, this can be $\ell_1 \in \left\{ 0,1,2,3 \right\}$. But now we explicitly see the aforementioned interesting novelty as compared to the one-dimensional case. If we consider the strictly vertical direction, $\ell_{1}=0$, then the above expansion continues indefinitely, pretty much as in the one-dimensional case. But if otherwise, \textit{i.e.}, if $\ell_1 \in \left\{ 1,2,3 \right\}$ mixing vertical and horizontal lattice motions, then the expansion will \textit{truncate}, which in the one-dimensional case only happened for \textit{backward motions}. It is without surprise that the same features we uncovered when discussing Stokes vectors now reappear when discussing Stokes automorphisms: as already explained, this just means we have reached the boundary of the $\BN^2_0$ semi-positive grid, and in this regard is somewhat similar to the one-dimensional truncation. Quantitatively, the $n$th term in the expansion above is proportional to the sector $\Phi_{\left(3-n\,\ell_{1},2+n\right)}$, which is non-zero only if $3-n\,\ell_{1}\ge0$. Let us understand this in terms of motions and data on the \textit{alien lattice} from figure~\ref{fig:sec5-2d-lattice}, using the ``statistical mechanical'' language we used back in section~\ref{sec:quartic}. Similar to what happened in the one-dimensional case, this should yield a ``physical'' explanation behind the coefficients (Borel residues) in \eqref{eq:sec5-forward-Stokes-expanded} above. In turns, let us consider a case where the expansion \eqref{eq:sec5-forward-Stokes-expanded} does \textit{not} truncate, and one where \textit{it does} truncate:
\begin{itemize}
\item \textit{What is the Stokes automorphism along the strictly vertical direction $\boldsymbol{\ell}=(0,1)$, when acting upon the node $\Phi_{(3,2)}$?} It is simple to compare the action of $\underline{\mathfrak{S}}_{\theta_{(0,1)}}$ on $\Phi_{(3,2)}$ from \eqref{eq:sec5-forward-Stokes-expanded}, with the allowed strictly-vertical resurgence motions out of the node $\Phi_{(3,2)}$ on the alien lattice of figure~\ref{fig:sec5-2d-lattice}. The coefficients of the several terms in the expansion above then have the following ``statistical mechanical'' origin:
\begin{itemize}
\item $\Phi_{(3,2)}\rightarrow\Phi_{(3,2)}$: this motion simply leaves the node invariant;
\item $\Phi_{(3,2)}\rightarrow\Phi_{(3,3)}$: there is a single path, of length
$\ell=1$ and weight\footnote{This is $w = (3,3) \cdot \boldsymbol{S}_{(0,1)} = 3 S^{(2)}_{\boldsymbol{e}_2}$. It is interesting to compare this case to the one following \eqref{stokes-zero-one-param-expanded}.} $w = 3 S^{(2)}_{\boldsymbol{e}_2}$, leading to a combinatorial factor $\mathrm{CF}=\frac{1}{1!}$ and nonperturbative contribution $\rme^{-\frac{A_2}{x}}$;
\item $\Phi_{(3,2)}\rightarrow\Phi_{(3,4)}$: there is a single path (made of two steps), with length $\ell=2$ and weight $w = 12 \left( S^{(2)}_{\boldsymbol{e}_2} \right)^2$, leading to $\mathrm{CF}=\frac{1}{2!}$ and nonperturbative contribution $\rme^{-2\frac{A_2}{x}}$;
\item $\Phi_{(3,2)}\rightarrow\Phi_{(3,5)}$: again there is a single path (with three steps), of length $\ell=3$ and weight $w = 60 \left( S^{(2)}_{\boldsymbol{e}_2} \right)^3$, leading to $\mathrm{CF}=\frac{1}{3!}$ and nonperturbative contribution $\rme^{-3\frac{A_2}{x}}$.
\end{itemize}
\item \textit{What is the Stokes automorphism along the diagonal ``northwest'' direction $\boldsymbol{\ell}=(-1,1)$, when acting upon the node $\Phi_{(3,2)}$?} The procedure is essentially the same, only now we compare the action of $\underline{\mathfrak{S}}_{\theta_{(-1,1)}}$ on $\Phi_{(3,2)}$, with the allowed lattice diagonal-motions out of $\Phi_{(3,2)}$ and along $\boldsymbol{\ell}=(-1,1)$. Again, our coefficients have a ``statistical mechanical'' origin:
\begin{itemize}
\item $\Phi_{(3,2)}\rightarrow\Phi_{\left(3,2\right)}$: this motion leaves the node invariant;
\item $\Phi_{(3,2)}\rightarrow\Phi_{\left(2,3\right)}$: there is one single path, of length $\ell=1$ and weight $w = \boldsymbol{S}_{(-1,1)} \cdot (2,3)$, leading to a combinatorial factor $\mathrm{CF}=\frac{1}{1!}$ and nonperturbative contribution $\rme^{-\frac{A_{2}-A_{1}}{x}}$;
\item $\Phi_{(3,2)}\rightarrow\Phi_{\left(1,4\right)}$: there is one single path (made of two steps), of length $\ell=2$ and weight $w = \left( \boldsymbol{S}_{(-1,1)} \cdot (2,3) \right) \left( \boldsymbol{S}_{(-1,1)} \cdot (1,4) \right)$, leading to $\mathrm{CF}=\frac{1}{2!}$ and nonperturbative contribution $\rme^{-2\frac{A_{2}-A_{1}}{x}}$;
\item $\Phi_{(3,2)}\rightarrow\Phi_{\left(0,5\right)}$: again there is one single path (with three steps), of length $\ell=3$ and weight $w = \prod_{m=1}^{3} \boldsymbol{S}_{(-1,1)} \cdot (3-m,2+m)$, leading to $\mathrm{CF}=\frac{1}{3!}$ and nonperturbative contribution $\rme^{-3\frac{A_{2}-A_{1}}{x}}$. Further, one cannot move diagonally ``northwest'' from this node, in which case the expansion necessarily truncates here.
\end{itemize}
\end{itemize}
\noindent
The general pattern should now be clear. Furthermore, we see how our initial class of \textit{forward motions}---motivated by \eqref{eq:sec5-Stokes-forward-direction-theta}---is actually best addressed if \textit{split in two}, where one class is closer to the one-dimensional example (where there is no truncation of the exponential action) and the other somewhat closer to the \textit{backward motions} due to the above truncation. Happily, this was precisely the original split we made when classifying $k$-orthants according to Stokes vectors.

Next, one can turn to the class of \textit{backward motions}. Let us now fix the lattice direction as $\boldsymbol{\ell}=\left(-\ell_{1},-\ell_{2}\right)$, corresponding to the singular direction $\theta_{\boldsymbol{\ell}} = \arg \left( \boldsymbol{\ell} \cdot \boldsymbol{A} \right)$ on the complex Borel plane. In order to understand what will the Stokes automorphism \eqref{Stokes-aut-as exponential-singularities-AGAIN} look like in this case, we only need to recall the discussion that led up to \eqref{eq:sec5-Stokes-forward-direction-theta}. One quickly realizes that now the Stokes automorphism for \textit{backward motions} must be very similar to \eqref{one-param-stokes-pi}: akin to back then, there may be several backward motions, some of which may even be iterated, but a truncation will always occur as soon as one hits the transseries-grid boundary. The straightforward $k$-dimensional generalization of \eqref{one-param-stokes-pi} is then, very simply\footnote{Recall that here, when selecting directions, $\boldsymbol{\ell}$ is lattice versor.},
\begin{equation}
\label{eq:sec5-Stokes-backward-direction-theta}
\underline{\mathfrak{S}}_{\theta_{\boldsymbol{\ell}}} \Phi_{\boldsymbol{n}} = \exp\left( \sum_{m=1}^{\textsf{n}} \rme^{-m\, \frac{\boldsymbol{\ell} \cdot \boldsymbol{A}}{x}}\, \Delta_{m\,\boldsymbol{\ell} \cdot \boldsymbol{A}} \right) \Phi_{\boldsymbol{n}}, \qquad \textsf{n} = \min \left\{ \left\lfloor \frac{n_i}{\ell_i} \right\rfloor \right\}_{i=1,\ldots,k},
\end{equation}
\noindent
where $\textsf{n}$ locates the transseries-grid boundary. Considering our usual ``example-sector'' $\boldsymbol{n} = (3,2)$, one may move backwards to $\left( -\ell_{1}, -\ell_{2} \right)$, with $\ell_{1} \in \left\{ 0,1,2,3 \right\}$ and $\ell_{2} \in \left\{ 0,1,2 \right\}$ (obviously excluding $(0,0)$ and $(3,2)$). Let us then understand the resulting Stokes automorphism using the familiar ``statistical mechanical'' language to describe motions and data on the alien lattice.
\begin{itemize}
\item \textit{What is the Stokes automorphism along the diagonal ``southwest'' direction $\boldsymbol{\ell}=(-1,-1)$, when acting upon the node $\Phi_{(3,2)}$?} In this case ${\textsf{n}}=2$ and $\boldsymbol{\ell} \cdot \boldsymbol{A} = - A_{1} - A_{2}$. Having said this, the procedure is essentially the same as in the previous case. First expand the exponential \eqref{eq:sec5-Stokes-backward-direction-theta}, and use the resurgence relations to compute:
\begin{eqnarray}
\underline{\mathfrak{S}}_{\theta_{\boldsymbol{\ell}}} \Phi_{(3,2)} &=& \left( 1 + \sum_{m=1}^{2} \rme^{-m\, \frac{\boldsymbol{\ell} \cdot \boldsymbol{A}}{x}}\, \Delta_{m\, \boldsymbol{\ell} \cdot \boldsymbol{A}} + \frac{1}{2!} \left( \sum_{m=1}^{2} \rme^{-m\, \frac{\boldsymbol{\ell} \cdot \boldsymbol{A}}{x}}\, \Delta_{m\, \boldsymbol{\ell} \cdot \boldsymbol{A}} \right)^2 + \right. \nonumber \\
&&
\left.
+ \frac{1}{3!} \left( \sum_{m=1}^{2} \rme^{-m\, \frac{\boldsymbol{\ell} \cdot \boldsymbol{A}}{x}}\, \Delta_{m\, \boldsymbol{\ell} \cdot \boldsymbol{A}} \right)^3 + \cdots \right) \Phi_{(3,2)} = \nonumber \\
&=& \Phi_{(3,2)} + \boldsymbol{S}_{(-1,-1)} \cdot \left(2,1\right) \rme^{\frac{A_{1}+A_{2}}{x}}\, \Phi_{\left(2,1\right)} + \left( \boldsymbol{S}_{(-2,-2)} \cdot \left(1,0\right) + \vphantom{\frac{1}{2!}} \right. \nonumber \\
&&
\label{eq:sec5-backwards-Stokes-expandded}
\left. + \frac{1}{2!} \Big( \boldsymbol{S}_{(-1,-1)} \cdot \left(2,1\right) \Big) \Big( \boldsymbol{S}_{(-1,-1)} \cdot \left(1,0\right) \Big) \right) \rme^{2\frac{A_{1}+A_{2}}{x}}\, \Phi_{\left(1,0\right)} \\
&=& \Phi_{(3,2)} - \mathsf{S}_{(3,2)\to(2,1)}\, \rme^{\frac{A_{1}+A_{2}}{x}}\, \Phi_{\left(2,1\right)} - \mathsf{S}_{(3,2)\to(1,0)}\, \rme^{2\frac{A_{1}+A_{2}}{x}}\, \Phi_{\left(1,0\right)}. \nonumber
\end{eqnarray}
\noindent
The result may be written with either Stokes constants or, equivalently, Borel residues, via\footnote{Again, just like in the one-dimensional case, if we did not know equation \eqref{kd-ANO-stokesS-to-borelS-GEN}, one could also take each exponential term in \eqref{eq:sec5-backwards-Stokes-expandded} as the \textit{defining equations} for the Borel residues as functions of the Stokes constants.} \eqref{kd-ANO-stokesS-to-borelS-1} and \eqref{kd-ANO-stokesS-to-borelS-2} (or, more generally, via \eqref{kd-ANO-stokesS-to-borelS-GEN}). It includes all the backward motions contributing to the Stokes automorphism along the direction defined by $\boldsymbol{\ell}=\left(-1,-1\right)$ and starting-off at node $\Phi_{(3,2)}$. Comparing this action of $\underline{\mathfrak{S}}_{\theta_{\boldsymbol{\ell}}}$ on $\Phi_{(3,2)}$ with the allowed \textit{backward} resurgence motions out of the $(3,2)$ node on the alien lattice of figure~\ref{fig:sec5-2d-lattice}, its ``physical'' interpretation easily follows:
\begin{itemize}
\item $\Phi_{(3,2)}\rightarrow\Phi_{(3,2)}$: this simply leaves the node invariant;
\item $\Phi_{(3,2)}\rightarrow\Phi_{(2,1)}$: there is a single path of length $\ell=1$ and weight $w = 2 S_{(-1,-1)}^{(1)} + S_{(-1,-1)}^{(2)}$, leading to $\mathrm{CF}=\frac{1}{1!}$ and nonperturbative contribution $\rme^{+\frac{A_{1}+A_{2}}{x}}$;
\item $\Phi_{(3,2)}\rightarrow\Phi_{(1,0)}$: there are now two paths. One consists of a single step, with length $\ell=1$, weight $w = S_{(-2,-2)}^{(1)}$ and $\mathrm{CF}=\frac{1}{1!}$; the other consists of two steps, with length $\ell=2$, weight $w = \left( 2 S_{(-1,-1)}^{(1)} + S_{(-1,-1)}^{(2)} \right) S_{(-1,-1)}^{(1)}$ and $\mathrm{CF}=\frac{1}{2!}$. They both lead to a nonperturbative contribution $\rme^{+2\frac{A_{1}+A_{2}}{x}}$.
\end{itemize}
\end{itemize}
\noindent
From these examples a general pattern should be easily recognizable, which we shall spell out in the two-dimensional case. In line with our earlier constructions, one may now define the Stokes discontinuities, $\disc_{\theta_{\boldsymbol{\ell}}}$, purely in terms of motions and data on the alien chain\footnote{One may also think of the statistical component below, $\mathrm{SF}_{(n \rightarrow m)}$, as a ``statistical-mechanical'' \textit{definition} of the Borel residues (up to a sign).}:

\begin{center}
\vspace{10pt}
\setlength{\fboxsep}{10pt}
\Ovalbox{
\parbox{14cm}{
\vspace{5pt}

\textbf{Classification of Stokes directions:} $\theta_{\boldsymbol{\ell}} = \arg \left( \boldsymbol{\ell}\cdot\boldsymbol{A} \right)$, labeled by $\boldsymbol{\ell}=\left(\ell_1,\ell_2\right)$ and thus naturally divided into three types:
\vspace{5pt}
\begin{itemize}
\item Solely forward: $\boldsymbol{\ell}=(1,0)$ or $\boldsymbol{\ell}=(0,1)$;  
\item Mixed forward/backward: $\boldsymbol{\ell}=(1,-\ell_2)$ or $\boldsymbol{\ell}=(-\ell_1,1)$, with $\ell_1,\ell_2 > 0$;
\item Solely backwards: $\boldsymbol{\ell}= m\, (-\ell_1,-\ell_2)$ or $\boldsymbol{\ell}= m\, (-1,0)$ or $\boldsymbol{\ell}= m\, (0,-1)$, with $\ell_1,\ell_2,m > 0$,  and $\ell_1/\ell_2$ an irreducible fraction.
\end{itemize}
\vspace{5pt}

\textbf{Stokes discontinuity on direction $\theta_{\boldsymbol{\ell}}$:} $\disc_{\theta_{\boldsymbol{\ell}}}\Phi_n$ is given by a \textit{sum} over \textit{all}  paths, linking to nodes along the same direction, $\Phi_{\boldsymbol{n} + m\,\boldsymbol{\ell}}$ (with $m>0$). 
\vspace{5pt}

For all cases, each term in this sum ($\Phi_{\boldsymbol{n}} \rightarrow \Phi_{\boldsymbol{n} + m\, \boldsymbol{\ell}}$) can be decomposed in two factors:
\begin{itemize}
\item Functional component, dictated solely by beginning and end nodes:
\begin{equation*}
-\rme^{-m\, \frac{\boldsymbol{\ell} \cdot \boldsymbol{A}}{x}}\, \Phi_{\boldsymbol{n} + m\, \boldsymbol{\ell}}.
\end{equation*}
\item Statistical component, sum over all the allowed paths $\mathcal{P}(\boldsymbol{n} \rightarrow \boldsymbol{n} + m\, \boldsymbol{\ell})$ linking the nodes as in figure~\ref{fig:sec5-2d-lattice}:
\begin{equation*}
\mathrm{SF}_{(\boldsymbol{n} \rightarrow \boldsymbol{n} + m\, \boldsymbol{\ell})} \equiv \sum_{\mathcal{P}(\boldsymbol{n} \rightarrow \boldsymbol{n} + m\, \boldsymbol{\ell})} \mathrm{CF} \left( \mathcal{P} \right) w \left(\mathcal{P}\right).
\end{equation*}
\end{itemize}
\vspace{-5pt}
}}
\vspace{10pt}
\par\end{center}

Having arrived at a complete description of all Stokes discontinuities, at this stage we shall not further pursue the general analysis of asymptotics and large-order relations within the context of multi-parameter transseries. While most formulae may be obtained somewhat straightforwardly\footnote{This is in itself a very good and challenging exercise for the reader.} by following our general reasonings in section~\ref{sec:quartic}---now based upon the above results for the Stokes discontinuities---a particular example of a multi-parameter transseries (further displaying resonance) will be thoroughly studied in the upcoming section~\ref{sec:elliptic}, and addressing it should be enough. As such, let us pause for a moment in order to address \textit{Stokes phenomena}.

We have mentioned Stokes phenomenon in the one-dimensional setting, \eqref{stokespheno}, but we are still short of a proof for such statement (another simple proof, based on the bridge equations, may be found in appendix~\ref{app:alien-calculus}). This is actually very simple, in light of the combinatorics of the Borel residues. Consider the one-dimensional setting with transseries \eqref{Phitransseries} and \eqref{Phi(n)},
\be
\Phi (x,\sigma) = \sum_{n=0}^{+\infty} \sigma^n\, \rme^{-n \frac{A}{x}}\, \Phi_n (x),
\ee
\noindent
and forward ($\theta=0$) Stokes automorphism \eqref{one-param-stokes-0}, or, equivalently for our present purposes, forward Borel residues \eqref{stokesS-to-borelS+GEN}---which we shall now write binomially as
\be
\label{stokesS-to-borelS+GEN-BINOMIAL}
\mathsf{S}_{n\to n+\ell} = - \binom{n+\ell}{\ell}\, S_1^\ell.
\ee
\noindent
Instead of focusing solely on the action of the Stokes automorphism upon a transseries \textit{node}, as we have been doing up to now, we may ask what is the action of the Stokes automorphism on the \textit{full transseries}. The calculation is elementary, but illustrative:
\bea
\underline{\mathfrak{S}}_0 \Phi (x,\sigma) &=& \Phi (x,\sigma) - \sum_{n=0}^{+\infty} \sigma^n\, \rme^{-n \frac{A}{x}}\, \disc_0 \Phi_n (x) = - \sum_{n=0}^{+\infty} \sum_{\ell=0}^{+\infty} \mathsf{S}_{n\to n+\ell}\, \sigma^n\, \rme^{- \left( n+\ell \right) \frac{A}{x}}\, \Phi_{n+\ell} \\
&&
\hspace{-30pt}
= \sum_{n=0}^{+\infty} \sum_{\ell=0}^{+\infty} \binom{n+\ell}{\ell}\, S_1^\ell\, \sigma^n\, \rme^{- \left( n+\ell \right) \frac{A}{x}}\, \Phi_{n+\ell} = \sum_{m=0}^{+\infty} \left( \sigma+S_1 \right)^m \rme^{-m \frac{A}{x}}\, \Phi_m (x) = \Phi \left( x,\sigma+S_1 \right), \nonumber
\eea
\noindent
where the key step occurs at the beginning of the second line, when we make use of the binomial formulation \eqref{stokesS-to-borelS+GEN-BINOMIAL}. It should be clear that this trivially generalizes to the $k$-dimensional setting, when \textit{and only when} in the forward-axes case: this is essentially due to the fact that the Stokes vector only has \textit{one} component, $\boldsymbol{S}_{\boldsymbol{e}_i} = S_{\boldsymbol{e}_i}^{(i)}\, \boldsymbol{e}_i$, in which case \eqref{stokesS-to-borelS+GEN} and \eqref{kd-FA-stokesS-to-borelS+GEN} are basically the same and the calculation above follows unchanged. In such a \textit{purely forward} direction, $\boldsymbol{e}_i$, one thus finds that the action of the Stokes automorphism \eqref{eq:sec5-Stokes-forward-direction-theta} upon the full multi-parameter transseries \eqref{eq:sec5-transseries-vec} is given by
\be
\label{stokespheno-kd-FA}
\underline{\mathfrak{S}}_{\theta_i} u \left( x, \boldsymbol{\sigma} \right) = u \left( x, \boldsymbol{\sigma} + \boldsymbol{S}_{\boldsymbol{e}_i} \right),
\ee
\noindent
generalizing \eqref{stokespheno} for rather broad transseries classes.

On the other hand, in a mixed forward/backward direction, while the combinatorics remain similar the action of the Stokes automorphism will truncate; and in a solely backwards direction the combinatorics get much more involved while the action of the Stokes automorphism keeps truncating. This results in much more intricate formulae (and although we briefly mention them in appendix~\ref{app:alien-calculus}, we also refer the reader to \cite{as13} for some such expressions). In this way, an alternative and perhaps easier path to address Stokes automorphisms along these \textit{backward} directions is to first notice the following key fact concerning the above, purely \textit{forward} direction: the action of the Stokes automorphism \eqref{stokespheno-kd-FA} is a \textit{translation} along $\boldsymbol{e}_i$ in $\boldsymbol{\sigma}$-space. In other words, we can write
\be
\underline{\mathfrak{S}}_{\theta_i} u \left( x, \boldsymbol{\sigma} \right) = u \left( x, \boldsymbol{\sigma} + \boldsymbol{S}_{\boldsymbol{e}_i} \right) \equiv \exp \left( \boldsymbol{S}_{\boldsymbol{e}_i} \cdot \frac{\partial}{\partial\boldsymbol{\sigma}} \right) u \left( x, \boldsymbol{\sigma} \right).
\ee
\noindent
But, from \eqref{eq:sec5-Stokes-forward-direction-theta}, the Stokes automorphism in this case is
\be
\underline{\mathfrak{S}}_{\theta_i} = \exp \left( \rme^{-\frac{\boldsymbol{e}_i\cdot\boldsymbol{A}}{x}} \Delta_{\boldsymbol{e}_i\cdot\boldsymbol{A}} \right),
\ee
\noindent
which implies that the alien derivative has a $\boldsymbol{\sigma}$-space \textit{vector-field representation} given by
\be
\label{a-BRIDGE-glimpse}
\rme^{-\frac{\boldsymbol{e}_i\cdot\boldsymbol{A}}{x}} \Delta_{\boldsymbol{e}_i\cdot\boldsymbol{A}} = \boldsymbol{S}_{\boldsymbol{e}_i} \cdot \frac{\partial}{\partial\boldsymbol{\sigma}}.
\ee
\noindent
The reader should recall that this is not the first time such an expression appears; we have seen this very same feature back in section~\ref{sec:quartic}, equations \eqref{stokes-zero-sigmas} and \eqref{stokes-pi-sigmas}, or \eqref{eq:Stokes-aut-zero-v2} and \eqref{eq:Stokes-aut-pi-v3}.

This vector-field representation is, essentially, the famous \textit{bridge equation} we already alluded to several times before. In fact, such representation establishes a \textit{bridge} between alien derivations and ordinary calculus---albeit in $\boldsymbol{\sigma}$-space. Recall that, \textit{e.g.}, in the ODE context, the transseries parameters encode the boundary conditions needed to specify a particular solution, which means that $\boldsymbol{\sigma}$-space is essentially describing the moduli space of solutions to this given ODE. But the take-home message from this discussion is that such vector-field representations, or bridge equations, may allow for a more geometrical viewpoint on the analysis of Stokes phenomena along more complicated (\textit{i.e.}, non-elementary) directions \cite{as13}. This is undissociated from the algebraic structures encoded in the alien derivations, to which we need to turn next.

\subsection*{Algebras of Alien Derivatives and Virasoro-Like Algebras}

The alert reader may have already noticed two ``teasers'' earlier in the text, pointing towards what is to come now. These were, first, the \textit{non-commutativity} of alien derivatives along the \textit{same} (one-dimensional) direction in \eqref{virasoro-1d-teaser}, and, second, the \textit{commutativity} of alien derivatives along \textit{orthogonal} ($k$-dimensional) directions in \eqref{eq:sec5-commutator-orthogonal-forward}. In particular, equation \eqref{virasoro-1d-teaser}, which we rewrite in here as
\be
\label{1dALIENisVIR?}
\left[ \frac{\Delta_{-mA}}{S_{-m}}, \frac{\Delta_{-nA}}{S_{-n}} \right] = \left( m-n \right) \frac{\Delta_{-(m+n)A}}{S_{-(m+n)}}.
\ee
\noindent
is already making a very strong case for the appearance of the $c=0$ (classical) Virasoro algebra in the present context,
\be
\label{c=0-VIR}
\left[ L_m, L_n \right] = \left( m-n \right) L_{m+n}.
\ee
\noindent
The similarity is striking, but one should proceed with caution in order to understand exactly how much of the algebra of alien derivatives is equatable to the Virasoro algebra (and, thus, to the use of eventual benefits from its well-studied representation theory). One immediately notices that there cannot be an isomorphism as not all Virasoro generators $L_n$---where one has $n \in \BZ$---will have an alien derivative counterpart $\Delta_{nA}$---where now $n \leq 1$ with $n \neq 0$. It is precisely this lack of a $L_0$ counterpart which prevents the use of the Virasoro representation theory in the present context. In order to delve on this a bit further, let us first make a very short detour in order to summarize basic ideas of conformal transformations and the representation theory of the Virasoro algebra (but see also, \textit{e.g.}, \cite{fqs84, g91}, which we follow below).

Within the setting of conformal field theory (CFT), the aforementioned Virasoro operators $L_{n}$ with $n \in \BZ$ (and their complex conjugates $\bar{L}_{n}$), generate \textit{local} conformal transformations acting on two-dimensional conformal fields, with given conformal weight $(h,\bar{h})$ (in particular implying that these conformal fields will have associated scaling dimension $\Delta=h+\bar{h}$ and spin $s=h-\bar{h}$). These operators $L_{n}$ are generators of the (full) Virasoro algebra, with central charge $c$, given by
\be
\label{c-VIR}
\left[ L_{m}, L_{n} \right] = \left(m-n\right) L_{m+n} + \frac{c}{12} \left(m^{3}-m\right) \delta_{m+n}.
\ee
\noindent
However, due to a possibly non-vanishing central charge $c$, \textit{global} conformal transformations are only generated by the subset $\left\{ L_{-1}, L_{0}, L_{1} \right\}$ (alongside their complex conjugates). 

Constructing representations of the Virasoro algebra follows standard steps\footnote{As we shall see in the following, the representation theory of the Virasoro algebra is quite similar to the one of $\text{SU}(2)$, with $L_0$ playing the role of $J_3$ and with the several $L_{n>0}$ and $L_{n<0}$ playing the role of an infinite number of copies of the (single) $\text{SU}(2)$ pair $J_{\pm}$.}. It is based upon the existence of eigenstates $\left|\psi\right\rangle$ of $L_0$,
\be
L_0 \left|\psi\right\rangle = h_\psi \left|\psi\right\rangle,
\ee
\noindent
for which the $L_n$ may act as either creation or annihilation operators. In fact, because $\left[ L_0, L_{n} \right] = -n\, L_{n}$, operators $L_n$ with $n>0$ will act as \textit{lowering} or annihilation operators for $L_0$,
\be
L_0 \left( L_n \left|\psi\right\rangle \right) = \left( h_\psi-n \right) \left( L_n \left|\psi\right\rangle \right).
\ee
\noindent
Likewise, operators $L_n$ with $n<0$ will act as \textit{raising} or creation operators for $L_0$. In this setting, it is then rather natural to define a \textit{highest-weight state}, denoted by $\left|h\right\rangle$, as an eigenvector of the $L_0$ operator (with eigenvalue $h$) and which is further annihilated by \textit{all} the \textit{lowering} operators $L_{n>0}$, \textit{i.e.},
\be
L_0 \left|h\right\rangle = h \left|h\right\rangle, \qquad L_{n>0} \left|h\right\rangle = 0.
\ee
\noindent
If one is to act instead with the \textit{raising} operators $L_{n<0}$ on the highest-weight state $\left|h\right\rangle$, this then creates a tower of \textit{descendent states} associated to $\left|h\right\rangle$. These descendent states will also be eigenstates of $L_0$, and one may thus group them into different levels in such a way that the $N$th level is spanned by the states
\be
L_{-k_1} L_{-k_2} \cdots L_{-k_q} \left|h\right\rangle \quad \text{ with } \quad k_1 \ge \cdots \ge k_q > 0 \quad \text{ and } \quad \sum_{i=1}^q k_i = N,
\ee
\noindent
with corresponding $L_0$ eigenvalue $h+N$. The set of descendant states constructed from a highest-weight state $\left|h\right\rangle$ is known as a \textit{Verma module}, which summarizes a Virasoro representation.

The conclusion is thus quite swift: due to the lack of a $L_0$ counterpart (\textit{i.e.}, $\Delta_0$ does not exist), there is no sense of a highest-weight state in the alien derivative context, and one cannot make use of the Virasoro representation theory in this context. Nonetheless, there is one thing that might be learnt from the (partial) match between \eqref{1dALIENisVIR?} and \eqref{c=0-VIR}: this is yet another (and, actually, rather non-standard) glimpse at the origin of the \textit{bridge equations}; further supporting and extending our previous \eqref{a-BRIDGE-glimpse}. The (classical) Virasoro algebra \eqref{c=0-VIR} has rather well-know representations using vector fields (this algebra describes diffeomorphisms of the circle), and thus so will \eqref{1dALIENisVIR?}. One such representation using homogeneous vector fields (of \textit{degree} or \textit{weight} $m$, $n$, and $m+n$, respectively) is precisely given by
\be
\left[ \left( -\sigma^{m+1}\, \frac{\partial}{\partial\sigma} \right), \left( -\sigma^{n+1}\, \frac{\partial}{\partial\sigma} \right) \right] = \left( m-n \right) \left( -\sigma^{m+n+1}\, \frac{\partial}{\partial\sigma} \right).
\ee
\noindent
Upon comparison with \eqref{1dALIENisVIR?}, this immediately implies:
\be
\label{a-BRIDGE-glimpse-BACKWARDS}
\Delta_{-nA} \propto S_{-n}\, \sigma^{n+1}\, \frac{\partial}{\partial\sigma}.
\ee
\noindent
Notice how this yields a \textit{standard derivative} differential representation for the \textit{alien derivative}. The identification of the present $\sigma$-variable with the transseries parameter is done based upon \eqref{a-BRIDGE-glimpse}, which unfolded a bridge equation for \textit{forward} motions. The above result, \eqref{a-BRIDGE-glimpse-BACKWARDS}, now unfolds a completion of the bridge equation \eqref{a-BRIDGE-glimpse} also for \textit{backward} motions. This \textit{bridge} between standard and alien calculus is herein established via the Virasoro-like vector-field representation of the alien-derivative algebra \eqref{1dALIENisVIR?}, although we also refer the reader to appendix~\ref{app:alien-calculus} for a more standard (and complete!) derivation of this equation.

Having started a ``CFT description'' of alien chains, based upon the Virasoro-like algebraic structure found in \eqref{1dALIENisVIR?}, we may ask whether this description extends to alien lattices (recall our standard example in figure~\ref{fig:sec5-2d-lattice}). In this case, the multi-dimensional resurgence relations \eqref{eq:sec5-bridge-eqs-COMPLETE} lead us to first introduce the more general operators $G_{\boldsymbol{n}} (\boldsymbol{v})$, depending on an arbitrary vector field, $\boldsymbol{v}$, and defined as
\be
\label{eq:sec5-def-operators-G}
G_{\boldsymbol{n}} (\boldsymbol{v})\, \Phi_{\boldsymbol{k}} = \boldsymbol{v} \cdot  \left( \boldsymbol{k}+\boldsymbol{n} \right) \Phi_{\boldsymbol{k}+\boldsymbol{n}}.
\ee
\noindent
Of course that when the vector field $\boldsymbol{v}$ is chosen to be $\boldsymbol{v} = \boldsymbol{S}_{\boldsymbol{n}}$ we recover the alien derivative operators (at least when $\boldsymbol{n}$ obeys its required constraints in \eqref{eq:sec5-bridge-eqs-COMPLETE}). This implies that the algebraic structure of the alien derivatives, acting on a multi-dimensional alien lattice, essentially sits inside the general algebraic structure of the above $G_{\boldsymbol{n}} (\boldsymbol{v})$ operators. This general structure is itself given by\footnote{Finding this result is a straightforward exercise which we leave for the reader. Just use \eqref{eq:sec5-def-operators-G} twice to compute the consecutive action of two $G_{\boldsymbol{n}}$-operators on a fixed sector,
\be
G_{\boldsymbol{n}} (\boldsymbol{v})\, G_{\boldsymbol{m}} (\boldsymbol{u})\, \Phi_{\boldsymbol{k}} = \boldsymbol{u} \cdot \left( \boldsymbol{k} + \boldsymbol{m} \right)\, \boldsymbol{v} \cdot \left( \boldsymbol{k} + \boldsymbol{m} + \boldsymbol{n} \right) \Phi_{\boldsymbol{k}+\boldsymbol{m}+\boldsymbol{n}},
\ee
\noindent
and then compute the commutator by expressing it back as the action of \eqref{eq:sec5-def-operators-G} on the chosen fixed sector.} the commutation relations\footnote{These may also be written component-wise, as
\be
\label{multi-dim-alien-chain-ALG-COMPONENTS}
\left[ G_{\boldsymbol{n}}^i, G_{\boldsymbol{m}}^j \right] = \left( m^i\, \delta^j_{\ell} - n^j\, \delta^i_{\ell} \right) G_{\boldsymbol{n}+\boldsymbol{m}}^{\ell}.
\ee
}
\be
\label{multi-dim-alien-chain-ALG}
\left[ G_{\boldsymbol{n}} (\boldsymbol{v}), G_{\boldsymbol{m}} (\boldsymbol{u}) \right] = G_{\boldsymbol{n}+\boldsymbol{m}} \left( \left( \boldsymbol{v} \cdot \boldsymbol{m} \right) \boldsymbol{u} - \left( \boldsymbol{u} \cdot \boldsymbol{n} \right) \boldsymbol{v} \right).
\ee
\noindent
The algebra of alien-lattice alien-derivatives then follows as the \textit{subalgebra} of \eqref{multi-dim-alien-chain-ALG} where we constrain ourselves to operators of the form $G_{\boldsymbol{\ell}} (\boldsymbol{S}_{\boldsymbol{\ell}}) \equiv \Delta_{\boldsymbol{\ell}\cdot\boldsymbol{A}}$. In this case, one finds:
\be
\label{eq:sec5-commutator-general-op-with stokes}
\left[ G_{\boldsymbol{n}} (\boldsymbol{S}_{\boldsymbol{n}}), G_{\boldsymbol{m}} (\boldsymbol{S}_{\boldsymbol{m}}) \right] = G_{\boldsymbol{n}+\boldsymbol{m}} \left( \left( \boldsymbol{S}_{\boldsymbol{n}} \cdot \boldsymbol{m} \right) \boldsymbol{S}_{\boldsymbol{m}} - \left( \boldsymbol{S}_{\boldsymbol{m}} \cdot \boldsymbol{n} \right) \boldsymbol{S}_{\boldsymbol{n}} \right).
\ee
\noindent
Two facts are immediately noticeable concerning this final result. The first is that this  multi-dimensional alien-lattice algebraic structure is a multi-dimensional \textit{generalization} of the Virasoro algebra (albeit, as we shall see in the following, the standard Virasoro algebra \eqref{c-VIR} also sits inside this structure). The second, more problematic issue, is that this multi-dimensional alien-lattice algebraic structure is generically \textit{open}, \textit{i.e.}, the commutator of two alien derivatives is \textit{not} an alien derivative. This is unlike the previous (more general) algebraic structure \eqref{multi-dim-alien-chain-ALG} (and, in fact, it is the reason why we started-off with that structure in the first place; because it is algebraically \textit{closed}). Closure of the alien-lattice algebraic structure may nonetheless be achieved when one has
\be
\label{closure-condition-stokes}
\left( \boldsymbol{S}_{\boldsymbol{n}} \cdot \boldsymbol{m} \right) \boldsymbol{S}_{\boldsymbol{m}} - \left( \boldsymbol{S}_{\boldsymbol{m}} \cdot \boldsymbol{n} \right) \boldsymbol{S}_{\boldsymbol{n}} \propto \boldsymbol{S}_{\boldsymbol{n}+\boldsymbol{m}}.
\ee
\noindent
This is a non-trivial requirement on Stokes vectors, and it would be interesting to classify what types of linear or nonlinear problems satisfy this constraint. In the following, however, we shall limit ourselves to a couple of simple exercises where the alien-lattice algebra does close, and where we may further rediscover Virasoro-like structures as in \eqref{1dALIENisVIR?}.

One simple exercise to return to is that of the commutator of two, \textit{orthogonal}, forward motions, as in \eqref{eq:sec5-commutator-orthogonal-forward}. Choosing $\boldsymbol{n} = \boldsymbol{e}_{n}$ and $\boldsymbol{m} = \boldsymbol{e}_{m}$, with $n \neq m$, we already know that this implies corresponding Stokes vectors $\boldsymbol{S}_{\boldsymbol{e}_n} = S_{\boldsymbol{e}_n}^{(n)}\, \boldsymbol{e}_n$ and $\boldsymbol{S}_{\boldsymbol{e}_m} = S_{\boldsymbol{e}_m}^{(m)}\, \boldsymbol{e}_m$. Then it immediately follows that $\left( \boldsymbol{S}_{\boldsymbol{e}_n} \cdot \boldsymbol{e}_{m} \right) \boldsymbol{S}_{\boldsymbol{e}_m} - \left( \boldsymbol{S}_{\boldsymbol{e}_m} \cdot \boldsymbol{e}_{n} \right) \boldsymbol{S}_{\boldsymbol{e}_n} = \boldsymbol{0}$; in which case, with $G_{\boldsymbol{\ell}} (\boldsymbol{0}) = \boldsymbol{0}$ due to linearity in the vectorial argument, one finds the expected result, \textit{i.e.}, \eqref{eq:sec5-commutator-general-op-with stokes} translates to
\be
\left[ G_{\boldsymbol{e}_n} (\boldsymbol{S}_{\boldsymbol{e}_n}), G_{\boldsymbol{e}_m} (\boldsymbol{S}_{\boldsymbol{e}_m}) \right] = G_{\boldsymbol{e}_n+\boldsymbol{e}_m} (\boldsymbol{0}) = 0.
\ee
\noindent
Without surprise, this recovers the commutation relation we had already found.

The other simple exercise to discuss is one of the cleanest scenarios where \eqref{closure-condition-stokes} is indeed verified and the alien-lattice algebra \eqref{eq:sec5-commutator-general-op-with stokes} \textit{closes}: this is the case where \textit{all} Stokes vectors are \textit{parallel}. Let us pick some direction, defined by the versor $\boldsymbol{u}$, and let us suppose that all Stokes vectors, $\boldsymbol{S}_{\boldsymbol{n}}$, $\boldsymbol{S}_{\boldsymbol{m}}$, and $\boldsymbol{S}_{\boldsymbol{m}+\boldsymbol{n}}$, are parallel, \textit{i.e.},
\be
\label{parallel-stokes-vectors}
\boldsymbol{S}_{\boldsymbol{n}} \parallel \boldsymbol{S}_{\boldsymbol{m}} \parallel \boldsymbol{S}_{\boldsymbol{m}+\boldsymbol{n}} \qquad \Rightarrow \qquad
\begin{cases}
\,\, \boldsymbol{S}_{\boldsymbol{n}} = S_{\boldsymbol{n}}\, \boldsymbol{u}, \\
\,\, \boldsymbol{S}_{\boldsymbol{m}} = S_{\boldsymbol{m}}\, \boldsymbol{u}, \\
\,\, \boldsymbol{S}_{\boldsymbol{m}+\boldsymbol{n}} = S_{\boldsymbol{m}+\boldsymbol{n}}\, \boldsymbol{u}.
\end{cases}
\ee
\noindent
This immediately implies that, in this case,
\be
\left( \boldsymbol{S}_{\boldsymbol{n}} \cdot \boldsymbol{m} \right) \boldsymbol{S}_{\boldsymbol{m}} - \left( \boldsymbol{S}_{\boldsymbol{m}} \cdot \boldsymbol{n} \right) \boldsymbol{S}_{\boldsymbol{n}} = \boldsymbol{u} \cdot \left( \boldsymbol{m} - \boldsymbol{n} \right)\, S_{\boldsymbol{n}} S_{\boldsymbol{m}}\, \boldsymbol{u},
\ee
\noindent
and thus the alien-lattice algebra \textit{closes} as
\be
\left[ \frac{G_{\boldsymbol{n}} (\boldsymbol{S}_{\boldsymbol{n}})}{S_{\boldsymbol{n}}}, \frac{G_{\boldsymbol{m}} (\boldsymbol{S}_{\boldsymbol{m}})}{S_{\boldsymbol{m}}} \right] = \boldsymbol{u} \cdot \left( \boldsymbol{m} - \boldsymbol{n} \right) \frac{G_{\boldsymbol{m}+\boldsymbol{n}} (\boldsymbol{S}_{\boldsymbol{m}+\boldsymbol{n}})}{S_{\boldsymbol{m}+\boldsymbol{n}}}.
\ee
\noindent
This is essentially the same result as in \eqref{1dALIENisVIR?}. In the present multi-dimensional case one thus finds underlying Virasoro-like algebraic structures along directions of \textit{parallel} Stokes vectors.

In short, the algebraic structure of multi-dimensional alien derivatives clearly extends beyond the realm of the Virasoro algebra. It is properly defined as the more general algebraic structure \eqref{multi-dim-alien-chain-ALG}, as the particular alien-derivative subset \eqref{eq:sec5-commutator-general-op-with stokes} does not necessarily form a subalgebra (\textit{i.e.}, does not necessarily close). In cases where it does close, such as the above \eqref{parallel-stokes-vectors}, then we see how Virasoro-like structures reappear along special directions in the space of Stokes vectors.

Let us finish this section with a multi-dimensional generalization of the rather non-standard (albeit partial) derivation of the \textit{bridge equations}, which we discussed earlier in the one-dimensional case \eqref{a-BRIDGE-glimpse-BACKWARDS} (as mentioned many times before, the standard derivation of the bridge equations is done in appendix~\ref{app:alien-calculus}). A differential representation of \eqref{multi-dim-alien-chain-ALG}, with homogeneous vector fields, may be achieved via (with no implicit summations)
\be
\left[ \left( - \boldsymbol{\sigma}^{-\boldsymbol{n}}\, \sigma_i\, \frac{\partial}{\partial\sigma_i} \right), \left( - \boldsymbol{\sigma}^{-\boldsymbol{m}}\, \sigma_j\, \frac{\partial}{\partial\sigma_j} \right) \right] = \left( m^i\, \delta_j^{\ell} - n^j\, \delta_i^{\ell} \right) \left( -\boldsymbol{\sigma}^{-\boldsymbol{n}-\boldsymbol{m}}\, \sigma_{\ell}\, \frac{\partial}{\partial\sigma_{\ell}} \right),
\ee
\noindent
where under $\sigma_i \to \lambda \sigma_i$ the \textit{degree} or \textit{weight} of each vector field is
\be
\boldsymbol{\sigma}^{\boldsymbol{n}}\, \sigma_j\, \frac{\partial}{\partial\sigma_j} \to \lambda^{n_i}\, \boldsymbol{\sigma}^{\boldsymbol{n}}\, \sigma_j\, \frac{\partial}{\partial\sigma_j}.
\ee
\noindent
Upon comparison with \eqref{multi-dim-alien-chain-ALG-COMPONENTS} and \eqref{eq:sec5-commutator-general-op-with stokes}, this immediately implies:
\be
\Delta_{\boldsymbol{\ell}\cdot\boldsymbol{A}} \propto \boldsymbol{\sigma}^{-\boldsymbol{\ell}}\, \sum_{i=1}^{k} S_{\boldsymbol{\ell}}^{(i)}  \sigma_i\, \frac{\partial}{\partial\sigma_i}.
\ee
\noindent
Again, this yields a standard-derivative differential representation for the alien derivative, with the expected identification of the present $\boldsymbol{\sigma}$-variables with the transseries parameters following from \eqref{a-BRIDGE-glimpse}. This equation gives further support to the existence of a multi-dimensional bridge equation, also building upon \eqref{a-BRIDGE-glimpse} and \eqref{a-BRIDGE-glimpse-BACKWARDS}, and appendix~\ref{app:alien-calculus} should make a good departure point into these technical issues for the interested reader.

\section{Resurgent Analysis of an Elliptic-Potential Integral}\label{sec:elliptic}

Having arrived this far, the reader has by now acquired a fairly complete knowledge on the main ideas of resurgence, transseries, and their asymptotics. There is, however, one final topic we still need to address in order to complete the basic toolset: this is the phenomenon of \textit{resonance}. Resonance is a new feature which only occurs within the realm of multi-parameter transseries, and which we shall introduce in the present section as it appears in a simple example (much along the spirit of what was done back in section~\ref{sec:quartic}, when introducing generalities on resurgence and transseries via the quartic potential). Note how on the road to resonance we will necessarily drive by familiar locations (\textit{e.g.}, the asymptotics of non-resonant resurgent-transseries roughly occupies the first half of the present section). When doing so we shall be less detailed than before (leaving the filling of the details as exercises), but obviously will still keep our higher pedagogical standards from previous sections as soon as any discussions involving resonance begin.

\begin{figure}[t!]
\begin{center}
\includegraphics[width=9cm]{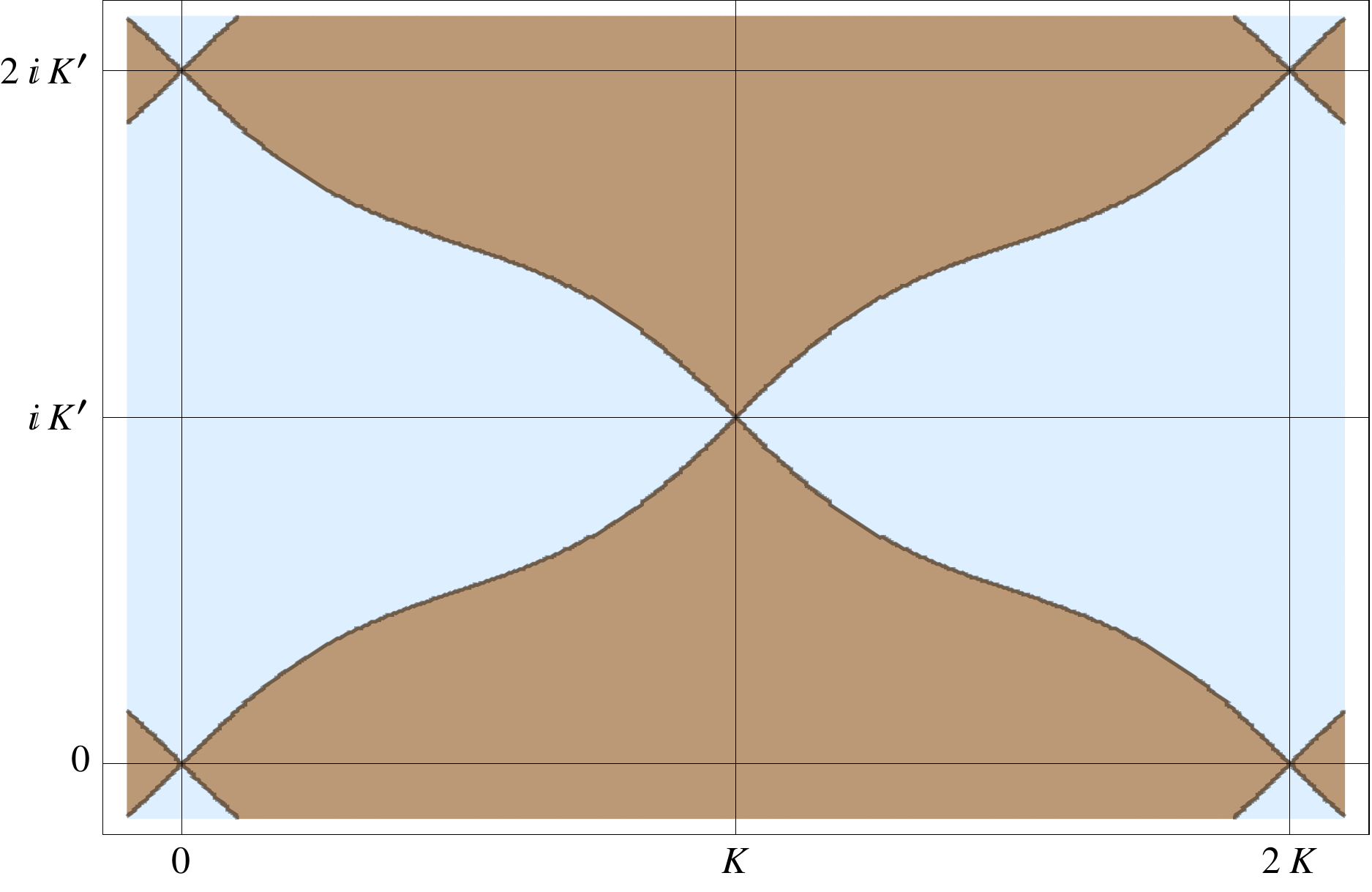}
\end{center}
\caption{The elliptic potential in \eqref{elliptic_Z} in the complex $z$-plane, with $\hbar \in \BR^+$ and $m=0.9$ in the plot. The brown (darker) region is where $\re\, V(z) > 0$, while the blue (lighter) region is where $\re\, V(z) < 0$.
}
\label{ellipticpotentialfig}
\end{figure}

In parallel with what was done in section~\ref{sec:quartic}, let us consider a partition-function toy-model described by a one-dimensional integral, as in \eqref{quarticintegral}, but where the potential is now given by an elliptic function. This ``elliptic partition function'' is given by
\be
\label{elliptic_Z}
Z (\hbar | m) = \frac{1}{\sqrt{\pi\hbar}} \int_\Gamma \rmd z\, \exp \left( - \frac{1}{\hbar}\, \mathrm{sd}^2 (z | m) \right),
\ee
\noindent
where $\mathrm{sd} (z | m)$ is a subsidiary Jacobian elliptic function with (squared) modulus $m$ (see, \textit{e.g.}, \cite{olbc10} for further details). The potential, $\mathrm{sd}^2 (z | m)$, is then a doubly-periodic complex function, along both real and imaginary axes with respective periods\footnote{In here, $\BK(m)$ denotes the complete elliptic-integral of the first kind, $\BK (m) = \int_{0}^{\frac{\pi}{2}} \frac{\rmd\theta}{\sqrt{1 - m \sin^2 \theta}}$.} $2\BK (m)$ and $2\rmi\, \BK^\prime (m) \equiv 2\rmi\, \BK (1-m)$. In particular, this elliptic potential has a zero at the origin and a double pole at $z=\BK+\rmi\, \BK^\prime$. Due to the double periodicity the potential is naturally defined over a torus, which we may usually take as the fundamental parallelogram $\left\{ 0, 2\BK, 2\rmi\, \BK^\prime, 2 \left( \BK+\rmi\, \BK^\prime \right) \right\}$ in $\BC$, with the appropriate congruent identifications (see figure~\ref{ellipticpotentialfig}). The (squared) modulus $m\in[0,1]$ sets the ratio of the orthogonal periods of this torus. In particular, at the limiting value $m=0$ the imaginary period diverges, whereas at $m=1$ it is the real period which diverges. For (standard) notational simplicity, we also define $m^\prime \equiv 1-m$. This example, \eqref{elliptic_Z}, was originally introduced in \cite{bdu13} as a toy model to illustrate the appearance of complex saddles leading to negative-real instanton actions within a simple\footnote{Such negative-real instanton actions first appeared within the harder Painlev\'e realm \cite{gikm10}; see also \cite{asv11, sv13}.} setting. The choice of such an elliptic potential also had a very natural physical justification, being the sole Jacobian-elliptic combination interpolating between the familiar sine-Gordon ($\sin^2 (z)$ as $m \to 0$) and sinh-Gordon ($\sinh^2 (z)$ as $m \to 1$) potentials. Herein, we will extend the large-order analysis of \cite{bdu13} into a complete resurgent analysis and further use this elliptic potential as a toy model to introduce the phenomenon of \textit{resonance}.

In order to specify the contours of integration, $\Gamma$, and understand what are their admissible choices on the present \textit{compact} domain (this is in contrast with what happened in section~\ref{sec:quartic}), let us begin with na\"\i ve steepest-descent analysis. There are three critical points of the potential
\be
\label{eq:ellipticsd2potential}
V(z) = \mathrm{sd}^2 ( z | m ),
\ee
\noindent
given by the zeroes of the $\mathrm{sd} (z)$, $\mathrm{cd} (z)$, and $\mathrm{nd} (z)$ subsidiary Jacobian elliptic functions. They are located at, respectively,
\be
z_0^* = 0, \qquad z_1^* = \BK, \quad \text{ and } \quad z_2^* = \rmi\, \BK^\prime.
\ee
\noindent
These critical points lead to $V(z^*_0) = 0$ and to the (real) instanton actions
\be
\label{eq:elliptic_critical}
V(z^*_1) = \frac{1}{m^\prime} \equiv A_1 \qquad \text{and} \qquad V(z^*_2) = - \frac{1}{m} \equiv A_2.
\ee
\noindent
In this way, the critical point at the origin, $z^*_0$, corresponds to the standard perturbative saddle; the critical point which sits on the positive \textit{real} axis, $z^*_1$, leads to a \textit{positive}-real instanton action; and the critical point which sits on the positive \textit{imaginary} axis, $z^*_2$, finally leads to a \textit{negative}-real instanton action (at least when $\hbar \in \BR^+$). Further, due to their dependence upon the (squared) modulus $m$, both these instanton actions are \textit{tunable} (\textit{i.e.}, they vary with $m$). We will later see how specific choices of $m$ will either display, or not, a resonant structure.

At this stage we may already guess the form of the partition-function transseries as (compare with \eqref{quartictransseries})
\bea
\CZ (\hbar, \sigma_{0}, \sigma_{1}, \sigma_{2}) &=&  \sigma_{0}\, \Phi_0 (\hbar) + \sigma_1\, \rme^{-\frac{A_1}{\hbar}}\, \Phi_1 (\hbar) + \sigma_2\, \rme^{-\frac{A_2}{\hbar}}\, \Phi_2 (\hbar) \\ 
&=& \sigma_0\, \Phi_0 (\hbar) + \sigma_1\, \rme^{-\frac{1}{\hbar m^\prime}}\, \Phi_1 (\hbar) + \sigma_2\, \rme^{+\frac{1}{\hbar m}}\, \Phi_2 (\hbar).
\label{ell_lin_transseries}
\eea
\noindent
The $\Phi_i (\hbar)$ are formal asymptotic series in $\hbar$ and the $\sigma_i$ are the transseries parameters. Recalling the discussion in section~\ref{sec:quartic}, it is natural to assume that each of these asymptotic series is associated to specific---and distinct---steepest-descent contours. In order to understand which are these admissible choices for $\Gamma$ in \eqref{elliptic_Z}, we will next follow the ideas introduced in subsection~\ref{subsec:4intphases}, and formalised in section~\ref{sec:lefschetz} in terms of Picard--Lefschetz theory, in order to construct the steepest-descent contours, or Lefschetz thimbles, associated to each saddle. This will further provide a geometrical perspective on the Stokes phenomenon.

\subsection{Phase Diagram from Stokes Phenomena}\label{sec:elliptic_thimbles}

\begin{figure}[t!]
\begin{center}
\includegraphics[width=3cm]{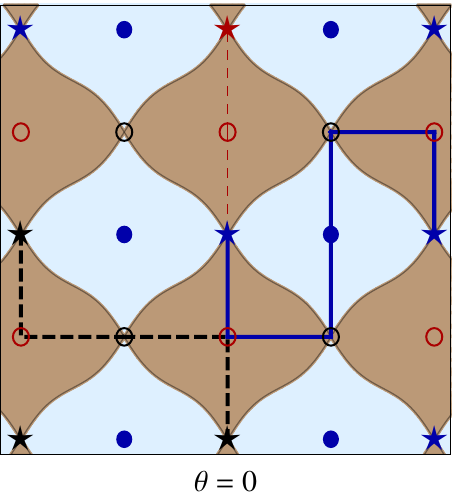}
$\qquad$
\includegraphics[width=3cm]{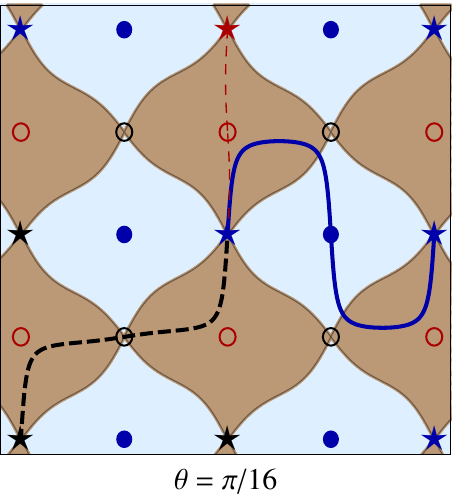}
$\qquad$
\includegraphics[width=3cm]{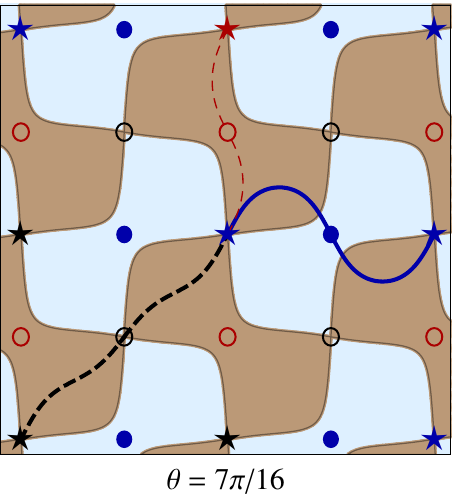}
$\qquad$
\includegraphics[width=3cm]{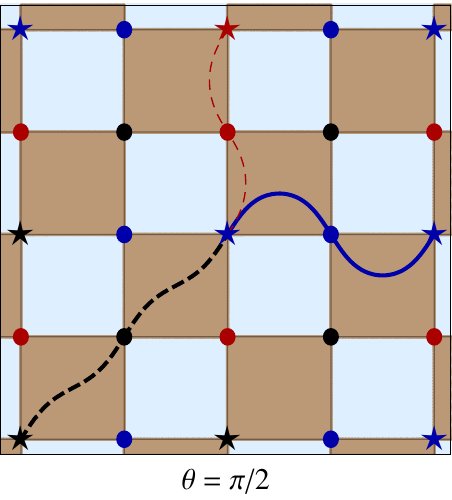}
\\
\includegraphics[width=3cm]{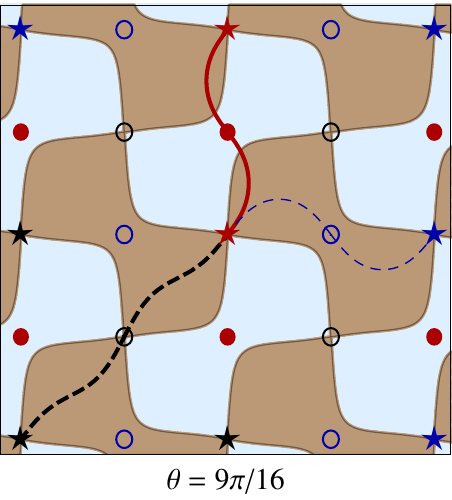}
$\qquad$
\includegraphics[width=3cm]{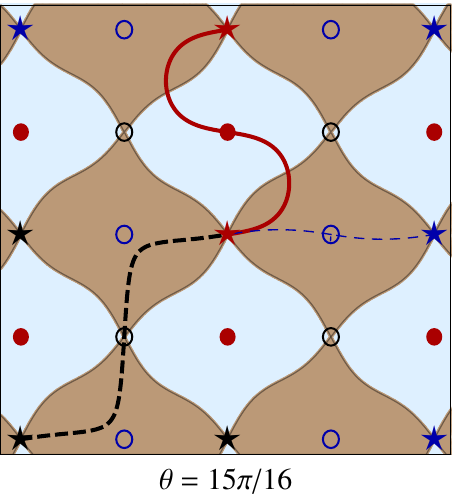}
$\qquad$
\includegraphics[width=3cm]{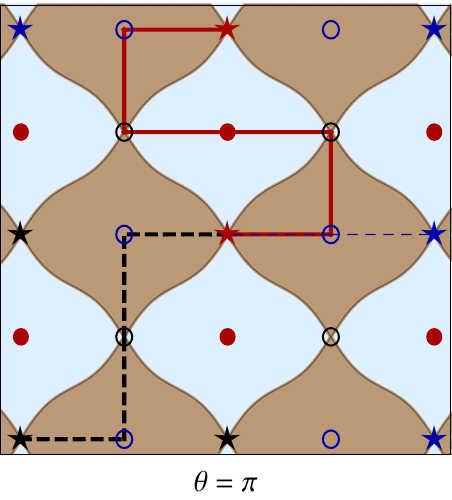}
$\qquad$
\includegraphics[width=3cm]{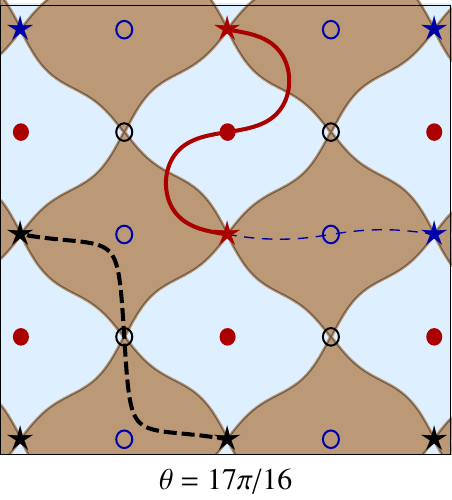}
\\
\includegraphics[width=3cm]{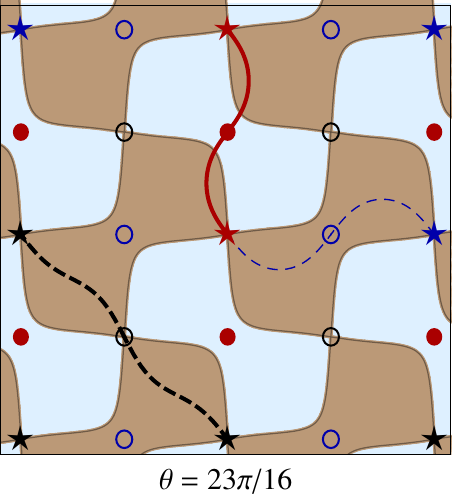}
$\qquad$
\includegraphics[width=3cm]{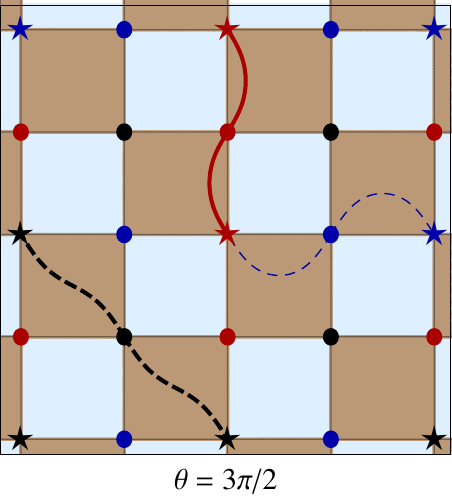}
$\qquad$
\includegraphics[width=3cm]{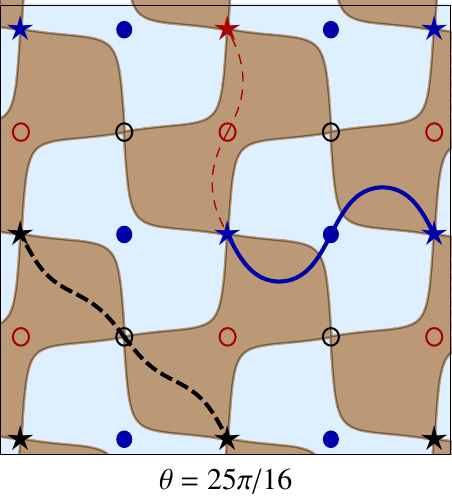}
$\qquad$
\includegraphics[width=3cm]{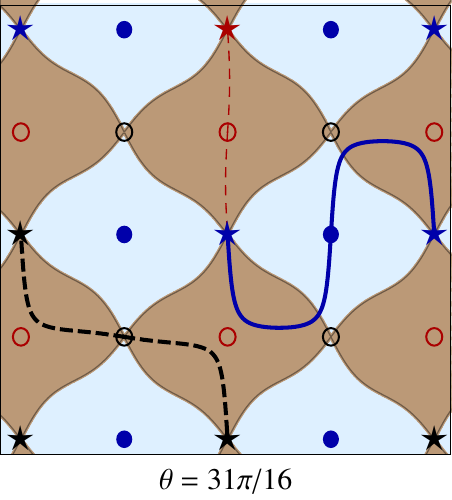}
\end{center}
\caption{Steepest-descent contours for the elliptic integral, through the several saddles, along with positivity of the potential, for different values of $\theta = \arg x$ ($|x|=1$, $m=0.9$ in the plot). The saddles and their respective thimbles are depicted as follows: $z^*_0$ in black, $z^*_1$ in red, and $z^*_2$ in blue. The leading saddles are plotted as solid lines/filled circles, while the subleading saddles are plotted as empty circles/dashed lines (the ``lighter'' dashed-lines being the most suppressed). The paths start and end at the double-poles, represented by filled stars. Stokes lines are at $\arg x = 0, \pi$ when the steepest-descent contours hit more than one saddle; anti-Stokes lines are at $\arg x = \frac{\pi}{2}, \frac{3\pi}{2}$. See the main text for a full discussion.
}
\label{steepest-elliptic}
\end{figure}

The Picard--Lefschetz theory outlined in section~\ref{sec:lefschetz} was built upon a polynomial potential, $V(z)$, which made it easy to identify the set of (steepest-descent) integration cycles over which the integral \eqref{eq:nd_integral} was convergent. In fact, therein, convergence was simply achieved by requiring $\re \left( - \frac{1}{\hbar} V(z) \right) \rightarrow -\infty$ as $|z| \rightarrow +\infty$ along some appropriate region within $\BC$. Identifying the set of integration cycles over which the integral \eqref{elliptic_Z} is convergent is now necessarily different as the potential is no longer polynomial (it is in fact meromorphic) neither is its domain unbounded (the torus $\BT^2$ is compact). Happily, however, the potential \eqref{eq:ellipticsd2potential} has a single (double) pole on $\BT^2$ (located at $z=\BK+\rmi\, \BK^\prime$ in the fundamental parallelogram), in which case it turns out that we can set-up Picard--Lefschetz theory in the present setting in a completely analogous fashion to what we did earlier in section~\ref{sec:lefschetz}. All one has to note is that the Lefschetz thimbles are now curves (cycles) which originate and end at this double-pole, wrapping around the torus and passing through the relevant critical point(s). In this particular example things are even simpler as the thimbles amount to oriented paths of steepest descent, just like in \eqref{steepestdescentcontour-def}; recall it:
\be
\label{steepestdescentcontour-def-elliptic}
\im \left( V(z) - V(z^*) \right) = 0.
\ee
\noindent
For the perturbative saddle at the origin, $z^*_0$, and for both positive-real and negative-real instanton saddles, $z^*_1$ and $z^*_2$, steepest-descent contours are depicted in figure~\ref{steepest-elliptic} for different values of $\theta = \arg \hbar$ (in the following we use $x \equiv \hbar$ in order to make it easier to compare present formulae with the ones in section~\ref{sec:quartic}). In this figure, we have also plotted the regions where the elliptic potential is positive (negative): comparing with figure~\ref{ellipticpotentialfig}, we now see how these regions change with $\theta$. Let us explain the plotting details of figure~\ref{steepest-elliptic}: for each value of $\theta$ we have plotted the leading saddle(s) with a solid disk, while the subleading saddle(s) were plotted with a circle. Regardless of dominance, the perturbative saddle is always plotted in black, with the positive-real instanton saddle in red and the negative-real instanton saddle in blue. In this way, solid contours are steepest-descent contours associated with the leading saddle(s), while dashed contours are steepest-descent contours associated with the subleading saddle(s). All steepest-descent contours originate and end at the double-poles, which are plotted with solid stars. Note how the plots extend over a couple of fundamental parallelograms, but this is for drawing clarity (\textit{i.e.}, so that the three steepest-descent contours may be plotted without falling on top of each other). As such, the period lattice must still be modded-out in order to obtain the corresponding final picture on $\BT^2$. Finally, as expected, generically a steepest-descent contour goes through a single saddle---but this is not always the case due to Stokes phenomenon.

In the above set-up, the analysis carried out in section~\ref{sec:lefschetz} holds. In such Picard--Lefschetz framework, let us first analyze the structure of steepest-descent contours for the elliptic partition-function \eqref{elliptic_Z} as the phase of $\hbar$ changes (subsequently denoted by $x$). Associated to the three critical points $z^*_i$ there will be three corresponding thimbles $\CJ_i$. For generic values of $\theta=\arg x$ all exponential arguments $\sim A/x$ from \eqref{eq:elliptic_critical} have different imaginary parts, in which case there is a well-defined thimble decomposition of the integration contour as
\be
\label{thimbledecompositionelliptic}
\Gamma = \sum_{i=0}^{2} n_i\, \CJ_i.
\ee
\noindent
However, when $\arg x = 0, \pi$, the imaginary parts of all the critical values coincide---in fact they all vanish. These values of $\theta$ describe the \textit{Stokes lines}, and the intersection numbers $n_i$ encounter monodromies as we cross them. These results are also summarised in figure~\ref{steepest-elliptic}, for different values of $\theta \in [0,2\pi)$ (where the Stokes lines are very distinctive, with different thimbles crossing more than one saddle). In-between this interchange of exponential dominance, one also identifies the corresponding \textit{anti-Stokes lines}, at $\theta = \frac{\pi}{2}, \frac{3\pi}{2}$; the directions where all saddles are of the same magnitude. These are the turning points where saddles which used to be leading become subleading, and vice-versa. Let us next discuss the Stokes jumps, at both Stokes lines $\theta=0, \pi$. 

\subsubsection*{Monodromy and Stokes Phenomenon at $\theta=0$}

Comparing the last and second plots in figure~\ref{steepest-elliptic} it is clear how some thimbles are jumping at the $\theta=0$ Stokes line. This may also be simply checked by evaluating \eqref{steepestdescentcontour-def-elliptic} at either $\theta=0^-$ or $\theta=0^+$. What is happening is that the asymptotic series in \eqref{ell_lin_transseries}, obtained by integrations along the corresponding cycles, must remain continuous across Stokes rays. As we already discussed back in section~\ref{sec:quartic}, this can only be guaranteed as long as each thimble has the same asymptotics for $\theta=0^-$ and $\theta=0^+$. This \textit{continuity condition} then implies that, as one moves counterclockwise from $\theta=0^-$ to $\theta=0^+$, the cycles $\CJ_i$ must acquire the monodromy (``jumps'')
\begin{equation}
\label{ell-monodromy-0}
\left(
\begin{array}{ccc}
\CJ_2 \\
\CJ_0 \\
\CJ_1
\end{array}
\right) \to S_{\circlearrowleft}^{(0)} \left(
\begin{array}{ccc}
\CJ_2 \\
\CJ_0 \\
\CJ_1
\end{array}
\right) = \left(
\begin{array}{ccc}
1 & -2 & -2 \\
0 &  1 &  2 \\
0 &  0 &  1
\end{array}
\right) \left(
\begin{array}{ccc}
\CJ_2 \\
\CJ_0 \\
\CJ_1
\end{array}
\right),
\end{equation}
\noindent
where $S_{\circlearrowleft}^{(0)}$ is the ``Stokes matrix'' associated to $\theta=0$.

\begin{figure}[t!]
\begin{center}
\includegraphics[width=0.3\textwidth]{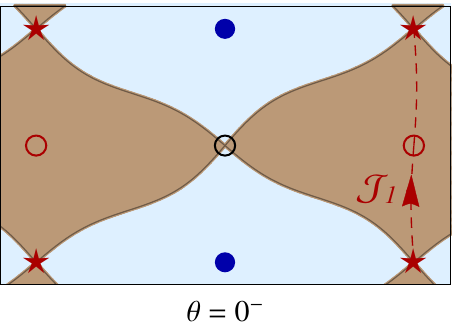}
$\qquad$
$\qquad$
\includegraphics[width=0.3\textwidth]{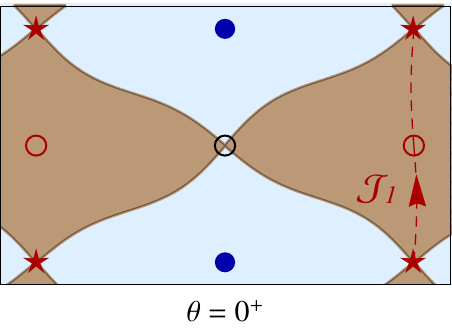}
\\
\includegraphics[width=0.3\textwidth]{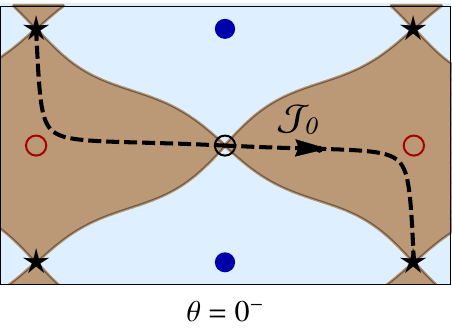}
$\qquad$
$\qquad$
\includegraphics[width=0.3\textwidth]{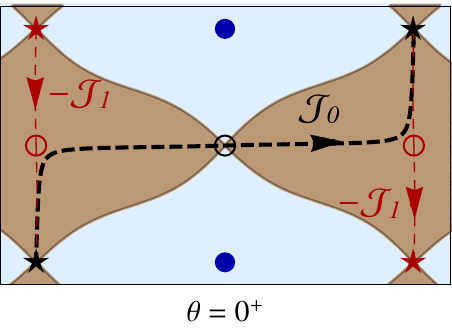}
\\
\includegraphics[width=0.3\textwidth]{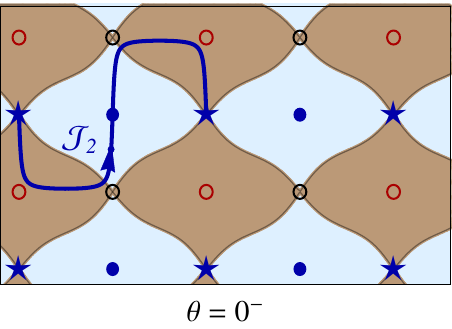}
$\qquad$
$\qquad$
\includegraphics[width=0.3\textwidth]{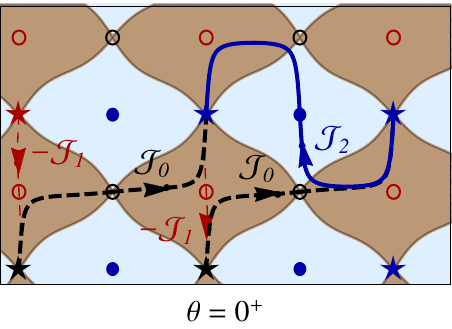}
\end{center}
\caption{Stokes phenomenon around $\theta=0$, from $\theta=0^-$ to $\theta=0^+$, and where $- A_2/x > 0 > - A_1/x $. At this particular angle, nothing really happens with the $\CJ_1$ thimble (top plots). On the other hand, as explained in the main text, both thimbles $\CJ_0$ and $\CJ_2$ do acquire discontinuities which occur at their tails. In order to preserve the continuity of the final integration contours, these discontinuities have to be balanced by addition of other thimbles, as pictured (middle and bottom plots, respectively).
}
\label{fig:thimbles-1}
\end{figure}

In more geometric terms, this non-trivial monodromy may also be understood as a result of the \textit{continuity} of the \textit{integration cycle} itself, $\Gamma$, across $\theta=0$. In other words, the thimble decomposition \eqref{thimbledecompositionelliptic} of a given cycle $\Gamma$ is \textit{different} for $\theta=0^{-}$ and $\theta=0^{+}$, and thus the \textit{same} cycle $\Gamma$ is expressed as \textit{different} linear combination of thimbles for $\theta=0^{\pm}$ (recall \eqref{eq:decomposition2}). Let us elucidate this point starting with the particular ``perturbative'' cycle $\Gamma=\CJ_0$, illustrated in the two middle plots of figure~\ref{fig:thimbles-1} (and start with the middle-left plot). As $\theta$ changes from $0^-$ to $0^+$, the tails of the cycle $\CJ_0$ (\textit{i.e.}, the neighbourhood of $\CJ_0$ around the pole) flip. This flip is necessary in order to keep the orientation of $\CJ_0$ around the saddle point, which itself is required to preserve continuity of integration along such contour. Now, at the level of the integration cycle $\Gamma$, continuity is kept as long as one compensates for this flip of $\CJ_0$ at $\theta=0^+$ by adding $-2 \CJ_1$ to $\CJ_0 (0^+)$, as illustrated in the middle-right plot of figure~\ref{fig:thimbles-1} (note how $\Gamma = \CJ_1$ is unchanged; see top plots of figure~\ref{fig:thimbles-1}). In fact, without this addition there would be an exponentially suppressed discontinuity (of order $\sim \rme^{-A_1/x}$; see \eqref{stokesPhi0toPhi0-2Phi-elliptic} below) in the integral over $\Gamma$. In short, the monodromy arises as a result of demanding continuity of the integration cycle $\Gamma$ across a Stokes line---which also amounts to a description of the Stokes phenomenon. Indeed, from the identification between the transseries and cycle decomposition, equations  \eqref{eq:decomposition2} and \eqref{eq:thimble_integral}, we can conclude that
\be
\int_{\CJ_0 (0^-)} \rmd z\, \rme^{-\frac{1}{x} V(z)} = \int_{\CJ_0 (0^+) - 2\CJ_1 (0^+)} \rmd z\, \rme^{-\frac{1}{x}V(z)}
\ee
\noindent
translates to
\be
\label{stokesPhi0toPhi0-2Phi-elliptic}
\lim_{\theta \to 0^-} \Phi_0 = \lim_{\theta \to 0^+} \left( \Phi_0 - 2\, \rme^{-\frac{A_1}{x}}\, \Phi_1 \right).
\ee

The case of $\CJ_2$ is similar, albeit slightly more intricate. This time around, in order to maintain the continuity of the ``instanton'' cycle $\Gamma = \CJ_2$, we will need the addition of terms proportional to both $\CJ_0$ and $\CJ_1$. This is illustrated in the bottom two plots of figure~\ref{fig:thimbles-1}. As expected from the structure of the transseries in \eqref{ell_lin_transseries}, both these extra terms are exponentially suppressed relative to the contribution from the critical point $z^*_2$. Following the contour plots in figure~\ref{fig:thimbles-1}, the reader should be able to arrive at the end result
\be
\int_{\CJ_2 (0^-)} \rmd z\, \rme^{-\frac{1}{x}V(z)} = \int_{\CJ_2 (0^+) + 2 \CJ_0 (0^+) - 2 \CJ_1 (0^+)} \rmd z\,\rme^{-\frac{1}{x}V(z)},
\ee
\noindent 
which translates to
\be
\lim_{\theta \to 0^-} \rme^{-\frac{A_2}{x}}\, \Phi_2 = \lim_{\theta \to 0^+} \left( \rme^{-\frac{A_2}{x}}\, \Phi_2 + 2\, \Phi_0 - 2\, \rme^{-\frac{A_1}{x}}\, \Phi_1 \right).
\ee
\noindent
Note how the exponential hierarchy at play, $-A_2/x > 0 > -A_1/x$, dictates the form of the monodromy matrix to be upper-triangular, since all the Stokes jumps have to be exponentially suppressed compared to the main contribution to the integral. 

\subsubsection*{Monodromy and Stokes Phenomenon at $\theta=\pi$}

The analysis of the monodromies and Stokes jumps at $\theta=\pi$ is analogous to what we just did at $\theta=0$. Since $x = -|x|$, the exponential hierarchy is now reversed as
\be
-\frac{A_1}{x} > 0 > -\frac{A_2}{x}, \qquad \text{ for } \quad \theta=\pi.
\ee
\noindent
Therefore there is no Stokes jump for the $\Gamma = \CJ_2$ cycle; the jump in the ``perturbative'' $\Gamma = \CJ_0$ cycle is now proportional to $\CJ_2$; and the jump in the ``instanton'' cycle $\Gamma = \CJ_1$ now has contributions from both $\CJ_0$ and $\CJ_1$. All these jumps are encoded in the ``Stokes matrix'' associated to $\theta=\pi$, $S_{\circlearrowleft}^{(\pi)}$, where:
\begin{equation}
\label{ell-monodromy-pi}
\left(
\begin{array}{ccc}
\CJ_2 \\
\CJ_0 \\
\CJ_1
\end{array}
\right) \to S_{\circlearrowleft}^{(\pi)} \left(
\begin{array}{ccc}
\CJ_2 \\
\CJ_0 \\
\CJ_1
\end{array}
\right) = \left(
\begin{array}{ccc}
 1 &  0 & 0 \\
 2 &  1 & 0 \\
-2 & -2 & 1
\end{array}
\right)  \left(
\begin{array}{ccc}
\CJ_2 \\
\CJ_0 \\
\CJ_1
\end{array}
\right).
\end{equation}
\noindent
Finally, we can write the \textit{monodromy matrix} $\mathfrak{M} = S_{\circlearrowleft}^{(\pi)} \cdot S_{\circlearrowleft}^{(0)}$, and check that $\mathfrak{M}^2 = \boldsymbol{1}$.

\subsubsection*{Complex Phase Diagram}

\begin{figure}[t!]
\begin{center}
\includegraphics[width=8cm]{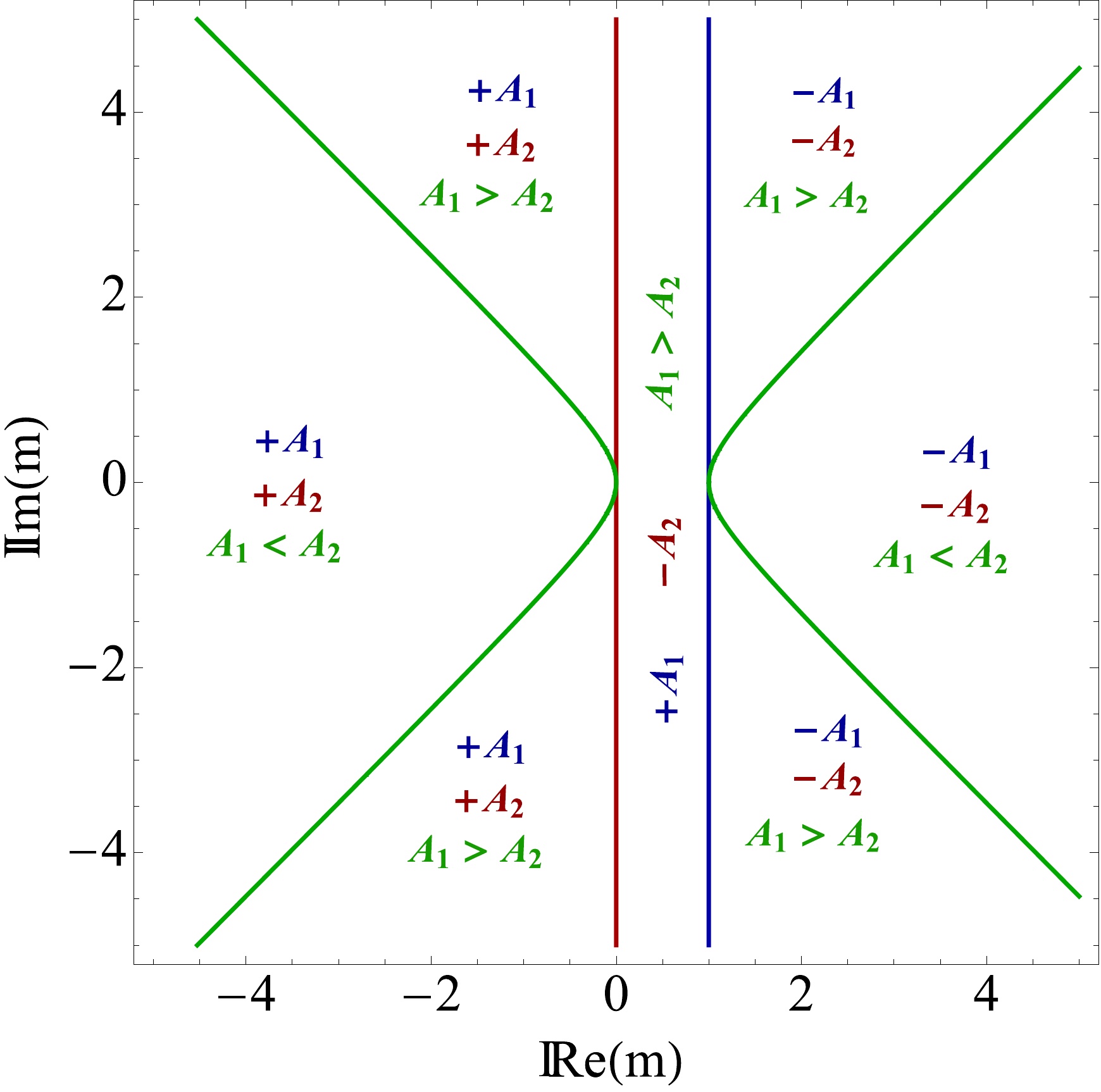}
\end{center}
\caption{Phase diagram for the elliptic partition function, in the complex $m$-plane. Only the anti-Stokes lines are plotted, satisfying $\re \left( A_i(m) - A_j(m) \right) = 0$ (we are now considering $x \in \BR^+$). Different colors represent different pairs of actions: in blue $i=1$, $j=0$; in red $i=2$, $j=0$; in green $i=1$, $j=2$. The relative sign of $\re \left( A_i(m) - A_j(m) \right)$ is also shown in the plots (recall that $A_0=0$).}
\label{fig:ell-phase}
\end{figure}

Before proceeding with the resurgence properties of this problem, let us briefly comment on the possible phase diagrams associated to the elliptic partition function \eqref{elliptic_Z}. If we set $x = |x|\, \rme^{\rmi\theta}$, and vary $\theta$ as we did earlier, then the phase diagram of our model is the \textit{very same} as the phase diagram for the quartic partition function \eqref{quarticintegral}, which was plotted back in figure~\ref{quarticphase} (herein holding true for any modulus $m \in \left(0,1\right)$, and leading to the structure of Stokes and anti-Stokes lines just discussed). On the other hand, as we shall see in the following (and in parallel with what was done back in subsection~\ref{subsec:basicresurgence}), the elliptic partition function satisfies a third-order linear ODE. From this standpoint, each of the three linearly-independent solutions to the said ODE is actually better thought-of as a one-parameter family of solutions, parametrized by the modulus $m$. As an ODE parameter, it is then natural to let $m$ be complex; $m = |m|\, \rme^{\rmi\varphi}$. In this way, as long as $\varphi \neq 0, \pi$, the actions $A_1$ and $A_2$ will also be complex and no longer collinear. One is then led to address \textit{another} phase diagram associated to the elliptic partition function, this time around keeping $\theta$ fixed and varying $\varphi$ instead---as shown in figure~\ref{fig:ell-phase}. This turns out to be an example of \textit{parametric}, or \textit{co-equational resurgence} (a topic not addressed in these lectures; see, \textit{e.g.}, \cite{afss18}) where the actions and the coefficients of the asymptotic series depend on a parameter, and where varying this parameter will change the position and shape of the Stokes lines.

\subsection{Basic Formulae for Resurgent Analysis}\label{sec:ell-basic-partfunc}

In order to explicitly evaluate the different asymptotic series appearing in \eqref{ell_lin_transseries}, it is simpler to proceed differentially rather than integrally. This is straightforward once one notices that the elliptic partition function \eqref{elliptic_Z} satisfies a \textit{third-order} linear ODE\footnote{Recall that we have defined $x\equiv\hbar$ and $m^\prime=1-m$.} \cite{bdu13, cku14}
\bea
&& 4\, m m^\prime x^4\, Z^{\prime\prime\prime} (x) - 4\, \Big( m - m^\prime - 6\, m m^\prime x \Big)\, x^2\, Z^{\prime\prime} (x) - \Big( 4 - \left( m - m^\prime \right) x + \nonumber \\
&& + 9 \left( m - m^\prime - 3\, m m^\prime x \right) x \Big)\, Z^{\prime}(x) - \Big( m - m^\prime - 3\, m m^\prime x \Big)\, Z (x) = 0.
\label{elliptic_linear_ODE}
\eea
\noindent
It is simple to check that this differential equation is invariant under the simultaneous interchange
\begin{equation}
m \rightarrow m^\prime=1-m \quad \text{and} \quad x \rightarrow -x.
\label{modular_sym}
\end{equation}
\noindent
This symmetry is a special case of the full modular symmetry associated with the underlying torus: it first exchanges the two periods of the torus, which accompanied with a change in the sign of $x$ ends up leaving the partition function unchanged. We shall use it later on.

Running in parallel with subsection~\ref{subsec:basicresurgence}, let us try to solve the above differential equation with the usual ``exponential times asymptotic-power-series'' \textit{ansatz},
\be
Z^{(i)} (x) = x^{\beta_i}\, \rme^{-\frac{A_i}{x}}\, \Phi_ i(x) \simeq x^{\beta_i}\,  \rme^{-\frac{A_i}{x}}\, \sum_{n=0}^{+\infty} Z^{(i)}_n\, x^{n}.
\label{eq:ell-partfunc-terms}
\ee
\noindent
This is now slightly more intricate than for the quartic partition function back in \eqref{quarticODE}. Indeed, plugging \eqref{eq:ell-partfunc-terms} into the ODE \eqref{elliptic_linear_ODE} leads to a more complicated recursion for the coefficients $Z^{(i)}_{n}$; namely
\bea
&&
4 A_i \left( 1 + \left( m-m^\prime \right) A_i - m m^\prime A_i^2 \right) Z^{(i)}_n + \nonumber \\
&&
+ 4 \left( \beta_i + n - 1 \right) \left( 1 + 2 \left( m-m^\prime \right) - 3\, m m^\prime A_i^2 \right) Z^{(i)}_{n-1} + \nonumber \\
&&
+ \left( m - m^\prime - 3\, m m^\prime A_i \right) \left( 1 + 4 \left( \beta_i + n - 1 \right) \left( \beta_i + n - 2 \right) \right) Z^{(i)}_{n-2} - \nonumber \\
&&
- m m^\prime \left( 3 + \left( \beta_i + n -3 \right) \left( 3 + 4 \left( \beta_i + n - 1 \right) \left( \beta_i + n - 2 \right) \right) \right) Z^{(i)}_{n-3} 
= 0,
\label{ell_lin_rec}
\eea
\noindent
where we set $Z^{(i)}_{n} \equiv 0$ whenever $i,n<0$ as usual. In particular, if we set $n=0$ and $n=1$ in \eqref{ell_lin_rec} we learn that there are three sectors, $Z^{(i)}(x)$, $i=0,1,2$, with, respectively,
\be
A_i \in \left\{ 0, \frac{1}{m^\prime}, -\frac{1}{m} \right\} \quad \text{ and } \quad \beta_i=0, \quad i=0,1,2.
\ee
\noindent
This matches our earlier saddle-point expectations, and is now a straightforward consequence of the fact that $Z(x)$ satisfies a linear, third-order ODE. As mentioned, we shall be particularly interested in the fact that $A_1$ and $A_2$ are such that they can be linearly combined to vanish, \textit{i.e.}, that specific nonperturbative sectors can end up with the same exponential weight as the perturbative sector, and subsequently mix with one another---one of the hallmarks of resonance. But we leave that discussion to the upcoming subsections~\ref{sec:nonlinear-resonance} and on.

Focusing for the moment in the non-resonant case, one can use the above recursion relation \eqref{ell_lin_rec} to compute the coefficients $Z^{(i)}_n$ associated to each sector. In spite of being a more intricate relation, akin to the results for quartic potential in \eqref{coeff-quartic} and \eqref{coeff-quartic-inst} one can herein also compute such coefficients in closed form. For the perturbative sector $Z^{(0)}(x)$, nonperturbative sector $Z^{(1)}(x)$ with action $A_1=1/m^\prime$, and nonperturbative sector $Z^{(2)}(x)$ with action $A_2=-1/m$, we find, respectively,
\bea
\Phi_0 (x) &\simeq& \sum_{n=0}^{+\infty} Z^{(0)}_n (m)\, x^{n}, \quad Z^{(0)}_n (m) = \frac{\Gamma^2\left(n+\frac{1}{2}\right)}{\pi\, \Gamma\left(n+1\right)}\, {}_2 F_{1} \left( -n, -n, \frac{1}{2}-n\, \right| \left. \vphantom{\frac{1}{2}} m \right),
\label{eq:elliptic_coeffs0} \\
\Phi_1 (x) &\simeq& \sum_{n=0}^{+\infty} Z^{(1)}_n (m)\, x^{n}, \quad Z^{(1)}_n (m) = Z^{(1)}_0\, \frac{(-1)^n}{A_1^n}\, \frac{\Gamma^2\left(n+\frac{1}{2}\right)}{\pi\, \Gamma\left(n+1\right)}\, {}_2 F_{1} \left( \frac{1}{2}, -n, \frac{1}{2}-n\, \right| \left. \vphantom{\frac{1}{2}} m \right),
\label{eq:elliptic_coeffs1} \\
\Phi_2 (x) &\simeq& \sum_{n=0}^{+\infty} Z^{(2)}_n (m)\, x^{n}, \quad Z^{(2)}_n (m) = Z^{(2)}_0\, \frac{(-1)^n}{A_2^n}\, \frac{\Gamma^2\left(n+\frac{1}{2}\right)}{\pi\, \Gamma\left(n+1\right)}\, {}_2 F_{1} \left( \frac{1}{2}, -n, \frac{1}{2}-n\, \right| \left. \vphantom{\frac{1}{2}} m^\prime \right).
\label{eq:elliptic_coeffs2}
\eea
\noindent
The instanton actions are evident in the growth of the several coefficients $\sim A^{-n} n!$. Further, the coefficients in the nonperturbative sectors relate to each other as $Z^{(2)}_n (m) = (-1)^n\, Z^{(1)}_n(m^\prime)$, which follows from the modular symmetry \eqref{modular_sym}. Finally, the residual coefficients were set to $Z^{(1)}_0 \equiv Z^{(1)}_0 (m) = \rmi\sqrt{m^\prime}$ and $Z^{(2)}_0 \equiv Z^{(2)}_0 (m) = \rmi\sqrt{m}$, which will be used throughout. 

Having constructed three linearly independent solutions---given by \eqref{eq:ell-partfunc-terms}, with asymptotic series \eqref{eq:elliptic_coeffs0}, \eqref{eq:elliptic_coeffs1}, and \eqref{eq:elliptic_coeffs2}---to the linear third-order ODE \eqref{elliptic_linear_ODE}, the general solution follows as their linear combination. This introduces three integration constants, parameterizing boundary conditions, and leading to the three-parameter transseries \eqref{ell_lin_transseries}, which we now rewrite as
\be
\label{ell_lin_transseries-v2}
\CZ (x, \sigma_0, \sigma_1, \sigma_2 ) = \sigma_0\, Z^{(0)}(x) + \sigma_1\, Z^{(1)}(x) + \sigma_2\, Z^{(2)}(x).
\ee

Proceeding to analyze the resurgent properties of this example, let us turn to the computation of the Borel transforms \eqref{boreltrans} for each asymptotic series above, $\Phi_{i}(x)$. Unlike in subsection~\ref{subsec:basicresurgence}, however, we shall now do this in two different steps: we will first address general resurgence expectations (based on what we have learnt so far), and only then will we compute explicit expressions. On very general grounds we expect the Borel transform of any given sector to have non-zero but finite radius of convergence. For instance, $\CB [\Phi_0] (s)$ will converge for $s < \min \left( |A_1|, |A_2| \right) = \min \left( 1/m^\prime, 1/m \right)$ (recall that the leading singularity on the Borel plane is determined by the closest-to-the-origin relative-action). Considering the simple representative, the singularities at $s=-1/m$ and at $s=1/m^\prime$ will be logarithmic branch-points (where the branch cuts emerging from these points can be tested numerically as accumulations of Pad\'e-approximant poles; recall, \textit{e.g.}, figures~\ref{fig:Pade-poles-quartic-pert-series} or \ref{fig:sec5-BP-Riccati}). When further considering the nonperturbative sectors, generically $\CB [\Phi_i] (s)$ will have singularities located at $\omega_{ij} = A_j-A_i$, for $i \ne j \in \left\{ 0, 1, 2 \right\}$ and with $A_0:=0$, and \textit{near each} of these singular points we find the \textit{simple singularity} structure \eqref{simpleBorelsingularities},
\be
\label{ellipticBPhiiSij}
\CB [\Phi_i] (s) = \mathsf{S}_{i\to j}\, \frac{Z^{(j)}_0}{2\pi\rmi \left( s-\omega_{ij} \right)} + \mathsf{S}_{i\to j} \times \CB [\Phi_j] \left(s-\omega_{ij}\right) \frac{\log \left(s-\omega_{ij}\right)}{2\pi\rmi} + \text{holomorphic}.
\ee
\noindent
The \textit{resurgence} of the sector $\Phi_j$ within the singular behaviour of the Borel transform for $\Phi_i$, $i\ne j$, is quite explicit as expected. The proportionality constant $\mathsf{S}_{i\to j}$ is the ``Borel residue'', which has already appeared back in, \textit{e.g.}, \eqref{Borel-transf-expanded-one-param} or \eqref{eq:sec5-Borel-transf}.

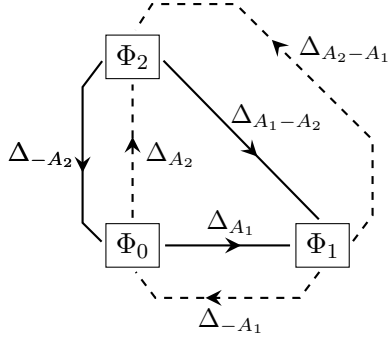
\begin{figure}[t!]
\begin{center}
\begin{tikzpicture}[>=latex,decoration={markings, mark=at position 0.6 with {\arrow[ultra thick]{stealth};}} ]
\begin{scope}[node distance=2.5cm]
  \node (Phi0) [draw] at (0,0) {$\Phi_0$};
  \node (Phi2) [above of=Phi0] [draw] {$\Phi_2$};
  \node (Phi1) [right of=Phi0] [draw] {$\Phi_1$};
\end{scope}
   \draw [thick,postaction={decorate},-,>=stealth,shorten <=2pt,shorten >=2pt]  (Phi0.east) --  (Phi1.west);
   \draw [dashed,thick,postaction={decorate},-,>=stealth,shorten <=2pt,shorten >=2pt] (Phi1.south) -- +(-0.3,-0.4) -- ++(-2.2,-0.4) -- (Phi0.south);
   \draw [thick,postaction={decorate},-,>=stealth,shorten <=2pt,shorten >=2pt]  (Phi2.west) -- +(-0.3,-0.4) -- ++(-0.3,-2.2) -- (Phi0.west);
   \draw [dashed,thick,postaction={decorate},-,>=stealth,shorten <=2pt,shorten >=2pt] (Phi0.north) -- (Phi2.south);
   \draw [thick,postaction={decorate},-,>=stealth,shorten <=2pt,shorten >=2pt] (Phi2.east) --  (Phi1.north);
   \draw [dashed,thick,postaction={decorate},-,>=stealth,shorten <=2pt,shorten >=2pt] (Phi1.east) -- +(0.3,0.4) -- ++(0.3,1.4) -- ++(-1.8,1.8) -- ++(-1.1,0)-- (Phi2.north);
  \node (+A1) at (1.3,0.3) {\small{$\Delta_{A_1}$}};
  \node (-A1) at (1.3,-1) {\small{$\Delta_{-A_1}$}};
  \node (-A2) at (-1.2,1.2) {\small{$\Delta_{-A_2}$}};
  \node (A2) at (0.5,1.2) {\small{$\Delta_{A_2}$}};
  \node (-A2) at (-1.2,1.2) {\small{$\Delta_{-A_2}$}};
  \node (A1-A2) at (1.9,1.7) {\small{$\Delta_{A_1-A_2}$}};
  \node (A2-A1) at (2.8,2.6) {\small{$\Delta_{A_2-A_1}$}};
\end{tikzpicture}
\end{center}
\caption{Algebraic structure associated to the behaviour of Borel transforms or alien derivatives for the $\Phi_i$ sectors, at their respective singularities. When $m \in \left(0,1\right)$ the motions fall along the real axis on the Borel plane: solid lines correspond to motions along the $\theta=0$ direction; dashed lines along $\theta=\pi$.}
\label{linear-algebraic-structure-elliptic}
\end{figure}

Let us next rewrite the above Borel expression at the usual resurgence algebraic-level-of-abstraction, based upon alien derivations $\Delta_\omega$. Given the three transseries sectors, there are six possible singularities, $\omega_{ij}$ and $\omega_{ji} = -\omega_{ij}$, of the form \eqref{ellipticBPhiiSij},
\begin{equation}
\label{ell-singularities}
\omega_{01} = A_1, \qquad \omega_{21} = A_1 - A_2 = A_1 + |A_2|, \qquad \omega_{02} = A_2 = - |A_2|. 
\end{equation}
\noindent
Each action $A_i$ is defined in \eqref{eq:elliptic_critical}; $A_0:=0$, $A_1>0$ and $A_2<0$. Continuing our parallel with subsection~\ref{subsec:basicresurgence}, we may now use the definition \eqref{def-linear-alien} appropriate for a linear problem---or even the general formalism discussed in appendix~\ref{app:alien-calculus}---in order to write all possible alien derivatives. One finds (compare with \eqref{Delta+A} and \eqref{Delta-A}):
\begin{eqnarray}
\label{ell-linear-aliender-1}
\Delta_{\omega_{01}} \Phi_0 = S_{\omega_{01}}\, \Phi_1, & & \qquad \Delta_{\omega_{02}} \Phi_0 = S_{\omega_{02}}\, \Phi_2, \\
\label{ell-linear-aliender-2}
\Delta_{\omega_{10}} \Phi_1 = S_{\omega_{10}}\, \Phi_0, & & \qquad  \Delta_{\omega_{12}} \Phi_1 = S_{\omega_{12}}\, \Phi_2, \\
\label{ell-linear-aliender-3}
\Delta_{\omega_{20}} \Phi_2 = S_{\omega_{20}}\, \Phi_0, & & \qquad \Delta_{\omega_{21}} \Phi_2 = S_{\omega_{21}}\, \Phi_1.
\end{eqnarray}
\noindent
In these expressions the $S_{\omega_{ij}}$ are the Stokes coefficients, related to the Borel residues $\mathsf{S}_{i\to j}$ appearing in the Borel transforms, and whose values may be determined, \textit{e.g.}, from an analysis of Stokes phenomena (as discussed later in the present subsection), or from an analysis of large-order behaviour (as discussed in the next subsection). The algebraic structure written above closes upon itself, much like what happened for the partition function of the quartic potential back in \eqref{linear-algebraic-structure}, albeit now with a few more allowed motions. The algebraic structure encoded in \eqref{ell-linear-aliender-1}--\eqref{ell-linear-aliender-3} is depicted in figure~\ref{linear-algebraic-structure-elliptic}, where we have assumed $m \in \left(0,1\right)$ and where we plot motions along the positive-real direction as solid lines, and motions along the negative-real direction as dashed lines. Note that while for $m \in \left(0,1\right)$ the three actions $A_i$ are all real and we will only have two singular directions, along $\theta=0$ and $\pi$, if one allows $m \in \BC$ the actions $A_1$ and $A_2$ become complex and might not fall along the same direction (this was already noticed in the discussion associated to figure~\ref{fig:ell-phase}). Also for this reason we are now plotting the sectors corresponding to each of these actions as \textit{orthogonal}; in order to depict the general independence between the two actions (in fact, as occurs for the completely general case; recall figure~\ref{fig:sec5-2d-lattice}).

The relation between the Stokes coefficients in \eqref{ell-linear-aliender-1}--\eqref{ell-linear-aliender-3} and the Borel residues in \eqref{ellipticBPhiiSij} can be easily determined following the example\footnote{This was dubbed the ``explicit (linear) example'' in appendix~\ref{app:alien-calculus}.} outlined in appendix~\ref{app:alien-calculus}; or else along the very general lines discussed earlier in sections~\ref{sec:quartic} and~\ref{sec:physics}. It is simple to find:
\begin{align}
\label{eq:nonres-Borel-stokes-01}
&\mathsf{S}_{0\to 1} = - S_{\omega_{01}}, 
&&\mathsf{S}_{1\to 0} = - S_{\omega_{10}}, \\
&\mathsf{S}_{0\to 2} = - S_{\omega_{02}}, 
&&\mathsf{S}_{2\to 0} = - S_{\omega_{20}}, \\
&\mathsf{S}_{1\to 2} = - S_{\omega_{12}} - \frac{1}{2}\, S_{\omega_{10}}\,S_{\omega_{02}}, 
&&\mathsf{S}_{2\to 1} = - S_{\omega_{21}} - \frac{1}{2}\, S_{\omega_{20}}\,S_{\omega_{01}}.
\label{eq:nonres-Borel-stokes-21}
\end{align}

The next step is to determine the Stokes automorphism, given by \eqref{stokesauto} and explicitly in \eqref{Stokes-aut-as exponential-singularities}, along all possible singular directions. This essentially follows from the alien derivatives \eqref{ell-linear-aliender-1}--\eqref{ell-linear-aliender-3}. Possible singular directions may depend on the value of the modular parameter $m$, and to make things simpler we will only have in mind the case of $m\in(0,1)$ already depicted in figure~\ref{linear-algebraic-structure-elliptic}. The Stokes automorphism at $\theta=0$ is
\be
\label{eq:disc-Z-0}
\underline{\mathfrak{S}}_0 = \exp \left( \rme^{-\frac{\omega_{01}}{x}} \Delta_{\omega_{01}} + \rme^{-\frac{\omega_{20}}{x}} \Delta_{\omega_{20}} + \rme^{-\frac{\omega_{21}}{x}} \Delta_{\omega_{21}} \right).
\ee
\noindent
As expected, only the alien derivatives whose associated relative actions are positive (\textit{i.e.}, $\omega_{01}$, $\omega_{20}$, $\omega_{21}$) contribute to the discontinuity along $\theta=0$, albeit not all will act non-trivially on each sector $\Phi_i$. Similarly, for $\theta=\pi$ we get
\be
\label{eq:disc-Z-pi}
\underline{\mathfrak{S}}_\pi = \exp \left( \rme^{-\frac{\omega_{10}}{x}} \Delta_{\omega_{10}} + \rme^{-\frac{\omega_{02}}{x}} \Delta_{\omega_{02}}+ \rme^{-\frac{\omega_{12}}{x}} \Delta_{\omega_{12}} \right),
\ee
\noindent
with only the negative relative actions $\omega_{10}$, $\omega_{02}$, $\omega_{12}$ contributing. Acting on each asymptotic sector $\Phi_i$, via expansion of the exponentials in \eqref{eq:disc-Z-0} and \eqref{eq:disc-Z-pi} (which collapse to finite sums), and further using \eqref{ell-linear-aliender-1}--\eqref{ell-linear-aliender-3}, it simply follows:
\begin{eqnarray}
\label{ell-linear-Stokes-00}
\underline{\mathfrak{S}}_0 \Phi_0 &=& \Phi_0 + S_{\omega_{01}}\, \mathrm{e}^{-\frac{A_1}{x}}\, \Phi_1 = \Phi_0 - \mathsf{S}_{0\to 1}\, \mathrm{e}^{-\frac{A_1}{x}}\, \Phi_1, \\
\label{ell-linear-Stokes-pi0}
\underline{\mathfrak{S}}_{\pi} \Phi_0 &=& \Phi_0 + S_{\omega_{02}}\, \mathrm{e}^{-\frac{A_2}{x}}\, \Phi_2 = \Phi_0 - \mathsf{S}_{0\to 2}\, \mathrm{e}^{-\frac{A_2}{x}}\, \Phi_2, \\
\label{ell-linear-Stokes-pi1}
\underline{\mathfrak{S}}_{\pi} \Phi_1 &=& \Phi_1 + S_{\omega_{10}}\, \mathrm{e}^{+\frac{A_1}{x}}\, \Phi_0 + \left( S_{\omega_{12}} + \frac{1}{2} S_{\omega_{02}} S_{\omega_{10}} \right) \mathrm{e}^{+\frac{A_1-A_2}{x}}\, \Phi_2 \nonumber \\
&=& \Phi_1 - \mathsf{S}_{1\to 0}\, \mathrm{e}^{+\frac{A_1}{x}}\, \Phi_0 - \mathsf{S}_{1\to 2}\, \mathrm{e}^{+\frac{A_1-A_2}{ x}}\, \Phi_2, \\
\label{ell-linear-Stokes-02}
\underline{\mathfrak{S}}_0 \Phi_2 &=& \Phi_2 + S_{\omega_{20}}\, \mathrm{e}^{+\frac{A_2}{x}}\, \Phi_0 + \left( S_{\omega_{21}} + \frac{1}{2} S_{\omega_{01}} S_{\omega_{20}} \right) \mathrm{e}^{-\frac{A_1-A_2}{x}}\, \Phi_1 \nonumber \\
&=& \Phi_2 - \mathsf{S}_{2\to 0}\, \mathrm{e}^{+\frac{A_2}{x}}\, \Phi_0 - \mathsf{S}_{2\to 1}\, \mathrm{e}^{-\frac{A_1-A_2}{x}}\, \Phi_1,
\end{eqnarray}
\noindent
where we made use of the relations \eqref{eq:nonres-Borel-stokes-01}--\eqref{eq:nonres-Borel-stokes-21} to display all results with either Stokes coefficients or Borel residues. These expressions encode Stokes phenomena across all Stokes lines. The lateral resummations across these lines, \eqref{stokesauto}, applied to the transseries \eqref{ell_lin_transseries-v2} can be easily written down, from where one obtains the \textit{Stokes transitions}
\begin{eqnarray}
\label{ell-stokes-transitions-0}
\mathcal{S}_{0^+} \mathcal{Z} \left( x, \sigma_0, \sigma_1, \sigma_2 \right) &=& \mathcal{S}_{0^-} \mathcal{Z} \left( x,\, \sigma_0-\mathsf{S}_{2\to 0}\,\sigma_2,\, \sigma_1-\mathsf{S}_{0\to 1}\,\sigma_0 - \mathsf{S}_{2\to 1}\,\sigma_2,\, \sigma_2 \right), \\
\label{ell-stokes-transitions-pi}
\mathcal{S}_{\pi^+} \mathcal{Z} \left( x, \sigma_0, \sigma_1, \sigma_2 \right) &=& \mathcal{S}_{\pi^-} \mathcal{Z} \left( x,\, \sigma_0-\mathsf{S}_{1\to 0}\,\sigma_1,\, \sigma_1,\, \sigma_2-\mathsf{S}_{0\to 2}\,\sigma_0-\mathsf{S}_{1\to 2}\,\sigma_1 \right).
\end{eqnarray}
\noindent
For simplicity, we have written the Stokes transitions in terms of the Borel residues. The same expressions in terms of the Stokes coefficients may be easily obtained from \eqref{eq:nonres-Borel-stokes-01}--\eqref{eq:nonres-Borel-stokes-21}.

By comparing the Stokes transitions with the jumps of the steepest-descent contours encoded in the Stokes matrices \eqref{ell-monodromy-0} and \eqref{ell-monodromy-pi}, one can easily determine the value of the Borel residues and Stokes coefficients. The jump of a given contour $\mathcal{J}_i$ across a Stokes ray will correspond to the jump of the corresponding asymptotic sector $Z^{(i)}$ in the Stokes transition. For $\theta=0$ one has (compare with \eqref{ell-monodromy-0})
\begin{equation}
\left(
\begin{array}{ccc}
Z^{(2)} \\
Z^{(0)} \\
Z^{(1)}
\end{array}
\right) \to S_{\circlearrowleft}^{(0)} \left(
\begin{array}{ccc}
Z^{(2)} \\
Z^{(0)} \\
Z^{(1)}
\end{array}
\right) = \left(
\begin{array}{ccc}
1 & -\mathsf{S}_{2\to 0} & - \mathsf{S}_{2\to 1} \\
0 & 1 & -\mathsf{S}_{0\to 1} \\
0 & 0 & 1
\end{array}
\right) \left(
\begin{array}{ccc}
Z^{(2)} \\
Z^{(0)} \\
Z^{(1)}
\end{array}
\right),
\end{equation}
\noindent
from where we can read the Borel residues $\mathsf{S}_{2\to 0} = \mathsf{S}_{2\to 1} = -\mathsf{S}_{0\to 1} = 2$. The exact same reasoning applied at $\theta=\pi$ yields $\mathsf{S}_{0\to 2} = -\mathsf{S}_{1\to 2} = -\mathsf{S}_{1\to 0} = -2$. Finally, using \eqref{eq:nonres-Borel-stokes-01}--\eqref{eq:nonres-Borel-stokes-21}, one obtains the complete set of Stokes coefficients,
\be
S_{\omega_{01}} = 2, 
\qquad 
S_{\omega_{10}} = -2, 
\qquad
S_{\omega_{02}} = 2, 
\qquad
S_{\omega_{20}} = -2, 
\qquad
S_{\omega_{12}} = 0 = S_{\omega_{21}}.
\label{eq:linear-stokes}
\ee

With these numbers in hand, computing monodromies is now straightforward. In fact, moving twice around the origin, \textit{i.e.}, acting with $\left( \underline{\mathfrak{S}}_{\pi} \underline{\mathfrak{S}}_0 \right)^2$ upon the transseries, the parameters $\sigma_i$ get mapped onto themselves. In other words,
\be
\left( \underline{\mathfrak{S}}_{\pi} \underline{\mathfrak{S}}_0 \right)^2 = \boldsymbol{1},
\ee
\noindent
which encompasses the monodromy encoded in the transseries solution for the elliptic partition function; once again leading to a multi-sheeted Riemann surface (compare with \eqref{quarticmonodromy}).

It is also very instructive to analyze the resurgent relations between the different sectors $\Phi_i$ purely from the perspective of the Picard--Lefschetz theory introduced in section~\ref{sec:lefschetz} (and further using the main discussion from section~\ref{sec:borel}, as will be clear momentarily). Each asymptotic series in \eqref{eq:elliptic_coeffs0} through \eqref{eq:elliptic_coeffs2} is basically dictated by the small $x\equiv\hbar$ expansion of the integral \eqref{elliptic_Z} evaluated on the appropriate thimble $\Gamma_i$, \textit{i.e.}, the one associated with the critical point $z^*_i$. The calculation is rather straightforward thanks to the elliptic identities
\bea
\frac{\rmd}{\rmd z} \mathrm{sd} (z | m) &=& \mathrm{nd} (z | m)\, \mathrm{cd} (z | m), \\
\mathrm{nd}^2 (z | m) &=& 1 + m\, \mathrm{sd}^2 (z | m), \\
\mathrm{cd}^2 (z | m) &=& 1 - m^\prime\, \mathrm{sd}^2 (z | m).
\eea
\noindent
Using these identities and the change of variables $s \equiv \mathrm{sd}^2 (z | m)$---which is of the type discussed in section~\ref{sec:borel}; recall \eqref{eq:Change-of-variables-to-Borel}---the integration over the cycle $\CJ_0 (\theta)$ may be expressed as 
\be
\label{JA}
\frac{1}{\sqrt{\pi x}} \int_{\CJ_0(\theta)} \rmd z\, \rme^{-\frac{1}{x}\, \mathrm{sd}^2 (z | m)} = \frac{1}{\sqrt{\pi x}} \int_0^{\rme^{\rmi\theta}\infty} \rmd s\, \frac{\rme^{-\frac{s}{x}}}{\sqrt{s \left( 1 - m^\prime s \right) \left( 1 + m s \right)}} \simeq \rme^{-\frac{A_0}{x}} \sum_{n=0}^{+\infty} Z^{(0)}_n (m)\, x^n,
\ee
\noindent
where, in the third step, we formally Taylor-expanded the denominator outside its radius of convergence and carried out the resulting $s$-integration, leading to the  asymptotic perturbative expansion $\Phi_0(x)$. In this way, it should not come as a great surprise that the coefficients of the above asymptotic series are found as 
\begin{eqnarray}
Z^{(0)}_n (m) &=& \frac{\Gamma\left(n+\frac{1}{2}\right)}{\pi^{\frac{3}{2}}}\, \sum_{k=0}^n (-1)^k\, \frac{\Gamma\left(k+\frac{1}{2}\right) \Gamma\left(n-k+\frac{1}{2}\right)}{\Gamma\left(k+1\right) \Gamma\left(n-k+1\right)}\, m^k\, m^{\prime\, n-k} \nonumber \\
&=& \frac{\Gamma^2\left(n+\frac{1}{2}\right)}{\pi\, \Gamma\left(n+1\right)}\, {}_2F_{1} \left( -n, -n, \frac{1}{2}-n\, \right| \left. \vphantom{\frac{1}{2}} m \right),
\end{eqnarray} 
\noindent
in precise agreement with expression \eqref{eq:elliptic_coeffs0}, which was derived from the recursion relations \eqref{ell_lin_rec}. Notice that, as expected from the discussion of section~\ref{sec:borel}, the form of the second integral in \eqref{JA} resembles the resummation of the Borel transform $\CB [\Phi_0] (s)$, up to the extra factor of $1/\sqrt{x}$ in front. To deal with these extra factors and ``odd-looking'' Borel transforms, in section~\ref{sec:borel} we introduced sectors of the form $\Phi_{i,[\alpha]}(x) = x^{\alpha}\, \Phi_{i,[0]}(x)$ (recall \eqref{Phi_[gamma]-DEF}) where herein $\Phi_{i,[0]}(x)$ is any of the original asymptotic sectors $\Phi_i (x)$ whose Borel transforms have (simple) logarithmic singularities. Form \eqref{JA} one can now immediately identify
\be
\Phi_{0,[1/2]}(x) = \frac{1}{\sqrt{\pi}} \int_{\CJ_0 (\theta)} \rmd z\, \rme^{-\frac{1}{x}\, \mathrm{sd}^2 (z | m)},
\ee 
\noindent
whose Borel transform is given by
\be
\CB [\Phi_{0,[1/2]}] (s) = \frac{1}{\sqrt{\pi s \left( 1 - m^\prime s \right) \left( 1 + m s \right)}}.
\ee

The Borel transforms associated to sectors $\Phi_{i,[\alpha]}(x)$ for different $\alpha$ are related by a fractional derivative, via \eqref{eq:finding-Borel-rep-from-other}. As discussed in section~\ref{sec:borel}, and due to the involved fractional derivatives, these Borel transforms are related in highly non-local fashions. Although alien calculus might give preference to the choice $\alpha=0$, we shall demonstrate below that it may also be illuminating to work with $\alpha=1/2$. In fact, herein we will find a surprisingly simple relation between the Borel transforms of the series associated with the different critical points.

Similarly to \eqref{JA}, the integration over the thimble $\CJ_1(\theta)$ may be expressed as 
\begin{eqnarray}
\frac{1}{\sqrt{\pi x}} \int_{\CJ_1(\theta)} \rmd z\, \rme^{-\frac{1}{x}\, \mathrm{sd}^2 (z | m)} &=& - \frac{1}{\sqrt{\pi x}} \int_{\frac{1}{m^\prime}}^{\rme^{\rmi\theta}\infty} \rmd s\, \frac{\rme^{-\frac{s}{x}}}{\sqrt{s \left( 1 - m^\prime s \right) \left( 1 + m s \right)}} \nonumber \\
&=& \rmi\, \rme^{-\frac{1}{m^\prime x}}\, \frac{\sqrt{m^\prime}}{\sqrt{\pi x}}  \int_0^{\rme^{\rmi\theta} \infty} \rmd s\, \frac{\rme^{-\frac{s}{x}}}{\sqrt{s \left( 1 + m m^\prime s \right) \left( 1 + m^\prime s \right)}} \nonumber \\
&\equiv& \rme^{-\frac{A_1}{x}}\, \frac{1}{\sqrt{x}}\, \CS_\theta \Phi_{1,[1/2]} (x). 
\end{eqnarray}
\noindent
Notice that the integrand of the integral on the right-hand-side of the first line is precisely the Borel transform of $\Phi_{0,[1/2]}(x)$ (up to a shift $1/m^\prime$), while in the last line we obtain instead a resummation of $\Phi_{1,[1/2]}(x)$. Furthermore, the shift in question is precisely the relative action $A_0-A_1$. As we shall see next, this is a manifestation of resurgence: the two asymptotic series, attached to different critical points, are related by a simple shift in the Borel plane, where the shift is precisely the difference between the actions at those critical points. We can moreover write
\be
\CB [\Phi_{1,[1/2]}] (s) = \frac{\rmi \sqrt{m^\prime}}{\sqrt{\pi s \left( 1 + m m^\prime s \right) \left( 1 + m^\prime s \right)}}.
\ee
\noindent
Finally, the  integration over the thimble $\CJ_2(\theta)$ can be expressed in a totally similar fashion as
\begin{eqnarray}
\frac{1}{\sqrt{\pi x}} \int_{\CJ_2(\theta)} \rmd z\, \rme^{-\frac{1}{x}\, \mathrm{sd}^2 (z | m)} &=& - \frac{1}{\sqrt{\pi x}} \int_{-\frac{1}{m}}^{\rme^{\rmi\theta} \infty} \rmd s\, \frac{\rme^{-\frac{s}{x}}}{\sqrt{s \left( 1 - m^\prime s \right) \left( 1 + m s \right)}} \nonumber \\
&=& \rmi\, \rme^{\frac{1}{m x}}\, \frac{\sqrt{m}}{\sqrt{\pi x}} \int_0^{\rme^{\rmi\theta} \infty} \rmd s\, \frac{\rme^{-\frac{s}{x}}}{\sqrt{s \left( 1 - m m^\prime s \right) \left( 1 - m s \right)}} \nonumber \\
&\equiv& \rme^{-\frac{A_2}{x}}\, \frac{1}{\sqrt{x}}\, \CS_\theta \Phi_{2,[1/2]} (x),
\end{eqnarray}
\noindent
where
\be
\CB [\Phi_{2,[1/2]}] (s) = \frac{\rmi \sqrt{m}}{\sqrt{\pi s \left( 1 - m m^\prime s \right) \left( 1 - m s \right)}}.
\ee

It is now clear that the Borel transforms of the series associated with each critical point are related to each other in a strikingly simple way,
\be
\CB [\Phi_{i,[1/2]}] \left(s\right) = \pm \CB [\Phi_{j,[1/2]}] \left(s-\left( A_j-A_i\right)\right), \qquad i,j=0,1,2,  
\ee
\noindent
where the plus-sign occurs when $j=0$. In words: on the Borel plane, and up to a sign\footnote{This sign is governed by the branch of the square-root one goes through when performing the shift.}, these Borel transforms are expressed by the exact \textit{same function}, only shifted by the relative action $A_j-A_i$. An immediate consequence of this relation is that, having access to, say, the perturbative asymptotic series around the vacuum, one can then reconstruct the expansions around \textit{all} other nonperturbative sectors without much effort. Further recalling the expected general relations between such Borel transforms, \eqref{eq:B_Psi_n:d=odd}, we can proceed to read-off the values of the Borel residues and then of the Stokes coefficients $S_{\omega_{ij}}$ in \eqref{ell-linear-aliender-1}--\eqref{ell-linear-aliender-3}. These numbers agree with the ones obtained in \eqref{eq:linear-stokes}. These very same values can also be obtained from a large-order analysis of the asymptotic behaviour of the coefficients $Z^{(i)_n}$, to which we turn next.

\subsection{Asymptotics and Large-Order Behaviour: Partition Function}\label{sec:linear_ode}

Let us now turn to the analysis of the large-order behaviour of the partition-function coefficients $Z^{(i)}_n$. For concreteness, let us start with the perturbative sector.  Using the connection formula for the hypergeometric function, one can write 
\be
\label{eq:linear-ell-lo}
Z^{(0)}_n (m) = \frac{\Gamma\left(n+1\right)}{\pi \left( n+\frac{1}{2} \right)} \left\{ \left( m^\prime \right)^{n+\frac{1}{2}}\, {}_2F_{1} \left( \frac{1}{2}, \frac{1}{2}, \frac{3}{2}+n\, \right| \left. \vphantom{\frac{1}{2}} m^\prime \right) + (-1)^n\, m^{n+\frac{1}{2}}\, {}_2F_{1} \left( \frac{1}{2}, \frac{1}{2}, \frac{3}{2}+n\, \right| \left. \vphantom{\frac{1}{2}} m \right) \right\}. 
\ee
\noindent
This decomposition of the perturbative coefficients already hints that they must somehow relate to nonperturbative content, since $A_1 = \left( m^\prime \right)^{-1}$ and $A_2 = - m^{-1}$. Of course such relations may be made very precise via resurgence. But before we do that, let us proceed by expanding the above hypergeometric functions in both $m^{\prime}$ and $m$,
\begin{eqnarray}
\label{ell_pertecomp}
Z^{(0)}_n (m) &\simeq& \frac{\Gamma\left(n+1\right)}{\pi \left( 2n+1 \right)} \left\{ \left(m^\prime\right)^{n+\frac{1}{2}} \left( 2 + \frac{m^\prime}{2n+3} +\frac{9 \left( m^\prime \right)^2}{4 \left( 2n+3 \right) \left( 2n+5 \right)} + \cdots \right) + \right. \\
&&
\hspace{65pt}
\left. + (-1)^n\, m^{n+\frac{1}{2}} \left( 2 + \frac{m}{2n+3} + \frac{9 \left( m \right)^2}{4 \left( 2n+3 \right) \left( 2n+5 \right)} + \cdots \right) \right\}. \nonumber
\end{eqnarray} 
\noindent
This expansion is valid for any $n$; in particular one can take the limit $n\gg1$ and reorganize the several terms as
\begin{eqnarray}
Z^{(0)}_n (m) &\simeq& \frac{\left( m^\prime \right)^{n+\frac{1}{2}}\, \Gamma\left(n\right)}{\pi} \left( 1 - \frac{1}{2n} + \frac{1}{4n^2} + \cdots + \frac{m^\prime}{4n} - \frac{m^\prime}{2n^2} + \cdots + \frac{9 \left(m^\prime\right)^2}{32n^2} + \cdots \right) + \nonumber \\
&+& \frac{(-1)^n\, m^{n+\frac{1}{2}}\, \Gamma\left(n\right)}{\pi} \left( 1 - \frac{1}{2n} + \frac{1}{4n^2} + \cdots + \frac{m}{4n} - \frac{m}{2n^2} + \cdots + \frac{9\, m^2}{32n^2} + \cdots \right) \\
&\approx& \frac{\left( m^\prime \right)^{n} \Gamma\left(n\right)}{\pi} \left( \sqrt{m^\prime} - \sqrt{m^\prime}\, \frac{m^\prime-2}{4 \left(n-1\right)} + \sqrt{m^\prime}\, \frac{9 \left(m^\prime\right)^2 - 24\, m^\prime + 24}{32 \left(n-1\right) \left(n-2\right)} + \cdots \right) + \nonumber \\
&+& \frac{(-1)^n\, m^{n}\, \Gamma\left(n\right)}{\pi} \left( \sqrt{m} - \sqrt{m}\, \frac{m-2}{4 \left(n-1\right)} + \sqrt{m}\, \frac{9\, m^2 - 24\, m + 24}{32 \left(n-1\right) \left(n-2\right)} + \cdots \right).
\end{eqnarray} 
\noindent
Interestingly enough, this form already is \textit{precisely} the expected large-order behaviour for the perturbative series, usually written as
\begin{eqnarray}
Z^{(0)}_n (m) &\simeq& \frac{2}{2\pi\rmi}\, \frac{\Gamma\left(n\right)}{A_1^n} \left( Z^{(1)}_0 + \frac{A_1}{n-1}\, Z^{(1)}_1 + \frac{A_1^2}{\left(n-1\right) \left(n-2\right)}\, Z^{(1)}_2 + \cdots \right) + \nonumber \\
&+& \frac{2}{2\pi\rmi}\, \frac{\Gamma\left(n\right)}{A_2^n} \left( Z^{(2)}_0 + \frac{A_2}{n-1}\, Z^{(2)}_1 + \frac{A_2^2}{\left(n-1\right) \left(n-2\right)}\, Z^{(2)}_2 + \cdots \right). 
\label{large_order_aA}
\end{eqnarray}

The standard approach---which in fact allows for a determination of the large-order relations associated to \textit{all} (perturbative and nonperturbative) sectors---is to follow the same steps as in subsection~\ref{subsec:Zasymptotics}. From Cauchy's theorem, 
\be
\Phi_i (x) = -\frac{1}{2\pi\rmi}\, \sum_{\theta=0,\pi}\, \int_0^{\rme^{\rmi\theta} \infty} \rmd w\, \frac{\disc_\theta \Phi_i(w)}{w-x},
\ee
\noindent
where the discontinuity is defined by \eqref{StokesDisc} and the relevant Stokes automorphisms are given in \eqref{ell-linear-Stokes-00}--\eqref{ell-linear-Stokes-02}. For $\Phi_0$, it follows
\be
\sum_{n=0}^{+\infty} Z^{(0)}_n\, x^n \simeq \frac{S_{\omega_{01}}}{2\pi\mathrm{i}}\, \int_0^{+\infty} \mathrm{d}w\, \frac{1}{w-x}\, \mathrm{e}^{-\frac{A_1}{w}}\, \sum_{k=0}^{+\infty} Z^{(1)}_k\, w^k + \frac{S_{\omega_{02}}}{2\pi\mathrm{i}}\, \int_0^{-\infty} \mathrm{d}w\, \frac{1}{w-x}\, \mathrm{e}^{-\frac{A_2}{w}}\, \sum_{k=0}^{+\infty}Z^{(2)}_k\,w^k,
\ee
\noindent
leading to
\be
\label{large_order_aA-v2}
Z^{(0)}_n \simeq \frac{S_{\omega_{01}}}{2\pi\mathrm{i}}\, \frac{\Gamma\left(n\right)}{A_1^n}\, \sum_{k=0}^{+\infty} \frac{\Gamma\left(n-k\right)}{\Gamma\left(n\right)}\, A_1^k\, Z^{(1)}_k + \frac{S_{\omega_{02}}}{2\pi\mathrm{i}}\, \frac{\Gamma\left(n\right)}{A_2^n}\, \sum_{k=0}^{+\infty} \frac{\Gamma\left(n-k\right)}{\Gamma\left(n\right)}\, A_2^k\, Z^{(2)}_k.
\ee
\noindent
This exactly matches \eqref{large_order_aA}, once the Stokes coefficients are identified as $S_{\omega_{01}}=S_{\omega_{02}}=2$.

Let us take another look at \eqref{large_order_aA} or \eqref{large_order_aA-v2}. It is clear that the large-order coefficients in the perturbative expansion are dominated at next-to-leading order by the nearest action to the origin, which, depending on whether $m<1/2$ or $m>1/2$, is either $A_1$ or $A_2$. However, at the special point where $m=1/2=m^\prime$ both instanton actions will contribute \textit{symmetrically}. In particular, this will cause all the \textit{odd} powers $Z^{(0)}_{2\BN+1}$ to \textit{vanish} (which can already be seen directly from \eqref{eq:linear-ell-lo}). This is a particular case of resonant behaviour, where $A_1+A_2=0$.

The large-order behaviours of the series $\Phi_1$ and $\Phi_2$ can be determined in the exact same way. Repeating such familiar procedure, the large-order growth of $\Phi_1$ is found to be
\begin{eqnarray}
\label{large_order_aB}
Z^{(1)}_n (m) &\simeq& \frac{S_{\omega_{10}}}{2\pi\mathrm{i}}\, \frac{\Gamma\left(n\right)}{\left(-A_1\right)^n} \left( Z^{(0)}_0 + \frac{-A_1}{n-1}\, Z^{(0)}_1 + \frac{\left(-A_1\right)^2}{\left(n-1\right) \left(n-2\right)}\, Z^{(0)}_2 + \cdots \right) + \\
&&
\hspace{-30pt}
+ \frac{S_{\omega_{12}}+\frac{1}{2}S_{\omega_{02}}S_{\omega_{10}}}{2\pi \mathrm{i}}\, \frac{\Gamma\left(n\right)}{\left(A_2-A_1\right)^n} \left( Z^{(2)}_0 + \frac{A_2-A_1}{n-1}\, Z^{(2)}_1 + \frac{\left(A_2-A_1\right)^2}{\left(n-1\right) \left(n-2\right)}\, Z^{(2)}_2 + \cdots \right).
\nonumber 
\end{eqnarray}
\noindent
Recall that $S_{\omega_{10}}=-2$ and $S_{\omega_{12}}+\frac{1}{2}S_{\omega_{02}}S_{\omega_{10}}=-2$. Further recall that $-A_1$ is actually the relative action $A_0-A_1$, where $A_0=0$ is the ``perturbative action''. Both relative actions $A_0-A_1$ and $A_2-A_1$ are negative, in which case $\Phi_1$ only has Borel singularities along the negative real axis. Similarly, the large-order growth of $\Phi_2$ is
\begin{eqnarray}
Z^{(2)}_n (m) &\simeq& \frac{S_{\omega_{20}}}{2\pi\mathrm{i}}\, \frac{\Gamma\left(n\right)}{\left(-A_2\right)^n} \left( Z^{(0)}_0 + \frac{-A_2}{n-1}\, Z^{(0)}_1 + \frac{\left(-A_2\right)^2}{\left(n-1\right) \left(n-2\right)}\, Z^{(0)}_2 + \cdots \right) + \\
&&
\hspace{-30pt}
+ \frac{S_{\omega_{21}}+\frac{1}{2}S_{\omega_{01}}S_{\omega_{20}}}{2\pi \mathrm{i}}\, \frac{\Gamma\left(n\right)}{\left(A_1-A_2\right)^n} \left( Z^{(1)}_0 + \frac{A_1-A_2}{n-1}\, Z^{(1)}_1 + \frac{\left(A_1-A_2\right)^2}{\left(n-1\right) \left(n-2\right)}\, Z^{(1)}_2 + \cdots \right).
\nonumber
\end{eqnarray}
\noindent
Herein one should recall that $S_{\omega_{20}}=-2$ and $S_{\omega_{21}}+\frac{1}{2}S_{\omega_{01}}S_{\omega_{20}}=-2$. Further, and in contrast to $\Phi_1$, the Borel singularities of $\Phi_2$ are now located along the positive real axis. In summary, we have made it rather explicit how the large-order behaviour of \textit{each} formal series $\Phi_i$ is controlled by the relative actions with respect to the \textit{other} critical points. The most dominant contribution always comes from the nearest critical point $j$ such that $|A_j-A_i|$ is the smallest. Furthermore, for $x\in\mathbb R^+$, if $A_j>A_i$ then the contribution of the critical point $j$ to the large-order growth of $i$ is non-alternating whereas for $A_j<A_i$ it is alternating.  

\begin{figure}[t!]
\begin{center}
\includegraphics[height=5.1cm]{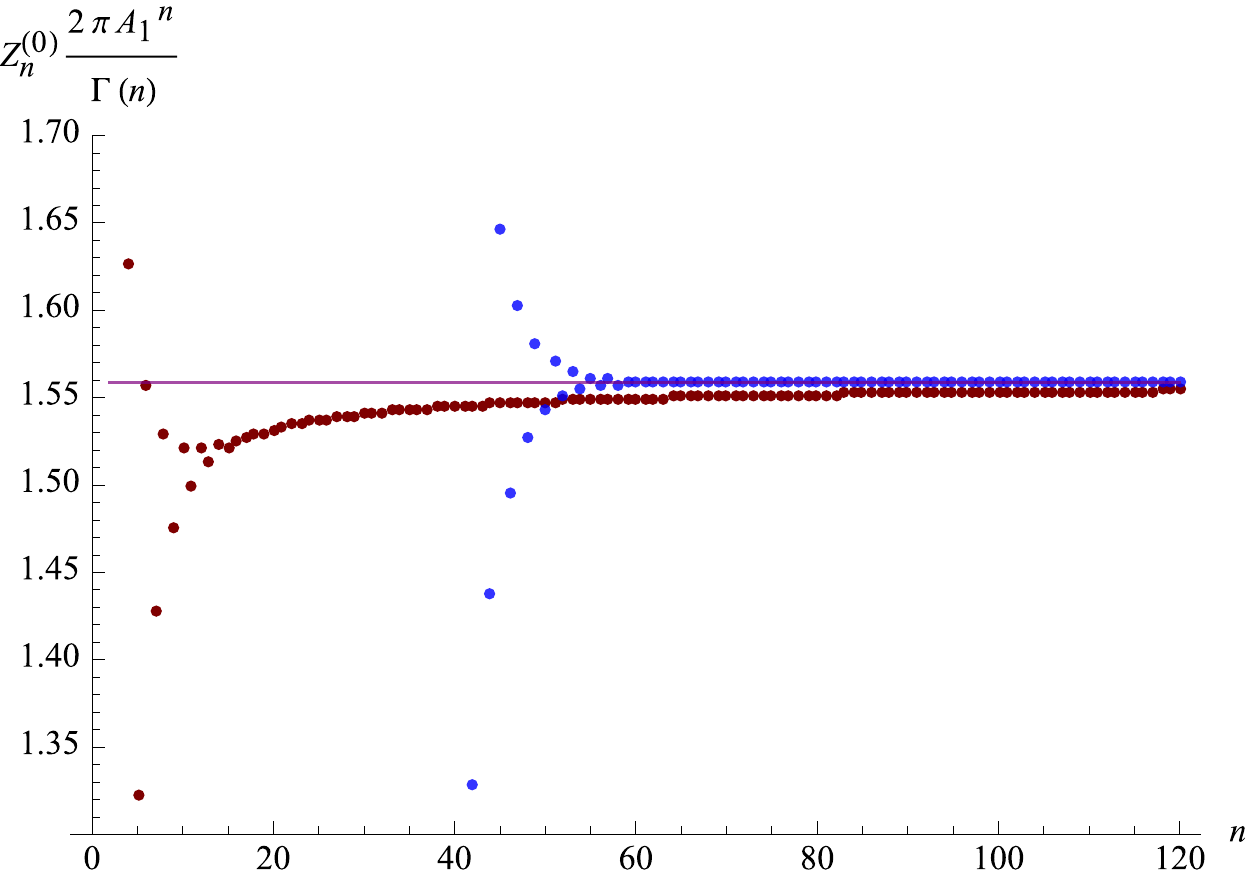}
$\qquad$
\includegraphics[height=5.1cm]{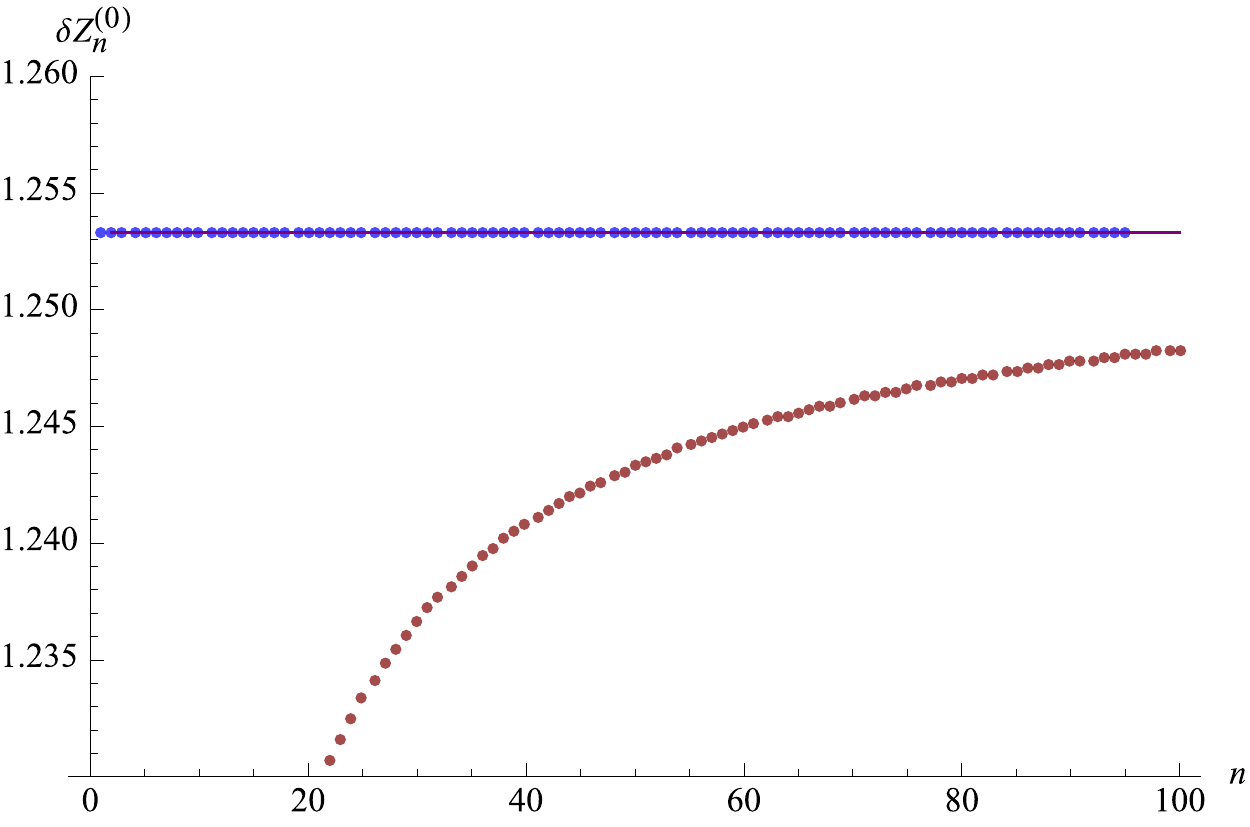}
\end{center}
\caption{Numerical plots of the asymptotic large-order growth of the perturbative series coefficients, $Z_n^{(0)}$ (left), and its subleading exponentially-suppressed growth (right), as predicted by \eqref{large_order_aA} (with $m=\pi/8$ in the plots). On the left we plot the coefficients $Z_n^{(0)}$ weighted by their growth factor (in red), together with the corresponding fifth Richardson transform (in blue). The sequence is seen to converge to the predicted value $-S_{\omega_{01}} \mathrm{i}\, Z^{(1)}_0 = 2\sqrt{1-m}$ (in purple). On the right we plot the weighted resummation \eqref{ell-partfunc-resummed} (in red), the corresponding fifth Richardson transform (in blue), and the predicted value of convergence, $-S_{\omega_{02}} \mathrm{i}\, Z^{(2)}_0 = 2\sqrt{m}$ (in purple). The errors associated with this numerical calculation are $\sim\mathcal{O}(10^{-13})$.}
\label{fig:large_order}
\end{figure}

In linear examples we are fortunate to have closed-form expressions for the asymptotic-series coefficients, from which one may analyze large-order behaviours without resorting to numerical calculations. In most nonlinear examples, however, this is not the case and asymptotic-series coefficients can only be computed recursively up to some fixed given order. Nevertheless, we have learnt in section~\ref{sec:quartic} how to fit the large-order behaviour predicted by resurgence to such recursive (numerical) data; and in this way extract the values of the Stokes coefficients. In the present (simple) linear scenario, figure~\ref{fig:large_order} illustrates such fit. On the left image, we plot the coefficients $Z^{(0)}_n$ weighted by the growth factor $2\pi A_1^n/\Gamma\left(n\right)$, and show the convergence towards the predicted value  \eqref{large_order_aA}, given by $-S_{\omega_{01}} \mathrm{i}\, Z^{(1)}_0 = 2\sqrt{1-m}$ (where in the plot we set $m=\pi/8$). For our choice of $m$, the large-order growth is dominated by the critical point $z^*_1$ since $A_1<|A_2|$. In order to numerically check the exponentially-suppressed contributions associated with $A_2$ in \eqref{large_order_aA}, we first need to resum the asymptotic large-order series\footnote{See subsection~\ref{sec:large-order-F} for definitions and a thorough introduction to this whole procedure.}
\be
\chi_{0\rightarrow1}(n) \simeq \sum_{k=0}^{+\infty} \frac{\Gamma\left(n-k\right)}{\Gamma\left(n\right)}\, A_1^k\, Z^{(1)}_k.
\ee
\noindent
With this result in hand one may then plot the weighted resummation
\be
\label{ell-partfunc-resummed}
\delta Z^{(0)}_n \equiv \left( Z^{(0)}_n\, \frac{2\pi A_1^n}{\Gamma\left(n\right)} - \left( -\mathrm{i} S_{\omega_{01}} \right) \mathcal{S}_0 \mathrm{BP}_{100} \left[\chi_{0\rightarrow1}(n)\right] \right) \left( \frac{A_2}{A_1} \right)^n,
\ee
\noindent
which is done on the right-hand plot of figure~\ref{fig:large_order} (and where $\mathrm{BP}_{100} \left[\chi_{0\rightarrow1}(n)\right]$ is the resummed Borel--Pad\'{e} approximant associated to the series $\chi_{0\rightarrow1}(n)$). We can very clearly see the convergence towards the expected value, $-S_{\omega_{02}} \mathrm{i}\, Z^{(2)}_0 = 2\sqrt{m}$.

\subsection{First Steps Towards Nonlinear Resurgent Analysis}\label{subsec:nonlinearFnonreso}

Having studied the linear problem associated with the partition function, let us turn to its closely related nonlinear problem---in parallel with subsection~\ref{subsec:Fnonlinear}---namely, the ``elliptic free energy''. Our motivation is dual. On the one hand this is a more intricate nonlinear problem than the one addressed in section~\ref{sec:quartic}, and it will help the reader gain further insight into the general structure of transseries with infinitely many instanton sectors; their asymptotics and their large-order behaviour. But, on the other hand and more importantly, the present non-resonant set-up will lay the ground to later address \textit{resonance} within nonlinear problems which is the main goal of the present section (to be addressed in the upcoming subsections~\ref{sec:nonlinear-resonance} and on).

The elliptic-potential free energy, $F = \log Z$, satisfies a \textit{nonlinear} third-order ODE (compare with the quartic-potential example back in \eqref{quarticNLODE}),
\begin{eqnarray}
&&
4 m m^\prime x^4 \left( F'''(x) + 3 F'(x) F''(x) + \left( F'(x) \right)^3 \right) - 4 x^2 \left( m - m^\prime - 6 m m^\prime x \right) \left( F''(x) + \left( F'(x) \right)^2 \right) - \nonumber \\
&&
- \left( 4 + 8 \left( m - m^\prime \right) x - 27 m m^\prime x^2 \right) F'(x) - \left( m - m^\prime - 3 m m^\prime x \right) = 0.
\label{eq:nl_ell_ode}
\end{eqnarray}
\noindent
Due to the nonlinearity of this equation, there will be infinitely many instanton sectors in the resulting transseries. Before diving into these nonperturbative sectors let us address the asymptotic perturbative solution, with standard \textit{ansatz}
%
\begin{figure}[t!]
\begin{center}
\includegraphics[width=7.4cm]{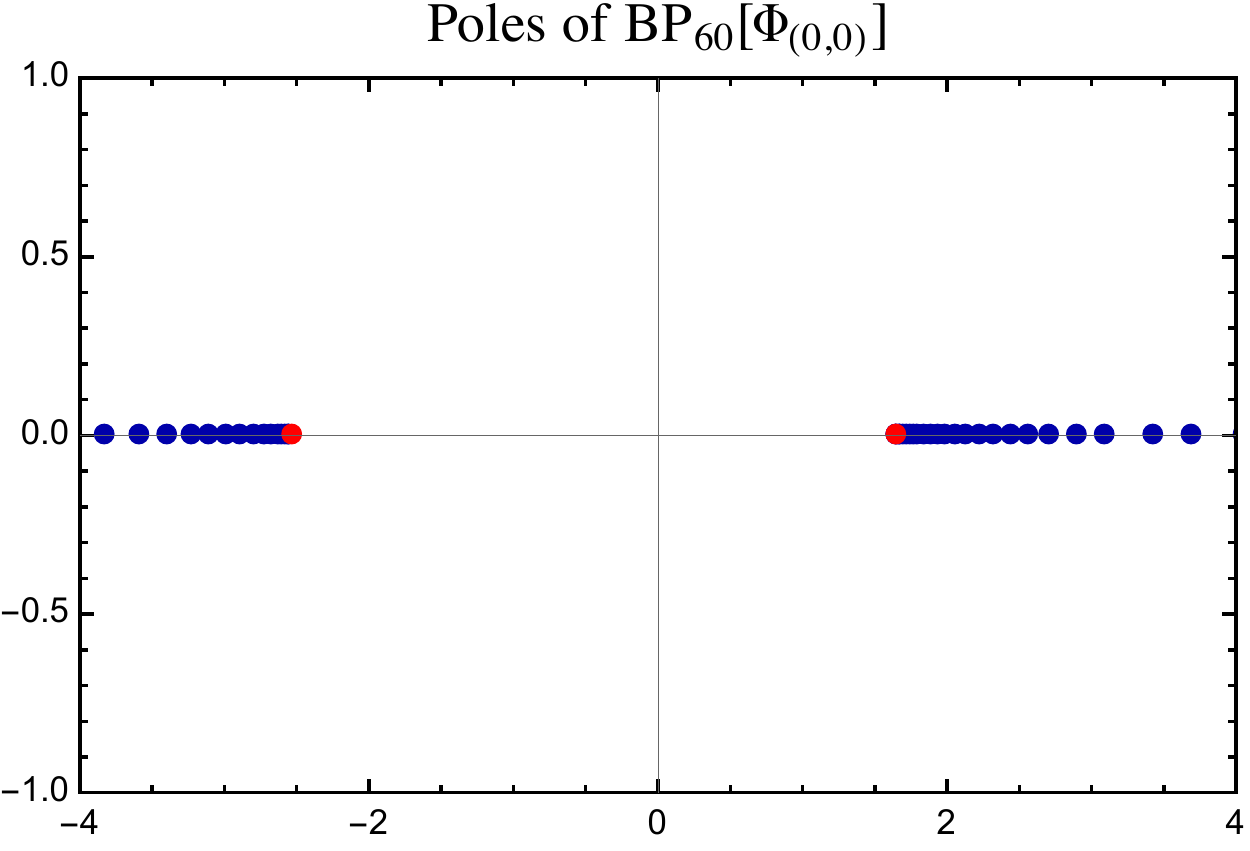}
$\qquad$
\includegraphics[width=7.4cm]{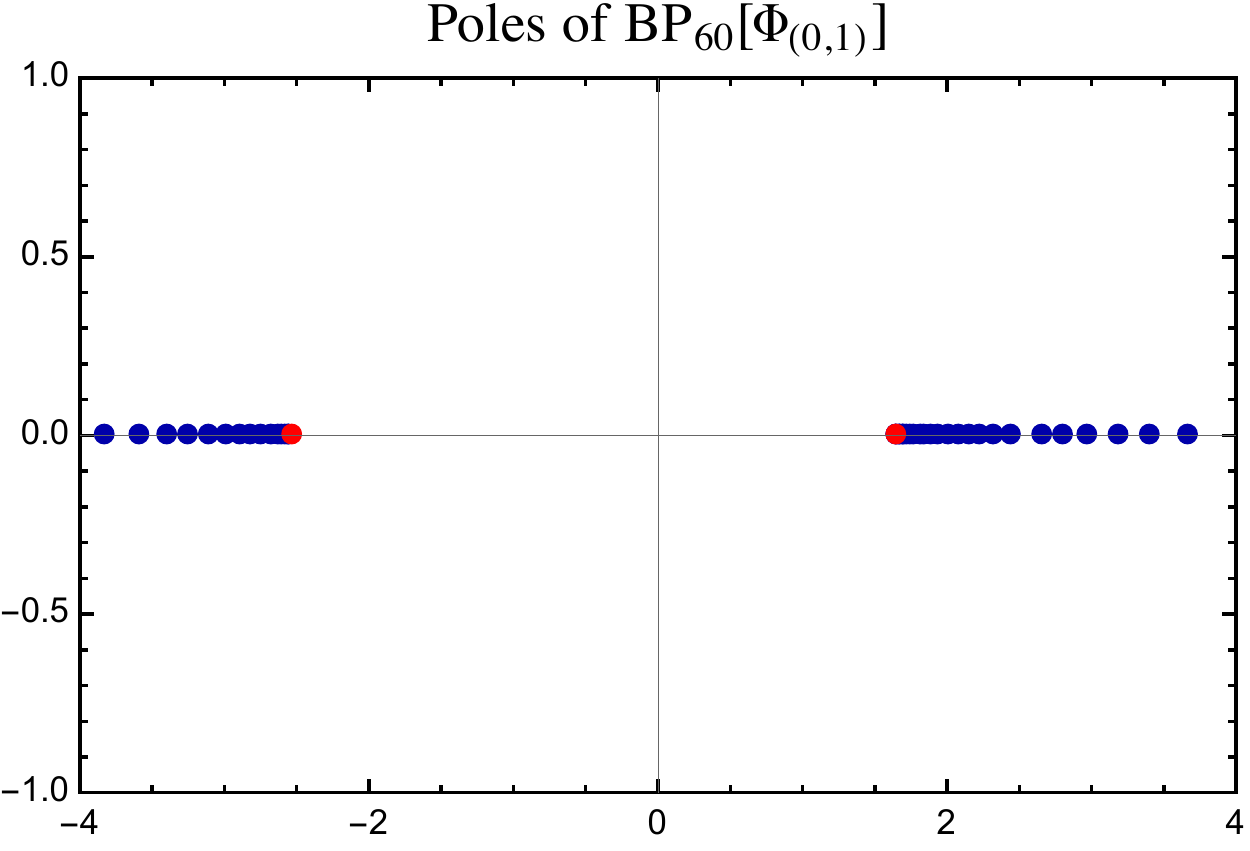}
\end{center}
\caption{Poles of the (diagonal) order-60 Pad\'e approximants for the Borel transforms of perturbative ($\mathrm{BP}_{60} [\Phi_{(0,0)}]$; left image) and $(0,1)$-instanton ($\mathrm{BP}_{60} [\Phi_{(0,1)}]$; right image) sectors in the free energy of the elliptic potential, with the asymptotic series truncated at $N=120$. The poles (shown in blue) are all on the positive real line, condensing into branch-cuts which start at either $A_1 = 1/(1-m)$ and $A_2 = -1/m$ (instanton actions shown in red). In these plots we have set $m=\pi/8$.}
\label{fig:borel-plane}
\end{figure}
%
\be
F(x) \simeq x^\beta\, \sum_{k=0}^{+\infty} F_k^{(0)} x^k.
\ee
\noindent
Plugging this perturbative \textit{ansatz} into \eqref{eq:nl_ell_ode} yields a recursion relation\footnote{This recursion relation can be found in appendix~\ref{app:recursive-free-en}, by simply setting $\boldsymbol{n} = \left( n_1, n_2 \right) = \boldsymbol{0}$ in the general recursion.} for the perturbative coefficients $F_k^{(0)}$. The reader may check that this leads to an asymptotic series whose Borel transform has singularities along the real axis. Without surprise, these singularities are branch points, located at $A_1 = 1/(1-m)$ and $A_2 = -1/m$. This is depicted in the left plot of figure~\ref{fig:borel-plane}, where we used a symmetric $N=60$ Pad\'e approximant (recall \eqref{pade-S}) for the Borel transform of $F(x)\equiv\Phi_{(0,0)}(x)$. The sequence of poles which accumulate at $A_1 \approx 1.65$ and $A_2 \approx -2.55$ reflects the underlying singularities of the Borel transform which are closest to the origin (we chose $m=\pi/8$ in the plots). The fact that there are two distinct branch points implies the existence of at least \textit{two} separate instanton sectors, with actions $A_1$ and $A_2$. Indeed, considering a slightly more general \textit{ansatz} for the solution of \eqref{eq:nl_ell_ode}, of the form (recall \eqref{oneparametertransseriesquartic})
\be
F(x) \simeq \mathrm{e}^{-\frac{\ell A}{x}}\, x^{\ell \beta}\, \sum_{n=0}^{+\infty} F_n^{(\ell)}\, x^n,
\ee
\noindent
one easily finds that a solution exists only if either $A = 0$, $A = 1/m^\prime$, or $A = -1/m$, and $\beta=0$. The existence of these three sectors---apparently one perturbative and two nonperturbative---is expected given that the free energy satisfies a third-order ODE, and at first one might expect that the general solution is a two-parameter transseries (\textit{i.e.}, one per nonperturbative sector). However, and in the exact same fashion as it happened for the quartic-potential starting in \eqref{two-param-quartic} through \eqref{one-point-five-transseries}, this is not the full story: the above $A = 0$ sector is \textit{not} quite the perturbative sector. In other words, in addition to the nonperturbative sectors associated with the non-zero instanton actions $A_1$ and $A_2$, there is one extra sector whose ``action'' vanishes, but which is \textit{distinct} from the perturbative sector. Furthermore, there are no perturbative fluctuations associated with this sector; it is merely a \textit{constant}. Of course that the reader is already familiar with this phenomenon, given the experience acquired in section~\ref{sec:quartic} and comparing \eqref{eq:nl_ell_ode} back to \eqref{quarticNLODE}. Also now, the existence of such a non-asymptotic, or ``silent'' sector---henceforth denoted by $\widetilde{\Phi}_{(0,0)}$---is associated with the fact that the third-order ODE \eqref{eq:nl_ell_ode} only depends on \textit{derivatives} of the free energy, and hence it can be trivially integrated into a \textit{second}-order ODE by a redefinition $\mathfrak{F} (x) \equiv F'(x)$. In this way, this new sector can then be identified with the integration constant of the redefinition. Consequently, one effectively has a ``two-and-a-half'' transseries: a three-parameter transseries where one of the sectors, $\widetilde{\Phi}_{(0,0)}$ associated to the ``instanton action'' $A=0$, is not asymptotic and essentially decouples from the asymptotic sectors associated to the non-zero instanton actions $A_1$ and $A_2$ (which themselves give rise to a \textit{bona fide} two-parameter transseries). Labelling these two truly nonperturbative sectors as $\Phi_{(0,n_1,n_2)}$ and denoting the constant sector by $\Phi_{(1,0,0)} \equiv \widetilde{\Phi}_{(0,0)}$, this structure leads to the transseries \textit{ansatz}
\be
\label{eq:nl_ell_ansatz-v0}
F (x, \sigma_0, \sigma_1, \sigma_2 ) = \sigma_0\, \Phi_{(1,0,0)} + \sum_{n_1=0}^{+\infty} \sum_{n_2=0}^{+\infty} \sigma_1^{n_1} \sigma_2^{n_2}\, \rme^{-\frac{n_1\,A_1+n_2\,A_2}{x}}\, \Phi_{(0,n_1,n_2)} (x),
\ee
\noindent
where
\be
\label{eq:ell-nl-sectors}
\Phi_{(0,n_1,n_2)} (x) \simeq \sum_{k=0}^{+\infty} F^{(n_1,n_2)}_k\, x^k.
\ee
\noindent
Note how this structure is completely analogous to the free energy of the quartic potential discussed in section~\ref{sec:quartic}. Because of the simple form of the ``third'' transseries sector (\textit{i.e.}, the one related to the transseries parameter $\sigma_0$), one is led to further denote $\Phi_{(0,n_1,n_2)}(x) \equiv \Phi_{(n_1,n_2)}(x)$, and finally rewrite the full transseries for the free energy as
\be
\label{eq:nl_ell_ansatz}
F \left( x, \sigma_0, \boldsymbol{\sigma} \right) = \sigma_0\, \widetilde{\Phi}_{(0,0)} + \sum_{\boldsymbol{n} \in \BN_0^2} \boldsymbol{\sigma}^{\boldsymbol{n}}\, \rme^{- \frac{\boldsymbol{n} \cdot \boldsymbol{A}}{x}}\, \Phi_{\boldsymbol{n}}(x) .
\ee
\noindent
We follow the multi-dimensional\footnote{Recall from section~\ref{sec:physics} that the sum over $\boldsymbol{n}\in\mathbb{N}_0^2$ corresponds to summing over $n_1\ge0$ and $n_2\ge0$.} notation of section~\ref{sec:physics}, defining vectors $\boldsymbol{A} = \left( A_1, A_2 \right)$, $\boldsymbol{\sigma} = \left( \sigma_1, \sigma_2 \right)$ and $\boldsymbol{n} = \left( n_1, n_2 \right)$, and  products $\boldsymbol{\sigma}^{\boldsymbol{n}} = \sigma_1^{n_1}\, \sigma_2^{n_2}$ and $\boldsymbol{n} \cdot \boldsymbol{A} = n_1\,A_1 + n_2\,A_2$.

Plugging the final transseries \textit{ansatz} \eqref{eq:nl_ell_ansatz} into the free-energy ODE \eqref{eq:nl_ell_ode} leads to recursion relations for \textit{all} asymptotic coefficients $F^{(n_1,n_2)}_k$ (as usual, by simply keeping track of equal powers of $\mathrm{e}^{-\frac{\boldsymbol n \cdot \boldsymbol A}{x}}$ and of $x^k$). Even though these recursion relations are relatively straightforward to obtain, they are rather lengthy and not particularly illuminating---as such, we have included them in appendix~\ref{app:recursive-free-en} rather than presenting them explicitly in the main body of the text. Note, however, that all the coefficients $F^{(\boldsymbol{n})}_k \equiv F^{(\boldsymbol{n})}_k (m)$ are \textit{polynomials} in the modular parameter $m$. 

The next step to address concerns resurgence relations in-between different sectors $\Phi_{\boldsymbol{n}}$. This is best done starting with the pictorial language of alien lattices introduced in section~\ref{sec:physics}. With the experience accumulated throughout section~\ref{sec:quartic}, where we were dealing with a ``1.5 dimensional'' alien chain as in figure~\ref{fig:first-5-sector-alien-one-point-five}, it is clear from \eqref{eq:nl_ell_ansatz} that we are now dealing with a ``2.5 dimensional'' alien lattice which should similarly extend figure~\ref{fig:sec5-2d-lattice}. One can then start by identifying each sector $\Phi_{\boldsymbol{n}}$ with a lattice node labeled by the two-dimensional lattice vector $\boldsymbol{n} = (n_1,n_2)$. The extra ``half dimension'' follows as an additional lattice-site which corresponds to the constant $\widetilde{\Phi}_{(0,0)}$ sector. Having this comparison between ``1.5'' and ``2.5'' transseries in mind---and recalling how in the ``1.5'' case the structure of Borel singularities \eqref{Borel-transf-expanded-one-point-five} was a straightforward extension of \eqref{Borel-transf-expanded-one-param}---, uncovering the local structure of Borel singularities in the ``2.5'' case should be equally straightforward. In parallel with \eqref{Borel-transf-expanded-one-point-five}, we expect only mild changes to \eqref{eq:sec5-Borel-transf} and find\footnote{Recall that $\widetilde{\Phi}_{(0,0)}$ is a constant and Borel transforms are not defined for constant terms. This caveat is not a big problem: one simply needs to define them as an identity on the Borel plane in order to conduct calculations.}
\be
\label{eq:elliptic-bridge-borel}
\CB [\Phi_{\boldsymbol{n}}] ( s + \boldsymbol{\ell} \cdot \boldsymbol{A} ) = \left( \mathsf{S}_{\boldsymbol{n} \to \boldsymbol{n}+\boldsymbol{\ell}} \times \CB [ \Phi_{\boldsymbol{n}+\boldsymbol{\ell}} ] (s) + \delta_{\boldsymbol{n}+\boldsymbol{\ell}}\, \mathsf{S}_{\boldsymbol{n} \to \widetilde{\boldsymbol{0}}} \times \CB [\widetilde{\Phi}_{(0,0)}] (s) \right) \frac{\log s}{2\pi\rmi}, \qquad \boldsymbol{\ell} \neq \boldsymbol{0}.
\ee

Moving towards the usual alien-derivative algebraic level of abstraction should by now also be quite straightforward, and the process is essentially the same which led to \eqref{eq:sec5-bridge-eqs} and \eqref{eq:sec5-bridge-eqs-COMPLETE}. The resurgence relations associated to \eqref{eq:elliptic-bridge-borel} are thus given by
\be
\label{eq:elliptic-bridge}
\Delta_{\boldsymbol{\ell} \cdot \boldsymbol{A}} \Phi_{\boldsymbol{n}} =  \boldsymbol{S}_{\boldsymbol{\ell}} \cdot \left(\boldsymbol{n}+\boldsymbol{\ell}\right) \Phi_{\boldsymbol{n}+\boldsymbol{\ell}} + \delta_{\boldsymbol{n}+\boldsymbol{\ell}}\, S^{(0)}_{\boldsymbol{\ell}}\, \widetilde{\Phi}_{(0,0)}.
\ee
\noindent
Let us make a couple of remarks about this equation. By definition, $\Delta_{\boldsymbol{\ell} \cdot \boldsymbol{A}} \Phi_{\boldsymbol{n}} = 0$ whenever any component is $\ell_i < - n_i$ (since $\Phi_{\boldsymbol{m}}$ vanishes whenever one $m_j$ is negative). Further, $\Delta_{\boldsymbol{\ell} \cdot \boldsymbol{A}} \widetilde{\Phi}_{(0,0)} = 0$ for any $\boldsymbol{\ell}$, as the sector $\widetilde{\Phi}_{(0,0)}$ is not asymptotic. The \textit{Stokes vectors} in \eqref{eq:elliptic-bridge} are \textit{two}-dimensional, $\boldsymbol{S}_{\boldsymbol{\ell}} = ( S_{\boldsymbol{\ell}}^{(1)}, S_{\boldsymbol{\ell}}^{(2)} )$, but---as thoroughly discussed in section~\ref{sec:physics}---many of their coefficients are zero. In particular, writing $\boldsymbol{\ell} =  ( \ell_1,\ell_2 )$,
\bea
\label{eq.S.coeff.a}
S^{(i)}_{\boldsymbol{0}} &=& 0, \qquad \text{ for }\, i=0,1,2, \\
\label{eq.S.coeff.b}
S^{(0)}_{\boldsymbol{\ell}} &=& 0, \qquad \text{ if any }\, \ell_i>0, \\
\label{eq.S.coeff.c}
S^{(1)}_{\boldsymbol{\ell}} &=& 0, \qquad \text{ if }\, \ell_1>1 \text{  or  } \ell_2>0, \\
\label{eq.S.coeff.d}
S^{(2)}_{\boldsymbol{\ell}} &=& 0, \qquad \text{ if }\, \ell_1>0 \text{  or  } \ell_2>1.
\eea
\noindent
This is a simple generalization\footnote{As usual, these results may be derived rather than just motivated by starting-off with the \textit{bridge equations}, and we included such technical alien-calculus results in appendix~\ref{app:alien-calculus}. For the reader who rather stay mid-way between motivation and derivation, let us just flash the argument. At the level of the full transseries \eqref{eq:nl_ell_ansatz}, the \textit{bridge equations} of this problem have the form
\be
\rme^{- \frac{\boldsymbol{\ell} \cdot \boldsymbol{A}}{x}} \Delta_{\boldsymbol{\ell} \cdot \boldsymbol{A}} F \left( x, \sigma_0, \boldsymbol{\sigma} \right) = \sum_{i=0}^2 S^{(i)}_{\boldsymbol{\ell}} (\sigma_0, \boldsymbol{\sigma})\, \frac{\partial F}{\partial \sigma_i} \left( x, \sigma_0, \boldsymbol{\sigma} \right),
\ee
\noindent
where, via the arguments of degree and homogeneity of appendix~\ref{app:alien-calculus} (also addressed in section~\ref{sec:physics}), the proportionality Stokes coefficients must be monomials of the form (and with a slight abuse of notation)
\begin{equation}
S^{(0)}_{\boldsymbol{\ell}} (\sigma_0, \boldsymbol{\sigma}) \equiv S^{(0)}_{\boldsymbol{\ell}}\, \sigma_1^{-\ell_1}\, \sigma_2^{-\ell_2},
\qquad
S^{(1)}_{\boldsymbol{\ell}} (\sigma_0, \boldsymbol{\sigma}) \equiv S^{(1)}_{\boldsymbol{\ell}}\, \sigma_1^{1-\ell_1}\, \sigma_2^{-\ell_2},
\qquad
S^{(2)}_{\boldsymbol{\ell}} (\sigma_0, \boldsymbol{\sigma}) \equiv S^{(2)}_{\boldsymbol{\ell}}\, \sigma_1^{-\ell_1}\, \sigma_2^{1-\ell_2}.
\end{equation}
\noindent
The allowed values of $\boldsymbol{\ell}$ are then such that the \textit{bridge equations} are \textit{regular} in $\boldsymbol{\sigma}$ (which is our result above).
\label{eq:elliptic-bridge0}} from the two-parameter case encoded in \eqref{eq:sec5-Stokes-const-condition} or \eqref{eq:sec5-bridge-eqs-COMPLETE} and illustrated in figure~\ref{fig:sec5-2d-lattice}. Of course that once again the Stokes vectors appearing in the resurgence relations \eqref{eq:elliptic-bridge} are directly related to the Borel residues appearing in the Borel transforms \eqref{eq:elliptic-bridge-borel}. This occurs in the exact same manner as was explained in section~\ref{sec:physics}. The extra residue, $\mathsf{S}_{\boldsymbol{n} \to \widetilde{\boldsymbol{0}}}$, simply appears due to the existence of the extra (non-asymptotic) sector $\widetilde{\Phi}_{(0,0)}$.

\begin{figure}[t!]
\begin{center}
\begin{tikzpicture}[>=latex,decoration={markings, mark=at position 0.6 with {\arrow[ultra thick]{stealth};}} ]
\begin{scope}[node distance=2.5cm]
  \node (Phi00) [draw] at (0,0) {$\Phi_{(0,0)}$};
  \node (Phi10) [right of=Phi00] [draw] {$\Phi_{(1,0)}$};
  \node (Phi20) [right of=Phi10] [draw] {$\Phi_{(2,0)}$};
  \node (Phi30) [right of=Phi20] [draw] {$\Phi_{(3,0)}$};
  \node (Phi40) [right of=Phi30] [draw] {$\Phi_{(4,0)}$};
  \node (Phi50) [right of=Phi40] [draw] {$\Phi_{(5,0)}$};
  \node (cdots50) [right of=Phi50]  {$\cdots$};
  \node (Phit00) [below of=Phi00] [draw] {$\widetilde{\Phi}_{(0,0)}$};
  \node (Phi01) [above of=Phi00] [draw] {$\Phi_{(0,1)}$};
  \node (Phi02) [above of=Phi01] [draw] {$\Phi_{(0,2)}$};
  \node (Phi03) [above of=Phi02] [draw] {$\Phi_{(0,3)}$};
  \node (Phi04) [above of=Phi03] [draw] {$\Phi_{(0,4)}$};
  \node (cdots05) [above of=Phi04]  {$\vdots$};
  \node (Phi11) [above of=Phi10] [draw] {$\Phi_{(1,1)}$};
  \node (Phi21) [above of=Phi20] [draw] {$\Phi_{(2,1)}$};
  \node (Phi31) [above of=Phi30] [draw] {$\Phi_{(3,1)}$};
  \node (Phi41) [above of=Phi40] [draw] {$\Phi_{(4,1)}$};
  \node (Phi51) [above of=Phi50] [draw] {$\Phi_{(5,1)}$};
  \node (cdots51) [above of=cdots50]  {$\cdots$};
  \node (Phi12) [above of=Phi11] [draw] {$\Phi_{(1,2)}$};
  \node (Phi22) [above of=Phi21] [draw] {$\Phi_{(2,2)}$};
  \node (Phi32) [above of=Phi31] [draw] {\color{red}$\Phi_{(3,2)}$};
  \node (Phi42) [above of=Phi41] [draw] {$\Phi_{(4,2)}$};
  \node (Phi52) [above of=Phi51] [draw] {$\Phi_{(5,2)}$};
  \node (cdots52) [above of=cdots51]  {$\cdots$};
  \node (Phi13) [above of=Phi12] [draw] {$\Phi_{(1,3)}$};
  \node (Phi23) [above of=Phi22] [draw] {$\Phi_{(2,3)}$};
  \node (Phi33) [above of=Phi32] [draw] {$\Phi_{(3,3)}$};
  \node (Phi43) [above of=Phi42] [draw] {$\Phi_{(4,3)}$};
  \node (Phi53) [above of=Phi52] [draw] {$\Phi_{(5,3)}$};
  \node (cdots53) [above of=cdots52]  {$\cdots$};
  \node (Phi14) [above of=Phi13] [draw] {$\Phi_{(1,4)}$};
  \node (Phi24) [above of=Phi23] [draw] {$\Phi_{(2,4)}$};
  \node (Phi34) [above of=Phi33] [draw] {$\Phi_{(3,4)}$};
  \node (Phi44) [above of=Phi43] [draw] {$\Phi_{(4,4)}$};
  \node (Phi54) [above of=Phi53] [draw] {$\Phi_{(5,4)}$};
  \node (cdots54) [above of=cdots53]  {$\cdots$};
  \node (cdots15) [above of=Phi14] {$\vdots$};
  \node (cdots25) [above of=Phi24] {$\vdots$};
  \node (cdots35) [above of=Phi34] {$\vdots$};
  \node (cdots45) [above of=Phi44] {$\vdots$};
  \node (cdots55) [above of=Phi54]  {$\vdots$};
  \node (cdotsbb) [above of=cdots54]  {$\iddots$};
\end{scope}
\draw [thick,postaction={decorate},-,>=stealth,shorten <=2pt,shorten >=2pt] (Phi32.east)  -- (Phi42.west);
\draw[thick,white!30!black,postaction={decorate},-,>=stealth,shorten <=2pt,shorten >=2pt] (Phi32.south) .. controls +(-1.3,-0.5) and +(0.6,0.5)  ..  (Phi21.east);
\draw[thick,white!70!black,postaction={decorate},-,>=stealth,shorten <=2pt,shorten >=2pt] (Phi32.west) .. controls +(-2.3,-1.8) and +(0.4,2.9)  ..  (Phit00.north);
\draw [dashed,thick,postaction={decorate},-,>=stealth,shorten <=2pt,shorten >=2pt] (Phi32.west)  -- (Phi22.east);
\draw[dashed,thick,white!40!black,postaction={decorate},-,>=stealth,shorten <=2pt,shorten >=2pt] (Phi32.west) .. controls +(-1.,-0.95) and +(0.3,2.5)  ..  (Phi01.north);
\draw [thick,blue!60!black,postaction={decorate},-,>=stealth,shorten <=2pt,shorten >=2pt] (Phi32.south)  -- (Phi31.north);
\draw[thick,blue!99!black,postaction={decorate},-,>=stealth,shorten <=2pt,shorten >=2pt] (Phi32.south) .. controls +(-0.9,-0.8) and +(1.7,0)  ..  (Phi10.east);
\draw[dashed,thick,blue!99!black,postaction={decorate},-,>=stealth,shorten <=2pt,shorten >=2pt] (Phi32.north) .. controls +(-1.3,0.7) and +(1.2,0.9)  ..  (Phi12.north);
\draw [dashed,thick,blue!60!black,postaction={decorate},-,>=stealth,shorten <=2pt,shorten >=2pt] (Phi32.north)  -- (Phi33.south);
\draw [thick,red!60!black,postaction={decorate},-,>=stealth,shorten <=2pt,shorten >=2pt] (Phi32.east) .. controls +(1.5,-0.1) and +(0,0.5)  ..  (Phi41.north);
\draw[thick,red!80!black,postaction={decorate},-,>=stealth,shorten <=2pt,shorten >=2pt] (Phi32.south) .. controls +(-1.1,-1.6) and +(1.3,1.3)  ..  (Phi20.east);
\draw[dashed,thick,red!80!black,postaction={decorate},-,>=stealth,shorten <=2pt,shorten >=2pt] (Phi32.north) .. controls +(-1.,1.3) and +(0.7,1.5)  ..  (Phi02.north);
\draw [dashed,thick,red!60!black,postaction={decorate},-,>=stealth,shorten <=2pt,shorten >=2pt] (Phi32.north) .. controls +(-0.1,1.7) and +(0.5,0)  ..  (Phi23.east);
\draw[thick,green!40!black,postaction={decorate},-,>=stealth,shorten <=2pt,shorten >=2pt] (Phi32.east) .. controls +(1.3,-0.7) and +(1.2,0.9)  ..  (Phi30.east);
\draw [dashed,thick,green!40!black,postaction={decorate},-,>=stealth,shorten <=2pt,shorten >=2pt] (Phi32.north) .. controls +(-0.2,1.5) and +(0.,-0.3)  ..  (Phi13.east);
\draw [dashed,thick,orange!80!black,postaction={decorate},-,>=stealth,shorten <=2pt,shorten >=2pt] (Phi32.north) .. controls +(-0.2,1.3) and +(0.1,-0.5)  ..  (Phi03.east);
\draw [thick,orange!80!black,postaction={decorate},-,>=stealth,shorten <=2pt,shorten >=2pt] (Phi32.east) .. controls +(1.5,-0.3) and +(-1.0,0.7)  ..  (Phi40.north);
\draw[dashed,thick,yellow!60!black,postaction={decorate},-,>=stealth,shorten <=2pt,shorten >=2pt] (Phi32.west) .. controls +(-1.,-0.9) and +(0.3,1.)  ..  (Phi11.north);
\node (m3-1) at (1.3,8.5) {\textcolor{orange!80!black}{\small{$w_{-3A_1 +A_2}$}}};
\node (1-arrow) at (1.3,7.7) {\textcolor{orange!80!black}{$\left\downarrow\rule{0cm}{0.4cm}\right.$}};
\node (m2-1) at (3.7,8.5) {\textcolor{green!40!black}{\small{$w_{-2A_1 + A_2}$}}};
\node (1-arrow) at (3.7,7.7) {\textcolor{green!40!black}{$\left\downarrow\rule{0cm}{0.4cm}\right.$}};
\node (m1-1) at (6.3,8.4) {\textcolor{red!60!black}{\small{$w_{-A_1 + A_2}$}}};
\node (1-arrow) at (6.3,7.7) {\textcolor{red!60!black}{$\left\downarrow\rule{0cm}{0.35cm}\right.$}};
\node (1-m2) at (11.1,1.3) {\textcolor{orange!80!black}{\small{$w_{A_1 - 2A_2}$}}};
\node (1-arrow) at (10.0,1.3) {\textcolor{orange!80!black}{$\longleftarrow$}};
\node (1-m1) at (11.5,3.6) {\textcolor{red!60!black}{\small{$w_{A_1 - A_2}$}}};
\node (1-arrow) at (10.4,3.6) {\textcolor{red!60!black}{$\longleftarrow$}};
\node (m3-0) at (0.4,6.4) {\textcolor{red!80!black}{\small{$w_{-3A_1 }$}}};
\node (0-1) at (8.1,6.7) {\textcolor{blue!60!black}{\small{$w_{A_2}$}}};
\node (1-0) at (8.9,5.4) {\small{$w_{A_1}$}};
\node (0-m2) at (8.3,1.4) {\textcolor{green!40!black}{\small{$w_{-2A_2}$}}};
\node (m1-m2) at (7.1,0.8) {\textcolor{red!80!black}{\small{$w_{-A_1 - 2A_2}$}}};
\node (m3-m1) at (0.1,4.25) {\textcolor{white!40!black}{\small{$w_{-3A_1- A_2}$}}};
\node (m2-m2) at (3.7,1.0) {\textcolor{blue!99!black}{\small{$w_{-2A_1 - 2A_2}$}}};
\node (0-m1) at (8.1,3.6) {\textcolor{blue!60!black}{\small{$w_{-A_2}$}}};
\node (m1-0) at (6.2,5.3) {\small{$w_{-A_1}$}};
\node (m2-0) at (4.0,5.6) {\textcolor{blue!99!black}{\small{$w_{-2A_1}$}}};
\node (m2-m1) at (1.95,3.4) {\textcolor{yellow!60!black}{\small{$w_{-2A_1 - A_2}$}}};
\node (m1-m1) at (5.25,3.1) {\textcolor{white!30!black}{\small{$w_{-A_1 - A_2}$}}};
\node (m3-m2) at (1.3,-1.2) {\textcolor{white!55!black}{\small{$w_{-3A_1 - 2A_2}$}}};
\end{tikzpicture}
\end{center}
\caption{The ``2.5d'' \textit{alien lattice}: a pictorial representation of the action of the alien derivative upon the $(3,2)$-sector (compare with figure~\ref{fig:sec5-2d-lattice}). Different single arrows correspond to different single steps, where the weight of each step is dictated by the right-hand-side of the resurgence relations \eqref{eq:elliptic-bridge}. This weight is written next to the corresponding arrow, and is given in terms of Stokes vectors as explained in the main text. The solid and dashed lines represent steps with $\boldsymbol{\ell} \cdot \boldsymbol{A} > 0$ and $\boldsymbol{\ell} \cdot \boldsymbol{A} < 0$, respectively, for the choice $m=\pi/8$ (and which will later correspondingly appear in the discontinuities along $\theta=0$ and $\theta=\pi$, respectively). Different colors represent different magnitudes $\left| \boldsymbol{\ell} \cdot \boldsymbol{A} \right|$ which dictate exponential dominance. The steps shown are \textit{all} the single steps which are allowed starting at the $\Phi_{(3,2)}$ lattice-site (and where \textit{paths} can then be constructed with different middle nodes, and corresponding products of weights).
}
\label{fig:chain-nonres}
\end{figure}
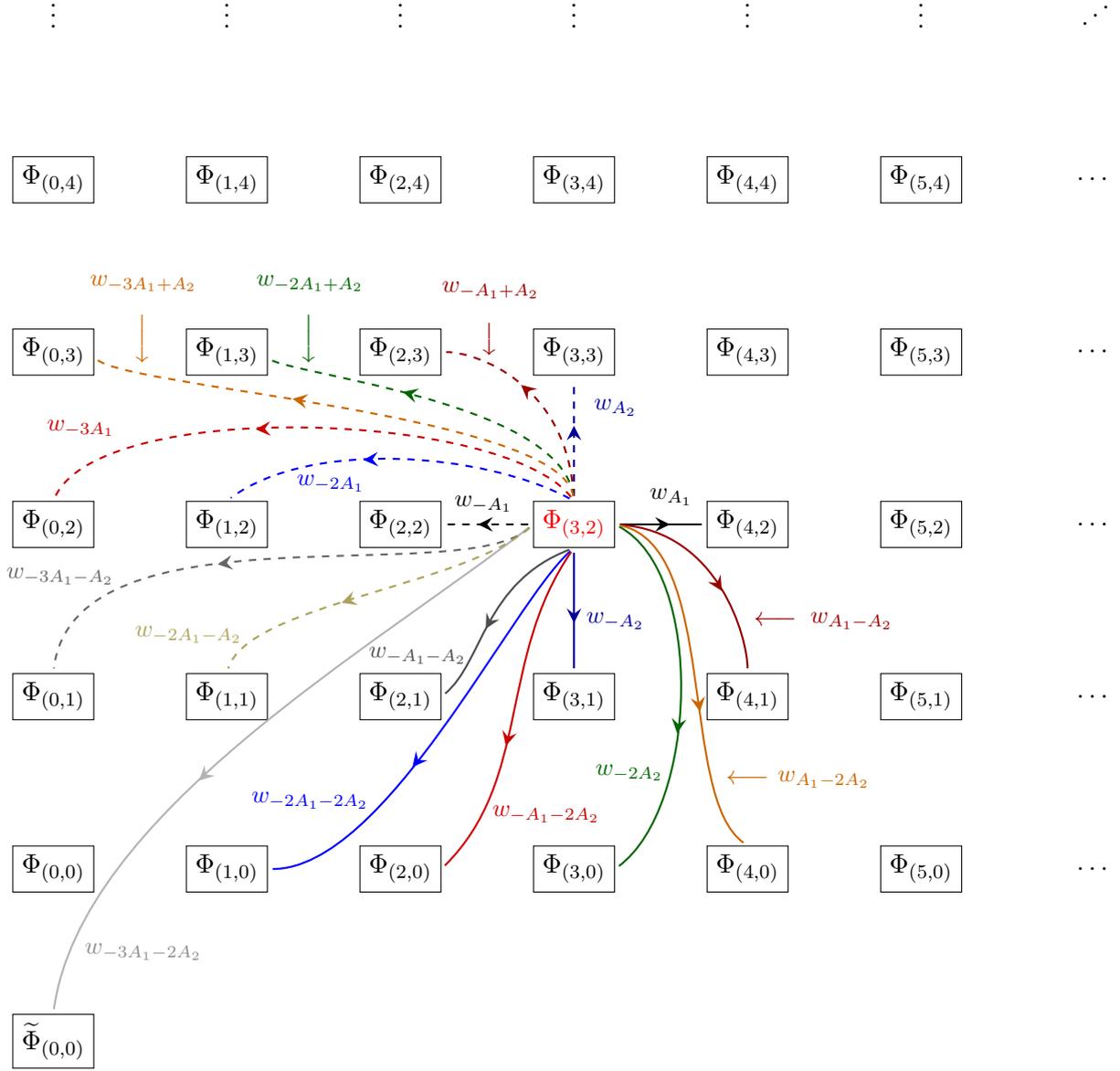

The ``2.5d'' alien lattice where the alien derivative \eqref{eq:elliptic-bridge} induces different motions is illustrated in figure~\ref{fig:chain-nonres} (in particular, it describes the action of $\Delta_{\boldsymbol{\ell} \cdot \boldsymbol{A}}$ on the same $(3,2)$-sector of figure~\ref{fig:sec5-2d-lattice}). The range of (single step) depicted motions follows directly from the discussion in section~\ref{sec:physics}. In particular, strictly-forward motions can only occur in \textit{unit} jumps, $(1,0)$ or $(0,1)$, implying that the only non-vanishing alien-derivative actions along these directions arise from either $\Delta_{A_1}$ or $\Delta_{A_2}$. This translates into the above vanishing of Stokes coefficients, \textit{i.e.}, $S^{(i)}_{\boldsymbol{\ell}} = 0$ for any $\ell_1>0$ and $\ell_2>0$ \textit{except} for $S^{(1)}_{(1,0)}$ and $S^{(2)}_{(0,1)}$. These two elementary forward motions are depicted in figure~\ref{fig:chain-nonres} with weights $w_{A_1}$ and $w_{A_2}$. Cases where $\boldsymbol{\ell}$ only has non-positive coefficients characterize strictly-backward motions: these are the links which connect $\Phi_{(3,2)}$ to any sector to the left and/or below it, \textit{e.g.}, the link in lighter red connecting to the sector $\Phi_{(2,0)}$ with weight $w_{-A_1-2A_2}$. Finally, if one of the components of $\boldsymbol{\ell}$ is negative and the other one positive (equal to $1$), we find mixed forward/backward motions. An example of such a mixed motion is the link in orange connecting to sector $\Phi_{(4,0)}$, with weight $w_{A_1-2A_2}$. The present lattice is the natural extension of the ``1.5d'' alien chain, associated to the free energy of the quartic potential and illustrated in figure~\ref{fig:first-5-sector-alien-one-point-five}, into the realm of 2d alien lattices illustrated in figure~\ref{fig:sec5-2d-lattice}. One final and important point to have in mind is the following: in this example, the lattice notion of \textit{forward} directions does \textit{not} correlate to singularities along the \textit{positive} real axis on the Borel plane. In fact, because $A_1>0$ and $A_2<0$, the projection map \eqref{projectionmapZkC} is such that, say, the lattice ``forward'' direction connecting $\Phi_{(3,2)}$ to $\Phi_{(4,2)}$ does fall on the \textit{positive} real axis, while, say, the lattice ``forward'' direction connecting $\Phi_{(3,2)}$ to $\Phi_{(3,3)}$ instead falls on the \textit{negative} real axis.

The ``statistical mechanical'' language of section~\ref{sec:quartic}, with its multi-dimensional generalization of section~\ref{sec:physics}, obviously still holds. The \textit{weight} $w$ associated to a \textit{step} $\CS$ is defined as in \eqref{k-dim-weight!}. For a step connecting lattice sites $\boldsymbol{n}$ and $\boldsymbol{n} + \boldsymbol{\ell}$, one has the familiar
\be
\label{eq:elliptic-weight-step}
w \left( \mathcal{S} \left( \boldsymbol{n} \rightarrow \boldsymbol{n} + \boldsymbol{\ell} \right) \right) = \left( \boldsymbol{n} + \boldsymbol{\ell} \right) \cdot \boldsymbol{S}_{\boldsymbol{\ell}},
\ee
\noindent
where this expression holds as long as $\Phi_{\boldsymbol{n} + \boldsymbol{\ell}} \ne \widetilde{\Phi}_{(0,0)}$. In the special case where a step links some sector $\Phi_{\boldsymbol{n}}$ to the sector $\widetilde{\Phi}_{(0,0)}$, which will be denoted by $\widetilde{\mathcal{S}} \left( \boldsymbol{n} \rightarrow \boldsymbol{0} \right)$, the corresponding weight is given by
\be
\label{eq:elliptic-weight-tilde}
w \big( \widetilde{\mathcal{S}} \left( \boldsymbol{n} \rightarrow \boldsymbol{0} \right) \big) = S^{(0)}_{-\boldsymbol{n}}.
\ee
\noindent
These weights are shown in figure~\ref{fig:chain-nonres}, where the departing node is  $\boldsymbol{n} = (3,2)$, and where we use the standard notation introduced for figure~\ref{fig:sec5-2d-lattice}, \textit{i.e.}, $w \left( \mathcal{S} \left( \boldsymbol{n} \rightarrow \boldsymbol{n} + \boldsymbol{\ell} \right) \right)) \equiv w_{\ell_1 A_1+\ell_2 A_2}$.

Equipped with a geometric picture and an associated ``statistical mechanical'' language, we may now turn to the standard discussion of Stokes discontinuities. Since there is one positive-real $(A_1)$ and one negative-real ($A_2$) instanton actions, the projection map \eqref{projectionmapZkC} implies that the expected singularities on the Borel plane are along $\theta=0$ and $\theta=\pi$. These lead to discontinuities which may be determined from \eqref{Stokes-aut-as exponential-singularities}---and which in the multi-dimensional case this can lead to more explicit equations, such as \eqref{eq:sec5-Stokes-forward-direction-theta} or \eqref{eq:sec5-Stokes-backward-direction-theta}---and \eqref{StokesDisc},
\begin{eqnarray}
\label{eq:nonlinear-disc-0}
\disc_0 \Phi_{\boldsymbol{n}} &=& \Phi_{\boldsymbol{n}} - \underline{\mathfrak{S}}_0 \Phi_{\boldsymbol n} = \left\{ \1 - \exp \left( \sum_{\,\boldsymbol{\ell} \in \BL^+} \mathrm{e}^{-\frac{\boldsymbol{\ell} \cdot \boldsymbol{A}}{x}} \Delta_{\boldsymbol{\ell} \cdot \boldsymbol{A}} \right) \right\} \Phi_{\boldsymbol{n}}, \\
\label{eq:nonlinear-disc-pi}
\disc_\pi \Phi_{\boldsymbol{n}} &=& \Phi_{\boldsymbol{n}} - \underline{\mathfrak{S}}_{\pi} \Phi_{\boldsymbol{n}} = \left\{ \1 - \exp \left( \sum_{\,\boldsymbol{\ell} \in \BL^-} \mathrm{e}^{-\frac{\boldsymbol{\ell} \cdot \boldsymbol{A}}{x}} \Delta_{\boldsymbol{\ell} \cdot \boldsymbol{A}} \right) \right\} \Phi_{\boldsymbol{n}},
\end{eqnarray}
\noindent
where $\BL^+ = \left\{ \boldsymbol{\ell} \in \mathbb{Z}^2 \,\,|\,\, \boldsymbol{\ell} \cdot \boldsymbol{A} >0 \right\}$ and $\BL^- = \left\{ \boldsymbol{\ell} \in \mathbb{Z}^2 \,\,|\,\, \boldsymbol{\ell} \cdot \boldsymbol{A} < 0 \right\}$ (of course that many of the lattice vectors chosen in this way lead to vanishing alien derivatives due to the constraints on the Stokes vectors, but, as long as the reader has this in mind, for the moment it is simpler to label the discontinuities in this way). With all this in hand, computing the discontinuities follows the same procedure we already used several times in these lectures, since back in subsection~\ref{subsec:Fnonlinear}. Let us quickly run through it as it applies to the present example. The ``statistical mechanical'' algorithmic procedure directly follows from a simple expansion of the above exponentials and subsequent use of the resurgence relations \eqref{eq:elliptic-bridge}. It leads to a sum over all allowed paths $\CP$, defined in \eqref{DEF-path}, that connect a given fixed node $\Phi_{\boldsymbol{n}}$ to all possible other nodes $\left\{ \Phi_{\boldsymbol{m}} \right\}$ such that $\left( \boldsymbol{m} - \boldsymbol{n} \right) \cdot \boldsymbol{A}$ is positive (negative) for $\disc_0$ ($\disc_\pi$). Each such path $\CP$ is made out of steps $\mathcal{S}_i$, defined in \eqref{DEF-stepS}, where each step corresponds to a lattice shift $\boldsymbol{\ell}_i$ (with $\boldsymbol{\ell}_i\cdot\boldsymbol{A}>0$ for $\disc_0$, and $\boldsymbol{\ell}_i\cdot\boldsymbol{A}<0$ for $\disc_\pi$) such that the set of these shifts builds\footnote{Each step links different intermediate nodes between $\boldsymbol{n}$ and $\boldsymbol{m}$. For example, consider a path contributing to the discontinuity along $\theta=0$, divided into 3 steps $\mathcal{S}_i$, $i=1,2,3$, and where step $\mathcal{S}_1$ connects $\boldsymbol{n} \rightarrow \boldsymbol{n} + \boldsymbol{\ell}_1$, step $\mathcal{S}_2$ connects $\boldsymbol{n} + \boldsymbol{\ell}_1 \rightarrow \boldsymbol{n} + \boldsymbol{\ell}_1 + \boldsymbol{\ell}_2$, and finally step $\mathcal{S}_3$ connects $\boldsymbol{n} + \boldsymbol{\ell}_1 + \boldsymbol{\ell}_2 \rightarrow \boldsymbol{m}$. As such, $\boldsymbol{m}-\boldsymbol{n} = \sum_{i=1}^3 \boldsymbol{\ell}_i$.} the corresponding path, \textit{i.e.}, $\sum_i \boldsymbol{\ell}_i = \boldsymbol{m}-\boldsymbol{n}$. The length of a path, $\ell (\mathcal{P})$, is then naturally defined as the number of its constituent steps, as in \eqref{DEF-ell-path}. In the aforementioned sum over all allowed paths, each contribution corresponds to the weight of each path, as defined in \eqref{DEF-w-path},
\begin{equation}
\label{eq:elliptic-weight-path}
w (\mathcal{P}) = \prod_{i=1}^{\ell (\CP)} \,w (\mathcal{S}_i),
\end{equation}
\noindent
and where the weights of each individual step were defined above; see \eqref{eq:elliptic-weight-step} and \eqref{eq:elliptic-weight-tilde}. This sum over all allowed paths is further weighted by adequate combinatorial factors; recall  \eqref{DEF-CF-path}
\begin{equation}
\label{eq:elliptic-CF}
\mathrm{CF} (\mathcal{P}) = \frac{1}{\left(\ell (\mathcal{P})\right)!}.
\end{equation}
\noindent
This is a straightforward generalization of the analogous procedure discussed in section~\ref{sec:physics}.

As concrete examples, let us now explicitly work out the $\disc_0$ and $\disc_\pi$ discontinuities for a couple of sectors. Our first example is the perturbative $(0,0)$ sector. Located at the boundary of the semi-positive transseries grid, any lattice motion which originates from this sector has to move forward. Therefore the only term that contributes to the sum inside the exponential in \eqref{eq:nonlinear-disc-0}, for the $\theta=0$ discontinuity, is $\Delta_{\boldsymbol{\ell} \cdot \boldsymbol{A}}$ with $\boldsymbol{\ell} = \boldsymbol{e}_1 \equiv (1,0)$. Expanding this exponential and using \eqref{eq:elliptic-bridge} one obtains
\be
\label{00nonreso-the0-compare}
\disc_0 \Phi_{(0,0)} = - \sum_{n=1}^{+\infty} \big( S^{(1)}_{\boldsymbol{e}_1} \big)^n\, \mathrm{e}^{-\frac{n A_1}{x}}\, \Phi_{(n,0)}.
\ee
\noindent
Similarly, for the $\theta=\pi$ discontinuity the only contributing term comes from $\boldsymbol{\ell} = \boldsymbol{e}_2 \equiv (0,1)$, which leads to
\be
\label{00nonreso-thepi-compare}
\disc_\pi \Phi_{(0,0)} = - \sum_{n=1}^{+\infty} \big( S^{(2)}_{\boldsymbol{e}_2} \big)^n\, \mathrm{e}^{\frac{n |A_2|}{x}}\, \Phi_{(0,n)}.
\ee

A slightly more complicated example is, \textit{e.g.}, the sector $\Phi_{(0,1)}$. In this case, and in addition to the standard forward motions, it is also possible to move strictly backwards to $\widetilde{\Phi}_{(0,0)}$. Taking $m=\pi/8$, the $\theta=0$ discontinuity will then include contributions such as $\boldsymbol{\ell} = (0,-1) = - \boldsymbol{e}_2$ and\footnote{If we choose a value $m<1/2$, then one has $A_1+A_2=(2m-1)/(m-m^2)<0$---which is the case for our numerical demonstrations below. But when $m>1/2$, the $\boldsymbol{\ell} = (-1,-1)$ term would show up in the discontinuity along $\theta=\pi$ instead of the one along $\theta=0$, since then $\boldsymbol{\ell} \cdot \boldsymbol{A} < 0$ for $m>1/2$.} $\boldsymbol{\ell} = (-1,-1) = - \boldsymbol{e}_1 - \boldsymbol{e}_2$. Expanding the exponential in \eqref{eq:nonlinear-disc-0} and using \eqref{eq:elliptic-bridge} leads to
\begin{eqnarray}
\label{eq:01-disc-0}
\disc_0 \Phi_{(0,1)} &=& - \left( S^{(0)}_{-\boldsymbol{e}_2}+\frac{1}{2} S^{(1)}_{\boldsymbol{e}_1} S^{(0)}_{-\boldsymbol{e}_1-\boldsymbol{e}_2} \right) \mathrm{e}^{-\frac{|A_2|}{x}}\, \widetilde \Phi_{(0,0)} - S^{(1)}_{\boldsymbol{e}_1-\boldsymbol{e}_2}\, \mathrm{e}^{-\frac{A_1+|A_2|}{x}}\, \Phi_{(1,0)} - \\
&& - S^{(1)}_{\boldsymbol{e}_1}\, \mathrm{e}^{-\frac{A_1}{x}}\, \Phi_{(1,1)} - \left( S^{(1)}_{\boldsymbol{e}_1} \right)^2 \mathrm{e}^{-\frac{2A_1}{x}}\, \Phi_{(2,1)} - \frac{1}{2!} \left( S^{(1)}_{\boldsymbol{e}_1} \right)^3 \mathrm{e}^{-\frac{3A_1}{x}}\, \Phi_{(3,1)} - \cdots \nonumber \\
&=& \mathsf{S}_{(0,1) \rightarrow \widetilde{\boldsymbol{0}}}\, \mathrm{e}^{-\frac{|A_2|}{x}}\, \widetilde{\Phi}_{(0,0)} + \mathsf{S}_{(0,1) \rightarrow (1,0)}\, \mathrm{e}^{-\frac{A_1+|A_2|}{x}}\, \Phi_{(1,0)} + \\
&& + \mathsf{S}_{(0,1) \rightarrow (1,1)}\, \mathrm{e}^{-\frac{A_1}{x}}\, \Phi_{(1,1)} + \mathsf{S}_{(0,1) \rightarrow (2,1)}\, \mathrm{e}^{-\frac{2A_1}{x}}\, \Phi_{(2,1)} + \mathsf{S}_{(0,1) \rightarrow (3,1)}\, \mathrm{e}^{-\frac{3A_1}{x}}\, \Phi_{(3,1)} + \cdots. \nonumber
\end{eqnarray}
\noindent
As usual, in the above expression we show the discontinuity written both in terms of Stokes coefficients as well as Borel residues. As for the discontinuity at $\theta=\pi$, a similar exercise yields
\begin{eqnarray}
\disc_\pi \Phi_{(0,1)} &=& - 2\, S^{(2)}_{\boldsymbol{e}_2}\, \mathrm{e}^{\frac{|A_2|}{x}}\, \Phi_{(0,2)} - 3 \left( S^{(2)}_{\boldsymbol{e}_2} \right)^2 \mathrm{e}^{\frac{2 |A_2|}{x}}\, \Phi_{(0,3)} - 4 \left( S^{(2)}_{\boldsymbol{e}_2} \right)^3 \mathrm{e}^{\frac{3 |A_2|}{x}}\, \Phi_{(0,4)} - \cdots 
\label{eq:01-disc-pi} \\
&=& \mathsf{S}_{(0,1) \rightarrow (0,2)}\, \mathrm{e}^{\frac{|A_2|}{x}}\, \Phi_{(0,2)} + \mathsf{S}_{(0,1) \rightarrow (0,3)}\, \mathrm{e}^{\frac{2 |A_2|}{x}}\, \Phi_{(0,3)} + \mathsf{S}_{(0,1) \rightarrow (0,4)}\, \mathrm{e}^{\frac{3 |A_2|}{x}}\, \Phi_{(0,4)} + \cdots. \nonumber
\end{eqnarray}

As one final example let us consider the sector $\Phi_{(3,2)}$, which was the sector used to illustrate figure~\ref{fig:chain-nonres} (recall, depicted in the case where $m=\pi/8$). The discontinuity at $\theta=0$ will involve lattice sites labeled by $(3,2) + \boldsymbol{\ell}$, with $\boldsymbol{\ell} \cdot \boldsymbol{A} > 0$ (further recall how steps between the $(3,2)$ node and these sites were plotted with solid lines back in figure~\ref{fig:chain-nonres}). The factor $\mathrm{e}^{-\frac{\boldsymbol{\ell} \cdot \boldsymbol{A}}{x}}$ in \eqref{eq:nonlinear-disc-0} determines the exponential magnitude of the contributions from different sectors, and different magnitudes were represented by different colors in figure~\ref{fig:chain-nonres}. The smaller $\boldsymbol{\ell} \cdot \boldsymbol{A}$ is, the more dominant (\textit{i.e.}, less exponentially suppressed) the contribution of the sector $\Phi_{(3,2) + \boldsymbol{\ell}}$ will be (of course, provided it remains $(3,2) + \boldsymbol{\ell} \ge \boldsymbol{0}$, since otherwise $\Delta_{\boldsymbol{\ell} \cdot \boldsymbol{A}} \Phi_{(3,2)} = 0$). The smallest such value is simply reached when $\boldsymbol{\ell} = - (3,2)$. Always assuming $m=\pi/8$, the ordering of dominances then goes as $\boldsymbol{\ell} \in \left\{ (-3,-2), (-1,-1), (1,0), (-2,-2), \cdots \right\}$, and the $\disc_0$ discontinuity is
\begin{eqnarray}
\disc_0\,\Phi_{(3,2)} &=& - \mathrm{SF}_{( (3,2)\rightarrow \widetilde{\boldsymbol{0}} )}\, \mathrm{e}^{\frac{3A_1+2A_2}{x}}\, \widetilde{\Phi}_{(0,0)} - \mathrm{SF}_{( (3,2)\rightarrow (2,1) )}\, \mathrm{e}^{\frac{A_1+A_2}{x}}\, \Phi_{(2,1)} - \nonumber \\
&& - \mathrm{SF}_{( (3,2)\rightarrow (4,2) )}\, \mathrm{e}^{-\frac{A_1}{x}}\, \Phi_{(4,2)} - \mathrm{SF}_{( (3,2)\rightarrow (1,0) )}\, \mathrm{e}^{\frac{2A_1+2A2}{x}}\, \Phi_{(1,0)} - \cdots.
\label{eq:disc-32-nonres-0}
\end{eqnarray}
\noindent
The statistical factors $\mathrm{SF}_{( \boldsymbol{n}\rightarrow\boldsymbol{m} )}$  appearing in \eqref{eq:disc-32-nonres-0} are just the Borel residues $\mathsf{S}_{\boldsymbol{n}\rightarrow\boldsymbol{m}} = - \mathrm{SF}_{( \boldsymbol{n}\rightarrow\boldsymbol{m} )}$. As thoroughly discussed, they are combinations of Stokes coefficients $S^{(i)}_{\boldsymbol \ell}$ which arise from all the different ways of decomposing the motion $(3,2) \rightarrow (3,2) + \boldsymbol{\ell}$ in different paths, \textit{i.e.}, different collections of steps\footnote{Arising from the expansion of the exponentials in \eqref{eq:nonlinear-disc-0} or \eqref{eq:nonlinear-disc-pi}, and applying $\cdots \Delta_{\boldsymbol{\ell}_2 \cdot \boldsymbol{A}} \Delta_{\boldsymbol{\ell}_1 \cdot \boldsymbol{A}}$ to $\Phi_{\boldsymbol{n}}$.} $\boldsymbol{\ell} = \sum_i \boldsymbol{\ell}_i$, with $\boldsymbol{\ell}_i \cdot \boldsymbol{A} > 0$. In other words, there are different ways to go from $\Phi_{(3,2)}$ to any other fixed sector of figure~\ref{fig:chain-nonres}. For instance, consider the contribution associated to connecting $\Phi_{(3,2)}\rightarrow \Phi_{(4,1)}$. There are three possible paths, with weights and combinatorial factors given by \eqref{eq:elliptic-weight-path} and \eqref{eq:elliptic-CF} (and where the step-weights are \eqref{eq:elliptic-weight-step} and \eqref{eq:elliptic-weight-tilde}):
\begin{itemize}
\item The first path, $\mathcal{P}_1$, consists of a single step connecting the two nodes directly, with path-length $\ell=1$, combinatorial factor $\mathrm{CF} = \frac{1}{1!}$, and weight $w = w_{A_1-A_2} = 4 S^{(1)}_{\boldsymbol{e}_1-\boldsymbol{e}_2} + S^{(2)}_{\boldsymbol{e}_1-\boldsymbol{e}_2}$; 
\item The second path, $\mathcal{P}_2$, consists of two steps, $\mathcal{S} \left( (3,2) \rightarrow (3,1) \right)$ followed by $\mathcal{S} \left( (3,1) \rightarrow (4,1) \right)$, with path-length $\ell=2$, combinatorial factor $\mathrm{CF} = \frac{1}{2!}$, and path-weight
\be
w (\mathcal{P}_2) = w \Big( \mathcal{S} \left( (3,2) \rightarrow (3,1) \right) \Big)\, w \Big( \mathcal{S} \left( (3,1) \rightarrow (4,1) \right) \Big) = \left( 3 S^{(1)}_{-\boldsymbol{e}_2} + S^{(2)}_{-\boldsymbol{e}_2} \right) \left( 4 S^{(1)}_{\boldsymbol{e}_1} + S^{(2)}_{\boldsymbol{e}_1} \right);
\ee
\item The final path, $\mathcal{P}_3$, also consists of two steps, which are now $\mathcal{S} \left( (3,2) \rightarrow (4,2) \right)$ followed by $\mathcal{S} \left( (4,2) \rightarrow (4,1) \right)$, with path-length $\ell=2$, combinatorial factor $\mathrm{CF} = \frac{1}{2!}$, and path-weight
\be
w (\mathcal{P}_3) = w \Big( \mathcal{S} \left( (3,2) \rightarrow (4,2) \right) \Big)\, w \Big( \mathcal{S} \left( (4,2) \rightarrow (4,1) \right) \Big) = \left( 4 S^{(1)}_{\boldsymbol{e}_1} + 2 S^{(2)}_{\boldsymbol{e}_1} \right) \left( 4 S^{(1)}_{-\boldsymbol{e}_2} + S^{(2)}_{-\boldsymbol{e}_2} \right). 
\ee
\end{itemize}
\noindent
The statistical factor $\mathrm{SF}_{( (3,2) \rightarrow (4,1) )}$ will then be
\be
\mathrm{SF}_{( (3,2) \rightarrow (4,1) )} = \sum_{i=1}^3 \mathrm{CF} (\mathcal{P}_i)\, w (\mathcal{P}_i).
\ee
\noindent
Similarly, the discontinuity at $\theta=\pi$ involves lattice sites labeled by $(3,2) + \boldsymbol{\ell}$, now with $\boldsymbol{\ell} \cdot \boldsymbol{A} < 0$ (and whose steps were plotted with dashed lines back in figure~\ref{fig:chain-nonres}). This time around, the exponential magnitude of the different contributions will be ordered starting with smaller $\left| \boldsymbol{\ell} \cdot \boldsymbol{A} \right|$, so as to correspond to successively more dominant (\textit{i.e.}, more exponentially enhanced) contributions, and such difference of magnitudes was again represented by different colors in figure~\ref{fig:chain-nonres}. Always assuming $m=\pi/8$, this ordering of dominances goes as $\boldsymbol{\ell} \in \left\{ (-2,-1), (-1,0), (-3,-1), (0,1), \cdots \right\}$. Repeating the same exercise as above now leads to
\begin{eqnarray}
\disc_\pi\,\Phi_{(3,2)} &=& - \mathrm{SF}_{( (3,2) \rightarrow (1,1) )}\, \mathrm{e}^{\frac{2A_1+A_2}{x}}\, \Phi_{(1,1)} - \mathrm{SF}_{( (3,2) \rightarrow (2,2) )}\, \mathrm{e}^{\frac{A_1}{x}}\, \Phi_{(2,2)} - \nonumber \\
&& - \mathrm{SF}_{( (3,2) \rightarrow (0,1) )}\, \mathrm{e}^{\frac{3A_1+A_2}{x}}\,  \Phi_{(0,1)} - \mathrm{SF}_{( (3,2) \rightarrow (3,3) )}\, \mathrm{e}^{\frac{|A_2|}{x}}\,  \Phi_{(3,3)} - \cdots,
\label{eq:disc-32-nonres-pi}
\end{eqnarray}
\noindent
where once again the statistical factors are identified with the Borel residues, $\mathsf{S}_{\boldsymbol{n}\rightarrow\boldsymbol{m}} = - \mathrm{SF}_{( \boldsymbol{n}\rightarrow\boldsymbol{m} )}$, and are given by the combinations of Stokes coefficients which arise as a result of decomposing each contribution labeled by $\boldsymbol{\ell}$ into all different ways (different paths) $\boldsymbol{\ell} = \sum_i \boldsymbol{\ell}_i$ (now with $\boldsymbol{\ell}_i \cdot \boldsymbol{A} < 0$), and then applying the associated alien derivatives to $\Phi_{\boldsymbol{n}}$.

Just like what happened throughout section~\ref{sec:quartic}, when dealing with the quartic-potential free energy, the results derived in the present subsection will now also allow us to perform very accurate numerical checks in upcoming subsections. These checks will verify that the transseries solution we are considering is indeed the \textit{complete} solution to the nonlinear problem of the elliptic-potential free energy. But, akin to what we did back in subsection~\ref{subsec:stokesZvsF}, before proceeding towards these large-order checks we shall first focus our attention on actually \textit{determining} the Stokes coefficients of the elliptic free energy---via a direct comparison with the ones already determined for the partition function---and this is what we shall turn to next.

\subsection{Stokes Constants of Partition Function versus Free Energy}\label{sec:stokes_z-vs-f}

Given that the free-energy \textit{nonlinear} problem may actually be derived from the partition-function \textit{linear} problem (\textit{i.e.}, the one for which we do know Stokes coefficients), then the Stokes coefficients for the nonlinear problem should be explicitly computable simply by formally expanding $\CF = \log \CZ$ (this is exactly the same strategy as the one used in subsection~\ref{subsec:stokesZvsF}). In order to do this, start by rewriting the transseries for the partition function \eqref{ell_lin_transseries-v2} as
\be
\CZ (x, \sigma_0, \sigma_1, \sigma_2) = \sigma_0\, \Phi_0 (x) \left( 1 + \zeta_1\, \mathrm{e}^{-\frac{A_1}{x}}\, \frac{\Phi_1 (x)}{\Phi_0 (x)} + \zeta_2\, \mathrm{e}^{-\frac{A_2}{x}}\, \frac{\Phi_2 (x)}{\Phi_0 (x)} \right),
\ee
\noindent
where we have introduced $\zeta_i \equiv \frac{\sigma_i}{\sigma_0}$, $i=1,2$. Next, as the free energy is the logarithm of the partition function, we determine it as the \textit{formal} expansion of the function $\log(1+z)$. It follows: 
\be
\label{ell_nl-from-lin}
\CF (x, \sigma_0, \sigma_1, \sigma_2) = \log \sigma_0 + \log \Phi_0 + \sum_{n_1=0}^{+\infty} \sum_{n_2=0}^{+\infty} \left( 1 - \delta_{n_1,0} \delta_{n_2,0} \right) \zeta_1^{n_1}\, \zeta_2^{n_2}\, \mathrm{e}^{-\frac{n_1 A_1+n_2 A_2}{x}}\, F^{(n_1,n_2)} (x),
\ee
\noindent
where the free-energy nonperturbative sectors $F^{(n_1,n_2)}$ may be written in terms of the original sectors in the partition function as
\be
\label{ell-free-en-from-partfunc}
F^{(n_1,n_2)} := (-1)^{n_1+n_2+1}\, \frac{(n_1+n_2-1)!}{n_1!\, n_2!}\left( \frac{\Phi_1}{\Phi_0} \right)^{n_1} \left( \frac{\Phi_2}{\Phi_0} \right)^{n_2}.
\ee
\noindent
On the other hand, from the discussions in the previous subsection, we know that the free energy will be given by a three-parameter\footnote{We have renamed the transseries parameters of the free energy as $\tau_i$, in order to avoid confusion with the transseries parameters of the partition function $\sigma_i$.} transseries, namely,
\be
F (x, \boldsymbol{\tau} ) = \sum_{\boldsymbol{n}\in\mathbb{N}_0^3} \boldsymbol{\tau}^{\boldsymbol{n}}\, \mathrm{e}^{-\frac{\boldsymbol{n}\cdot\boldsymbol{A}}{x}}\, \Phi_{\boldsymbol{n}} (x).
\ee
\noindent
Here we used the notation introduced in section~\ref{sec:physics}, in particular defined\footnote{Note that, for the purposes of this subsection, we are using three-dimensional vectors within a three-parameter transseries, although we already know that $\tau_0$ is associated to a single (non-asymptotic) sector.} $\boldsymbol{\tau} = (\tau_0, \tau_1, \tau_2)$, $\boldsymbol{n} = (n_0, n_1, n_2)$ and $\boldsymbol{A} = (0, A_1, A_2)$; and the free-energy nonperturbative sectors $\Phi_{\boldsymbol{n}}$ may be directly read from \eqref{eq:nl_ell_ansatz-v0} and \eqref{eq:ell-nl-sectors}. Comparing this expression for the free energy with the previous one, \eqref{ell_nl-from-lin}, which follows directly from the partition function, we immediately identify
\begin{eqnarray}
\label{ell-relation-parameters}
\tau_0 &=& \log \sigma_0, \qquad \tau_i = \frac{\sigma_i}{\sigma_0},\,\, i=1,2, \\
\label{ell-relation-phi000}
\Phi_{(0,0,0)} &=& \log \Phi_0, \\
\label{ell-relation-phis}
\Phi_{(0,n_1,n_2)} &=& F^{(n_1,n_2)}, \\
\label{ell-relation-phitilde}
\Phi_{(1,0,0)} &=& 1, \\
\Phi_{(n_1,n_2,n_3)} &=& 0, \qquad \text{if }\, n_1>1 \vee \left( n_1=1 \wedge (n_2,n_3)\ne(0,0) \right).
\end{eqnarray}
\noindent
In particular, this identification is now validating our ``2.5''-parameter transseries \textit{ansatz} in \eqref{eq:nl_ell_ansatz}.

Following subsection~\ref{subsec:stokesZvsF}, let us next see if we can make similar identifications at the level of resurgence relations and, eventually, at the level of Stokes data. In particular, let us follow the discussion at the very end of subsection~\ref{subsec:stokesZvsF}, where we were able to relate the Stokes coefficients of the partition function to those of the free energy via a direct comparison of Stokes \textit{automorphisms}. Recall that the main idea behind that discussion was to use the fact that $\Delta_\omega$ is a derivation, in order to then show how the Stokes automorphisms for the free energy have to be the \textit{exact same} operators as those acting at the level of the partition function, \textit{once} the identifications \eqref{ell-relation-parameters} between transseries parameters are taken into account. For the partition function, the Stokes coefficients\footnote{Akin to subsection~\ref{subsec:stokesZvsF}, also in this subsection we shall label Stokes coefficients associated with the partition function with a superscript $Z$, while those associated with the free energy will be labeled with a superscript $F$.} were already determined in \eqref{eq:linear-stokes}; recall
\be
S_{\omega_{01}}^Z = 2, 
\qquad 
S_{\omega_{10}}^Z = - 2, 
\qquad 
S_{\omega_{02}}^Z = 2, 
\qquad 
S_{\omega_{20}}^Z = - 2, 
\qquad 
S_{\omega_{12}}^Z = 0 = S_{\omega_{21}}^Z.
\ee
\noindent
With these coefficients, one may completely determine the Stokes automorphisms \eqref{eq:disc-Z-0} and \eqref{eq:disc-Z-pi}, and subsequently the Stokes transitions for the partition function, \eqref{ell-stokes-transitions-0} and \eqref{ell-stokes-transitions-pi}. But in order to carry through our present goal, one must first rewrite all these expressions with a slightly different flavor (albeit one we have already seen back in subsection~\ref{subsec:stokesZvsF}).

Starting along $\theta=0$, one easily finds\footnote{The last line can be simply checked by expanding the exponential and applying the required derivatives. In this process, it is interesting to note how the \textit{alien} derivatives (acting upon asymptotic series) from the first line ended up being translated into \textit{regular} derivatives (now acting upon transseries \textit{parameters}) in the last line. As repeated many times before, this is the essence of the \textit{bridge equations}, which is now being made explicit: to establish a ``bridge'' between alien and ordinary calculus---see appendix~\ref{app:alien-calculus} for further details.}
\begin{eqnarray}
\underline{\mathfrak{S}}_0 \mathcal{Z} (x, \sigma_0, \sigma_1, \sigma_2) &=& \exp\left( \mathrm{e}^{-\frac{\omega_{01}}{x}} \Delta_{\omega_{01}} + \mathrm{e}^{-\frac{\omega_{21}}{x}} \Delta_{\omega_{21}} + \mathrm{e}^{-\frac{\omega_{20}}{x}} \Delta_{\omega_{20}} \right) \mathcal{Z} (x, \sigma_0, \sigma_1, \sigma_2) = \nonumber \\
&=& \mathcal{Z} \left( x, \sigma_0 + S_{\omega_{20}}^Z \sigma_2, \sigma_1 + S_{\omega_{01}}^Z \sigma_0 + \left( S_{\omega_{21}}^Z + \frac{1}{2} S_{\omega_{01}}^Z  S_{\omega_{20}}^Z \right) \sigma_2, \sigma_2 \right) = \nonumber \\
&\equiv& \exp \left\{ S_{\omega_{20}}^Z \sigma_2\, \frac{\partial}{\partial\sigma_0} + \left( S_{\omega_{01}}^Z \sigma_0 + S_{\omega_{21}}^Z \sigma_2 \right) \frac{\partial}{\partial\sigma_1} \right\} \mathcal{Z} (x, \sigma_0, \sigma_1, \sigma_2).
\end{eqnarray}
\noindent
If one now uses the relations \eqref{ell-relation-parameters}, between partition-function and free-energy transseries parameters, the above differential-representation of the Stokes automorphism may be rewritten for the $\tau_i$ variables rather than the $\sigma_i$ variables. Using
\begin{eqnarray}
\sigma_2\, \frac{\partial}{\partial\sigma_0} &=& \tau_2 \left( \frac{\partial}{\partial\tau_0} - \tau_1\, \frac{\partial}{\partial\tau_1} - \tau_2\, \frac{\partial}{\partial\tau_2} \right), \\
\sigma_0\, \frac{\partial}{\partial\sigma_1} &=& \frac{\partial}{\partial\tau_1}, \\
\sigma_2\, \frac{\partial}{\partial\sigma_1} &=& \tau_2\, \frac{\partial}{\partial\tau_1},
\end{eqnarray}
\noindent
it immediately follows
\be
\label{eq:stokes-l_nl-0}
\underline{\mathfrak{S}}_0 = \exp \left\{ S_{\omega_{20}}^Z\, \tau_2\, \frac{\partial}{\partial\tau_0} + \left( S_{\omega_{01}}^Z + S_{\omega_{21}}^Z\, \tau_2 - S_{\omega_{20}}^Z\, \tau_1 \tau_2 \right) \frac{\partial}{\partial\tau_1} - S_{\omega_{20}}^Z\, \tau_2^2\, \frac{\partial}{\partial\tau_2} \right\}.
\ee
\noindent
When acting upon the free energy, this operator must exactly match the one we already computed when addressing the $\theta=0$ discontinuity, \textit{i.e.}, \eqref{eq:stokes-l_nl-0} must exactly match the Stokes automorphism in \eqref{eq:nonlinear-disc-0}. In other words, the (differential operator) argument of the above exponential must equal $\sum_{\boldsymbol{\ell}} \mathrm{e}^{-\frac{\boldsymbol{\ell} \cdot \boldsymbol{A}}{x}} \Delta_{\boldsymbol{\ell} \cdot \boldsymbol{A}}$ for $\boldsymbol{\ell} \cdot \boldsymbol{A} > 0$. This simple relation between the Stokes automorphisms of linear and nonlinear problems essentially stems from the fact that the free energy ends up inheriting the Borel singularities of the partition function, as explained in detail in subsection~\ref{subsec:stokesZvsF}. Now, to match \eqref{eq:stokes-l_nl-0} and \eqref{eq:nonlinear-disc-0} we need to dig further into the \textit{bridge equations}, relating alien to regular calculus. In the present case it can be shown that they imply\footnote{This is basically the content of footnote~\ref{eq:elliptic-bridge0}---or else we refer the reader to the discussion in appendix~\ref{app:alien-calculus}.}
\be
\label{eq:stokes-tau}
\mathrm{e}^{-\frac{\boldsymbol{\ell} \cdot \boldsymbol{A}}{x}} \Delta_{\boldsymbol{\ell} \cdot \boldsymbol{A}} = S_{\boldsymbol{\ell}}^{F(0)}\, \tau_1^{-\ell_1} \tau_2^{-\ell_2}\, \frac{\partial}{\partial\tau_0} + S_{\boldsymbol{\ell}}^{F(1)}\, \tau_1^{1-\ell_1} \tau_2^{-\ell_2}\, \frac{\partial}{\partial\tau_1} + S_{\boldsymbol{\ell}}^{F(2)}\, \tau_1^{-\ell_1} \tau_2^{1-\ell_2}\, \frac{\partial}{\partial\tau_2},
\ee
\noindent
where $\boldsymbol{\ell} = (\ell_1,\ell_2)$ satisfies $\boldsymbol{\ell} \cdot \boldsymbol{A} > 0$. Comparing \eqref{eq:stokes-tau} with the argument of the exponential in \eqref{eq:stokes-l_nl-0} results in that only three steps, namely $\boldsymbol{\ell} = (1,0)$, $\boldsymbol{\ell} = (1,-1)$, and $\boldsymbol{\ell} = (0,-1)$, respectively related to $\omega_{01}$, $\omega_{21}$, and $\omega_{20}$, contribute to the Stokes automorphism. Further, this immediately leads to the Stokes identifications:
\begin{align}
\label{ell-stokes-freeen-1-1}
S^{F(1)}_{(1,0)} &= S_{\omega_{01}}^Z &(\, &= 2\,\, ), \\
\label{ell-stokes-freeen-1-2}
S^{F(1)}_{(1,-1)} &= S_{\omega_{21}}^Z &(\, &= 0\,\, ), \\
\label{ell-stokes-freeen-1-3}
S^{F(0)}_{(0,-1)} &= S_{\omega_{20}}^Z &(\, &= - 2\,\, ), \\
\label{ell-stokes-freeen-1-4}
S^{F(1)}_{(0,-1)} &= - S_{\omega_{20}}^Z &(\, &= 2\,\, ), \\
\label{ell-stokes-freeen-1-5}
S^{F(2)}_{(0,-1)} &= - S_{\omega_{20}}^Z &(\, &= 2\,\, );
\end{align}
\noindent
all other coefficients vanishing.

The exact same procedure may be applied along $\theta=\pi$, where one now simply starts from \eqref{eq:disc-Z-pi} rather than \eqref{eq:disc-Z-0}. In this case, the Stokes transition \eqref{ell-stokes-transitions-pi} implies
\begin{eqnarray}
\underline{\mathfrak{S}}_\pi \mathcal{Z} (x, \sigma_0, \sigma_1, \sigma_2) &=& \exp \left( \mathrm{e}^{-\frac{\omega_{02}}{x}} \Delta_{\omega_{02}} + \mathrm{e}^{-\frac{\omega_{10}}{x}} \Delta_{\omega_{10}} + \mathrm{e}^{-\frac{\omega_{12}}{x}} \Delta_{\omega_{12}} \right) \mathcal{Z} (x, \sigma_0, \sigma_1, \sigma_2) = \nonumber \\
&=& \mathcal{Z} \left( x, \sigma_0 + S_{\omega_{10}}^{Z} \sigma_1, \sigma_1, \sigma_2 + S_{\omega_{02}}^{Z} \sigma_0 + \left( S_{\omega_{12}}^{Z} + \frac{1}{2} S_{\omega_{02}}^{Z} S_{\omega_{10}}^{Z} \right) \sigma_1 \right) = \nonumber \\
&\equiv& \exp \left\{ S_{\omega_{10}}^{Z} \sigma_1\, \frac{\partial}{\partial\sigma_0} + \left( S_{\omega_{02}}^{Z} \sigma_0 + S_{\omega_{12}}^{Z} \sigma_1 \right) \frac{\partial}{\partial\sigma_2} \right\} \mathcal{Z} (x, \sigma_0, \sigma_1, \sigma_2).
\end{eqnarray}
\noindent
Using
\begin{eqnarray}
\sigma_1\, \frac{\partial}{\partial\sigma_0} &=& \tau_1 \left( \frac{\partial}{\partial\tau_0} - \tau_1\, \frac{\partial}{\partial\tau_1} - \tau_2\, \frac{\partial}{\partial\tau_2} \right), \\
\sigma_0\, \frac{\partial}{\partial\sigma_2} &=& \frac{\partial}{\partial\tau_2}, \\
\sigma_1\, \frac{\partial}{\partial\sigma_2} &=& \tau_1\, \frac{\partial}{\partial\tau_2},
\end{eqnarray}
\noindent
it immediately follows
\be
\label{eq:stokes-l_nl-pi}
\underline{\mathfrak{S}}_\pi = \exp \left\{ S_{\omega_{10}}^Z\, \tau_1\, \frac{\partial}{\partial\tau_0} - S_{\omega_{10}}^Z\, \tau_1^2\, \frac{\partial}{\partial\tau_1} + \left( S_{\omega_{02}}^Z - S_{\omega_{10}}^Z\, \tau_1 \tau_2 - S_{\omega_{12}}^Z\, \tau_1 \right) \frac{\partial}{\partial\tau_2} \right\}.
\ee
\noindent
Comparing the (differential operator) argument of the exponential in \eqref{eq:stokes-l_nl-pi} above with \eqref{eq:stokes-tau} (where now $\boldsymbol{\ell} = (\ell_1,\ell_2)$ satisfies $\boldsymbol{\ell} \cdot \boldsymbol{A} < 0$), results in that only three steps, namely $\boldsymbol{\ell} = (0,1)$, $\boldsymbol{\ell} = (-1,0)$, and $\boldsymbol{\ell} = (-1,1)$, respectively related to $\omega_{02}$, $\omega_{10}$, and $\omega_{12}$, contribute to the Stokes automorphism. Further, this immediately leads to the Stokes identifications:
\begin{align}
\label{ell-stokes-freeen-2-1}
S^{F(2)}_{(0,1)} &= S_{\omega_{02}}^Z &(\, &= 2\,\, ), \\
\label{ell-stokes-freeen-2-2}
S^{F(2)}_{(-1,1)} &= -S_{\omega_{12}}^Z &(\, &= 0\,\, ), \\
\label{ell-stokes-freeen-2-3}
S^{F(0)}_{(-1,0)} &= S_{\omega_{10}}^Z &(\, &= - 2\,\, ), \\
\label{ell-stokes-freeen-2-4}
S^{F(1)}_{(-1,0)} &= - S_{\omega_{10}}^Z &(\, &= 2\,\, ), \\
\label{ell-stokes-freeen-2-5}
S^{F(2)}_{(-1,0)} &= - S_{\omega_{10}}^Z &(\, &= 2\,\, );
\end{align}
\noindent
all other coefficients vanishing.

\subsection{Asymptotics and Large-Order Behaviour: Free Energy}\label{sec:large-order-nonres}

Having arrived at this point, it is immediate to acknowledge our construction of a transseries solution for the elliptic-potential free energy, as well as the computation of its complete set of Stokes data. This means we can access (at least in principle, and recursively) all the free-energy (multi) instanton nonperturbative sectors, as well as explicitly write down the associated Stokes discontinuities for all these sectors. However, some assumptions were made along the way, and short of a rigorous proof that we have achieved the correct and complete solution, the next-best thing to do (and in complete analogy with the procedure of section~\ref{sec:quartic}; in particular of subsection~\ref{sec:large-order-F}) is to \textit{test} our constructions via resurgent large-order (numerical) analyses.

These upcoming tests (which will completely vindicate our constructions and calculations), will follow very closely what was done in subsection~\ref{sec:large-order-F} for the quartic free-energy, and to where we refer the reader if in need of a reminder on numerical methods and techniques. Without further ado, first note that---similarly to the case of the quartic potential---also in the current example the constant part of the transseries \eqref{eq:nl_ell_ansatz}, $F_0^{(0,0)} + \sigma_0\, \widetilde{\Phi}_{(0,0)}$, is an integration constant which does not get fixed by the recursion relations (and does not change the numerical analyses performed below). Of course it is fixed by a choice of initial conditions, corresponding to some particular solution of the ODE \eqref{eq:nl_ell_ode}. It may also be fixed upon comparison of free-energy against partition-function; \textit{i.e.}, taking \eqref{ell-relation-phi000} and \eqref{ell-relation-phitilde} into account it is natural to set $F_0^{(0,0)}=0$ and $\widetilde{\Phi}_{(0,0)}=1$. But this is not all: as \eqref{eq:nl_ell_ode} is a third-order ODE, there will still be two other coefficients which need to be fixed from initial conditions, namely, $F_0^{(1,0)}$ and $F_0^{(0,1)}$. Comparing again with the partition function, this time via \eqref{ell-relation-phis}, we easily find\footnote{Recall from subsection~\ref{sec:ell-basic-partfunc} that $Z_0^{(0)}=1$, $Z_0^{(1)}(m)=\mathrm{i}\sqrt{1-m}$, and  $Z_0^{(2)}(m)=\mathrm{i}\sqrt{m}$.}
\be
F_0^{(1,0)} = \frac{Z_0^{(1)}}{Z_0^{(0)}} = \mathrm{i}\sqrt{1-m}, \qquad F_0^{(0,1)} = \frac{Z_0^{(2)}}{Z_0^{(0)}} = \mathrm{i}\sqrt{m}.
\ee
\noindent
We shall use these values in the numerical analyses addressed below.

Let us begin by discussing the general large-order pattern we find, and how it may be succinctly written-down based upon ``statistical mechanical'' rules (this is essentially the higher-dimensional analogue of \eqref{large-order-factor} and \eqref{large-order-one-point-five}). The generalization of \eqref{large-order-factor}, the \textit{large-order factor} for a path on the alien lattice connecting the nodes $\Phi_{\boldsymbol{n}}$ and $\Phi_{\boldsymbol{m}}$, is straightforward\footnote{Up to the ``multi-instanton-number factorization'' which is now no longer possible.}
\be
\label{ell-large-order-factor}
\chi_{(\boldsymbol{n} \rightarrow \boldsymbol{m})} (k) \equiv \sum_{h=0}^{+\infty} \frac{\Gamma(k-h)}{\Gamma(k)}\, F_{h}^{(\boldsymbol{m})} \left( \left( \boldsymbol{m} - \boldsymbol{n} \right) \cdot \boldsymbol{A} \right)^h.
\ee
\noindent
In here, we are assuming that the starting node can be any node except the non-asymptotic sector $\widetilde{\Phi}_{(0,0)}$. The end node, on the other hand, can be any node except the perturbative sector $\Phi_{(0,0)}$. As both these sectors have $\boldsymbol{m} = (0,0)$, let us distinguish them by introducing the notation $\boldsymbol{m} = \widetilde{\boldsymbol{0}}$ for the sector $\Phi_{\widetilde{\boldsymbol{0}}} \equiv \widetilde{\Phi}_{(0,0)}$, while $\boldsymbol{m} = \boldsymbol{0}$ naturally remains attached to $\Phi_{(0,0)}$ (recall\footnote{If one were to use the notation of a full three-parameter transseries, as in \eqref{eq:nl_ell_ansatz-v0}, one should really have $\boldsymbol{m} = (0,0,0) \equiv \boldsymbol{0}$ and $\boldsymbol{m} = (1,0,0) \equiv \widetilde{\boldsymbol{0}}$; and with all other sectors labeled by $\boldsymbol{m} = (0,m_1,m_2)$.} \eqref{eq:nl_ell_ansatz}). In this way, the starting node will have $\boldsymbol{n} \ge \boldsymbol{0}$, while the end node will obey either $\boldsymbol{m} \equiv (m_1,m_2) > \boldsymbol{0}$ (\textit{i.e.}, $m_i\ge 0$ and $\boldsymbol{m} \ne (0,0)$) or $\boldsymbol{m} = \widetilde{\boldsymbol{0}}$. Finally, recall that $ \widetilde{\boldsymbol{0}}$ is not asymptotic: we have only one non-zero coefficient $F_{0}^{(\widetilde{\boldsymbol{0}})}=1$ (all others vanish; $F_{h>0}^{(\widetilde{\boldsymbol{0}})} = 0$). Having all this in mind, a straightforward re-run of the analysis in subsection~\ref{sec:large-order-F} allows us to write down generic large-order relations for the ``two-and-a-half'' transseries of the elliptic potential, purely in terms of motions and data on the alien lattice:

\vspace{10pt}
\setlength{\fboxsep}{10pt}
\Ovalbox{%
\parbox{14cm}{%
\vspace{5pt}

\textbf{Large-order relations:}  The large-order (large $k$) behaviour of the coefficients $F_k^{(\boldsymbol{n})}$, associated to the node $\Phi_{\boldsymbol{n}}$, is given by a \textit{sum} over \textit{all} paths, linking to nodes $\Phi_{\boldsymbol{m}}$ with $\boldsymbol{m} > \boldsymbol{0}$, \textit{in addition} to a sum over paths to the extra ``orthogonal'' node in the lattice, $\widetilde{\Phi}_{(0,0)}$ (when $\boldsymbol{m} = \boldsymbol{0}$).
      
\vspace{10pt}        
Each term in this sum ($\Phi_{\boldsymbol{n}} \rightarrow \Phi_{\boldsymbol{m}}$) can be decomposed into three factors:
\begin{itemize}
\item Leading growth-factor:
\begin{equation*}
\frac{\Gamma(k)}{2\pi\rmi \left( \left( \boldsymbol{m}-\boldsymbol{n} \right) \cdot \boldsymbol{A} \right)^k}.
\end{equation*}
\noindent
\item Large-order factor \eqref{ell-large-order-factor}, dictated by beginning and end nodes:
\begin{equation*}
\chi_{(\boldsymbol{n} \rightarrow \boldsymbol{m})}( k).
\end{equation*}
\noindent
\item Statistical factor, sum over all the allowed paths $\mathcal{P}(\boldsymbol{n} \rightarrow \boldsymbol{m})$ linking the nodes as in figure~\ref{fig:chain-nonres}:\label{ell-stat-fact}
\begin{equation*}
\mathrm{SF}_{(\boldsymbol{n} \rightarrow \boldsymbol{m})} \equiv \sum_{\mathcal{P}(\boldsymbol{n} \rightarrow \boldsymbol{m})} \mathrm{CF} (\CP)\, w (\mathcal{P}).
\end{equation*}
\end{itemize}
\vspace{-5pt}
}}
\vspace{15pt}

\noindent
Spelled out in writing, the result is:
\be
F_k^{(\boldsymbol{n})} \simeq \frac{\Gamma(k)}{2\pi\rmi \left( \left( \boldsymbol{m}-\boldsymbol{n} \right) \cdot \boldsymbol{A} \right)^k}\, \sum_{\boldsymbol{m} \ne \boldsymbol{n}} \mathrm{SF}_{(\boldsymbol{n} \rightarrow \boldsymbol{m})}\, \chi_{(\boldsymbol{n} \rightarrow \boldsymbol{m})}(k).
\ee

Further proceeding along the lines of subsection~\ref{sec:large-order-F}, let us next address the (resurgent) large-order behaviour of the perturbative sector, $\Phi_{(0,0)}$. The analogue of \eqref{large-order-pert-quartic} will now be
\begin{eqnarray}
\label{eq:00-nonres}
F_k^{(0,0)} &\simeq& \frac{S^{(1)}_{\boldsymbol{e}_1}}{2\pi\mathrm{i}}\, \frac{\Gamma(k)}{A_1^k} \left( F^{(1,0)}_0 + \frac{A_1}{k-1}\, F^{(1,0)}_1 + \frac{A_1^2}{(k-1)(k-2)}\, F^{(1,0)}_2 + \cdots \right) + \\
&& + \frac{\left( S^{(1)}_{\boldsymbol{e}_1} \right)^2}{2\pi\mathrm{i}}\, \frac{\Gamma(k)}{\left( 2A_1 \right)^k} \left( F^{(2,0)}_0 + \frac{2A_1}{k-1}\, F^{(2,0)}_1 + \frac{\left( 2A_1 \right)^2}{(k-1)(k-2)}\, F^{(2,0)}_2 + \cdots \right) + \cdots \nonumber \\
&& + \frac{S^{(2)}_{\boldsymbol{e}_2}}{2\pi\mathrm{i}}\, \frac{\Gamma(k)}{A_2^k} \left( F^{(0,1)}_0 + \frac{A_2}{k-1}\, F^{(0,1)}_1 + \frac{A_2^2}{(k-1)(k-2)}\, F^{(0,1)}_2 + \cdots \right) + \nonumber \\
&& + \frac{\left( S^{(2)}_{\boldsymbol{e}_2} \right)^2}{2\pi\mathrm{i}}\, \frac{\Gamma(k)}{\left( 2A_2 \right)^k} \left( F^{(0,2)}_0 + \frac{2A_2}{k-1}\, F^{(0,2)}_1 + \frac{\left( 2A_2 \right)^2}{(k-1)(k-2)}\, F^{(0,2)}_2 + \cdots \right) + \cdots, \nonumber
\end{eqnarray} 
\noindent
where the \textit{two} instanton actions $A_1$ and $A_2=-|A_2|$ \textit{both} come into play at large order, albeit generically with distinct weights (a novelty as compared to the quartic potential). In particular, these relative contributions get exponentially-suppressed as $(A_1/|A_2|)^k$ when $A_1<|A_2|$, and $(|A_2|/A_1)^k$ when $A_1>|A_2|$; and such that the sector with smallest action is always the dominant one. We shall see how to ``disentangle'' them in the following. One similarity with the quartic potential is that contributions to the large-order growth of the perturbative series are exclusively associated to forward motions (either along $\boldsymbol{e}_1$, $\Phi_{(0,0)} \rightarrow \Phi_{(1,0)} \rightarrow \Phi_{(2,0)} \rightarrow \Phi_{(3,0)} \cdots$, or along $\boldsymbol{e}_2$, $\Phi_{(0,0)} \rightarrow \Phi_{(0,1)} \rightarrow \Phi_{(0,2)} \rightarrow \Phi_{(0,3)} \cdots$). However, since $A_2<0$, the motion along the $\boldsymbol{e}_2$-direction now yields oscillatory contributions to the large-order behaviour, since $A_2^k = (-1)^k |A_2|^k$. 

\begin{figure}[t!]
\begin{center}
\includegraphics[height=5.1cm]{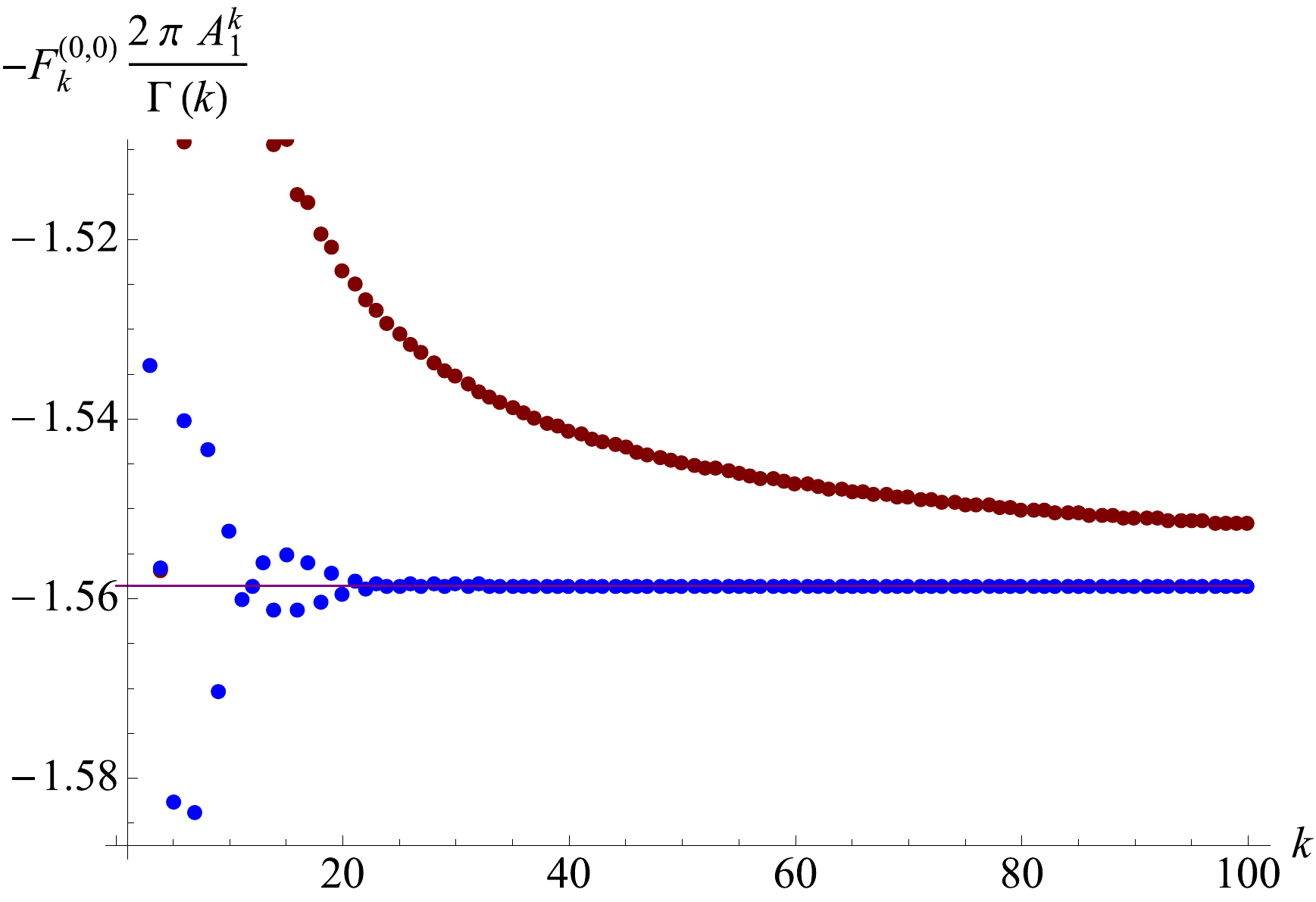}
$\,$
\includegraphics[height=5.1cm]{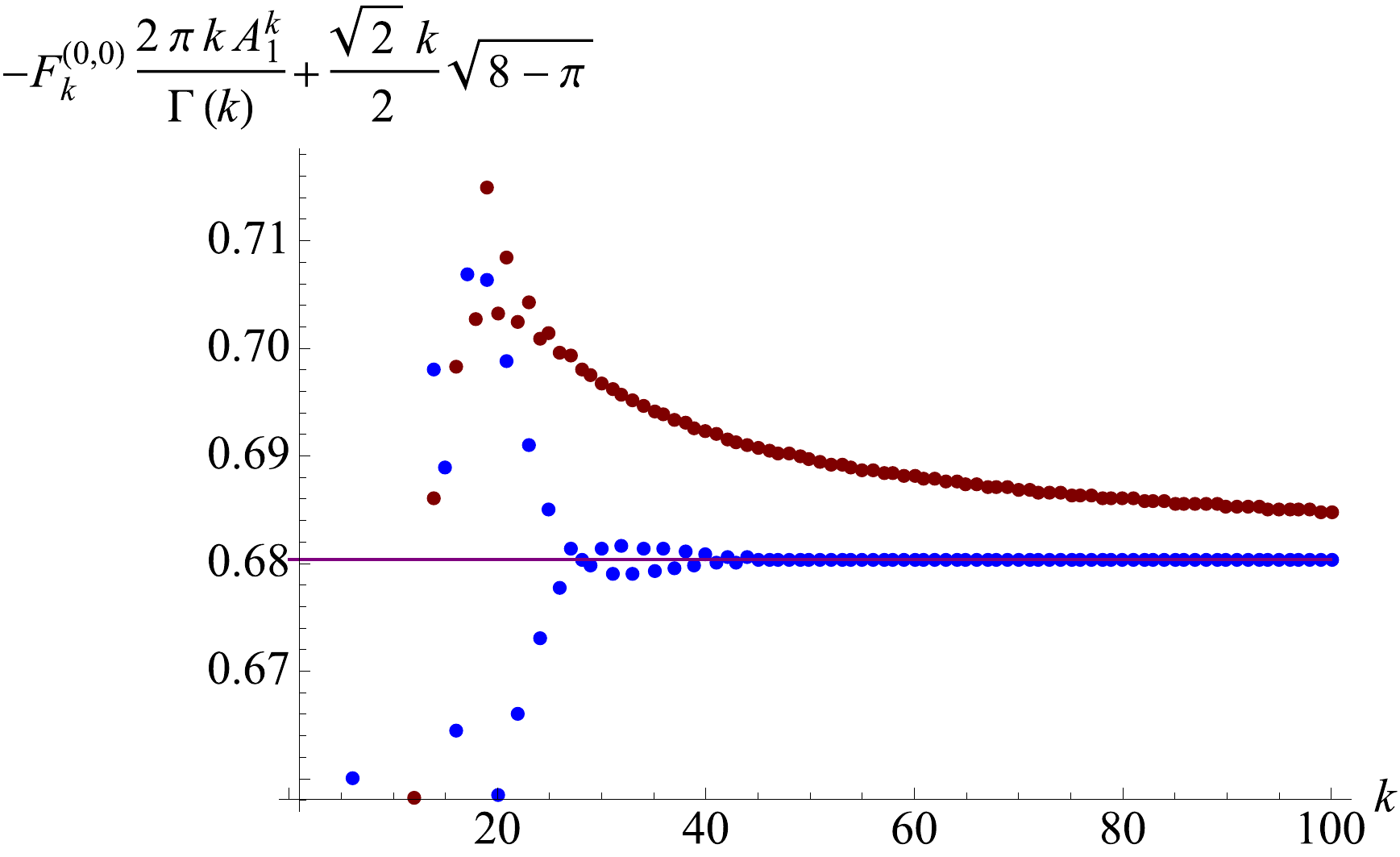}
\end{center}
\caption{The leading (left) and subleading $1/k$ (right) large-order behaviour of the perturbative sector $F^{(0,0)}_k$, weighted by the leading growth-factor. In red we plot the original sequences, while in blue we plot the Richardson transforms $\mathrm{RT}_{(0,0)}(r,k,5)$ ($r=0$ for the left plot and $r=1$ for the right plot). The purple lines show the constant values to which these sequences converge: on the left, $\mathrm{i} S^{(1)}_{\boldsymbol{e}_1} F^{(1,0)}_0 = - 2 \sqrt{1-m}$; while on the right, $\mathrm{i} s_1 A_1 S^{(1)}_{\boldsymbol{e}_1} F^{(1,0)}_1 = \frac{2}{\sqrt{1-m}}$ (and where we have set $m=\frac{\pi}{8}$).}
\label{fig:00-nonres}
\end{figure}

Let us next illustrate how one can decode nonperturbative information (say, the one-instanton sector $\Phi_{(1,0)}$) solely out of the above large-order behaviour of the perturbative coefficients. As usual, we set the non-resonant value of $m=\pi/8$. For this choice of $m$, $A_1<|A_2|$ and the dominant contribution to the large-order growth of the perturbative sector thus comes from the $\Phi_{(1,0)}$ nonperturbative sector, \textit{i.e.}, the first line of equation \eqref{eq:00-nonres}. At leading (exponential) order we have
\begin{eqnarray}
\label{eq:00-nonres-leading}
F_{k}^{(0,0)}\, \frac{2\pi\mathrm{i} A_1^k}{\Gamma(k)} &\simeq& S^{(1)}_{\boldsymbol{e}_1} \left( F^{(1,0)}_0 + \frac{F^{(1,0)}_1 A_1}{k} + \frac{F^{(1,0)}_1 A_1 + F^{(1,0)}_2 A_1^2}{k^2} + \cdots \right) + \mathcal{O}(2^{-k}) \\
&\equiv&
\sum_{r=0}^{+\infty} \frac{s_r}{k^r} + \mathcal{O}(2^{-k}),
\end{eqnarray}
\noindent
where the coefficients $\{ s_r \}$ in the second line are simply defined to match the expansion of the first line. If we plot the left-hand side of \eqref{eq:00-nonres-leading}, it should asymptote to $s_0 \equiv S^{(1)}_{\boldsymbol{e}_1} F^{(1,0)}_0 = 2\mathrm{i}\sqrt{1-m}$. Figure~\ref{fig:00-nonres} (left) shows how this is indeed the case, albeit rather slowly. In order to increase the rate of convergence, let us use Richardson extrapolation akin to what was done in subsection~\ref{sec:large-order-F} (recall that the Richardson transform effectively removes the subleading $1/k$ corrections that appear on the right-hand side of \eqref{eq:00-nonres-leading}, leading to much faster convergence). As such, in the left image of figure~\ref{fig:00-nonres} we have plotted the left-hand side of \eqref{eq:00-nonres-leading} (multiplied by $\mathrm{i}$) in red, whereas in blue we have plotted its corresponding fifth Richardson transform ($5$-RT) $\mathrm{RT}_{(0,0)}(0,k,5)$. The solid purple line shows the constant value $s_0$ to which this Richardson transform is converging to. Note that if at this stage we did not know the value of the Stokes coefficient $S^{(1)}_{\boldsymbol{e}_1}$, but only of $F^{(1,0)}_0$, then this large-order result would allow us to further determine\footnote{This procedure of computing Stokes coefficients might seem somehow redundant since we already know their values from the previous subsection. However, in general we do not have the luxury of independent computations of Stokes coefficients and this is essentially the only way to determine them. Of course that if neither Stokes coefficient nor the first term of the instanton series are known, then the best one can do is to determine their product; see \cite{asv11} on this point.} $S^{(1)}_{\boldsymbol{e}_1} = 2$. Comparing the exact value of $s_0$ with the value of the Richardson transform $\mathrm{RT}_{(0,0)}(0,100,5)$, one finds the usual, remarkably small relative errors
\be
\frac{\mathrm{RT}_{(0,0)} (0,100,5) - \mathrm{i} s_0}{\mathrm{i} s_0} \approx 2.7372 \times 10^{-10}.
\ee

Having computed the value of $F^{(1,0)}_0$ from the partition function, large order then allowed us to determine (or, more precisely, \textit{to check}) the value of the Stokes coefficient $S^{(1)}_{\boldsymbol{e}_1}$. But from here on one can proceed to compute all subleading terms $F^{(1,0)}_r$, with $r\ge1$, without any further input. In the right image of figure~\ref{fig:00-nonres} we have plotted the sequence $- F_k^{(0,0)}\, \frac{2\pi k A_1^k}{\Gamma(k)} - \mathrm{i} s_0 k$ in red, whereas in blue we have plotted its corresponding $5$-RT, $\text{RT}_{(0,0)}(1,k,5)$. These two sequences converge to $\mathrm{i} s_1 = \mathrm{i} A_1 S^{(1)}_{\boldsymbol{e}_1} F^{(1,0)}_1 = \frac{2}{\sqrt{1-m}} \approx 0.6804$ (whose value is shown as the solid purple line), in complete agreement with the value of $F^{(1,0)}_1$ which we directly computed from the recursion relations. One may then continue this process and subtract $s_0 + \frac{s_1}{k}$ from \eqref{eq:00-nonres-leading}, multiply the result by $k^2$, and thus obtain $F^{(1,0)}_2$. Iterate the procedure and one obtains $F^{(1,0)}_3$, $F^{(1,0)}_4$, and so on. At all stages, comparing exact coefficients with the numerical Richardson extrapolations one finds the familiar extremely small relative errors; \textit{e.g.},
\be
\frac{\mathrm{RT}_{(0,0)}(1,100,5) - \mathrm{i} s_1}{\mathrm{i} s_1} \approx -1.3120 \times 10^{-11}.
\ee

\begin{figure}[t!]
\begin{center}
\includegraphics[height=7cm]{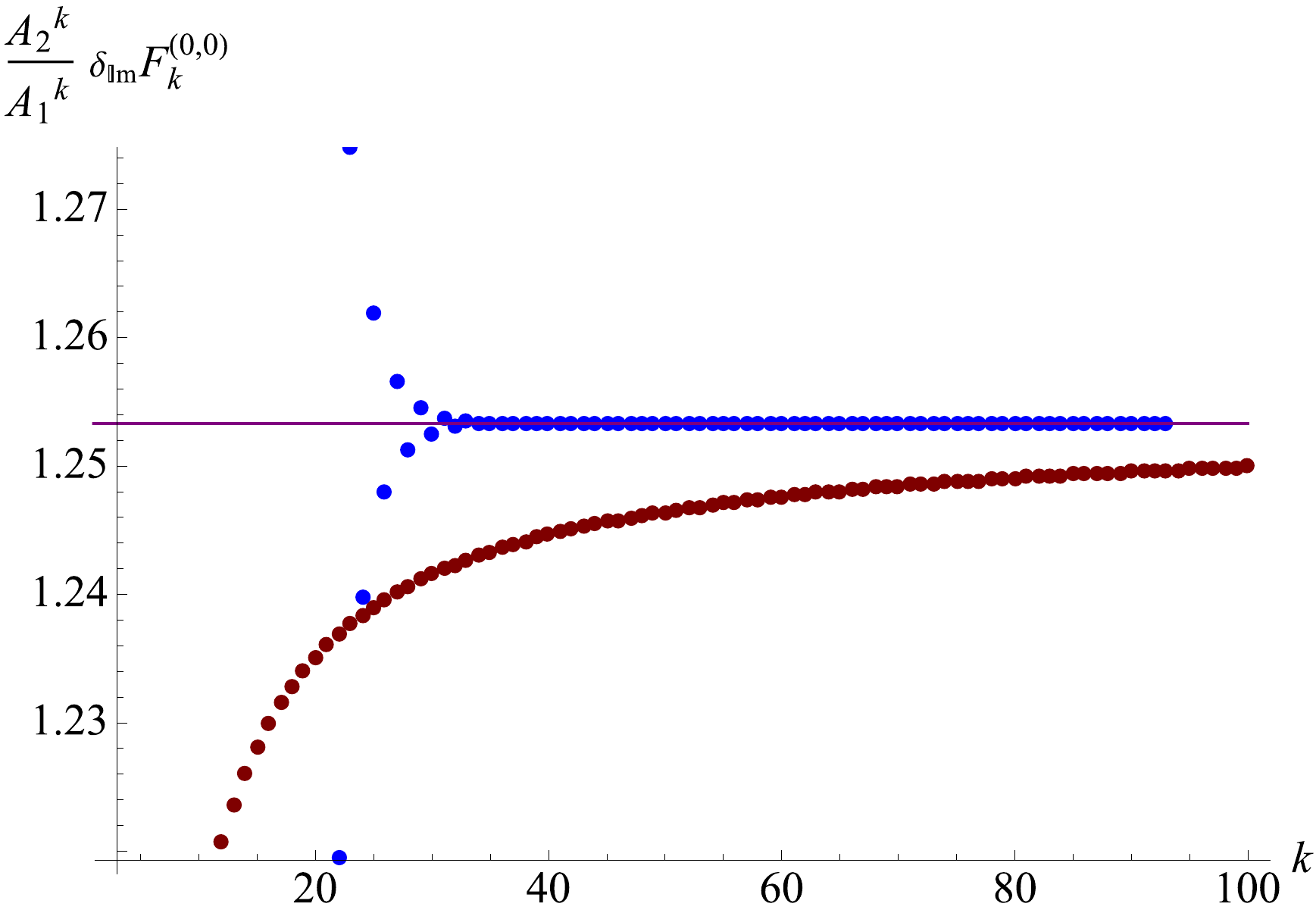}
\end{center}
\caption{Illustration of the subleading, exponentially-suppressed contributions to the large-order behaviour of the perturbative sector. The plot shows the sequence $\left( \frac{A_2}{A_1} \right)^k \delta_{\im} F^{(0,0)}_k$, defined in \eqref{eq:00-subleading-def}, and corresponding to a numerical BP resummation of order $N=50$. The original sequence is plotted in red, while in blue we plot its corresponding $\mathrm{RT}_{(0,0),\frac{A_2}{A_1}} (0,k,5)$ Richardson transform. The purple line shows the constant value that this sequence is converging to, $-\mathrm{i} S^{(2)}_{\boldsymbol{e}_2} F^{(0,1)}_0 = \sqrt{\pi/2} \approx 1.2533$.}
\label{fig:00-subexp}
\end{figure}

Moving on, let us focus on the terms which are exponentially suppressed, and for starts let us focus on the second and third lines of \eqref{eq:00-nonres}. In particular, note that since $|A_2| \approx 2.54 < 2A_1 \approx 3.29$ the leading exponentially-suppressed-sector contributing to the large-order growth of the perturbative series will be $\Phi_{(0,1)}$. In order to isolate the contribution of this sector, the contribution from the leading sector $\Phi_{(1,0)}$ needs to be removed, as a whole. Such procedures were thoroughly discussed in subsection~\ref{sec:large-order-F} via the use of Borel--Pad\'e (BP) resummation (\textit{e.g.}, recall \eqref{pade-S}), which may now be applied to the asymptotic series $\chi_{( (0,0) \rightarrow (1,0) )} (k)$, that may then be subtracted-out from the original sequence $F_k^{(0,0)}\, \frac{2\pi\mathrm{i} A_1^k}{\Gamma(k)}$. From \eqref{eq:00-nonres}, it then simply follows\footnote{Recall that $\mathrm{SF}_{( (0,0) \rightarrow (1,0) )} = S_{\boldsymbol{e}_1}^{(1)}$ and that $\mathrm{SF}_{( (0,0) \rightarrow (0,1) )} = S_{\boldsymbol{e}_2}^{(2)}$.} (compare with \eqref{Fk00BP-1})
\bea
\label{eq:ell-pert-subtraction}
F_k^{(0,0)}\, \frac{2\pi\mathrm{i} A_1^k}{\Gamma(k)} - \mathrm{SF}_{( (0,0) \rightarrow (1,0) )} \times \mathcal{S}_{0^{-}} \mathrm{BP}_N [\chi_{( (0,0) \rightarrow (1,0) )}] (k) &\simeq& \\
&&
\hspace{-250pt}
\simeq \left( \frac{A_1}{A_2} \right)^k \mathrm{SF}_{( (0,0) \rightarrow (0,1) )} \left( F^{(0,1)}_0 + \frac{F^{(0,1)}_1 A_2}{k} + \frac{F^{(0,1)}_1 A_2 + F^{(0,1)}_2 A_2^2}{k^2} + \cdots \right) + \mathcal{O} (2^{-k}). \nonumber
\eea
\noindent
The above Laplace transform yielding the resummed series, $\CS_{0^-} \mathrm{BP}_N [\chi_{( (0,0) \rightarrow (1,0) )}] (k)$, is a \textit{lateral} resummation evaluated along $\theta=0^{-}$ (\textit{i.e.}, just below the real line). This is essentially the same formulae that we have already seen in subsection~\ref{sec:large-order-F}---\textit{e.g.}, recall the discussion following \eqref{Fk00BP-1}---and such lateral resummation is needed in order to avoid the poles of the BP approximant to $\chi_{( (0,0) \rightarrow (1,0) )}$ which lie on the positive real line. Exactly as it happened in subsection~\ref{sec:large-order-F}, this will produce a result which has both real and imaginary parts (compare with \eqref{S-BP=Re+Im-compare})
\be
\label{lateralS-BPchi00-01}
\CS_{0^{-}} \mathrm{BP}_{N} [\chi_{( (0,0) \rightarrow (1,0) )}] (k) = \CS_{\re} \mathrm{BP}_{N} [\chi_{( (0,0) \rightarrow (1,0) )}] (k) + \rmi\, \CS_{\im} \mathrm{BP}_{N} [\chi_{( (0,0) \rightarrow (1,0) )}] (k).
\ee
\noindent
We find that the imaginary component is the leading one, of the same magnitude as the original sequence; while the real component is exponentially suppressed, as $\CO (2^{-k})$. Once the subtraction in \eqref{eq:ell-pert-subtraction} is carried through (do recall that the perturbative coefficients $F_k^{(0,0)}$ are real), one finds the following magnitudes\footnote{As $|A_2| < 2A_1$ it follows $|A_2|/A_1 < 2$ and the imaginary contribution is the exponentially \textit{leading} one. This is different from what happened back with the quartic potential, wherein the imaginary part was the exponentially \textit{suppressed} one---recall, \textit{e.g.}, figure~\ref{fig:large-order-pert-series-1inst-resum}.}:
\begin{eqnarray}
\label{eq:00-subleading-def}
\mathrm{i}\, \delta_{\im} F^{(0,0)}_k &\equiv& F_k^{(0,0)}\, \frac{2\pi\mathrm{i} A_1^k}{\Gamma(k)} - \mathrm{i} \mathrm{SF}_{( (0,0) \rightarrow (1,0) )}\, \CS_{\im} \mathrm{BP}_{N} [\chi_{( (0,0) \rightarrow (1,0) )}] (k) \simeq \CO ( (A_2/A_1)^{-k} ), \qquad \\
\label{eq:00-subleading-def-im}
\delta_{\re} F^{(0,0)}_k &\equiv& - \mathrm{SF}_{( (0,0) \rightarrow (1,0) )}\, \CS_{\re} \mathrm{BP}_{N} [\chi_{( (0,0) \rightarrow (1,0) )}] (k) \simeq \CO (2^{-k}).
\end{eqnarray}
\noindent
In this way, the large-order sequence, and corresponding behaviour, we wish to address is, finally,
\be
\label{eq:00-subleading-exp}
\left(\frac{A_2}{A_1} \right)^k \delta_{\im} F^{(0,0)}_k \simeq - \mathrm{i} \mathrm{SF}_{( (0,0) \rightarrow (0,1) )} \left( F^{(0,1)}_0 + \frac{F^{(0,1)}_1 A_2}{k} + \frac{F^{(0,1)}_1 A_2 + F^{(0,1)}_2 A_2^2}{k^2} + \cdots \right).
\ee
\noindent
The overall factor $\left( \frac{A_2}{A_1} \right)^k$ on the left-hand side of \eqref{eq:00-subleading-exp} removes any exponential dependence, and we are left with a simple sequence which should converge to $-\mathrm{i} S^{(2)}_{\boldsymbol{e}_2} F^{(0,1)}_0 = \sqrt{\pi/2}$ (where we are using $S^{(2)}_{\boldsymbol{e}_2} = 2$ and $F^{(0,1)}_0 = \mathrm{i}\frac{\sqrt{\pi}}{2\sqrt{2}}$). Indeed, such convergence is very explicitly seen in figure~\ref{fig:00-subexp} (corresponding to a numerical BP resummation of order $N=50$). Let us recall the notation $\mathrm{RT}_{\boldsymbol{n},\ell}(0,k,N)$, used to denote the $N^{\text{th}}$-RT associated to the $\ell^{-k}$ exponentially-suppressed behaviour of the large-order relation obeyed by the $\boldsymbol{n}$-instanton coefficients $F_k^{(\boldsymbol{n})}$. For our current case, \textit{i.e.}, for the sequence $\left( \frac{A_2}{A_1} \right)^k \delta_{\im} F^{(0,0)}_k$ obtained via BP resummation, we find the usual extremely-small convergence\footnote{Had we not previously known the separate values of $S^{(2)}_{\boldsymbol{e}_2}$ and $F^{(0,1)}_0$, then we could now use this convergence to compute their product, $S^{(2)}_{\boldsymbol{e}_2} F^{(0,1)}_0$, to such small error.} error
\be
\frac{\mathrm{RT}_{(0,0),\frac{A_2}{A_1}}(0,95,5) - \sqrt{\frac{\pi}{2}}}{\sqrt{\frac{\pi}{2}}} \approx 4.0376 \times 10^{-10}.
\ee
\noindent
Via continued BP resummations of the subleading large-order contributions, one could go on analysing the subsequent exponentially-suppressed sequences within \eqref{eq:00-nonres}, much along the spirit of what was done in subsection~\ref{sec:large-order-F}---a laborious exercise left for the interested reader.

Let us now turn to the analysis of the large-order behaviour of higher instanton sectors, in particular let us focus upon the $\Phi_{(0,1)}$ sector (which in itself will include most relevant features of multi-instantonic large-orders). Using the expressions for both Stokes discontinuities \eqref{eq:01-disc-0} and \eqref{eq:01-disc-pi} into Cauchy's theorem \eqref{Cauchy-two-disc}, one obtains the large-order relation\footnote{The ordering of the contributions is organized according to their exponential dominance, which follows from the ordered sequence $1 < |A_2|/A_1 \approx 1.546 < 2 < (A_1+|A_2|)/A_1 \approx 2.546\, ( = |A_2| ) < 3 < \cdots$.}
\begin{eqnarray}
\label{eq:01-nonres}
F_k^{(0,1)}\, \frac{2\pi\mathrm{i}A_1^k}{\Gamma(k)} &\simeq& S^{(1)}_{\boldsymbol{e}_1} \left( F^{(1,1)}_0 + \frac{A_1}{k-1}\, F^{(1,1)}_1 + \frac{A_1^2}{(k-1)(k-2)}\, F^{(1,1)}_2 + \cdots \right) + \\
&&
\hspace{-50pt}
+ \left( -\frac{A_1}{|A_2|} \right)^k 2 S^{(2)}_{\boldsymbol{e}_2} \left( F^{(0,2)}_0 + \frac{-|A_2|}{k-1}\, F^{(0,2)}_1 + \frac{|A_2|^2}{(k-1)(k-2)}\, F^{(0,2)}_2 + \cdots \right) + \nonumber \\
&& 
\hspace{-50pt}
+ \left( \frac{A_1}{|A_2|} \right)^k \left( S^{(0)}_{-\boldsymbol{e}_2} + \frac{1}{2} S^{(1)}_{\boldsymbol{e}_1} S^{(0)}_{-\boldsymbol{e}_1-\boldsymbol{e}_2} \right) F_0^{(\widetilde{\boldsymbol{0}})} + \nonumber \\
&&
\hspace{-50pt}
+ \frac{1}{2^k} \left( S^{(1)}_{\boldsymbol{e}_1} \right)^2 \left( F^{(2,1)}_0 + \frac{2A_1}{k-1}\, F^{(2,1)}_1 + \frac{(2A_1)^2}{(k-1)(k-2)}\, F^{(2,1)}_2 + \cdots \right) + \nonumber \\
&&
\hspace{-50pt}
+ \left( \frac{A_1}{A_1+|A_2|} \right)^k S^{(1)}_{\boldsymbol{e}_1-\boldsymbol{e}_2} \left( F^{(1,0)}_0 + \frac{A_1+|A_2|}{k-1}\, F^{(1,0)}_1 + \frac{(A_1+|A_2|)^2}{(k-1)(k-2)}\, F^{(1,0)}_2 + \cdots \right) + \nonumber \\
&&
\hspace{-50pt}
+ \mathcal{O} (3^{-k}). \nonumber
\end{eqnarray} 
\noindent
This large-order relation is particularly interesting as it displays the many different types of resurgence which appear within this problem. In particular, the third line above corresponds to the ``orthogonal'' resurgence motion in the alien lattice, akin to the analogous term which appeared back in \eqref{large-order-1-inst-quartic}. The large-order relation \eqref{eq:01-nonres} may be equivalently written with slightly simpler combinatorial factors, using Borel residues instead of Stokes coefficients. One has:
\begin{eqnarray}
\label{eq:01-nonres-BOREL}
F_k^{(0,1)}\, \frac{2\pi\mathrm{i}A_1^k}{\Gamma(k)} &\simeq& - \mathsf{S}_{(0,1) \rightarrow (1,1)} \left( F^{(1,1)}_0 + \frac{A_1}{k-1}\, F^{(1,1)}_1 + \frac{A_1^2}{(k-1)(k-2)}\, F^{(1,1)}_2 + \cdots \right) - \\
&&
\hspace{-50pt}
- \left( -\frac{A_1}{|A_2|} \right)^k \mathsf{S}_{(0,1) \rightarrow (0,2)} \left( F^{(0,2)}_0 + \frac{-|A_2|}{k-1}\, F^{(0,2)}_1 + \frac{|A_2|^2}{(k-1)(k-2)}\, F^{(0,2)}_2 + \cdots \right) - \nonumber \\
&& 
\hspace{-50pt}
- \left( \frac{A_1}{|A_2|} \right)^k \mathsf{S}_{(0,1) \rightarrow \widetilde{\boldsymbol{0}}}\, F_0^{(\widetilde{\boldsymbol 0})} - \nonumber \\
&&
\hspace{-50pt}
- \frac{1}{2^k}\, \mathsf{S}_{(0,1) \rightarrow (2,1)} \left( F^{(2,1)}_0 + \frac{2A_1}{k-1}\, F^{(2,1)}_1 + \frac{(2A_1)^2}{(k-1)(k-2)}\, F^{(2,1)}_2 + \cdots \right) - \nonumber \\
&&
\hspace{-50pt}
- \left( \frac{A_1}{A_1+|A_2|} \right)^k \mathsf{S}_{(0,1) \rightarrow (1,0)} \left( F^{(1,0)}_0 + \frac{A_1+|A_2|}{k-1}\, F^{(1,0)}_1 + \frac{(A_1+|A_2|)^2}{(k-1)(k-2)}\, F^{(1,0)}_2 + \cdots \right) - \nonumber \\
&&
\hspace{-50pt}
- \mathcal{O} (3^{-k}). \nonumber
\end{eqnarray} 

\begin{figure}[t!]
\begin{center}
\includegraphics[height=5.2cm]{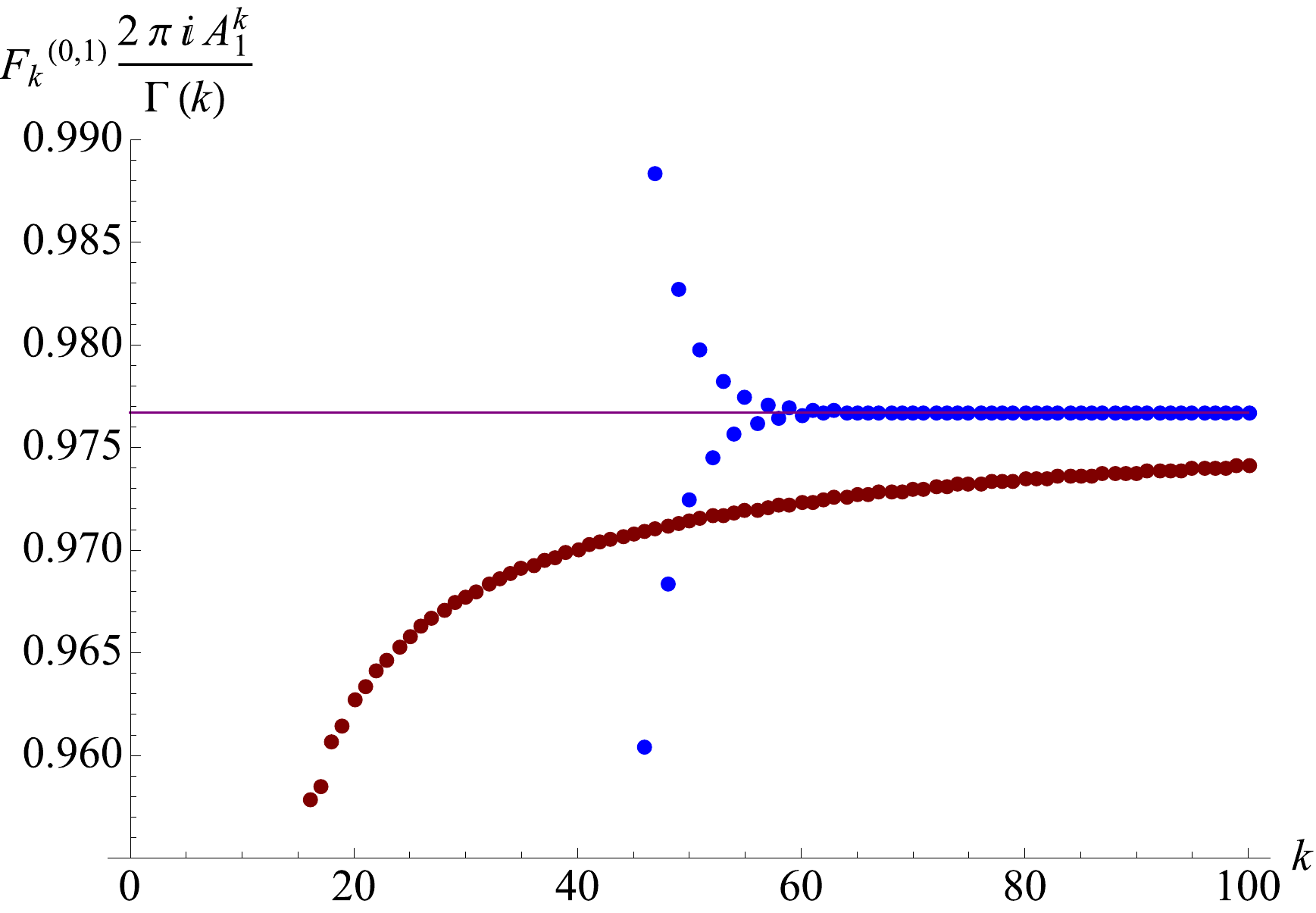}
$\qquad$
\includegraphics[height=5.2cm]{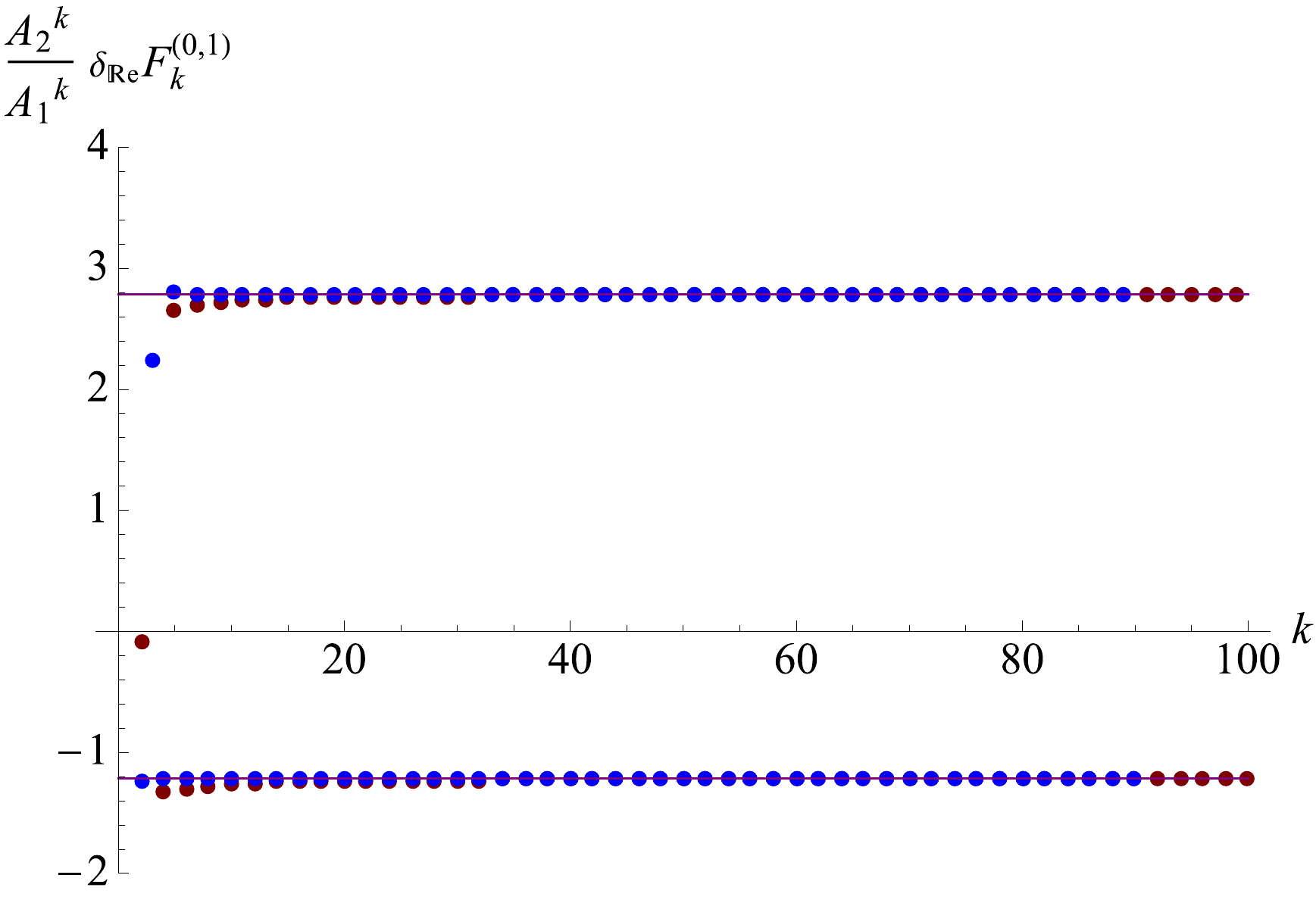}
\end{center}
\caption{Leading (left) and exponentially suppressed (right; via BP resummation of order $N=50$) large-order behaviour of the coefficients $F^{(0,1)}_k$, weighted by the leading growth factor. In red we plot the original sequences, while their corresponding $5$-RTs are plotted in blue (these are $\mathrm{RT}_{(0,1)}(0,k,5)$ for the left plot and $\mathrm{RT}_{(0,1),\frac{A_2}{A_1},\text{even/odd}}(0,k,5)$ for the right plot). Purple (solid) lines denote the constants to which the sequences converge: $S^{(1)}_{\boldsymbol{e}_1} F^{(1,1)}_0 = 2 \sqrt{\frac{\pi}{8}\big(1-\frac{\pi}{8}\big)}$ (left) and  $2 S^{(2)}_{\boldsymbol{e}_2} F^{(0,2)}_0 \pm S^{(0)}_{-\boldsymbol{e}_2} F_0^{(\widetilde{\boldsymbol{0}})} = \frac{\pi}{4} \mp 2$ (right).}
\label{fig:01-nonres}
\end{figure}

Large-order analysis now follows as usual. Let us start with a quick check of the leading sequence (the first line above). On the left image of figure~\ref{fig:01-nonres} we plot the sequence $F_k^{(0,1)}\, \frac{2\pi\mathrm{i} A_1^k}{\Gamma(k)}$ (red) alongside its fifth Richardson transform (blue), $\mathrm{RT}_{(0,1)}(0,k,5)$, and compare these with the expected result $S^{(1)}_{\boldsymbol{e}_1} F_0^{(1,1)}$ (purple) (using the explicit values $S^{(1)}_{\boldsymbol{e}_1} = 2$ and $F_0^{(1,1)} = \sqrt{m-m^2} \approx 0.488$). The convergence is excellent, as always, with a relative error given by
\be
\frac{\mathrm{RT}_{(0,1)}(0,95,5) - 2 \sqrt{\frac{\pi}{8} \big(1-\frac{\pi}{8}\big)}}{2 \sqrt{\frac{\pi}{8} \big(1-\frac{\pi}{8}\big)}} \approx 5.0594 \times 10^{-10}.
\ee

Moving towards exponentially-suppressed contributions (the second and following lines above), let us first address how to ``disentangle'' the second and third lines of \eqref{eq:01-nonres}. These correspond to contributions from the sectors $\Phi_{(0,2)}$ and $\widetilde{\Phi}_{(0,0)}$, which have the same weight but are exponentially suppressed as compared to the contribution of the $\Phi_{(1,1)}$ sector. As such, in order to isolate the contribution of these subleading sectors, the whole contribution of $\Phi_{(1,1)}$ has to be resummed and subtracted from the large-order growth of the coefficients $F^{(0,1)}_k$. This is basically the same which was just done within the analysis of the perturbative sector: the contribution to be subtracted is in itself an asymptotic series, and one first needs to calculate its BP resummation. This resummation will once again have both real and imaginary parts, and given that the coefficients $F^{(0,1)}_k$ are purely imaginary, one ends up working with the combinations\footnote{The statistical factor here is $\mathrm{SF}_{( (0,1) \rightarrow (1,1)) } = S_{\boldsymbol{e}_1}^{(1)}$.}
\begin{eqnarray}
\label{eq:01-subleading-def}
\delta_{\re} F^{(0,1)}_k &\equiv& F_k^{(0,1)}\, \frac{2\pi\rmi A_1^k}{\Gamma(k)} - \mathrm{SF}_{( (0,1) \rightarrow (1,1) )}\, \CS_{\re} \mathrm{BP}_{N} [\chi_{( (0,1) \rightarrow (1,1) )}] (k) \simeq \CO ( (A_2/A_1)^{-k} ), \qquad \\
\label{eq:01-subleading-def-IM}
\mathrm{i}\, \delta_{\im} F^{(0,1)}_k &\equiv& - \rmi \mathrm{SF}_{( (0,1) \rightarrow (1,1) )}\, \CS_{\im} \mathrm{BP}_{N} [\chi_{( (0,1) \rightarrow (1,1) )}] (k) \simeq \CO (2^{-k}).
\end{eqnarray}
\noindent
Clearly the relevant sequence is $\delta_{\re} F^{(0,1)}_k$, whose leading large-order growth will be dictated by the second and third lines of \eqref{eq:01-nonres} (while $\delta_{\im} F^{(0,1)}_k$ is exponentially suppressed with respect to $\delta_{\re} F^{(0,1)}_k$). This large order is:
\begin{eqnarray}
\left( \frac{A_2}{A_1} \right)^k \delta_{\re} F^{(0,1)}_k &\simeq& \mathrm{SF}_{( (0,1) \rightarrow (0,2) )} \left( F^{(0,2)}_0 + \frac{F^{(0,2)}_1 A_2}{k} + \frac{F^{(0,2)}_1 A_2 + F^{(0,2)}_2 A_2^2}{k^2} + \cdots \right) + \nonumber \\
&&
+ \left(-1\right)^k \mathrm{SF}_{( (0,1) \rightarrow \widetilde{\boldsymbol{0}} )}\, F_0^{(\widetilde{\boldsymbol{0}})}.
\label{eq:01-subleading}
\end{eqnarray} 
\noindent
The statistical factors in the above expression are given by $\mathrm{SF}_{( (0,1) \rightarrow (0,2) )} = 2 S^{(2)}_{\boldsymbol{e}_2} = 4$ and $\mathrm{SF}_{( (0,1) \rightarrow \widetilde{\boldsymbol{0}} )} = S^{(0)}_{-\boldsymbol{e}_2} + \frac{1}{2}S^{(1)}_{\boldsymbol{e}_1} S^{(0)}_{-\boldsymbol{e}_1-\boldsymbol{e}_2} = -2$ (using the values for the Stokes coefficients previously determined in \eqref{ell-stokes-freeen-2-1}, \eqref{ell-stokes-freeen-1-3}, and \eqref{ell-stokes-freeen-1-1}). Further, we have $F_0^{(0,2)} = \frac{1}{2}m$ and $F_0^{(\widetilde{\boldsymbol{0}})} = 1$. Now, the leading large-order behaviour of the sequence on the left-hand side of \eqref{eq:01-subleading} has a fixed contribution (coming from the first line) alongside an alternating contribution (coming from the second line). As such, this sequence is composed of two (even and odd) sub-sequences, which respectively converge to either $\mathrm{SF}_{( (0,1) \rightarrow (0,2) )} F^{(0,2)}_0 + \mathrm{SF}_{( (0,1) \rightarrow \widetilde{\boldsymbol{0}} )} F_0^{(\widetilde{\boldsymbol{0}})} = 2 m - 2 \approx -1.215$ or $\mathrm{SF}_{( (0,1) \rightarrow (0,2) )} F^{(0,2)}_0 - \mathrm{SF}_{( (0,1) \rightarrow \widetilde{\boldsymbol{0}} )} F_0^{(\widetilde{\boldsymbol{0}})} = 2 m + 2 \approx 2.785$. These two sub-sequences, together with their respective 5-RTs, are plotted on the right image of figure~\ref{fig:01-nonres} (the label even/odd on the RTs denotes the Richardson transform of the even/odd sequence, respectively). Comparing the limits of these sub-sequences---obtained from their $5$-RT---with the exact coefficients---obtained from the recursion relations---we find a relative numerical error of the order $\sim 10^{-10}$. As always, the numerical precision obtained from large-order analysis is quite remarkable. 

This concludes our analysis of the resurgent structure of the transseries solution we constructed for the elliptic-potential free energy. The numerical checks we performed thoroughly verified the large-order predictions which were obtained from resurgence, and in this way completely vindicated our earlier transseries constructions and calculations. It is worth noting that up to now the analysis was quite similar to the case of the quartic potential, once one upgrades the underlying structure from the one-dimensional alien chain to a two-dimensional alien lattice. This in itself is an important extension to learn and have in mind, but, as alluded to earlier, there is more to this example than just a higher-dimensional generalization of section~\ref{sec:quartic}. This is in fact the final topic we wish to discuss, as we finally turn to the study of \textit{resonance}.

\subsection{Nonlinear Resonance and Transseries Structures}\label{sec:nonlinear-resonance}

As already pointed out in section~\ref{sec:physics}, multiple instanton actions may give rise to new phenomena. Back then, we limited ourselves to discuss what we dubbed as the ``non-resonant case'', where the instanton actions in $\boldsymbol{A} = \left( A_1, \ldots, A_k \right)$ were $\BZ$-linearly independent; recall equation \eqref{non-resonant-case}
\be
\left. \nexists \,\, \boldsymbol{n} \neq \boldsymbol{0} \in \BZ^k \,\, \right| \,\, \boldsymbol{n} \cdot \boldsymbol{A} = 0.
\ee
\noindent
In particular, this condition implied that the projection map \eqref{projectionmapZkC} had vanishing kernel, \textit{i.e.}, $\ker \mathfrak{P} = \boldsymbol{0}$. It so happens that in our present example the instanton actions \eqref{eq:elliptic_critical} do depend on a parameter (they are ``tunable'') and, depending on this choice of $m$, the above condition may be violated. In this case, we have to deal with \textit{resonance}, to which we turn in the following.

One way to understand the origin of this phenomenon is to think of the \textit{inverse} of the coupling-constant variable, $1/x$, as a (imaginary) ``time variable''; to think of the instanton actions, $A_i$, as ``frequencies''; and to think of the transmonomials as ``plane-wave contributions''. In such a familiar physical scenario, resonance usually occurs for special choices of the different frequencies at play (\textit{e.g.}, at special values of the driving frequencies). It turns out that the relevant concept of (nonlinear) resonance within the transseries setting finds its origin within perturbed Hamiltonian dynamics and its associated KAM theory; see, \textit{e.g.}, \cite{gh83, p09}. In the following we shall discuss resonance within the present transseries language, but will follow such discussion in parallel with the one one might have if discussing Hamiltonian dynamics.

Consider the scenario of section~\ref{sec:physics}, where we are dealing with a multi-parameter (specifically, $k$-dimensional) transseries of the form \eqref{eq:sec5-transseries-vec} and \eqref{eq:sec5-asympt-sectors-vec}, and where the $k$ instanton actions assemble vectorially as $\boldsymbol{A} = \left( A_1, \ldots, A_k \right)$. The instanton actions are said to be \textit{rationally independent} or \textit{non-resonant}, when they are linearly independent over the rationals:
\be
\forall \,\, \boldsymbol{n} \in \BZ^k, \qquad \boldsymbol{n} \cdot \boldsymbol{A} = 0 \,\, \Rightarrow \,\, \boldsymbol{n}=\boldsymbol{0}.
\ee
\noindent
Otherwise, the instanton actions are said to be \textit{rationally dependent} or \textit{resonant}, and one has:
\be
\label{rationally-dependent-resonant-actions}
\left. \exists \,\, \boldsymbol{\mathfrak{n}} \neq \boldsymbol{0} \in \BZ^k \,\, \right| \,\, \boldsymbol{\mathfrak{n}} \cdot \boldsymbol{A} = 0.
\ee
\noindent
Each such vector $\boldsymbol{\mathfrak{n}}$ is denoted as an \textit{action correlation}. The set of all action correlations, $\left. \left\{ \boldsymbol{\mathfrak{n}}_i \neq \boldsymbol{0} \in \BZ^k \,\,\right|\,\, \boldsymbol{\mathfrak{n}}_i \cdot \boldsymbol{A} = 0 \right\}$, alongside the standard vector addition, is a subgroup of $\left( \BZ^k, + \right)$. This group of action-correlations can generically have $r$ generators, with $1 \leq r \leq k-1$, which are also linearly-independent $\BZ^k$-vectors. This number of generators equals the number of truly independent correlations one can write down, in-between the $k$ instanton actions.

In the Hamiltonian dynamics formulation of integrable systems, set-up in action-angle variables, the above instanton actions correspond to the frequencies of the periodic motion taking place in phase space. More precisely, the trajectory of motion in phase space generically takes place over a $k$-dimensional torus, $\BT^k$. But if there are, say, $r$ frequency correlations, then the trajectory ``collapses'' to having support on the smaller torus $\BT^{k-r} \subset \BT^k$. As long as the frequencies are rationally independent (non-resonant), the phase-curves on the torus $\BT^k$ are everywhere dense. But when they are rationally dependent (resonant), each orbit is only dense in the lower-dimensional torus $\BT^{k-r}$ (\textit{i.e.}, on the original torus $\BT^k$ the trajectories are now only closed). This split between resonant and non-resonant tori\footnote{Note how resonant tori are the analogue of the rational numbers, while non-resonant tori are the analogue of the irrational numbers; which extends to their respective cardinalities.} is crucial for KAM theory. Start off with an integrable system, whose motion is confined to an invariant torus. Upon an arbitrarily small but nonlinear perturbation, resonant tori are generically destroyed by this perturbation (eventually giving rise to chaotic behaviour). However, non-resonant tori (which end-up being the majority) do survive (thus constraining the possible chaotic motions near each resonance). This fascinating story is developed at length within the setting of KAM theory, to which we refer the reader to \cite{gh83, p09} and references therein---the goal of this paragraph was simply to point out how there are countless interesting links between the theory of resurgence and that of dynamical systems.

\begin{figure}[t!]
\begin{center}
\includegraphics[width=7.5cm]{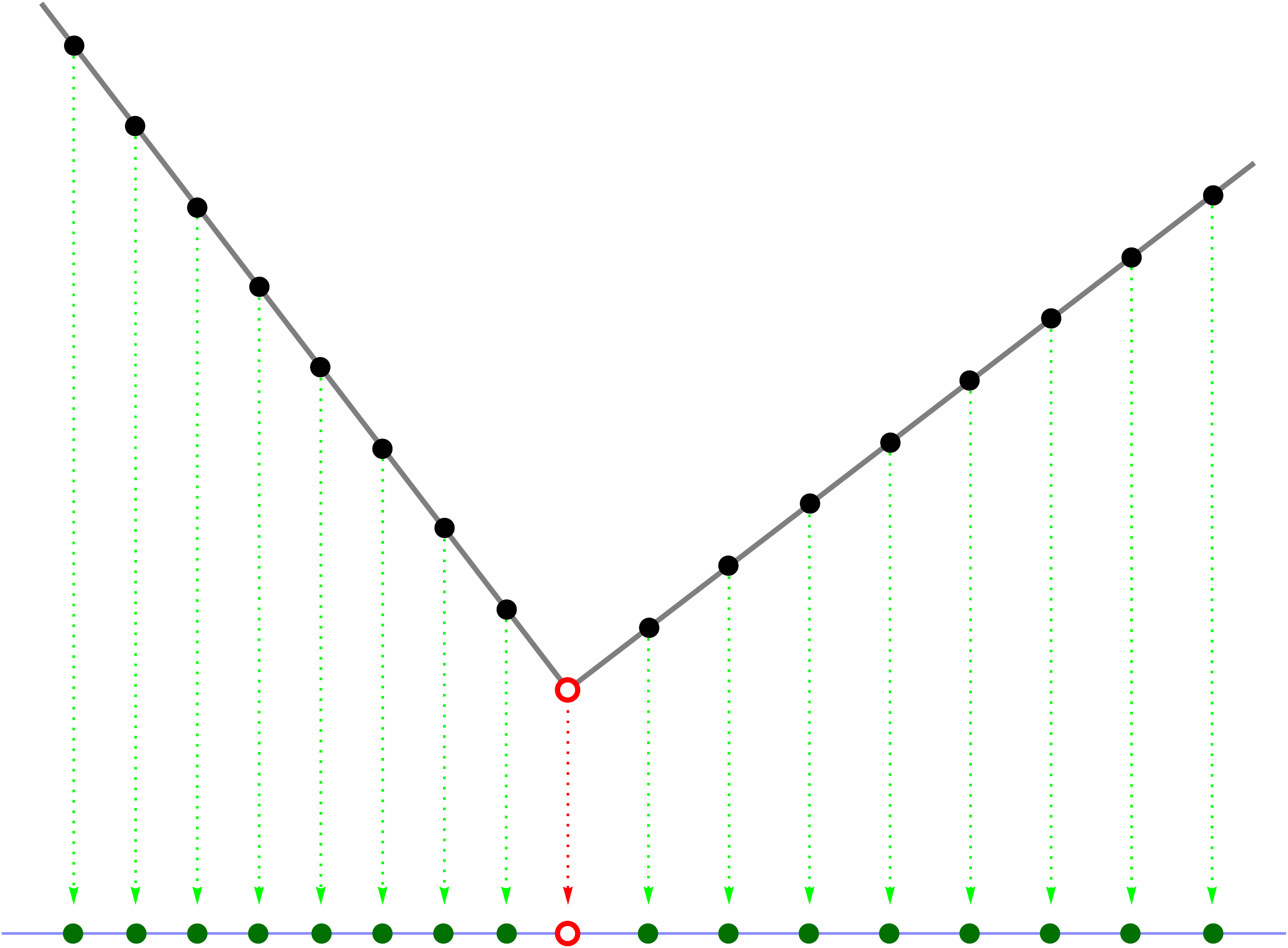}
$\qquad$
\includegraphics[width=7.5cm]{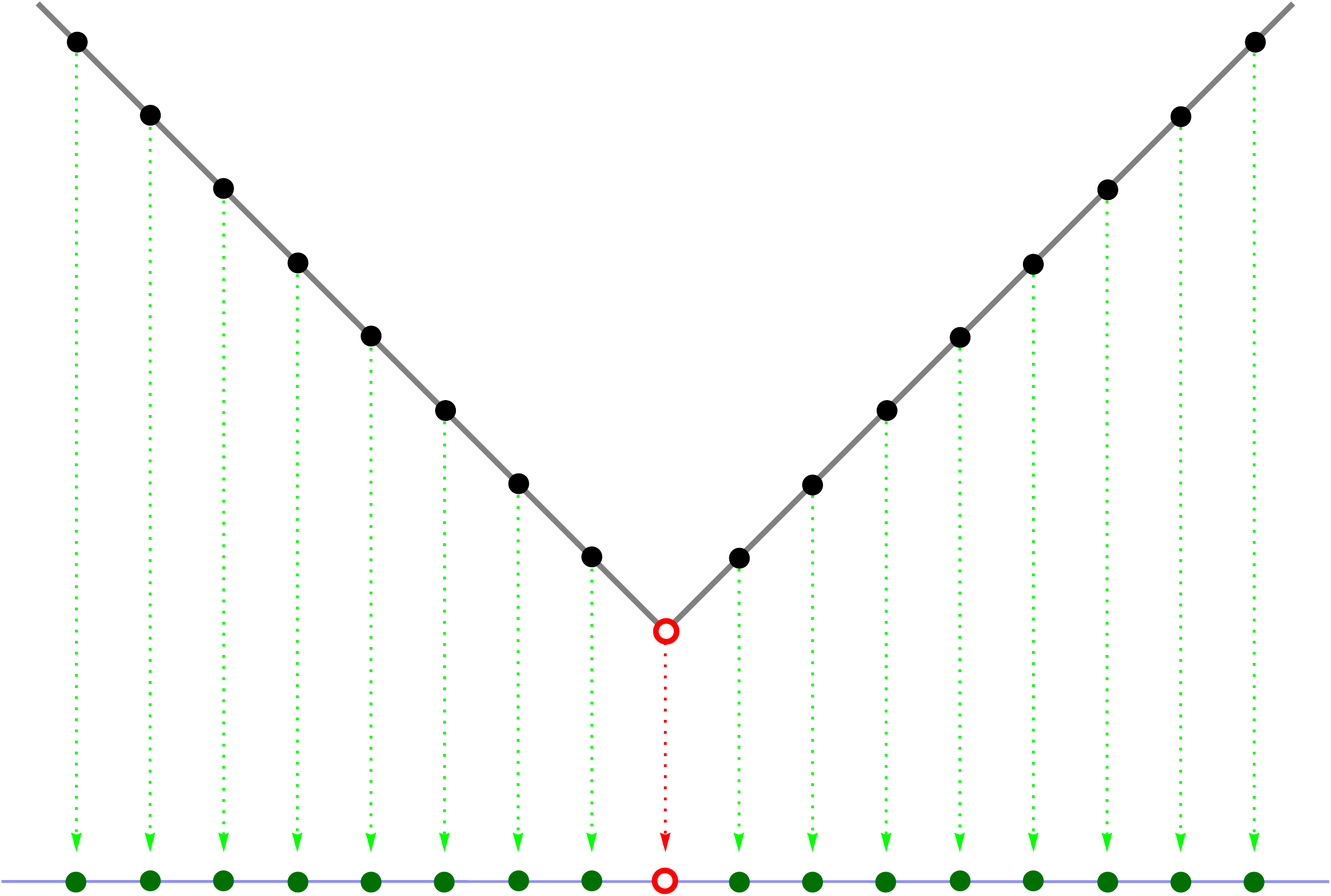}
\end{center}
\caption{Non-vanishing nodes of the $(0,0)$ alien lattice, and their projection upon the complex Borel plane. The left plot shows a non-resonant example, where $A_1$ and $A_2$ are rationally independent (but both still real and symmetric), while the right plot shows the resonant example we are considering, with $A_2 = - A_1$. For the perturbative sector, nothing special really happens as we approach the resonant case.
}
\label{fig:sec6-00resonance}
\end{figure}

\begin{figure}[t!]
\begin{center}
\includegraphics[width=7.5cm]{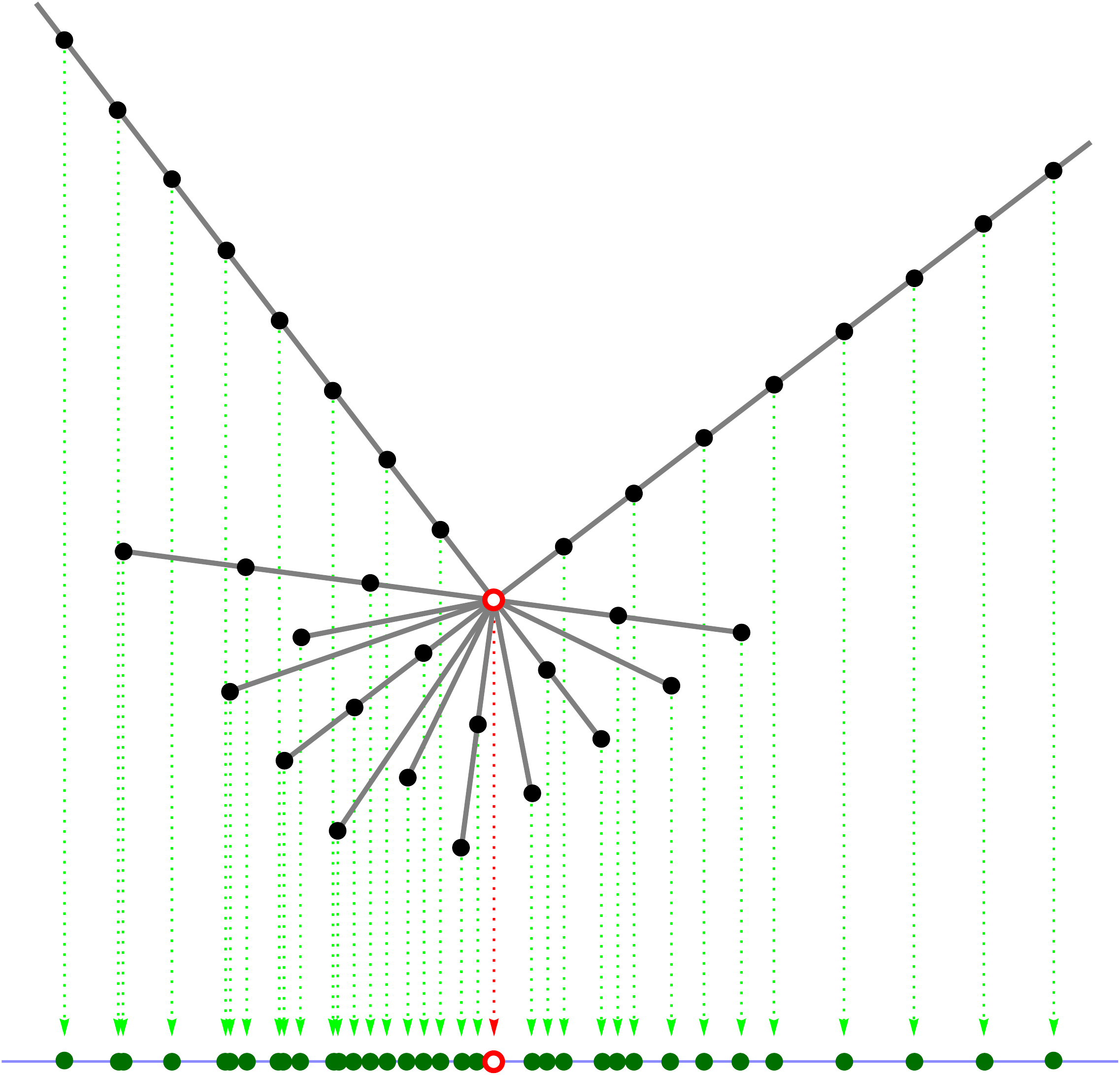}
$\qquad$
\includegraphics[width=7.5cm]{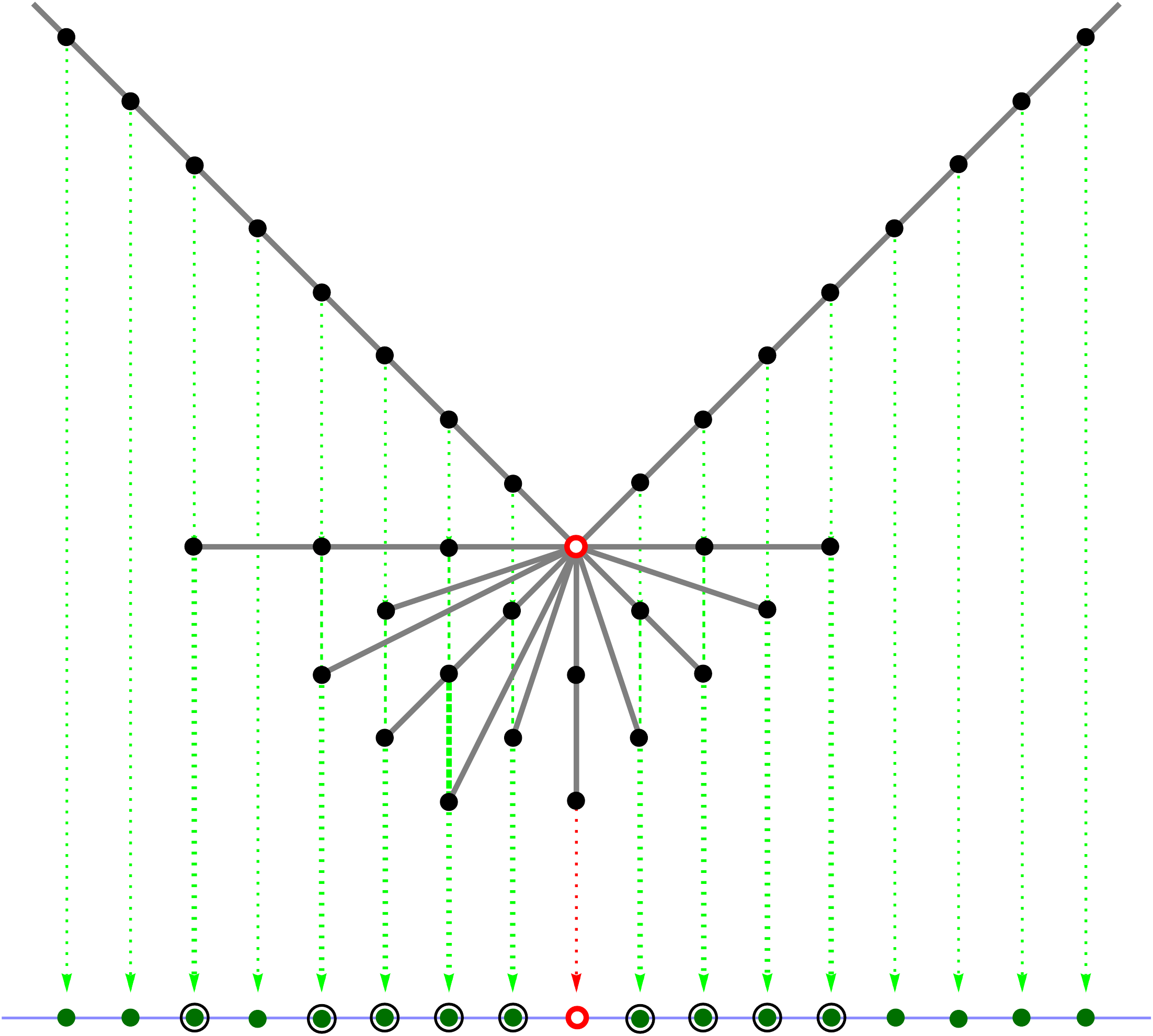}
\end{center}
\caption{Non-vanishing nodes of the $(3,2)$ alien lattice, and their projection upon the complex Borel plane. The left plot shows a non-resonant example, where the projection map is one-to-one, while the right plot shows the resonant example we are considering, with $\boldsymbol{\mathfrak{n}}_0 = \left( 1,1 \right)$ and where the projection map is now many-to-one. It is clear that when considering multi-instanton sectors, the approach to resonance creates new phenomena implying new non-trivial relations between Stokes vectors and Borel residues.
}
\label{fig:sec6-32resonance}
\end{figure}

Perhaps one of the simplest possible examples to consider\footnote{This is the relevant set-up for addressing resonance in both Painlev\'e~I \cite{gikm10, asv11} and Painlev\'e~II \cite{sv13} examples.} is when $k=2$ and the instanton actions are such that $\boldsymbol{A} = \left( A, -A \right)$, with $A \in \BR^+$. In this case, the group of action correlations is  basically $\left\{ \left( n,n \right), + \right\}$, with $n \in \BZ$, and it is thus generated by $\boldsymbol{\mathfrak{n}}_0 = \left( 1,1 \right)$ (\textit{i.e.}, there is a single \textit{independent} correlation between the actions). Let us analyze how the transseries and its resurgence relations are affected, as compared to the discussion in section~\ref{sec:physics}. The first change occurs when considering the projection map \eqref{projectionmapZkC}, as now $\ker \mathfrak{P} = \CL \left\{ \boldsymbol{\mathfrak{n}}_i \right\}$ is a $r$-dimensional linear space, and the projection is no longer one-to-one. What will happen to Stokes vectors and Borel residues in this case? Clearly, Stokes vectors remain unchanged as long as we think of them as living on alien-lattice sites. But their projection into Borel residues must necessarily change. Let us illustrate how this occurs remaining within the ``Painlev\'e-like'' example and focusing on the perturbative and $(3,2)$ transseries-nodes of figures~\ref{fig:sec5-2d-lattice} and~\ref{fig:sec5-BP-Riccati}, as usual. In our present example, the projection \eqref{projectionmapZkC} results on a line on the complex Borel plane; more specifically its real axis. In figures~\ref{fig:sec6-00resonance} and~\ref{fig:sec6-32resonance} we illustrate the corresponding alien lattices (and their projections) for perturbative and $(3,2)$ transseries-nodes, both at resonance and slightly way from resonance (where we perturb our present example with a small irrational non-resonant content). It is clear how nothing significant really changes for the perturbative sector, but how the projection is now many-to-one for the instanton sectors. In fact, whenever they exist, lattice nodes of the form $\boldsymbol{n} + \CL_{\BZ} \left\{ \boldsymbol{\mathfrak{n}}_0 \right\}$ are all projected to the same point. It is this new aspect that changes the relations between Stokes vectors and Borel residues which were discussed at length in section~\ref{sec:physics}. In the following subsections, we shall discuss the exact nature of these required changes.

It is important to realize that resonance is a property of the particular equation we might be dealing with (\textit{e.g.}, the equations for partition function \eqref{elliptic_linear_ODE} and free energy \eqref{eq:nl_ell_ode} in the present section are resonant; while the equations for partition function \eqref{quarticODE} and free energy \eqref{quarticNLODE} back in section~\ref{sec:quartic} were not). Of course it is always possible to consider special solutions which themselves do not display resonance, \textit{e.g.}, solutions where a single instanton action has been turned on. Nonetheless, in order to construct the whole respective transseries, resonance must \textit{already} have been taken into consideration. In fact, when using power-series and transseries to solve \textit{differential} equations, the latter are turned into \textit{recursion} equations for the coefficients in these power-series or transseries \textit{ans\"atze} (see appendix~\ref{app:recursive-free-en}). At the level of these recursion equations, resonance then translates into the cancelation of the \textit{leading} (corresponding to the \textit{unknown}) coefficient term, at each step in the recursion. But if such a cancelation occurs, then the recursion is \textit{not} solving for any unknown coefficients, and is thus \textit{not} allowing for an iterative construction of the solution in the first place. One way around this problem is when resonance is ``tunable'', as in the example in the present section. In this case, one first solves the recursion for \textit{arbitrary} values of the ``tunable parameter'' (which is $m$, in this case) and only sets for its resonant values at the very end (this was done in appendix~\ref{app:recursive-free-en}). Generically, however, resonance need not be ``tunable''. For instance, it may be associated to special integrability properties, as in the Painlev\'e-type examples addressed in \cite{gikm10, asv11, sv13}. In these cases, the aforementioned cancelation of unknown coefficients is simply telling us that the (power-series or transseries) solution \textit{ansatz} we are using is \textit{inconsistent}. Instead, it must be \textit{modified}, in such a way that the resulting recursion will work again. For instance, in the Painlev\'e-type examples addressed in \cite{gikm10, asv11, sv13}, the transseries needed to be modified so as to include \textit{logarithmic sectors}. Once this is done, the recursion equations get modified just enough as to no longer display cancelation of the unknown coefficients and, instead, allow for a fully iterative construction of the transseries solution. It should be clear that, in general, more intricate modifications of the transseries \textit{ansatz} may be required, than the aforementioned simple inclusion of logarithms.

\subsection{Linear Resonant-Asymptotics: Partition Function}\label{sec:linear-resonance}

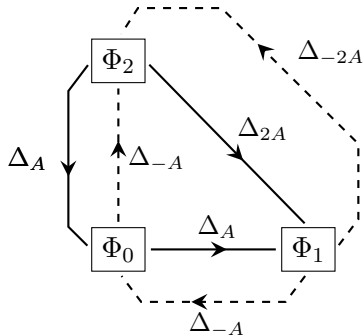
\begin{figure}[t!]
\begin{center}
\begin{tikzpicture}[>=latex,decoration={markings, mark=at position 0.6 with {\arrow[ultra thick]{stealth};}} ]
\begin{scope}[node distance=2.5cm]
  \node (Phi0) [draw] at (0,0) {$\Phi_0$};
  \node (Phi2) [above of=Phi0] [draw] {$\Phi_2$};
  \node (Phi1) [right of=Phi0] [draw] {$\Phi_1$};
\end{scope}
   \draw [thick,postaction={decorate},-,>=stealth,shorten <=2pt,shorten >=2pt]  (Phi0.east) --  (Phi1.west);
   \draw [dashed,thick,postaction={decorate},-,>=stealth,shorten <=2pt,shorten >=2pt] (Phi1.south) -- +(-0.3,-0.4) -- ++(-2.2,-0.4) -- (Phi0.south);
   \draw [thick,postaction={decorate},-,>=stealth,shorten <=2pt,shorten >=2pt]  (Phi2.west) -- +(-0.3,-0.4) -- ++(-0.3,-2.2) -- (Phi0.west);
   \draw [dashed,thick,postaction={decorate},-,>=stealth,shorten <=2pt,shorten >=2pt] (Phi0.north) -- (Phi2.south);
   \draw [thick,postaction={decorate},-,>=stealth,shorten <=2pt,shorten >=2pt] (Phi2.east) --  (Phi1.north);
   \draw [dashed,thick,postaction={decorate},-,>=stealth,shorten <=2pt,shorten >=2pt] (Phi1.east) -- +(0.3,0.4) -- ++(0.3,1.4) -- ++(-1.8,1.8) -- ++(-1.1,0)-- (Phi2.north);
  \node (2) at (1.3,0.3) {\small{$\Delta_{A}$}};
  \node (-2) at (1.3,-1) {\small{$\Delta_{-A}$}};
  \node (2) at (-1.2,1.2) {\small{$\Delta_{A}$}};
  \node (-2) at (0.5,1.2) {\small{$\Delta_{-A}$}};
  \node (2) at (-1.2,1.2) {\small{$\Delta_{A}$}};
  \node (4) at (1.9,1.6) {\small{$\Delta_{2A}$}};
  \node (-4) at (2.8,2.6) {\small{$\Delta_{-2A}$}};
\end{tikzpicture}
\end{center}
\caption{Algebraic structure associated to the behaviour of Borel transforms or alien derivatives for the $\Phi_i$ sectors, at their respective singularities, for the resonant case with $m=\frac{1}{2}$. Motions fall along the real axis on the Borel plane: solid lines correspond to motions along the $\theta=0$ direction; dashed lines along $\theta=\pi$. The steps which connect either $\Phi_1$ or $\Phi_2$ to the perturbative sector are of equal size, $A=2$, which implies that their singularities are equidistant to the origin on the Borel plane.}
\label{fig:linear-algebraic-structure-elliptic-resonant}
\end{figure}

Our simplest illustration of resonance occurs within the linear problem, \textit{i.e.}, when considering the case of the partition function of the elliptic potential. The alien lattice of this system is so simple (recall figure~\ref{linear-algebraic-structure-elliptic}) that it is almost blind to any resonant effects---except in the case where $m = \frac{1}{2}$. For this value of the modulus, the manifestation of resonance actually becomes rather clear. The magnitude of positive and negative instanton actions becomes the same, $A_1 = - A_2 \equiv A = 2$; with the associated alien lattice illustrated in figure~\ref{fig:linear-algebraic-structure-elliptic-resonant}. In particular, this causes all the \textit{odd} $n$ terms in the perturbative expansion $Z^{(0)}_n$ to \textit{cancel}, and the resulting \textit{resonant} perturbative series is in \textit{even} powers, $x^2$, alone (instead of all powers of $x$, as for any other value of $m$).

\begin{figure}[t!]
\begin{center}
\includegraphics[height=7.3cm]{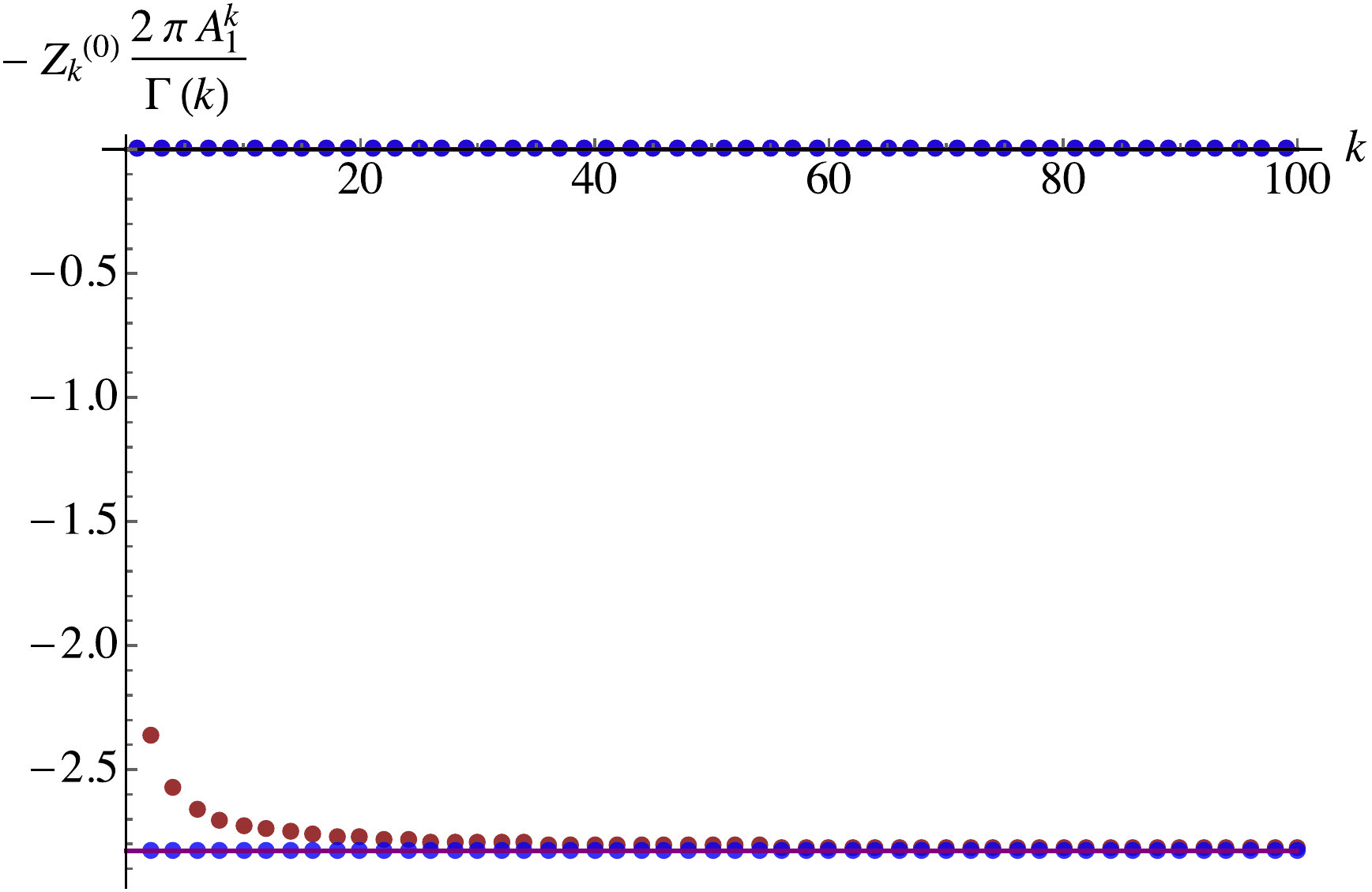}
\end{center}
\caption{Illustration of the asymptotic large-order behaviour of the perturbative-series coefficients, $Z^{(0)}_k$; via \eqref{large_order_aA} for the resonant value $m=1/2$. In red we plot the original weighted sequence $- Z^{(0)}_k\, \frac{2\pi A_1^k}{\Gamma(k)}$, while in blue we plot its corresponding $5$-RT. The solid purple lines denote the  constant values to which these sequences converge, $2 \mathrm{i} Z^{(1)}_0 \pm 2 \mathrm{i} Z^{(2)}_0$. For odd values of $k$ the coefficients vanish as a result of resonance since $Z^{(1)}_0=Z^{(2)}_0$; for even values of $k$ the sequence converges to $-2\sqrt{2}$ as expected.}
\label{fig:large_order-res}
\end{figure}

This can also be seen from the large-order behaviour of the perturbative coefficients in \eqref{large_order_aA} (and illustrated in figure~\ref{fig:large_order-res}). Recall that, due to the modular symmetry \eqref{modular_sym}, the coefficients in the partition-function nonperturbative sectors relate to each other as $Z^{(2)}_n (m) = (-1)^n\, Z^{(1)}_n(m^\prime)$. Thus, for $m=m^\prime=\frac{1}{2}$ and consequentially $A_1 = -A_2$, one can immediately see that all the odd $n$ terms cancel in \eqref{large_order_aA}. This phenomenon is a straightforward consequence of the fact that both nonperturbative saddles contribute to the large-order growth of the perturbative series with the \textit{same} magnitude but with \textit{symmetric} values (\textit{i.e.}, the contribution of saddle $(2)$ is alternating while that of saddle $(1)$ is not). This leads to the ``destructive interference'' which causes all odd terms in the perturbative series to vanish. Note that this is an extremely simple yet quite important phenomenon: given an asymptotic perturbative-series in even powers of the coupling $x^2$, one might na\"\i vely assume that its associated nonperturbative terms should be of the form $\sim \exp \left( - A/x^2 \right)$. However, our example illustrates how this need not be the case. When our system is \textit{resonant}, nonperturbative terms of the form $\sim \exp \left( - A/x \right)$ do lead to perturbative series in even powers $x^2$. For example, these resonance phenomena occur in the asymptotics of Painlev\'e~I and Painlev\'e~II equations \cite{gikm10, asv11, sv13}, describing non-critical strings; as well as in topological string theory \cite{ps09, cesv13, cesv14, c15}. This seems to point to the fact that resonance might play a central role in the asymptotics of string theory, as it explains how open and closed string sectors play with each other at large order\footnote{A more technical discussion of this point may be found in subsection~4.3 of \cite{asv11}.}: the closed-string perturbative expansion is asymptotic, in powers of string-coupling-\textit{squared}, but, \textit{via resonance}, its large-order behaviour is still controlled by D-branes (open strings) with nonperturbative weight $\sim \exp \left( - 1/\text{string-coupling} \right)$ \cite{s90, p95}.

\subsection{First Steps Towards Nonlinear Resonant Analysis}\label{sec:nonlinear-resonance-res}

Having discussed the simpler aspects of linear resonance, let us turn to resonance within our free-energy \textit{nonlinear} problem. Without surprise, the nonlinear case will turn out to be much richer than the linear one. Let us first recall that our two non-zero instanton actions, $A_1 = \frac{1}{1-m} > 0$ and $A_2 = - \frac{1}{m} < 0$, depend on a modulus $m \in (0,1)$. This implies, as per definition \eqref{rationally-dependent-resonant-actions}, that these actions are \textit{resonant} for any \textit{rational} value of $m$. In fact, given any $m \in \BQ$, one can always find a non-vanishing $\boldsymbol{n} = (n_1,n_2) \in \BZ^2$ satisfying $n_1/n_2=(1-m)/m$, which will thus imply $\boldsymbol{n} \cdot \boldsymbol{A} = 0$. For concreteness, let us tune $m$ to a particular fixed resonant value for the remainder of this section; say:
\begin{equation}
m = \frac{1}{3}.
\end{equation}
\noindent
With this choice of $m$, we now have an explicit realization of a resonant transseries, with two rationally dependent instanton actions,
\begin{equation}
A_1 = \frac{3}{2} \equiv A \quad \text{ and } \quad A_2 = -2A = -3,
\end{equation}
\noindent
defining the action-vector $\boldsymbol{A} = (A,-2A)$. Following the discussion in subsection~\ref{sec:nonlinear-resonance}, in this case the group of action correlations is generated by $\nres = (2,1)$ (there is a single independent correlation between the actions), and the kernel of the projection map  \eqref{projectionmapZkC} is thus given by
\be
\label{ell-res-vector}
\ker \mathfrak{P} = \mathcal{L} \left\{ \nres \right\}.
\ee
\noindent
One canonical choice we shall use for these generators is simply to pick the ``smallest'' ones, \textit{i.e.}, a basis of the kernel whose vector-entries are the smallest-possible non-negative integers.

As discussed in section~\ref{sec:physics}, generic transseries $\boldsymbol{\CT}$ are defined as \textit{ordered} formal-sums of transmonomials. For a $k$-dimensional transseries with exponential transmonomials \eqref{eq:sec5-transseries-vec} this ordering is simply organized according to the respective exponential weights (either growing or decaying). The arguments of these ordered exponential-transmonomials then determine the distinct locations of Borel singularities, leading---upon projection \eqref{projectionmapZkC}---to the singular structure \eqref{eq:sec5-Borel-transf}. But for \textit{resonant} systems---and as explained in subsection~\ref{sec:nonlinear-resonance}; in particular illustrated in figure~\ref{fig:sec6-32resonance}---many \textit{different} alien-lattice nodes may be associated to the \textit{same} exponential weight and thus project to the \textit{same} point on the Borel plane, contributing simultaneously. We shall thus need to reorganize the non-resonant transseries ordering in \eqref{eq:sec5-transseries-vec} in order to better understand how does resonance change the singular structure encoded in \eqref{eq:sec5-Borel-transf} (and, eventually, how does it change the resulting resurgence relations, \eqref{eq:sec5-bridge-eqs} or \eqref{eq:sec5-bridge-eqs-COMPLETE}).

Consider two distinct nonperturbative sectors, $\boldsymbol{n}$ and $\boldsymbol{n}^\prime$, which differ by an element in the projection kernel \eqref{ell-res-vector}, \textit{i.e.},
\be
\label{equivalence-relation-for-quotient}
\boldsymbol{n} - \boldsymbol{n}^\prime \in \ker \mathfrak{P}.
\ee
\noindent
Both these sectors will have the \textit{same} exponential grading in the transseries, and, in fact, generically there will be infinitely many sectors in these conditions. The required reordering of the transseries should then be naturally organized according to the equivalence classes in the resulting quotient defined by \eqref{equivalence-relation-for-quotient}, \textit{i.e.}, in $\BN^2 / \ker \mathfrak{P}$. In our on-going example, there is an infinite number of such equivalence classes which may be neatly organized into four basic families: the ``perturbative'' contribution $\BF_0 := [(0,0)]$, the ``positive instanton'' family $\BF_{+} (n) := [(n,0)]$, the ``negative-even instanton'' family $\BF_{-,\text{e}} (n) := [(0,n)]$, and the ``negative-odd instanton'' family $\BF_{-,\text{o}} (n) := [(1,n)]$. The exponential grading associated to this decomposition is, respectively, $0$, $nA$, $-2nA$, and $-(2n-1)A$ (herein we are always considering $n \in \BN^+$). The choice of representative element $\boldsymbol{\rho} (n) \in \BN^2 / \ker \mathfrak{P}$ for each family follows our canonical choice of picking the ``smallest'' vectors, \textit{i.e.}, the ones whose entries are the smallest-possible non-negative integers. In this way, any transseries alien-lattice node $\boldsymbol{n} \in \mathbb{Z}^2$ can be decomposed as the sum of a representative in one of the four families $\boldsymbol{\rho}_\alpha (n) \in \BF_\alpha (n)$ together with an element of the kernel of the form $\CL_{\BN^+} \{\nres\}$. The end result is that the \textit{resonant} free-energy transseries may be finally reorganized\footnote{Although, as we shall see next, this transseries reordering is key to understanding the resonant structure of Borel singularities, it need not be the most efficient way to do resonant-transseries analytic resummation; see \cite{asv18a}.} as (compare with \eqref{eq:nl_ell_ansatz})
\be
\label{eq:nl_ell_ansatz-res}
F \left( x, \sigma_0, \boldsymbol{\sigma} \right) = \sigma_0\, \widetilde{\Phi}_{(0,0)} + \sum_{\boldsymbol{\rho}\, \in\, \BN^2/\ker \mathfrak{P}} \mathrm{e}^{- \frac{\boldsymbol{\rho} \cdot \boldsymbol{A}}{x}} \sum_{\ell=0}^{+\infty} \boldsymbol{\sigma}^{\boldsymbol{\rho} + \ell \nres}\, \Phi_{\boldsymbol{\rho} + \ell \nres}(x).
\ee
\noindent
This reordering has implications for the large-order behaviour, as we shall see in subsection~\ref{sec:nonlinear-resonance-lo}.

The $\Phi_{\boldsymbol{\rho} + \ell \nres}$ sectors in \eqref{eq:nl_ell_ansatz-res} are labeled by an element $\boldsymbol{\rho}$ in the quotient $\BN^2 / \ker \mathfrak{P}$ plus an element in the positive-integer kernel $\CL_{\BN^+} \{\nres\}$, and thus continue living on the original $2$-dimensional (semi-positive) lattice. But if their Borel singularities will still be determined by the argument of their corresponding exponential-transmonomials, then their locations are at $s = \boldsymbol{\rho} \cdot \boldsymbol{A}$. These locations \textit{only} depend on $\boldsymbol{\rho} \in \BN^2 / \ker \mathfrak{P}$, in which case the whole (integer) kernel $\CL \{\nres\}$ must thus contribute \textit{in the same way}, \textit{at the same point}. As such, the natural resonant generalization of \eqref{eq:sec5-Borel-transf} becomes
\begin{equation}
\label{eq:sec6-Borel-transf-RESONANT}
\CB [\Phi_{\boldsymbol{n}}] (s) \Big|_{s=\boldsymbol{\rho}\cdot\boldsymbol{A}} = \left\{ \sum_{\ell \in \BZ}\, \mathsf{S}_{\boldsymbol{n} \to \boldsymbol{n}+\boldsymbol{\rho}+\ell\nres} \times \CB [\Phi_{\boldsymbol{n}+\boldsymbol{\rho}+\ell\nres}] (s-\boldsymbol{\rho}\cdot\boldsymbol{A}) \right\} \frac{\log \left(s-\boldsymbol{\rho}\cdot\boldsymbol{A}\right)}{2\pi\rmi}, \qquad \boldsymbol{\rho} \neq \boldsymbol{0}.
\end{equation}
\noindent
Notice, however, that the above sum generically \textit{truncates} and becomes \textit{finite}, as most $k$-orthants are empty of Stokes vectors (one might recall the discussion in section~\ref{sec:physics}, but this is actually very clearly illustrated in figure~\ref{fig:sec6-32resonance}). For our precise transseries \eqref{eq:nl_ell_ansatz-res}, the structure of Borel singularities is slightly more intricate than just \eqref{eq:sec6-Borel-transf-RESONANT}, and one finds (compare with \eqref{eq:elliptic-bridge-borel}):
\be
\label{eq:elliptic-bridge-borel-res}
\CB [\Phi_{\boldsymbol{n}}] ( s + \boldsymbol{\rho} \cdot \boldsymbol{A} ) = \sum_{\{ \boldsymbol{\ell} \in \BZ^2 | \boldsymbol{\ell} = \boldsymbol{\rho} + \ker \mathfrak{P} \}} \left( \mathsf{S}_{\boldsymbol{n} \to \boldsymbol{n}+\boldsymbol{\ell}} \times \CB [ \Phi_{\boldsymbol{n}+\boldsymbol{\ell}} ] (s) + \delta_{\boldsymbol{n}+\boldsymbol{\ell}}\, \mathsf{S}_{\boldsymbol{n} \to \widetilde{\boldsymbol{0}}} \times \CB [\widetilde{\Phi}_{(0,0)}] (s) \right) \frac{\log s}{2\pi\mathrm{i}}.
\ee

Addressing the familiar alien-derivative algebraic level of abstraction is by-now straightforward. Let us first readdress the arguments which led to \eqref{eq:nl_ell_ansatz-res} and \eqref{eq:elliptic-bridge-borel-res} from the standpoint of (alien-derivative induced) motions on the alien lattice. Due to resonance, the infinitely many nodes which lie along each foliation of the $\BZ^2$-lattice by the action correlation $\nres$ have the exact same exponential grading---these are just the equivalence classes \eqref{equivalence-relation-for-quotient}. As such, the alien derivatives should induce no motions whatsoever along these directions, \textit{i.e.}, if $\boldsymbol{n} - \boldsymbol{n}^\prime \in \ker \mathfrak{P}$ then
\begin{equation}
\Delta_{\boldsymbol{n} \cdot \boldsymbol{A}} = \Delta_{\boldsymbol{n}^\prime \cdot \boldsymbol{A}}.
\end{equation} 
\noindent
All one is left with are motions in-between distinct equivalence classes, which in this case imply that the original two-dimensional \textit{lattice} effectively collapses to a one-dimensional alien \textit{chain}. The resulting one-dimensional forward or backward motions, jumping from equivalence class to equivalence class, are organized by our aforementioned four basic families, $\BF_\alpha (n)$. In particular, nodes in the ``perturbative'' contribution $\BF_0$ project to the origin of the Borel plane and must have vanishing alien derivative, \textit{i.e.}, are associated with no motions on the alien chain. On the other hand, vectors in the ``positive instanton'' family $\BF_{+} (n)$ will induce forward motions, while vectors in both ``negative instanton'' families, even $\BF_{-,\text{e}} (n)$ and odd $\BF_{-,\text{o}} (n)$, will induce all possible backward motions. Furthermore, all our steps have the exact same size, given by $A$. On the Borel plane, singularities are thus located at positions $s = n A$, with $n \in \BZ \setminus \{ 0 \}$; falling upon both positive ($n>0$) and negative ($n<0$) real axes. One would thus be tempted to think that the alien derivative should be only labeled by a non-zero integer $n$, as $\Delta_{n A}$, and equivalent to the one-dimensional alien chain case \eqref{Delta-kA-one-param}. But this problem is more intricate than that one-dimensional case. In fact, the one subtlety still to take into consideration is that the many different nodes on the two-dimensional alien lattice belonging to the same equivalence class have ended-up being projected into the exact same position on the Borel plane. This infinite degeneracy of each step, and which is the hallmark of resonance, must still be present somewhere---and the information about this resonant degeneracy precisely appears at the level of a rearrangement of the \textit{Stokes data}, much like in \eqref{eq:elliptic-bridge-borel-res}. The \textit{resonant} resurgence relations generalizing \eqref{eq:elliptic-bridge} are then given by
\be
\label{eq:bridgeres}
\Delta_{\boldsymbol{\rho} \cdot \boldsymbol{A}} \Phi_{\boldsymbol{n}} = \sum_{\{ \boldsymbol{\ell} \in \BZ^2 | \boldsymbol{\ell} = \boldsymbol{\rho} + \ker \mathfrak{P} \}} \left( \boldsymbol{S}_{\boldsymbol{\ell}} \cdot \left(\boldsymbol{n}+\boldsymbol{\ell} \right) \Phi_{\boldsymbol{n}+\boldsymbol{\ell}} + \delta_{\boldsymbol{n}+\boldsymbol{\ell}}\, S_{\boldsymbol{\ell}}^{(0)}\, \widetilde{\Phi}_{(0,0)} \right).
\ee
\noindent
In this expression, $\boldsymbol{\rho}$ is a (canonical) representative of one of the equivalence classes, and elements in the right-hand side vanish whenever $\boldsymbol{n}+\boldsymbol{\ell}$ has a negative entry. Moreover, the above sum further truncates (and is in fact generically finite) as most $k$-orthants are empty of Stokes vectors. The Stokes coefficients obey the same conditions as in the non-resonant case, \eqref{eq.S.coeff.a} through \eqref{eq.S.coeff.d}, which we reproduce in here for completeness:
\bea
S^{(i)}_{\boldsymbol{0}} &=& 0, \qquad \text{ for }\, i=0,1,2, \\
S^{(0)}_{\boldsymbol{\ell}} &=& 0, \qquad \text{ if any }\, \ell_i>0, \\
S^{(1)}_{\boldsymbol{\ell}} &=& 0, \qquad \text{ if }\, \ell_1>1 \text{  or  } \ell_2>0, \\
S^{(2)}_{\boldsymbol{\ell}} &=& 0, \qquad \text{ if }\, \ell_1>0 \text{  or  } \ell_2>1.
\eea

\begin{figure}[t!]
\begin{center}
\begin{tikzpicture}[>=latex,decoration={markings, mark=at position 0.6 with {\arrow[ultra thick]{stealth};}} ]
\begin{scope}[node distance=2.5cm]
  \node (Phi00) [draw] at (0,0) {$\Phi_{(0,0)}$};
  \node (Phi10) [right of=Phi00] [draw] {$\Phi_{(1,0)}$};
  \node (Phi20) [right of=Phi10] [draw] {$\Phi_{(2,0)}$};
  \node (Phi30) [right of=Phi20] [draw] {$\Phi_{(3,0)}$};
  \node (Phi40) [right of=Phi30] [draw] {$\Phi_{(4,0)}$};
  \node (Phi50) [right of=Phi40] [draw] {$\Phi_{(5,0)}$};
  \node (cdots50) [right of=Phi50]  {$\cdots$};
  \node (Phit00) [below of=Phi00] [draw] {$\widetilde{\Phi}_{(0,0)}$};
  \node (Phi01) [above of=Phi00] [draw] {$\Phi_{(0,1)}$};
  \node (Phi02) [above of=Phi01] [draw] {$\Phi_{(0,2)}$};
  \node (Phi03) [above of=Phi02] [draw] {$\Phi_{(0,3)}$};
  \node (Phi04) [above of=Phi03] [draw] {$\Phi_{(0,4)}$};
  \node (cdots05) [above of=Phi04]  {$\vdots$};
  \node (Phi11) [above of=Phi10] [draw,dotted] {$\Phi_{(1,1)}$};
  \node (Phi21) [above of=Phi20] [draw] {$\Phi_{(2,1)}$};
  \node (Phi31) [above of=Phi30] [draw] {$\Phi_{(3,1)}$};
  \node (Phi41) [above of=Phi40] [draw] {$\Phi_{(4,1)}$};
  \node (Phi51) [above of=Phi50] [draw] {$\Phi_{(5,1)}$};
  \node (cdots51) [above of=cdots50]  {$\cdots$};
  \node (Phi12) [above of=Phi11] [draw] {$\Phi_{(1,2)}$};
  \node (Phi22) [above of=Phi21] [draw] {$\Phi_{(2,2)}$};
  \node (Phi32) [above of=Phi31] [draw] {\color{red}$\Phi_{(3,2)}$};
  \node (Phi42) [above of=Phi41] [draw] {$\Phi_{(4,2)}$};
  \node (Phi52) [above of=Phi51] [draw] {$\Phi_{(5,2)}$};
  \node (cdots52) [above of=cdots51]  {$\cdots$};
  \node (Phi13) [above of=Phi12] [draw] {$\Phi_{(1,3)}$};
  \node (Phi23) [above of=Phi22] [draw] {$\Phi_{(2,3)}$};
  \node (Phi33) [above of=Phi32] [draw] {$\Phi_{(3,3)}$};
  \node (Phi43) [above of=Phi42] [draw] {$\Phi_{(4,3)}$};
  \node (Phi53) [above of=Phi52] [draw,dotted] {$\Phi_{(5,3)}$};
  \node (cdots53) [above of=cdots52]  {$\cdots$};
  \node (Phi14) [above of=Phi13] [draw] {$\Phi_{(1,4)}$};
  \node (Phi24) [above of=Phi23] [draw] {$\Phi_{(2,4)}$};
  \node (Phi34) [above of=Phi33] [draw] {$\Phi_{(3,4)}$};
  \node (Phi44) [above of=Phi43] [draw] {$\Phi_{(4,4)}$};
  \node (Phi54) [above of=Phi53] [draw] {$\Phi_{(5,4)}$};
  \node (cdots54) [above of=cdots53]  {$\cdots$};
  \node (cdots15) [above of=Phi14] {$\vdots$};
  \node (cdots25) [above of=Phi24] {$\vdots$};
  \node (cdots35) [above of=Phi34] {$\vdots$};
  \node (cdots45) [above of=Phi44] {$\vdots$};
  \node (cdots55) [above of=Phi54]  {$\vdots$};
  \node (cdotsbb) [above of=cdots54]  {$\iddots$};
\end{scope}
  \draw [thick,dashed,orange!80!black,postaction={decorate},-,>=stealth,shorten <=2pt,shorten >=2pt] (Phi32.north) .. controls +(-0.2,1.3) and +(0.1,-0.5)  ..  (Phi03.east);
  \draw [thick,orange!80!black,postaction={decorate},-,>=stealth,shorten <=2pt,shorten >=2pt] (Phi32.east) .. controls +(1.5,-0.3) and +(-1.0,0.7)  ..  (Phi40.north);
  \draw [thick,dashed,green!40!black,postaction={decorate},-,>=stealth,shorten <=2pt,shorten >=2pt] (Phi32.north) .. controls +(-0.2,1.5) and +(0.,-0.3)  ..  (Phi13.east);
  \draw[thick,green!40!black,postaction={decorate},-,>=stealth,shorten <=2pt,shorten >=2pt] (Phi32.east) .. controls +(1.3,-0.7) and +(1.2,0.9)  ..  (Phi30.east);
  \draw[thick,dashed,red!70!black,postaction={decorate},-,>=stealth,shorten <=2pt,shorten >=2pt] (Phi32.north) .. controls +(-1.,1.3) and +(0.7,1.5)  ..  (Phi02.north);
  \draw [thick,dashed,red!70!black,postaction={decorate},-,>=stealth,shorten <=2pt,shorten >=2pt] (Phi32.north) .. controls +(-0.1,1.7) and +(0.5,0)  ..  (Phi23.east);
  \draw [thick,red!70!black,postaction={decorate},-,>=stealth,shorten <=2pt,shorten >=2pt] (Phi32.east) .. controls +(1.5,-0.1) and +(0,0.5)  ..  (Phi41.north);
  \draw[thick,red!70!black,postaction={decorate},-,>=stealth,shorten <=2pt,shorten >=2pt] (Phi32.south) .. controls +(-1.1,-1.6) and +(1.3,1.3)  ..  (Phi20.east);
  \draw[thick,dashed,blue!80!black,postaction={decorate},-,>=stealth,shorten <=2pt,shorten >=2pt] (Phi32.north) .. controls +(-1.3,0.7) and +(1.2,0.9)  ..  (Phi12.north);
  \draw [thick,dashed,blue!80!black,postaction={decorate},-,>=stealth,shorten <=2pt,shorten >=2pt] (Phi32.north)  -- (Phi33.south);
  \draw [thick,blue!80!black,postaction={decorate},-,>=stealth,shorten <=2pt,shorten >=2pt] (Phi32.south)  -- (Phi31.north);
  \draw[thick,blue!80!black,postaction={decorate},-,>=stealth,shorten <=2pt,shorten >=2pt] (Phi32.south) .. controls +(-0.9,-0.8) and +(1.7,0)  ..  (Phi10.east);
  \draw[thick,postaction={decorate},-,>=stealth,shorten <=2pt,shorten >=2pt] (Phi32.south) .. controls +(-1.3,-0.5) and +(0.6,0.5)  ..  (Phi21.east);
  \draw[thick,dashed,postaction={decorate},-,>=stealth,shorten <=2pt,shorten >=2pt] (Phi32.west) .. controls +(-1.1,-0.95) and +(0.3,2.5)  ..  (Phi01.north);
  \draw [thick,postaction={decorate},-,>=stealth,shorten <=2pt,shorten >=2pt] (Phi32.east)  -- (Phi42.west);
  \draw [thick,dashed,postaction={decorate},-,>=stealth,shorten <=2pt,shorten >=2pt] (Phi32.west)  -- (Phi22.east);
  \draw[thick,postaction={decorate},-,>=stealth,shorten <=2pt,shorten >=2pt] (Phi32.west) .. controls +(-2.1,-1.9) and +(0.4,2.9)  ..  (Phit00.north);
\node (m3-1) at (1.3,8.4) {\textcolor{orange!80!black}{\small{$w_{\footnotesize{-5A,0}}$}}};
\node (1-arrow) at (1.3,7.7) {\textcolor{orange!80!black}{$\left\downarrow\rule{0cm}{0.4cm}\right.$}};
\node (1-m2) at (10.8,1.25) {\textcolor{orange!80!black}{\small{$w_{\footnotesize{5A,0}}$}}};
\node (1-arrow) at (10.0,1.3) {\textcolor{orange!80!black}{$\longleftarrow$}};
\node (m2-1) at (3.7,8.4) {\textcolor{green!40!black}{\small{$w_{\footnotesize{-4A,0}}$}}};
\node (1-arrow) at (3.7,7.7) {\textcolor{green!40!black}{$\left\downarrow\rule{0cm}{0.4cm}\right.$}};
\node (0-m2) at (8.2,0.8) {\textcolor{green!40!black}{\small{$w_{\footnotesize{4A,0}}$}}};
\node (m1-1) at (6.3,8.25) {\textcolor{red!70!black}{\small{$w_{\footnotesize{-3A,1}}$}}};
\node (1-arrow) at (6.3,7.7) {\textcolor{red!70!black}{$\left\downarrow\rule{0cm}{0.35cm}\right.$}};
\node (1-m1) at (11.25,3.55) {\textcolor{red!70!black}{\small{$w_{\footnotesize{3A,1}}$}}};
\node (1-arrow) at (10.4,3.6) {\textcolor{red!70!black}{$\longleftarrow$}};
\node (m3-0) at (0.3,6.2) {\textcolor{red!70!black}{\small{$w_{\footnotesize{-3A,0}}$}}};
\node (m1-m2) at (6.75,0.8) {\textcolor{red!70!black}{\small{$w_{\footnotesize{3A,0}}$}}};
\node (0-1) at (8.25,6.3) {\textcolor{blue!80!black}{\small{$w_{\footnotesize{-2A,1}}$}}};
\node (m2-m2) at (4.2,1.0) {\textcolor{blue!80!black}{\small{$w_{\footnotesize{2A,0}}$}}};
\node (0-m1) at (8.15,3.7) {\textcolor{blue!80!black}{\small{$w_{\footnotesize{2A,1}}$}}};
\node (m2-0) at (3.95,5.55) {\textcolor{blue!80!black}{\small{$w_{\footnotesize{-2A,0}}$}}};
\node (1-0) at (8.8,5.3) {\small{$w_{\footnotesize{A,2}}$}};
\node (m3-m1) at (0.05,4.0) {\small{$w_{\footnotesize{-A,0}}$}};
\node (m1-0) at (6.25,5.3) {\small{$w_{\footnotesize{-A,1}}$}};
\node (m1-m1) at (5.7,3.4) {\small{$w_{\footnotesize{A,1}}$}};
\node (m3-m2) at (0.8,-1.2) {\small{$w_{\footnotesize{A,0}}$}};
\end{tikzpicture}
\end{center}
\caption{The \textit{resonant alien lattice}: a pictorial representation of the action of the \textit{resonant} alien derivative upon the $(3,2)$-sector (compare with the non-resonant figure~\ref{fig:chain-nonres}). Different single arrows correspond to different single steps, where the weight of each step is dictated by the right-hand-side of the \textit{resonant} resurgence relations \eqref{eq:bridgeres} (and where we are using the $w_{\boldsymbol{\rho} \cdot \boldsymbol{A},\, \kappa}$ labelling explained in the text). Solid and dashed lines represent steps with positive and negative actions, respectively. Different colors represent different magnitudes in exponential grading. In particular, note how \textit{equal} colors are organized according to foliations of the lattice by the resonance $\nres = (2,1)$. For example, nodes $(1,1)$ and $(5,3)$ are in the same equivalence class of our $(3,2)$-sector (they have been singled-out by their dotted squares), and project to the origin of the Borel plane. As a result, there are no steps between them or with anyone else. The steps shown are \textit{all} the single steps which are allowed starting at the $\Phi_{(3,2)}$ lattice-site (and where \textit{paths} can then be constructed with different middle nodes, and corresponding products of weights).
}
\label{fig:chain-res}
\end{figure}
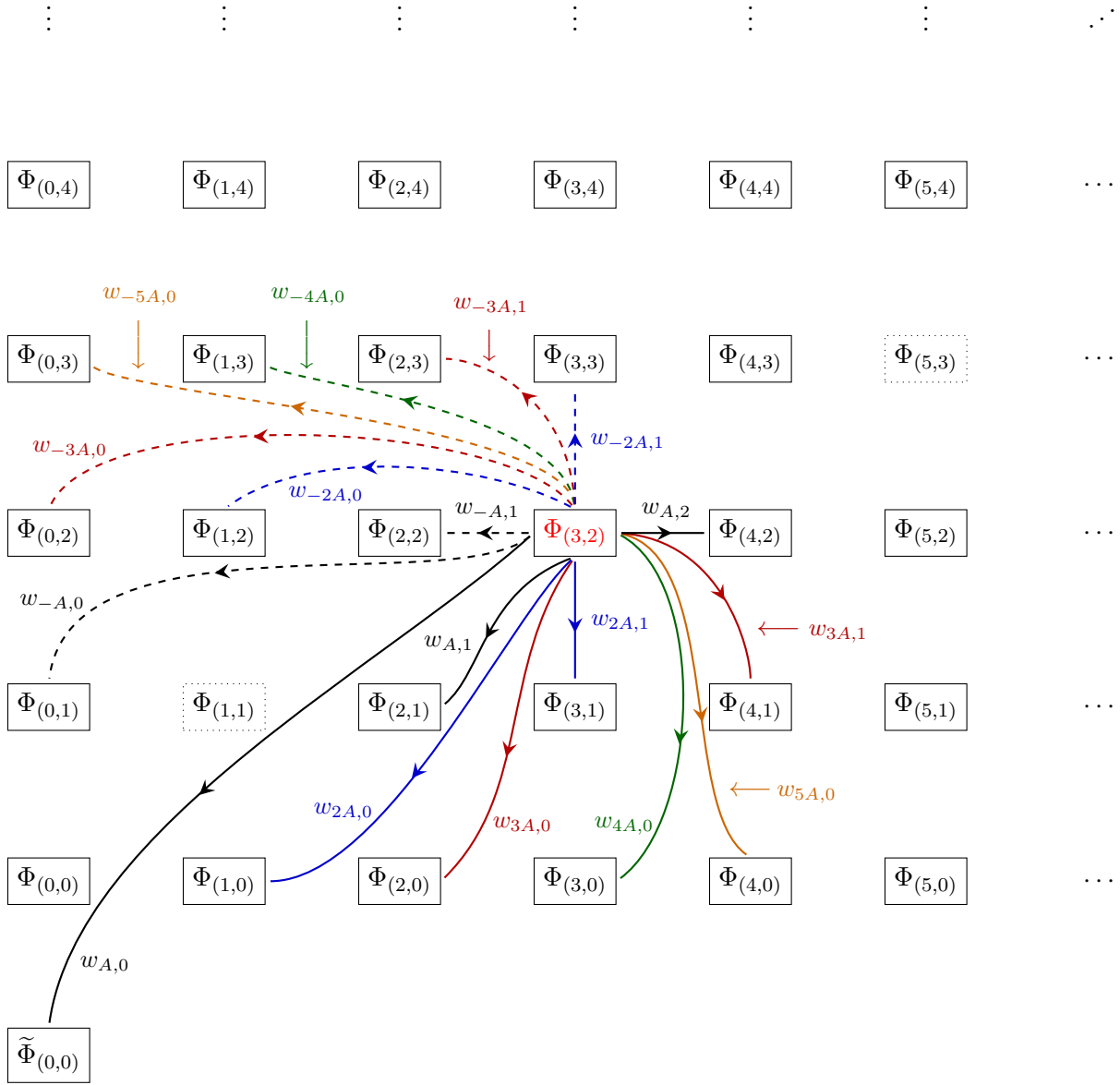

The resonant alien lattice where the alien derivative \eqref{eq:bridgeres} induces different motions is illustrated in figure~\ref{fig:chain-res} (it describes the action of $\Delta_{\boldsymbol{\ell} \cdot \boldsymbol{A}}$ on the same $(3,2)$-sector of figures~\ref{fig:sec5-2d-lattice} and~\ref{fig:chain-nonres}). It is particularly illuminating to compare the present resonant case of figure~\ref{fig:chain-res} to the exact same but non-resonant problem of figure~\ref{fig:chain-nonres}. As usual, solid lines indicate motions associated to singularities which lie on the positive real axis of the Borel plane, while dashed lines indicate motions associated to singularities which fall upon the negative real axis. Throughout, we have been labelling weights $w$ with subscript $s = \boldsymbol{\ell} \cdot \boldsymbol{A}$, but in the resonant case this is no longer enough. It is simple to see from figure~\ref{fig:chain-res} how weights with this very same label organize according to each foliation of the alien-lattice by $\nres = (2,1)$. In particular the foliation associated to the $(3,2)$-node itself \textit{cannot} have any associated motions (it projects to the origin on the Borel plane). In the figure this includes nodes $(1,1)$ and $(5,3)$, which are ``dotted out''. But along other foliations, the label is always the same although the weight might be distinct; depending on the end-nodes according to \eqref{eq:bridgeres}. This is simply solved by adding one extra label as in\footnote{Note that when addressing the $\BZ^2$ lattice rather than the $\BN^2$ lattice, \textit{i.e.}, when addressing motions starting-off at some transseries node $\boldsymbol{n}$ rather than the transseries itself which starts at $\boldsymbol{0}$, our canonical choice for $\boldsymbol{\rho}$ needs to be adapted (so as to remain in the same ``minimal'' spirit). This choice is now the vector with the most negative entries, such that it \textit{still} ends-up in an allowed (\textit{i.e.}, non-vanishing) node.} $\boldsymbol{\ell} = \boldsymbol{\rho} + \kappa\, \nres$ and $w_{\boldsymbol{\ell} \cdot \boldsymbol{A}} \equiv w_{\boldsymbol{\rho} \cdot \boldsymbol{A},\, \kappa}$ (for instance, the $(3,2)$-node itself is in this way decomposed as $(3,2) = (1,1) + \nres$). In this set-up, the ``statistical mechanical'' rules for Stokes data are basically the same as in the non-resonant case starting in \eqref{eq:elliptic-weight-step} and \eqref{eq:elliptic-weight-tilde}. The \textit{weight} $w$ associated to the \textit{step} $\CS$ connecting lattice sites $\boldsymbol{n}$ and $\boldsymbol{n} + \boldsymbol{\ell}$ is given by the standard expressions
\be
\label{eq:ell-res-weights}
w \left( \mathcal{S} \left( \boldsymbol{n} \rightarrow \boldsymbol{n} + \boldsymbol{\ell} \right) \right) = \left( \boldsymbol{n} + \boldsymbol{\ell} \right) \cdot \boldsymbol{S}_{\boldsymbol{\ell}}
\ee
\noindent
and
\be
\label{eq:ell-res-weightstilde}
w \big( \widetilde{\mathcal{S}} \left( \boldsymbol{n} \rightarrow \boldsymbol{0} \right) \big) = S^{(0)}_{-\boldsymbol{n}}.
\ee
\noindent
These weights are shown in figure \ref{fig:chain-res}, where the departing node is $\boldsymbol{n} = (3,2)$, and where we use the notation explained above, \textit{i.e.}, $w \left( \mathcal{S} \left( \boldsymbol{n} \rightarrow \boldsymbol{n} + \boldsymbol{\ell} \right) \right)) \equiv w_{\boldsymbol{\rho} \cdot \boldsymbol{A},\, \kappa}$.

Equipped with the (resonant) resurgence relations \eqref{eq:bridgeres} and their geometrical picture, and with the associated ``statistical mechanical'' language essentially unchanged from what was used in subsection~\ref{subsec:nonlinearFnonreso}, we may now turn to the discussion of resonant Stokes discontinuities. Let us start from definition \eqref{Stokes-aut-as exponential-singularities}. For the discontinuity at $\theta=0$, the relevant singularities on the Borel plane are those which lie along the positive real axis, \textit{i.e.}, one needs to consider the forward steps generated by $\Delta_{nA}$ with $n>0$. For the discontinuity at $\theta=\pi$, we instead need to consider all the backward steps generated by $\Delta_{-nA}$ with $n>0$. This leads to the standard picture where these discontinuities arise as sums over all paths $\CP$ which connect a given fixed sector $\Phi_{\boldsymbol{n}}$ to any other possible sectors $\Phi_{\boldsymbol{n}^{\prime}}$ such that $\left( \boldsymbol{n}^{\prime}-\boldsymbol{n} \right) \cdot \boldsymbol{A} = \pm nA$; positive or negative for either $\disc_0$ or $\disc_\pi$, respectively. Each such path $\CP$ is a collection of steps $\CS_i$, where in this case each step is associated to a jump $\pm n_i A$, such that $\sum_i n_i = n$. Further, each path $\CP$ contributes with an associated weight, given by \eqref{eq:elliptic-weight-path}, and an associated combinatorial factor, given by \eqref{eq:elliptic-CF}. For the perturbative sector, it is very easy to check that we get the following discontinuities:
\begin{eqnarray}
\label{ell-pert-disc-0}
\disc_0 \Phi_{(0,0)} &=& - \sum_{n=1}^{+\infty} \big( S^{(1)}_{\boldsymbol{e}_1} \big)^n\, \mathrm{e}^{-\frac{n A}{x}}\, \Phi_{(n,0)}, \\
\label{ell-pert-disc-pi}
\disc_\pi \Phi_{(0,0)} &=& - \sum_{n=1}^{+\infty} \big( S^{(2)}_{\boldsymbol{e}_2} \big)^n\, \mathrm{e}^{\frac{2 n A}{x}}\, \Phi_{(0,n)}.
\end{eqnarray}
\noindent
These are none others than \eqref{00nonreso-the0-compare} and \eqref{00nonreso-thepi-compare}, once we appropriatly select $A_1$ and $A_2$. No surprise.

For a more non-trivial comparison with the non-resonant case, addressed earlier in subsection~\ref{subsec:nonlinearFnonreso}, let us next work out the $\Phi_{(0,1)}$ sector. The resurgence relations \eqref{eq:bridgeres} for this sector read
\begin{eqnarray}
\label{DAP01=P11}
\Delta_A \Phi_{(0,1)} &=& S^{(1)}_{\boldsymbol{e}_1}\, \Phi_{(1,1)}, \\
\label{ell-delta-pi-sec01}
\Delta_{-2A} \Phi_{(0,1)} &=& 2 S^{(2)}_{\boldsymbol{e}_2}\, \Phi_{(0,2)}, \\
\label{D2AP01=tP00}
\Delta_{2A} \Phi_{(0,1)} &=& S^{(0)}_{-\boldsymbol{e}_2}\, \widetilde{\Phi}_{(0,0)}, \\
\Delta_{3A} \Phi_{(0,1)} &=& S^{(1)}_{\boldsymbol{e}_1-\boldsymbol{e}_2}\, \Phi_{(1,0)}.
\end{eqnarray}
\noindent
These are all the allowed single-step motions starting-out on sector $\Phi_{(0,1)}$. Up to now, other than appropriate selecting $A_1$ and $A_2$ there are still no novelties due to resonance. The discontinuities for $\theta=0,\pi$ follow from definition, \eqref{Stokes-aut-as exponential-singularities} and \eqref{StokesDisc}. Let us start with the $\theta=0$ discontinuity,
\begin{eqnarray}
\disc_0 \Phi_{(0,1)} &=& \left\{ \1 - \exp \left( \sum_{n=1}^{+\infty} \mathrm{e}^{-\frac{nA}{x}} \Delta_{nA} \right) \right\} \Phi_{(0,1)} = \\
&=& - \left( \mathrm{e}^{-\frac{A}{x}} \Delta_{A} + \mathrm{e}^{-\frac{2A}{x}} \Delta_{2A} + \frac{1}{2!} \left( \mathrm{e}^{-\frac{A}{x}} \Delta_{A}\right)^2 \right) \Phi_{(0,1)} + \mathcal{O}(\mathrm{e}^{-3A/x}). \nonumber
\end{eqnarray}
\noindent
Having seen how the single-step motions bear no novelties due to resonance, let us consider multi-step paths. Up to the exponential order we are considering, there is a single factor, $\left( \Delta_A \right)^2$, associated to such a multi-step path, as can be seen in the expression above. Let us compare its evaluation in both resonant and non-resonant scenarios. One begins with \eqref{DAP01=P11}, which is the same is both cases \eqref{eq:elliptic-bridge} and \eqref{eq:bridgeres} (up to the appropriate identification of $A_1 = A$),
\begin{align}
\label{DA01reso}
\Delta_A \Phi_{(0,1)} &= S^{(1)}_{\boldsymbol{e}_1}\, \Phi_{(1,1)}, &\text{ (resonant) } \\
\Delta_{A_1} \Phi_{(0,1)} &= S^{(1)}_{\boldsymbol{e}_1}\, \Phi_{(1,1)}. &\text{ (non-resonant) }
\end{align}
\noindent
Resonance then manifests itself when considering the \textit{second} alien-derivative, as
\begin{align}
\label{DA11reso}
\Delta_A \Phi_{(1,1)} &= 2 S^{(1)}_{\boldsymbol{e}_1}\, \Phi_{(2,1)} + S^{(0)}_{-\boldsymbol{e}_1-\boldsymbol{e}_2}\, \widetilde{\Phi}_{(0,0)}, &\text{ (resonant) } \\
\Delta_{A_1} \Phi_{(1,1)} &= 2 S^{(1)}_{\boldsymbol{e}_1}\, \Phi_{(2,1)}. &\text{ (non-resonant) }
\end{align}
\noindent
In the resonant case it is clear how the $\Phi_{(2,1)}$ and $\widetilde{\Phi}_{(0,0)}$ sectors fall upon the same equivalence class, and thus both appear in the right-hand side of \eqref{eq:bridgeres}. But in the non-resonant case there seems to be no trace of the $\widetilde{\Phi}_{(0,0)}$ contribution, even after the appropriate identification of $A_1 = A$. Of course even in the non-resonant case one may reach $\widetilde{\Phi}_{(0,0)}$ out of $\Phi_{(1,1)}$, only with a \textit{different} motion:
\be
\label{D-A1-A211nonreso}
\Delta_{-A_1-A_2} \Phi_{(1,1)} = S^{(0)}_{-\boldsymbol{e}_1-\boldsymbol{e}_2}\, \widetilde{\Phi}_{(0,0)}. \qquad\qquad \text{ (non-resonant) }
\ee
\noindent
This apparent mismatch is now made clear: upon appropriately selecting $A_1=A$ and $A_2=-2A$, then \textit{also} $-A_1-A_2=A$ and the two non-resonant contributions are joint together upon resonance (as it happens!). The resonant resurgence relations \eqref{eq:bridgeres} precisely take this into account.

Having understood the different resonant contributions, alongside their comparison to the earlier non-resonant case, let us proceed with the discontinuity $\disc_0 \Phi_{(0,1)}$. At leading exponential order, $\mathrm{e}^{-\frac{A}{x}}$, only one sector may be reached, and this occurs via the single forward step generated by $\Delta_A$. This is $\Phi_{(1,1)}$ as in \eqref{DAP01=P11}, corresponding to the step $\mathcal{S} \left( (0,1) \rightarrow (1,1) \right)$. The corresponding path has length $\ell=1$, weight $w = S^{(1)}_{\boldsymbol{e}_1}$, and combinatorial factor $\mathrm{CF}=\frac{1}{1!}$. This resulting resonant-contribution is essentially the same as in the non-resonant case \eqref{eq:01-disc-0}. At next-to-leading exponential order, $\mathrm{e}^{-\frac{2A}{x}}$, there are two sectors within reach: $\Phi_{(2,1)}$ and $\widetilde{\Phi}_{(0,0)}$, as discussed above. The first of these, sector $\Phi_{(2,1)}$, is reached via the two-step path $\Phi_{(0,1)} \rightarrow \Phi_{(1,1)} \rightarrow \Phi_{(2,1)}$ (with length $\ell=2$), which is generated by the $\left( \Delta_A \right)^2$ factor via \eqref{DA01reso} and \eqref{DA11reso}, has weight $w = 2 \left( S^{(1)}_{\boldsymbol{e}_1} \right)^2$, and combinatorial factor $\mathrm{CF}=\frac{1}{2!}$. Again, the resulting resonant-contribution is essentially the same as in the non-resonant case \eqref{eq:01-disc-0}. As to the second sector, $\widetilde\Phi_{(0,0)}$, it is reached by two distinct paths: a single step path (length $\ell=1$) $\Phi_{(0,1)} \rightarrow \widetilde{\Phi}_{(0,0)}$, generated by $\Delta_{2A}$ via \eqref{D2AP01=tP00}, with weight $w = S^{(0)}_{-\boldsymbol{e}_2}$ and combinatorial factor $\mathrm{CF}=\frac{1}{1!}$; and a two-step path $\Phi_{(0,1)} \rightarrow \Phi_{(1,1)} \rightarrow \widetilde{\Phi}_{(0,0)}$ (with length $\ell=2$), which is generated by the $\left( \Delta_A \right)^2$ factor via \textit{the very same} \eqref{DA01reso} and \eqref{DA11reso} which led us to $\Phi_{(2,1)}$, with weight $w = S^{(1)}_{\boldsymbol{e}_1} S^{(0)}_{-\boldsymbol{e}_1-\boldsymbol{e}_2}$ and combinatorial factor $\mathrm{CF}=\frac{1}{2!}$. As should be now clear, its resulting resonant-contribution is essentially the same as in the non-resonant case \eqref{eq:01-disc-0}, but it was now reached with \textit{just} the two $\Delta_A$ motions \eqref{DA01reso} and \eqref{DA11reso}, while the non-resonant case \textit{further required} a distinct motion \eqref{D-A1-A211nonreso}. This is a straightforward consequence of the resonant reorganization \eqref{eq:bridgeres}. All in all, this results in the discontinuity
\begin{eqnarray}
\disc_0 \Phi_{(0,1)} &=& - S^{(1)}_{\boldsymbol{e}_1}\, \mathrm{e}^{-\frac{A}{x}}\, \Phi_{(1,1)} - \left( S^{(1)}_{\boldsymbol{e}_1} \right)^2 \mathrm{e}^{-\frac{2A}{x}}\, \Phi_{(2,1)} - \left( S^{(0)}_{-\boldsymbol{e}_2}+\frac{1}{2} S^{(1)}_{\boldsymbol{e}_1} S^{(0)}_{-\boldsymbol{e}_1-\boldsymbol{e}_2} \right) \mathrm{e}^{-\frac{2A}{x}}\, \widetilde{\Phi}_{(0,0)} - \cdots \nonumber \\
\label{eq:01-disc-0-res}
&=& \mathsf{S}_{(0,1) \rightarrow (1,1)}\, \mathrm{e}^{-\frac{A}{x}}\, \Phi_{(1,1)} + \mathrm{e}^{-\frac{2A}{x}} \left( \mathsf{S}_{(0,1) \rightarrow (2,1)}\, \Phi_{(2,1)} + \mathsf{S}_{(0,1) \rightarrow \widetilde{\boldsymbol{0}}}\, \widetilde{\Phi}_{(0,0)} \right) + \cdots.
\end{eqnarray}
\noindent
Comparing with \eqref{eq:01-disc-0}, we see that upon identification of the actions $A_1$ and $A_2$ the result is the same. The difference lies in the resonant organization of the alien-derivative-induced motions \eqref{eq:bridgeres}, which made us arrive at the same result via a different route. Nevertheless, because of the resonant structure of the actions, this result will have a very clear impact on the large-order behaviour, as we shall see in a subsequent subsection.

The discontinuity at $\theta=\pi$ follows via a completely similar reasoning (compare with \eqref{eq:01-disc-pi})
\begin{eqnarray}
\disc_\pi \Phi_{(0,1)} &=& - 2\, S^{(2)}_{\boldsymbol{e}_2}\, \mathrm{e}^{\frac{2A}{x}}\, \Phi_{(0,2)} - 3 \left( S^{(2)}_{\boldsymbol{e}_2} \right)^2 \mathrm{e}^{\frac{4A}{x}}\, \Phi_{(0,3)} - 4 \left( S^{(2)}_{\boldsymbol{e}_2} \right)^3 \mathrm{e}^{\frac{6A}{x}}\, \Phi_{(0,4)} - \cdots \nonumber \\
\label{eq:01-disc-pi-res}
&=& \mathsf{S}_{(0,1) \rightarrow (0,2)}\, \mathrm{e}^{\frac{2A}{x}}\, \Phi_{(0,2)} + \mathsf{S}_{(0,1) \rightarrow (0,3)}\, \mathrm{e}^{\frac{4A}{x}}\, \Phi_{(0,3)} + \mathsf{S}_{(0,1) \rightarrow (0,4)}\, \mathrm{e}^{\frac{6A}{x}}\, \Phi_{(0,4)} + \cdots. \qquad\quad
\end{eqnarray}

Our final example in this subsection is, of course, to consider our favourite nonperturbative sector, namely, $\Phi_{(3,2)}$ (illustrated in figure~\ref{fig:chain-res}). In particular, we shall sketch the resonant resurgence-relations based upon this sector, using \eqref{eq:bridgeres}, which would then allow us to compute the resonant versions of \eqref{eq:disc-32-nonres-0} and \eqref{eq:disc-32-nonres-pi} (which we leave as a straightforward exercise for the reader). Along the $\theta=0$ direction, we find the contributions
\begin{eqnarray}
\Delta_A \Phi_{(3,2)} &=& w_{A,0}\, \widetilde{\Phi}_{(0,0)} + w_{A,1}\, \Phi_{(2,1)} + w_{A,2}\, \Phi_{(4,2)}, \\
\Delta_{2A} \Phi_{(3,2)} &=& w_{2A,0}\, \Phi_{(1,0)} + w_{2A,1}\, \Phi_{(3,1)}, \\
\Delta_{3A} \Phi_{(3,2)} &=& w_{3A,0}\, \Phi_{(2,0)} + w_{3A,1}\, \Phi_{(4,1)}, \\
\Delta_{4A} \Phi_{(3,2)} &=& w_{4A,0}\, \Phi_{(3,0)}, \\
\Delta_{5A} \Phi_{(3,2)} &=& w_{5A,0}\, \Phi_{(4,0)}.
\end{eqnarray}
\noindent
These are all the allowed (forward) \textit{single}-step motions starting out at our $(3,2)$-node, which are also represented with appropriate matching colors in figure~\ref{fig:chain-res}. It is simple to see how the right-hand side of the above expressions is arranged according to the foliations of the alien-lattice by $\ker \mathfrak{P}$. The values of the different weights in these equations are:
\begin{align}
w_{A,0} &= S^{(0)}_{(-3,-2)}, &w_{A,1} &= 2 S_{(-1,-1)}^{(1)} + S_{(-1,-1)}^{(2)}, &w_{A,2} &= 4 S^{(1)}_{(1,0)}, \\
w_{2A,0} &= S^{(1)}_{(-2,-2)}, &w_{2A,1} &= 3 S^{(1)}_{(0,-1)} + S^{(2)}_{(0,-1)}, \\
w_{3A,0} &= 2 S^{(1)}_{(-1,-2)}, &w_{3A,1} &= 4 S^{(1)}_{(1,-1)} + S^{(2)}_{(1,-1)}, \\
w_{4A,0} &= 3 S^{(1)}_{(0,-2)}, \\
w_{5A,0} &= 4 S^{(1)}_{(1,-2)}.
\end{align}
\noindent
For completeness we also list all the allowed backward (along the $\theta=\pi$ direction) \textit{single}-step motions starting out at our $(3,2)$-node, now represented by dashed lines in figure~\ref{fig:chain-res}. One finds:
\begin{eqnarray}
\Delta_{-A} \Phi_{(3,2)} &=& w_{-A,0}\, \Phi_{(0,1)} + w_{-A,1}\, \Phi_{(2,2)}, \\
\Delta_{-2A} \Phi_{(3,2)} &=& w_{-2A,0}\, \Phi_{(1,2)} + w_{-2A,1}\, \Phi_{(3,3)}, \\
\Delta_{-3A} \Phi_{(3,2)} &=& w_{-3A,0}\, \Phi_{(0,2)} + w_{-3A,1}\, \Phi_{(2,3)}, \\
\Delta_{-4A} \Phi_{(3,2)} &=& w_{-4A,0}\, \Phi_{(1,3)}, \\
\Delta_{-5A} \Phi_{(3,2)} &=& w_{-5A,0}\, \Phi_{(0,3)},
\end{eqnarray}
\noindent
where the weights are given by
\begin{align}
w_{-A,0} &= S^{(2)}_{(-3,-1)}, &w_{-A,1} &= 2 S_{(-1,0)}^{(1)} + 2S_{(-1,0)}^{(2)}, \\
w_{-2A,0} &= S^{(1)}_{(-2,0)} + 2 S^{(2)}_{(-2,0)}, &w_{-2A,1} &= 3 S^{(2)}_{(0,1)}, \\
w_{-3A,0} &= 2 S^{(2)}_{(-3,0)}, &w_{-3A,1} &= 2 S^{(1)}_{(-1,1)} + 3 S^{(2)}_{(-1,1)}, \\
w_{-4A,0} &= S^{(1)}_{(-2,1)} + 3 S^{(2)}_{(-2,1)}, \\
w_{-5A,0} &= 3 S^{(2)}_{(-3,1)}.
\end{align}

So far we have seen how resonance ended-up yielding a reorganization of transseries structures \eqref{eq:nl_ell_ansatz-res}, Borel singularities \eqref{eq:elliptic-bridge-borel-res}, and, equivalently, alien-derivative resurgence relations \eqref{eq:bridgeres}. These apparently mild reorganizations will have dire consequences in the large-order behaviour, as we shall see in the following. But first, let us take our usual pause to consider the relation between partition-function and free-energy Stokes coefficients, which will be required for the numerical verifications in the upcoming asymptotic large-order checks.

\subsection{Stokes Constants of Partition Function versus Free Energy: Resonance}

In the present resonant case, Stokes data for the free energy may also be equated with that for the partition function. Indeed, such relations follow in the exact same way as in the non-resonant case, earlier addressed in subsection~\ref{sec:stokes_z-vs-f}. This is essentially because the fact that the free-energy inherits the fundamental Borel singularities of the partition-function still holds in the presence of resonance. Given the similarity of the argument, we will not repeat the full calculation. Instead, we shall limit ourselves to pointing out the differences due to resonance, and then list the values for the free-energy Stokes coefficients as they relate to partition-function Stokes-data.

The analysis in subsection~\ref{sec:stokes_z-vs-f} is valid up to the determination of  the Stokes automorphism along $\theta = 0$ with respect to the free-energy transseries parameters $\tau_i$, starting in \eqref{eq:stokes-l_nl-0} (equivalently, along $\theta = \pi$, starting in \eqref{eq:stokes-l_nl-pi}). As stated many times before, the \textit{bridge equations} implement a bridge between alien-operators $\Delta$ and regular-derivatives $\partial$, which in the non-resonant case is implemented by the map \eqref{eq:stokes-tau}. Applying the exact same such procedure to the present resonant case, this implies we now need to match \eqref{eq:stokes-l_nl-0} with\footnote{Recall that the entries $\ell_i$ are assumed to be such that all powers of the $\tau_i$ are non-negative.}
\be
\label{eq:stokes-tau-res}
\mathrm{e}^{-\frac{n A}{x}} \Delta_{n A} = \sum_{\{ \boldsymbol{\ell} \in \BZ^2 | \boldsymbol{\ell} \cdot \boldsymbol{A} = n A \}} \left( S_{\boldsymbol{\ell}}^{F(0)}\, \tau_1^{-\ell_1} \tau_2^{-\ell_2}\, \frac{\partial}{\partial\tau_0} + S_{\boldsymbol{\ell}}^{F(1)}\, \tau_1^{1-\ell_1} \tau_2^{-\ell_2}\, \frac{\partial}{\partial\tau_1} + S_{\boldsymbol{\ell}}^{F(2)}\, \tau_1^{-\ell_1} \tau_2^{1-\ell_2}\, \frac{\partial}{\partial\tau_2} \right).
\ee
\noindent
The sum is over $\boldsymbol{\ell} \cdot \boldsymbol{A} = n A$ (where $n \in \BN^+$ for the $\theta = 0$ direction), which implies that the vector-entries of $\boldsymbol{\ell} = (\ell_1,\ell_2)$ obey the constraint $\ell_1 = n + 2 \ell_2$. The differential-operator in the right-hand side of \eqref{eq:stokes-tau-res} may thus be rewritten as\footnote{Where $\ell_2$ must still satisfy the $\tau_i$-regularity requirements, which are now $n$-dependent.}
\be
\label{rewritepartiald}
S_{( n + 2 \ell_2, \ell_2)}^{F(0)}\, \tau_1^{- n - 2 \ell_2} \tau_2^{-\ell_2}\, \frac{\partial}{\partial\tau_0} + S_{( n + 2 \ell_2, \ell_2)}^{F(1)}\, \tau_1^{1 - n - 2 \ell_2} \tau_2^{-\ell_2}\, \frac{\partial}{\partial\tau_1} + S_{( n + 2 \ell_2, \ell_2)}^{F(2)}\, \tau_1^{- n - 2 \ell_2} \tau_2^{1-\ell_2}\, \frac{\partial}{\partial\tau_2}.
\ee
\noindent
The required match now easily follows. Using \eqref{eq:stokes-tau-res} with \eqref{rewritepartiald}, and then comparing against \eqref{eq:stokes-l_nl-0} is quite straightforward. For instance, consider the terms which are proportional to $\frac{\partial}{\partial \tau_0}$. Their matching requires
\be
\sum_{n=1}^{+\infty} \sum_{\ell_2=-\infty}^{-\left\lfloor{\frac{n}{2}}\right\rfloor} S_{(n+2\ell_2,\ell_2)}^{F(0)}\, \tau_1^{-n-2\ell_2}\, \tau_2^{-\ell_2} = S_{\omega_{20}}^Z \tau_2.
\ee
\noindent
This implies there is a single (free energy) non-zero Stokes coefficient, associated to the $\tau_0$-action. One must set $\ell_2=-1$---and subsequently $n=2$---in which case this Stokes coefficient is
\be
\label{eq:ell-res-Stokes-pred-0-1}
S_{(0,-1)}^{F(0)} = S_{\omega_{20}}^Z \quad (\, = - 2\, ).
\ee
\noindent
As another example, let us consider the terms which are proportional to $\frac{\partial}{\partial \tau_1}$. Their matching now requires
\be
\label{ell-comp-Stokes-1}
\sum_{n=1}^{+\infty} \sum_{\ell_2=-\infty}^{-\left\lfloor{\frac{n-1}{2}}\right\rfloor} S_{(n+2\ell_2,\ell_2)}^{F(1)}\, \tau_1^{1-n-2\ell_2}\, \tau_2^{-\ell_2} = S_{\omega_{01}}^Z + S_{\omega_{21}}^Z\, \tau_2 - S_{\omega_{20}}^Z\, \tau_1 \tau_2.
\ee
\noindent
For example, if $\ell_2=0$ we must further have $n=1$ in which case one obtains the first term on the right-hand-side, resulting in $S_{(1,0)}^{F(1)} = S_{\omega_{01}}^Z$. Repeating the same procedure for the other terms in \eqref{ell-comp-Stokes-1}, and for the matching associated to terms proportional to $\frac{\partial}{\partial \tau_2}$, yields:
\begin{align}
\label{eq:ell-res-Stokes-pred-0-2-a}
S_{(1,0)}^{F(1)} &= S_{\omega_{01}}^Z &(\, &= 2\,\, ), \\
\label{eq:ell-res-Stokes-pred-0-2-b}
S_{(1,-1)}^{F(1)} &= S_{\omega_{21}}^Z &(\, &= 0\,\, ), \\
\label{eq:ell-res-Stokes-pred-0-2-c}
S_{(0,-1)}^{F(1)} &= - S_{\omega_{20}}^Z &(\, &= 2\,\, ), \\
\label{eq:ell-res-Stokes-pred-0-3}
S_{(0,-1)}^{F(2)} &= - S_{\omega_{20}}^Z &(\, &= 2\,\, ).
\end{align}
\noindent
Interestingly enough, these are the exact same formulae as in subsection~\ref{sec:stokes_z-vs-f}.

Having computed all non-vanishing Stokes coefficients associated to the free-energy Stokes discontinuities along $\theta=0$, let us address the same problem along $\theta=\pi$. The matching is now done against \eqref{eq:stokes-l_nl-pi}, and the procedure is completely analogous to the one above (just recall that in \eqref{eq:stokes-tau-res} $n$ now is $n<0$). The final result is:
\begin{align}
\label{eq:ell-res-Stokes-pred-pi-a}
S^{F(2)}_{(0,1)} &= S_{\omega_{02}}^Z &(\, &= 2\,\, ), \\
\label{eq:ell-res-Stokes-pred-pi-b}
S^{F(2)}_{(-1,1)} &= -S_{\omega_{12}}^Z &(\, &= 0\,\, ), \\
\label{eq:ell-res-Stokes-pred-pi-c}
S^{F(0)}_{(-1,0)} &= S_{\omega_{10}}^Z &(\, &= - 2\,\, ), \\
\label{eq:ell-res-Stokes-pred-pi-d}
S^{F(1)}_{(-1,0)} &= - S_{\omega_{10}}^Z &(\, &= 2\,\, ), \\
\label{eq:ell-res-Stokes-pred-pi-e}
S^{F(2)}_{(-1,0)} &= - S_{\omega_{10}}^Z &(\, &= 2\,\, ).
\end{align}
\noindent
Again, these end-up being the exact same formulae as in subsection~\ref{sec:stokes_z-vs-f}.

Having matched Stokes data from the partition function with that for the resonant free-energy associated to the elliptic potential, all we have left to do is to proceed with the numerical checks of \textit{nonlinear resonant resurgence}, via our usual large-order analyses.

\subsection{Nonlinear Resonant-Asymptotics: Free Energy}\label{sec:nonlinear-resonance-lo}

Our final task in these lectures is to address the \textit{nonlinear resonant asymptotics} of the elliptic free-energy, and, in particular, to compare against its non-resonant asymptotics discussed in subsection~\ref{sec:large-order-nonres}. Equipped with the Stokes data computed in the previous subsection, obtaining the associated discontinuities and subsequent large-order relations is by-now straightforward. Most of the construction occurs in parallel with what was done in subsection~\ref{sec:large-order-nonres} and, as such, herein we shall mostly focus on the manifestations of resonance.

As usual, let us start with the perturbative sector. Its large-order behaviour can be derived from \eqref{ell-pert-disc-0} and \eqref{ell-pert-disc-pi}, such that the resonant version of \eqref{eq:00-nonres} is now
\begin{eqnarray}
\label{eq:00-res}
F_k^{(0,0)}\, \frac{2\pi\mathrm{i} A^{k}}{\Gamma(k)} &\simeq& S^{(1)}_{\boldsymbol{e}_1} \left( F^{(1,0)}_0 + \frac{A}{k-1}\, F^{(1,0)}_1 + \frac{A^2}{(k-1)(k-2)}\, F^{(1,0)}_2 + \cdots \right) + \\
&& + 2^{-k} \left( S^{(1)}_{\boldsymbol{e}_1} \right)^2 \left( F^{(2,0)}_0 + \frac{2A}{k-1}\, F^{(2,0)}_1 + \frac{(2A)^2}{(k-1)(k-2)}\, F^{(2,0)}_2 + \cdots \right) + \nonumber \\
&& + (-2)^{-k}\, S^{(2)}_{\boldsymbol{e}_2} \left( F^{(0,1)}_0 + \frac{-2A}{k-1}\, F^{(0,1)}_1 + \frac{(-2A)^2}{(k-1)(k-2)}\, F^{(0,1)}_2 + \cdots \right) + \mathcal{O}(3^{-k}). \nonumber
\end{eqnarray} 
\noindent
As compared to the non-resonant case in \eqref{eq:00-nonres}, we see that now the instanton action $A$ may be factored-out to the left-hand-side (similarly to what happened in the one-dimensional case \eqref{large-order-pert-quartic}). But, other than that, at \textit{leading} exponential order---the first line of \eqref{eq:00-res}---the resonant large-order growth is exactly the \textit{same} as the non-resonant one\footnote{In fact, given that it is the same, we shall obviously not readdress it in the present subsection.}. However, at \textit{next-to-leading} exponential-order resonance kicks-in and the resonant large-order growth becomes rather \textit{distinct} from the non-resonant one. In the non-resonant case of subsection~\ref{sec:large-order-nonres}, with $m=\pi/8$, the large-order relation \eqref{eq:00-nonres} shows how the contribution of the $\Phi_{(2,0)}$-sector was exponentially-suppressed by $(A_2/(2A_1))^k$ as compared to $\Phi_{(0,1)}$, and these two sectors could not play with each other at large-order. But in the present resonant case the contributions of sectors $\Phi_{(2,0)}$ and $\Phi_{(0,1)}$ are now of the \textit{same order}, and, furthermore, the $\Phi_{(0,1)}$ contribution is \textit{oscillatory}. 

\begin{figure}[t!]
\begin{center}
\includegraphics[height=5cm]{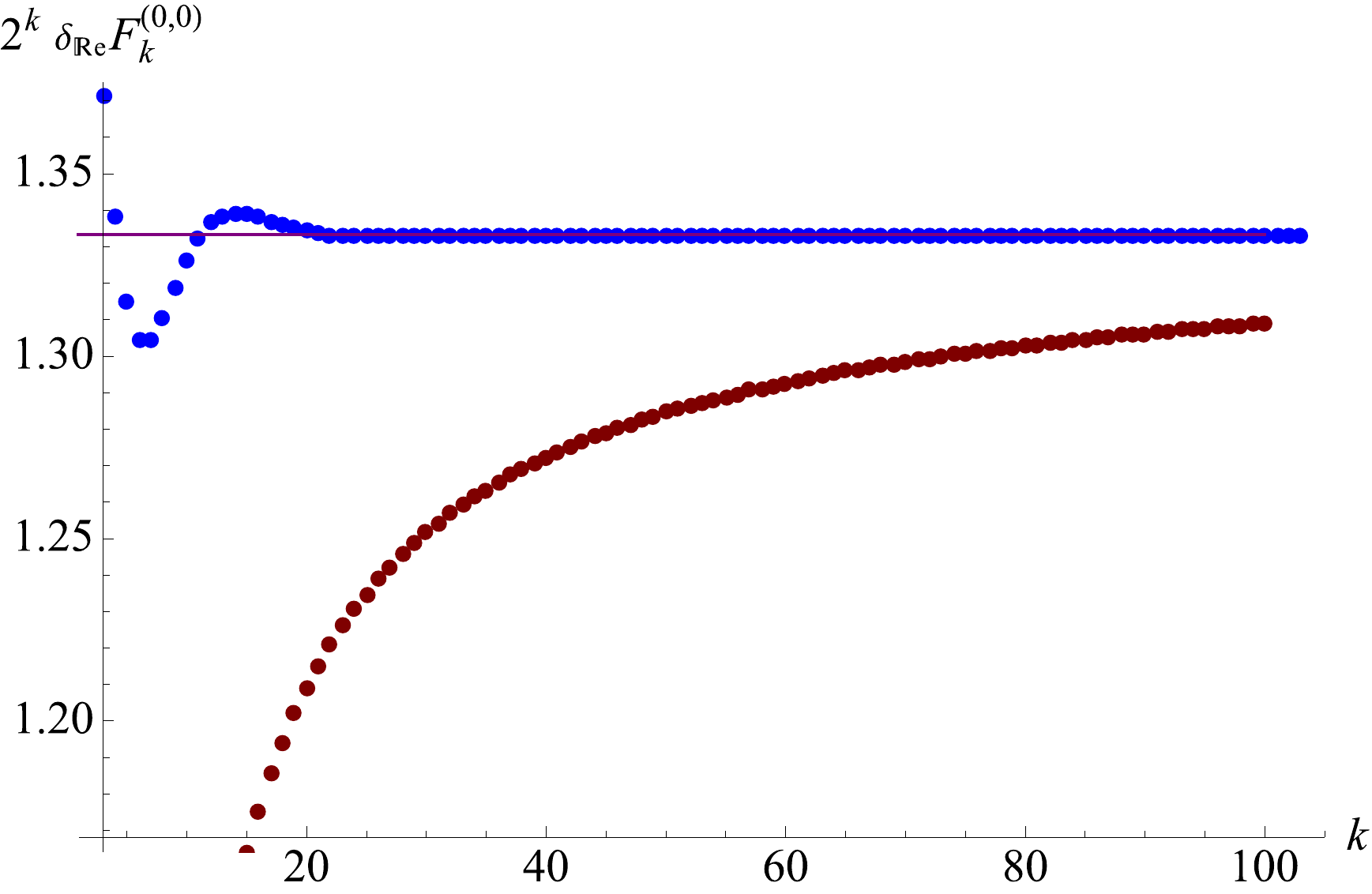}
$\quad$
\includegraphics[height=5cm]{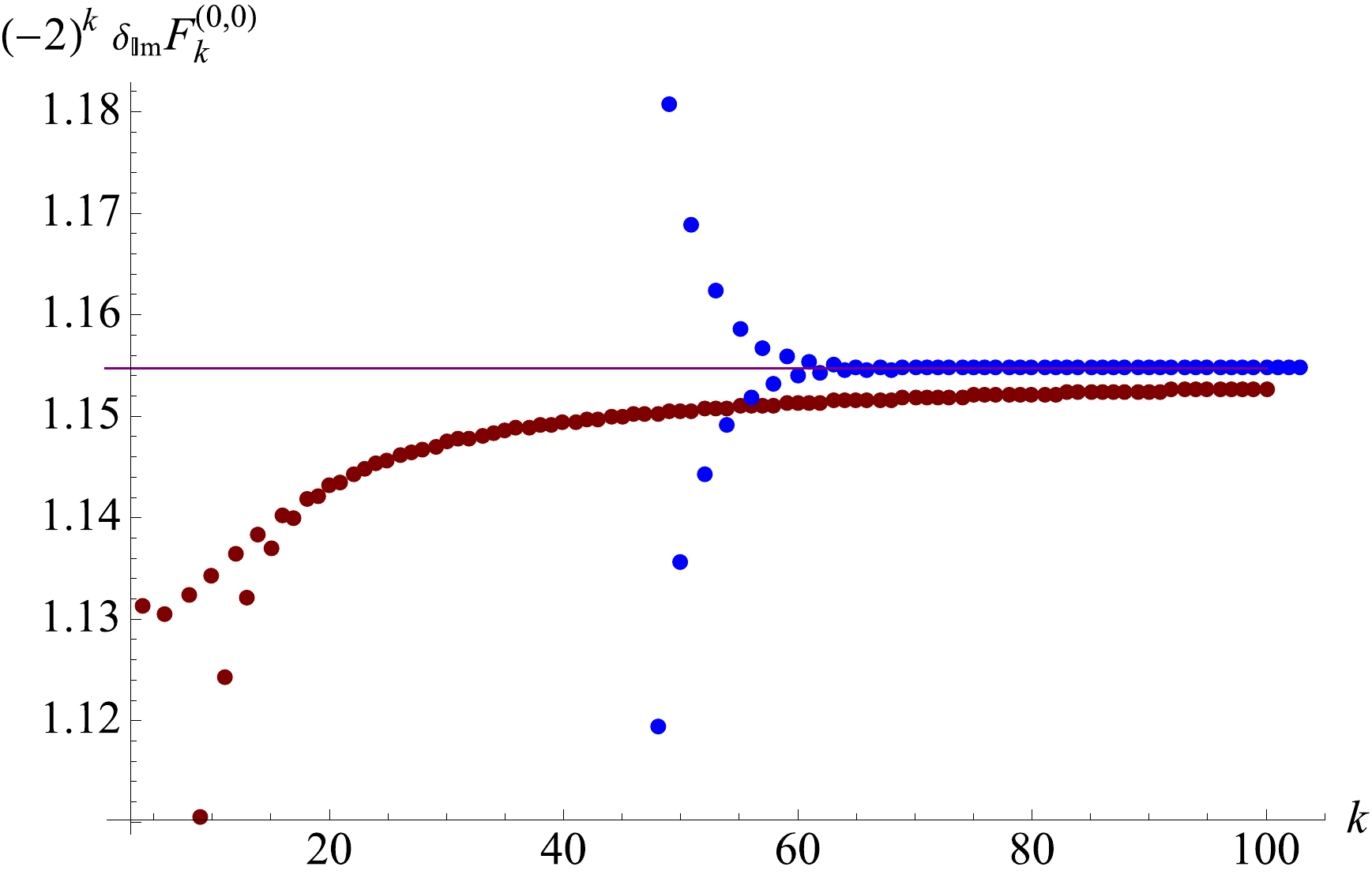}
\end{center}
\caption{Exponentially-suppressed contributions to the large-order behaviour of the resonant perturbative sector. On the left we plot the leading behaviour of the real sequence $2^k\, \delta_{\re} F^{(0,0)}_k$, defined in \eqref{eq:res-00-subl-re}, while on the right we plot the leading behaviour of the imaginary sequence $(-2)^k\, \delta_{\im} F^{(0,0)}_k$, defined in \eqref{eq:res-00-subl-im}. In both cases the original sequences are plotted in red, while in blue we plot their $5$-RTs. Furthermore, both cases involve a BP resummation of order $N=50$. Purple (solid) lines denote the expected (exact) convergence values: $4 F^{(2,0)}_0 = \frac{4}{3} \approx 1.3333$ (left) and $-2 \mathrm{i} F_0^{(0,1)} = \frac{2}{\sqrt{3}} \approx 1.1547$ (right).}
\label{fig:00-res-sub}
\end{figure}

Let us then illustrate the specificities of this resonant behaviour at order $\sim \mathcal{O}(2^{-k})$, and how it differs from what was discussed back in subsection~\ref{sec:large-order-nonres}. Making use of our standard methods, subleading exponential growths are isolated by resumming and then subtracting any leading-order growth. Recalling the definition of large-order factor in \eqref{ell-large-order-factor}, the exponentially-subleading large-order growth inside \eqref{eq:00-res} may then be numerically observed by plotting the sequence (note the differences from \eqref{eq:ell-pert-subtraction})
\begin{eqnarray}
F_k^{(0,0)}\, \frac{2\pi\mathrm{i} A^{k}}{\Gamma(k)} - \mathrm{SF}_{( (0,0) \rightarrow (1,0) )} \times \mathcal{S}_{0^{-}} \mathrm{BP}_N [\chi_{( (0,0) \rightarrow (1,0)}] (k) &\simeq& \\
&&
\hspace{-250pt}
\simeq 2^{-k} \left( \mathrm{SF}_{( (0,0) \rightarrow (2,0) )} \times \chi_{( (0,0) \rightarrow  (2,0) )} + (-1)^k\, \mathrm{SF}_{( (0,0) \rightarrow (0,1) )} \times \chi_{( (0,0) \rightarrow (0,1)) } \right) + \mathcal{O}(3^{-k}). \nonumber
\end{eqnarray}
\noindent
The notation is always the same. $\CS_{0^-} \mathrm{BP}_N [\chi_{( (0,0) \rightarrow (1,0) )}] (k)$ denotes the lateral resummation of the large-order factor along $\theta = 0^-$ (avoiding the poles which lie on the positive real axis). All present statistical factors were defined in the box of page~\pageref{ell-stat-fact} (and are now evaluated with respect to the likes of figure~\ref{fig:chain-res}). Finally, weights and combinatorial factors of each \textit{path} were defined in \eqref{eq:elliptic-weight-path} and \eqref{eq:elliptic-CF}; while the weights of each \textit{step} are now given by \eqref{eq:ell-res-weights} and \eqref{eq:ell-res-weightstilde}. In particular, making use of the results in the previous subsection, $\mathrm{SF}_{( (0,0) \rightarrow (n,0) )} = \left( S^{(1)}_{\boldsymbol{e}_1} \right)^n = 2^n$, with $n\ge1$, and $\mathrm{SF}_{( (0,0) \rightarrow (0,1) )} = S^{(2)}_{\boldsymbol{e}_2} = 2$. The lateral resummation follows the exact same calculation as in subsection~\ref{sec:large-order-nonres}, recall \eqref{lateralS-BPchi00-01}, with both real and imaginary parts:
\be
\CS_{0^{-}} \mathrm{BP}_{N} [\chi_{( (0,0) \rightarrow (1,0) )}] (k) = \CS_{\re} \mathrm{BP}_{N} [\chi_{( (0,0) \rightarrow (1,0) )}] (k) + \rmi\, \CS_{\im} \mathrm{BP}_{N} [\chi_{( (0,0) \rightarrow (1,0) )}] (k).
\ee
\noindent
The first imprint of resonance appears once the above subtraction is carried out, as both real and imaginary components are now of the same order (unlike in the non-resonant case, recall \eqref{eq:00-subleading-def} and \eqref{eq:00-subleading-def-im}),
\begin{eqnarray}
\label{eq:res-00-subl-im}
\mathrm{i}\, \delta_{\im} F^{(0,0)}_k &\equiv& F_k^{(0,0)}\, \frac{2\pi\mathrm{i}A^{k}}{\Gamma(k)} - \mathrm{i} \mathrm{SF}_{( (0,0) \rightarrow (1,0) )}\, \CS_{\im} \mathrm{BP}_{N} [\chi_{( (0,0) \rightarrow (1,0) )}] (k) \simeq \CO (2^{-k}), \\
\label{eq:res-00-subl-re}
\delta_{\re} F^{(0,0)}_k &\equiv& - \mathrm{SF}_{( (0,0) \rightarrow (1,0) )}\, \CS_{\re} \mathrm{BP}_{N} [\chi_{( (0,0) \rightarrow (1,0) )}] (k) \simeq \CO (2^{-k}).
\end{eqnarray}
\noindent
Having the same magnitude, it would be natural to expect these two contributions to intertwine, with the large-order matching then occurring against the growth of the sum of second and third lines in \eqref{eq:00-res}---with possibly yet another intertwine of their own. This particular case actually turns out to be simpler than that, as\footnote{This can be seen directly, \textit{e.g.}, from the relation of free-energy sectors $\Phi_{(2,0)}$ and $\Phi_{(0,1)}$ with the original sectors in the partition function, as related via \eqref{ell-free-en-from-partfunc}.} all coefficients $F_m^{(2,0)}$ are real, and all $F_m^{(0,1)}$ are purely imaginary. As such, one finds the simpler identifications (in particular, the first line below is essentially the same as \eqref{eq:00-subleading-exp}):
\begin{eqnarray}
(-2)^k\, \delta_{\im} F^{(0,0)}_k &\simeq& - \mathrm{i} \mathrm{SF}_{( (0,0) \rightarrow (0,1) )} \left( F_0^{(0,1)} - \frac{2A}{k}\, F_1^{(0,1)} + \cdots \right), \\
2^k\, \delta_{\re} F^{(0,0)}_k &\simeq& \mathrm{SF}_{( (0,0) \rightarrow (2,0) )} \left( F_0^{(2,0)} + \frac{2A}{k}\, F_1^{(2,0)} + \cdots \right).
\end{eqnarray}
\noindent
The imaginary-component of our (exponentially-weighted) sequence, $(-2)^k\, \delta_{\im} F^{(0,0)}_k$, is predicted to converge to $-\mathrm{i} \mathrm{SF}_{( (0,0) \rightarrow (0,1) )} F_0^{(0,1)} \equiv 2\sqrt{m}$, whereas\footnote{From \eqref{ell-relation-phis} and \eqref{ell-free-en-from-partfunc} we can read-off $F_0^{(0,1)} = \mathrm{i}\sqrt{m}$ and $F_0^{(2,0)}=\frac{1}{2} (1-m)$.} the real component, $2^k\, \delta_{\re} F^{(0,0)}_k$, is predicted to converge to $\mathrm{SF}_{( (0,0) \rightarrow (2,0) )} F_0^{(2,0)} = 2(1-m)$. Note how the contribution from the $\Phi_{(0,1)}$-sector to the imaginary sequence is alternating, thus accounting for the minus sign in the exponential growth. These convergences are shown very explicitly in figure~\ref{fig:00-res-sub} (recall we have fixed $m=\frac{1}{3}$ to illustrate the resonant example). Comparing the fifth Richardson transforms of the corresponding sequences with their expected exact results, we find the typical vanishingly-small relative errors\footnote{We have added labels $\re$ and $\im$ to the RTs, in order to distinguish real and imaginary sequences.}
\begin{eqnarray}
\frac{\mathrm{RT}_{(0,0),2,\re}(0,95,5) - \frac{4}{3}}{\frac{4}{3}} &\approx& 1.4877 \times 10^{-8}, \\
\frac{\mathrm{RT}_{(0,0),2,\im}(0,95,5) - \frac{2}{\sqrt{3}}}{\frac{2}{\sqrt{3}}} &\approx& 5.425 \times 10^{-9}.
\end{eqnarray}

\begin{figure}[t!]
\begin{center}
\includegraphics[height=7cm]{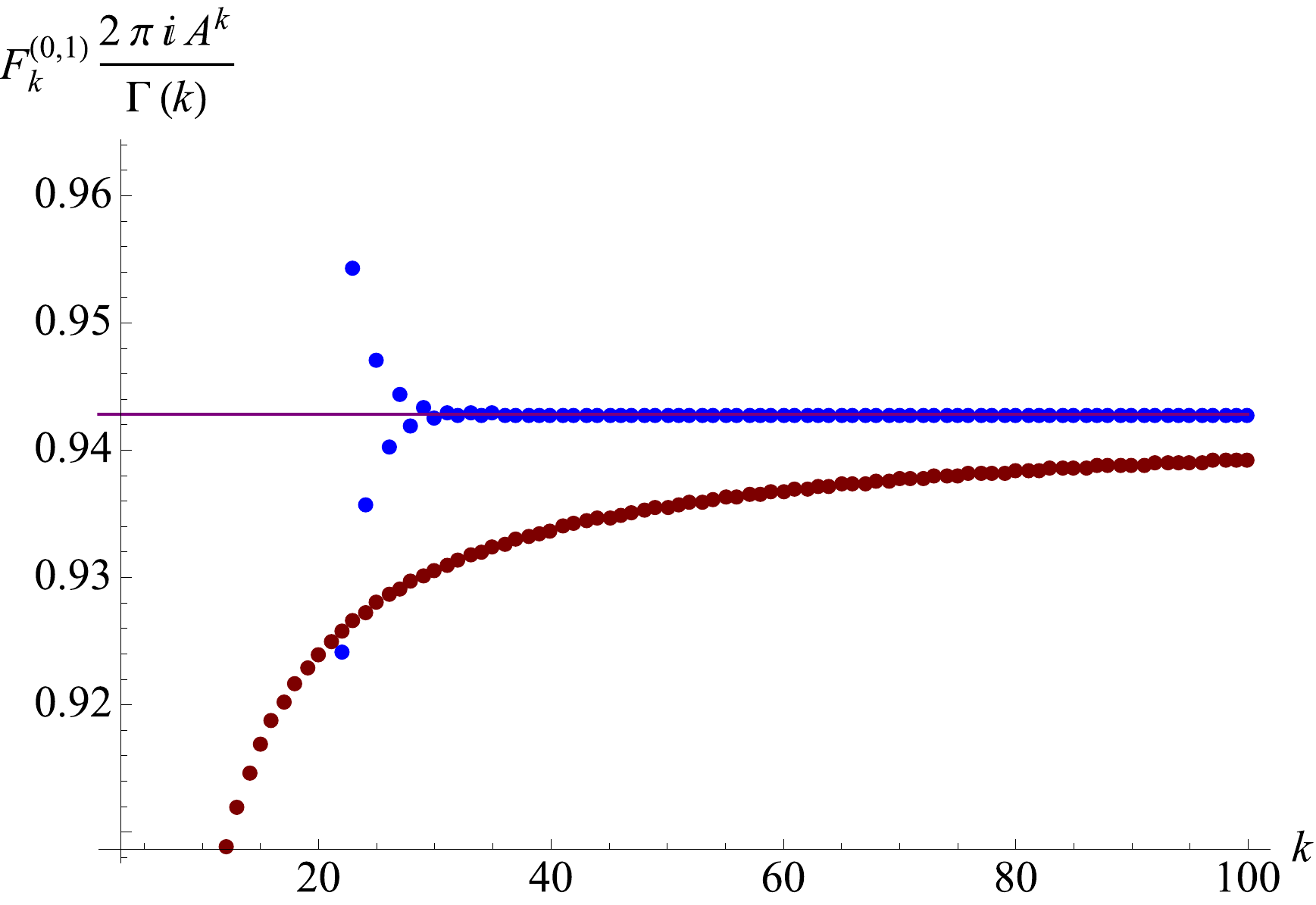}
\end{center}
\caption{Leading large-order growth of the $\Phi_{(0,1)}$-sector in the resonant free-energy. The original weighted-sequence $F_k^{(0,1)}\, \frac{2\pi\mathrm{i} A^k}{\Gamma(k)}$ is plotted in red, while in blue we plot its corresponding $\mathrm{RT}_{(0,1)}(0,k,5)$. The purple line shows the value to which these sequences converge, $S^{(1)}_{\boldsymbol{e}_1} F^{(1,1)}_0 = \frac{2\sqrt{2}}{3} \approx 0.9428$.
}
\label{fig:01-res}
\end{figure}

As the final example, let us consider an instanton sector; \textit{e.g.}, let us revisit the $\Phi_{(0,1)}$-sector (also analyzed in the non-resonant case). Its large-order follows directly from \eqref{eq:01-disc-0-res} and \eqref{eq:01-disc-pi-res},
\begin{eqnarray}
\label{eq:01-res}
F_k^{(0,1)}\, \frac{2\pi\mathrm{i} A^k}{\Gamma(k)} &\simeq& S^{(1)}_{\boldsymbol{e}_1} \left( F^{(1,1)}_0 + \frac{A}{k-1}\, F^{(1,1)}_1 + \frac{A^2}{(k-1)(k-2)}\, F^{(1,1)}_2 + \cdots \right) + \\
&&
+ \frac{1}{(-2)^k}\, 2 S^{(2)}_{\boldsymbol{e}_2} \left( F^{(0,2)}_0 + \frac{-2A}{k-1}\, F^{(0,2)}_1 + \frac{(-2A)^2}{(k-1)(k-2)}\, F^{(0,2)}_2 + \cdots \right) + \nonumber \\
&&
+ \frac{1}{2^k} \left( S^{(0)}_{-\boldsymbol{e}_2} + \frac{1}{2} S^{(1)}_{\boldsymbol{e}_1} S^{(0)}_{-\boldsymbol{e}_1-\boldsymbol{e}_2} \right) F_0^{(\widetilde{\boldsymbol{0}})} + \nonumber \\
&&
+ \frac{1}{2^k} \left( S^{(1)}_{\boldsymbol{e}_1} \right)^2 \left( F^{(2,1)}_0 + \frac{2A}{k-1}\, F^{(2,1)}_1 + \frac{(2A)^2}{(k-1)(k-2)}\, F^{(2,1)}_2 + \cdots \right) + \nonumber \\
&&
+ \mathcal{O}( 3^{-k}). \nonumber
\end{eqnarray} 
\noindent
Upon comparison with the non-resonant case \eqref{eq:01-nonres}, it is clear that the overall structure is the same. Furthermore, at leading exponential-order the result is \textit{exactly} the same. However, at next-to-leading exponential-order, resonance kicks-in as many sectors will now talk to each other at large order, and the resonant asymptotics is now \textit{distinct} from the non-resonant one. It is interesting to realize how, so far, resonance has not lead to any qualitative changes in the exponentially-\textit{leading} large-order growth: similarly to what happened for the perturbative sector in \eqref{eq:00-res}, there is now only one sector which contributes at leading exponential-order, $\Phi_{(1,1)}$, and this is the exact same as in the non-resonant case \eqref{eq:01-nonres}. This leading behaviour is illustrated in figure~\ref{fig:01-res}, where the leading growth of the (weighted) sequence $F_k^{(0,1)}\, \frac{2\pi\mathrm{i} A^k}{\Gamma(k)}$ is shown to converge to its predicted value, $S^{(1)}_{\boldsymbol{e}_1} F^{(1,1)}_0 = \frac{2\sqrt{2}}{3}$, with remarkable accuracy (and this is in fact basically the same as in the non-resonant case, in the left image of figure~\ref{fig:01-nonres}). 

\begin{figure}[t!]
\begin{center}
\includegraphics[height=4.9cm]{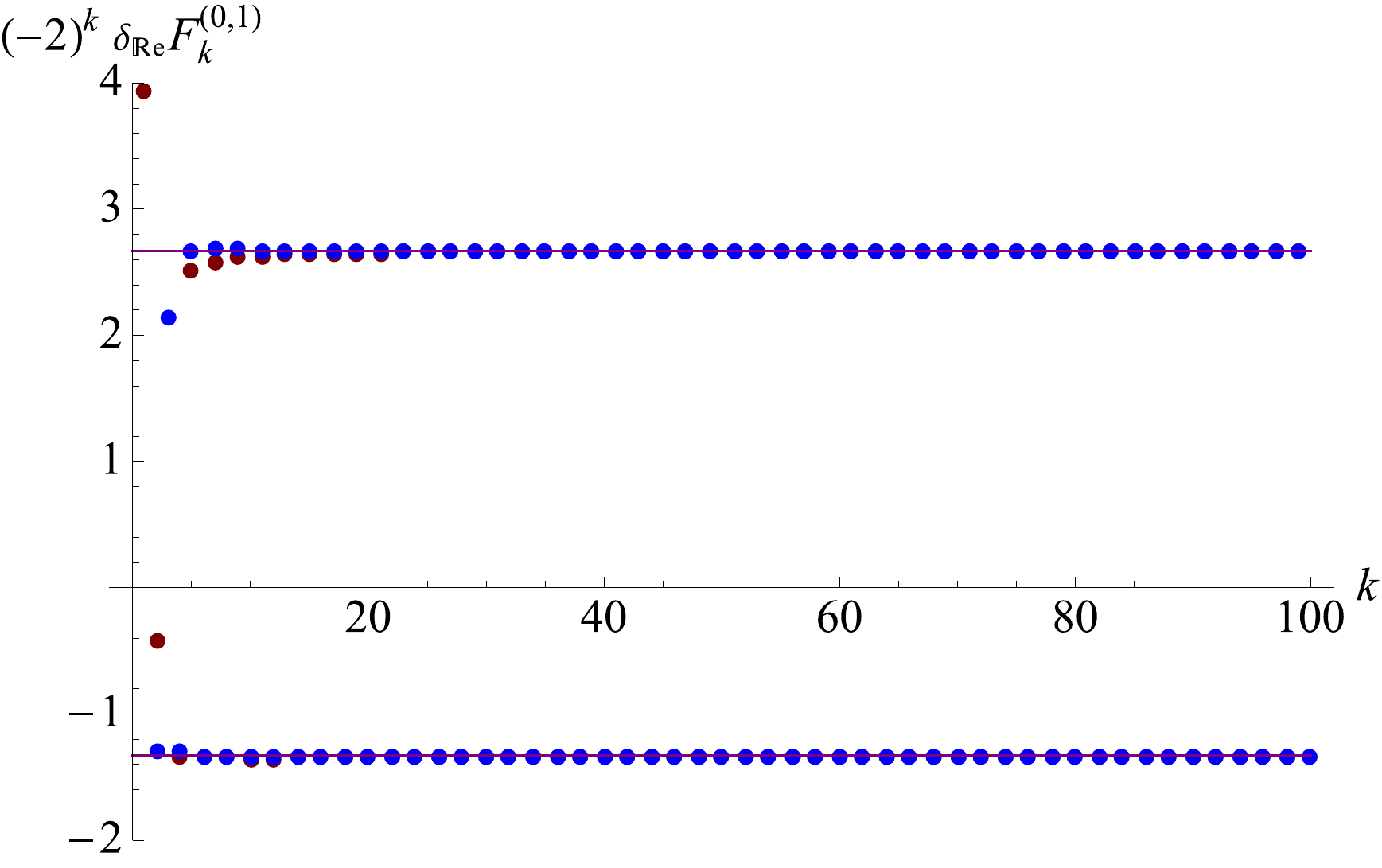}
$\quad$
\includegraphics[height=4.9cm]{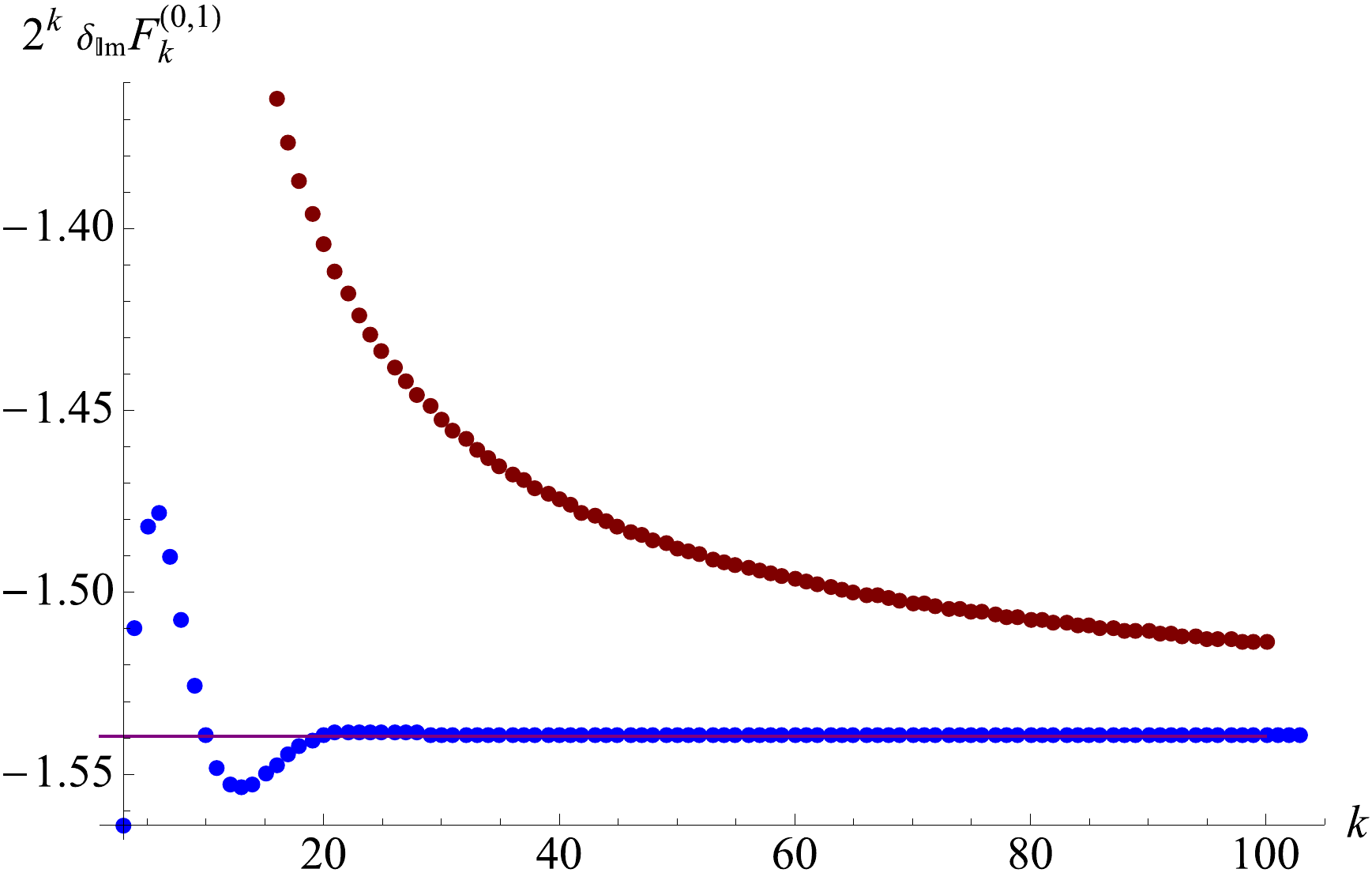}
\end{center}
\caption{Exponentially-suppressed contributions to the large-order behaviour of the resonant $\Phi_{(0,1)}$ sector. We plot the leading behaviour of both real $(-2)^k\, \delta_{\re} F^{(0,1)}_k$ (left) and imaginary $2^k\, \delta_{\im} F^{(0,1)}_k$ (right) sequences, as defined in \eqref{eq:res-01-subl-re} and \eqref{eq:res-01-subl-im}, respectively. In red we plot the original sequences, with their $5$-RTs in blue ($\mathrm{RT}_{(0,1),2,\re,\text{even}/\text{odd}}(0,k,5)$ for the left plot and $\mathrm{RT}_{(0,1),2,\im}(0,k,5)$ for the right plot). Purple (solid) lines denote the values to which these sequences converge, namely, $\pm \mathrm{SF}_{( (0,1) \rightarrow \widetilde{\boldsymbol{0}}}\,F _0^{(\widetilde{\boldsymbol{0}})} + \mathrm{SF}_{( (0,1) \rightarrow (0,2) )}\, F_0^{(0,2)} = \frac{2}{3} \mp 2$ (left) and $- \mathrm{i} \mathrm{SF}_{( (0,1) \rightarrow (2,1) )}\, F_0^{(2,1)} = - \frac{8}{3\sqrt{3}}$ (right).}
\label{fig:01-res-sub}
\end{figure}

At next-to-leading exponential-order, the effects of resonance start materializing in the asymptotics. Sectors $\widetilde{\Phi}_{(0,0)}$ (third line of \eqref{eq:01-res}) and $\Phi_{(2,1)}$ (fourth line of \eqref{eq:01-res}) contribute to the large-order growth on equal footing, \textit{as they are both reached via forward motions of the same length} (whereas in the non-resonant case the contribution of the $\Phi_{(2,1)}$ sector was exponentially-suppressed\footnote{At least for our choice of non-resonant modulus $m$.} with respect to that of $\widetilde{\Phi}_{(0,0)}$). This can be understood pictorially: figure~\ref{fig:chain-res} illustrates the same phenomenon (albeit now centered on sector $\Phi_{(3,2)}$), where in the plot it is sectors $\Phi_{(3,1)}$ and $\Phi_{(1,0)}$ which are both reached by a forward motion of length $2A$. Back to our $\Phi_{(0,1)}$-sector; the large-order contribution of $\Phi_{(0,2)}$ (second line of \eqref{eq:01-res}) also contributes at the same exponential-order, altough it arises from a backwards path (in the same way as in the non-resonant case), thus its alternating-sign within the growth. In order to numerically analyse all these exponentially-suppressed contributions, we perform the usual BP resummation of the leading growth (first line of \eqref{eq:01-res}), split it into real and imaginary parts, and define (as in \eqref{eq:01-subleading-def} and \eqref{eq:01-subleading-def-IM})
\begin{eqnarray}
\label{eq:res-01-subl-re}
\delta_{\re} F^{(0,1)}_k &\equiv& F_k^{(0,1)}\, \frac{2\pi\mathrm{i} A^{k}}{\Gamma(k)} - \mathrm{SF}_{( (0,1) \rightarrow (1,1) )}\, \CS_{\re} \mathrm{BP}_{N} [\chi_{( (0,1) \rightarrow (1,1) )}] (k) \simeq \CO (2^{-k}), \\
\label{eq:res-01-subl-im}
\mathrm{i}\, \delta_{\im} F^{(0,1)}_k &\equiv& - \mathrm{i} \mathrm{SF}_{( (0,1) \rightarrow (1,1) )}\, \CS_{\im} \mathrm{BP}_{N} [\chi_{( (0,1) \rightarrow (1,1) )}] (k) \simeq \CO (2^{-k}).
\end{eqnarray}
\noindent
In these expressions we used the fact that the coefficients $F_k^{(0,1)}$ are purely imaginary. As expected, both real and imaginary sequences are of the same exponential order. Finally, because the coefficients $F_k^{(0,2)}$ and $F_0^{(\widetilde{\boldsymbol{0}})} \equiv 1$ are real, while the $F_k^{(2,1)}$ are purely imaginary, we can write the respective predicted growths as
\begin{eqnarray}
(-2)^k\, \delta_{\re} F^{(0,1)}_k &\simeq& (-1)^k \mathrm{SF}_{( (0,1) \rightarrow \widetilde{\boldsymbol{0}} )}\, F_0^{(\widetilde{\boldsymbol{0}})} + \mathrm{SF}_{( (0,1) \rightarrow (0,2) )} \left( F_0^{(0,2)} - \frac{2A}{k}\, F_1^{(0,2)} + \cdots \right), \qquad \\
2^k\, \delta_{\im} F^{(0,1)}_k &\simeq& -\mathrm{i} \mathrm{SF}_{( (0,1) \rightarrow  (2,1) )} \left( F_0^{(2,1)} + \frac{2A}{k}\, F_1^{(2,1)} + \cdots \right).
\end{eqnarray}
\noindent
Using the values for the Stokes coefficients from the previous subsection, namely, \eqref{eq:ell-res-Stokes-pred-0-2-a}, \eqref{eq:ell-res-Stokes-pred-0-1}, and \eqref{eq:ell-res-Stokes-pred-pi-a}, the required statistical-factors\footnote{These are just the Borel residues associated to the singularities of $\CB [\Phi_{(0,1)}] (s)$; $\mathrm{SF}_{( (0,1) \rightarrow \boldsymbol{m} )} = - \mathsf{S}_{(0,1) \rightarrow \boldsymbol{m}}$.} above are:
\begin{eqnarray}
\mathrm{SF}_{( (0,1) \rightarrow (1,1) )} &\equiv& S^{(1)}_{\boldsymbol{e}_1} = 2, \\
\mathrm{SF}_{( (0,1) \rightarrow \widetilde{\boldsymbol{0}} )} &\equiv& S^{(0)}_{-\boldsymbol{e}_2} + \frac{1}{2} S^{(1)}_{\boldsymbol{e}_1} S^{(0)}_{-\boldsymbol{e}_1-\boldsymbol{e}_2} = - 2, \\
\mathrm{SF}_{( (0,1) \rightarrow (0,2) )} &\equiv& 2 S^{(2)}_{\boldsymbol{e}_2} = 4, \\
\mathrm{SF}_{( (0,1) \rightarrow (2,1) )} &\equiv& \left( S^{(1)}_{\boldsymbol{e}_1} \right)^2 = 4.
\end{eqnarray}
\noindent
The large-order tests are shown in figure~\ref{fig:01-res-sub}. The left image plots the leading growth of the real sequence, $(-2)^k\, \delta_{\re} F^{(0,1)}_k$, defined in \eqref{eq:res-01-subl-re}, and its convergence to the predicted alternating values\footnote{Using $F_0^{(\widetilde{\boldsymbol{0}})} = 1$ and $F_0^{(0,2)} = \frac{1}{6}$.} $\pm \mathrm{SF}_{( (0,1) \rightarrow \widetilde{\boldsymbol{0}} )}\, F_0^{(\widetilde{\boldsymbol{0}})} + \mathrm{SF}_{( (0,1) \rightarrow (0,2) )}\, F_0^{(0,2)} = \frac{2}{3} \mp 2$. The right image plots the leading growth of the imaginary sequence, $2^k\, \delta_{\im} F^{(0,1)}_k$, defined in \eqref{eq:res-01-subl-im}, and its convergence to the predicted value\footnote{Using $F_0^{(2,1)} = - \frac{2\mathrm{i}}{3\sqrt{3}}$.} $- \mathrm{i} \mathrm{SF}_{( (0,1) \rightarrow (2,1) )}\, F_0^{(2,1)} = - \frac{8}{3\sqrt{3}}$. The fifth-order Richardson transforms of both these real and imaginary sequences (the blue sequences in figure~\ref{fig:01-res-sub}) converge very clearly to their predicted values, with vanishingly-small relative errors of the order of $\sim 10^{-10}$ for the imaginary sequence, and $\sim 10^{-8}$ for the real\footnote{Once again, due to its alternating nature, one needs to separate and independently address even and odd terms in order to perform the Richardson transforms for the real sequence.} one.

This concludes our analysis of the \textit{resonant} resurgent-structure of the transseries solution we constructed for the elliptic-potential free energy. The many numerical checks clearly validate all large-order predictions, which were now obtained from \textit{resonant} resurgence. This example was particularly illuminating as resonance was \textit{tunable}, and, in this way, allowed us to compare differences and similarities between resonant and non-resonant cases. In general, in harder problems, resonance need not be tunable and the asymptotics may be much more intricate than illustrated here. Recent such examples include different string-theoretic scenarios, as are the cases of the first Painlev\'e equations describing 2d (super) gravity \cite{gikm10, asv11, sv13} and topological string theory \cite{ps09, cesv13, cesv14, c15}. This introduction to resonance should hopefully allow the interested reader to easily understand the resurgent properties of those examples.

As we reach the end, hopefully we have contributed to helping the reader enlarge their toolbox of methods for addressing nonperturbative questions in modern theoretical and mathematical physics. It should be clear that resurgence and transseries have a built-in very natural power to tackle truly nonperturbative questions, both at numerical \textit{and} analytical levels. Further, they are explicitly constructed in a physically-intuitive \textit{semiclassical} language. The range of problems where they may be applicable---from quantum mechanics, field theory, gauge theory, through string theory---is enormous, and we hope this contribution will help opening new nonperturbative windows into so many fascinating problems. Closer to our own research interests, while these methods by themselves may not tell us \textit{why} string perturbation theory diverges (we think of the ``why'' as a question concerning the physics), very likely they will tell us exactly \textit{how} string perturbation theory diverges (as these lectures hopefully made clear, the ``how'' is, in fact, a very sharp and answerable mathematical question): seemingly in a very specific mixture of both \textit{resonant} and \textit{parametric}/\textit{co-equational} resurgences. While the former has, at this stage, become well-known to the reader, the latter will have to wait for future work.

\acknowledgments
We would like to thank Aleksey Cherman, Santiago Codesido, Ovidiu Costin, Ricardo Couso-Santamar\'\i a, Gerald Dunne, Jos\'e Edelstein, Fr\'ed\'eric Fauvet, Tiago Dinis da Fonseca, Stavros Garoufalidis, Alba Grassi, Chris Howls, Marcos Mari\~no, Sara Pasquetti, Jorge Russo, David Sauzin, Mithat \"Unsal, Ricardo Vaz, Marcel Vonk, Andr\'e Voros, Marlene Weiss, Niclas Wyllard, and Szabolcs Zakany, for very valuable collaborations, discussions, comments and/or correspondence; since the early days we started working on these topics, and then over the slow process across the years that it took for these lectures to finally become concluded. IA would further like to thank the Kavli Institute for Theoretical Physics for extended hospitality, where parts of this work were conducted. GB would further like to thank the Kavli Institute for Theoretical Physics for hospitality, where parts of this work were conducted. RS would further like to thank CERN TH-Division, the Kavli Institute for Theoretical Physics, and the University of Geneva, for extended hospitality, where parts of this work were conducted. IA's research was partially supported by the NCN grant 2012/06/A/ST2/00396. GB is supported by the U.S. Department of Energy under Contract No.~DE~FG02-01ER41195. The research of RS was partially supported by the FCT-Portugal grants PTDC/MAT/119689/2010, EXCL/MAT-GEO/0222/2012, and UID/MAT/04459/2013; and the Swiss-NSF grant NCCR~51NF40-141869 ``The Mathematics of Physics'' (SwissMAP). This research was supported in part by the National Science Foundation under Grant No.~NSF~PHY-1125915.

\newpage

\appendix

\section{Further Details on Alien Calculus}\label{app:alien-calculus}

The goal of this appendix is to present a mathematical (albeit much shorter) introduction to \textit{alien calculus}. This will hopefully lay down a notational bridge between the contents of the main text and the mathematical literature; thus guiding the interested reader within the general framework described in, \textit{e.g.}, \cite{e81, cnp93a, cnp93b, c95, c98, c98b, dp99, ss03, s07, s14, d2014, ms16, lr16, d16, p16}, and from then on. We shall begin with general definitions and results, and then proceed to an example involving two instanton actions but with no multi-instanton corrections, \textit{i.e.}, a strictly ``linear'' example\footnote{This is, of course, an example inspired by a partition function already computed in section~\ref{sec:elliptic}.}. This example will nonetheless have enough structure so as to illustrate all the required material.

Note that, when compared against the main text, the asymptotic expansions in this appendix are in the variable (coupling constant) $z=1/x$, with $z \sim +\infty$. The analogue of \eqref{divergent} is now simply
\be
\label{divergent-app}
\Phi^{(0)} (z) \simeq \sum_{g=0}^{+\infty} \frac{\Phi^{(0)}_g}{z^{g+1}}.
\ee
\noindent
While this is the standard notation within the mathematical literature---and we shall follow it in this appendix---, as one compares against formulae in the main text---and, in particular, large-order formulae---it is important to notice that there might be some overall minus-sign differences.

\subsection*{Asymptotic Gevrey Series}

Up to this point we have been dealing with asymptotic series whose coefficients grow as $\Phi_g \sim g!$, but there are many other possibilities. One very useful classification of different possible types of asymptotic behaviour is due to Gevrey (see, \textit{e.g.}, \cite{j14} for an introduction). Herein, one says that the formal series \eqref{divergent-app} is a \textit{Gevrey-$p$ series}, iff $\exists\, M,C \in \BR^+$ such that
\be
| \Phi_g | \le M\, C^g\, \left( g! \right)^p
\ee
\noindent
In particular, a convergent series is Gevrey-$0$ while the asymptotic series we have been considering in the main text are Gevrey-$1$. The set of all Gevrey-$p$ formal series, which we shall denote by $\BC_{(p)} [[1/z]]$, is a differential algebra. One can make sense out of these general, divergent series by following a rather straightforward generalization of the Borel resummation procedure outlined at the beginning of section~\ref{sec:quartic}. Let $\Phi (z)$ in \eqref{divergent-app} be a Gevrey-$p$ series, \textit{i.e.}, $\Phi (z) \in \BC_{(p)} [[1/z]]$. Further, define the index-$m$ Borel transform as
\be
\CB_m \left[ \frac{1}{z^{\alpha+1}} \right](s) = \frac{s^{\alpha}}{\Gamma \left( \frac{\alpha}{m}+1 \right)},
\ee
\noindent
which, in particular, generalizes \eqref{boreltrans}. In this set-up, the index-$\frac{1}{p}$ Borel transform
\be
\CB_{1/p} [ \Phi ](s)
\ee
\noindent
has \textit{positive} radius of convergence\footnote{Hence, is a germ of an analytic function.} and may thus by analytically extended throughout $s \in \BC$, albeit with the resulting function having singularities and branch cuts, as expected. Once this is done, the index-$\frac{1}{p}$ Borel resummation\footnote{This resummation is defined via the index-$m$ Laplace transform, itself given by
\be
\CL_m \left[ f(s) \right] (z) := z^{m-1} \int_0^{+\infty} \rmd s\, f ( s^{1/m} )\, \rme^{- z^m s}.
\ee} (generalizing the standard Borel resummation \eqref{borelresum})
\be
\label{borel-resum-app}
\CS_{1/p} \Phi (z) = z^{\frac{1}{p}-1} \int_0^{+\infty} \rmd s\, \CB_{1/p} [\Phi](s^{p})\, \rme^{- z^{\frac{1}{p}} s}
\ee
\noindent
finally defines a function out from the Gevrey-$p$ series \eqref{divergent-app} (see, \textit{e.g.}, \cite{j14} for further details). Of course that, as usual, this integral only makes sense as long as the integration contour does not cross any singularities of the integrand. If it does, Stokes automorphisms must come into play and the resurgence framework necessarily makes its way onto the realm of general Gevrey series. Albeit very interesting, we shall not further pursue the question of multi-summability in this appendix, but refer the interested reader to, \textit{e.g.}, \cite{lr16}. Having briefly unveiled the general set-up of Borel multi-summability, we shall next return to our analysis of Gevrey-$1$ series.

\subsection*{Resurgent Functions and Simple Resurgent Functions}

Under the above set-up, let us introduce the definition of resurgence as due to \'Ecalle \cite{e81}. The Gevrey-$1$ formal series \eqref{divergent-app} is said to be \textit{resurgent}, or, more generically, to define a \textit{resurgent function}, if its Borel transform has \textit{endless analytic continuation}. This means that, for any chosen path emanating from some neighborhood of the origin, and which avoids a fixed \textit{countable} set on the complex plane, the Borel transform must have analytic continuation along this path. This ``forbidden'' set is, of course, the set of Borel singularities. This definition leads to a very general class of functions, from which a particularly useful subclass is that of \textit{simple} resurgent functions (also as already discussed in section~\ref{sec:borel}). This subclass is still rather large so as to include many examples, but a bit more under control on what concerns Borel singularities. Recall from section~\ref{sec:borel} that a \textit{simple resurgent function} is a resurgent function whose Borel singularities restrict to simple poles and logarithmic branch-points. If we recall \eqref{simpleBorelsingularities}, the precise statement is\footnote{In this appendix we are dropping the simple pole in \eqref{simpleBorelsingularities}, associated to a residual coefficient, without loss of generality. It should be straightforward for the reader to include it back at any stage in the present analysis.}
\be
\label{simple-app}
\CB [\Phi] (s) = \mathsf{S}_\omega \times \CB [\Psi_\omega] \left( s-\omega \right) \frac{\log \left( s-\omega \right)}{2\pi\rmi} + \text{holomorphic},
\ee
\noindent
near each singular point $\omega$, with $\mathsf{S}_\omega \in \BC$ and $\Psi_\omega$ some other asymptotic sector (but such that $\CB [\Psi_\omega]$ is holomorphic near the origin, \textit{i.e.}, it introduces no further singularities near $s \approx \omega$).

For simple resurgent functions, one may actually use the above explicit expression \eqref{simple-app} in order to be precise upon what happens as the Borel resummation \eqref{borel-resum-app} crosses a Stokes line (recall figure~\ref{stokescrossingfig} in section~\ref{sec:quartic}). Using the lateral Borel resummations
\be
\CS_{\theta^\pm} \Phi (z) = \int_0^{\rme^{\rmi\theta^\pm}\infty} \rmd s\, \CB[\Phi](s)\, \rme^{-z s},
\ee
\noindent
and labeling the (possibly infinite) singular points on the Stokes line along $\theta$ by $\left\{ \omega_n \right\}$ (\textit{i.e.}, $\forall\, n \in \BN,\, \arg \omega_n = \theta$), it is straightforward to obtain
\be
\CS_{\theta^+} \Phi (z) = \CS_{\theta^-} \Phi (z) - \sum_{\left\{ \omega_n \right\}} \mathsf{S}_{\omega_n}\, \rme^{-\omega_n z}\, \CS_{\theta^-} \Psi_{\omega_n} (z).
\ee
\noindent
This expression encodes the \textit{Stokes automorphism} $\underline{\mathfrak{S}}_\theta$ in \eqref{stokesauto}, $\CS_{\theta^+} = \CS_{\theta^-} \circ \underline{\mathfrak{S}}_\theta$, or, equivalently, the \textit{discontinuity} across the Stokes line,
\be
\CS_{\theta^+} - \CS_{\theta^-} = - \CS_{\theta^-} \circ \disc_\theta
\ee
\noindent
where we recall from \eqref{StokesDisc} that $\disc_\theta = \1 - \underline{\mathfrak{S}}_\theta$. One concludes that, for simple resurgent functions with Borel singularities \eqref{simple-app}, the Stokes discontinuities take the general form\footnote{This also shows how $\Phi$ alone cannot be the complete answer to whatever problem one is addressing: all $\left\{ \Psi_{\omega_n} \right\}$ contributions found by probing the singularity structure of $\CB[\Phi]$ must also be appropriately taken into account.}
\be
\label{sum-disc-app}
\disc_\theta \Phi (z) = \sum_{\left\{ \omega_n \right\}} \mathsf{S}_{\omega_n}\, \rme^{-\omega_n z}\, \Psi_{\omega_n} (z).
\ee
\noindent
This \textit{directional} discontinuity is easily decomposable as $\disc_\theta \Phi = \sum_{\left\{ \omega_n \right\}} \disc_{\omega_n} \Phi$, where the discontinuity at each specific singularity $\omega$ is given by
\be
\label{disc-alone-app}
\disc_\omega \Phi \equiv \mathsf{S}_{\omega}\, \rme^{-\omega z}\, \Psi_{\omega}.
\ee
\noindent
This ``discontinuity operator at $\omega$'' is \textit{linear} by construction. It may be equally defined as either acting on the asymptotic series or on the corresponding Borel transforms. It also relates in a completely direct way to the Borel singularity structure in \eqref{simple-app}. As we shall see next, this state of affairs may be further improved by introducing a closely related operator: the alien derivation.

\subsection*{Alien Derivatives and Alien Calculus}

The alien derivative at $\omega$, denoted by $\Delta_\omega \Phi$, improves upon the  discontinuity operator, $\disc_\omega \Phi$, precisely by further being a \textit{derivation} (and thus, a vector field). One way to approach it, starting from \eqref{simple-app}, is to define the \textit{alien derivative} of $\CB [\Phi]$ at the singularity $\omega$ (of $\Phi$, at $\omega$), denoted by $\Delta_\omega \CB [\Phi]$ (by $\Delta_\omega \Phi$), as a specific weighted-average of analytic continuations of $\CB [\Psi_\omega]$ along every possible path from the origin up to $\omega$ along $\theta$ (where $\arg \omega = \theta$), which circumvent the \textit{previous} singularities on the Stokes line in every possible way (\textit{e.g.}, recall figure~\ref{stokesweirdcrossingfig}). The alien derivative thus constructed is linear and may be equally defined as either acting on the Borel transforms or on the corresponding asymptotic series. But, in contrast with the discontinuity, it is also a \textit{derivation} in the sense that it satisfies the Leibniz rule with respect to either the convolution product of Borel transforms or the standard product of asymptotic series. Thus its usefulness over \eqref{disc-alone-app}. The precise definition is too technical for this short appendix and we shall next approach it in a slightly detoured way (see, \textit{e.g.}, \cite{e81, s07, s14} for rigorous definitions).

Within the usual picture of figures~\ref{stokescrossingfig} and~\ref{stokesweirdcrossingfig} in section~\ref{sec:quartic}, and our on-going discussion, let us consider a Stokes line along $\theta$ with singularities $\left\{ \omega_n \right\}$. Let us further introduce the \textit{pointed} (or dotted) alien derivative, simply defined by the combination $\dot{\Delta}_\omega := \rme^{-\omega z} \Delta_\omega$. Then, akin to \eqref{sum-disc-app}, one defines the \textit{directional} pointed alien-derivative as
\be
\label{pointed-dir-app}
\dot{\underline{\Delta}}_\theta = \sum_{\left\{ \omega_n \right\}} \dot{\Delta}_{\omega_n}.
\ee
\noindent
The main relation between all these concepts (and which, for the purposes of this appendix, we shall take as the \textit{definition} of alien derivative) is now \cite{e81}
\be
\label{Delta-vs-Disc-app}
\dot{\underline{\Delta}}_\theta := \log \underline{\mathfrak{S}}_\theta = \log \left( \1 - \disc_\theta \right) = - \disc_\theta - \frac{1}{2} \disc_\theta \circ \disc_\theta - \frac{1}{3} \disc_\theta \circ \disc_\theta \circ \disc_\theta - \cdots.
\ee
\noindent
In this way, via the logarithm, the proof that $\Delta_{\omega}$ is a derivation becomes equivalent to the proof that $\underline{\mathfrak{S}}_\theta$ is an automorphism  \cite{e81}. As it stands, \eqref{Delta-vs-Disc-app} is a surprisingly simple relation. But once we insert \eqref{pointed-dir-app}, and iteratively \eqref{sum-disc-app}, it will result in intricate\footnote{Except, of course, if there is a single singularity $\omega$ along $\theta$, in which case $\dot{\Delta}_\theta = 
- \disc_\theta$ as $\disc_\theta^2 \Phi \sim \disc_\theta \Psi_\omega = 0$.} expressions\footnote{See \cite{s07} for some very explicit expressions and examples, obtained by following this sort of analysis.} for the alien derivatives $\Delta_{\omega_n}$ (resulting in the aforementioned weighted averages). Nevertheless, the main point is that although $\disc_\omega$ connects to the Borel transform in a very direct way (as we have seen above), it is \textit{not} a derivation---thus the usefulness of introducing $\Delta_\omega$ instead.

\subsection*{Transseries and the Bridge Equation}

The natural question which follows from the above discussion is: is there an alternative (meaning simpler) way to compute $\Delta_\omega$? The answer is ``yes'', and it will be achieved by the use of the \textit{bridge equation}. One of the cleanest ways to introduce it is to follow the historical development of resurgence, going back to its origins within the context of (first order) nonlinear ODEs. Herein, one constructs \textit{formal solutions} to the said ODE via the use of \textit{transseries}, just like the one back in \eqref{Phitransseries},
\be
\label{Phitransseries-app}
\Phi (z,\sigma) = \sum_{n=0}^{+\infty} \sigma^n\, \Phi^{(n)} (z).
\ee
\noindent
Under appropriate conditions, $\Phi^{(0)} (z)$ is a Gevrey-$1$ series of the form \eqref{divergent-app} and the $\Phi^{(n)} (z)$ are $n$-instanton contributions\footnote{If by no other way, all sectors in the transseries \eqref{Phitransseries-app} may be found by probing the singularity structure of $\CB[\Phi^{(0)}]$, as in \eqref{sum-disc-app} (and, if necessary, successively the singularity structures of the thus uncovered $\CB[\Phi^{(n)}]$).}, analogues of \eqref{Phi(n)},
\be
\label{Phi(n)-app}
\Phi^{(n)} (z) \simeq \rme^{-n A z}\, \sum_{g=1}^{+\infty} \frac{\Phi_g^{(n)}}{z^{g + \beta_n}} \equiv \rme^{-n A z}\, \Phi_n (z),
\ee
\noindent
with $\Phi_n$ also Gevrey-$1$ series. The \textit{transseries parameter} $\sigma$ parameterizes different choices of initial (boundary) conditions, and one can write an infinite sequence of \textit{linear} ODEs (homogeneous for $n=1$; inhomogeneous for $n\ge2$) for the $\Phi_n$, starting from the original nonlinear equation.

The construction of the bridge equation now follows from a simple observation. The \textit{pointed} alien derivative which we introduced earlier, $\dot{\Delta}_\omega$, turns out to commute with the \textit{ordinary} derivative, \textit{i.e.}, $[ \dot{\Delta}_\omega, \partial_z ] = 0$ (in particular, this implies that the \textit{standard} alien derivative $\Delta_\omega$ does \textit{not} commute with the ordinary derivative), see, \textit{e.g.}, \cite{e81, s07, s14}. Further, a partial derivative with respect to the transseries parameter, $\partial_\sigma$, obviously also commutes with $\partial_z$. This implies that both $\dot{\Delta}_\omega \Phi (z,\sigma)$ and $\partial_\sigma \Phi (z,\sigma)$ will satisfy the very \textit{same}, first-order ODE. In particular, this will be a \textit{linear}, homogeneous ODE in the variable $z$, which implies that these two quantities must be \textit{proportional} to each other. In formulae:
\be
\label{bridge-equation-app}
\dot{\Delta}_\omega \Phi (z,\sigma) = S_\omega(\sigma)\, \frac{\partial\Phi}{\partial\sigma} (z,\sigma),
\ee
\noindent
where $S_\omega(\sigma)$ is the ($z$-independent) proportionality coefficient, and this equation is precisely the \textit{bridge equation}, establishing a ``bridge'' between \textit{alien} and \textit{ordinary} derivatives. The argument may be generalized to higher-order ODEs, to PDEs, to finite-difference equations, multi-dimensional integrals and so on, which is to say that the bridge equation is quite general and somewhat independent from the context where it was derived \cite{e81, s07, s14}.

One may be more specific about the proportionality coefficient $S_\omega(\sigma)$. The transmonomial building blocks of \eqref{Phitransseries-app} are of the form $\sigma^n \rme^{m A z}$, to which we may assign weight $\mathsf{w} = n+m$. Now, the derivative $\dot{\Delta}_{k A}$ acts on these transmonomials by lowering\footnote{For positive $k$ and $A$, which is implicitly assumed when saying ``lowering''.} their weight as $m \to m-k$ and the derivative $\partial_\sigma$ acts upon them by lowering weight as $n \to n-1$. Since the transseries \eqref{Phitransseries-app} has overall weight $\mathsf{w}=0$, the bridge equation \eqref{bridge-equation-app} can only hold\footnote{These concepts of degree and homogeneity were further discussed, in a slightly different context, in section~\ref{sec:physics}.} at $\omega = k A$ if $S_{k A} (\sigma)$ has weight $\mathsf{w} = 1-k$, \textit{i.e.},
\be
S_{k A} (\sigma) \equiv S_k\, \sigma^{1-k}
\ee
\noindent
is a monomial in $\sigma$. Notice that one must further have $k\leq1$, as both sides of \eqref{bridge-equation-app} are regular. In this case, the bridge equation is finally written as
\be
\label{final-bridge-equation-app}
\dot{\Delta}_{k A} \Phi (z,\sigma) = S_k\, \sigma^{1-k}\, \frac{\partial\Phi}{\partial\sigma} (z,\sigma), \qquad k \leq 1, \quad k \neq 0,
\ee
\noindent
up to the as-yet-unknown complex numbers $\left\{ S_{1}, S_{-1}, S_{-2}, \cdots \right\}$; the \'Ecalle analytic invariants\footnote{These numbers are invariants of the ODE in the following sense: two (apparently distinct) ODEs with the \textit{same} set of analytic invariants, must in fact be the \textit{same} equation up to a (possibly complicated) change of variables.} for the original ODE. This bridge equation that the transseries formal-solution satisfies, \eqref{final-bridge-equation-app}, translates to an infinite sequence of resurgence relations for the transseries components $\Phi_n$ (by simple replacement) as
\be
\label{bridge-app}
\Delta_{k A} \Phi_n = S_k \left( n+k \right) \Phi_{n+k}, \qquad k \leq 1, \quad k \neq 0.
\ee
\noindent
These expressions compute \textit{all} alien derivatives, up to the analytic invariants. It is also simple to realize that, via \eqref{Delta-vs-Disc-app}, the invariants $S_\omega$ are determined by the ``Borel residues'' $\mathsf{S}_\omega$ in \eqref{simple-app}.

This discussion also illustrates the advantage of working with the alien operator $\Delta_\omega$ rather than the discontinuity $\disc_\omega$: as $\Delta_\omega$ is a derivation, it is also a vector field, and it is precisely such a vector field realization that the bridge equation is implementing.

\subsection*{Stokes Phenomena and its Stokes Constants}

The framework described above now allows for a very elegant reformulation of Stokes phenomena (which we first explained back in \eqref{stokespheno}). Let us consider the action of the Stokes automorphism $\underline{\mathfrak{S}}_\theta$ on the transseries \eqref{Phitransseries-app}, along $\theta=0$. In this direction, the bridge equation \eqref{final-bridge-equation-app} yields a single singularity at $\omega = A$, in which case it follows
\be
\label{ecalle-conn-0}
\underline{\mathfrak{S}}_0 \Phi (z,\sigma) = \rme^{\dot{\underline{\Delta}}_0} \Phi (z,\sigma) = \rme^{\dot{\Delta}_A} \Phi (z,\sigma) = \rme^{S_1 \frac{\partial}{\partial\sigma}} \Phi (z,\sigma) = \Phi (z, \sigma+S_1 ),
\ee
\noindent
where we made use of the relation between the Stokes automorphism and the alien derivative \eqref{Delta-vs-Disc-app} in the first equality, and of the bridge equation \eqref{final-bridge-equation-app} in the third equality. This expression shows why the Stokes automorphism is indeed an \textit{automorphism}: it acts upon transseries (via the bridge equation) as inducing a \textit{translation} along the $\sigma$-direction. It also shows why this automorphism is named \textit{Stokes}. But, interestingly enough, this is not the only type of Stokes phenomena that must be considered. The bridge equation \eqref{final-bridge-equation-app} also predicts a Stokes line along $\theta=\pi$, where we now find nonlinear Stokes phenomena. In fact, along this direction there is an infinite number of singularities, and the Stokes automorphism acts as
\bea
\underline{\mathfrak{S}}_\pi \Phi (z,\sigma) &=& \rme^{\dot{\underline{\Delta}}_\pi} \Phi (z,\sigma) = \exp \left( \sum_{n=1}^{+\infty} \dot{\Delta}_{- n A} \right) \Phi (z,\sigma) = \\
&=& \exp \left( \sum_{n=1}^{+\infty} S_{-n}\, \sigma^{n+1}\, \frac{\partial}{\partial\sigma} \right) \Phi (z,\sigma) \equiv \rme^{\underline{S}_\pi (\sigma) \frac{\partial}{\partial\sigma}} \Phi (z,\sigma) = \Phi (z, \underline{\BS}_\pi (\sigma) ),
\label{ecalle-conn-pi}
\eea
\noindent
where $\underline{\BS}_\pi (\sigma)$ essentially generalizes the above translation $\sigma \mapsto \underline{\BS}_0 (\sigma) \equiv \sigma + S_1$ to a more generic automorphism $\sigma \mapsto \underline{\BS}_\pi (\sigma)$ generated by the one-parameter flow of the formal vector field $\underline{S}_\pi (\sigma) \frac{\partial}{\partial\sigma}$. Expressions \eqref{ecalle-conn-0} and \eqref{ecalle-conn-pi} are also known as the \'Ecalle connection formulae. A more complete discussion of Stokes phenomena and the Stokes automorphism, along different types of Stokes lines and for transseries with multiple parameters, may be found in, \textit{e.g.}, \cite{as13}.

The above discussion further shows how the analytic invariants are in fact the \textit{Stokes constants} of the problem; which themselves must relate to the \textit{Borel residues}. In this way, one may equally denote the $S_k$ constants as \'Ecalle analytic invariants, Stokes constants, or \'Ecalle--Stokes invariants. But perhaps the most interesting point here is that, because they relate to the Borel residues $\mathsf{S}_\omega$ in \eqref{simple-app} or \eqref{disc-alone-app} via \eqref{Delta-vs-Disc-app}, such relation gives us, at least in principle, a way to compute them. Let us understand how this occurs in our present setting.

Rewriting \eqref{simple-app} for any given sector of our specific transseries \eqref{Phitransseries-app}, one finds
\be
\label{borel-gen-app}
\CB [\Phi_n] (s) \Big|_{s=kA} = \mathsf{S}_{n\to n+k} \times \CB [\Phi_{n+k}] \left( s-kA\right)\, \frac{\log \left( s-kA \right)}{2\pi\rmi} + \text{holomorphic}, \qquad k\neq0,
\ee
\noindent
where the enlarged labeling of the Borel residues, including departure and arrival nodes, will become clear shortly. Further rewriting \eqref{sum-disc-app} for any of the transseries sectors, one finds
\be
\label{disc-gen-app}
\disc_\theta \Phi_n = \sum_{ \left\{ k \right\} } \mathsf{S}_{n\to n+k}\, \rme^{-kAz}\, \Phi_{n+k},
\ee
\noindent
where the sum in taken over all singularities $\omega = kA$ appearing along\footnote{For the two directions of interest in the one-parameter transseries setting, $\theta=0$ and $\theta=\pi$, this results in:
\bea
\disc_0 \Phi_n &=& \sum_{k=1}^{+\infty} \mathsf{S}_{n\to n+k}\, \rme^{-kAz}\, \Phi_{n+k}, \\
\disc_\pi \Phi_n &=& \sum_{k=1}^{n} \mathsf{S}_{n\to n-k}\, \rme^{+kAz}\, \Phi_{n-k}.
\eea} $\theta$. We may now expand \eqref{Delta-vs-Disc-app} in the present setting, and compare equal exponential terms in order to obtain the alien derivative acting on a sector $\Phi_n$ as a combination of Borel residues. Along $\theta=0$ one finds (with $k>0$)
\be
\label{alien-borel-app}
\Delta_{kA} \Phi_n = - \sum_{s=1}^{k} \frac{1}{s}\, \sum_{\substack{\left\{ \ell_1,\ldots,\ell_s\ge 1 \right\} \\ \vphantom{\frac{1}{2}} \sum \ell_i=k}} \mathsf{S}_{n\to n+\ell_1}\, \mathsf{S}_{n+\ell_1\to n+\ell_1+\ell_2}\, \cdots\, \mathsf{S}_{n+k-\ell_s\to n+k}\, \Phi_{n+k}.
\ee
\noindent
But now the constraint $k\leq1$ in \eqref{bridge-app} implies that there is only one non-vanishing term, \textit{i.e.}, $\Delta_{kA} \Phi_n=0$ for $k>1$. Comparing \eqref{alien-borel-app} with the resurgence relations \eqref{bridge-app}, this non-vanishing term yields
\be
\label{1-Borel-res-app}
S_1 = - \frac{1}{n+1}\, \mathsf{S}_{n\to n+1},
\ee
\noindent
which relates the $k=1$ Stokes constant with the Borel residues of a ``single-step'' (and vice-versa). All the remaining vanishing terms in \eqref{alien-borel-app} then yield a series of constraints which essentially build the ``multi-step'' Borel residues as products of ``single-step'' ones. One finds
\be
\mathsf{S}_{n\to n+k} = \frac{(-1)^{k-1}}{k!}\, \prod_{\ell=0}^{k-1} \mathsf{S}_{n+\ell\to n+\ell+1},
\ee
\noindent
or, explicitly\footnote{This also serves as a consistency check on the $n$-independence of $S_1$ in \eqref{1-Borel-res-app}.},
\be
\mathsf{S}_{n\to n+k} = - \frac{1}{k!}\, \frac{(n+k)!}{n!}\, S_1^k.
\ee
\noindent
Along $\theta=\pi$ the story is more intricate. One now finds a very similar result to the case of $\theta=0$ (with $k>0$)
\be
\label{alien-borel-2-app}
\Delta_{-kA} \Phi_n = - \sum_{s=1}^{k} \frac{1}{s}\, \sum_{\substack{\left\{ \ell_1,\ldots,\ell_s\ge 1 \right\} \\ \vphantom{\frac{1}{2}} \sum \ell_i=k}} \mathsf{S}_{n\to n-\ell_1}\, \mathsf{S}_{n-\ell_1\to n-\ell_1-\ell_2}\, \cdots\, \mathsf{S}_{n-k+\ell_s\to n-k}\, \Phi_{n-k},
\ee
\noindent
but where there are no vanishing constraints. Comparing \eqref{alien-borel-2-app} with the resurgence relations \eqref{bridge-app} finally yields a closed-form expression for the Stokes constants in terms of the Borel residues\footnote{Note that although the right-hand sides of these relations have an explicit $n$-dependence, the specific combinations which appear in these formulae are such that the left-hand sides turn out to be $n$-\textit{independent}. This relates back to the resurgence relations \eqref{bridge-app}, where the $n$-dependence of their proportionality coefficient $S_k \left( n+k \right)$ appears only (and explicitly) in the \textit{second} factor $\left( n+k \right)$. In other words, the Stokes constants $S_k$ do \textit{not} depend on departure or arrival transseries nodes; they are ``alien-chain translational-invariant'' (unlike the Borel residues).} \cite{s07}
\be
\label{neg-Borel-res-app}
S_{-k} = - \frac{1}{n-k}\, \sum_{s=1}^{k} \frac{1}{s}\, \sum_{\substack{\left\{ \ell_1,\ldots,\ell_s\ge 1 \right\} \\ \vphantom{\frac{1}{2}} \sum \ell_i=k}} \mathsf{S}_{n\to n-\ell_1}\, \mathsf{S}_{n-\ell_1\to n-\ell_1-\ell_2}\, \cdots\, \mathsf{S}_{n-k+\ell_s\to n-k}, \qquad 1\le k < n.
\ee
\noindent
Before finding the (equivalent) inverse mapping, let us first notice a simple consequence of the resurgence relations \eqref{bridge-app}. They yield $\Delta_{-nA}\Phi_n=0$, which, in terms of Borel residues, translates to\footnote{This restriction was lifted in the example studied in section~\ref{sec:quartic}, as it dealt with a two-parameter transseries.}
\be
\mathsf{S}_{n\to0}=0, \qquad n\ge1.
\ee
\noindent
Now, the complete inverse mapping corresponding to \eqref{neg-Borel-res-app} can be determined for each $k\ge 1$. It is given by\footnote{This type of expressions may be written as sums over partitions, and many such closed-form formulae may be found in \cite{asv11, as13}. However, the vectorial generalization of this type of expressions---which follows straightforwardly from the above reasoning and was presented in section~\ref{sec:physics}---is perhaps more illuminating.}:
\be
\mathsf{S}_{n\to n-k} = - \sum_{s=1}^{k} \frac{1}{s!}\, \sum_{\substack{\left\{ \ell_1,\ldots,\ell_s\ge 1 \right\} \\ \vphantom{\frac{1}{2}} \sum \ell_i=k}}\, \prod_{a=1}^{s} \left( n - \sum_{j=1}^{a} \ell_j \right)\, \prod_{b=1}^{s} S_{-\ell_b}, \qquad  1\le k < n.
\ee

The above formulae also show how the analytic invariants $S_k$ are in some sense more ``fundamental'' than the Borel residues $\mathsf{S}_{n\to n+k}$. In fact, the analytic invariants only depend on the ``distance vector'' $k$ between transseries nodes $n$ and $n+k$ (they are ``alien-chain translational invariant''), while the Borel residues need to have specified ``departure'', $n$, and ``arrival'', $n+k$, nodes in their labels (thus also implicitly depending on this ``distance vector'' $k$). In this way, the analytic invariants depend on less data, and there are less of them. On the other hand, as we shall see next, in spite of less ``fundamental'' the Borel residues do turn out to appear in simpler combinations than the Stokes constants in most formulae, thus still being rather useful.

\subsection*{An Explicit (Linear) Example}

In order to make the above mathematical framework more explicit, let us consider an example based on the partition function addressed in section~\ref{sec:elliptic}, \textit{i.e.}, a linear example involving a ``perturbative'' sector and \textit{two}, distinct ``instanton'' sectors. The transseries for such partition function will be chosen of the form
\be
\label{3-Z-app}
\CZ (z, \boldsymbol{\sigma}) = \sigma_0\, \Phi_0 (z) + \sigma_1\, \rme^{-A_1 z}\, \Phi_1 (z) + \sigma_2\, \rme^{+A_2 z}\, \Phi_2 (z),
\ee
\noindent
with $A_1, A_2 > 0$, seemingly leading to Borel singularities located at $+A_1$ and $-A_2$ (being a linear problem, there will be no multi-instanton Borel singularities). Each sector above is, of course, asymptotic (as in \eqref{divergent-app} and \eqref{Phi(n)-app}),
\be
\label{3-Z-Phi-app}
\Phi_n (z) \simeq \sum_{g=1}^{+\infty} \frac{\Phi_g^{(n)}}{z^{g + \beta_n}}.
\ee

Let us start by computing the alien derivatives of this example. The bridge equation \eqref{bridge-equation-app} in this problem is
\be
\label{bridge-Z-app}
\dot{\Delta}_\omega \CZ (z, \boldsymbol{\sigma}) = \boldsymbol{S}_\omega (\boldsymbol{\sigma}) \cdot \frac{\partial\CZ}{\partial\boldsymbol{\sigma}} (z,\boldsymbol{\sigma}),
\ee
\noindent
where $\boldsymbol{S}_\omega (\boldsymbol{\sigma})$ is now a ($z$-independent) proportionality \textit{vector}, which in principle may be built out of monomials of the form $\sigma_0^n \sigma_1^m \sigma_2^k$. However, in this linear example things turn out to be much simpler. In fact, due to linearity, the right-hand side of \eqref{bridge-Z-app} includes the very simple
\be
\boldsymbol{S}_\omega (\boldsymbol{\sigma}) \cdot \frac{\partial\CZ}{\partial\boldsymbol{\sigma}} = \boldsymbol{S}_\omega (\boldsymbol{\sigma}) \cdot \left( \Phi_0, \rme^{-A_1 z}\, \Phi_1, \rme^{+A_2 z}\, \Phi_2 \right),
\ee
\noindent
whereas the left-hand side may be written as
\be
\dot{\Delta}_\omega \CZ = \boldsymbol{\sigma} \cdot \left( \rme^{-\omega z}\, \Delta_\omega \Phi_0, \rme^{- \left( \omega+A_1 \right) z}\, \Delta_\omega \Phi_1, \rme^{- \left( \omega-A_2 \right) z}\, \Delta_\omega \Phi_2 \right).
\ee
\noindent
This immediately implies that
\be
\boldsymbol{S}_\omega (\boldsymbol{\sigma}) = \left(
\begin{array}{ccc}
S_\omega^{00} & S_\omega^{01} & S_\omega^{02} \\
S_\omega^{10} & S_\omega^{11} & S_\omega^{12} \\
S_\omega^{20} & S_\omega^{21} & S_\omega^{22}
\end{array}
\right) \boldsymbol{\sigma} \equiv \text{\textbf{S}}_\omega \cdot \boldsymbol{\sigma},
\ee
\noindent
where we have organized the Stokes analytic-invariants into a matrix. Matching equal powers of the transseries parameters in \eqref{bridge-Z-app} leads to the three equations
\be
\rme^{- \left( \omega+A_\ell \right) z}\, \Delta_\omega \Phi_\ell = S_\omega^{0\ell}\, \Phi_0 + S_\omega^{1\ell}\, \rme^{-A_1 z}\, \Phi_1 + S_\omega^{2\ell}\, \rme^{+A_2 z}\, \Phi_2,
\ee
\noindent
with $A_\ell = 0, A_1, -A_2$, respectively, and from where the resurgence relations may be found (alongside the vanishing Stokes constants). Indeed, now matching equal powers of the transmonomials $\rme^{-z}$ finally leads to the resurgence relations\footnote{Recall that Borel transforms cannot have singularities at the origin; this would contradict their definition based on holomorphicity around some neighborhood of the origin.}
\begin{alignat}{2}
\Delta_{A_1} \Phi_0 &= S_{A_1}^{10}\, \Phi_1, \qquad\qquad & \Delta_{-A_2} \Phi_0 &= S_{-A_2}^{20}\, \Phi_2, \\
\Delta_{-A_1} \Phi_1 &= S_{-A_1}^{01}\, \Phi_0, \qquad\qquad & \Delta_{-A_1-A_2} \Phi_1 &= S_{-A_1-A_2}^{21}\, \Phi_2, \\
\Delta_{A_2} \Phi_2 &= S_{A_2}^{02}\, \Phi_0, \qquad\qquad & \Delta_{A_1+A_2} \Phi_2 &= S_{A_1+A_2}^{12}\, \Phi_1,
\end{alignat}
\noindent
or, equivalently, to the analytic-invariant matrices\footnote{Because each matrix has a single non-vanishing entry, this shows very explicitly how the Stokes constants only depend on their subscript (the ``distance vector'' between nodes) and not on their superscripts (labeling ``departure'' and ``arrival'' transseries nodes), as discussed earlier. In the following we will drop these superscripts and label the Stokes constants only by their ``translational invariant'' subscripts.}
\begin{alignat}{3}
\text{\textbf{S}}_{A_1} &= \left(
\begin{array}{ccc}
0 & 0 & 0 \\
S_{A_1}^{10} & 0 & 0 \\
0 & 0 & 0
\end{array}
\right),
\qquad &
\text{\textbf{S}}_{A_2} &= \left(
\begin{array}{ccc}
0 & 0 & S_{A_2}^{02} \\
0 & 0 & 0 \\
0 & 0 & 0
\end{array}
\right),
\qquad &
\text{\textbf{S}}_{A_1+A_2} &= \left(
\begin{array}{ccc}
0 & 0 & 0 \\
0 & 0 & S_{A_1+A_2}^{12} \\
0 & 0 & 0
\end{array}
\right), \\
\text{\textbf{S}}_{-A_1} &= \left(
\begin{array}{ccc}
0 & S_{-A_1}^{01} & 0 \\
0 & 0 & 0 \\
0 & 0 & 0
\end{array}
\right),
\qquad &
\text{\textbf{S}}_{-A_2} &= \left(
\begin{array}{ccc}
0 & 0 & 0 \\
0 & 0 & 0 \\
S_{-A_2}^{20} & 0 & 0
\end{array}
\right),
\qquad &
\text{\textbf{S}}_{-A_1-A_2} &= \left(
\begin{array}{ccc}
0 & 0 & 0 \\
0 & 0 & 0 \\
0 & S_{-A_1-A_2}^{21} & 0
\end{array}
\right).
\end{alignat}

\begin{figure}[t!]
\begin{center}
\includegraphics[width=15.5cm]{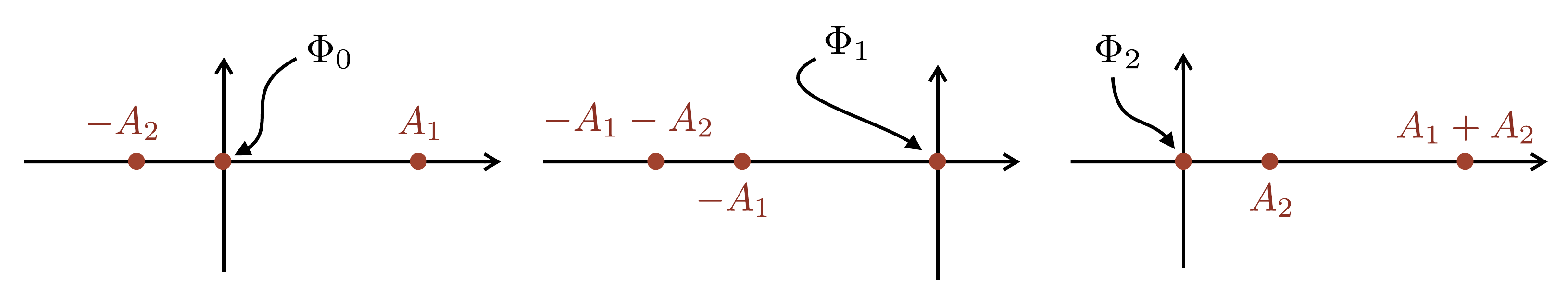}
\end{center}
\caption{Singularities on the Borel plane, for each of the three sectors $\Phi_{n}$.
}
\label{fig:Borel-plane-3-sectors}
\end{figure}

Besides having computed all alien derivatives in the problem (up to the Stokes constants), this shows how there are indeed two Stokes lines: one along $\theta=0$ and another along $\theta=\pi$. The Borel singularity structure is slightly more intricate than one had na\"\i vely guessed from the original transseries, and is fully illustrated in figure~\ref{fig:Borel-plane-3-sectors}. Furthermore, Stokes phenomena is now straightforward to implement. Proceeding stepwise, one first writes the directional alien derivatives $\dot{\underline{\Delta}}_\theta$ along $\theta=0$,
\be
\dot{\underline{\Delta}}_0 = \rme^{- A_1 z} \Delta_{A_1} + \rme^{- A_2 z} \Delta_{A_2} + \rme^{- \left( A_1+A_2 \right) z} \Delta_{A_1+A_2},
\ee
\noindent
and along $\theta=\pi$,
\be
\dot{\underline{\Delta}}_\pi = \rme^{A_1 z} \Delta_{-A_1} + \rme^{A_2 z} \Delta_{-A_2} + \rme^{\left( A_1+A_2 \right) z} \Delta_{-A_1-A_2}.
\ee
\noindent
Next, these expansions lead to their corresponding Stokes automorphisms\footnote{Note how there is only one type of second-order alien derivative along each direction, but no higher-order ones.}
\bea
\underline{\mathfrak{S}}_0 \Phi_0 &=& \Phi_0 + S_{A_1}\, \rme^{- A_1 z}\, \Phi_1, \\
\underline{\mathfrak{S}}_0 \Phi_1 &=& \Phi_1, \\
\underline{\mathfrak{S}}_0 \Phi_2 &=& \Phi_2 + S_{A_2}\, \rme^{- A_2 z}\, \Phi_0 + \left( S_{A_1+A_2} + \frac{1}{2!}\, S_{A_1} S_{A_2} \right) \rme^{- \left( A_1+A_2 \right) z}\, \Phi_1,
\eea
\noindent
and
\bea
\underline{\mathfrak{S}}_\pi \Phi_0 &=& \Phi_0 + S_{-A_2}\, \rme^{A_2 z}\, \Phi_2, \\
\underline{\mathfrak{S}}_\pi \Phi_1 &=& \Phi_1 + S_{-A_1}\, \rme^{A_1 z}\, \Phi_0 + \left( S_{-A_1-A_2} + \frac{1}{2!}\, S_{-A_2} S_{-A_1} \right) \rme^{\left( A_1+A_2 \right) z}\, \Phi_2, \\
\underline{\mathfrak{S}}_\pi \Phi_2 &=& \Phi_2.
\eea
\noindent
These operators act on the full partition-function transseries as
\be
\underline{\mathfrak{S}}_0 \CZ (z, \boldsymbol{\sigma}) = \CZ (z, \boldsymbol{\underline{\BS}_0} \cdot \boldsymbol{\sigma}) \qquad \text{and} \qquad \underline{\mathfrak{S}}_\pi \CZ (z, \boldsymbol{\sigma}) = \CZ (z, \boldsymbol{\underline{\BS}_\pi} \cdot \boldsymbol{\sigma})
\ee
\noindent
where, finally, the flows acting on the transseries parameters $\boldsymbol{\sigma} \mapsto \boldsymbol{\underline{\BS}_{\theta}} \cdot \boldsymbol{\sigma}$ are implemented by the Stokes matrices
\be
\boldsymbol{\underline{\BS}_0} = \left(
\begin{array}{ccc}
1 & \,0\, & S_{A_2} \\
S_{A_1} & \,1\, & S_{A_1+A_2} + \frac{1}{2} S_{A_1} S_{A_2}\vphantom{\frac{\frac{1}{2}}{\frac{1}{2}}} \\
0 & \,0\, & 1
\end{array}
\right)
\qquad \text{and} \qquad
\boldsymbol{\underline{\BS}_\pi} = \left(
\begin{array}{ccc}
1\, & S_{-A_1} & \,0 \\
0\, & 1 & \,0 \\
S_{-A_2}\, & S_{-A_1-A_2} + \frac{1}{2} S_{-A_2} S_{-A_1}\vphantom{\frac{\frac{1}{2}}{\frac{1}{2}}} & \,1
\end{array}
\right).
\ee

It is also interesting to rewrite these matrices using the Borel residues instead of the analytic invariants, via \eqref{neg-Borel-res-app}, where one finds the much neater:
\be
\boldsymbol{\underline{\BS}_0} = \left(
\begin{array}{ccc}
1 & \,0\, & -\mathsf{S}_{2\to0} \\
-\mathsf{S}_{0\to1} & \,1\, & -\mathsf{S}_{2\to1} \\
0 & \,0\, & 1
\end{array}
\right),
\qquad
\boldsymbol{\underline{\BS}_\pi} = \left(
\begin{array}{ccc}
1\, & -\mathsf{S}_{1\to0} & \,0 \\
0\, & 1 & \,0 \\
-\mathsf{S}_{0\to2}\, & -\mathsf{S}_{1\to2} & \,1
\end{array}
\right).
\ee

\subsection*{Large-Order Behaviour and Transseries Asymptotics}

The one question that remains is: how do the above concepts relate to asymptotics?, which was a central topic throughout these lectures. As discussed at length in the main body of this text, the connection arises via the (Stokes) discontinuity \eqref{sum-disc-app}, and how it may be used within the Cauchy theorem. This was well explained in the main text; here we shall limit ourselves to very briefly readdressing this issue within the alien-calculus mathematical framework outlined above.

As first explained back in subsection~\ref{subsec:Zasymptotics}---but now within the example of the partition function \eqref{3-Z-app} with Stokes lines along $\theta=0, \pi$---, straightforward application of the Cauchy theorem to each of its asymptotic components \eqref{3-Z-Phi-app} yields (this is very similar to \eqref{Cauchy-two-disc})
\be
\Phi (z) = \frac{1}{2\pi\rmi} \int_{0}^{+\infty} \rmd w\, \frac{\disc_{0}\Phi(w)}{w-z} + \frac{1}{2\pi\rmi} \int_{0}^{-\infty} \rmd w\, \frac{\disc_{\pi}\Phi(w)}{w-z}.
\ee
\noindent
where $\disc_\theta = \1 - \underline{\mathfrak{S}}_\theta$ and $\underline{\mathfrak{S}}_\theta$ was computed earlier for each $\Phi_n$-sector via alien calculus. Using the asymptotic expansions \eqref{3-Z-Phi-app} and expanding the above integrands around large $z$, it follows:
\begin{itemize}
\item Discontinuities of the perturbative sector $\Phi_0$:
\bea
\disc_0 \Phi_0 (w) &=& - S_{A_1}\, \rme^{- A_1 w}\, \Phi_1 (w), \\
\disc_\pi \Phi_0 (w) &=& - S_{-A_2}\, \rme^{A_2 w}\, \Phi_2 (w).
\eea
\item Asymptotics of the perturbative sector $\Phi_0$ (with analytic invariants):
\bea
\Phi_{g\gg1}^{(0)} &\simeq& \frac{S_{A_1}}{2\pi\rmi}\, \frac{\Gamma \left( g-\beta_1 \right)}{A_1^{g-\beta_1}}\, \sum_{h=1}^{+\infty} \frac{\Gamma \left( g-\beta_1-h+1 \right)}{\Gamma \left( g-\beta_1 \right)}\, \Phi_h^{(1)}\, A_1^{h-1} + \\
&&
\hspace{50pt}
+ \frac{S_{-A_2}}{2\pi\rmi}\, \frac{\Gamma \left( g-\beta_2 \right)}{\left(-A_2\right)^{g-\beta_2}}\, \sum_{h=1}^{+\infty} \frac{\Gamma \left( g-\beta_2-h+1 \right)}{\Gamma \left( g-\beta_2 \right)}\, \Phi_h^{(2)} \left(-A_2\right)^{h-1}. \nonumber
\eea
\item Asymptotics of the perturbative sector $\Phi_0$ (with Borel residues):
\bea
\Phi_{g\gg1}^{(0)} &\simeq& - \frac{\mathsf{S}_{0\to1}}{2\pi\rmi}\, \frac{\Gamma \left( g-\beta_1 \right)}{A_1^{g-\beta_1}}\, \sum_{h=1}^{+\infty} \frac{\Gamma \left( g-\beta_1-h+1 \right)}{\Gamma \left( g-\beta_1 \right)}\, \Phi_h^{(1)}\, A_1^{h-1} - \\
&&
\hspace{50pt}
- \frac{\mathsf{S}_{0\to2}}{2\pi\rmi}\, \frac{\Gamma \left( g-\beta_2 \right)}{\left(-A_2\right)^{g-\beta_2}}\, \sum_{h=1}^{+\infty} \frac{\Gamma \left( g-\beta_2-h+1 \right)}{\Gamma \left( g-\beta_2 \right)}\, \Phi_h^{(2)} \left(-A_2\right)^{h-1}. \nonumber
\eea
\item Leading large-order behaviour:
\bea
\Phi_{g\gg1}^{(0)} &\simeq& \frac{S_{A_1}}{2\pi\rmi}\, \frac{\Gamma \left( g-\beta_1 \right)}{A_1^{g-\beta_1}} \left( \Phi_1^{(1)} + \frac{A_1}{g-\beta_1-1}\, \Phi_2^{(1)} + \cdots \right) + \\
&&
\hspace{50pt}
+ \frac{S_{-A_2}}{2\pi\rmi}\, \frac{\Gamma \left( g-\beta_2 \right)}{\left(-A_2\right)^{g-\beta_2}} \left( \Phi_1^{(2)} - \frac{A_2}{g-\beta_2-1}\, \Phi_2^{(2)} + \cdots \right). \nonumber
\eea
\item Discontinuities of the nonperturbative sector $\Phi_1$:
\bea
\disc_0 \Phi_1 (w) &=& 0, \\
\disc_\pi \Phi_1 (w) &=& - S_{-A_1}\, \rme^{A_1 w}\, \Phi_0 (w) - \left( S_{-A_1-A_2} + \frac{1}{2}\, S_{-A_2} S_{-A_1} \right) \rme^{\left( A_1+A_2 \right) w}\, \Phi_2 (w).
\eea
\item Asymptotics of the nonperturbative sector $\Phi_1$ (with analytic invariants):
\bea
\Phi_{g\gg1}^{(1)} &\simeq& \frac{S_{-A_1}}{2\pi\rmi}\, \frac{\Gamma \left( g+\beta_1 \right)}{\left(-A_1\right)^{g+\beta_1}}\, \sum_{h=0}^{+\infty} \frac{\Gamma \left( g+\beta_1-h-1 \right)}{\Gamma \left( g+\beta_1 \right)}\, \Phi_h^{(0)} \left(-A_1\right)^{h+1} + \\
&&
\hspace{-50pt}
+ \frac{S_{-A_1-A_2} + \frac{1}{2}\, S_{-A_2} S_{-A_1}}{2\pi\rmi}\, \frac{\Gamma \left( g+\beta_1-\beta_2 \right)}{\left(-A_1-A_2\right)^{g+\beta_1-\beta_2}}\, \sum_{h=1}^{+\infty} \frac{\Gamma \left( g+\beta_1-\beta_2-h \right)}{\Gamma \left( g+\beta_1-\beta_2 \right)}\, \Phi_h^{(2)} \left(-A_1-A_2\right)^h. \nonumber
\eea
\item Asymptotics of the nonperturbative sector $\Phi_1$ (with Borel residues):
\bea
\Phi_{g\gg1}^{(1)} &\simeq& -\frac{\mathsf{S}_{1\to0}}{2\pi\rmi}\, \frac{\Gamma \left( g+\beta_1 \right)}{\left(-A_1\right)^{g+\beta_1}}\, \sum_{h=0}^{+\infty} \frac{\Gamma \left( g+\beta_1-h-1 \right)}{\Gamma \left( g+\beta_1 \right)}\, \Phi_h^{(0)} \left(-A_1\right)^{h+1} - \\
&&
\hspace{20pt}
- \frac{\mathsf{S}_{1\to2}}{2\pi\rmi}\, \frac{\Gamma \left( g+\beta_1-\beta_2 \right)}{\left(-A_1-A_2\right)^{g+\beta_1-\beta_2}}\, \sum_{h=1}^{+\infty} \frac{\Gamma \left( g+\beta_1-\beta_2-h \right)}{\Gamma \left( g+\beta_1-\beta_2 \right)}\, \Phi_h^{(2)} \left(-A_1-A_2\right)^h. \nonumber
\eea
\item Leading large-order behaviour:
\bea
\Phi_{g\gg1}^{(1)} &\simeq& \frac{S_{-A_1}}{2\pi\rmi}\, \frac{\Gamma \left( g+\beta_1 \right)}{\left(-A_1\right)^{g+\beta_1}} \left( - \frac{A_1}{g+\beta_1-1}\, \Phi_0^{(0)} + \cdots \right) + \\
&&
+ \frac{S_{-A_1-A_2} + \frac{1}{2}\, S_{-A_2} S_{-A_1}}{2\pi\rmi}\, \frac{\Gamma \left( g+\beta_1-\beta_2 \right)}{\left(-A_1-A_2\right)^{g+\beta_1-\beta_2}} \left( - \frac{A_1+A_2}{g+\beta_1-\beta_2-1}\, \Phi_1^{(2)} + \cdots \right). \nonumber
\eea
\item Discontinuities of the nonperturbative sector $\Phi_2$:
\bea
\disc_0 \Phi_2 (w) &=& - S_{A_2}\, \rme^{- A_2 w}\, \Phi_0 (w) - \left( S_{A_1+A_2} + \frac{1}{2}\, S_{A_1} S_{A_2} \right) \rme^{- \left( A_1+A_2 \right) w}\, \Phi_1 (w), \\
\disc_\pi \Phi_2 (w) &=& 0.
\eea
\item Asymptotics of the nonperturbative sector $\Phi_2$ (with analytic invariants):
\bea
\Phi_{g\gg1}^{(2)} &\simeq& \frac{S_{A_2}}{2\pi\rmi}\, \frac{\Gamma \left( g+\beta_2 \right)}{A_2^{g+\beta_2}}\, \sum_{h=0}^{+\infty} \frac{\Gamma \left( g+\beta_2-h-1 \right)}{\Gamma \left( g+\beta_2 \right)}\, \Phi_h^{(0)} A_2^{h+1} + \\
&&
\hspace{-20pt}
+ \frac{S_{A_1+A_2} + \frac{1}{2}\, S_{A_1} S_{A_2}}{2\pi\rmi}\, \frac{\Gamma \left( g+\beta_2-\beta_1 \right)}{\left(A_1+A_2\right)^{g+\beta_2-\beta_1}}\, \sum_{h=1}^{+\infty} \frac{\Gamma \left( g+\beta_2-\beta_1-h \right)}{\Gamma \left( g+\beta_2-\beta_1 \right)}\, \Phi_h^{(1)} \left(A_1+A_2\right)^h. \nonumber
\eea
\item Asymptotics of the nonperturbative sector $\Phi_2$ (with Borel residues):
\bea
\Phi_{g\gg1}^{(2)} &\simeq& -\frac{\mathsf{S}_{2\to0}}{2\pi\rmi}\, \frac{\Gamma \left( g+\beta_2 \right)}{A_2^{g+\beta_2}}\, \sum_{h=0}^{+\infty} \frac{\Gamma \left( g+\beta_2-h-1 \right)}{\Gamma \left( g+\beta_2 \right)}\, \Phi_h^{(0)} A_2^{h+1} - \\
&&
- \frac{\mathsf{S}_{2\to1}}{2\pi\rmi}\, \frac{\Gamma \left( g+\beta_2-\beta_1 \right)}{\left(A_1+A_2\right)^{g+\beta_2-\beta_1}}\, \sum_{h=1}^{+\infty} \frac{\Gamma \left( g+\beta_2-\beta_1-h \right)}{\Gamma \left( g+\beta_2-\beta_1 \right)}\, \Phi_h^{(1)} \left(A_1+A_2\right)^h. \nonumber
\eea
\item Leading large-order behaviour:
\bea
\Phi_{g\gg1}^{(2)} &\simeq& \frac{S_{A_2}}{2\pi\rmi}\, \frac{\Gamma \left( g+\beta_2 \right)}{A_2^{g+\beta_2}} \left( \frac{A_2}{g+\beta_2-1}\, \Phi_0^{(0)} + \cdots \right) + \\
&&
+ \frac{S_{A_1+A_2} + \frac{1}{2}\, S_{A_2} S_{A_1}}{2\pi\rmi}\, \frac{\Gamma \left( g+\beta_2-\beta_1 \right)}{\left(A_1+A_2\right)^{g+\beta_2-\beta_1}} \left( \frac{A_1+A_2}{g+\beta_2-\beta_1-1}\, \Phi_1^{(1)} + \cdots \right). \nonumber
\eea
\end{itemize}
\noindent
Note how the above formulae become invalid and must necessarily change under conditions for resonance, \textit{i.e.}, if $A_1+A_2=0$. This issue was discussed at length in section~\ref{sec:elliptic}. Many other closed-form large-order relations, both resonant and non-resonant, may be found in \cite{asv11}.

\section{Stokes Lines Within the WKB Expansion}\label{app:stokes}

Both Stokes and anti-Stokes lines were discussed at length in the main body of this work, illustrated via, \textit{e.g.}, figures~\ref{steepest} and~\ref{quarticphase} for the example discussed in section~\ref{sec:quartic}, and figures~\ref{steepest-elliptic}, \ref{fig:thimbles-1} and~\ref{fig:ell-phase} for the example discussed in section~\ref{sec:elliptic}. These examples were associated to one-dimensional integrals for (simplified) partition functions. For the sake of completeness, let us also illustrate how Stokes or anti-Stokes lines appear when considering problems based upon differential equations. Addressing a simple and standard setting (but perhaps also one of the most popular ones), let us briefly discuss the WKB analysis\footnote{WKB is textbook material. For modern introductions see, \textit{e.g.}, \cite{kt05, kp09} and references therein.} of quantum mechanical potentials.

Start with the one-dimensional time-independent Schr\"odinger equation,
\be
\left( - \frac{\hbar^2}{2m}\, \frac{\rmd^2}{\rmd q^2} + V(q) \right) \psi(q) = E\, \psi (q),
\ee
\noindent
and rewrite it (adequately for what follows) as
\be
\label{gen-schrodinger-app}
\hbar^2\, \psi''(q) + p^2(q)\, \psi(q) = 0,
\ee
\noindent
where $p$ is the ``classical momentum'' $p = \sqrt{2m \left( E - V(q) \right)}$ (at least when this result is real). WKB analysis now follows via its familiar \textit{ansatz} for the wavefunction solution,
\be
\label{WKB-app}
\psi(q) = \exp \left( \frac{\rmi}{\hbar} \int_{q_0}^q \rmd z\, \mathsf{V} (z,\hbar) \right),
\ee
\noindent
where $\mathsf{V} (q,\hbar)$ satisfies the Riccati equation $\mathsf{V}^2 - \rmi\hbar\, \mathsf{V}' = p^2$ via \eqref{gen-schrodinger-app} and should be thought of as a power series in $\hbar$,
\be
\label{V-exp-app}
\mathsf{V} (q,\hbar) = \mathsf{V}_0 (q) + \hbar\, \mathsf{V}_1 (q) + \hbar^2\, \mathsf{V}_2 (q) + \cdots.
\ee
\noindent
In other words, a (perturbative) WKB solution to the above one-dimensional time-independent Schr\"odinger equation is determined by the recursion:
\bea
\label{pmp-app}
\mathsf{V}_0 &=& \pm p, \\
\mathsf{V}_n &=& \frac{1}{2 \mathsf{V}_0} \left( \rmi \frac{\rmd}{\rmd q} \mathsf{V}_{n-1} - \sum_{k=1}^{n-1} \mathsf{V}_{k}\, \mathsf{V}_{n-k} \right).
\eea

A simplification of the WKB \textit{ansatz} \eqref{WKB-app} occurs when one splits even and odd powers of $\hbar$ in the expansion \eqref{V-exp-app}, as $\mathsf{V}_{\text{even}} (q,\hbar) = \mathsf{V}_0 (q) + \cdots$ and $\mathsf{V}_{\text{odd}} (q,\hbar) = \hbar\, \mathsf{V}_1 (q) + \cdots$. One then immediately notices that the odd coefficients are determined by the even ones,
\be
\mathsf{V}_{\text{odd}} = \frac{\rmi\hbar}{2}\, \frac{\rmd}{\rmd q} \log \mathsf{V}_{\text{even}}.
\ee
\noindent
Replacing this in the WKB \textit{ansatz} \eqref{WKB-app} and considering the two options implied by the recursion in \eqref{pmp-app}, one finally obtains the expected two linearly-independent solutions (in fact forming a basis for the space of solutions)
\be
\psi_{\pm} (q) = \frac{1}{\sqrt{\mathsf{V}_\text{even} (q,\hbar)}}\, \exp \left( \pm \frac{\rmi}{\hbar} \int_{q_0}^q \rmd z\, \mathsf{V}_\text{even} (z,\hbar) \right).
\ee
\noindent
Of course that the WKB solution was perturbative to start-off with, in which case one should still expand these wavefunctions as (asymptotic) series (see, \textit{e.g.}, \cite{cm16} for some explicit expressions)
\be
\label{WKBwf-app}
\psi_{\pm} (q) \simeq \exp \left( \pm \frac{\rmi}{\hbar} \int_{q_0}^q \rmd z\, p (z) \right) \sum_{k=0}^{+\infty} \hbar^k \psi_{\pm,k} (q).
\ee
\noindent
These WKB solutions are local, \textit{i.e.}, valid in specific regions of the complex plane. In order to construct a global solution for the wavefunction one needs ``connection formulae'', in order to stepwise match WKB solutions of the type \eqref{WKBwf-app}, valid in adjacent regions. Further, this match only works if the corresponding WKB solutions have similar (exponential) magnitudes, and thus the relevance of Stokes and anti-Stokes lines in the present framework.

In this way, our focus of interest in this appendix is precisely the above exponential pre-factor. Let us normalize it by setting $q_0$ to be a classical \textit{turning point}, \textit{i.e.}, separating a classically allowed region from a classically forbidden one. These turning points are the zeroes of the ``classical momentum'', $p(q)$, which also signal the associated square-root branch points. In this context, at real $\hbar$, Stokes lines of \eqref{WKBwf-app} are defined as\footnote{Recall the extra factor of $\rmi$ in the exponential \eqref{WKBwf-app}, as compared to what was discussed in the main text.}
\be
\re \int_{q_0}^q \rmd z\, \sqrt{2m \left( E - V(z) \right)} = 0,
\ee
\noindent
while anti-Stokes lines are (where WKB solutions will have similar exponential weights)
\be
\label{WKBantistokes-app}
\im \int_{q_0}^q \rmd z\, \sqrt{2m \left( E - V(z) \right)} = 0.
\ee
\noindent
Note how at simple zeroes of the argument of the square-root, one has $\int \sqrt{z} \sim z^{3/2}$, in which case (anti) Stokes lines will form networks with \textit{trivalent} nodes at the turning points. 

\begin{figure}[t!]
\begin{center}
\includegraphics[height=5.25cm]{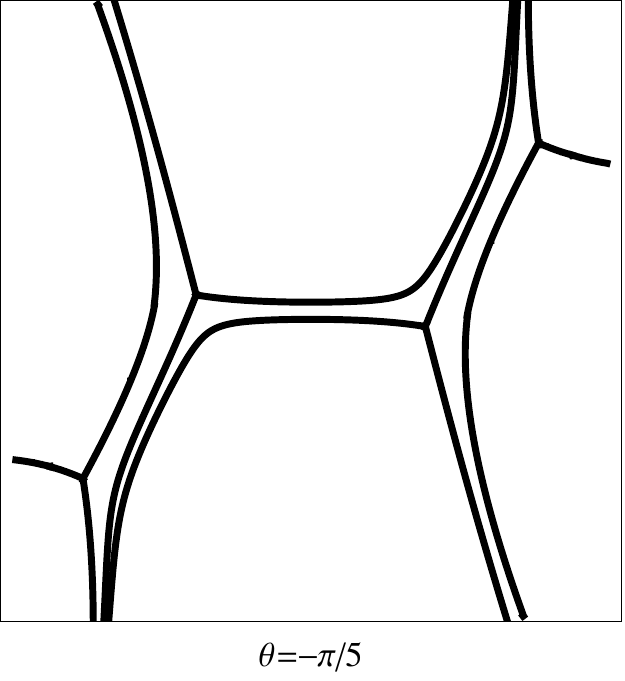}
$\quad$
\includegraphics[height=5.25cm]{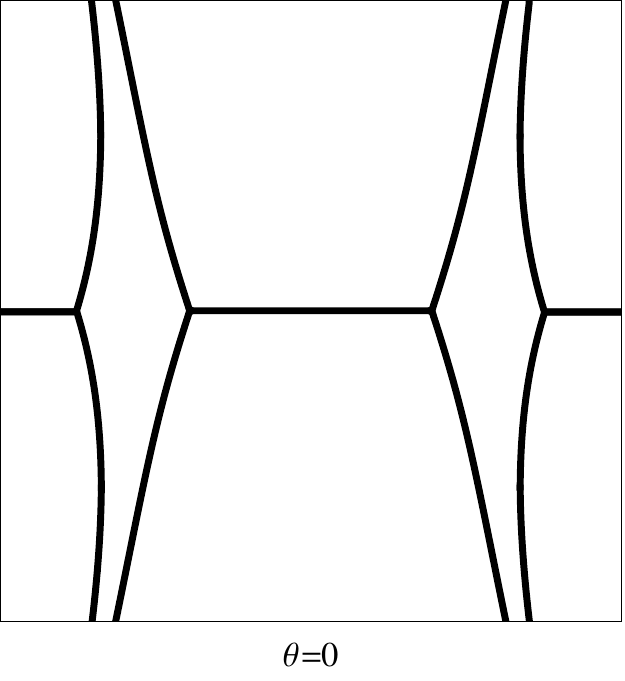}
$\quad$
\includegraphics[height=5.25cm]{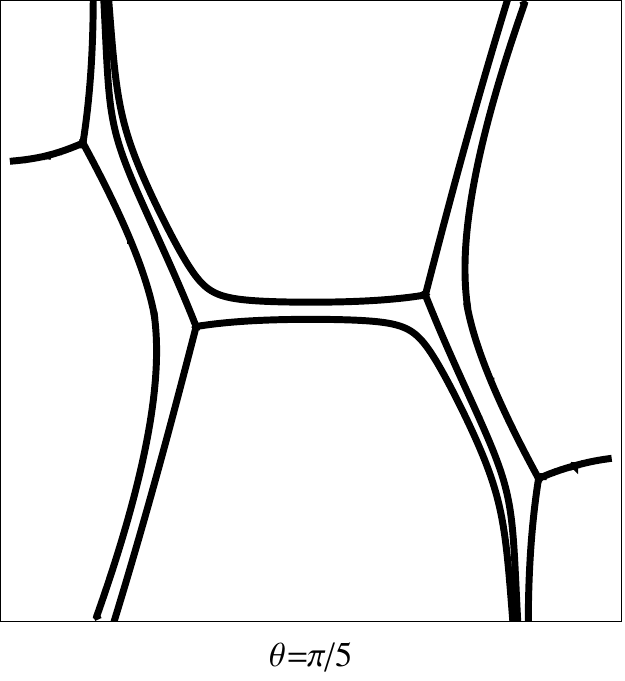}
\end{center}
\caption{Anti-Stokes lines for the quartic quantum-mechanical potential \eqref{appBquarticQMpot}. Wall-crossing is very clear across $\theta = 0$, when the topology of the network effectively changes.}
\label{fig:WKB-anti-Stokes-app}
\end{figure}

Let us illustrate how this WKB matching might work by plotting anti-Stokes lines \eqref{WKBantistokes-app} for the quartic potential (which may then be compared to figure~\ref{steepest} in section~\ref{sec:quartic}). Recall that in our conventions this potential is \eqref{quarticpotential},
\be
\label{appBquarticQMpot}
V(z) = \frac{1}{2} z^2 - \frac{\lambda}{24} z^4.
\ee
\noindent
Then, at fixed energy $E$, the four turning points are $E=V(q_0)$, \textit{i.e.},
\be
q_0^4 - \frac{12}{\lambda} q_0^2 + \frac{24}{\lambda} E = 0 \qquad \Leftrightarrow \qquad q_0^2 = \frac{2\sqrt{3}}{\lambda} \left( \sqrt{3} \pm \sqrt{3 - 2 \lambda E}\right)
\ee
\noindent
The interesting aspect here is that, at fixed energy $E$, these turning points (and, subsequently, the structure of anti-Stokes lines) still depend on the quartic coupling $\lambda$ (a phenomenon which was well addressed in the main text). As such, the whole structure of anti-Stokes lines changes as this parameter is changed. This is illustrated in figure~\ref{fig:WKB-anti-Stokes-app} for our quartic potential. In particular, it is quite clear how at $\theta = \arg \lambda = 0$ the \textit{topology} of the network ``jumps'', a phenomenon sometimes going by the names of ``wall crossing'' or ``parametric Stokes phenomena''. See, \textit{e.g.}, \cite{kt05, gmn09} for excellent overviews of these WKB spectral networks.

Let us finally mention that these techniques are not exclusive to quantum mechanics, and have found applications in many other problems. For instance, they have even been applied in the study of black hole physics (see, \textit{e.g.}, \cite{mn03, n03, cns04, ns04, hns07}) and, in fact, it would be very interesting to further apply resurgent analysis to this large class of gravitational problems.

\section{On the Nature of Strong-Coupling Expansions}\label{app:strongcoupling}

Our main concern in these lectures dealt with making perturbation theory complete, in the sense of extracting exact information out of initially divergent or asymptotic perturbative expansions. But, of course, not all perturbative expansions need to be divergent or asymptotic. One very interesting feature of many quantum theoretic problems is that while, on the one hand, their \textit{weak}-coupling perturbative expansions are indeed asymptotic, one finds that instead, on the other hand, their \textit{strong}-coupling perturbative expansions lead to \textit{convergent} expansions. Note that this is not a solution to the problems of perturbation theory, as it is not common to have access to such strong-coupling expansions. But, when such a regime is indeed available, this is a rather interesting phenomena which we shall now illustrate.

Let us consider our original example of the ``quartic partition function'' from section~\ref{sec:quartic}, as written in \eqref{quarticintegralX}. In the main text we have addressed the calculation of this partition function using a weak-coupling (small $x\ll1$) perturbative analysis. The (opposite) strong-coupling regime corresponds to large $x\gg1$ and proceeds as follows\footnote{This is rather standard in the literature. See, \textit{e.g.}, \cite{chps16} for a recent very similar analysis.}. Let us first change variables in the integrand by a simple rescaling $z \to y=\frac{z}{x^{1/4}}$, and let us momentarily work with the now more convenient coupling $\mathsf{x}=x^{1/4}$. In this case, the partition function becomes (compare with \eqref{quarticintegralX})
\begin{equation}
\label{quarticintegralX-app}
Z (\mathsf{x}) = \frac{\sqrt{\hbar}}{2\pi\mathsf{x}} \int_{\Gamma} \rmd y\, \exp \left( - \frac{1}{2\mathsf{x}^2} y^2 + \frac{1}{24} y^4 \right).
\end{equation}
\noindent
In the limit where $\mathsf{x} \to +\infty$, one can use the (convergent) power-series expansion of the exponential to write
\begin{equation}
Z_{\infty} (\mathsf{x}) = \frac{\sqrt{\hbar}}{2\pi\mathsf{x}}\, \sum_{n=0}^{+\infty} \frac{(-1)^n}{2^n n!}\, \frac{1}{\mathsf{x}^{2n}} \int_{\Gamma} \rmd y\, y^{2n}\, \rme^{\frac{1}{24}y^{4}}.
\end{equation}
\noindent
Of course that the path of integration $\Gamma$ still needs to be chosen such that all integrals above converge. This simply amounts to selecting $\Gamma = \left( - \infty\, \rme^{\rmi\frac{\pi}{4}}, + \infty\, \rme^{\rmi \frac{\pi}{4}} \right)$, in which case we can evaluate these integrals and obtain
\begin{equation}
Z_{\infty} (\mathsf{x}) = \frac{\sqrt{\hbar}}{2\pi\mathsf{x}} \left(-\frac{3}{2}\right)^{1/4}\, \sum_{n=0}^{+\infty} \frac{\Gamma \left( \frac{2n+1}{4} \right)}{n!} \left( - \frac{\left( -24 \right)^{1/2}}{2\mathsf{x}^{2}} \right)^{n}.
\end{equation}
\noindent
The main point now is that this is a \textit{convergent} expansion for $\mathsf{x}\gg1$. In fact, one may write
\begin{equation}
\frac{Z_{\infty} (\mathsf{x})}{\frac{\sqrt{\hbar}}{2\pi\mathsf{x}} \left( - \frac{3}{2} \right)^{1/4}} = \sum_{n=0}^{+\infty} \frac{\Gamma \left( \frac{2n+1}{4} \right)}{n!} \left(-w\right)^n \equiv \sum_{n=0}^{+\infty} \zeta_n\, w^n,
\end{equation}
\noindent
with $w = \frac{\left(-24\right)^{1/2}}{2\mathsf{x}^2}$, where the radius of convergence of this power series around $w=0$ ($\mathsf{x}=\infty$) is actually \textit{infinite}, \textit{i.e.}, the above resulting function has no singularities on the complex plane except perhaps at $w=\infty$ ($\mathsf{x}=0$). These results are not a big surprise, as the sum of this series is actually well-known (see, \textit{e.g.}, \cite{olbc10}):
\begin{equation}
\sum_{n=0}^{+\infty} \zeta_n\, w^n = \rme^{\frac{w^2}{8}}\, \sqrt{w}\, K_{\frac{1}{4}} \left( \frac{w^2}{8} \right),
\end{equation}
\noindent
where $K_{\nu} (z)$ is a modified Bessel function of the second kind. Going back to the original variable $x$, this implies that\footnote{As already briefly mentioned in the discussion surrounding \eqref{bessel-mentioned}.}
\begin{eqnarray}
Z_{\infty} (x) &=& \rme^{\rmi\frac{\pi}{2}} \sqrt{\frac{\hbar}{2\pi}}\, \sqrt{\frac{3}{2\pi x}}\, \rme^{-\frac{3}{4x}}\, K_{\frac{1}{4}} \left(\rme^{\rmi\pi} \frac{3}{4x} \right) = \\
&=& \rme^{\rmi\frac{\pi}{4}} \sqrt{\frac{\hbar}{2}}\, \sqrt{\frac{3}{4x}}\, \rme^{-\frac{3}{4x}}\, \left( I_{-\frac{1}{4}} \left( \frac{3}{4x} \right) - \rmi\, I_{\frac{1}{4}} \left( \frac{3}{4x} \right) \right).
\label{eq:Zinf-as-S0p}
\end{eqnarray}
\noindent
It is interesting to see how exponentials with the ``instanton action'' naturally appear in our end result even though there is no notion of a Borel plane in this calculation (recall that, in the appropriate regime, there is still an exponential piece inside the Bessel function which, together with the exponential in the expression above, will give rise to the instanton action \eqref{instactions}).

In summary, we have managed to compute the partition function \eqref{quarticintegralX-app} by \textit{summing} the \textit{convergent} expansion at strong coupling $x\gg1$. But a couple of remarks on how to arrive at this final result are still in order. First, note that the function $K_{1/4} (w)$ has two branch points, located at $w=0$ and $w=\infty$. It is the factor $\sqrt{w}$ multiplying the Bessel function which guarantees a convergent behaviour at $w=0$, as $K_{1/4}(w)$ behaves like $\log w$ close to the origin. Finally, to obtain the end result above, we used the familiar connection formula/analytic continuation for the modified Bessel functions\footnote{In particular, for $\nu=\frac{1}{4}$ we have: 
\begin{equation}
K_{\frac{1}{4}} \left( \rme^{\rmi\pi m} \cdot z \right) = \frac{\pi}{\sqrt{2}}\, \rme^{ - \rmi m \frac{\pi}{4}} \left( I_{-\frac{1}{4}} (z) - \rme^{\rmi m \frac{\pi}{2}}\, I_{\frac{1}{4}} (z) \right).
\end{equation}
} (see, \textit{e.g.}, \cite{olbc10}):
\begin{eqnarray}
K_{\nu} (z) &=& \frac{\pi}{2 \sin (\nu\pi)}\, \Big(\, I_{-\nu} (z) - I_{\nu} (z)\, \Big), \\
K_{\nu}\left( \rme^{\rmi\pi m} \cdot z \right) &=& \rme^{-\rmi\nu m\pi} K_{\nu} (z) - \rmi\pi\, \frac{\sin (m\nu\pi)}{\sin (\nu\pi)}\, I_{\nu} (z), \qquad m\in\mathbb{Z}.
\end{eqnarray}

\begin{figure}[t!]
\begin{center}
\includegraphics[height=7cm]{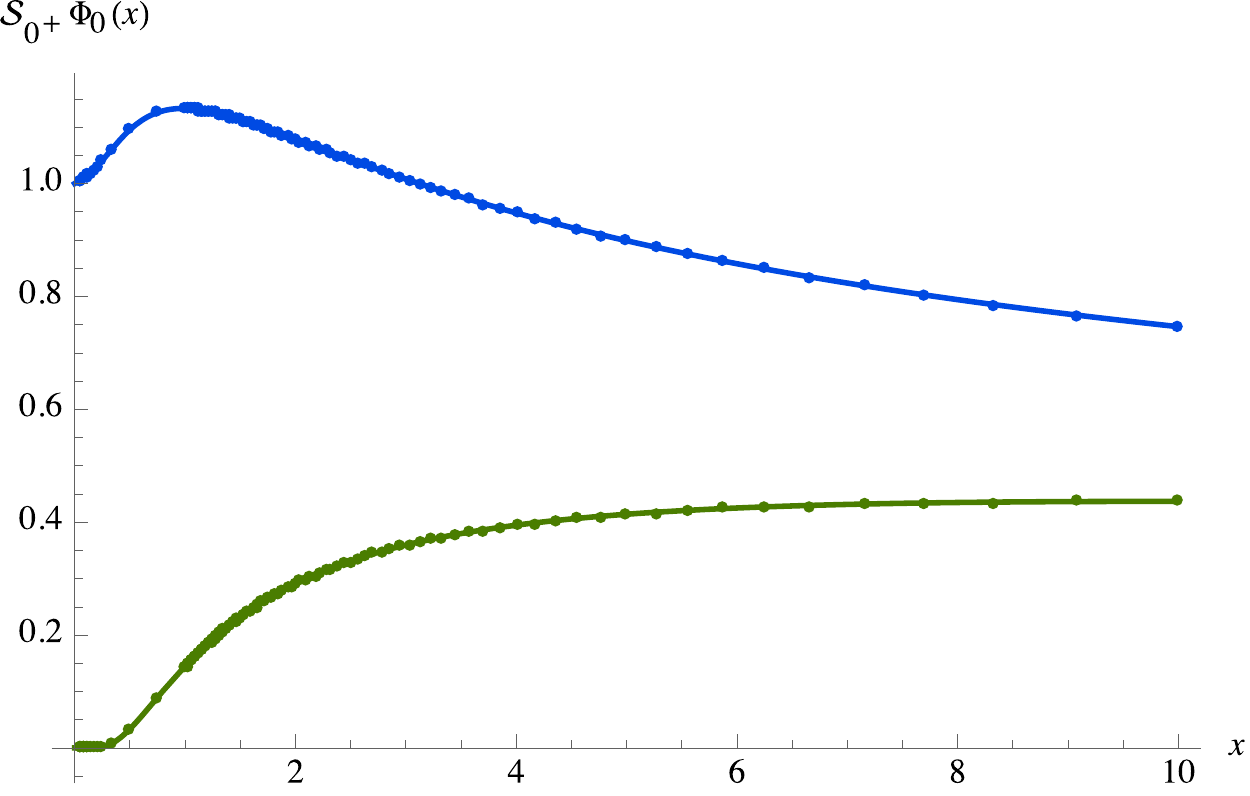}
\end{center}
\caption{Strong-weak coupling interpolation out from the weak-coupling asymptotic series $\Phi_{0}$, herein Borel resummed along the $\theta=0^+$ direction. The numerical Borel resummation is depicted by the blue (real part) and green dots (imaginary part). As is clearly seen in the plot, this matches to great accuracy against the corresponding analytic value (solid lines) determined from the strong-coupling result $Z_{\infty}(x)$.
}
\label{fig:Strong-weak-interp}
\end{figure}

As a consistency check, it would be interesting to compare this result against the one that follows from the transseries solution to the quartic partition-function \eqref{quartictransseries} addressed in the main text (for some fixed values of the transseries parameters, $\sigma_{0}$, $\sigma_{1}$). This is particularly interesting in the light of the fact that while the above results were obtained via strong-coupling \textit{convergent} power-series expansions, the ones in the main text were obtained by working with weak-coupling \textit{asymptotic} expansions. If we perform this comparison along the positive real axis $\arg x=0$, we must first recall that this line is a Stokes line. Nevertheless, performing the Borel resummation above the real axis (in the direction $\theta=0^{+}$), or, alternatively, choosing the steepest-descent contour passing through the origin and with coupling with a small, positive imaginary part, we can easily obtain the result \eqref{eq:Zinf-as-S0p}. Thus, for $x = \left|x\right| \rme^{\rmi\theta}$, with $0<\theta<\pi$, one finds
\noindent
\begin{equation}
Z_{\infty} (x) = \CS_\theta \Phi_0 (x).
\end{equation}
\noindent
In the specified angular region this result is valid for \textit{all} values of $\left|x\right|$, \textit{i.e.}, the function $Z_{\infty} (x)$ may be analytically continued throughout the entire upper-half plane, always matching against the Borel resummation $\CS_\theta \Phi_0 (x)$. We have checked this match numerically by choosing values for the coupling ranging from small $\left|x\right|\ll1$ up to large $\left|x\right|\apprge10$. This is actually rather simple to implement numerically in the present case, as there is no need for Pad\'e approximants in order to perform a numerical Borel resummation. In fact, we know the \textit{exact} Borel transform associated to the perturbative sector, earlier computed in \eqref{borel-quartic-int}. Focusing on the ray just above the real axis, with $\theta=0^+$, we have computed the corresponding Borel resummation for a sample of values $\left|x\right| = \frac{1}{19}, \cdots, 10$. This is depicted by the dots in figure~\ref{fig:Strong-weak-interp} (where the real part of the resummation is plotted in blue, while the imaginary part is plotted in green). These values are then compared against the (blue and green) solid lines, which correspond to the analytical value of $Z_{\infty} (x)$. The resulting perfect match\footnote{Note that were we to plot the difference between this Borel resummation and the corresponding value of $Z_{\infty} (x)$, we would obtain an identically vanishing result up to the error of the numerical calculation. It so happens that in this case this error is extremely small, of order $\sim 10^{-200}$. In general, for nonlinear problems, this will not be the case as such a difference can then become of the order of higher, exponentially suppressed contributions.} is very clear in the figure. Effectively, this exercise also shows how to perform a strong-weak coupling (numerical) interpolation, along a particular direction $\theta$, starting \textit{only} from data associated to the \textit{asymptotic} expansion at \textit{weak} coupling. 

\begin{figure}[t!]
\begin{center}
\includegraphics[height=7.5cm]{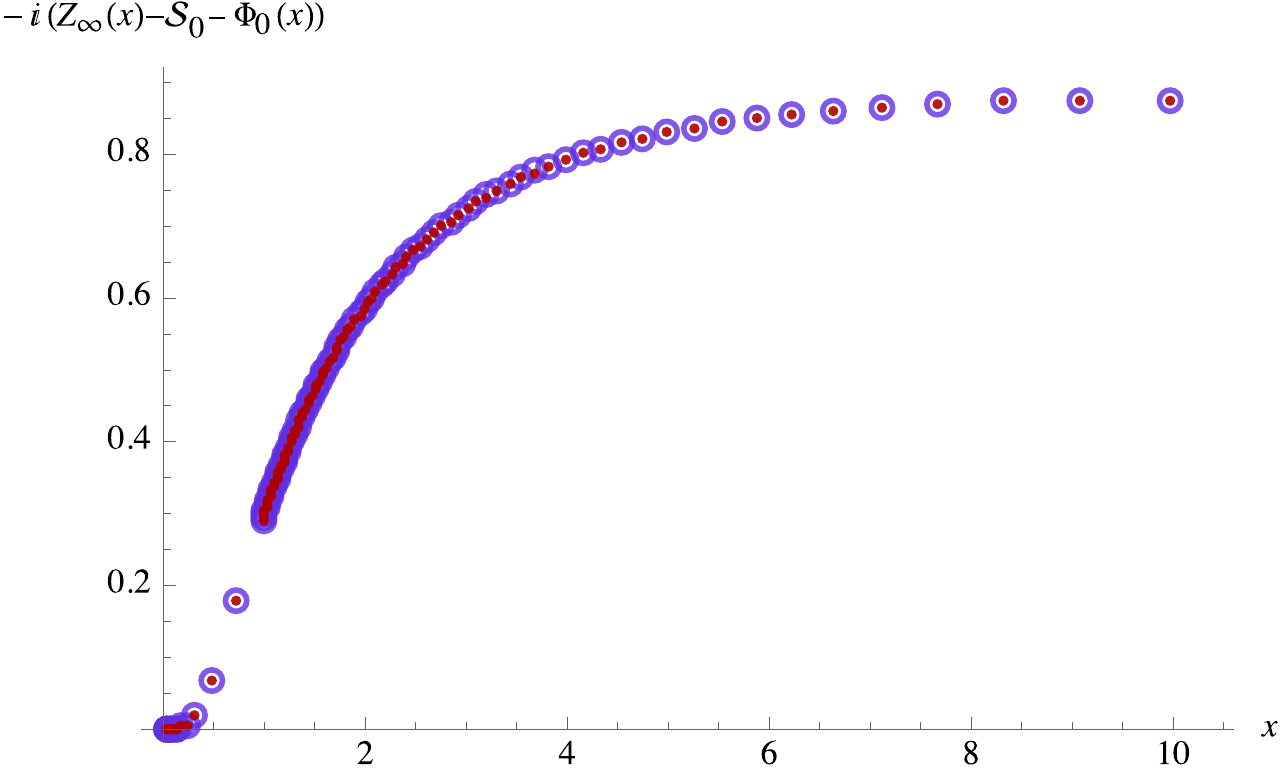}
\end{center}
\caption{Comparison of the resummation $-\rmi \left( Z_{\infty} (x) - \CS_{0^-} \Phi_0 (x) \right)$, Borel resummed along $\theta=0^-$, against the Borel resummation $2\rmi\, \CS_0 \Phi_1 (x)$ along $\theta=0$. The former is plotted as the large purple circles, while the latter is plotted as the (inner) small red dots. As usual, they show perfect agreement.
}
\label{fig:Strong-weak-interp-comp}
\end{figure}

One may next ask what will happen were we to consider $\theta$ outside of the range $(0,\pi)$? The answer is again quite straightforward: all one has to do is to properly take Stokes phenomena into account on the transseries side, and then compare against the strong-coupling result $Z_{\infty} (x)$. In order to illustrate how this works, consider for example the ray just below the positive real line, $\theta=0^-$. Computing the Borel resummations $\CS_{0^-} \Phi_0 (x)$ and $\CS_0 \Phi_1 (x)$ is again quite easy to implement (recall that we have previously computed the \textit{exact} Borel transforms of both perturbative and one-instanton sectors, in equations \eqref{borel-quartic-int} and \eqref{borel-quartic-int-inst}), and we have done so for the same sample of values of the coupling as above, $\left|x\right| = \frac{1}{19}, \cdots, 10$. One finds
\begin{equation}
Z_{\infty} (x) = \CS_{0^-} \Phi_0 (x) - 2 \CS_0 \Phi_1 (x),
\end{equation}
\noindent
as expected from the Stokes transition at $\theta=0$. This agreement is shown in figure~\ref{fig:Strong-weak-interp-comp}, where we have plotted\footnote{Once again, given the simplicity of this linear problem, the difference between the analytic result and the resummed one will vanish up to the extremely small error of the numerical calculation, of order $\sim 10^{-200}$.} both numerical resummations, $-\rmi \left( Z_{\infty} (x) - \CS_{0^-} \Phi_0 (x) \right)$ and $2\rmi\, \CS_0 \Phi_1 (x)$, for different values of the coupling in the ranges $0<x\ll1$ and $x\apprge1$. The match is spot on.

In this appendix we discussed the strong-coupling analysis for the partition function of section~\ref{sec:quartic}. Doing the same analysis for the free energy, \textit{i.e.}, for a nonlinear problem, would be much more involved: in nonlinear problems it is much harder to have direct access to the strong-coupling region. We have also shown how resummation may interpolate from weak to strong coupling, and such resurgence methods have recently been used in many different problems in order to explicitly connect weak to strong coupling regimes; see, \textit{e.g.}, \cite{m08, gmz14, cku14, csv15, a15, cms16, cms17}.

\section{Recurrence Relations for Free Energies}\label{app:recursive-free-en}

In the main text we dealt with a few nonlinear differential equations, and transseries \textit{ans\"atze} were used in order to solve them recursively, \textit{i.e.}, incorporating both (perturbative) analytical and (nonperturbative) non-analytical contributions. This led to intricate recurrence equations for the multi-loop multi-instanton coefficients in the asymptotic expansions building up the transseries, and we present such recursive relations in this appendix. The set-up was already discussed in the main text, herein we shall restrict ourselves to presenting the recurrence relations.

\subsection*{Recurrence Relations for the Quartic Free Energies}

The one-parameter transseries \eqref{one-param-transseries},
\be
F (x, \sigma) = \sum_{\ell=0}^{+\infty} \sigma^\ell\, \rme^{-\frac{\ell A}{x}}\, \Phi_{\ell} (x),
\ee
\noindent
is built out of asymptotic expansions \eqref{one-param-inst-asympt},
\be
\Phi_{\ell} (x) \simeq \sum_{n=0}^{+\infty} F_n^{(\ell)}\, x^n,
\ee
\noindent
with multi-loop multi-instanton coefficients $F_n^{(\ell)}$. If one is to plug-in this transseries \textit{ansatz} \eqref{one-param-transseries} into the quartic nonlinear differential equation \eqref{quarticNLODE}, while keeping track of equal powers of  $\rme^{-\frac{n A}{x}}$ and $x^{k}$ (recall that $A=\frac{3}{2}$), one immediately obtains a set of nonlinear, coupled recursive relations for these multi-loop multi-instanton coefficients $F_n^{(\ell)}$. The nonlinearity of the original equation implies that there will be an infinite number of both multi-loop \textit{and} multi-instanton coefficients.

For $n=0$ and $k \ge 1$,
\begin{eqnarray}
F_1^{(0)} &=& \frac{1}{8}, \\
F_2^{(0)} &=& \frac{1}{12}, \\
F_{k+1}^{(0)} &=& \frac{2}{3} \left( k\, F_k^{(0)} + \frac{1}{k+1} \sum_{\ell=1}^{k-1} \left( k-\ell \right) \ell\, F_{k-\ell}^{(0)}\, F_{\ell}^{(0)} \right), \qquad k\ge2.
\end{eqnarray}
\noindent
For $n=1$ and $k \ge 1$,
\begin{eqnarray}
F_1^{(1)} &=& -\frac{1}{4} F_{0}^{(1)}, \\
F_2^{(1)} &=& \frac{1}{32} F_{0}^{(1)}, \\
F_{k+1}^{(1)} &=& -\frac{1}{\left(k+1\right)A} \left( k \left(k+1\right) F_k^{(1)} + 2 \sum_{\ell=1}^{k-1} \left(k-\ell\right) \ell\, F_{k-\ell}^{(1)}\, F_{\ell}^{(0)} 
+ 2 A \sum_{\ell=0}^{k} \left(k-\ell+1\right) F_{k-\ell+1}^{(0)}\, F_{\ell}^{(1)} \right). \nonumber \\
&&
\end{eqnarray}
\noindent
For $n \ge 2$ and $k\ge0$,
\begin{eqnarray}
F_{k}^{(n)} &=& -\frac{1}{n \left(n-1\right)} \sum_{m=1}^{n-1} \left(n-m\right)m\,  \sum_{\ell=0}^{k} F_{k-\ell}^{(n-m)}\, F_{\ell}^{(m)}, \qquad k=0,1, \\
F_{k}^{(n)} &=& - \frac{1}{n \left(n-1\right) A^{2}} \left( A^{2} \sum_{m=1}^{n-1} \left(n-m\right) m\, \sum_{\ell=0}^{k} F_{k-\ell}^{(n-m)}\, F_{\ell}^{(m)} + \left(2n-1\right) \left(k-1\right) A\, F_{k-1}^{(n)} + \right. \nonumber \\
&& 
+ \left(k-2\right) \left(k-1\right) F_{k-2}^{(n)} + 2 A \sum_{m=1}^{n} \sum_{\ell=0}^{k-2} m \left(k-\ell-1\right) F_{k-\ell-1}^{(n-m)}\, F_{\ell}^{(m)} + \nonumber \\
&&
\left. + \sum_{m=0}^{n} \sum_{\ell=1}^{k-3} \ell \left(k-\ell-2\right) F_{k-\ell-2}^{(n-m)}\, F_{\ell}^{(m)} \right), \qquad k\ge2.
\end{eqnarray}
\noindent
Do note that the coefficients $F_{0}^{(0)}$ and $F_{0}^{(1)}$ are not fixed by any recurrence relations; they are unknown integration constants. Further recall that throughout this paper, we are using conventions where if the upper limit of a sum is less than its lower limit, that sum has no contribution.

\subsection*{Recurrence Relations for the Elliptic Free Energies}

For the case of the elliptic nonlinear differential equation \eqref{eq:nl_ell_ode} one deals with a ``two-and-a-half'' transseries \eqref{eq:nl_ell_ansatz}. This may be turned into an effective two-parameter transseries by simply incorporating the sectors $\widetilde{\Phi}_{(0,0)} \equiv \Phi_{(1,0,0)}$ and $\Phi_{(0,0,0)}$ into a new, perturbative sector which, with some abuse of notation, we shall denote in this appendix by $\Phi_{(0,0)}$. As such, this effective two-parameter transseries is written as
\be
F(x,\boldsymbol{\sigma}) = \sum_{\boldsymbol{n} \in \mathbb{N}^2_{0}}\, \boldsymbol{\sigma}^{\boldsymbol{n}}\, \rme^{-\frac{\boldsymbol{n} \cdot \boldsymbol{A}}{x}}\, \Phi_{\boldsymbol{n}} (x),
\ee
\noindent
where we are using the usual notation from sections~\ref{sec:physics} and \ref{sec:elliptic}, and it is built out of the asymptotic expansions 
\begin{eqnarray}
\Phi_{(0,0)}(x) &=& \sigma_0\, \Phi_{(1,0,0)} + \Phi_{(0,0,0)} \simeq \sum_{k=0}^{+\infty} F_k^{(0,0)} x^k, \\
\Phi_{\boldsymbol{n}}(x) &=& \Phi_{(0,\boldsymbol{n})} \simeq \sum_{k=0}^{+\infty} F_k^{(\boldsymbol{n})} x^k, \qquad \boldsymbol{n}\ne(0,0).
\end{eqnarray}
\noindent
Just as in the quartic case, if one is to plug-in the above transseries \textit{ansatz} into the elliptic nonlinear differential equation \eqref{eq:nl_ell_ode}, while keeping track of equal powers of $\rme^{-\frac{\boldsymbol{n} \cdot \boldsymbol{A}}{x}}$ and $x^k$ (recall that $\boldsymbol{A} = \left( \frac{1}{1-m}, -\frac{1}{m} \right)$), one immediately obtains a set of nonlinear, coupled recursive relations for the multi-loop multi-instanton coefficients $F_k^{(\boldsymbol{n})}$. The nonlinearity of the original equation implies that there will be an infinite number of both multi-loop \textit{and} multi-instanton coefficients.

Let us start by analyzing the non-resonant case. Define the useful combination:
\begin{equation}
c^{(\boldsymbol{n})}_{k} \equiv \left( \boldsymbol{n} \cdot \boldsymbol{A} \right) F^{(\boldsymbol{n})}_k + \left(k-1\right) F^{(\boldsymbol{n})}_{k-1},
\end{equation}
\noindent
where we are assuming, as usual, that $F^{(\boldsymbol{n})}_k=0$ when $k<0$. Then, these recursion relations may be written in a more-or-less compact form as
\begin{eqnarray}
0 &=& -\left( m-m^\prime \right) \delta_{k,2}\, \delta_{\boldsymbol{n},\boldsymbol{0}} + 3\, m\, m^\prime\, \delta_{k,3}\, \delta_{\boldsymbol{n},\boldsymbol{0}} + 4 \left( m\, m^\prime \left( \boldsymbol{n} \cdot \boldsymbol{A} \right)^2 - \left( m-m^\prime \right) \left( \boldsymbol{n} \cdot \boldsymbol{A} \right) -1 \right) c^{(\boldsymbol{n})}_k + \nonumber \\
&&
+ 4 \left( k-1 \right) \left( 2\, m\, m^\prime \left( \boldsymbol{n} \cdot \boldsymbol{A} \right) - \left( m-m^\prime \right) \right) c^{(\boldsymbol{n})}_{k-1} + m\, m^\prime \left( 3 + 4 \left(k-1\right) \left(k-2\right) \right) c^{(\boldsymbol{n})}_{k-2} - \nonumber \\
&&
- 4 \left( m-m^\prime \right) \sum_{n_1^\prime=0}^{n_1}\, \sum_{n_2^\prime=0}^{n_2}\, \sum_{k^\prime=0}^{k}\, c^{(\boldsymbol{n} - \boldsymbol{n}^\prime)}_{k-k^\prime}\, c^{(\boldsymbol{n}^\prime)}_{k^\prime} + \nonumber \\
&&
+ 12\, m\, m^\prime\, \sum_{n_1^\prime=0}^{n_1}\, \sum_{n_2^\prime=0}^{n_2}\, \sum_{k^\prime=0}^{k} \left( \left( \boldsymbol{n}^\prime \cdot \boldsymbol{A} \right) c^{(\boldsymbol{n}^\prime)}_{k^\prime} + \left( k^\prime-1 \right) c^{(\boldsymbol{n}^\prime)}_{k^\prime-1} \right) c^{(\boldsymbol{n}-\boldsymbol{n}^\prime)}_{k-k^\prime} + \nonumber \\
&&
+ 4\, m\, m^\prime \left( \sum_{n_1^\prime=0}^{n_1}\, \sum_{n_1^{\prime\prime}=0}^{n_1^\prime}\, \sum_{n_2^\prime=0}^{n_2}\, \sum_{n_2^{\prime\prime}=0}^{n_2^\prime}\, \sum_{k^\prime=0}^k\, \sum_{k^{\prime\prime}=0}^{k^\prime}\, c^{(\boldsymbol{n}-\boldsymbol{n}^\prime)}_{k-k^\prime}\, c^{(\boldsymbol{n}^\prime-\boldsymbol{n}^{\prime\prime})}_{k^\prime-k^{\prime\prime}}\, c^{(\boldsymbol{n}^{\prime\prime})}_{k^{\prime\prime}} \right),
\label{eq:ell_nl_recursion}
\end{eqnarray}
\noindent
where we set $k\ge 0$ and used $\boldsymbol{n}-\boldsymbol{n}^{\prime} = \left( n_1-n_1^{\prime}, n_2-n_2^{\prime} \right)$. Note that for $\boldsymbol{n} = \boldsymbol{0}$ the coefficients are $c^{(\boldsymbol{0})}_k=0$ for $k=0,1$, which means that $F^{(\boldsymbol{0})}_0$ remains indeterminate from the recursion. Furthermore, the coefficients $c^{(1,0)}_0$ and $c^{(0,1)}_0$ also cannot be determined from the recursion relations.

In the resonant case, there are some extra subtleties. As discussed at length in section~\ref{sec:elliptic}, resonance implies that different sectors $\boldsymbol{n}$ will end-up leading to the same exponential factor $\rme^{-\frac{\boldsymbol{n} \cdot  \boldsymbol{A}}{x}}$. For instance, when $m=1/3$ the exponential depends only on an element $\boldsymbol{\rho} \in \BN^2/\ker \mathfrak{P}$. In this case, defining $\boldsymbol{n} = \left(n_1, n_2\right) \equiv \boldsymbol{\rho}_\alpha + \kappa\, \nres$, where $\nres = \left(2,1\right)$, will result in the same exponential factor for any $\kappa \in \BN$. This sort of effects may naturally change the above recursion relations. One example where this phenomenon of resonance is particularly visible is when one tries to write the recursion relations for the sector $\boldsymbol{n} = \left(2,2\right) \equiv \boldsymbol{\rho}_\alpha + \nres$ (or any other sectors with $\kappa \neq 0$). Let us illustrate what happens by considering the equations for $c^{(2,2)}_0$ and $c^{(2,2)}_1$, as a function of the parameter $m$. They are:
\begin{eqnarray}
0 &=& 4 \left( 9 \left(m-1\right) m + 2 \right) c^{(2,2)}_0 + 12 \left(2m-1\right) \left(3m-2\right) \left(3m-1\right), \\
0 &=& 4 \left( 9 \left(m-1\right) m + 2 \right) c^{(2,2)}_1 + 18 \left( 9 \left(m-1\right) m + 2 \right) \left(2m-1\right)^2.
\end{eqnarray}
\noindent
Setting $m=1/3$ in both these equations makes their right-hand side identically vanish, and we cannot solve for the unknown coefficients. An analogous situation may be found in the realm of Painlev\'e equations, \textit{e.g.}, the analysis of resonance for the Painlev\'e~I \cite{gikm10, asv11} and Painlev\'e~II equations \cite{sv13}. In those cases, resonance implied that the transseries itself needed to be modified in order to include logarithmic sectors (which lifted the indeterminacy created by the resonant behaviour). In the present case this is not necessary: one can study solutions to the elliptic nonlinear ODE as functions\footnote{This is \textit{not} an option in the  Painlev\'e cases, where resonance is not ``tunable'' as in the present example.} of the free parameter $m$. In this way, the solutions to the above (apparently indetermined) equations are simply $c^{(2,2)}_0 = - 3 \left( m-m^\prime \right)$ and $c^{(2,2)}_1 = - \frac{9}{2} \left( m-m^\prime \right)^2$, both of which have well-defined limits when $m=1/3$. In this way it is possible to determine \textit{every} resonant sector, simply from the recursion relations for arbitrary $m$.

\section{Further Details on Borel Transforms}\label{app:borel}

This appendix presents a formal derivation of the relation between Borel transforms within the same resurgence class; equation \eqref{eq:finding-Borel-rep-from-other} in the main text. Let us consider a family of asymptotic series in the same resurgence class---with the class specified by the constants $\alpha$ and $\beta$ (different sets yielding different classes) and the family $\{ \Phi_{[\gamma]} \}$ parameterized by $\gamma$---which is given by
\be
\label{eq:Family-res-funct-app}
\Phi_{[\gamma]} (\lambda) = \lambda^\gamma\, \Phi_{[0]} (\lambda) \simeq \sum_{k=0}^{+\infty} \Phi_k\, \lambda^{\alpha k + \beta + \gamma},
\ee
\noindent
and where we are taking $\Phi_{[0]}$ to be the standard representative. This essentially means that the coefficients $\Phi_k$ have the precise (large-order) factorial growth of
\be
\Phi_k \sim \left( \alpha k + \beta - 1 \right)!,
\ee
\noindent
and that the Borel transform $\CB [\Phi_{[0]}] (s)$ will have the singular structure \eqref{simpleBorelsingularities}. All other Borel transforms, associated with each other  element in this family of functions, will be defined as
\be
\CB [ \widehat{\Phi}_{[\gamma]} ] (s) = \sum_{k=n_0(\gamma)}^{+\infty} \Phi_k\, \frac{s^{\alpha k + \beta + \gamma - 1}}{\Gamma \left( \alpha k + \beta + \gamma \right)}.
\ee
\noindent
Let us explain the notation. In order to obtain the Borel transform of the generic element $\Phi_{[\gamma]} (\lambda)$, one first needs to remove the first $k<n_{0}(\gamma)$ terms within its asymptotic expansion, in such a way that $\alpha\, n_{0}(\gamma)+\beta+\gamma>0$ and the Borel transform is thus well-defined. This ``truncated'' asymptotic series is then denoted by $\widehat{\Phi}_{[\gamma]} (\lambda)$. As to the terms which get removed, they are handled separately, and can be added back-in at the very end, once we have Borel resummed and obtained $\CS_\theta \widehat{\Phi}_{[\gamma]}$ following the standard procedures. In order to simplify our present discussion, we will not consider these terms as they are always straightforward to add at any stage.

For integer values of $\gamma$, the straightforward relation (with $n,m \in \BN^+$)
\be
\frac{s^n}{\Gamma \left( n+1 \right)} = \frac{\rmd^m}{\rmd s^m}\, \frac{s^{n+m}}{\Gamma \left( n+m+1 \right)}
\ee
\noindent
directly implies the relation between Borel transforms\footnote{As an historical aside, note that this sort of relations was probably first noticed in \cite{bpv78b}, but, unfortunately, the role of the standard representative---with its (simple) logarithmic ramified structure, and where resurgent analysis becomes most straightforward---was not emphasized enough to spark subsequent further analysis.}
\be
\label{eq:relation-between-borels-der}
\CB [ \widehat{\Phi}_{[\gamma']} ] (s) = D_s^{\gamma-\gamma'} \CB [ \widehat{\Phi}_{[\gamma]} ] (s).
\ee
\noindent
Assuming that $\gamma-\gamma'$ is an integer and that $\gamma>\gamma'$, the operator $D_s^n$ is herein defined as the $n$th derivative with respect to $s$. The relation \eqref{eq:relation-between-borels-der} also holds if $\gamma<\gamma'$, in which case one simply defines $D_s^n$ as the $n$th \textit{primitive}\footnote{Each primitive yields an integration constant, and upon multiple primitivation one obtains a polynomial in $s$ with maximum degree $\gamma'-\gamma$. This polynomial is directly related to the first $n_0(\gamma)$ coefficients which were removed from the original asymptotic series $\Phi_{[\gamma]} (\lambda)$ in order to obtain the ``Borel transformable'' asymptotic series $\widehat{\Phi}_{[\gamma]} (\lambda)$.} with respect to $s$. If $\gamma=\gamma'$ then $D_s^0$ is simply the identity.

The assumption that $\gamma-\gamma'$ is an integer may be lifted, allowing \eqref{eq:relation-between-borels-der} to be valid for a \textit{continuous} parameter $\gamma \in \BR$. Let us show how this is done when\footnote{The case where $\gamma<\gamma'$ follows in complete analogy by simply inverting \eqref{eq:relation-between-borels-der} and having the operator $D_s^{\gamma'-\gamma}$, which is now a \textit{derivative}, acting on $\CB [ \widehat{\Phi}_{[\gamma']} ] (s)$.} $\gamma>\gamma'$. In this case we have to deal with a \textit{fractional derivative}, where $D_s^\alpha$ is defined, when $0<\alpha<1$, as (see, \textit{e.g.}, \cite{lot76})
\begin{equation}
D_s^\alpha f(s) = \frac{1}{\Gamma \left(1-\alpha\right)}\, \frac{\rmd}{\rmd s} \int_0^s \rmd t\, \frac{f(t)}{\left( s-t \right)^{\alpha}}.
\end{equation}
\noindent
This definition naturally extends to any real order $\gamma$ by setting $\gamma = \alpha + m$, with $\gamma \in \BR$, $m \in \BN_+$, and $\alpha$ as above, in which case
\be
D_s^\gamma f(s) = D_s^\alpha D_s^m f(s) = D_s^\alpha\, \frac{\rmd^m f}{\rmd s^m} (s).
\ee
\noindent
Applying the fractional derivative to each term in the definition of a Borel transform is straightforward; one only needs to know that, for $f(s) = s^\rho$, $\rho>-1$, one has
\begin{equation}
D_s^\alpha s^\rho = \frac{\Gamma \left(\rho+1\right)}{\Gamma \left(\rho+1-\alpha\right)}\, s^{\rho-\alpha}.
\end{equation}
\noindent
The action of the fractional derivative $D_s^{\gamma-\gamma'}$ on the Borel transform $\CB [ \widehat{\Phi}_{[\gamma]} ] (s)$ in \eqref{eq:relation-between-borels-der} can then be easily computed making use of the above expression order-by-order in the series definition of $\CB [ \widehat{\Phi}_{[\gamma]} ] (s)$. One obtains
\be
D_s^{\gamma-\gamma'} \CB [ \widehat{\Phi}_{[\gamma]} ] (s) = \sum_{k=n_0(\gamma)}^{+\infty} \Phi_k\, \frac{s^{\alpha k + \beta + \gamma' - 1}}{\Gamma \left( \alpha k + \beta + \gamma' \right)} \approx \CB [ \widehat{\Phi}_{[\gamma']} ] (s).
\ee
\noindent
We have used the symbol $\approx$ as a reminder that this equality only holds up to the removal/addition of $\gamma-\gamma'$ initial terms in the expansions (residual coefficients \textit{et al}).

This shows how to relate Borel transforms of asymptotic series within the same resurgence class. Note that the derivation above was done at the level of the series expansions of the respective Borel transforms, around $s \sim 0$. One natural question is whether such relation still holds should we instead expand the Borel transforms around their singularities---and thus if generic resurgence structures may always be mapped back to those associated to the standard representative. In the main text we have illustrated how this holds in the example of the quartic partition-function, where this relation is upheld by a semi-derivative, $D_s^{1/2}$. This should be a generic feature, with the relation \eqref{eq:relation-between-borels-der} holding around singularities of Borel transforms.

\newpage

\bibliographystyle{plain}

\end{document}